\documentclass[11pt,twoside,a4paper]{report} 
\usepackage{geometry} 

\pdfoutput=1

\usepackage{amsmath,color,multicol}
\usepackage{amsfonts}
\usepackage{amssymb}
\usepackage{graphicx}
\usepackage{mathrsfs}
\usepackage{epsfig}
\usepackage{latexsym}
\usepackage{graphicx}
\usepackage{color}
\usepackage{cite}
\usepackage{slashed, cancel}
\usepackage{hyperref}
\usepackage{datetime}
\usepackage{cancel}
\usepackage{multirow}
\usepackage[T1]{fontenc}

\numberwithin{equation}{chapter}

\def\circa#1{\,\raise.3ex\hbox{$#1$\kern-.75em\lower1ex\hbox{$\sim$}}\,}

\newcommand{\beq}{\begin{equation}}
\newcommand{\eeq}{\end{equation}}
\newcommand{\beqn}{\begin{eqnarray}}
\newcommand{\eeqn}{\end{eqnarray}}

\newcommand{\TeV}{{\rm \,TeV}}

\newcommand{\LSM}{$\mathcal{L}_{SM}$}

\newcommand{\V}{\mathbf{V}}
\newcommand{\T}{\mathbf{T}}

\newcommand{\F}{\mathcal{F}}

\newcommand{\cL}{\mathcal{L}}
\newcommand\cO{\mathcal{O}}

\newcommand\cP{\mathcal{P}}
\newcommand\cT{\mathcal{T}}

\def\Tr{{\rm Tr}}

 \def\be   {\begin{equation}}   \def\ee   {\end{equation}}
 \def\ba   {\begin{array}}      \def\ea   {\end{array}}
 \def\bea  {\begin{eqnarray}}   \def\eea  {\end{eqnarray}}
 \def\bean {\begin{eqnarray*}}  \def\eean {\end{eqnarray*}}

\def\eq#1{eq.~(\ref{#1})}

\usepackage{fancyhdr}
\fancypagestyle{plain}{%
\fancyhead{} % usuń p. górne na stronach pozbawionych numeracji (plain)
}

\makeatletter
\renewcommand{\maketitle}{\begin{titlepage}
    \begin{center}\normalsize
    \textit{Institute of Theoretical Physics \\
    Faculty of Physics \\
    University of Warsaw}
    \end{center}
    \vspace{4cm}
    \begin{center}
      \LARGE \@title
      \\
        \vspace{2cm}
        \begin{large}
        \@author
        \end{large}
         \end{center}
    \vspace{7cm}
    \begin{flushright}
    \begin{minipage}{7cm}
    \begin{flushright}
    \normalsize Doctoral dissertation\\ prepared under the supervision of
\\ prof. dr hab. Stefan Pokorski
    \end{flushright}
    \end{minipage}
     \end{flushright}
    \vspace*{\stretch{8}}
    \begin{center}
   Warsaw 2019
    \end{center}
  \end{titlepage}
}
\makeatother
\author{\textbf{Pawe{\l} Koz{\'o}w}}
\title{\textbf{The W and Z scattering as a probe of physics beyond the Standard Model:  Effective Field Theory approach}}

\newcommand{\podziekowania}{\begin{titlepage}
	\begin{center}
	    \vspace*{\stretch{9}}
		\textbf{Acknowledgements}
			\end{center}
	First, I would like to thank prof. Stefan Pokorski for his priceless mentorship, support and for organising wonderful research opportunities. In particular I would like to thank for his care, truly individual approach and for his constant and solid supervision on this thesis. His help has been inevitable practically at any stage of my doctoral studies.
\vspace{0.7cm}

I am especially indebted to dr hab. Micha{\l} Szleper, dr Luca Merlo, dr S{\l}awomir Tkaczyk, prof. dr hab. Jan Kalinowski, prof. dr hab. Janusz Rosiek, dr Manjit Kaur, Kaur Sandeep and Geetanjali Chaudhary for very valuable and fruitful collaborations. 
\vspace{0.7cm}

I am grateful to dr Adam Falkowski for valuable physics discussions.
\vspace{0.7cm}

I am especially thankful to Pawe{\l} Szczerbiak for the long-term collaboration on a project of which the results are, in the end, not the topic of this thesis, and his help on different stages of the doctoral studies in general.  
In the latter aspect I am also thankful to Zosia Fabisiewicz, dr Tomasz Krajewski, dr Olga Czerwi{\'n}ska, Pawe{\l} Olszewski, dr Marek Lewicki, dr Katarzyna Grzelak, dr Arkadiusz Trawi{\'n}ski and dr Bogumi{\l}a {\'S}wie{\.z}ewska.
\vspace{0.7cm}

Last but not least, I want to thank my parents whose support was crucial. \\\mbox{} \\

This work is supported by National Science Centre, Poland, the PRELUDIUM project under contract 2018/29/N/ST2/01153. \\ \mbox{} \\

The author also acknowledges support by the Polish National Science Centre (NCN) under the grant DEC-2015/18/M/ST2/00054.

		\vspace*{\stretch{10}}
			\vspace{7cm}
\end{titlepage}
}

\numberwithin{equation}{chapter}
\begin{document}
\newgeometry{tmargin=2cm, bmargin=3cm, lmargin=3cm, rmargin=3cm} 
\maketitle
\clearpage
\mbox{}
\thispagestyle{empty}
\clearpage
\podziekowania
\thispagestyle{empty}
\clearpage
\mbox{}
\clearpage
\newgeometry{tmargin=2cm, bmargin=3cm, lmargin=2.5cm, rmargin=2.5cm} 
 \begin{abstract}
There are strong arguments in favour of the point of view that the Standard Model (SM) of elementary interactions
is only an effective theory, a low-energy approximation of a more fundamental one. One of the very important processes that may
shed light on a deeper theory is the scattering of the electroweak vector bosons, $W$ and $Z$. It is indirectly accessible
in the LHC experiments, particularly with its high luminosity (HL-LHC) or high energy (HE-LHC) phase, through the process $pp\rightarrow2
\mathrm{jets} + 2 \mathrm{lepton\ pairs}$. The process of vector bosons scattering is the most direct test of the mechanism of the electroweak symmetry breaking and also is sensitive to the existence of new particles interacting electroweakly. 

There are
two general approaches to the search for a beyond the SM theory. One is to propose and investigate explicit models,
the other one is based on the so-called Effective Field Theory (EFT) approach. The strength and the effectiveness of the
EFT approach follows from the fact that one can study the discovery potential of the physics beyond the SM without knowing complete theories.
EFT is particularly useful as long as no new particles are directly discovered experimentally but their
existence could manifest itself at energies much lower than their masses $\Lambda$ ($E<<\Lambda$) as corrections to the SM
predictions suppressed as $(E/\Lambda)$. These corrections can be parametrized by non-renormalizable
operators added to the SM Lagrangian. 
%Every
%concrete complete model, after decoupling of the heavy degrees of freedom predicts some series, in principle
%infinite, of operators with concrete coefficients, called Wilson coefficients (being functions of the coupling
%constants $g$ and the ratios $g/\Lambda$). 
The EFT approach relies on the justified assumption that a limited number of certain operators provides a good approximation to certain classes of complete
models when $E<<\Lambda$. Each choice of a limited set of operators, with their (Wilson) coefficients taken at some (arbitrary) fixed
values, defines an EFT ''model'' to be tested for its discovery potential. At this point expected magnitude of deviations with respect to the SM predictions can be estimated as well as the effects of different non-renormalizable operators on kinematic distributions studied. Once the future data are available and feature significant deviations from the SM, the analysis in the framework of EFT ''models'' shall serve as a guide towards concrete deeper models. The general novel element of this work, important for both discussed above aspects of using EFT, is determination of the region of validity of the EFT approach to gauge boson scattering. The latter is strongly constrained by the partial wave unitarity bound applied
in presence of non-renormalizable operators.

Two main classes of effective Lagrangians can be
considered, depending on how the electroweak symmetry breaking is assumed to be realized: linearly for
elementary Higgs particle embedded as a component of the SM scalar doublet (Standard Model Effective Field Theory, SMEFT) or non-linearly, on the three Goldstone bosons constituting the longitudinal components of the $W^\pm$ and $Z$
gauge fields (Higgs Effective Field Theory, HEFT). In the latter case the physical Higgs field is a singlet of the symmetry group. The non-linear EFT is particularly suitable for low-energy effective description of models in which the Higgs particle is a fundamental or composite (pseudo) Nambu-Goldstone bosons arising from the spontaneous breaking of some global symmetry in the deeper theory. 

In this work the vector boson scattering process is investigated through the reaction $pp\rightarrow 2
\mathrm{jets} + W^\ast W^\ast \rightarrow 2
\mathrm{jets} + l\nu_l l'\nu_l'$, where $W^\ast$ denote in general off-shell $W^+$, in the EFT approach with the HL-LHC and HE-LHC experiments in mind. We have investigated the discovery potential of certain classes of the EFT ''models'' of both the SMEFT and HEFT bases with the
particular emphasis on using the EFT ''models'' in their region of validity. A novel method
has been proposed for determining the discovery regions of physics beyond the SM, described by the EFT
''models''. Independent of the basis chosen, 
the discovery regions are found to be non-empty. We then compare differences in experimental signatures between SMEFT and HEFT, which is an important step for distinguishing between the two hypotheses in the future data. Finally, we investigated what the effect on the discovery regions is when increasing the $pp$ collision energy.
\end{abstract}

%\newpage
%\thispagestyle{empty} 
%\mbox{}
\renewcommand{\abstractname}{Streszczenie}
\begin{abstract} 
Istniej\k{a} silne argumenty za punktem widzenia, {\.z}e Model Standardowy (MS) oddzia{\l}ywa{\'n} elementarnych jest tylko teori\k{a} efektywn\k{a}, niskoenergetycznym przybli{\.z}eniem jakiej{\'s} bardziej fundamentalnej teorii. Jednym z bardzo wa{\.z}nych proces{\'o}w, kt{\'o}ry mo{\.z}e rzuci{\'c} {\'s}wiat{\l}o na teori\k{e} g{\l}\k{e}bsz\k{a}, jest rozpraszanie elektros{\l}abych bozon{\'o}w wektorowych, $W$ oraz $Z$. Proces ten jest po{\'s}rednio dost\k{e}pny w eksperymencie LHC, szczeg{\'o}lnie w fazie wysokiej {\'s}wietlno{\'s}ci (HL-LHC), przez reakcj\k{e} $pp\rightarrow2
\text{\ d{\.z}ety} + 2 \text{\ pary lepton{\'o}w}$. Proces rozpraszania bozon{\'o}w wektorowych jest najbardziej bezpo{\'s}rednim testem mechanizmu naruszenia symetrii elektros{\l}abej, a tak{\.z}e czu{\l}y jest na istnienie nowych cz\k{a}stek oddzia{\l}uj\k{a}cych elektros{\l}abo.

Istniej\k{a} dwa generalne podej{\'s}cia do poszukiwa{\'n} teorii wychodz\k{a}cej poza MS. Jedno to proponowanie i badanie konkretnych modeli, drugie bazuje na tak zwanej Efektywnej Teorii Pola (Effective Field Theory, EFT). Si{\l}a i skuteczno{\'s}{\'c} podej{\'s}cia przez EFT wynika z faktu, {\.z}e mo{\.z}liwe jest badanie potencja{\l}u odkrywczego fizyki wychodz\k{a}cej poza MS, bez znajomo{\'s}ci pe{\l}nych teorii. EFT jest szczeg{\'o}lnie u{\.z}yteczne tak d{\l}ugo, jak {\.z}adne nowe cz\k{a}stki nie s\k{a} bezpo{\'s}rednio odkryte do{\'s}wiadczalnie, ale ich istnienie mog{\l}oby si\k{e} przejawia{\'c} przy energiach du{\.z}o ni{\.z}szych ni{\.z} ich masy $\Lambda$ ($E<<\Lambda$),
 jako poprawki do przewidywa{\'n} MS t{\l}umione jak $(E/\Lambda)$. Te poprawki mog\k{a} by{\'c} parametryzowane przez nierenormalizowalne operatory dodane do Lagran{\.z}janu MS. Podej{\'s}cie przez EFT bazuje na uzasadnionym za{\l}o{\.z}eniu, {\.z}e ograniczona liczba zadanych operator{\'o}w stanowi dobre przybli{\.z}enie pewnych klas kompletnych modeli, gdy $E<<\Lambda$. Ka{\.z}dy wyb{\'o}r ograniczonego zbioru operator{\'o}w, z ich wsp{\'o}{\l}czynnikami (Wilsona) wzi\k{e}tymi z jakimi{\'s} (dowolnymi) ustalonymi warto{\'s}ciami, definiuje ''model'' EFT, kt{\'o}ry ma by{\'c} testowany wzgl\k{e}dem jego potencja{\l}u odkrywczego. Na tym etapie mo{\.z}e by{\'c} oszacowana przewidywana skala odchyle{\'n} wzgl\k{e}dem MS, tak samo jak efekty od r{\'o}{\.z}nych nierenormalizowalnych operator{\'o}w, w badanych rozk{\l}adach kinematycznych. Kiedy przysz{\l}e dane b\k{e}d\k{a} dost\k{e}pne i poka{\.z}\k{a} istotne odchylenia od MS, analiza w ramach ''modeli'' EFT pos{\l}u{\.z}y jako wskaz{\'o}wka w kierunku konkretnych g{\l}\k{e}bszych modeli. Og{\'o}lny nowy element tej pracy, wa{\.z}ny w obu powy{\.z}szych aspektach stosowania EFT, to wyznaczanie rejon{\'o}w stosowalno{\'s}ci podej{\'s}cia przez EFT do rozpraszania bozon{\'o}w cechowania. Te ostatnie s\k{a} silnie ograniczone przez perturbacyjn\k{a} unitarno{\'s}{\'c} fal parcjalnych w obecno{\'s}ci operator{\'o}w nierenormalizowalnych.

Mo{\.z}na rozwa{\.z}a{\'c} dwie g{\l}{\'o}wne klasy efektywnych Lagran{\.z}jan{\'o}w w zale{\.z}no{\'s}ci od za{\l}o{\.z}enia, jak realizowane jest {\l}amanie symetrii elektros{\l}abej: liniowo, z elementarn\k{a} cz\k{a}stk\k{a} Higgsa jako sk{\l}adnik skalarnego dubletu MS (Standard Model Effective Field Theory, SMEFT) lub nieliniowo, przez trzy bozony Goldstona stanowi\k{a}ce pod{\l}u{\.z}ne sk{\l}adowe p{\'o}l cechowania $W^\pm$ oraz $Z$ (Higgs Effective Field Theory, HEFT). W drugim przypadku fizyczne pole Higgsa jest singletem grupy symetrii. Nieliniowa EFT jest szczeg{\'o}lnie odpowiednia w niskoenergetycznym efetywnym opisie modeli, w kt{\'o}rych cz\k{a}stka Higgsa jest elementarnym, b\k{a}d{\'z} z{\l}o{\.z}onym (pseudo) bozonem Nambu-Goldstona, wy{\l}aniaj\k{a}cym si\k{e} ze spontanicznego naruszenia pewnej symetrii globalej w g{\l}\k{e}bszej teorii.

 W tej pracy zbadano rozpraszanie bozon{\'o}w wektorowych przez reakcj\k{e} $pp\rightarrow 2
\mathrm{jets} + W^\ast W^\ast \rightarrow 2
\mathrm{jets} + l\nu_l\, l'\nu_l'$, gdzie  $W^\ast$ oznacza w og{\'o}lno{\'s}ci $W^+$ poza pow{\l}ok\k{a} masy, w podej{\'s}ciu przez EFT, pod k\k{a}tem eksperyment{\'o}w HL-LHC i HE-LHC. Zbadano potencja{\l} odkrywczy pewnych klas ''modeli'' EFT obu baz SMEFT i HFET, ze szczeg{\'o}lnym podkre{\'s}leniem u{\.z}ywania ''modeli'' EFT w ich rejonach stosowalno{\'s}ci. Zaproponowano now\k{a} metod\k{e} w celu okre{\'s}lania rejon{\'o}w odkrywczych fizyki wychodz\k{a}cej poza MS, opisanej przez ''modele'' EFT. Niezale{\.z}nie od wyboru bazy, znalezione rejony odkrywcze s\k{a} niepuste.  Nast\k{e}pnie por{\'o}wnano r{\'o}{\.z}nice sygnatur eksperymentalnych pomi\k{e}dzy SMEFT i HEFT, co jest wa{\.z}nym krokiem dla odr{\'o}{\.z}nienia tych dw{\'o}ch hipotez w przysz{\l}ych danych. Na koniec zbadano, jaki jest efekt na rejony odkrywcze gdy zwi\k{e}kszy{\'c} energi\k{e} zderze{\'n} $pp$. 
\end{abstract}
%\newpage
%\thispagestyle{plain} % empty
%\mbox{}
%spis treści
\newgeometry{twoside, tmargin=2cm, bmargin=3cm, inner=3cm, outer=3cm} 
\tableofcontents

%\chapter{{Introduction}
%
%\chapter{{Foundations}
	%
	%\chapter{Quantization of the scalar field in curved spacetime}
%
	%\chapter{Notion of vacuum in curved spacetime}
%
		%\section{Production without quantum corrections}
			%
		%\section{Instant preheating}
%
%\newpage
%
%\begin{center}
%\textbf{Summary of Chapter}
%\end{center} 
%\begin{itemize}
%\item 
%\item
%\end{itemize}
%
%\chapter{{Summary}
%
%\appendix
%
%\chapter{{Yang-Feldman equation \label{app:YF}}

%\addcontentsline{toc}{Chapter}{Bibliography}
%\bibliographystyle{JHEP}
%\bibliography{inflation}
\chapter{Introduction}

There are strong arguments in favour of the point of view that the Standard Model (SM) of elementary interactions~\cite{Glashow:1961tr,Weinberg:1967tq,Salam:1968rm,tHooft:1972tcz},
is only an effective theory, a low-energy approximation of a more fundamental one. The search for an extension
of the SM that would address the open questions left unanswered by the present theory is now the main goal of
particle physics, both in the experimental and theoretical research. One of the very important processes that may
shed light on a deeper theory is the scattering of the electroweak vector bosons, $W$ and $Z$. It is indirectly accessible
in the LHC experiments, particularly with its high luminosity phase (HL-LHC), through the process $pp\rightarrow2
\mathrm{jets} + 2 \mathrm{lepton\ pairs}$. The process of vector bosons scattering is the most direct test of the mechanism of the electroweak symmetry breaking~\cite{Englert:1964et,Higgs:1964ia,Higgs:1964pj,Guralnik:1964eu,Higgs:1966ev,Kibble:1967sv}
and also is sensitive to the existence of new particles interacting electroweakly.

On the theoretical side, there are
two general approaches to the search for a beyond the SM theory. One is to propose and investigate explicit models,
the other one is based on the so-called Effective Field Theory (EFT) approach~\cite{Manohar:2018aog,Pich:2015lkh}. The strength and the effectiveness of the
EFT approach follows from the fact that one can study the discovery potential of the physics beyond the SM without knowing complete theories.
EFT is particularly useful as long as no new particles are directly discovered experimentally but their
existence could manifest itself at energies much lower than their masses $\Lambda$ ($E<<\Lambda$) as corrections to the SM
predictions suppressed as $(E/\Lambda)$. These corrections can be parametrized by non-renormalizable
operators added to the SM Lagrangian. Every
concrete complete model, after decoupling of the heavy degrees of freedom predicts some series, in principle
infinite, of operators with concrete coefficients, called Wilson coefficients (being functions of the coupling
constants $g$ and the ratios $g/\Lambda$). The EFT approach relies on the justified assumption that a limited number of certain operators provides a good approximation to certain classes of complete
models when $E<<\Lambda$. Each choice of a limited set of operators, with their (Wilson) coefficients taken at some (arbitrary) fixed
values, defines an EFT ''model'' to be tested for its discovery potential. At this point expected magnitude of deviations with respect to the SM predictions can be estimated as well as the effects of different non-renormalizable operators on kinematic distributions studied. Once the future data are available and feature significant deviations from the SM, the analysis in the framework of EFT ''models'' shall serve as a guide towards concrete deeper models. The general novel element of this work, important for both discussed above aspects of using EFT, is determination of the region of validity of the EFT approach to gauge boson scattering. The latter is strongly constrained by the partial wave unitarity bound applied
in presencen of non-renormalizable operators.%tutajX

Two main classes of effective Lagrangians can be
considered, depending on how the electroweak symmetry breaking is assumed to be realized: linearly for
elementary Higgs particle embedded as a component of the SM scalar doublet (Standard Model Effective Field Theory, SMEFT)~\cite{Buchmuller:1985jz,Grzadkowski:2010es} or non-linearly, on the three Goldstone bosons constituting the longitudinal components of the $W^\pm$ and $Z$
gauge fields (Higgs Effective Field Theory, HEFT)~\cite{Feruglio:1992wf,Grinstein:2007iv,Contino:2010mh,Alonso:2012px,Alonso:2012pz,Buchalla:2013rka,Brivio:2013pma,Brivio:2014pfa,Gavela:2014vra,Gavela:2014uta,Eboli:2016kko,Brivio:2016fzo,deFlorian:2016spz,Merlo:2016prs,Buchalla:2017jlu,Alonso:2017tdy,Pich:2015kwa,Pich:2016lew}. In the latter case the physical Higgs field is a singlet of the symmetry group. The non-linear EFT is particularly suitable for low-energy effective description of models in which the Higgs particle is a fundamental or composite (pseudo) Nambu-Goldstone bosons arising from the spontaneous breaking of some global symmetry in the deeper theory. The non-linear EFT is the so-called Chiral EFT: the operators in the effective Lagrangian are organized according to the
number of derivatives of a certain dimensionless unitary matrix describing the Goldstone degrees of freedom.

In this work we shall investigate the Vector Boson Scattering (VBS) process through the reaction $pp\rightarrow 2
\mathrm{jets} + W^\ast W^\ast \rightarrow 2
\mathrm{jets} + 2 \mathrm{lepton\ pairs}$, where $W^\ast$ denote in general off-shell $W^+$ gauge bosons, in the EFT approach with the HL-LHC and HE-LHC experiments in mind. This thesis is based on the results of [a-d] (listed separately at the end of this Chapter). 

Focusing first on the linear realization of the electroweak symmetry breaking, quantitative EFT description
is based on expansion in higher dimensional operators in the the SMEFT basis, formed with operators
invariant under the gauge symmetries of the SM. In [a]
we have investigated the discovery potential at the HL-LHC of a certain class of the EFT ''models'', defined in the
SMEFT basis, applied to the $W^+ W^+$ scattering in the process $pp\rightarrow2
\mathrm{jets} + 2 \mathrm{lepton\ pairs}$, with the
particular emphasis on using the EFT ''models'' in their region of validity. A novel method
has been proposed for determining the discovery regions of physics beyond the SM, described by the EFT
''models'' in the regions where the EFT description is sensible.

For effective theories originating from complete models with the Higgs boson a (pseudo) Goldstone boson of some
new strong interactions the HEFT basis is more appropriate. It is a very interesting
question about the discovery potential of the physics beyond the SM depending on the EFT basis used. Therefore,
following the developed methods mentioned above, in [b] we have investigated the discovery potential for physics beyond
the SM in the $WW$ scattering at the HL-LHC, using the HEFT parametrization. The comparison between the results
obtained for the operators of the SMEFT and HEFT bases has been also addressed and it helps to understand the differences in the
expected experimental signatures in these two cases.

Therefore, the main results of the thesis are:
\begin{itemize}
\item determination of the region of validity of the EFT approach to same-sign $WW$ scattering,
	\item determination of the discovery potential of the EFT approach \underline{in its region of validity} to same-sign $WW$ scattering at HL-LHC both in SMEFT and HEFT case,
	\item in both cases the discovery regions are found to be non-empty,
	\item comparison of differences in experimental signatures between SMEFT and HEFT hypothesis,
	\item Finally, in [c-d] we addressed the question what the effect on the discovery regions is when increasing the $pp$ collision energy. For this reason we studied the SMEFT discovery regions in the HE-LHC experiment and compared them with the HL-LHC case. 
\end{itemize}

The structure of the thesis is as follows: we start with overview of the SM in Sec. 1, where we focus both on the theoretical principles as well as sum up its very impressive experimental success. In Sec. 2 we however briefly discuss the reasons for its extension, providing a list of examples where the SM fails. Here, we also briefly argue about the potential usefulness of VBS processes as a probe of the (expected) beyond the SM effects and discuss their experimental accessibility. In Sec. 3 we discuss in detail perturbative partial wave unitarity bounds. Then, in Sec. 4 we start the investigation on VBS process by discussing its features in the SM. In particular, we focus on the aspect of perturbative unitarity bounds fulfilment by the SM amplitudes. In Sec. 5 we discuss general features of the effective field theory approach. The latter is 
used in Sec. 6 in the form of SMEFT and HEFT to parametrize the beyond the SM effects in same-sign $WW$ scattering. The proposition for data analysis strategy in the EFT approach is discussed and the discovery potential presented. The conclusions and outlook concerning further research objectives are presented in Sec. 7. An important element of the thesis is also justification that qualitative influence of the non-renormalizable operators on the scattering amplitudes can be inferred from studying the on-shell vector boson scattering, e.g. $W^+W^+\rightarrow W^+W^+$. The full list of numerical results characterizing the scattering is lengthy because of the number of non-renormalizable operators that must be investigated. The on-shell scattering is illustrated in the main body of the thesis on small number of examples, for clarity. The full set of results is presented in Appendices but their content is summarized in the main body.\\ \mbox{} \\

[a] J.~Kalinowski, P.~Koz{\'o}w, S.~Pokorski, J.~Rosiek, M.~Szleper and S.~Tkaczyk,
  ``Same-sign WW scattering at the LHC: can we discover BSM effects before discovering new states?,''
  Eur.\ Phys.\ J.\ C 78, no. 5, 403 (2018)
  doi:10.1140/epjc/s10052-018-5885-y
  [arXiv:1802.02366 [hep-ph]].\\ 
	
[b] P.~Koz{\'o}w, L.~Merlo, S.~Pokorski and M.~Szleper,
  ``Same-sign WW Scattering in the HEFT: Discoverability vs. EFT Validity,'' 
	[arXiv:1905.03354 [hep-ph]], \emph{accepted to JHEP}.\\
	
[c]  G.~Chaudhary, J.~Kalinowski, M.~Kaur, P.~Koz{\'o}w, S.~Pokorski, J.~Rosiek, K.~Sandeep, M.~Szleper, S.~Tkaczyk, 
``Higgs Physics at the HL-LHC and HE-LHC'', Report from Working Group 2 on the Physics of the HL-LHC, and Perspectives at the HE-LHC,	[arXiv:1902.00134 [hep-ph]].\\

[d] G.~Chaudhary, J.~Kalinowski, M.~Kaur, P.~Koz{\'o}w, K.~Sandeep, M.~Szleper and S.~Tkaczyk,
  ``EFT triangles in the same-sign WW scattering process at the HL-LHC and HE-LHC'', [arXiv:1906.10769 [hep-ph]].
%%\vspace{10mm}

\chapter{The Standard Model}
\section{General designing principles}

The current theory of particles and their interactions is the Standard Model. It merges the two fundamental principles of XX century physics: quantum mechanics (1920') and special relativity (1905). Hence it is a Quantum Field Theory (QFT) with the Poincare invariance imposed. The principles of general relativity (1915) are not accounted for - gravity is not among the forces the SM describes. Further basic principles the local QFT Lagrangian of the SM \LSM \ obeys, are:
\begin{enumerate}
	\item \emph{the gauge principle}: both the electromagnetic and weak interactions are described by the gauge group $SU(2)_L\times U(1)_Y$. Global symmetries are not imposed at the fundamental level. It may however happen than the $\mathcal{L}$ features some global symmetries as a consequence of certain local symmetry and the particle content. In fact the SM has \emph{accidental} global symmetries. As a consequence lepton and baryon numbers are conserved.
	\item concerning the fermion sector, the \LSM is constructed with the two-component anti-commuting (Grassman) fields that transform in the two inequivalent two-dimensional irreducible representations, 2 and $\bar{2}$, of the Poincar{\'e} group (Weyl spinors). The Dirac fermion field $\Psi$ that represents the electron and the positron in the Quantum Electrodynamics (QED), consists of a pair of such fields; one transforming as $2$ and the other as $\bar{2}$. The QED is \emph{vector-like} -- the $\mathcal{L}_{QED}$ can be constructed with the four-component Dirac fermion solely. The SM, though describes the electromagnetic interactions of the electron, is \emph{chiral} at the fundamental level of $SU(2)_L\times U(1)_Y$ -- the building blocks are the $2$ and $\bar{2}$ spinors. Of course, the Weyl spinors can be obtained from the four-component field by chiral projections $P_L\Psi\equiv\psi_L$, $P_R\Psi\equiv\psi_R$, where $P_L,P_R$ are the left (L) and right (R) projection operators.  In this framework $\psi_L,\psi_R$ are technically written in the four-component notation. This notation will be used throughout the entire text. The fields $\psi_L,\psi_R$ are chiral eigenstates. For massless particles or in the massless (relativistic) limit $\psi_L$, $\psi_R$ are also helicity left- and right-handed eigenstates, respectively. Hence chirality is helicity in this limit. For a reference on spinors see~\cite{Dreiner:2008tw}%
	\item the \LSM\ allows for arbitrarily precise predictions (at any order in perturbation theory) after a finite, number of measurements are conducted to determine the fundamental parameters in \LSM. This feature goes under the name of \emph{renormalizability}.
	\end{enumerate} 
Throughout the text we will work in the natural units where the speed of light $c$ and the Planck constant $\hbar$ are set to 1. Helicity projections in these units are simply $\pm1/2(\pm1)$ for fermions (the photon). The electromagnetic fine structure constant $\alpha\equiv\frac{e^2}{4\pi}$ is dimensionless. In these units all fields have dimension of mass. Spin-1 vector fields $A_\mu(x)$ and scalars $\phi(x)$ have dimension 1 and the spinor spin-1/2 fields dimension 3/2, in the mass units. The derivative is mass dimension 1. The mass dimension is equivalent to the energy dimension, e.g. electronvolt (eV). Gauge invariant QFT are renormalizable if the mass dimension ($D$) of its terms is $\leq 4$~\cite{tHooft:1972tcz}. Higher dimension operators spoil renormalizability. We will refer to such operators as non-renormalizable operators.
%gauge principle 
\subsection{Gauge invariance}
\textbf{A. The abelian case}\\ \mbox{}\\
We consider the free Dirac Langrangian
\begin{equation}
\mathcal{L}_0=i\bar{\Psi}(x)\gamma^\mu\partial_\mu\Psi(x) - m\bar{\Psi}(x)\Psi(x).
\label{eq:3}
\end{equation}
Summation over repeated indices (also non-relativistic) is understood throughout the entire text.
It is invariant under the global $U(1)$ transformation:
\begin{equation}
\Psi(x)\stackrel{U(1)}{\rightarrow} \Psi'(x) \equiv \exp\{iQ\theta\}\Psi(x)
\label{eq:4}
\end{equation}
where $Q$ is an arbitrary constant and $\theta$ is the $U(1)$ rotation angle. The transformation changes the complex phase of all the quantas $\Psi(x), x\in \mathcal{R}^4$ simultaneously.
Since the phase of each quantum $\Psi(x)$ at fixed x has no physical meaning the theory should be symmetric under the local phase rotations, i.e. $\theta\equiv\theta(x)$. However it is not the case, since
\begin{equation}
\partial_\mu\Psi(x)\rightarrow \Psi'(x) \equiv \exp\{iQ\theta\}(\partial_\mu+iQ\partial_\mu\theta)\Psi(x).
\label{eq:4p}
\end{equation}
The \emph{gauge principle} is the requirement that the Lagrangian is invariant under local phase transformations. To fix this issue in the context of the above example, one has to add an extra field $A_\mu(x)$ to the partial derivative $\partial_\mu$:
\begin{equation}
\partial_\mu\rightarrow \partial_\mu+ieQA_\mu(x)
\label{eq:5}
\end{equation}
and assume its transformation under the local $U(1)$ is:
\begin{equation}
A_\mu(x)\rightarrow A'_\mu(x)\equiv A_\mu(x)-\frac{1}{e}\partial_\mu\theta.
\label{eq:6}
\end{equation}
The partial derivative modification in eq.~\eqref{eq:5} defines the \emph{covariant} derivative $D_\mu$. It transforms in the same way under the local rotations, as the $\Psi$ field:
\begin{equation}
\begin{array}{c}
	D_\mu\Psi(x)\equiv [\partial_\mu+ieQA_\mu(x)]\Psi(x), \\
	D_\mu\Psi(x)\rightarrow(D_\mu\Psi)'(x)\equiv \exp\{iQ\theta(x)\}D_\mu\Psi(x).
\end{array}
\label{eq:7}
\end{equation}
The term that governs the $A_\mu$ propagation is build from the stress-tensor $F_{\mu\nu}\equiv\partial_\mu A_\nu - \partial_\nu A_\mu$ and reads:
\begin{equation}
\mathcal{L}_{\mathrm{Kin}}^A\equiv -\frac{1}{4}F_{\mu\nu}F^{\mu\nu}.
\label{eq:8}
\end{equation}
The $A_\mu$ field mass term $\frac{1}{2}m^2A_\mu A^{\mu}$ is forbidden by the gauge principle. The quantum $A_\mu(x)$ is a spin-1 massless particle that has two possible spin projections $\pm1$ on the spin quantization axis. The total Lagrangian
\begin{equation}
\mathcal{L}_{QED} \equiv \mathcal{L}_{\mathrm{kin}}^A + i\bar{\Psi}(x)\gamma^\mu D_\mu\Psi(x)  = \mathcal{L}_{\mathrm{kin}}^A + \mathcal{L}_0 - eQA_\mu(x)\bar{\Psi}(x)\gamma^\mu\Psi(x)
\label{eq:9}
\end{equation}
is the QED Lagrangian with $A_\mu$ the photon; in the second step in eq.~\eqref{eq:9} electron-phonton interaction was explicity separated. $Q$ denotes the charge of $\Psi$ in the units of the electron charge $e$, $\frac{e^2}{4\pi}\approx \frac{1}{137}$. After the global symmetry is \emph{gauged}, the photon emerges.

Instead of a single fermion $\Psi$, N fermions $\Psi_i$ can be introduced, where $i=1,2,\ldots, N$. Each $\Psi_i$ corresponds then to its own 
\begin{equation}
\mathcal{L}_{0,i}\equiv i\bar{\Psi}_i(x)\gamma^\mu\partial_\mu\Psi_i(x) - m_i\bar{\Psi}_i(x)\Psi_i
\label{eq:10}
\end{equation} 
and its own interaction term with the photon
\begin{equation}
\mathcal{L}_{\mathrm{int},i}\equiv-eQ_iA_\mu(x)\bar{\Psi}_i(x)\gamma^\mu\Psi_i(x),
\label{eq:11}
\end{equation}
i.e. $Q_i$ is arbitrary for each species $i$.
%SSB
The Lagrangian describing the electromagnetic interactions between all the species $i$ 
\begin{equation}
\mathcal{L}\equiv\mathcal{L}_{\mathrm{kin}}^A+ \sum_i\left(\mathcal{L}_{0,i}+\mathcal{L}_{\mathrm{int},i}\right)
\label{eq:12}
\end{equation}
is invariant under the local $U(1)_{EM}$, too. We shall refer to the different species as different flavors. The Lagrangian~\eqref{eq:12} could describe electromagnetic interactions of different quark flavors $u,d,s,c,b,t$ together with different charged lepton flavors $e,\mu,\tau$. \\\mbox{} \\
\textbf{B. The non-abelian case}\\\mbox{}\\
The non-abelian generalization of eq.~\eqref{eq:3} is obtained if $\Psi$ is promoted to a multiplet (vector) of N fields $\Psi^T\equiv(\Psi^1,\Psi^2,\ldots,\Psi^N)$ and we require that it is invariant under arbitrary global $U(N)$ transformations:
\begin{equation}
\Psi\stackrel{U(N)}{\rightarrow}\Psi'\equiv U\Psi\qquad U^\dagger U=UU^\dagger=1.
\label{eq:13}
\end{equation} 
The free Lagrangian looks the same in the vector notation:
\begin{equation}
\mathcal{L}_0=i\bar{\Psi}(x)\gamma^\mu\partial_\mu\Psi(x) - m\bar{\Psi}(x)\Psi(x),
\label{eq:33}
\end{equation}
where $m$ denotes now a $N\times N$ matrix, proportional to identity; similarly, the proper form of $\gamma^\mu$ is implicit. Notice also that $\mathcal{L}_0$ is vector-like, i.e. both the left $\Psi_L$ and right $\Psi_R$ fields are in the same representation of the symmetry group.
Since $U(N)=SU(N)\times U(1)$ we consider in this example only the global $SU(N)$ to be gauged. The extra condition is $\det U=1$. The exponential parametrization of the $U$ reads
\begin{equation}
U=\exp\left\{i T^a\theta_a\right\},
\label{eq:14}
\end{equation}
where $T^a$ ($a=1,2\ldots,N$) are the generators of the fundamental representation of $SU(N)$, traceless and Hermitian, and $\theta_a$ are arbitrary real numbers, the rotation angles. The structure constants are denoted by $f_{abc}$:
\begin{equation}
\left[T^a,T^b\right]=if^{abc}T^c,
\label{eq:15}
\end{equation}
where $[\cdot,\cdot]$ denotes the commutator; $f^{abc}$ is real for all $a,b,c$ and totally antisymmetric in these indices.

By analogy to the abelian case we require the symmetry under local transformations $U\equiv \exp\{iT^a\theta_a(x)\}$. Similarly the $\partial^\mu$ requires modification with an extra field $G^\mu(x)$, which is now a matrix
\begin{equation}
D^\mu\Psi\equiv\left[\partial^\mu +igG^\mu(x)\right]\Psi,
\label{eq:16}
\end{equation}
with the following transformation properties
\begin{equation}
G^\mu\rightarrow (G^\mu)'= UG^\mu U^\dagger +\frac{i}{g}U^\dagger (\partial^\mu U).
\label{eq:17}
\end{equation}
Eq.~\eqref{eq:17} assures covariance:
\begin{equation}
D^\mu\rightarrow (D^\mu)'=U D^\mu U^\dagger,
\label{eq:18}
\end{equation}
equivalently
\begin{equation}
D^\mu\Psi\rightarrow(D^\mu\Psi)'=U (D^{\mu}\Psi).
\label{eq:17p}
\end{equation} 
The following decomposition of the $G^\mu(x)$ matrix will prove convenient:
\begin{equation}
G^\mu(x)\equiv T^a G^\mu_a(x),
\label{eq:19}
\end{equation}
Using~\eqref{eq:19} and~\eqref{eq:17} with infinitesimal rotation angles $\theta_a=\delta\theta_a$, the transformation properties of $G^\mu_a(x)$ can be found:
\begin{equation}
G^\mu_a\rightarrow (G^\mu_a)'=G^\mu_a-\frac{1}{g}\partial^\mu(\delta\theta_a)-f^{abc}\delta\theta_b G^\mu_c.
\label{eq:20}
\end{equation} 
Propagation of $G^\mu_a$ is governed by
\begin{equation}
\begin{array}{l}
G^{\mu\nu}(x)\equiv -\frac{i}{g}[D^\mu,D^\nu]=\partial^\mu G^\nu-\partial^\nu G^\mu +i g [G^\mu,G^\nu]\equiv T^a G^{\mu\nu}_a(x),\\
G^{\mu\nu}_a(x)=\partial^\mu G^\nu_a-\partial^\nu G^\mu_a-gf^{abc}G^\mu_b G^\nu_c.
\end{array}
\label{eq:21}
\end{equation}
The $G^{\mu\nu}(x)$ stress-tensor is covariant by construction. Hence its trace $\mathrm{Tr}\,(G^{\mu\nu}G_{\mu\nu})=\frac{1}{2}G^{\mu\nu}_a G_{\mu\nu}^a$ is invariant; its canonical normalization is equal to $-\frac{1}{4}G^{\mu\nu}_a G^a_{\mu\nu}$.

Eq.~\eqref{eq:21} implies self-interactions of these gauge fields. The latter are in the form of triple and quartic gauge vertices -- the expansion of the kinetic term $\mathrm{Tr}\,(G^{\mu\nu}G_{\mu\nu})$ yields:
	\begin{equation}
	-\frac{1}{4}(\partial^\mu G^\nu_a-\partial^\nu G^\mu_a)^2 +\frac{g}{2}f^{abc}(\partial^\mu G^\nu_a-\partial^\nu G^\mu_a)G^b_\mu G^c_\nu - \frac{g^2}{2}f^{abc}f_{ade}G^\mu_b G^\nu_c G^d_\mu G^e_\nu.
	\label{eq:26}
	\end{equation}

The transformation properties under global non-abelian transformations are
	\begin{equation}
	G^\mu_a\rightarrow (G^\mu_a)'=G^\mu_a-f^{abc}\theta_b G^\mu_c
	\label{eq:24}
	\end{equation}
	In the vector notation ($\mu$ index is ommited) $G^T=(G^1, G^2 ,\ldots G^N)$ it is written as
	\begin{equation}
	G\rightarrow G - f^b G,
	\label{eq:25}
	\end{equation}
	where $f^b$ denotes an $N\times N$ matrices with the following prescription for $(a,b)$ matrix elements $(f^b)^a_{\ c}=f^{abc}$. The $if^b$ satisfy the same commutation relations as $T^a$. The corresponding representation is called the adjoint representation. The generators in this representation are totally antisymmetric, purely imaginary; the representation is real.

So far a single $\Psi$ multiplet (single flavor) was considered in the non-abelian case. We could consider more than one flavor, charged under the same $SU(N)$ interactions, to be present in our Lagrangian. Now, in the abelian case different flavors $i$ of the same gauge group can have arbitrary interaction strengths $eQ_i$ with the gauge field $A_\mu$. The situation in the non-abelian case is different -- the interaction strength of each flavor must be the same; in the next paragraph, we justify it. 

Let's assume, for concreteness, that we have two flavors in the non-abelian case, and that the interaction strength of the first flavor with $G^\mu_a$ is governed by the coupling $g$. Technically, to account for a difference of the interaction strength of the second flavor, one has to rescale, by an appropriate factor $Q$, the generators $T^a$ occurring in the covariant derivative corresponding to that flavor:
	\begin{equation}
	T^a\rightarrow Q T^a.
	\label{eq:23}
	\end{equation}
	Then, the non-linear relations between generators (eq.~\eqref{eq:15}) would imply that this covariant derivative term in the Lagrangian stays invariant only if the transformation properties of $G_\mu^a$ occuring in that covariant derivative would be modified (by rescaling $f^{abc}\rightarrow Qf^{abc}$ in the transformation rule eq.~\eqref{eq:20}). But, since we assume a single gauge group, i.e. a single multiplet of $G^a_\mu$, these cannot be altered, as these transformations are fixed by interaction strength of the first flavor.
	
	Depending on the choice of $N$, different theories are obtained. e.g. for $SU(2)$ the generators are $T^a=\frac{\tau^a}{2}$ $(a=1,2,3)$, the $\tau^a$ being the Pauli matrices; for $SU(3)$, $T^a=\frac{\lambda^a}{2}$ $(a=1,2,\ldots,8)$, the $\lambda^a$ being the Gell-Mann matrices. For $N=2$ the gauge fields form an triplet (3) of $SU(2)$ and in case of $N=3$, an octet (8) of $SU(3)$. The latter choice corresponds to the Lagrangian of Quantum Chromodynamics $QCD$ where the gauge fields $G^a_\mu(x) $ are the gluon fields mediating the strong interactions. The requirement for a single non-abelian gauge symmetry in QCD implies the interaction strength with the gluons are equal for all the flavors $\Psi_f=u,d,s,c,b,t$. Hence the exact form of $D_\mu$ corresponding to each $\Psi_f$ is the same for all $f$ and the QCD Lagrangian for all the flavors simply reads:
	\begin{equation}
\mathcal{L}_{QCD}=-\frac{1}{4}G^{\mu\nu}_a G^a_{\mu\nu}+\sum_f\left( i\bar{\Psi}_f(x)\gamma^\mu D_\mu\Psi_f(x) - m_f\bar{\Psi}_f(x)\Psi_f(x)\right).
\label{eq:22}
\end{equation}

%SSB higgs mechanizm
\subsection{Spontaneous Symmetry Breaking (SSB)}
\label{subsub:SSB}
The classic example where SSB occurs is the physics of a ferromagnet. On one hand, the Hamiltonian $\mathcal{H}$ is invariant under rotations in the 3-dimensional space. Below a certain temperature the ferromagnet magnetizes -- the electrons spins get correlated at macroscopic distances leading to a macroscopic magnetic moment, to which each electron contributes. This quantum state of a ferromagnet is the ground state. It is degenerate -- the ferromagnet has freedom to choose the magnetic moment direction. Any such choice violates the symmetry of $\mathcal{H}$ -- the ground state transforms (the magnetization change its direction) under rotations. Moreover, all the states that are finite excitations around the ground state will share this asymmetry. The symmetry is broken spontaneously in such set of quantum states by the choice of magnetization direction of the ground state. In QFT the ground state is called \emph{the vacuum} $\left|0\right>$. Particle states are excitations around the vacuum. Hence SSB occurs if a certain symmetry of $\mathcal{L}$ is not shared by the vacuum. Below we illustrate on examples physics of SSB in QFT that is triggered by a scalar potential $V$, $\mathcal{L}\supset -V$.  The $\mathcal{L}$ is assumed to be invariant under unitary transformations. The latter symmetry is the equivalent, in the context of SSB, of space rotations in the ferromagnet example. \\\mbox{} \\
\textbf{A. U(1) global} \\\mbox{}\\
We consider the $U(1)$ symmetric renormalizable Lagrangian of a complex scalar field $\Phi(x)$:
\begin{equation}
\mathcal{L}= \partial_\mu \Phi^\dagger \partial^\mu \Phi - V(\Phi),\qquad V(\Phi)=\mu^2\Phi^\dagger\Phi + \lambda \left(\Phi^\dagger\Phi\right)^2.
\label{eq:27}
\end{equation}
The $\Phi$ symmetry transformations read:
\begin{equation}
\Phi(x)\rightarrow \Phi'(x) \equiv \exp\{i\theta\}\Phi(x).
\label{eq:28}
\end{equation}
The vacuum $\left|0\right>$ is the state that minimizes the Hamiltonian expectation value. At tree-level the $V$ vacuum expectation value $\left<0\right|V\left|0\right>$ is equal to the minimum of $V$ treated classically, i.e. as a function of the scalar fields. The classical scalar field values at the minimum of $V$ are equal to vacuum expectation values of the scalar field operators. Hence the following quantity 
\begin{equation}
\left<0\right|V\left|0\right> = \mu^2\left|\left<\Phi\right>\right|^2+\lambda\left|\left<\Phi\right>\right|^4
\label{eq:30}
\end{equation}
is to be minimized with respect to $\left<\Phi\right>$. The minimum is, in general, a function of $\mu^2$, $\lambda$. For the stability of the vacuum $\lambda>0$ is required. There is no such restriction of the sign of $\mu^2$. It leads to two physically distinct cases:
\begin{itemize}
	\item $\mu^2>0$: there is a non-degenerate minimum at $\left<\Phi\right>=0$. Such $\mathcal{L}$ describes a particle of mass $\mu$ (and its antiparticle). The minimum is unaffected by the $U(1)$ rotations. The symmetry is not spontaneously broken.
	\begin{figure} 
\begin{center}	
\includegraphics[width=0.7\textwidth]{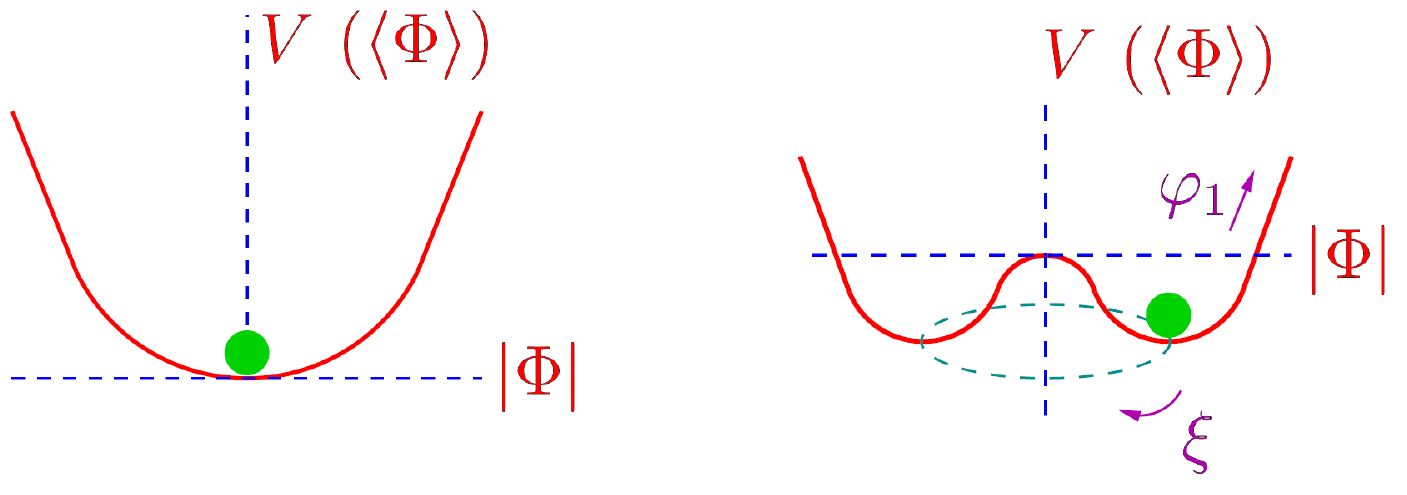}
\caption{Shape of the scalar potential for $\mu^2 > 0$ (left) and $\mu^2 < 0$ (right). In the second case there is
a continuous set of degenerate vacua, corresponding to different phases $\theta$, connected through a massless field
excitation $\xi$. For details see the text.}
\label{fig:mexicanHat}
\end{center}
\end{figure}
	
	%tutajKurwa
	\item $\mu^2<0$: $\left<V\right>$ as a function of $\left<\Phi\right>$ is shown in Fig.~\ref{fig:mexicanHat}. The condition for the vev is 
	\begin{equation}
	\left|\left<\Phi\right>\right| = \sqrt{\frac{-\mu^2}{2\lambda}}\equiv\frac{v}{\sqrt{2}}
	\label{eq:31}
	\end{equation}
	Any choice of the vev $\left<\Phi\right>=\exp\{i\theta\}\frac{v}{\sqrt{2}}$ is equivalent. The vacuum is degenerate, which reflects the symmetry of $\mathcal{L}$. It is characterized by a point in the complex plain. The point would transform non-trivially under the $U(1)$ rotations, i.e. it would go around the circle. Hence any vev choice of the physical system will break the symmetry $U(1)$. The symmetry gets spontaneously broken. For the remaining discussion we choose $\theta=0$ as our vacuum. The excitations $\phi_1(x)$, $\phi_2(x)$ (real fields) over the vev are then parametrized as follows:
	\begin{equation}
	\Phi(x)\equiv\frac{1}{\sqrt{2}}\left[v + \phi_1(x) +i \phi_2(x)\right]
	\label{eq:32}
	\end{equation}
	and the potential reads:
	\begin{equation}
	V = \left<V\right> +
	\lambda v^2 \phi_1^2 
	+ \lambda v \phi_1\left(\phi_1^2+\phi_2^2\right) 
	+ \frac{1}{4} \lambda  \left(\phi_1^2+\phi_2^2\right)^2.
	\label{eq:32p}
	\end{equation}
\end{itemize}
After the SSB one particle is massive $\phi_1$, its mass equal to $\sqrt{2\lambda} v$. The other, $\phi_2$, is massless. To gain further insight we introduce an equivalent parametrization for the field $\Phi(x)$:
\begin{equation}
\Phi(x) = \exp\left\{i\frac{\xi(x)}{v}\right\}(v+\varphi_1(x)).
\label{eq:33p}
\end{equation}
In this parametrization the $\xi$ dependence disappears from the potential $V$. Practically, one is left with the $\phi_1$ part of eq.~\eqref{eq:32p} (up to trivial difference in normalization). Hence the imaginary part of $\Phi(x)$, $\phi_2$, is explicitly massless. After the reparametrization, the $\xi(x)$ dependence occurs in the $\partial_\mu\Phi\partial^\mu\Phi$ term of $\mathcal{L}$. It implies the $\xi$ interacts necessarily through its derivatives, i.e. its vertices are momentum dependent and in the low energy limit $\xi$ interactions vanish. Since the fields $\varphi_2(x)$ and $\xi(x)$ are equivalent descriptions of the massless modes, the latter statement must be true also for $\varphi_2$ interactions. This is not obvious based on eq.~\eqref{eq:32p}; cancellations of the momentum-independent parts must take place. The mode $\xi(x)$ is called the Nambu-Goldstone or Goldstone (mode). The existence of massless derivative-interacting modes in theories with SSB, is stated in the Goldstone theorem. Its formulation in the more general non-abelian case will be followed by an illustrative $SO(3)$ example.\\\mbox{} \\
\textbf{B. SO(3) global}\\\mbox{} \\
We consider now $SO(3)$ symmetric Lagrangian of a triplet of (real) fields $\phi_i$ $(i=1,2,3)^T$. In the vector notation $\Phi=(\phi_1,\phi_2,\phi_3)$ the Lagrangian form is the same as in eq.~\eqref{eq:27} (with Hermitian conjugation replaced by transposition). The explicit form of the $SO(3)$ transformations read:
\begin{equation}
\begin{array}{c}
\Phi(x)\rightarrow \Phi'(x)=\exp\left\{i\theta_a T^a\right\}\Phi, \\ \mbox{}\\
T^1= \left(\begin{array}{ccc}
0 & 0 & 0 \\ 
0 & 0 & +i \\
0 & -i & 0 
\end{array}\right),
T^2 = \left(\begin{array}{ccc}
	0 & 0 & -i \\ 
0 & 0 & 0 \\
+i & 0 & 0 
\end{array}\right),
T^3 = \left(\begin{array}{ccc}
	0 & +i & 0 \\ 
-i & 0 & 0 \\
0 & 0 & 0 
\end{array}\right).
\end{array}
\label{eq:34}
\end{equation} 
The case $\mu^2<0$ implies SSB. The condition on the $\phi_i$ vev's is 
\begin{equation}
\left<\phi_1\right>^2+\left<\phi_2\right>^2+\left<\phi_3\right>^2 = v^2 \equiv \sqrt{\frac{-\mu^2}{2\lambda}},
\label{eq:35}
\end{equation}
i.e. the minimum is degenerate and in a form of the two-dimensional sphere $S_2$. 

Any choice of a point on $S_2$, i.e. of the vacuum, is physically equivalent to any other. We adopt the following choice 
$\left<\phi_1\right>=\left<\phi_2\right>=0$ and $\left<\phi_3\right>=v$. The excitations around the ground state are then parametrized as $\Phi(x)=(\varphi_1(x),\varphi_2(x), v + \varphi_3(x))^T$ and the potential in the broken phase reads:
\begin{equation}
V = -4  \lambda  v^2 \varphi_3^2 -\lambda \varphi_3^4 -4 \lambda v  \varphi_3^3 
-4  \lambda  v \varphi_3 \left(\varphi_{1}^2+\varphi_{2}^2\right)
-2 \lambda \varphi_3^2   \left(\varphi_{1}^2+\varphi_{2}^2\right)
-\lambda  \left(\varphi_{1}^2+\varphi_{2}^2\right)^2.
\label{eq:36}
\end{equation}
The particle $\phi_3$ is massive with the mass $2\sqrt{2\lambda} v$ and the particles $\varphi_1$, $\varphi_2$ are massless. Similarly as in the previous abelian example, reparametrizing $\Phi$ as follows 
\begin{equation}
\Phi(x) \equiv \exp\{i\xi_1(x) T^1 +i\xi_2(x) T^2\}(0 , 0 , v+\varphi_3)^T
\label{eq:37}
\end{equation}
shows both masslessness and derivative interactions of $\xi_1$ and $\xi_2$, explicitly. They are the Nambu-Goldstone modes. 
In eq.~\eqref{eq:37} we made use of the fact that $T^1$ and $T^2$ are ''broken'' by $\left<\Phi\right>$, i.e. $T^1\left<\Phi\right>,T^2\left<\Phi\right>\neq0$ -- there exists a Goldstone for each broken generator.
From eq.~\eqref{eq:37} it is to be noted that V, and hence $\mathcal{L}$, have a remnant symmetry described by the following transformations
\begin{equation}
\Phi(x) \rightarrow \Phi'(x) = \exp\{i\theta_3 T^3\}\Phi(x) = 
\left(\begin{array}{ccc}
	\cos\theta_3 & -\sin\theta_3 & 0 \\
	\sin\theta_3 & \cos\theta_3 & 0 \\
	0 & 0 & 1 
\end{array}\right)
\left(\begin{array}{c}
\varphi_1(x)\\
\varphi_2(x)\\
v+\varphi_3(x)
\end{array}\right)
\label{eq:38}
\end{equation}  
The remnant symmetry is $SO(2)$ and the SSB pattern in this example is $SO(3)\rightarrow SO(2)$. The fact the $\mathcal{L}$ in the broken phase is symmetric under $SO(2)$ corresponds to the fact that the generator $T^3$ remains unbroken by the vacuum, i.e. $T^3 \left<\Phi\right>= 0 $. Again, it is not true for the remaining two generators. 

This example illustrates a general result known as the Nambu-Goldstone theorem~\cite{Goldstone:1961eq}: if a Lagrangian is invariant under a continuous
symmetry group $G$, but the vacuum is only invariant under a subgroup $H\supset G$, then there must exist as
many massless spin-0 particles (Nambu--Goldstone bosons) as broken generators (i.e., generators of $G$
which do not belong to $H$). The vectors $iT^i\left<\Phi\right>$, where $i$ runs over the broken generators, 
are eigenvectors of the scalar mass-matrix with 0 eigenvalues.

In our example vectors $iT^1 \left<\Phi\right>$ , $iT^2 \left<\Phi\right>$ read
\begin{equation}
iT^1 \left<\Phi\right> = \left(\begin{array}{c} v \\ 0 \\ 0 \end{array}\right),\qquad iT^2 \left<\Phi\right> =\left(\begin{array}{c} 0 \\ -v \\ 0 \end{array}\right),
\label{eq:39}
\end{equation} 
which is consistent with the last statement.\\\mbox{} \\
\textbf{C. U(1) local}\\\mbox{}\\
We consider now the local $U(1)$ symmetry of a scalar field:
\begin{equation}
\begin{array}{l}
\mathcal{L}= (D_\mu \Phi)^\dagger D^\mu \Phi - V(\Phi),\qquad V(\Phi)=\mu^2\Phi^\dagger\Phi + \lambda \left(\Phi^\dagger\Phi\right)^2, \\
D_\mu = \partial_\mu+ i e A_\mu.
\end{array}
\label{eq:40}
\end{equation}
$\mathcal{L}$ is, in particular, invariant under the global $U(1)$ symmetry. We assume the latter is spontaneously broken, hence $\mu^2 < 0$. The field $\Phi$ can be parametrized as in eq.~\eqref{eq:33p} and, due to the gauge $U(1)$ symmetry, the Goldstone mode $\xi$ can be rotated away:
\begin{equation}
\Phi(x) =\exp\left\{i\frac{\xi(x)}{v}\right\}\frac{1}{\sqrt{2}}(v+h(x))\rightarrow \frac{1}{\sqrt{2}}(v+h(x)).
\label{eq:41}
\end{equation} 
This particular choice of gauge is called the unitary gauge. Simultaneously transformed is the gauge field $A_\mu(x)\rightarrow A'_\mu(x)= A_\mu(x)-\frac{1}{e}\partial_\mu\xi(x)/v$. We will reabsorb $\xi(x)$ into $A_\mu(x)$ by field redefinition $A_\mu(x)\equiv A'(x)$. In the unitary gauge, the kinetic term $ (D_\mu \Phi^\dagger) D^\mu \Phi $ reads:
\begin{equation}
\frac{1}{2}(\partial_\mu h)\partial^\mu h+ \frac{1}{2}e^2 (v+h)^2 A_\mu A^\mu .
\label{eq:42}
\end{equation}
After the SSB the gauge field acquired mass equal to $ e^2 v^2$. The gauge symmetry is lost, however the theory is renormalizable since its original formulation eq.~\eqref{eq:40} is gauge invariant. The Nambu-Goldstone field $
\xi(x)$ becomes a would-be Goldstone boson, ''eaten'' be the gauge field: since $A_\mu$ becomes massive it has three spin degrees of freedom $0,\pm1$ as opposed to the the massless case where it has only $\pm1$. The extra longitudinal degree of freedom, 0,  is nothing else but the would-be Goldstone scalar field. The gauge bosons mass generation through gauging of the global symmetry, that is spontaneously broken, is a mechanism that goes under the name of \emph{the Brout-Englert-Higgs mechanism}~\cite{Englert:1964et,Higgs:1964ia,Higgs:1964pj,Guralnik:1964eu,Higgs:1966ev,Kibble:1967sv}. We illustrate details of the mechanism in the following, relevant in the context of the SM physics, non-abelian example.\\ \mbox{} \\
\textbf{D. $SU(2)\times U(1)$ local }\\ \mbox{}\\ 
We consider $SU(2)\times U(1)$ group as the gauge symmetry. 
This example is special in the sense that the same group with the same SSB pattern appears in the SM. In the SM the $SU(2)$ is chiral, i.e. the left and right fermion components, $\psi_L,\psi_R$ transform in different representations under these transformations. We shall use results of this example in the discussion of the SM later on.
Going back to our example where no fermion sector is introduced, the field content is a single scalar doublet $\Phi$ of $SU(2)$, charged with Q=1/2 under the $U(1)$ group. The couplings of $SU(2)$ and $U(1)$ are denoted as $g,g'$, respectively. Hence the Lagrangian reads:
\begin{equation}
\begin{array}{l}
\mathcal{L}= -\frac{1}{4}B_{\mu\nu}B^{\mu\nu}-\frac{1}{4}W^a_{\mu\nu}W_a^{\mu\nu}+ (D_\mu \Phi)^\dagger D^\mu \Phi - V(\Phi),\qquad V(\Phi)=\mu^2\Phi^\dagger\Phi + \lambda \left(\Phi^\dagger\Phi\right)^2,\mu^2<0, \\\mbox{}\\
D^\mu = \partial^\mu+ \frac{i}{2} g W^\mu_a \tau^a + \frac{i}{2}g' B^\mu,\\\mbox{}\\
B_{\mu\nu} = \partial_{\mu}B_{\nu}-\partial_{\nu}B_{\mu},\\\mbox{}\\
W^{\mu\nu}_a =  \partial^{\mu}W_a^{\nu}-\partial^{\nu}W_a^{\mu} - g\epsilon^{abc}W^{\mu}_{b}W^\nu_c, \qquad\qquad\qquad\qquad\qquad (a=1,2,3),\qquad \epsilon^{123} = 1,\\\mbox{}\\
\Phi = \left(\begin{array}{c}
\phi_1	\\ 
\phi_2
\end{array}\right)\equiv \left(\begin{array}{c} 
\varphi_1+i\varphi_2\\
\varphi_3+i\varphi_4
\end{array}\right).
\end{array}
\label{eq:43}
\end{equation}
Due to the $SU(2)$ invariance, any choice of the vev direction in the $\left<\varphi_i\right>$ space is equivalent. Only the module of $\Phi$ is fixed: $|\left<\Phi\right>|^2=\frac{-\mu^2}{2\lambda}$. We adopt the following choice: $\left<\varphi_3\right> = \sqrt{\frac{-\mu^2}{2\lambda}}\equiv\frac{v}{\sqrt{2}}$,
\begin{equation}
\left<\Phi\right>=\left(\begin{array}{c} 0 \\ v/\sqrt{2} \end{array}\right).
\label{eq:44}
\end{equation}
There is a single combination of generators left unbroken 
\begin{equation}
\frac{\tau^3}{2}+\frac{I}{2},
\label{eq:45}
\end{equation}
where $I$ is the 2-by-2 identity matrix.
The remaining generators 
\begin{equation}
\frac{\tau^1}{2},\qquad \frac{\tau^2}{2},\qquad \frac{\tau^3}{2} - \frac{I}{2}
\label{eq:46}
\end{equation}
are broken. Hence the $\mathcal{L}$ in eq.~\eqref{eq:43} describes the $SU(2)\times U(1)\rightarrow U'(1)$ SSB pattern. There are three Goldstone bosons. Their directions in the $\Phi$ multiplet are: 
\begin{equation}
(i/2)\tau^1\left<\Phi\right>=\left(\begin{array}{c}iv/2 \\ 0 \end{array}\right),\qquad (i/2)\tau^2\left<\Phi\right>=\left(\begin{array}{c} v/2 \\ 0 \end{array}\right),\qquad 
\left((i/2)\tau^3-(i/2)I\right)\left<\Phi\right>=\left(\begin{array}{c} 0 \\ -iv \end{array}\right)
\label{eq:46pp}
\end{equation}
Therefore the $\varphi_1,\varphi_2,\varphi_3$ of eq.~\eqref{eq:43} are to be identified with the Goldstone modes. The scalar fluctuations around $\left<\Phi\right>$ read
\begin{equation}
\Phi(x) = \left(\begin{array}{c} 
\varphi_1(x)+i\varphi_2(x)\\
v + h(x) +i\varphi_4(x)
\end{array}\right) \equiv \exp\{i/2\xi^a(x)\tau^a/v -iI\xi^3(x)/v\}\left(\begin{array}{c}
	0 \\ v + h(x)
\end{array}\right), 
\label{eq:45p}
\end{equation}
In the last step the Goldstones were factorized in analogy to eq.~\eqref{eq:33p}. Again, these modes can be rotated away due to the $SU(2)\times U(1)$ gauge invariance. They are eaten by gauge bosons. To see the mass spectrum of the gauge fields we compute $D_\mu\left<\Phi\right>D^\mu\left<\Phi\right>$:
\begin{equation}
(D_\mu\left<\Phi\right>)^\dagger D^\mu\left<\Phi\right> = \frac{1}{4} v^2 \left(g'^2 B^\mu B_\mu -2  g g' B^\mu W^3_\mu+g^2
   \left(W^1_\mu \left.W^1\right.^\mu+W^2_\mu \left.W^2\right.^\mu+W^3_\mu \left.W^3\right.^\mu\right)\right)
\label{eq:46p}
\end{equation}
which implies the following the gauge bosons mass matrix
\begin{equation}
\frac{v^2}{4}\left(
\begin{array}{cccc}
 g^2 & 0 & 0 & 0 \\
 0 & g^2 & 0 & 0 \\
 0 & 0 & g^2 & -g g' \\
 0 & 0 & -g g' & g'^2 \\
\end{array}
\right).
\label{eq:47}
\end{equation}
Indices $1,2$ of the matrix correspond to $W^1$, $W^2$ fields respectively; indices $3,4$ to $W^3,B$ respectively. In particular the fields $W^1,W^2$ are mass eigenstates. The lower $2\times 2$ matrix is diagonalized by the following rotation:
\begin{equation}
\left(\begin{array}{c} Z_\mu \\ A_\mu\end{array}\right) = \left(\begin{array}{cc} \cos\theta & -\sin\theta \\ 
\sin\theta & \cos\theta \end{array}\right) \left(\begin{array}{c} W^3_\mu \\ B_\mu\end{array}\right),
\label{eq:48}
\end{equation}
where 
\begin{equation}
\cos\theta \equiv \frac{g}{\sqrt{g^2+\left.g'\right.^2}},\qquad \sin\theta\equiv \frac{g'}{\sqrt{g^2+\left.g'\right.^2}}.
\label{eq:49}
\end{equation}
The following definitions $W^\pm\equiv \frac{1}{\sqrt{2}}(W^1\mp iW^2)$ introduce states that are both mass eigenstates and the unbroken $U'(1)$ charge eigenstates, with charges $\pm$. The masses are:
\begin{equation}
m_{W^\pm}=\frac{1}{2}gv,\qquad m_{Z}=\frac{1}{2}\sqrt{g^2+\left.g'\right.^2}v,\qquad m_A=0.
\label{eq:50}
\end{equation}
After the SSB, three gauge bosons become massive. The three would-be Goldstone bosons constitute the longitudinal degrees of freedom of the gauge fields. Single gauge boson is massless. It corresponds to the unbroken $U'(1)$ local symmetry. There is a single massive scalar field $h$ with mass 
\begin{equation}
m_h=\sqrt{2\lambda}v.
\label{eq:51}
\end{equation}
The number of degrees of freedom before and after the SSB is, of course, the same. The above was a non-abelian illustration of the Higgs mechanism: if a global symmetry that exhibits SSB is gauged then the Goldstone bosons of the global symmetry are not physical anymore in the sense that there exist no massless asymptotic states corresponding to these modes and the latter only constitute the longitudinal polarizations of the gauge fields. Hence there are as many massive gauge fields as broken generators. The theory is renormalizable. Although we illustrated the Higgs mechanism at the level of classical Lagrangian, it holds at the quantum level as well (in particular after loop corrections are accounted for in perturbative calculations).
\section{SM structure and its experimental verification}
Upon the three general model-building features listed at the very begining of this Chapter, the SM is designed as follows:
\begin{enumerate}
	\item its (gauge) symmetry group is $G_{SM}=SU(3)_c\times SU(2)_L\times U(1)_Y$. The $SU(3)_c$ corresponds to QCD; QED and the weak interactions are described within the remaining subgroup. 
	\item there are three fermion generations $(i=1,2,3)$ consisting of five representations of $G_{SM}$. These are: left-handed quark $SU(2)_L$ doublets $Q_{Li}$, right-handed up $U_{Ri}$ and right-handed down $D_{Ri}$ quarks, left-handed lepton $SU(2)_L$ doublets $L_{Li}$ and the right-handed charged leptons $E_{Ri}$. The right-handed fields are singles under $SU(2)_L$.  It introduces parity violation in the weak interactions. QED interactions are however parity conserving. There is also a single scalar $SU(2)_L$ doublet $\Phi$. We denote each gauge representation of $Q, U, D, L, E$ and $\Phi$ by $(c,L)_Y$, where $c,L$ denote the $SU(3)_c$ and $SU(2)_L$ representations, respectively and $Y$ is the $U(1)_Y$ charge (the hypercharge). These are:
	\begin{equation}
	\begin{array}{c}
		Q_{Li}(3,2)_{+1/6},\ \ U_{Ri}(3,1)_{+2/3},\ \  D_{Ri}(3,1)_{-1/3},\ \  L_{Li}(1,2)_{-1/2},\ \   E_{Ri}(1,1)_{-1},\ \ \ (i=1,2,3)
		\\ \mbox{}\\
		\Phi(1,2)_{+1/2}.
		\end{array}
	\label{eq:1}
	\end{equation}
	All the three generations are in the same representations of the gauge group. Hence all necessarily have equal interaction strength to the non-abelian gauge bosons of $SU(3)_c$ and $SU(2)_L$.
\item The scalar content and its quantum numbers are the same as in our example \textbf{D} in Sec.~\ref{subsub:SSB}. Hence it triggers the following pattern of SSB $SU(2)_L\times U(1)_Y\rightarrow U(1)_{EM}$ where $U(1)_{EM}$ has now the interpretation of the gauge symmetry of electromagnetic interactions. The $SU(3)_c$ is left unbroken. The QED and QCD gauge interactions are hence unbroken and the SM SSB pattern reads:
\begin{equation}
SU(3)_c\times SU(2)_L\times U(1)_Y\rightarrow SU(3)_c\times U(1)_{EM}.
\label{eq:eq2}
\end{equation}	
\end{enumerate}

Naturally, the SM is a vast subject. In the remaining of this Chapter we focus on EW processes emerging already at tree-level in the SM, report the experimental results and compare. An exception is a brief characterization of electroweak precision measurements. The experimental values are taken from the Particle Data Group database~\cite{Nakamura:2010zzi}, unless stated explicitly. For overviews of the SM see~\cite{Nir,Grossman:2017thq,Pich:2012sx,Pich:2015lkh}.%

The SM is the most general Lagrangian $\mathcal{L}_{SM}$ that can be build with the the particle content (2.) allowed by the symmetry (1.) -- all terms that are allowed, are present. Obviously the SM is co-defined by concrete values of its otherwise free, physical parameters that are fixed by measurements.  $\mathcal{L}_{SM}$ can be split in three pieces:
\begin{equation}
\mathcal{L}_{SM} = \mathcal{L}_{\mathrm{kin}}+\mathcal{L}_{\mathrm{Yuk}}+\mathcal{L}_{\Phi}.
\label{eq:52}
\end{equation}
$\mathcal{L}_{\mathrm{kin}}$ denotes the bilinear, covariant derivative dependent terms; $\mathcal{L}_{\mathrm{Yuk}}$ denotes the scalar-fermion Yukawa interactions; $\mathcal{L}_{\Phi}$ denotes the scalar potential that triggers the SSB.

The $SU(3)$, $SU(2)_{L}$ and $U(1)_Y$ gauge fields form the following multiplets, respectively: 
\begin{equation}
G^\mu_a(8,1)_0,\qquad W^\mu_a(1,3)_0,\qquad B^\mu(1,1)_0.
\label{eq:53}
\end{equation}
The stress-tensors read explicitly:
\begin{equation}
\begin{array}{l}
G^{\mu\nu}_a=\partial^\mu G^\nu_a-\partial^\nu G^\mu_a-g_s f^{abc}G^\mu_b G^\nu_c, \\
W^{\mu\nu}_a=\partial^\mu W^\nu_a-\partial^\nu W^\mu_a-g \epsilon^{abc}W^\mu_b W^\nu_c, \\
B^{\mu\nu}=\partial^\mu B^\nu-\partial^\nu B^\mu,
\end{array}
\label{eq:53p}
\end{equation}
where $f^{abc}$ and $\epsilon^{abc}$ are the $SU(3)$, $SU(2)$ structure constants, respectively; the $g_s$ and $g$ are the $SU(3)$, $SU(2)$ gauge couplings, respectively.

In accord with~\eq{eq:1}, the explicit form of the covariant derivatives read:
\begin{equation}
\begin{array}{l}
D^\mu\Phi = (\partial^\mu+\frac{i}{2}gW^\mu_a\tau_a+\frac{i}{2}g'B^\mu)\Phi, \\\mbox{}\\
D^\mu Q_{Li} = (\partial^\mu+\frac{i}{2}g_sG^\mu_a\lambda_a+ \frac{i}{2}gW^\mu_a\tau_a+\frac{i}{6}g'B^\mu)Q_{Li}, \\\mbox{}\\
D^\mu U_{Ri} = (\partial^\mu+\frac{i}{2}g_sG^\mu_a\lambda_a+\frac{i}{6}g'B^\mu)U_{Ri},\\\mbox{}\\
D^\mu D_{Ri} = (\partial^\mu+\frac{i}{2}g_sG^\mu_a\lambda_a+\frac{2i}{3}g'B^\mu)D_{Ri}, \\\mbox{}\\
D^\mu L_{Li} = (\partial^\mu+\frac{i}{2}gW^\mu_a\tau_a-\frac{i}{2}g'B^\mu)L_{Li}, \\\mbox{}\\
D^\mu E_{Ri} = (\partial^\mu-\frac{i}g'B^\mu)E_{Ri}.
\end{array}
\label{eq:54}
\end{equation}
Except for the $E_{Ri}$ case, the proper tensor product form of the generators representations in eq.~\eqref{eq:54} is implicit. 
%More precisely, except for the $E_{Ri}$ case, the $B_\mu$ field is implicitly multiplied by an $n\times n$ identity matrix: $n=2,3,6$ for $\phi$, $U_{Ri}$, $D_{Ri}$ and $Q_{Li}$ respectively; $\lambda_a$ denotes the $3\times 3$ Gell-Mann matrices in case of $U_{Ri},D_{Ri}$; the $\tau_a$ denotes the $2\times 2$ Pauli matrices in case of $\Phi$ and $L_{Li}$; $\lambda_a$ in $Q_{Li}$ are $6\times 6$ block diagonal matrices with the $3\times 3$ Gell-Mann matrices constituting the blocks; the $\tau_a$ denote modified Pauli matrices with each entry replaced by that entry times a $3\times 3$ identity matrix. 
$\mathcal{L}_{\mathrm{kin}}$ reads
\begin{eqnarray}
\mathcal{L}_{\mathrm{kin}}& = &-\frac{1}{4}G^{\mu\nu}_a G^a_{\mu\nu} -\frac{1}{4}W^{\mu\nu}_a W^a_{\mu\nu} -\frac{1}{4}B^{\mu\nu} B_{\mu\nu} \nonumber \\ &&\mbox{}\nonumber\\ 
&&+i\bar{Q}_{Li}\slashed{D} Q_{Li}+i\bar{U}_{Ri}\slashed{D} U_{Ri}+i\bar{U}_{Di}\slashed{D} U_{Di}+i\bar{L}_{Li}\slashed{D} L_{Li}+i\bar{E}_{Ri}\slashed{D} E_{Ri} \nonumber\\&&\mbox{}\nonumber\\ 
&& +(D^\mu\Phi)^\dagger D_{\mu}\Phi. 
\label{eq:55}
\end{eqnarray}
$\mathcal{L}_{\mathrm{Yuk}}$ reads 
\begin{eqnarray}
\mathcal{L}_{\mathrm{Yuk}} & = & Y^d_{ij}\bar{Q}_{Li}\Phi D_{Rj}+Y^u_{ij}\bar{Q}_{Li}\tilde{\Phi} U_{Rj} + Y^e_{ij}\bar{L}_{Li}\Phi E_{Rj}+ h.c.,
\label{eq:56}
\end{eqnarray}
where $\tilde{\Phi}\equiv i\tau_2\Phi^\dagger$ and $Y^f$ are general $3\times 3$ complex matrices.

$\mathcal{L}_{\Phi}$ denotes the same scalar potential as in Sec. \ref{subsub:SSB} with $\mu^2<0$ triggering the same SSB pattern:
\begin{equation}
\mathcal{L}_{\Phi}=-\mu^2\Phi^\dagger\Phi-\lambda(\Phi^\dagger\Phi)^2,\qquad \left<\Phi\right> = \sqrt{\frac{-\mu^2}{2\lambda}}\equiv\frac{v}{\sqrt{2}}.
\label{eq:57}
\end{equation}
%\subsection{Gauge boson mass spectrum}
%The $W^\pm$ and $Z$ from that example are to be now identified with the massive weak force mediators while $A$ is the massless photon.  
%
%
%The upper bound on the photon mass:
%\begin{equation}
%m_A < 10^{-18}\ \mathrm{eV}.
%\label{eq:60}
%\end{equation}
%Hence the SM predicts the $\theta_W$ angle to be:
%\begin{equation}
%\frac{m_W}{m_Z}= 0.88147 \pm 0.00015\qquad \Longrightarrow \qquad\sin^2\theta_W = 0.2230\pm 0.0003.
%\label{eq:61}
%\end{equation}
%This prediction is confirmed experimentally which is a strong indication that the scenario of SSB triggered by a doublet of $SU(2)_L\times U(1)_Y$ is realized in Nature. Experimental verification of the relation between $m_W$ and $m_Z$ will be discussed later.
\subsection{Gauge boson and Fermion mass spectrum}
The $SU(2)\times U(1)\rightarrow U(1)'$ SSB pattern with a scalar doublet was discussed already in Sec.~\ref{subsub:SSB} in example \textbf{D}. Hence, the tree-level mass formulae for the SM vector bosons $W^\pm$, $Z$ are written in eq.~\eqref{eq:50}; $A$ is now the photon. Experimentally: 
\begin{equation}
m_{W^\pm} = 80.379 \pm 0.012 \ \mathrm{GeV}, \qquad m_Z = 91.1876 \pm 0.0021\  \mathrm{GeV},
\label{eq:59}
\end{equation}
while the experimental upper bound on the photon mass:
\begin{equation}
m_A < 10^{-18}\ \mathrm{eV}.
\label{eq:60}
\end{equation} 
The scalar boson mass is described by~\eqref{eq:51} and the particle shall be referred to as the Higgs boson. The weak and electromagnetic forces mediated by the four gauge bosons are unified in the SM within the $SU(2)_L\times U(1)_Y$ gauge group -- only two gauge couplings $g,g'$ are introduced to describe the couplings of the forces mediated by $W^\pm,Z,A$. The mixing angle $\theta$, that occurred in eq.~\eqref{eq:48}, is called the Weinberg angle and will be denoted by $\theta_W$.

In $\mathcal{L}_{SM}$ there is a single dimensionful parameter, $\mu^2$, which can be replaced by $v$ in the broken phase. In all occurrences of the $\Phi$ doublet, the scale $v$ will occur. In particular the SSB also generates fermion masses by the Yukawa interactions in $\mathcal{L}_Y$ (the Dirac mass terms $m\bar{\psi}_L\psi_R$ are otherwise forbidden by the chiral nature of the SM $SU(2)_L$ gauge group).

One has freedom to redefine the fields $Q_{Li}$, $D_{Ri}$, $U_{Ri}$ in eq.~\eqref{eq:56} by arbitrary unitary rotations $U(3)$ in the $i$ index, i.e. in the flavor space. Each definition means a particular interaction basis, in which the Yukawa matrices $Y^d,Y^u,Y^e$ take certain form. In general, two bases are physically important: the interaction basis and the mass basis. As a convenient first step of going from the former to the latter, one can start with one of the following two bases: (a) the one in which the $Y^d$ and $Y^e$ matrices are diagonal (b) the one in which the $Y^u$ and $Y^e$ are diagonal. Since in general $Y^d \neq Y^u$, the requirements (a) and (b) lead in general to two different bases.  

The $Y^e$ is diagonalized independent of $Y^d$ or $Y^u$. Certain unitary rotations in the flavor space applied to the fields $L_{Li}, E_{Ri}$ results in bi-unitary diagonalization of $Y^e$:
\begin{equation}
\begin{array}{l} 
L_{L}\rightarrow U^\dagger_{eL} L_{L}, \\ 
E_{R}\rightarrow U_{eR} E_{R}
\end{array}\qquad \Longrightarrow\qquad Y^e\rightarrow \hat{Y}^e \equiv U_{eL}Y^e U^\dagger_{eR}.
\label{eq:62}
\end{equation}
Vector notation in the flavor space, e.g. $E_R\equiv (E_{R1},E_{R2},E_{R3})^T$, was used; $U_{eL},U_{eR}$ belong to the flavor $U(3)$ rotations. The matrix $Y^e$ is diagonal and real:
\begin{equation}
\hat{Y}^e=\mathrm{diag}(y_e,y_\mu,y_\tau).
\label{eq:63}
\end{equation}
In the basis eq.~\eqref{eq:63} the components of the left lepton $SU(2)_L$ doublets and the right lepton $SU(2)_L$ singlets shall be denoted as follows:
\begin{equation}
\left(\begin{array}{c}
	\nu_{eL}\\ e_{L}
\end{array}\right),\qquad \left(\begin{array}{c}
	\nu_{\mu L}\\ \mu_{L}
\end{array}\right), \qquad
\left(\begin{array}{c}
	\nu_{\tau L}\\ \tau_{L}
\end{array}\right); \qquad e_R,\qquad\mu_R,\qquad\tau_R.
\label{eq:64}
\end{equation} 
The $e,\mu,\tau$ families are three different flavors, ordered by $y_e,y_\mu,y_\tau$ hierarchy from smallest to the largest. 

Concerning quarks, we first consider the case (a) where $Y^d$ is diagonalized:
\begin{equation}
\begin{array}{l}
Q_{Li}\rightarrow V^\dagger_{dL} Q_{Li},\\
D_{R}\rightarrow V_{dR} D_{R}
\end{array} \qquad \Longrightarrow \qquad Y^d\rightarrow \hat{Y}^d \equiv V_{dL}Y^d V^\dagger_{dR},
\label{eq:65}
\end{equation}
where $\hat{Y}^d$ is diagonal and real:
\begin{equation}
\hat{Y}^d = \mathrm{diag}(y_d,y_s,y_b).
\label{eq:66}
\end{equation}

In the basis eq.~\eqref{eq:66} the components of the left quark $SU(2)_L$ doublets and the right down quarks $SU(2)_L$ singlets shall be denoted as follows:
\begin{equation}
\left(\begin{array}{c}
	u_{dL}\\ d_{L}
\end{array}\right),\qquad 
\left(\begin{array}{c}
	u_{sL}\\ s_{L}
\end{array}\right), \qquad
\left(\begin{array}{c}
	u_{bL}\\ b_{L}
\end{array}\right); \qquad d_R,\qquad s_R,\qquad b_R.
\label{eq:68}
\end{equation} 
The $d,s,b$ are the down quark flavors.

We now consider the case where $Y^u$ is diagonalized:
\begin{equation}
\begin{array}{l}
Q_{Li}\rightarrow V^\dagger_{uL} Q_{Li},\\
U_{R}\rightarrow V_{uR} U_{R}\end{array}\qquad \Longrightarrow \qquad Y^u\rightarrow \hat{Y}^u \equiv V_{uL}Y^u V^\dagger_{uR},
\label{eq:69}
\end{equation}
where $\hat{Y}^u$ is diagonal and real:
\begin{equation}
\hat{Y}^u = \mathrm{diag}(y_u,y_c,y_t).
\label{eq:70}
\end{equation}
In the basis eq.~\eqref{eq:70} the components of the left quark $SU(2)_L$ doublets and the right up quarks, $SU(2)_L$ singlets, shall be denoted as follows:
\begin{equation}
\left(\begin{array}{c}
	u_{L}\\ d_{uL}
\end{array}\right),\qquad 
\left(\begin{array}{c}
	c_{L}\\ d_{cL}
\end{array}\right), \qquad
\left(\begin{array}{c}
	t_{L}\\ d_{tL}
\end{array}\right); \qquad u_R,\qquad c_R,\qquad t_R.
\label{eq:71}
\end{equation} 
The $u,c,t$ are the down quark flavors.

The flavors $d_L,s_L,b_L$ cannot be, in general, identified with the fields $d_{uL},d_{cL},d_{tL}$ because the bases eq.~\eqref{eq:70} and eq.~\eqref{eq:66} require, generally, different $V_{dL}$ and $V_{uL}$ rotations of $Q_{Li}$. Analogous remark is true for the up quark states. The two interaction bases are different.

In the case where $Y^u$ is diagonal, the $Y^d$ is related to the $\hat{Y}^d$ as follows
\begin{equation}
Y^u=\hat{Y}^u,\qquad Y^d=V \hat{Y}^d,
\label{eq:72}
\end{equation}
where 
\begin{equation}
V = V_{uL}V^\dagger_{dL}.
\label{eq:73}
\end{equation}
In the case where $Y^d$ is diagonal the $Y^u$ is related to the $\hat{Y}^u$ as follows
\begin{equation}
Y^d=\hat{Y}^d,\qquad Y^u=V^\dagger \hat{Y}^u.
\label{eq:74}
\end{equation}
In eq.~\eqref{eq:72} and~\eqref{eq:74}, the $Y^d$ and $Y^u$ forms assume the $V_{dR}$ and $V_{uR}$ transformations are applied, respectively. While the rotation matrices $V_{uR},V_{dR},V_{uL},V_{dL}$ depend on the basis we start with in eq.~\eqref{eq:56}, the combination $V$ does not. It is physical, as will be discussed later.

All in all, the terms in eq.~\eqref{eq:56} generate Dirac up, down quark and lepton masses after SSB and the masses read:
\begin{equation}
m_e=\frac{y_e v}{\sqrt{2}},\qquad m_\mu=\frac{y_\mu v}{\sqrt{2}}, \qquad m_\tau=\frac{y_\tau v}{\sqrt{2}},
\label{eq:75}
\end{equation}
for the leptons,
\begin{equation}
m_d=\frac{y_d v}{\sqrt{2}},\qquad m_s=\frac{y_s v}{\sqrt{2}}, \qquad m_b=\frac{y_b v}{\sqrt{2}},
\label{eq:75b}
\end{equation}
for the down quarks, and 
\begin{equation}
m_u=\frac{y_u v}{\sqrt{2}},\qquad m_c=\frac{y_c v}{\sqrt{2}}, \qquad m_t=\frac{y_t v}{\sqrt{2}},
\label{eq:75p}
\end{equation}
for the up quarks. Hence while the fermions are in chiral representations of $G_{SM}$, they are in vector representation of the unbroken group $SU(3)_c\times U(1)_{EM}$:
\begin{itemize}
	\item the left and right charged leptons $e,\mu,\tau$ are in $(1)_{-1}$ representation,
	\item the left and right charged up quarks  $u,c,t$ are in $(3)_{+2/3}$ representation and 
	\item the left and right charged down quarks  $d,s,b$ are in $(3)_{-1/3}$ representation.
\end{itemize} 
The only massless fermions in the SM are neutrinos:
\begin{equation}
m_{\nu_e} = m_{\nu_\mu} = m_{\nu_\tau} = 0.
\label{eq:76}
\end{equation}
From the classical Lagrangian it is obvious - with the absence of the right neutrinos $\nu_R$ in the SM, no bilinear mass terms are allowed. However, since the left neutrinos are singles $(1)_0$ of the unbroken $SU(3)_c\times U(1)_{EM}$, Majorana mass terms could in principle be generated radiatively. Nevertheless, it does not happen due to accidental global symmetries in the SM which correspond to conservation of the lepton flavor quantum numbers. 

In reality neutrinos are massive. It is implied by the neutrino oscillation phenomenon. While the absolute mass scale of neutrinos is unknown, known are the mass differences. The experimental bound on the neutrino $\nu_e,\nu_\mu,\nu_\tau$ masses is
\begin{equation}
m_{\nu}<2\ \mathrm{eV},
\label{eq:78}
\end{equation}
while the mass differences squared read:
\begin{equation}
\begin{array}{l}
 \Delta m_{21}^2 (7.53\pm 0.18) \times 10^{-5}\ 
\mathrm{eV}^2, \\
\Delta m_{32}^2= (2.51 \pm 0.05) \times 10^{-3}\ 
\mathrm{eV}^2.
\end{array}
\label{eq:79}
\end{equation}
Therefore, the physics of neutrino mass generation is beyond the SM description. Whether the neutrinos are Dirac or Majorana is still an open question. If three flavors of right-handed neutrinos $\nu_{Ri}$, singles under the $G_{SM}$, i.e. $(1)_0$, were introduced, Dirac mass terms would emerge from the Yukawa part. The result of such SM extension is an extra set of parameters in the form of a matrix -- a lepton sector direct analogue of the quark $V$ matrix. Again, such extension is not considered the SM -- there are no $\nu_{Ri}$ degrees of freedom in the SM.

The experimental values of the fermion masses are:
\begin{equation}
\begin{array}{c|c|c|c}
	m_{e,\mu,\tau}=& 0.5109989461(31)\ \mathrm{MeV} & 105.6583745(24)\ \mathrm{MeV} & 1776.86 \pm 0.12\ \mathrm{MeV}\\ \hline
	m_{d,s,b} = &4.7^{+ 0.5}_{−0.3}\  \mathrm{MeV} & 95^{+ 9}_{-3}\ \mathrm{MeV} &  4.18^{+ 0.04}_{-0.03}\ \mathrm{GeV} \\ \hline
	m_{u,c,t} = & 2.2^{+0.5}_{-0.4}\ \mathrm{MeV} &  1.275^{+ 0.025}_{
-0.035}\ \mathrm{GeV} &  173.0 \pm 0.4\ \mathrm{GeV}. \\ 
\end{array}
\label{eq:77}
\end{equation}

Except for the three left-handed neutrinos, all fermions are massive. Masses of fermions and weak bosons are consequence of the SSB. If it were not for the SSB, the masslessness of the former is protected by their chiral nature, while of the latter by the gauge symmetry. On the other hand, the mass of the scalar particle is not protected by any mechanism. This situation is typical for scalars and is a potential source of the so called hierarchy problem:  if there is a sufficiently large gap between the electroweak scale and a scale of some heavy sector beyond the SM (BSM), then the Higgs boson mass requires unnaturally exact tuning of the BSM model parameters (for an interesting discussion in this subject see the Appendix in~\cite{Manohar:2018aog}).
All masses are proportional to the only mass parameter in the model, the vev $v$, or equivalently to $\sqrt{-\mu^2}$. 
% interactions
\subsection{The charge current weak interactions}
As already pointed out, there exist, in general, no quark sector interaction basis in which both the down and up Yukawa matrices $Y^u$and $Y^d$ are diagonal simultaneously. However, one of them can always be diagonalized with bi-unitary transformations eq.~\eqref{eq:65},~\eqref{eq:69}. The quark kinetic terms in eq.~\eqref{eq:55} are invariant under the unitary rotations in the flavor space. Without loss of generality for physics conclusions we start the discussion of the $W^\pm$ mediated quark interactions in the basis eq.~\eqref{eq:66}, that correspond to diagonal $Y^d$. The relevant part of~\eqref{eq:55} read
\begin{equation}
\mathcal{L}_{\mathrm{kin}}\supset-\frac{g}{\sqrt{2}}(\bar{u}_{dL}\slashed{W}^+ d_{L}+ \bar{u}_{sL}\slashed{W}^+ s_{L} + \bar{u}_{bL}\slashed{W}^+ b_{L} +\mathrm{h.c.}).
\label{eq:80}
\end{equation}

It is clear from the unitary gauge (eq.~\eqref{eq:45p} with $\xi_i=0$) that the first (second) term in eq.~\eqref{eq:56} governs the down (up) quark interactions with $\Phi$. The up quarks Yukawa interactions are described by the following piece 
\begin{equation}
(\bar{u}_{dL}, \bar{u}_{sL} , \bar{u}_{bL})V^\dagger \hat{Y}^u\left(\begin{array}{c}
	u_R\\ c_R\\ t_R 
\end{array}\right).
\label{eq:81}
\end{equation}
This quark piece can be diagonalized by applying the following transformation of the left up quarks
\begin{equation}
\left(\begin{array}{c}
	u_{dL}\\ u_{sL}\\ u_{bL}
\end{array}\right)\rightarrow V \left(\begin{array}{c}
	u_{dL}\\ u_{sL}\\ u_{bL}
\end{array}\right) = \left(\begin{array}{c}
	u_{L}\\ c_{L}\\ t_{L}
\end{array}\right).
\label{eq:82}
\end{equation}
The $Q_{iL}$ kinetic term is not invariant under eq.~\eqref{eq:82} . The form of~\eqref{eq:80} read
\begin{equation}
-\frac{g}{\sqrt{2}}(\bar{u}_{L}, \bar{d}_{L} , \bar{t}_{L})V \slashed{W}^+\left(\begin{array}{c}
	d_L\\ s_L\\ b_L 
\end{array}\right) + \mathrm{h.c.}.
\label{eq:83}
\end{equation}
the flavor dependence of the $W$ mediated interactions are governed by the matrix $V$; the matrix is physical. Indeed one can check that $V$ is invariant under different choices of the quark interaction basis one starts with. In the general case the $W$ quark couplings are not universal, i.e. $V$ is not proportional to the identity matrix; nor is it diagonal. This case defines the SM. Only left-handed quarks couple to $W$ -- parity is maximally violated in $W$ emissions from quarks.

The case of leptons is more straightforward -- there exists an interaction basis that is also the mass basis~\eqref{eq:71}. The $W^\pm$ interactions in this basis read simply (compare with eq.\eqref{eq:80}):
\begin{equation}
\mathcal{L}_{\mathrm{kin}}\supset-\frac{g}{\sqrt{2}}(\bar{\nu}_{eL}\slashed{W}^+ e_{L}+ \bar{\nu}_{\mu L}\slashed{W}^+ \mu_{L} + \bar{\nu}_{\tau L}\slashed{W}^+ \tau_{L} +\mathrm{h.c.}).
\label{eq:84}
\end{equation}
Only left-handed leptons interact with $W^\pm$. It implies maximal parity-violation. The interactions are universal in the flavor space: the couplings to $W$ is the same for all three pairs $\tau\bar{\nu}_{\tau}$, $\mu\bar{\nu}_{\mu}$ and $e\bar{\nu}_{e}$, and flavor-off-diagonal couplings are forbidden, i.e. $W$ does not couple to $e\bar{\nu}_{\mu}$ etc.

Universality of $W$-$l\bar{\nu_l}$ interactions has been verified experimentally:
\begin{equation}
\begin{array}{c}
\mathrm{BR}(W^+\rightarrow e^+\nu) = (10.71 \pm 0.16) \%, \\	
\mathrm{BR}(W^+\rightarrow \mu^+\nu) = (10.63 \pm 0.15) \%, \\
\mathrm{BR}(W^+\rightarrow \tau^+\nu) = (11.38 \pm 0.21) \%. \\
\end{array}
\label{eq:85}
\end{equation}
%The lepton average is
%\begin{equation}
%\mathrm{BR}(W^+\rightarrow l^+\nu) =  (10.86\pm 0.09)\%. \\
%\label{eq:86}
%\end{equation}

The quark flavor mixing matrix $V$ is called the CKM matrix. The entries shall be denoted as follows: 
\begin{equation}
V = \left(\begin{array}{ccc}
	V_{ud} & V_{us} & V_{ub} \\
	V_{cd} & V_{cs} & V_{cb} \\
	V_{td} & V_{ts} & V_{tb}
\end{array}
\right).
\label{eq:87}
\end{equation}

Comparizon of quark and lepton couplings in eq.~\eqref{eq:80} and~\eqref{eq:84} implies the following prediction for the $W$ decay widths
\begin{equation}
\begin{array}{l}
\Gamma (W^+\rightarrow l^+ \nu_l) \propto 1,\\
\Gamma (W^+\rightarrow u_i \bar{d}_j) \propto 3|V_{ij}|^2,
\end{array}
\label{eq:88}
\end{equation}
where the index $i$ runs only over first two generations of up-quarks $u,c$, the index $j$ over all three down-quarks $d,s,b$. The limitation is due to kinematics. Common factors were omitted and differences in phase space factors neglected. The matrix $V$ is unitary, in particular
\begin{equation}
|V_{ud}|^2+|V_{us}|^2 + |V_{ub}|^2 = |V_{cd}|^2+|V_{cs}|^2 + |V_{cb}|^2 = 1.
\label{eq:89}
\end{equation}
Quark decays mean in practice decays to hadrons, then eq.~\eqref{eq:89} and~\eqref{eq:88} implies
\begin{equation}
\Gamma(W\rightarrow \mathrm{hadrons})\approx 2\Gamma(W\rightarrow \mathrm{leptons}).
\label{eq:90}
\end{equation}
Experimentally
\begin{equation}
BR(W\rightarrow\mathrm{hadrons}) = (67.41\pm 0.27)\%.
\label{eq:91}
\end{equation}

Then from eq.~\eqref{eq:85}:
\begin{equation}
\Gamma(W\rightarrow \mathrm{hadrons})/\Gamma(W\rightarrow\mathrm{leptons})= 2.06\pm 0.02
\label{eq:92}
\end{equation}
in good agreement with the SM. 

Another prediction of eq.~\eqref{eq:89} is that half of the hadronic $W$ decays are through the $c$ quark:
\begin{equation}
\Gamma(W\rightarrow uX) = \Gamma(W\rightarrow cX) = \frac{1}{2}\Gamma(W\rightarrow \mathrm{hadrons}).
\label{eq:93}
\end{equation} Experimentally,
\begin{equation}
\Gamma(W\rightarrow cX) = (33.3 \pm 2.6)\%,
\label{eq:94}
\end{equation}
which implies
\begin{equation}
\Gamma(W\rightarrow cX)/\Gamma(W\rightarrow\mathrm{hadrons})=0.49\pm0.04,
\label{eq:95}
\end{equation}
again in good agreement with the SM. 
\subsection{Neutral currents weak interactions}
Fermion interactions with the $Z$ boson are flavor diagonal and flavor universal. It is a consequence of the fact that all fermions of the same chirality and charge come from the same representation of $SU(2)_L\times U(1)_{Y}$. The relevant part of~\eqref{eq:55} in the mass basis reads
\begin{eqnarray}
\mathcal{L}_{\mathrm{kin}}&\supset&\frac{gg'/\sqrt{g^2+\left.g'\right.^2}}{s_W c_W}\left[ -\left(\frac{1}{2}-s^2_W\right)\bar{e}_{Li}\slashed{Z}e_{Li}+s^2_W\bar{e}_{Ri}\slashed{Z}e_{Ri}+\frac{1}{2}\bar{\nu}_{Li}\slashed{Z}\nu_{Li}\right. \nonumber\\
&&+\left.\left(\frac{1}{2}-\frac{2}{3}s^2_W\right)\bar{u}_{Li}\slashed{Z}u_{Li}-\frac{2}{3}s^2_W\bar{u}_{Ri}\slashed{Z}u_{Ri}-\left(\frac{1}{2}-\frac{1}{3}s^2_W\right)\bar{d}_{Li}\slashed{Z}d_{Li}+\frac{1}{3}s^2_W\bar{d}_{Ri}\slashed{Z}d_{Ri}\right],\nonumber\\
\label{eq:96}
\end{eqnarray}
where $\nu_{Li}=\nu_e,\nu_\mu,\nu_\tau$ and $e_{Li}=e_L,\mu_L,\tau_L$ and $u_{Li}=u_L,c_L,t_L$ etc.

Universality has been confirmed experimentally by flavor-diagonal branching rations:
\begin{equation}
\begin{array}{c}
\mathrm{BR}(Z\rightarrow e^+ e^-) = (3.3632\pm 0.0042)\%, \\
\mathrm{BR}(Z\rightarrow \mu^+ \mu^-) =  (3.3662\pm 0.0066)\%, \\
\mathrm{BR}(Z\rightarrow \tau^+ \tau^-) =  (3.3696\pm 0.0083)\% \\
\end{array}
\label{eq:97}
\end{equation}
and bounds set on flavor-off-diagonal interactions:
\begin{equation}
\begin{array}{c}
	\mathrm{BR}(Z\rightarrow e^\pm\mu^\mp) < 7.5 \times 10^{−7}, \\
	\mathrm{BR}(Z\rightarrow e^\pm\tau^\mp) < 9,8 \times 10^{−7}, \\
	\mathrm{BR}(Z\rightarrow \mu^\pm\tau^\mp) < 1.2 \times 10^{−7}. \\
\end{array}
\label{eq:98}
\end{equation}
Moreover eq.~\eqref{eq:96} implies the following $Z$ decay partial widths into a single fermion-pair generation of each type:
\begin{equation}
\begin{array}{l}
\Gamma(Z\rightarrow \nu\bar{\nu})\propto 1,\\
\Gamma(Z\rightarrow l\bar{l})\propto 1 - 4s_W^2+8s^4_W,\\
\Gamma(Z\rightarrow u\bar{u})\propto 3\left(1-\frac{8}{3}s^2_W+\frac{32}{9}s^4_W\right),\\
\Gamma(Z\rightarrow d\bar{d})\propto 3\left(1-\frac{4}{3}s^2_W+\frac{8}{9}s^4_W\right),
\end{array}
\label{eq:99}
\end{equation}
where again common factors are omitted and phase space differences neglected. The ratios are governed by a single parameter $s_W$. Substituting $\sin\theta_W$ with the formula~\eqref{eq:48} one obtains the prediction:
\begin{equation}
1:0.51:1.75:2.24.
\label{eq:100}
\end{equation}
% number of physical parameters
Experimentally,%
\begin{equation}
\begin{array}{l}
\mathrm{BR}(Z\rightarrow \nu\bar{\nu})=(6.67 \pm 0.02)\%, \\
\mathrm{BR}(Z\rightarrow l\bar{l})=(3.37 \pm 0.01)\%, \\
\mathrm{BR}(Z\rightarrow u\bar{u})=(11.6 \pm 0.6)\%, \\
\mathrm{BR}(Z\rightarrow d\bar{d})=(15.6 \pm 0.4)\%, \\
\end{array}
\label{eq:101}
\end{equation}
which yield the following rations
\begin{equation}
1:0.505:1.74:2.34.
\label{eq:102}
\end{equation}
Again, in good agreement with the SM prediction.
\subsection{QED and QCD interactions}
The SM Lagrangian is invariant under the $SU(3)_c\times U(1)_{EM}$ by construction. QED interactions are described by 
\begin{equation}
-\frac{2e}{3}\bar{u}_i\slashed{A} u_i+\frac{e}{3}\bar{d}_i\slashed{A} d_i+e\bar{l}_i\slashed{A} l_i,
\label{eq:103}
\end{equation}
where $u_{i}=u,c,t$, $d_i=d,s,b$, $l_i=e,\mu,\tau$ and $e$ in eq.~\eqref{eq:103} is the electron charge. In terms of $SU(2)_L\times U(1)_Y$ it is
\begin{equation}
e \equiv \frac{gg'}{\sqrt{g^2+\left.g'\right.^2}}=g\sin\theta_W.
\label{eq:103p}
\end{equation}
This relation is a consequence of the SM unification of the weak and electromagnetic forces. The QED couplings are vector-like. Parity is conserved QED interactions. Except for neutrinos all fermions interact with photons. The interaction is diagonal and universal in the flavor space, in particular the photon does not couple to $e^+\mu-$, etc.

The electron anomalous magnetic moment measurements~\cite{Hanneke:2008tm} provide the most accurate determination of the fine structure constant:
\begin{equation}
\alpha^{-1} \equiv \frac{e^2}{4\pi} = 137.035 999 084 \pm 0.000 000 051.
\label{eq:103b}
\end{equation}
Five loop QED order calculations are necessarily to match the experimental accuracy. Naturally, QED is to be understood as low-energy approximation of the electroweak interactions. In particular, there are corrections to the anomalous magnetic moment from virtual $W$ and $Z$. These are however of order $O(\alpha \frac{m_e^2}{m_{W/Z}^2})\sim 10^{-13}$, beyond the accuracy in~\eqref{eq:103b}. Alternatively one can use the measurement to test QED. This test is the most stringent available and QED as a low-energy approximation to the SM electroweak interactions, is consistent with the experiment. 

The QCD part of  the SM reads
\begin{equation}
-\frac{g_s}{2}\bar{q}\lambda_a\slashed{G}_a q,
\label{eq:104}
\end{equation}
where the $g_s$ denotes the QCD coupling. 

The $e^-e^-$ scattering provide a test of the number of QCD colors $N_C$ via the following processes
\begin{equation}
\begin{array}{c}
	e^+e^-\rightarrow \mathrm{hadrons}, \\
	e^+e^-\rightarrow \mu^+\mu^-,
\end{array}
\label{eq:105}
\end{equation}
by measuring the cross sections ratio
\begin{equation}
R_{e^+e^-}\equiv\frac{\sigma(e^+e^-\rightarrow \mathrm{hadrons})}{\sigma(e^+e^-\rightarrow\mathrm{\mu^+\mu^-}}
\label{eq:106}
\end{equation}
at different energies (invariant mass) of the $e^+e^-$ system.
The tree-level diagrams are those of s-channel $e^+e^-$ annihilation to a pair of quark-antiquark or the $\mu^+\mu^-$ through the photon and the $Z$ boson. At energies well below the $Z$ mass, the photon exchange amplitude dominates the process. The ratio reads
\begin{equation}
N_C\sum_{f=1}^{N_f}Q_f^2=\left\{\begin{array}{l}
	\frac{2}{3}N_C=2,\qquad (N_f=3:u,d,s),\\
	\frac{10}{9} N_C=3\frac{1}{3},\qquad (N_f=4:u,d,s,c),\\
		\frac{11}{9}N_C=3\frac{2}{3},\qquad (N_f=5:u,d,s,c,b).\\
\end{array}\right.
\label{eq:107}
\end{equation}
\begin{figure} 
\begin{center}	
\includegraphics[width=1.\textwidth]{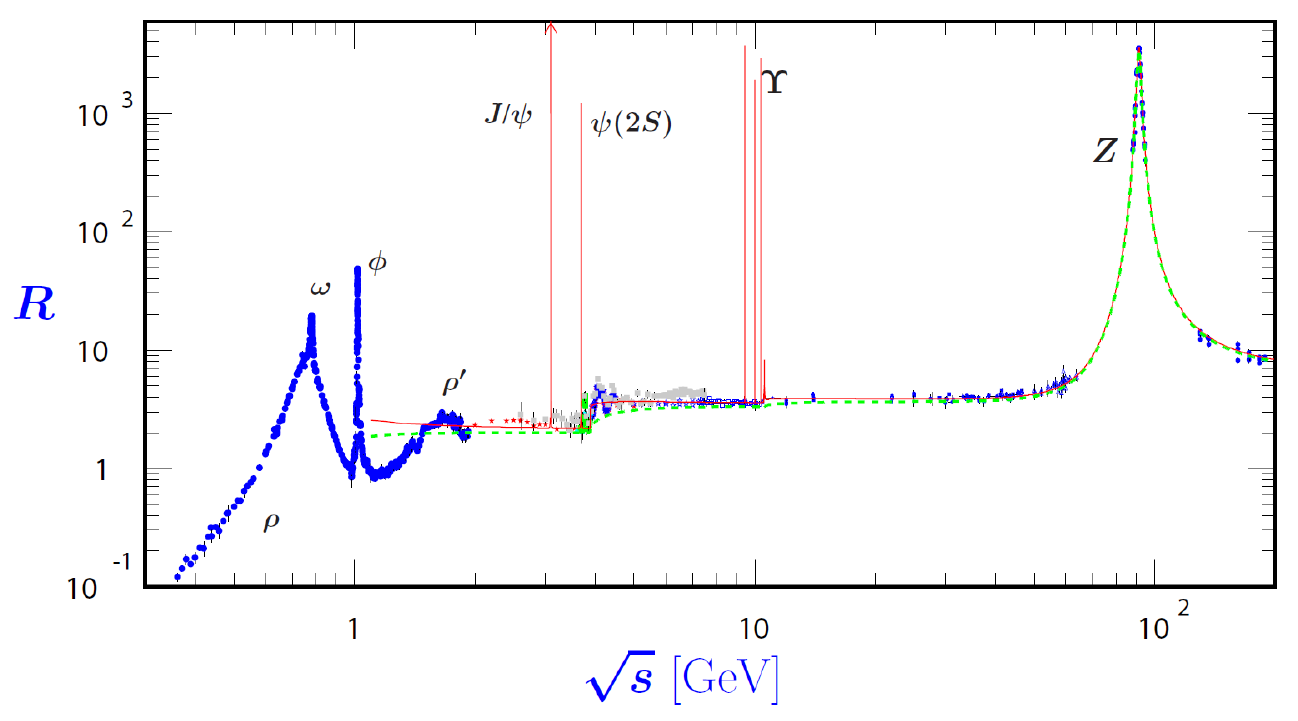}
\caption{World data on the ratio $R_{e^+e^-}$~\cite{Nakamura:2010zzi}. The broken lines show the naive quark model approximation with $N_C = 3$. The solid curve is the 3-loop perturbative QCD prediction.}
\label{fig:Ree}
\end{center}
\end{figure}
Different ratios correspond to different $e^+e^-$ energies, where only certain flavors can be produced. The measurement is shown in Fig.~\ref{fig:Ree}. The complicated resonance structure visible in the plot, is irrelevant for inferring $N_C$ -- the formulae~\eqref{eq:107} hold away from the resonant productions, i.e. they are valid in the regions where hadrons are produced in the form of a pair of jets. The resonances $J/\psi$, $\Upsilon$ are bound states of $c\bar{c}$ and $b\bar{b}$, respectively. Good agreement is reached for $N_C=3$.

A simple estimate on the value of $g_s$ can be made studying the ratio between 3-jet and 2-jet $e^+e^-$ annihilation -- a quark can emit a gluon that can hadronize to a jet. This estimate results in
\begin{equation}
\alpha_s\equiv \frac{g^2_s}{4\pi}\sim 0.12.
\label{eq:108}
\end{equation}
Accounting for QCD loop corrections result in~\cite{Bethke:2011tr}:
\begin{equation}
\alpha_s(m_{Z})=0.1183\pm 0.0010.
\label{eq:109}
\end{equation}
The result is to be understood as at the $m_Z$ scale. Moreover, the coupling $g_s$ is non-perturbative below the so called $\Lambda_{QCD}\sim 300\ \mathrm{MeV}$.  

The same remarks as for QED, are true for QCD, concerning flavor features. Flavor universality of QED and QCD is guaranteed by the fact that the corresponding gauge symmetries are left unbroken -- the kinetic terms in eq.~\eqref{eq:55} are invariant under any rotations in flavor space. 
\subsection{Interactions of the Higgs boson}
The Higgs boson has self-interactions, weak interactions, and Yukawa interactions:
\begin{eqnarray}
\mathcal{L}_{SM}& \supset &  -\frac{m_h^2}{2v}h^3-\frac{m_h^2}{8v^2}h^4\\ &&\mbox{}\nonumber\\ 
&& + m_W^2 W^-_\mu W^{+\mu}\left(\frac{2h}{v}+\frac{h^2}{v^2}\right)+\frac{1}{2}m_Z^2 Z_\mu Z^\mu\left(\frac{2h}{v}+\frac{h^2}{v^2}\right)\nonumber\\&&\mbox{}\nonumber\\ 
&& -\frac{h}{v}(m_l \bar{l}_L l_R + m_q \bar{q}_L q_R + h.c.),\qquad (l=e,\mu,\tau),\  (q=u,c,t,d,s,b)\nonumber
\label{eq:109p}
\end{eqnarray}
Higgs couplings to the fermion mass eigenstates  are diagonal. The reason Higgs boson couples diagonally to the quark mass eigenstates is that the Yukawa couplings determine both the masses and the Higgs couplings to the fermions. Thus, in the mass basis the Yukawa interactions are also diagonal. The couplings are non-universal, as they are proportional to the fermion masses: the heavier the fermion, the stronger the coupling; the factor of proportionality is $m_\psi/v$. Experimental verification of the SM Higgs particle prediction is briefly discussed in Sec.~\ref{sec:higgsDiscovery}.

\subsection{Electroweak precision measurements}
The SM predicts the following relation between the weak boson masses and the weak couplings at tree-level:
\begin{equation}
\frac{m^2_W}{m^2_Z}=\frac{g^2}{g^2+g'^2}=\cos^2\theta_W.
\label{eq:58}%
\end{equation}
%More generally, eq.~\eqref{eq:49} is a consequence of the SSB being triggered by $SU(2)$ doublets. 
It could be verified experimentally: the left hand side can be measured directly from the measured mass spectrum and the right hand side can be determined by measuring weak interaction rates. 

More specifically, in the gauge and scalar sectors there are four free physical parameters: $g,g',\mu^2$ and $\lambda$. Equivalently, one could choose as the free parameters some electroweak observables, e.g.: $m_Z,m_W,\alpha$ and $m_h$. We shall choose the following:
\begin{equation}
\begin{array}{l}
G_F = (1.166 378 8 \pm 0.000 000 7)\times 10^{-5}\mathrm{GeV}^{-2}, \\
\alpha^{-1} = 137.035 999 084 \pm 0.000 000 051, \\
m_Z = 91.1876 \pm 0.0021 \mathrm{GeV}^{-2},
\end{array} 
\label{eq:110}
\end{equation}
and the Higgs boson mass $m_h$. $G_F$ is the Fermi constant and is determined by the muon life-time $\tau_\mu$ measurement from the $\mu^-\rightarrow e^-\bar{\nu}_e\nu_\mu$ decay:
\begin{equation}
\frac{1}{\tau_\mu}=\Gamma_\mu\equiv \frac{G_F^2 m_\mu^5}{192\pi^3}f(m_e^2/m_\mu^2)(1+\delta_{RC}), \qquad f(x)\equiv 1-8x+8x^3 - x^4 - 12x^2\log x,
\label{eq:111}
\end{equation}
where $\delta_{RC}$ denotes QED radiative corrections to order $\mathcal{O}(\alpha^2)$. The explicit correspondence of $G_F$ to the electroweak parameters is the following: the decay is realized through a virtual $W^-$ emission (that subsequently decays to $e^-\bar{\nu}_e$). The momentum transfer $q^2$ carried by the $W$ propagator is much smaller than the $m_W$ scale
\begin{equation}
q^2=(p_\mu-p_{\nu_\mu})^2 = (p_e+p_{\nu_e})^2 \lesssim m_\mu^2 << m_W^2.
\label{eq:112}
\end{equation}
and one can approximate:
\begin{equation}
\frac{g^2}{m_W^2-q^2}\approx \frac{g^2}{m_W^2}=\frac{4\pi\alpha}{\sin^2\theta_W m_W^2}= 4\sqrt{2}G_F.
\label{eq:113}
\end{equation}

The relation~\eqref{eq:113} together with eq.~\eqref{eq:58} ($\sin^2\theta_W=1-\frac{m_W^2}{m_Z^2}$) then determine
\begin{equation}
\sin\theta_W^2 = 0.2122, \qquad m_W=80.94.
\label{eq:114}
\end{equation}
Eq.~\eqref{eq:114} is to be understood as SM prediction based on tree-level calculation. Concerning first the $W$ mass, already at tree-level, the prediction for $m_W$ agrees reasonably well with the experimental value~\eqref{eq:59}. 
In order to meet the experimental result, radiative corrections have to be accounted for. After loop corrections are included the $m_W$ mass prediction of SM is in very good agreement with the measurement~\cite{ALEPH:2010aa}. In general, both left and right hand side of the tree-level relation~\eqref{eq:58} acquire corrections. The experimental verification of the SM prediction for $\sin\theta_W$ at loop level, is discussed below.

\begin{figure} 
\begin{center}	
\includegraphics[width=0.5\textwidth]{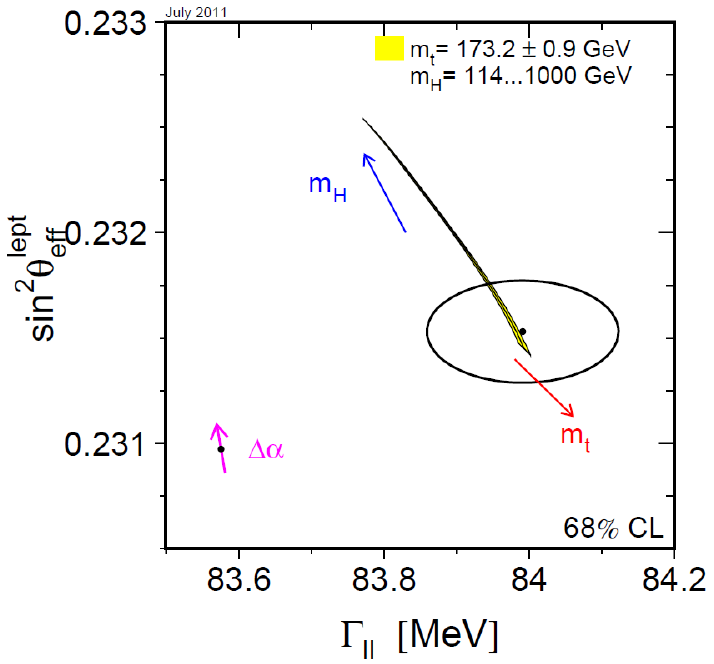}
\caption{Combined LEP and SLD measurements of $\sin2 \theta_{eff}$ (averaged over leptons $e,\mu,\tau$); $\Gamma_l$ denotes the width $\Gamma(Z\rightarrow l\bar{l})$ (averaged over the three leptonic
modes). The shaded region shows the SM prediction. The arrows point in the
direction of increasing values of $m_t$ and $m_h$. The point shows the predicted values if, among the electroweak
radiative corrections, only the photon vacuum polarization is included. Its arrow indicates the variation induced by
the uncertainty in $\alpha(m^2_
Z)$~\cite{ALEPH:2010aa}.}
\label{fig:precisionEW}
\end{center}
\end{figure}

The process that is particularly sensitive to $\sin^2\theta_W$ is $e^+e^-$ annihilation to leptons $e^+e^-\rightarrow l\bar{l}$. The sensitivity is obtained through the measurements of forward-backward asymmetry and polarization asymmetry of the final state leptons. The precision obtained in the LEP and SLD measurements~\cite{ALEPH:2010aa}  required accounting for virtual corrections, in particular of the Higgs boson $h$ and the top quark $t$. The latter particle was discovered at the Tevatron with mass $m_t=173.2 \pm 0.9$~\cite{Lancaster:2011wr}. In Fig.~\ref{fig:precisionEW} shown are the combined LEP and SLD measurements vs. SM prediction of $\sin^2\theta_{eff}$, which is an effective quantity corresponding to the measured charged lepton weak interaction rates, expressible in the SM as $\sin^2\theta_W$ plus the radiative corrections. 
In the plot the SM prediction is presented with the fixed $t$ mass as a function of the Higgs mass -- at the time of the electroweak precision and Tevatron experiments the scalar particle had not yet been discovered. It allows for indirect bounds for the Higgs mass determination. The lower bound on the $m_h$ considered was dictated by 95\%CL direct exclusion of the Higgs boson mass at LEP in the region
\begin{equation}
m_h<114.4\ \mathrm{GeV},
\label{eq:116}
\end{equation} 
The upper limit considered has theoretical motivation -- if $m_h\gtrsim 1000\mathrm{TeV}$ perturbative unitary is violated in weak boson scattering (see Chapter~\ref{sec:VVinSM} for details). The  analysis of Fig.~\ref{fig:precisionEW} indicated that the Higgs mass would be in the lower part of the considered region 114 -- 1000 GeV. In.~\cite{Baak:2011ze} the following indirect upper bound was established: 
\begin{equation}
m_h<169\ \mathrm{GeV}\qquad (95\% C.L.),
\label{eq:117}
\end{equation}
from the requirement of consistency of the data with the SM predictions at the level of two standard deviations (2$\sigma$). 
\subsection{Gauge weak self interactions}
Also pure gauge interactions were studied in $e^+e^-$ collisions. Such interactions are consequence of non-abelian type of the gauge symmetry. 

\begin{figure} 
\begin{center}	
\includegraphics[width=1\columnwidth]{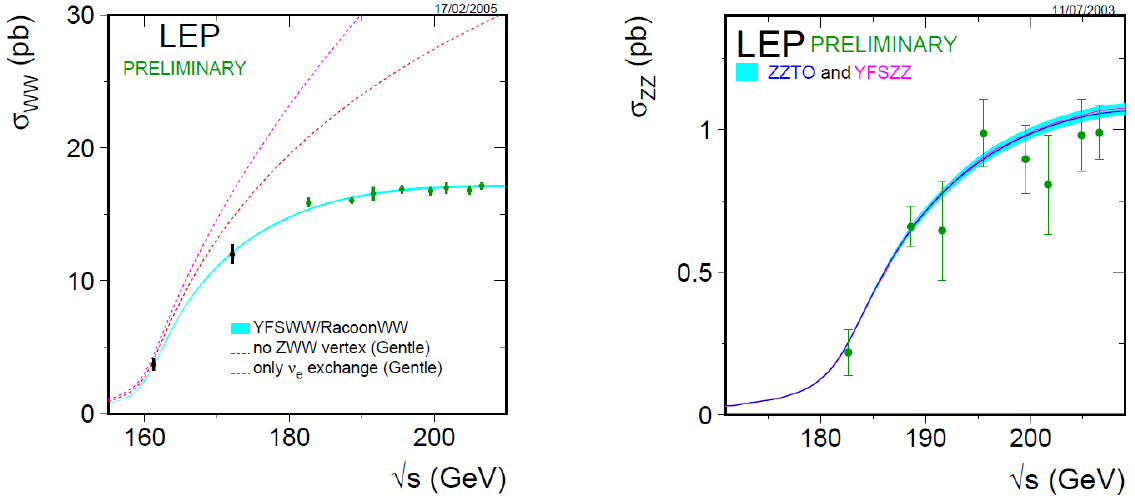}
\caption{Measured energy dependence of cross sections: $e^+e^− \rightarrow W^+W^−$ (left) and $e^+e^−\rightarrow ZZ$ (right). The three
curves shown for the $W$-pair production cross-section correspond to only the $\nu_e$-exchange contribution (upper
curve), $\nu_e$ exchange plus photon exchange (middle curve) and all contributions including also the $ZWW$ vertex
(lower curve). Only the $e$-exchange mechanism contributes to $Z$ pair production,~\cite{ALEPH:2010aa}}
\label{fig:precisionEW1}
\end{center}
\end{figure}

According to the SM there are three types of tree-level diagrams contributing to the process $e^+e^-\rightarrow W^+W^-$: 
\begin{itemize}
	\item[i)] through the exchange of $\nu_e$ in the $t$ channel, 
	\item[ii)] through the exchange of $A$ in the $s$ channel,
	\item[iii)] through the exchange of $Z$ in the $s$ channel.
\end{itemize}
The last two are due to the existence of $AWW$ and $ZWW$ vertices. Fig.~\ref{fig:precisionEW1} shows the $e^+e^-\rightarrow W^+W^-$ cross sections as functions of energy assuming various diagrams would contribute, and the data. 
Clearly the data can be explained only if all three contributions are added. Moreover omitting any of the three diagrams leads to unphysical behavior of the cross section -- it grows with energy, violating unitarity.

On the other hand, the process $e^+e^-\rightarrow ZZ$ can be used to show that there is no evidence for local $\gamma ZZ$ and $ZZZ$ interactions. The SM predicts only one type of diagram that contributes at tree-level: the $t$ channel exchange of the electron. The SM cross section as function of energy and the data are shown in Fig.~\ref{fig:precisionEW1}. The agreement is good.

The pure $SU(2)_L\times U(1)_Y$ gauge interactions of the SM, has been confirmed experimentally.
\section{Higgs discovery}
\label{sec:higgsDiscovery}
In 2012 a new particle, consistent with the SM Higgs boson, has been discovered~\cite{Aad:2012tfa,Chatrchyan:2012xdj} in proton-proton (pp) collisions at 7-8 TeV center of mass (c.o.m) collision energy at the Large Hadron Collider (LHC) by the ATLAS~\cite{Aad:2008zzm} and CMS~\cite{Chatrchyan:2008aa} experiments with mass in the window 125-126 GeV. The first results provided strong indication for a neutral boson with spin-1 hypothesis strongly disfavoured. The combined ATLAS and CMS analysis of the full run 1 (2011-2012) dataset resulted in the following mass determination~\cite{Aad:2015zhl}: 
\begin{equation}
125.09\pm 0.24\ \mathrm{GeV}.
\label{eq:higgsdis1}
\end{equation}
Based on the above dataset, the properties of this particle were studied and found to be consistent with the Standard Model expectations:
\begin{itemize}
	\item all tested alternatives to the $0^+$ spin-parity assignment were rejected in favour of the $0^+$ hypothesis~\cite{
Aad:2015mxa,Khachatryan:2014kca},
	\item the couplings were found to be consistent with the SM predictions with accuracies reaching around $10\%$ in the most favourable cases~\cite{Khachatryan:2016vau}. In particular, the decays into final $ZZ^\ast$, $WW^\ast$, $\gamma\gamma$ and $\tau^+\tau^−$ have been established. The measured signal strengths $\mu$, that are defined as proportionality factors multiplying the SM rate prediction (fitted in the measurement) read (from combined ATLAS and CMS analysis~\cite{Khachatryan:2016vau}):
	\begin{equation}
	\begin{array}{l}
	\mu^{\gamma\gamma} = 1.14^{+0.19}_{-0.18},\nonumber \\ \mbox{} \\
	\mu^{ZZ} = 1.29^{+0.26}_{−0.23}, \nonumber\\ \mbox{} \\
	\mu^{WW} = 1.09^{+0.18}_{−0.16}, \nonumber\\ \mbox{} \\
	\mu^{\tau\tau} = 1.11^{+0.24}_{−0.22},
	\end{array}
	\label{eq:higgsdis2}
	\end{equation}
\end{itemize}
where $\mu=1$ is the SM prediction. Hence, all are consistent with the SM.

As concerns quark flavor changing Higgs couplings, these have been searched for in $t \rightarrow qh$ decays $(q = c, u)$,~\cite{
Aad:2015pja}:
\begin{equation}
\begin{array}{l}
	\mathrm{BR}(t \rightarrow ch) < 4.0 \times 10^{-3},\nonumber \\
\mathrm{BR}(t \rightarrow uh) < 4.5 \times 10^{-3}.
\end{array}
\label{eq:higgsdis3}
\end{equation}
The first direct searches for the lepton-flavour violating Higgs decays were carried out in~\cite{Aad:2016blu} yielding the upper bounds:
\begin{equation}
\begin{array}{l}
	\mathrm{BR}(h \rightarrow \tau \mu) < 2.5 \times 10^{-3},\nonumber  \\
\mathrm{BR}(h \rightarrow \tau e) < 6.1 \times 10^{-3},\nonumber \\
\mathrm{BR}(h \rightarrow \mu e) < 3.4 \times 10^{-4}. \\
\end{array}
\label{eq:higgsdis4}
\end{equation}

The run 2 data taking period (13 TeV c.o.m. $pp$ energy) resulted, in particular, in discovery of the $b\bar{b}$ Higgs decay mode with the following signal strength result:
\begin{equation}
\mu^{b\bar{b}} =  1.02 \pm 0.15,
\label{eq:higgsdis5}
\end{equation} 
again consistent with the SM. In general, all measured properties of the scalar particle are consistent with the SM predictions. 

The discovery of the last remaining piece of the SM -- the Higgs boson -- ended a certain era in particle physics. Confirmed has been that the fundamental laws of physics are based on symmetry principle, violated by the vacuum properties. Despite the extreme experimental success of the SM, in particular in the context of the 13 TeV LHC data, it is widely expected that there is physics beyond the SM, with some new characteristic mass scale(s). We briefly discuss this issue in Chapter~\ref{beyond} where we also argue that vector boson scattering is a promising probe in the search for new physics.

\chapter{Why to go beyond}
\label{beyond}
The current situation in elementary particles is intriguing(see also discussion in~\cite{Pokorski:2005fb}):
on one side, there is the SM which is a renormalizable QFT based on gauge symmetries and spontaneous symmetry breaking. It is a very plausible theory due to its simplicity and high predictive power. It provides a very accurate description of the data. On the other side, there are both certain empirical facts and theoretical questions that the SM does not explain.
From the empirical side, one has for example:
\begin{itemize}
	\item neutrino masses,
	\item the existence of Dark Matter,
	\item matter-antimatter asymmetry,
	\item the acceleration (Dark Energy), 
	\item homogeneity and isotropy of the Universe (inflation),
\end{itemize}
while the theoretical questions the SM does not answer are for example the following:
\begin{itemize}
	\item what is the mechanism stabilizing the SM vacuum?
	\item what is responsible for no $\mathcal{CP}$ breaking in strong interactions?
	\item what explains quarks and leptons hierarchy?
	\item what is the mechanism explaining inflation in the framework of elementary interactions?
	\item what is the relation between elementary
interactions and gravity?
\end{itemize}
These are strong arguments to claim that the SM is only an effective description, a low-energy approximation of a more fundamental theory. The search for an extension of the SM that would address the open questions left unanswered by the SM is now the main goal of particle physics research.

In general, there are two ways how to search for a more
fundamental theory. One is to build concrete deeper theories models and test their predictions against the data. The other is the Effective Field Theory approach. It is a well-developed technique to investigate potential extensions of the SM without explicitly referring to a particular model. The standard technique is to investigate the departures from the SM
predictions in the presence of effective, non-renormalizable operators, that parametrize effects of heavy beyond the SM particles, as a function of the operators coefficients, to either put constraints
on new physics effects or estimate their discovery potential in as much as possible model independent way (for discussion on the EFT Lagrangians see Chapter~\ref{sec:EFT}). This technique is being widely used e.g. in flavor physics~\cite{Fajfer:2012vx} and in the Higgs physics~\cite{Pomarol:2013zra}. The EFT approach is also used to investigate theoretically potential departures from the SM predictions in the gauge boson scattering -- to
parametrize the potential deviations from those predictions. This approach is taken in this thesis. In the case of the gauge boson scattering, one aspect
of the EFT approach is particularly striking. This is the problem of using the EFT approach in its region of validity. There are two main aspects of this issue. One, already mentioned in the introduction, corresponds to perturbative unitarity violation in the EFT approach (for details on unitarity bounds see Chapter~\ref{unitarity}). Another is due to the way the vector boson scattering is accessible experimentally. Both issues are the subject of Chapter~\ref{WWscatEFT} where we apply the EFT approach to same-sign $WW$ scattering in the context of LHC experiment.

\section{Why vector boson scattering as a probe of New Physics} In particular the Higgs mechanism with a single
elementary Higgs boson that triggers the spontaneous EW symmetry breaking, although provides a very successful description of the gauge boson sector, is most likely a simple, effective parametrization of some larger sector. In fact, in
every proposed complete extension of the SM, such as supersymmetric models, composite Higgs models, little
Higgs models etc., that addresses the above mentioned experimental or theoretical issues, the Higgs mechanism involves more scalar bosons and/or non-elementary scalars, and there are
more particles interacting electroweakly. In consequence, also the predictions for the gauge boson interaction are
modified and those effects should manifest themselves at high enough energies. 

At the LHC, vector boson scattering is among the processes most sensitive to the electroweak and the Higgs sectors. In the
SM, the Feynman amplitudes grow with energy, but cancellations among diagrams involving quartic gauge boson couplings, trilinear gauge boson couplings and Higgs exchange occur, and lead to a
total amplitude that does not grow at large energies (for details see Chapter~\ref{sec:VVinSM}). If
modifications from physics beyond the SM exist, they are likely to spoil these cancellations and lead to sizeable cross section increases.

After the discovery of a scalar resonance at the LHC, VBS
received a renewed attention both from the experimental collaborations (see below) and from
theorists who analysed possible signals of NP in these processes by means of the SMEFT%tutajKURWA
Lagrangian~\cite{Kilian:2014zja,Brass:2018hfw,Gomez-Ambrosio:2018pnl}, or of the HEFT one~\cite{Espriu:2012ih,Espriu:2013fia,Delgado:2013loa,Delgado:2013hxa,Espriu:2014jya,Delgado:2014jda,Delgado:2017cls}, without however proper account of the problem of validity of the EFT approach.

\section{How is it accessible experimentally}
\begin{figure} 
\begin{center}	
\includegraphics[width=1.\textwidth]{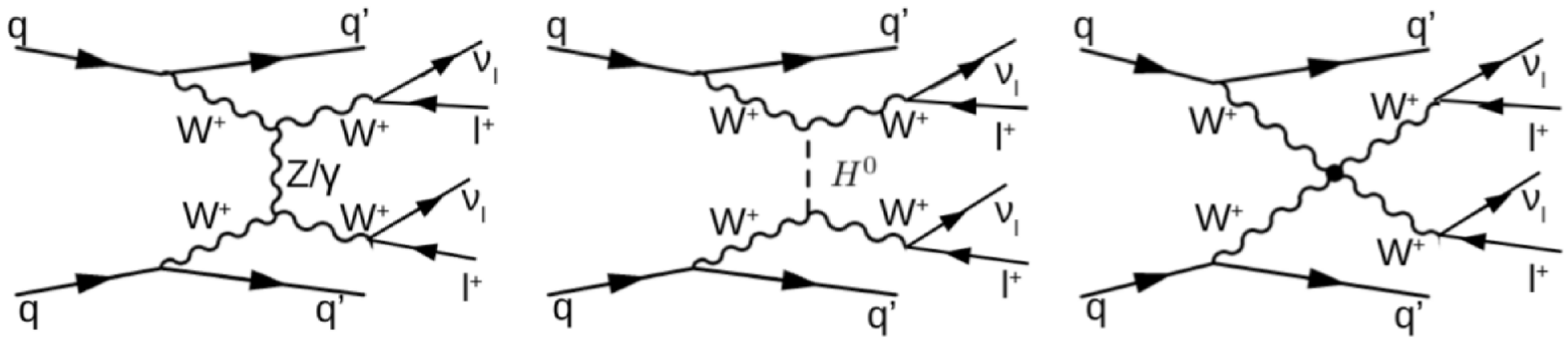}
\caption{Representative Feynman diagrams for single, triple, and quartic gauge couplings of
the EW-induced same-sign W boson pair production (vectro boson scattering).}
\label{fig:EWWWprod}
\end{center}
\end{figure}

This reaction is not directly accessible experimentally but one
can use the following reaction at the LHC:
\begin{equation}
pp \rightarrow 2j + V^\ast V^\ast \rightarrow 2j + 4l,
\label{eq:beyond1}
\end{equation}
where $j$ denotes a quark jet; $V$ denotes a weak gauge boson; $4l$ denotes 2 lepton pairs, decays of $V$ which has which are virtual particles in this reaction ($V^\ast$). A subset of Feynman diagrams representatives is shown in Fig.\ref{fig:EWWWprod}. 

First experimental results for electroweak same-sign W boson pair production searches were reported by the ATLAS and CMS based on data collected at $\sqrt{s} = 8$ TeV~\cite{Aad:2014zda,Khachatryan:2014sta}. The observed significance was 3.6 (2.0) standard
deviations ($\sigma$) for the ATLAS (CMS) study, where a significance of 2.8 (3.1) $\sigma$ was expected based on the SM prediction. The first observation, at the level of $5.5 \sigma$, was reported by CMS~\cite{Sirunyan:2017ret} based on 13 TeV dataset. Also, pioneering measurements of the $ZW^\pm jj$~\cite{Aad:2016ett} 
and $ZZ jj$~\cite{Sirunyan:2017fvv}
processes exist. For more material on experimental searches for vector boson scatterings see~\cite{ATLAS-CONF-2018-030,Aad:2012twa,Khachatryan:2016poo,Khachatryan:2016tgp}. For a review of QGC measurements at the LHC, see~\cite{Green:2016trm}.%,Aaboud:2016ffv,Aad:2012twa}%,Khachatryan:2016poo}%, Khachatryan:2016tgp,Aaboud:2016yus,Aaboud:2016ffv}. 

The lack of direct indications for the presence of new physics (NP) makes indirect searches more interesting. 
The HL-LHC upgrade will eventually collect an integrated luminosity of 3 ab$^{-1}$ 
of data in $pp$ collisions 
at c.o.m. energy of 14 TeV, which should maximize the LHC potential to uncover new phenomena.
It may however well be that the NP degrees of freedom are at  higher masses  making it difficult  at the LHC  to identify experimentally  
 new particles, or  new paradigms.
These considerations have been driving, in the last few years, intense activity worldwide to assess the future of 
collider experiments beyond the HL-LHC. Several proposals and studies have been performed. 
The prospects of pushing the LHC program  further with the  LHC tunnel and the whole CERN infrastructure, 
together with future magnet technology, is an exciting possibility that could push the energy up into an 
unexplored region with the 27 TeV (HE-LHC), that could collect an integrated luminosity of 15 ab$^{-1}$.

\chapter{Constraints on scattering amplitudes from unitarity}
\label{unitarity}
The purpose of this Section is to derive and discuss bounds on tree-level scattering amplitudes that are consequence of unitary evolution of states in Quantum Mechanics. The results of this Section shall be extensively used in the phenomenological analyses of Sec.~\ref{WWscatEFT}.

We shall begin with discussing consequences of unitarity that do not rely on perturbative expansion of the $S$ matrix, the latter being a unitary operator:
\begin{equation}
\hat{S}^\dagger \hat{S} = \hat{1}.
\label{eq:uni1}
\end{equation}
It is customary to decompose $\hat{S}$ into the ''interacting'' and ''non-interacting'' parts as follows:
\begin{equation}
\hat{S} \equiv \hat{1} - i \hat{T},
\label{eq:uni2}
\end{equation}
where $\hat{T}$ governs the interaction and $\hat{1}$ is the identity operator. At the level of matrix elements the four-momentum Dirac delta is factored explicitly from $\hat{T}$ which introduces the scattering amplitude $\mathcal{M}$:
\begin{equation}
\left<\beta\right|\hat{T}\left|\alpha\right> \equiv (2\pi)^2\delta^{(4)}(P_\alpha-P_\beta)\mathcal{M}_{\beta\alpha},
\label{eq:uni3}
\end{equation}
where $\left|\alpha\right>,\left|\beta\right>$ denote, in general, multiparticle states; $P_\alpha$ ($P_\beta$) is the total four-momentum of the state $\left|\alpha\right>$ ($\left|\beta\right>$).  At the level of matrix elements, eq.~\eqref{eq:uni1} then implies:
\begin{equation}
-i(\mathcal{M}^\ast_{\alpha\beta}-\mathcal{M}_{\beta\alpha}) = \int{d\gamma(2\pi)^4\delta^{(4)}(P_\gamma - P_\alpha)\mathcal{M}_{\gamma\beta}^\ast\mathcal{M}_{\gamma\alpha}}, 
\label{eq:uni4}
\end{equation}
in which in both sides $P_\beta=P_\alpha$ is implicitly assumed. The integration over $d\gamma$ involves in particular summation over different number of particles in the states $\left|\gamma\right>$; the factors $1/n_i!$ for each set of $n_i$ identical particles in $\left|\gamma\right>$ are implicit.
The existence of the symmetry factors is a consequence of the explicit form of the completeness
relation that we work with, and which reads:
\begin{eqnarray}
\hat{1}=\sum_{N=0}^{\infty}\left(\sum_{N1}\sum_{N1}\ldots\right)\delta_{N,(N_1+N_2+\ldots)}\frac{1}{N_1!N_2!\ldots} \nonumber \\
\sum_{\sigma_1,\ldots,\sigma_N}\int\ \Pi_{i=1}^N\ d\mathrm{\Gamma}_{\mathbf{p}_i}\left|\mathbf{p}_1\sigma_1,\ldots,\mathbf{p}_N\sigma_N\right>\left<\mathbf{p}_1\sigma_1,\ldots,\mathbf{p}_N\sigma_N\right|,
\label{eq:uni16p}
\end{eqnarray}
where $\sigma_i$ denotes a set of quantum numbers of particle $(i)$.

Eq.~\eqref{eq:uni4} is to be understood as a unitarity condition on the scattering matrix $\mathcal{M}$. In particular it implies the well known optical theorem. Indeed, taking $\beta=\alpha$ one obtains the following relation:
\begin{equation}
-2\mathrm{Im}\mathcal{M}_{\alpha\alpha} = \int{d\gamma(2\pi)^4\delta^{(4)}(P_\gamma - P_\alpha)|\mathcal{M}_{\gamma\alpha}|^2},
\label{eq:uni5}
\end{equation}
in which, for $\left|\alpha\right>$ a two-particle state, the right side is proportional to the total cross section of the $\alpha$ particles scattering. Hence the imaginary part of forward scattering is proportional to the total cross section:
\begin{equation}
\sigma_{tot}(\alpha\rightarrow\text{anything}) \propto -2\mathrm{Im}\mathcal{M_{\alpha\alpha}},
\label{eq:uni6}
\end{equation}
which is the more familiar formulation of the optical theorem.

In order to explore further consequences of the unitarity condition~\eqref{eq:uni4}, from now on we shall assume that both $\left|\alpha\right>$ and $\left|\beta\right>$ in eq.~\eqref{eq:uni4} are two-particle states and explore unitarity conditions for binary reactions. To this end, we shall also apply partial wave decomposition of these states. One can assign:
\begin{equation}
\begin{array}{l}
\left|\alpha\right> = \left|\mathbf{P},\mathbf{p},\lambda_{1},\lambda_{2}\right>\equiv \left|\mathbf{P},\sqrt{s},\hat{\mathbf{p}},\lambda_{1},\lambda_{2}\right>\\ 
\left|\beta\right> = \left|\mathbf{P'},\mathbf{p'},\lambda'_{1},\lambda'_{2}\right>\equiv \left|\mathbf{P'},\sqrt{s'},\hat{\mathbf{p}}',\lambda'_{1},\lambda'_{2}\right>,
\end{array} 
\label{eq:uni7}
\end{equation}
where $\mathbf{P}$ denotes total three-momentum of the corresponding state; $\mathbf{p}$ denotes three-momentum of the first particle in the center of mass frame of the state; $\lambda_i,\lambda'_i$ denote helicities of corresponding particles. The second equalities in each line corresponds to an equivalent labelling of the states: $\hat{\mathbf{p}}$ denotes the versor pointing the $\mathbf{p}$ direction and $\sqrt{s} = \sqrt{\mathbf{p}^2+m_1^2} + \sqrt{\mathbf{p}^2+m_2^2}$. For details on construction, normalization of multi-particle states introduced in this Section and detailed discussion on the $S$-matrix theory, see~\cite{chankowski}. 

We shall now decompose $\left|\alpha\right>$ into states of definite total angular momentum $j$ and its projection $m_j$ onto a quantization axis in the two-particle center of mass frame:
\begin{equation}
\left|\mathbf{P},\sqrt{s},\hat{\mathbf{p}},\lambda_1,\lambda_2\right> = \sum^\infty_j\sum_{m_j = -j}^{+j}\sqrt{\frac{2j+1}{4\pi}}\left|\mathbf{P},\sqrt{s},\lambda_1,\lambda_2,j,m_j\right>D^{(j)}_{m_j,\lambda_1-\lambda_2}(\Omega_{\mathbf{p}}),
\label{eq:uni8}
\end{equation}
where $D^{(j)}_{m_j,\lambda_1-\lambda_2}$ denotes a $(m_j,\lambda_1-\lambda_2)$ matrix element of the Wigner matrices $D^{(j)}(\Omega_{\mathbf{p}})$ and $\Omega_{\mathbf{p}}$ denotes the $\varphi,\vartheta$ angles that specify $\hat{\mathbf{p}}$. Later on, we shall make use of the following property of $D^{(j)}$:
\begin{equation}
\int{d\Omega_{\mathbf{p}}D^{(j')}_{m_j',\lambda}(\Omega_{\mathbf{p}})D^{(j)\ast}_{m_j,\lambda}(\Omega_{\mathbf{p}})=\frac{4\pi}{2j+1}\delta_{j'j}\delta_{m'_jm_j}}.
\label{eq:uni8p}
\end{equation}
The $\left|\beta\right>$ state can be decomposed accordingly which implies in turn the following decomposition of the $\left<\beta\right|\hat{T}\left|\alpha\right>$ matrix element:
\begin{eqnarray}
\lefteqn{\left<\mathbf{P'},\mathbf{p'},\lambda_1',\lambda_2'\right|\hat{T}\left|\mathbf{P},\mathbf{p},\lambda_1,\lambda_2\right> =  \sum_j^\infty\sum_{m_j}\sum_{j'}^\infty\sum_{m_j'}^\infty D^{(j')\ast}_{m'_j,\lambda_1'-\lambda_2'}(\Omega_{\mathbf{p}'}) D^{(j)\ast}_{m_j,\lambda_1-\lambda_2}(\Omega_{\mathbf{p}})\times} \nonumber \\ 
 && \frac{\sqrt{(2j'+1)(2j+1)}}{4\pi}\left<\mathbf{P'},\sqrt{s'},\lambda_1',\lambda_2',j',m_j'\right|\hat{T}\left|\mathbf{P},\sqrt{s},\lambda_1,\lambda_2,j,m_j\right>. 
\label{eq:uni9}
\end{eqnarray}
The matrix elements in the second line can be further factorized accounting for energy-momentum, total angular momentum $j$ and its projection on the quantization axis $m_j$ conservations:
\begin{eqnarray}
\lefteqn{\left<\mathbf{P'},\sqrt{s'},\lambda_1',\lambda_2',j',m_j'\right|\hat{T}\left|\mathbf{P},\sqrt{s},\lambda_1,\lambda_2,j,m_j\right> = } \nonumber \\
&& = (2\pi)^4\delta^{(3)}(\mathbf{P'}-\mathbf{P})\delta(P^{0'}-P^{0})64\pi^2\mathcal{T}^{(j)}_{\lambda_1',\lambda_2';\lambda_1,\lambda_2}(s)\delta_{j'j}\delta_{m'_j m_j}, 
\label{eq:uni10}
\end{eqnarray}
where $\mathcal{T}^{(j)}_{\lambda_1',\lambda_2';\lambda_1,\lambda_2}(s)$ are the partial wave amplitudes; the factor $64\pi^2$ is introduced for later convenience. Comparing~\eqref{eq:uni9} and~\eqref{eq:uni10} with~\eqref{eq:uni4} then implies:
\begin{equation}
\mathcal{M}_{\beta\alpha}=16\pi\sum_j^\infty \sum_{m_j}(2j+1)\mathcal{T}^{(j)}_{\lambda_1',\lambda_2';\lambda_1,\lambda_2}(s)D^{(j)\ast}_{m_j,\lambda_1' - \lambda_2'}(\Omega_{\mathbf{p}'}) D^{(j)}_{m_j,\lambda_1 - \lambda_2}(\Omega_{\mathbf{p}}).
\label{eq:uni11}
\end{equation}
The implications of unitarity from eq.~\eqref{eq:uni4} to binary reactions requires explicit partial wave expanded form of $\mathcal{M}_{\alpha\beta}^\ast$, $\mathcal{M}_{\beta\alpha}$, $\mathcal{M}_{\gamma\beta}^\ast$ and $\mathcal{M}_{\gamma\alpha}$ matrix elements. The decomposed matrix elements, that shall be subsequently substituted into eq.~\eqref{eq:uni4}, shall be marked below with a bullet for the reader convenience. The $\left|\gamma\right>$ states that we shall single out from~\eqref{eq:uni4} and expand into partial waves are two-particle states $\gamma = \left|\tilde{\mathbf{P}},\mathbf{\tilde{p}},\lambda_a,\lambda_b\right>$. Moreover, since four-momentum conservation is implicitly assumed between $\left|\beta\right>$ and $\left|\alpha\right>$ in~\eqref{eq:uni4} and we consider scattering in the center of mass of the $\left|\alpha\right>$ system, the $\left|\alpha\right>,\left|\beta\right>,\left|\gamma\right>$ in eq.~\eqref{eq:uni4} can be identified as follows:
\begin{equation}
\begin{array}{c}
	\left|\alpha\right> = \left|\mathbf{0},\mathbf{p},\lambda_1,\lambda_2\right>, \\
	\left|\beta\right> = \left|\mathbf{0},\mathbf{p'},\lambda_1',\lambda_2'\right>, \\
	\left|\gamma\right> = \left|\mathbf{0},\mathbf{\tilde{p}},\lambda_a,\lambda_b\right>.
\end{array}
\label{eq:uni13p}
\end{equation}

The choice of the angular momentum quantization axis in the direction of the momentum $\mathbf{p}$ simplifies $D^{(j)}_{m_j\lambda_1 - \lambda_2}(\Omega_{\mathbf{p}})$ to $\delta_{m_j\lambda_1-\lambda_2}$ in eq.~\eqref{eq:uni11} and consequently 
\begin{equation}
	\bullet\qquad \mathcal{M}_{\beta\alpha}=16\pi\sum_j^\infty (2j+1)\mathcal{T}^{(j)}_{\lambda_1',\lambda_2';\lambda_1,\lambda_2}(s)D^{(j)\ast}_{\lambda_1-\lambda_2, \lambda_1' - \lambda_2'}(\Omega_{\mathbf{p}'}).
	\label{eq:uni12}
	\end{equation}
	Similarly
	\begin{equation}
	\bullet\qquad \mathcal{M}_{\gamma\alpha}=16\pi\sum_j^\infty (2j+1)\mathcal{T}^{(j)}_{\lambda_a,\lambda_b;\lambda_1,\lambda_2}(s)D^{(j)\ast}_{\lambda_1-\lambda_2, \lambda_a - \lambda_b}(\Omega_{\mathbf{\tilde{p}}}),
	\label{eq:uni12p}
	\end{equation}
where the corresponding partial waves $\mathcal{T}^{(j)\ast}_{\lambda_a,\lambda_b;\lambda_1',\lambda_2'}(s)$ were introduced -- though the notation does not distinguish it, the $\mathcal{T}^{(j)}$ in eq.~\eqref{eq:uni12} and~\eqref{eq:uni12p} are different entities. One can keep track on which transition the partial waves correspond to, by looking at the indices $\lambda_1\lambda_2$, $\lambda_1'\lambda_2'$, $\lambda_a\lambda_b$.

Formula~\eqref{eq:uni12} already implies:
\begin{equation}
\bullet\qquad \mathcal{M}_{\gamma\beta}^\ast=16\pi\sum_j^\infty \sum_{m_j}(2j+1)\mathcal{T}^{(j)\ast}_{\lambda_a,\lambda_b;\lambda_1',\lambda_2'}(s)D^{(j)}_{m_j,\lambda_a - \lambda_b}(\Omega_{\mathbf{\tilde{p}}}) D^{(j)\ast}_{m_j,\lambda_1' - \lambda_2'}(\Omega_{\mathbf{p'}}),
\label{eq:uni13}
\end{equation}
In turn eq.~\eqref{eq:uni13} implies:
\begin{equation}
\bullet\qquad \mathcal{M^
\ast}_{\alpha\beta}=16\pi\sum_j^\infty (2j+1)\mathcal{T}^{(j)\ast}_{\lambda_1,\lambda_2;\lambda_1',\lambda_2'}(s)D^{(j)\ast}_{\lambda_1'-\lambda_2', \lambda_1 - \lambda_2}(\Omega_{\mathbf{p}'}).
\label{eq:uni14}
\end{equation}
The two-body $(ab)$ integrals in eq.~\eqref{eq:uni4} read:
\begin{multline}
\sum_{\lambda_a,\lambda_b} N_{\lambda_a,\lambda_b}\int{d\mathrm{\Gamma}_{\mathbf{\tilde{p}}_a}}\int{d\mathrm{\Gamma}_{\mathbf{\tilde{p}}_b}(2\pi)^4\delta(\tilde{E}_a+\tilde{E}_b-\sqrt{s})\delta^{(3)}(\mathbf{\tilde{p}}_a+\mathbf{\tilde{p}}_b)\mathcal{M}_{\gamma\beta}^\ast\mathcal{M}_{\gamma\alpha} = }  \\
 = \frac{1}{32\pi^2s}\lambda^{1/2}(s,m_a^2,m_b^2)\sum_{\lambda_a,\lambda_b}N_{\lambda_a\lambda_b}\int{d\mathrm{\Omega_{\mathbf{\tilde{p}}}}\mathcal{M}^\ast_{\gamma\beta}\mathcal{M}_{\gamma\alpha}},
\label{eq:uni15}
\end{multline}
where $N_{\lambda_a,\lambda_b}$ corresponds to the $1/{n_i!}$ discussed below eq.~\eqref{eq:uni4}: 
\begin{displaymath}
	\begin{array}{lll}
		N_{\lambda_a\lambda_b} = 1/2,&\qquad& \text{if the particles are identical, i.e. } a=b, \text{ and if } \lambda_a = \lambda_b, \\
		N_{\lambda_a\lambda_b} = 1,&\qquad& \text{otherwise},
	\end{array}
\end{displaymath}
and 
\begin{equation}
\lambda(x,y,z) = x^2 + y^2 + z^2 - 2xy - 2xz - 2yz.
\label{eq:uni16}
\end{equation}
After plugging appropriate ''bullet'' formulas into~\eqref{eq:uni15} and~\eqref{eq:uni4}, performing the $d\Omega_{\mathbf{\tilde{p}}}$ integral in the former formula (with the help of property~\eqref{eq:uni8p}), eq.~\eqref{eq:uni4} acquires the following form:
\begin{eqnarray}
\lefteqn{-i\sum_j{(2j+1)D^{(j)\ast}_{\lambda_1-\lambda_2,\lambda_1'-\lambda_2'}(\Omega_{\mathbf{p'}})\left[\mathcal{T}^{(j)\ast}_{\lambda_1\lambda_2;\lambda_1'\lambda_2'}(s)- \mathcal{T}^{(j)}_{\lambda_1'\lambda_2';\lambda_1\lambda_2}(s)\right]}} \nonumber\\
&\qquad\qquad=& \sum_{(ab)}\sum_j{(2j+1)D^{(j)\ast}_{\lambda_1-\lambda_2,\lambda_1'-\lambda_2'}(\Omega_{\mathbf{p'}})}\qquad \qquad\qquad\qquad\qquad\qquad\qquad\qquad\qquad\nonumber\\
&&\times \sum_{\lambda_a\lambda_b}{\frac{2}{s}\lambda^{1/2}(s,m_a^2,m_b^2)N_{\lambda_a\lambda_b}\mathcal{T}^{(j)\ast}_{\lambda_a\lambda_b;\lambda_1'\lambda_2'}(s)\mathcal{T}^{(j)}_{\lambda_a\lambda_b;\lambda_1\lambda_2}(s)} \nonumber\\
&&+\frac{1}{16\pi}\int{d\gamma(2\pi)^4\delta^{(4)}(P_\gamma-P_\alpha)\mathcal{M}^\ast_{\gamma\beta}\mathcal{M}_{\gamma\alpha}},
\label{eq:uni17}
\end{eqnarray}
where the sum over $(ab)$ means sum over all two-particle states that are kinematically allowed at $s$; the integral over $d\gamma$ includes the sum over three-, four-, etc. particle $\left|\gamma\right>$ states. Similar  factors to $D^{(j)\ast}_{\lambda_1-\lambda_2,\lambda_1'-\lambda_2'}$, present on the left side, and on the right side in the term involving the sum over two-particle states, can be factored in the last line, as well. Indeed, it is enough to expand into partial waves the states $\left|\alpha\right>$ and $\left|\beta\right>$; the $\mathcal{M}_{\gamma\beta}^\ast$ and $\mathcal{M}_{\gamma\alpha}$ in the last line of eq.~\eqref{eq:uni17} can then be generically written as:
\begin{eqnarray}
\mathcal{M}_{\gamma\beta}^\ast = \sum_{j'}^\infty\sum_{m'_j}\sqrt{\frac{2j'+1}{4\pi}} \mathcal{T}^{(j')}_{\gamma;\lambda_1',\lambda_2',m'_j}(\gamma;s)D^{(j')}_{m'_j,\lambda_1'-\lambda_2'}(\Omega_{\mathbf{p'}}), \nonumber \\
\mathcal{M}_{\gamma\alpha} = \sum_{j''}^\infty\sqrt{\frac{2j''+1}{4\pi}}\mathcal{T}^{(j'')}_{\gamma;\lambda_1,\lambda_2,\lambda_1-\lambda_2}(\gamma;s),\qquad\qquad\qquad
\label{eq:uni18}
\end{eqnarray}
where the amplitudes $\mathcal{T}^{(j')}_{\gamma;\lambda_1',\lambda_2',m'_j}(\gamma;s)$ are defined by the equation:
\begin{equation}
\left<\gamma\right|\hat{T}\left|\mathbf{0},\sqrt{s},\lambda_1',\lambda_2',j',m_j'\right> \equiv (2\pi)^4\delta^{(3)}(\mathbf{P}_\gamma)\delta(P^0_\gamma-\sqrt{s}) \mathcal{T}^{(j')}_{\gamma;\lambda_1',\lambda_2',m'_j}(\gamma;s).
\label{eq:uni19}
\end{equation}
The symbol $\gamma$ used above in the subscript of $\mathcal{T}^{(j)}$ is to remind that this amplitude
depends, apart from s, also on the variables needed to specify the multiparticle state $\left|\gamma\right>$. Similarly as in~\eqref{eq:uni12}, the equality $D^{(j)}_{m_j\lambda_1 - \lambda_2}(\Omega_{\mathbf{p}})=\delta_{m_j\lambda_1-\lambda_2}$ was used in the second line of eq.~\eqref{eq:uni18}. After the substitutions~\eqref{eq:uni18}, the last line in eq.~\eqref{eq:uni17} reads:
\begin{eqnarray}
\frac{1}{16\pi}\sum_{j''}^\infty\sum_{j'}^\infty\sum_{m_j'}\frac{\sqrt{(2j''+1)(2j'+1)}}{4\pi} D^{(j')\ast}_{m_j',\lambda_1'-\lambda_2'}(\Omega_{\mathbf{p'}})\qquad\qquad\qquad\qquad \nonumber \\
\times \int{d\gamma(2\pi)^4\delta^{(4)}(P_\gamma-P_\alpha)\mathcal{T}^{(j')}_{\gamma;\lambda_1',\lambda_2',m_j'}(\gamma;s)\mathcal{T}^{(j'')}_{\gamma;\lambda_1,\lambda_2,\lambda_1-\lambda_2}(\gamma;s)}.
\label{eq:uni20}
\end{eqnarray}
A factor $D^{(j')\ast}_{m_j',\lambda_1'-\lambda_2'}(\Omega_{\mathbf{p'}})$ occurred as anticipated.

The next step is to project both sides of eq.~\eqref{eq:uni17} on definite $j$ component by first multiplying both sides by $D^{(j)}_{\lambda_1-\lambda_2,\lambda_1'-\lambda_2'}(\Omega_\mathbf{p'})$ and then integrating over $d\Omega_{\mathbf{p'}}$ using again the property~\eqref{eq:uni8p}. The term~\eqref{eq:uni20} after the projection reads:
\begin{eqnarray}
\frac{1}{16\pi}\sum_{j''}^\infty\sqrt{\frac{2j''+1}{2j+1}}\int{d\gamma}(2\pi)^4\delta^{(4)}(P_\gamma-P_\alpha)\qquad\qquad\qquad\qquad \nonumber \\
\times \mathcal{T}^{(j)\ast}_{\gamma;\lambda_1',\lambda_2',\lambda_1-\lambda_2}(\gamma;s)\mathcal{T}^{(j'')}_{\gamma;\lambda_1,\lambda_2,\lambda_1-\lambda_2}(\gamma;s).
\label{eq:equni21}
\end{eqnarray}
The angular momentum conservation implies that only $j''=j$ can contribute in eq.~\eqref{eq:equni21}. All in all, after the projection eq.~\eqref{eq:uni17} reads:
\begin{eqnarray}
-i\left[\mathcal{T}^{(j)\ast}_{\lambda_1\lambda_2;\lambda_1'\lambda_2'}(s)-\mathcal{T}^{(j)}_{\lambda_1'\lambda_2';\lambda_1\lambda_2}(s)\right]= \nonumber\qquad\qquad\qquad\qquad\qquad\qquad\qquad\qquad\qquad\\
%=\frac{2}{s}\lambda^{1/2}(s,m_1^2,m_2^2)\sum_{\tilde{\lambda}_1,\tilde{\lambda}_2}N_{\tilde{\lambda}_1\tilde{\lambda}_2}\mathcal{T}^{(j)\ast}_{\tilde{\lambda}_1\tilde{\lambda}_2;\lambda_1'\lambda_2'}(s)\mathcal{T}^{(j)}_{\tilde{\lambda}_1\tilde{\lambda}_2;\lambda_1\lambda_2}(s) \qquad\qquad \nonumber\\
=\sum_{(ab)}\frac{2}{s}\lambda^{1/2}(s,m_a^2,m_b^2)\sum_{\lambda_a,\lambda_b}N_{\lambda_a\lambda_b} \mathcal{T}^{(j)\ast}_{\lambda_a\lambda_b;\lambda_1'\lambda_2'}(s)\mathcal{T}^{(j)}_{\lambda_a\lambda_b;\lambda_1\lambda_2}(s) \qquad\qquad\nonumber \\
+\frac{1}{16\pi}\int{d\gamma(2\pi)^4\delta^{(4)}(P_{\gamma}-P_{\alpha})\mathcal{T}^{(j)\ast}_{\gamma;\lambda_1',\lambda_2',\lambda_1-\lambda_2}(\gamma;s) \mathcal{T}^{(j)}_{\gamma;\lambda_1,\lambda_2,\lambda_1-\lambda_2}(\gamma;s)},
\label{eq:equni22}
\end{eqnarray}
where ${\displaystyle\sum_{(ab)}}$ denotes summation over different two-particle states; it is understood $(ab)=(ba)$. 

The formula in eq.~\eqref{eq:equni22} can now be used to generate different unitarity bounds for binary reactions. First, we assume that the particle content of $\left|\alpha\right>$ and $\left|\beta\right>$ is the same. Then the sum over $(ab)$ in~\eqref{eq:equni22} can be split into the elastic part, i.e. where the particles $(ab)$ are the same as $(12)$, and the remaining, inelastic part; more explicitly:
\begin{eqnarray}
-i\left[\mathcal{T}^{(j)\ast}_{\lambda_1\lambda_2;\lambda_1'\lambda_2'}(s)-\mathcal{T}^{(j)}_{\lambda_1'\lambda_2';\lambda_1\lambda_2}(s)\right] \nonumber\qquad\qquad\qquad\qquad\qquad\qquad\qquad\qquad\\
=\frac{2}{s}\lambda^{1/2}(s,m_1^2,m_2^2)\sum_{\tilde{\lambda}_1,\tilde{\lambda}_2}N_{\tilde{\lambda}_1\tilde{\lambda}_2}\mathcal{T}^{(j)\ast}_{\tilde{\lambda}_1\tilde{\lambda}_2;\lambda_1'\lambda_2'}(s)\mathcal{T}^{(j)}_{\tilde{\lambda}_1\tilde{\lambda}_2;\lambda_1\lambda_2}(s) \qquad\qquad \nonumber\\
+\sum_{(ab)\neq(12)}\frac{2}{s}\lambda^{1/2}(s,m_a^2,m_b^2)\sum_{\lambda_a,\lambda_b}N_{\lambda_a\lambda_b} \mathcal{T}^{(j)\ast}_{\lambda_a\lambda_b;\lambda_1'\lambda_2'}(s)\mathcal{T}^{(j)}_{\lambda_a\lambda_b;\lambda_1\lambda_2}(s) \nonumber \\
+\frac{1}{16\pi}\int{d\gamma(2\pi)^4\delta^{(4)}(P_{\gamma}-P_{\alpha})\mathcal{T}^{(j)\ast}_{\gamma;\lambda_1',\lambda_2',\lambda_1-\lambda_2}(\gamma;s) \mathcal{T}^{(j)}_{\gamma;\lambda_1,\lambda_2,\lambda_1-\lambda_2}(\gamma;s)},
\label{eq:equni22p}
\end{eqnarray}
where the first two lines correspond to the splitting between the elastic and inelastic parts of two-particle $\left|\gamma\right>$ states. We further assume that $\lambda_1'=\lambda_1$ and $\lambda_2'=\lambda_2$. In particular, then on the left side of eq.~\eqref{eq:equni22p} both $\mathcal{T}^{(j)}$ are partial wave amplitudes of the same elastic scattering with no helicity flip. Moreover, the last line in~\eqref{eq:equni22p} simplifies to
\begin{equation}
\frac{1}{16\pi}\int{d\gamma(2\pi)^4\delta^{(4)}(P_\gamma-P_\alpha)\left|\mathcal{T}^{(j)}_{\gamma;\lambda_1,\lambda_2,\lambda_1-\lambda_2}(\gamma;s)\right|^2}
\label{eq:equni23}
\end{equation}
and~\eqref{eq:equni22p} can be rewritten as 
\begin{equation}
\left[\mathrm{Re}\mathcal{T}^{(j)}_{\lambda_1\lambda_2;\lambda_1\lambda_2}(s)\right]^2+\left[\mathrm{Im}\mathcal{T}^{(j)}_{\lambda_1\lambda_2;\lambda_1\lambda_2}(s)+\frac{s\lambda_{12}^{-1/2}}{2N_{\lambda_1\lambda_2}}\right]^2 = \frac{s^2\lambda_{12}^{-1}}{4N^2_{\lambda_1\lambda_2}}-R^2_j(s),
\label{eq:uni24}
\end{equation}
where $R^2_j(s)$ is positive-definite and reads
\begin{eqnarray}
R^2_j(s) = \sum_{(\lambda_1'\lambda_2')\neq(\lambda_1\lambda_2)}\frac{N_{\lambda_1'\lambda_2'}}{N_{\lambda_1\lambda_2}}\left|\mathcal{T}^{(j)}_{\lambda_1'\lambda_2';\lambda_1\lambda_2}(s)\right|^2 \qquad\qquad\qquad\nonumber \\
+\sum_{(ab)\neq(12)}\sum_{(\lambda_a\lambda_b)}\frac{N_{\lambda_a\lambda_b}}{N_{\lambda_1\lambda_2}}\frac{\lambda_{ab}^{1/2}}{\lambda_{12}^{1/2}}\left|\mathcal{T}^{(j)}_{\lambda_a\lambda_b;\lambda_1\lambda_2}(s)\right|^2\qquad \nonumber \\\mbox{} \nonumber\\
+\left(\text{a term proportional to~\eqref{eq:equni23}}\right)\qquad\qquad\qquad.
\label{eq:uni25}
\end{eqnarray}
The eq.~\eqref{eq:uni24} implies that the amplitude $\mathcal{T}^{(j)}_{\lambda_1,\lambda_2;\lambda_1,\lambda_2}(s)$ of the
elastic scattering with no change of helicities must lie on a circle, the so-called Argand circle, whose radius in not grater than $s\lambda_{12}^{-1/2}/2N_{\lambda_1\lambda_2}$ and the center is
at the point $(0,-s\lambda_{12}^{-1/2}/2N_{\lambda_1,\lambda_2})$ in the complex plane, as shown graphically
\begin{figure} 
\begin{center}	
\includegraphics[width=0.8\textwidth]{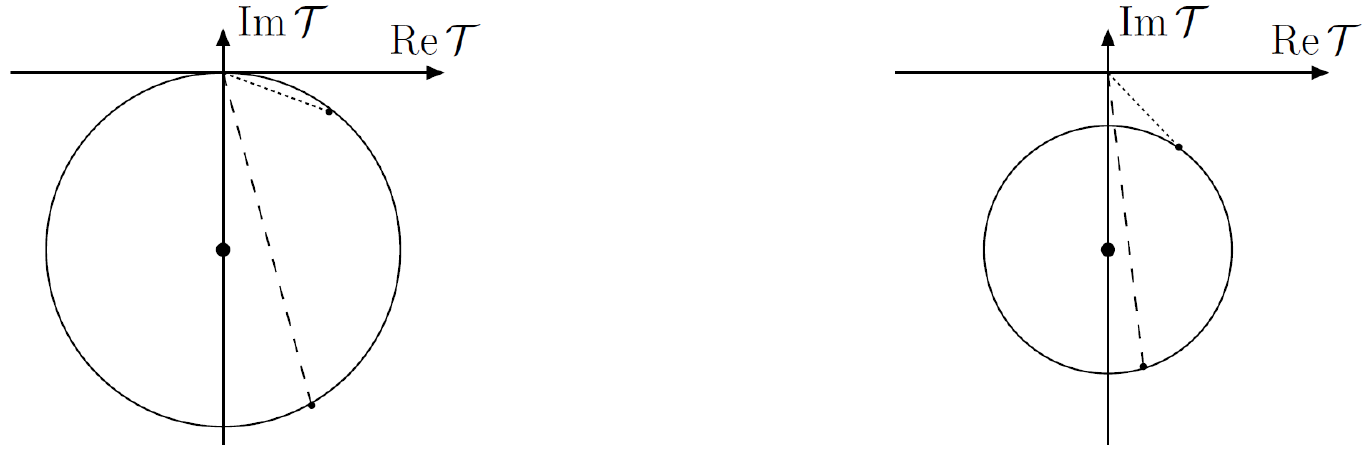}
\caption{Argand circles: if inelastic channels are closed, i.e. if $R^2_j (s) = 0$,
(left) the radius is $s\lambda^{−1/2}/2N_{\lambda_1\lambda_2}$; if inelastic channels are open (right) the
radius is smaller. Partial amplitudes of the elastic scattering must lie on
the Argand circle. Short-dashed lines show possible partial elastic scattering
amplitudes in a weakly coupled theory (small corrections in the perturbative
expansion) whereas the long-dashed ones illustrate elastic scattering amplitudes
typical for a strongly coupled (nonperturbative) theory;~\cite{chankowski}.}
\label{fig:argand}
\end{center}
\end{figure}
in Figure~\ref{fig:argand}. This shows, that the elastic scattering amplitude must have a nonzero imaginary part, which grows as more and more inelastic channels open up with increasing $\sqrt{s}$. Hence, at high energies elastic scattering amplitudes, at least with no helicity flip, are typically predominantly imaginary. In particular eq.~\eqref{eq:uni24} implies the following unitarity bounds on the partial waves for each $j$ separately:
\begin{eqnarray}
N_{\lambda_1\lambda_2}\left|\mathcal{T}^{(j)}_{\lambda_1\lambda_2;\lambda_1\lambda_2}(s)\right|\leq\frac{s}{\lambda^{1/2}(s,m_1^2,m_2^2)}, \nonumber\\
N_{\lambda_1\lambda_2}\left|\mathrm{Re}\mathcal{T}^{(j)}_{\lambda_1\lambda_2;\lambda_1\lambda_2}(s)\right|\leq\frac{s}{2\lambda^{1/2}(s,m_1^2,m_2^2)}, \nonumber\\
-\frac{s}{\lambda^{1/2}(s,m_1^2,m_2^2)}\leq N_{\lambda_1\lambda_2}\,\mathrm{Im}\mathcal{T}^{(j)}_{\lambda_1\lambda_2;\lambda_1\lambda_2}(s)\leq 0. 
\label{eq:uni26}
\end{eqnarray}
Moreover, since $R^2_j(s)$ cannot exceed $s^2\lambda_{12}^{-1}/4N^2_{\lambda_1\lambda_2}$ (the right hand side of~\eqref{eq:uni24} must be positive), one also obtains the bounds on partial wave amplitudes of any two-body (not necessarily elastic) scattering:
\begin{equation}
\sqrt{N_{\lambda_a\lambda_b}N_{\lambda_1\lambda_2}}\left|\mathcal{T}^{(j)}_{\lambda_a\lambda_b;\lambda_1\lambda_2}(s)\right|\leq\frac{s}{2\lambda^{1/4}(s,m_a^2,m_b^2)\lambda^{1/4}(s,m_1^2,m_2^2)}.
\label{eq:uni27}
\end{equation}
Interestingly, at the reaction threshold, where $\lambda^{1/2}(s,m^2_a,m^2_b) = 0$, the bounds disappear. On the other hand, if $\sqrt{s}$ is much greater than any of the masses involved, which is the case we shall assume in the context of later analysis, the unitarity bounds become:
\begin{eqnarray}
N_{\lambda_1\lambda_2}\left|\mathcal{T}^{(j)}_{\lambda_1\lambda_2;\lambda_1\lambda_2}(s)\right|\leq1, \label{eq:mod}\\
N_{\lambda_1\lambda_2}\left|\mathrm{Re}\mathcal{T}^{(j)}_{\lambda_1\lambda_2;\lambda_1\lambda_2}(s)\right|\leq\frac{1}{2},\label{eq:re} \\
-1\leq N_{\lambda_1\lambda_2}\mathrm{Im}\mathcal{T}^{(j)}_{\lambda_1\lambda_2;\lambda_1\lambda_2}(s)\leq 0, \label{eq:im}\\
 \sqrt{N_{\lambda_a\lambda_b}N_{\lambda_1\lambda_2}}\left|\mathcal{T}^{(j)}_{\lambda_a\lambda_b;\lambda_1\lambda_2}(s)\right|\leq\frac{1}{2}.
\label{eq:uni28}
\end{eqnarray}
The bounds~\eqref{eq:mod}-~\eqref{eq:uni28} have been derived assuming only that
the evolution of the quantum system is unitary. In particular they do not rely on any perturbative expansion. Scattering amplitudes derived from QFT models which (are believed to) give rise to unitary $S$-matrices should in principle, respect these bounds. Interestingly however, since elastic scattering partial wave amplitudes computed in the lowest order of the perturbative
expansion in QFT are (usually) real (i.e. lie on the horizontal axis in Fig.~\ref{fig:argand}), they cannot satisfy the unitarity relation~\eqref{eq:uni24}. Higher order contributions must therefore bring elastic amplitudes back on the Argand circle. Two cases can be then distinguished:
\begin{enumerate}
	\item If the absolute value of the real part of the lowest order amplitude is bounded by $1/2N_{\lambda_1\lambda_2}$ (compare with eq.~\eqref{eq:re}), loop contributions required to restore unitarity can be relatively small and the perturbative expansion is likely to be reliable.
	\item In contrast, if the real part of the lowest order amplitude exceeds $1/2N_{\lambda_1\lambda_2}$, the necessary higher order contributions must be comparable or even larger than the lowest order term and the perturbative expansion evidently fails.
\end{enumerate}	

Therefore the value $1/2N_{\lambda_1\lambda_2}$ emerges as a characteristic value above which elastic scattering partial waves with no spin flip violate \emph{perturbative unitarity}. If the tree-level amplitude is real, then the bound~\eqref{eq:re} is just stronger than~\eqref{eq:mod}, and~\eqref{eq:im} does not play any role. Similarly, one can argue that the characteristic scale of perturbative unitarity violation by partial waves in case of non-elastic scattering (or elastic with spin flip) can be identified with $1/(2\sqrt{N_{\lambda_a\lambda_b}N_{\lambda_1\lambda_2}})$ (see eq.\eqref{eq:uni28}).

In practice, the magnitude of the tree-level amplitudes depends usually on the energy $\sqrt{s}$. In renormalizable theories the lowest order amplitudes
are bounded in the asymptotic region $\sqrt{s} \rightarrow \infty$ by some constants and reliability of the perturbation
expansion depends on the magnitude of such a limiting value, i.w. whether such a constant is smaller or bigger than $1/2\sqrt{N_{\lambda_a\lambda_b}N_{\lambda_1\lambda_2}}$. Interestingly in the SM, which is a renormalizable QFT, the gauge boson scattering amplitude is limited in the asymptotic region by a constant proportional to the Higgs boson mass -- the larger the $m_h$ the closer it is to the unitarity bound; see Sec.~\ref{sec:VVinSM} for the corresponding discussion.

In non-renormalizable theories the amplitudes usually grow with $\sqrt{s}$ and above some critical energy the perturbation expansion unavoidably breaks down; EFT Lagrangians (non-renormalizable operators) typically generate amplitudes that grow with the energy.
 
Therefore, since we shall investigate gauge boson elastic scattering at tree-level with real three-level amplitudes, we shall apply the perturbative unitarity bound in the form of~\eqref{eq:re}, when examining various effective phenomenological (non-remormalizable) models later on.

The bounds~\eqref{eq:uni26},~\eqref{eq:uni27} and~\eqref{eq:mod}~--~\eqref{eq:uni28} can be further optimized. To see this we first go back to eq.~\eqref{eq:equni22p}, i.e. we assume that the two-particle states $\left|\alpha\right>$ and $\left|\beta\right>$  have the same particle content but both $\lambda_1'=\lambda_1$ and $\lambda_2'=\lambda_2$ are not assumed any more. We promote $\mathcal{T}^{(j)}_{\lambda_1\lambda_2;\lambda_1'\lambda_2'}$ entities to matrix elements of a matrix $\mathcal{T}^{(j)}$:
\begin{equation}
[\mathcal{T}^{(j)}]_{\lambda_1\lambda_2,\lambda_1'\lambda_2'} = \mathcal{T}^{(j)}_{\lambda_1\lambda_2;\lambda_1'\lambda_2'}
\label{eq:uni29}
\end{equation} $(\lambda_1\lambda_2)$ is the first and $(\lambda_1'\lambda_2')$ is the second index. We make this identification in each line of eq.~\eqref{eq:equni22p}. Hence we look at eq.~\eqref{eq:equni22p} as a matrix equation of $(2s_1+1)\times(2s_2+1)$ matrices, where $s_1$ and $s_2$ denote spin quantum number of particles (1) and (2) in $\left|\alpha\right>$. The factors $N_{\tilde{\lambda}_1\tilde{\lambda}_2}$ are also promoted to squares of diagonal matrices $N$:
\begin{equation}
\begin{array}{c}
 \qquad [N]_{\lambda_1\lambda_2,\tilde{\lambda}_1\tilde{\lambda}_2} \equiv \delta_{\lambda_1\lambda_2,\tilde{\lambda}_1\tilde{\lambda}_2} \sqrt{N_{\tilde{\lambda}_1\tilde{\lambda}_2}}\ , \\
\qquad [N_{ab}]_{\lambda_1\lambda_2,\lambda_a\lambda_b} \equiv \delta_{\lambda_1\lambda_2,\lambda_a\lambda_b} \sqrt{N_{\lambda_a\lambda_b}}.
\end{array}
\label{eq:uni30}
\end{equation}
The matrix form of eq.~\eqref{eq:equni22p} reads (the superscript $(j)$ shall be implicit in all matrix equations):
\begin{eqnarray}
-i\left[ \mathcal{T}^\dagger-\mathcal{T}\right]= \qquad\qquad \nonumber\\
=\frac{2}{s}\lambda^{1/2}_{12} \mathcal{T}^\dagger NN\mathcal{T} \nonumber\\
+\sum_{(ab)}\frac{2}{s}\lambda^{1/2}_{ab} \mathcal{T}_{ab}^\dagger N_{ab}^2\mathcal{T}_{ab} \nonumber\\
+\frac{1}{16\pi} \mathcal{T}_\gamma^\dagger \mathcal{T}_\gamma,
\label{eq:uni31}
\end{eqnarray}
where in the last line the matrix $\mathcal{T}_\gamma$ has continuous row index or equivalently the allowed integration region is discretized and the row index runs also over the discretized integration points; the subscripts $(ab)$, $\gamma$ were used to explicitly distinguish different matrix identifications in each line. Now,  we multiply eq.~\eqref{eq:uni31} both from left and right by the matrix $N$. Diagonalizability of the matrix $N\mathcal{T} N$ is implied by diagonalizability of the unitary matrix $S$. We denote the unitary matrix that diagonalizes $N\mathcal{T} N$ by $U_N$:
\begin{equation}
\mathcal{T}_{ND}\equiv U_N^\dagger N\mathcal{T}N U_N,\qquad \mathcal{T}_{ND}\text{ is diagonal}.
\label{eq:uni32}
\end{equation}
Hence, we multiply~\eqref{eq:uni31} from left by $U_N^\dagger$ and from right by $U_N$. The diagonal matrix elements of the last two terms, after the multiplications in eq.~\eqref{eq:uni31}, are positive definite, since they still can be written as matrix multiplications of the type $A^\dagger A$. The left side and the first term on the right side are now diagonal matrices with entries $\mathcal{T}^{(j),k}$, where $k=1,2,\ldots,(2s_1+1)\times(2s_2+1)$. The equation can be rewritten in the form analogous to that of eq.~\eqref{eq:uni24} for the diagonal entries of $\mathcal{T}_{ND}$:
\begin{equation}
\left[\mathrm{Re}\mathcal{T}_{ND}^{(j),k}(s)\right]^2+\left[\mathrm{Im}\mathcal{T}_{ND}^{(j),k}(s)+\frac{s\lambda_{12}^{-1/2}}{2}\right]^2 = \frac{s^2\lambda_{12}^{-1}}{4}-R^2_j(s),
\label{eq:uni33}
\end{equation}
where now the $R^2_j$ involves only the inelastic part, i.e. contributions from the last two lines in eq.~\eqref{eq:uni31}. As previously (eq.~\eqref{eq:uni26},~\eqref{eq:uni27} and eq.~\eqref{eq:mod}~-~\eqref{eq:uni28}), unitarity bounds emerge, where the difference with the previous bounds is that $\mathcal{T}^{(j)}_{\lambda_1\lambda_2;\lambda_1\lambda_2}$ are replaced with $\mathcal{T}_{ND}^{(j),k}$ and there are no explicit $N_{\lambda_1\lambda_2}$ etc. factors; in the limit $s\rightarrow\infty$:
\begin{eqnarray}
\left|\mathcal{T}_{ND}^{(j),k}(s)\right|\leq1,\label{eq:uniDiagMod} \\
\left|\mathrm{Re}\mathcal{T}_{ND}^{(j)}(s)\right|\leq\frac{1}{2},\label{eq:uniDiagRe} \\
-1\leq \mathrm{Im}\mathcal{T}_{ND}^{(j)}(s)\leq 0.
\label{eq:uni28p}
\end{eqnarray}

Compared to~\eqref{eq:uni24} now the $R^2_j$ term in eq.~\eqref{eq:uni33} does not involve elastic part with spin flip, since this part has been diagonalized. Hence, the diagonalization absorbed the elastic part of $R^2_j$. In~\eqref{eq:uni24} where no diagonalization was performed, the spin flip part contributes to $R^2_j$ as visible in eq.~\eqref{eq:uni25}. The larger the $R^2_j$ the more stringent the bounds from eq.~\eqref{eq:uni24} would be, but the bounds in eq.~\eqref{eq:uni26}~-~\eqref{eq:uni28} assumed $R^2_j = 0$. Hence accounting for the elastic spin flip part in the non-diagonalized case, would make the bounds~\eqref{eq:uni26}~-~\eqref{eq:uni28} stronger. Again, since after diagonalization this elastic part disappears, it must accounted for in the entries $\mathcal{T}_{ND}^{(j),k}$. It suggests therefore that the unitarity bounds~\eqref{eq:uniDiagMod}~-~\eqref{eq:uni28p} are stronger than~\eqref{eq:mod}~-~\eqref{eq:im}. We checked that it is indeed true in the same and opposite sign on-shell $WW$ scattering (Sec.~\ref{onshellSMEFT},~\ref{onshellSMEFT} and the Appendices~\ref{app:VVonshellSMEFT},~\ref{app:VVonshellHEFT}).

The bounds on non-elastic partial waves, i.e.~\eqref{eq:uni27} and~\eqref{eq:uni28} can also be further optimized by appropriate diagonalization. To this end we go back to eq.~\eqref{eq:equni22}, i.e. we do not assume that the particle content in $\left|\beta\right>$ and $\left|\alpha\right>$ is the same. We promote both $\mathcal{T}^{(j)}_{\lambda_1\lambda_2;\lambda_1'\lambda_2'}$ and $\mathcal{T}^{(j)}_{\lambda_a\lambda_b;\lambda_1'\lambda_2'}$ to the matrix elements of the same (large) square matrix $\mathcal{T'}^{(j)}$ of dimension: ${\displaystyle \sum_{(ab)}(2s_a+1)(2s_b+1)}$, i.e. the matrix indices run over all two particle states and the particles' helicities. The factors $N_{\lambda_a\lambda_b}$ are also promoted to a square of a diagonal matrix $N'$, as previously. The factors $\frac{1}{s}\lambda^{1/2}(s,m_a^2,m_b^2)$ are to be thought of as absorbed in the matrix $N'$. The integral $d\gamma$ term shall be promoted to a multiplication of a matrix and its Hermitian conjugate, as before; the latter matrix shall be denoted by $\mathcal{T}_\gamma'$. All in all, the  matrix form of eq.~\eqref{eq:equni22} reads:
\begin{eqnarray}
-i\left[ \mathcal{T'}^\dagger-\mathcal{T'}\right]= \qquad\qquad\qquad \nonumber\\
=2 \mathcal{T'}^\dagger N'N'\mathcal{T'} \nonumber\\
+\frac{1}{16\pi} {\mathcal{T}'_\gamma}^{\dagger} \mathcal{T}'_\gamma,
\label{eq:uni34}
\end{eqnarray}
As previously, we multiply eq.~\eqref{eq:uni34} from both left and right by $N'$, from left by ${U'_N}^\dagger$ and from right by $U'_N$, where $U'_N$ is a unitary matrix such that 
\begin{equation}
\mathcal{T'}_{ND}\equiv {U'_N}^\dagger N\mathcal{T'}N' U'_N,\qquad \mathcal{T'}_{ND}\text{ is diagonal}.
\label{eq:uni35}
\end{equation}
obtaining in turn the following equations for the (diagonal) matrix elements $\mathcal{T'}^{(j),k}_{ND}$ of $\mathcal{T'}^{(j)}_{ND}$:
\begin{equation}
\left[\mathrm{Re}\mathcal{T'}_{ND}^{(j),k}(s)\right]^2+\left[\mathrm{Im}\mathcal{T'}_{ND}^{(j),k}(s)+\frac{1}{2}\right]^2 = \frac{1}{4}-R^2_j(s), \qquad k=1,2,\ldots, \sum_{(ab)}(2s_a+1)(2s_b+1).
\label{eq:uni36}
\end{equation}
Again, unitarity bounds emerge: 
\begin{eqnarray}
\left|\mathcal{T'}_{ND}^{(j),k}(s)\right|\leq1, \\
\left|\mathrm{Re}\mathcal{T'}_{ND}^{(j)}(s)\right|\leq\frac{1}{2}, \\
-1\leq \mathrm{Im}\mathcal{T'}_{ND}^{(j)}(s)\leq 0.
\label{eq:uni35p}
\end{eqnarray}
Now, the $R^2_j(s)$ term in eq.~\eqref{eq:uni35p} includes only the inelastic part corresponding to three or more particle states $\left|\gamma\right>$, i.e. diagonalization in the space of all binary reactions and the helicities further optimizes both the unitarity bounds~\eqref{eq:uniDiagMod}~-~\eqref{eq:uni28p} and~\eqref{eq:uni28}. 

In principle, one could proceed further and ultimately diagonalize each $(j)$ whole partial wave. In practice any explicit step further that what presented above is cumbersome, as it requires explicit decomposition of states with more that two particles into partial waves, which is highly non-trivial.
\chapter{Vector boson scattering in the SM}
\label{sec:VVinSM}
% W+ W+ ------------------------------------------------%
In this Section we discuss tree-level on-shell scattering of gauge bosons $VV\rightarrow VV$ in the SM. We focus on the scattering of massive fields: while the massless gauge boson interactions (Yang-Mills) are renormalizable and in particular the perturbative unitarity holds, the mass can be introduced in a renormalizable way, only via a specific mechanism (in the case of SM it is the Higgs mechanism). It implies a special role of the Higgs particle fluctuating in  Feynman diagrams. We illustrate how cancellations between Feynman diagrams of scattering amplitudes take place, leading to fulfilment of perturbative unitarity bounds in the SM. The exchange of $h$ is essential to these cancellations. Also, we study the energy dependence of the on-shell $W^+W^+$ elastic scattering total unpolarized cross section and its polarized components. The latter reaction will be studied phenomenologically in Chapter~\ref{WWscatEFT}. 

A massive gauge boson, has 3 spin states. We choose to work in the helicity basis, hence the polarizations are: right (+), left (-) and longitudinal (0). Therefore one considers an amplitude $iM(ij\rightarrow kl)$ corresponding to a certain helicity configuration $(ijkl)$ of the four V system, $i,j,k,l\in\{+,-,0\}$. There are $3^4=81$ such amplitudes. Tree-level VV scattering implies that fermions do not enter the diagrams. Since the only source of $\mathcal{C}$, $\mathcal{P}$, $\mathcal{T}$ violation are the Yukawa interactions, no violation of the discrete symmetries is present in VV scattering at LO. In general, in VBS the 81 polarizations can be divided into classes that yield the same (polarized) cross sections due to relations between various helicity amplitudes guaranteed by some of these discrete symmetries and/or Bose statistics~\cite{deRham:2017zjm}. Which symmetries can be applied to reduce the number of amplitudes, depends on the reaction. In case of elastic same-sign $WW$ scattering one can apply $\mathcal{P},\mathcal{T}$ and Bose statistics. The number of polarization classes is 13. We choose representatives of each class as follows:
%\begin{table}
\begin{center}
\begin{tabular}{ccccccc}
 $\text{- - - - } $ & $ \text{- - - 0 } $ & $ \text{- - - + } $ & $ \text{- - 0 0 } $ & $ \text{- - 0 + } $ & $ \text{- - + + } $ & $ \text{- 0 - 0 } $ \\
$\text{- 0 - + } $ & $ \text{- 0 0 0 } $ & $ \text{- 0 0 + } $ & $ \text{- + - + } $ & $ \text{- + 0 0 } $ & $ \text{0 0 0 0 } $ & $ \mbox{}$
\end{tabular}
\end{center}
%\end{table}
Each class has a certain number of helicity configurations. For the above classes these numbers read 
%\begin{table}
\begin{center}
\begin{tabular}{ccccccc}
 2 & 8 & 8 & 4 & 8 & 2 & 8 \\ 
16 & 8 & 8 & 4 & 4 & 1, & \mbox{}
\end{tabular}
\end{center}
%\end{table}
respectively. The corresponding multiplicity factors must be accounted for, while computing the cross section.
In case of the opposite-sign $WW$ scattering there are 17 independent helicity configurations. Here, applicable are $\mathcal{C}$,$\mathcal{P}$ and $\mathcal{T}$. We choose the following representatives:
%\begin{table}%
\begin{center}\begin{tabular}{ccccccccc}
 \text{- - - - } & \text{- - - 0 } & \text{- - - + } & \text{- - 0 0 } & \text{- - 0 + } & \text{- - + + } & \text{- 0 - 0 } & \text{- 0 - + } & \text{- 0 0 - } \\
	\text{- 0 0 0 } & \text{- 0 0 + } & \text{- 0 + - } & \text{- 0 + 0 } & \text{- + - + } & \text{- + 0 0 } & \text{- + + - } & \text{0 0 0 0 }. & \mbox{} \\
\end{tabular}\end{center}
%\caption{}
%\label{}
%\end{table}
The corresponding multiplicities read:
%\begin{table}%
\begin{center}\begin{tabular}{ccccccccc}
 2 & 8 & 8 & 4 & 8 & 2 & 4 & 8 & 4 \\
8 & 4 & 8 & 4 & 2 & 4 & 2 & 1. & \mbox{}\\
\end{tabular}\end{center}
%\caption{}
%\label{}
%\end{table}

%\caption{}
%\label{}
%\end{table}
In case of the $W^+W^-\rightarrow ZZ$ scattering, one can apply $\mathcal{C},\mathcal{P}$ and Bose symmetry for the $ZZ$ pair. There are 20 independent helicity configurations:
\begin{center}
\begin{tabular}{ccccccccc}
 \text{- - - - } & \text{- - - 0 } & \text{- - - + } & \text{- - 0 0 } & \text{- - 0 + } & \text{- - + + } & \text{- 0 - - } & \text{- 0 - 0 } & \text{- 0 - + } \\ 
\text{- 0 0 0 }  & \text{- 0 0 + } & \text{- 0 + + } & \text{- + - - } & \text{- + - 0 } & \text{- + - + } & \text{- + 0 0 } & \text{0 0 - - } & \text{0 0 - 0 } \\
 \text{0 0 - + } & \text{0 0 0 0 }.
\end{tabular}
\end{center}
The corresponding multiplicities read:
%\begin{table}%
\begin{center}\begin{tabular}{cccccccccc}
2 & 4 & 4 & 2 & 4 & 2 & 4 & 8 & 8 \\ 
4  & 8 & 4 & 4 & 8 & 4 & 2 & 2 & 4 \\
 2 & 1. 
\end{tabular}
\end{center}
In case of the $W^+Z\rightarrow W^+Z$ scattering, one can apply $\mathcal{P},\mathcal{T}$. There are 25 independent helicity configurations:
\begin{center}
\begin{tabular}{ccccccccc}
 \text{- - - - } & \text{- - - 0 } & \text{- - - + } & \text{- - 0 - } & \text{- - 0 0 } & \text{- - 0 + } & \text{- - + - } & \text{- - + 0 } & \text{- - + + } \\ 
\text{- 0 - 0 } & \text{- 0 - + } & \text{- 0 0 - } & \text{- 0 0 0 } & \text{- 0 0 + } & \text{- 0 + - } & \text{- 0 + 0 } & \text{- + - + } & \text{- + 0 - } \\ 
\text{- + 0 0 } & \text{- + 0 + } & \text{- + + - } & \text{0 - 0 - } & \text{0 - 0 0 } & \text{0 - 0 + } & \text{0 0 0 0 }.
\end{tabular}
\end{center}
The corresponding multiplicities read:
\begin{center}
\begin{tabular}{ccccccccc}
 2 & 4 & 4 & 4 & 4 & 4 & 4 & 4 & 2 \\ 
2 & 4 & 4 & 4 & 4 & 4 & 2 & 2 & 4 \\ 
4 & 4 & 2 & 2 & 4 & 2 & 1 .
\end{tabular}
\end{center}
All the relations between different helicity amplitudes, guaranteed by $\mathcal{C}$ and/or $\mathcal{P}$ and/or $\mathcal{T}$ and/or Bose statistics, have been confirmed by explicit analytical calculations in all the four $VV\rightarrow VV$ reactions. 

Polarizations for a massive vector boson read:
\begin{equation}
\begin{array}{llll}
\epsilon^\mu_- &=& \frac{1}{\sqrt{2}}(0,+1-i,0) &\qquad \mathrm{(left),}\\
\epsilon^\mu_+ &=& \frac{1}{\sqrt{2}}(0,1-i,0) &\qquad \mathrm{(right),}\\
\epsilon^\mu_0 &=& (k,0,0,E)/m &\qquad \mathrm{(longitudinal),}\\
\end{array}
\label{eq:VVscattering}
\end{equation}
in the frame in which the tree-momentum of $k_\mu$ is in the $z$ direction. Hence, each longitudinal polarization introduces $\sqrt{s}$ energy factor in the relativistic limit; $s\equiv (p+p')^2$ is the Mandelstam variable. Hence, if there is no momentum dependence in all the vertices constituting a diagram, then the longitudinal scattering amplitude, i.e. $V_L V_L\rightarrow V_L V_L$, factors as $s^2$ in that diagram. If there are several diagrams for the tree-level amplitude and if there were no cancellations between diagrams, it would lead to perturbative unitarity violation above a certain scale $s^U$ in that process. 
In fact, there are 7 diagrams contributing to $W^+W^+\rightarrow W^+W^+$ at LO: $t$ and $u$ channel exchange of $Z_\mu$, $A_\mu$, $h$ and the contact gauge diagram. 
In particular the contact diagram has no momentum dependence (eq.\eqref{eq:22}) in the SM. Cancellations must take place, which is guaranteed in models with gauge symmetry (also exhibiting SSB) that are constructed with renormalizable operators only~\cite{Cornwall:1974km,LlewellynSmith:1973yud}. In the end, the SM amplitudes behave at most as constants asymptotically in energy, i.e as $\sim s^0$. The case of $W^+W^+\rightarrow W^+W^+$ is illustrated in Tab.~\ref{tab:WWonshell1} in the Appendix, for all the independent helicity combinations.
In the second column shown are would-be amplitudes where only the $t$,$u$ channel $Z_\mu$ and $A_\mu$ exchange diagrams are summed. In the fourth column the contact gauge diagram is in addition accounted for. Only the leading terms in the asymptotic limit $s\rightarrow\infty$ are shown for each helicity amplitude. 
In column third (fifth) shown are the exponents $n$ of the asymptotic behavior $iM\sim s^n$ from column second (fourth), for the reader's convenience.
 
One can see that in the second column many amplitudes growth with energy, at most as $s^2$. The $s^2$ energy dependence is however only the case of pure longitudinal scattering. This growth with energy is considerably reduced after the contact diagram is added, leading to all of the amplitudes to behave at most as constant, except for the longitudinal scattering which still grows as $s^1$. In the last column shown is the behavior after the two Higgs exchange diagrams are added. It further reduces the asymptotic behavior in several helicity configurations and regularizes the longitudinal amplitude behavior. The analytical formulae for the asymptotically leading terms corresponding to the full tree-level amplitudes are presented in Tab.~\ref{tab:WWonshell2}. 

The tables with analogous results for the $W^+W^-\rightarrow W^+W^-$ process are Tab.~\ref{tab:WWonshell3} and~\ref{tab:WWonshell4}. The latter reaction was determined with the amplitude for $W^+W^+$ crossed and the situation in $W^+W^-$ is very similar to the $W^+W^+$ case. Results for $W^+W^-\rightarrow ZZ$ are presented in Tab.~\ref{tab:WWonshell7} and~\ref{tab:WWonshell8}. Similarly, the reaction $W^+Z\rightarrow W^+Z$ was determined via $W^+W^-\rightarrow ZZ$ amplitude crossing and Tab.~\ref{tab:WWonshell9} and~\ref{tab:WWonshell10} correspond to this case.

In all four reactions the perturbative unitarity is preserved. Any shifts of SM couplings would result in spoiling these cancellations, leading to growth of some of the helicity amplitudes with energy. At a certain energy scale perturbative unitarity would be violated, which corresponds to strong interaction regime, i.e. sizeable cross section deviations from the SM predictions. The origin of such anomalous couplings would be new physics effects. This is why VBS is considered sensitive to BSM.

Strictly speaking the fact that in the SM the gauge bosons scattering helicity amplitudes behave at most as constant in the asymptotic region $\sqrt{s}\rightarrow\infty$, does not necessarily imply fulfilment of tree-level unitarity -- in principle one has to check the unitarity bounds, e.g.~\eqref{eq:re}. Since the amplitudes depend on the Higgs mass which appears in numerators of $i\mathcal{M}$ (see tables~\ref{tab:WWonshell2},~\ref{tab:WWonshell4},~\ref{tab:WWonshell8},~\ref{tab:WWonshell10}), the tree-level unitarity can be in principle violated, if the Higgs mass was large enough. To find the corresponding upper bound on $m_h$ we shall analyse now partial wave amplitudes and account for the perturbative unitarity bound from eq.~\eqref{eq:re}. In order to compute for the partial waves it is enough to make use of the partial wave expansion of $\mathcal{M}$ eq.~\eqref{eq:uni12} by performing projection using the property~\eqref{eq:uni8p}:
\begin{equation}
\mathcal{T}^{(j)}_{\lambda_1'\lambda_2';\lambda_1\lambda_2} = \frac{1}{32\pi}\int_0^\pi{d\vartheta\sin\vartheta\mathcal{M}_{\beta\alpha}D^{(j)}_{\lambda_1-\lambda_2;\lambda_1'-\lambda_2'}(\varphi=0,\vartheta)},
\label{eq:VVscattering1p}
\end{equation}
in which the azimuth $d\varphi$ integration has been already accounted for.
The partial waves for the minimal value of $j$ for all helicity combinations in: $W^+W^+\rightarrow W^+W^+$, $W^+W^-\rightarrow W^+W^-$, $W^+Z\rightarrow W^+Z$, $W^+W^-\rightarrow ZZ$ are presented in Tab.~\ref{tab:WWonshell11},~\ref{tab:WWonshell12},~\ref{tab:WWonshell13} and~\ref{tab:WWonshell14}, respectively. Again, only the leading in $\sqrt{s}$ in the limit $s\rightarrow \infty$ terms are shown. Moreover, the results on the partial waves correspond to subtraction, prior to the integration, of the Coulomb singularities if and only if such singularity would result in singular integration -- infinite range interactions cannot be expanded into partial waves; formally each partial wave is infinite in such case. However the subtraction is justified -- one could turn off the electromagnetic interactions by going to $g'\rightarrow 0$ and examine perturbative unitarity fulfilment by the other forces; the subtraction of the Coulomb poles is an alternative to considering the $g'\rightarrow0$ limit. 

From the tables one can see that in all four reactions the fully longitudinal partial waves approach a constant at $\sqrt{s}\rightarrow\infty$ and simultaneously have $m_h$ in their numerators. Hence at large enough Higgs mass, for which the tree-level unitarity is already violated, perturbative calculations stop making sense (at any $\sqrt{s}$). Using the perturbative unitarity bounds from eq.~\eqref{eq:re} and~\eqref{eq:uni28} one obtains the following upper bounds on the Higgs mass:
\begin{equation*}
\begin{array}{|c|c|c|c|c|}\hline
&W^+W^+& W^+W^-& W^+W^-\rightarrow ZZ & W^+Z\rightarrow W^+Z \\ \hline
m_h\lesssim M,\qquad M\mathrm{[GeV]} = & 1236. & 869. & 1462.  & 1235. \\\hline
\end{array}
\end{equation*} 
The common bound is around 1 TeV, which determined the investigated Higgs boson mass range in the electroweak precision indirect Higgs searches (Fig.~\ref{fig:precisionEW}). Since $m_h= \sqrt{2\lambda}v$, one can translate the above bounds into the bounds of the scalar self-interaction coupling $\lambda$; these read: 
\begin{equation*}
\begin{array}{|c|c|c|c|c|}\hline
&W^+W^+& W^+W^-& W^+W^-\rightarrow ZZ & W^+Z\rightarrow W^+Z \\ \hline
\lambda \lesssim X,\qquad X = & 12.64 & 6.25 & 17.66  & 12.62 \\\hline
\end{array}
\end{equation*} 
Interestingly the common upper bound from tree-level unitarity is around $O(10)$ which is $>>1$. 

\begin{figure} 
\begin{center}	
\includegraphics[width=0.8\textwidth]{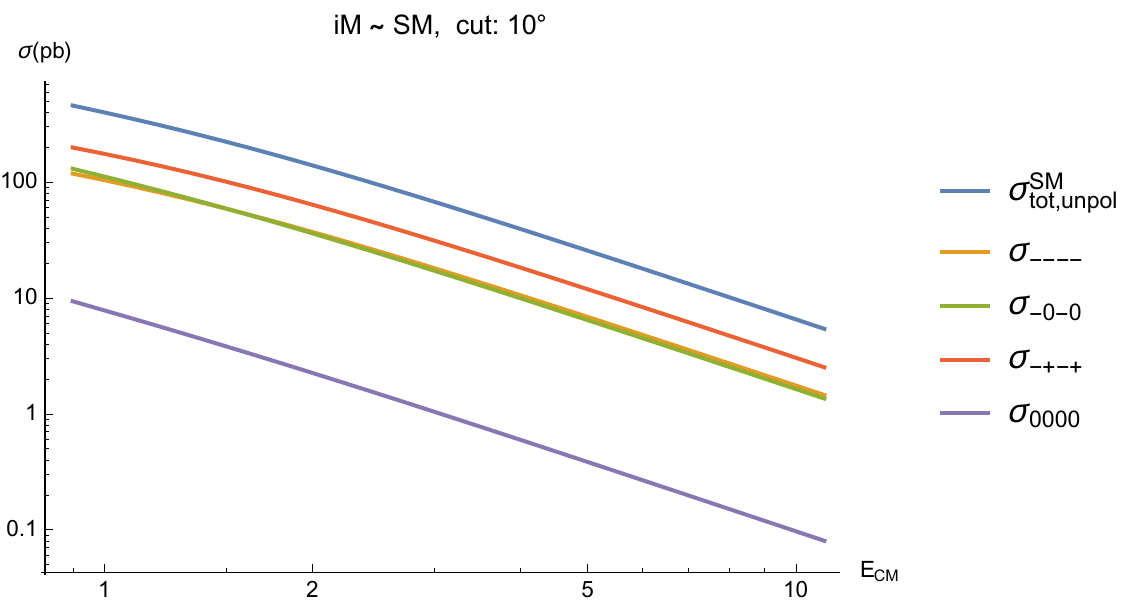}
\caption{Contributions of different helicities (multiplicity taken into account) to the total unpolarized cross section as a function of the center-of-mass collision energy ($E_{CM} \equiv \sqrt{s}$, in TeV) in the SM.  The total cross section is shown in blue.}
\label{fig:plotlistPlot10DegreeParticPolsSMOnly}
\end{center}
\end{figure}

The total unpolarized on-shell $VV\rightarrow VV$ cross section reads schematically:
\begin{equation}
\sigma\sim\frac{1}{9}\ \sum_{i,j,k,l}\ \ \left|A(ij\rightarrow kl)\right|^2,
\label{eq:VVscattering1}
\end{equation}
where $A$ is the scattering amplitude.
There are orders of magnitude differences concerning contributions of different helicities to~\eqref{eq:VVscattering1}. Only a few helicity configurations contribute non-negligibly at high VV energy ($M_{VV}$). We refer to such helicities as saturating helicities. The total unpolarized cross section decomposition into the saturating helicities in the case of the SM for the $W^+W^+\rightarrow W^+W^+$ reaction is shown in Fig.~\ref{fig:plotlistPlot10DegreeParticPolsSMOnly}. All cross sections are computed with a $10^{\circ}$ cut in the forward and backward scattering regions, in order to regularize the Coulomb singularity. The four saturating helicity configurations visible in Fig.~\ref{fig:plotlistPlot10DegreeParticPolsSMOnly} are the only ones whose scattering amplitude is asymptotically constant in energy. The remaining helicity configurations behave asymptotically at most as $1/s$, hence their suppression at large $M_{WW}$. The contributions of various helicity combinations to the total unpolarized $W^+W^+$ cross section at $\sqrt{s}=1$ TeV are shown numerically in the last column in Tab.~\ref{tab:WWonshell2} (in pb). The cross section is dominated by fully transverse configurations $W_TW_T\rightarrow W_T W_T$, while the fully longitudinal component constitutes about $2\%$. 
\clearpage
%----------------------------------------------------------------%

\chapter{Effective Field Theory approach}
\label{sec:EFT}
In this Chapter we discuss the most important features of the EFT approach by discussing the two different EFTs that have been constructed to parametrize possible BSM effects. They correspond to two different directions of thoughts on the EW symmetry breaking mechanism (EWSB), that have emerged in the last decades. According to the first one, the EWSB mechanism is due to the linearly realised
dynamics of the Higgs sector such as in the SM. The alternative consists in a non-linearly realised dynamics of the Higgs sector, typically occurring in the so-called Composite Higgs (CH) models. Both approaches are however based on the SM gauge symmetry $G_{SM}$ and the gauge symmetry breaking pattern in eq.~\eqref{eq:eq2}. These two distinct EFT approaches to indirect BSM searches are introduced in Sec.~\ref{SMEFT} and~\ref{HEFT}.

We put the EFT approach into better known perspective by discussing first the interesting analogy with the relation between QED and SM:
\begin{displaymath}
\begin{array}{ccc}
	\mathrm{QED} & \longrightarrow & \mathrm{SM} \\
	\mathrm{SM} &\longrightarrow & \mathrm{?}
\end{array}
\end{displaymath}
QED is a renormalizable QFT which gives precise predictions for electromagnetic processes (at low energy). But it is only an effective theory, low energy approximation to the SM. Hence we know that its predictions disagree with experiment at the level $\sim O(E/m_W)$, where $E$ is the characteristic energy for a given process. Starting from SM we can derive effective QED by decoupling heavy degrees of freedom, i.e. $W$ and $Z$ bosons, obtaining:
\begin{itemize}
	\item the renormalizable $\mathcal{L}_{QED}$, 
	\item and corrections: non-renormalizable operators, which give corrections to electromagnetic processes of order $O(E/m_{W,Z})$.
\end{itemize} 
Importantly, these corrections are invariant under the local $U(1)_{EM}$. 

\begin{figure} 
\begin{center}	
\includegraphics[width=0.4\textwidth]{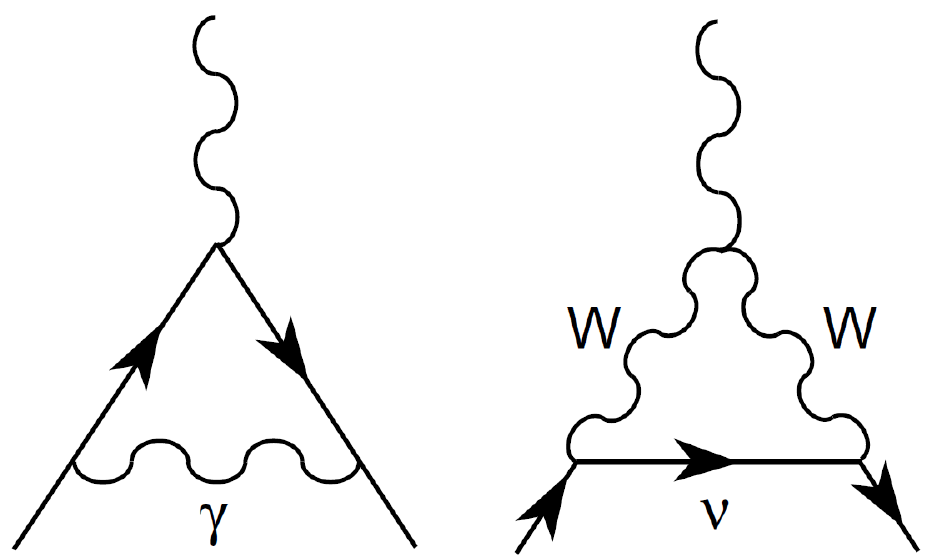}
\caption{One loop contributions to the anomalous lepton magnetic moment in the SM.}
\label{fig:EFT1}
\end{center}
\end{figure}

For example, one can consider the lepton magnetic moment. It gets one loop contributions from the diagrams depicted in Fig.~\ref{fig:EFT1}. 
Thus, for the nonrelativistic effective interaction with the magnetic moment one obtains:
\begin{equation}
{\cal H}_{eff} = \frac{e}{2m_l} {\boldsymbol\sigma}\cdot\mathbf{B}(1+\frac{\alpha}{2\pi}+{\cal O}\left(\alpha\frac{m_l^2}{M_W^2}\right)+\ldots),
\label{eq:EFT1}
\end{equation}
where the role of the energy scale is played by the lepton mass $m_l$. The weak correction is calculable in the full electroweak theory, but at the level of QED as an effective theory, it has to be added as a new non-renormalizable interaction:
\begin{equation}
{\cal L}_{QED}^{eff} = {\cal L}_{QED}+ {\frac{m_l}{m_W^2} \bar{e}\,\sigma_{\mu\nu}\,e\, F^{\mu\nu}} + \ldots
\label{eq:EFT2}
\end{equation}
This would have been a way to discover weak interactions (and to measure the weak scale) in purely electromagnetic processes: one extends QED to a non-renormalizable theory by adding higher dimension operators and looks for their experimental manifestation in purely electromagnetic processes once the experimental precision is high enough. Such approach shall be referred to as the bottom-up approach. The example illustrates the true usefulness and aim of the bottom-up EFT approach: to learn something about the BSM physics, i.e. its couplings and its scale $\Lambda$ prior to discovering heavy particles directly, in case of a considerable mass gap. The effective couplings (usually dimensionless) in front of the operators parametrize then our ignorance about the high-energy dynamics in a model-independent way, i.e. only certain symmetries are postulated in the effective theory. 

In general, one can distinguish two types of higher dimension operators constituting the effective QED: the ones that respect (e.g.~\eqref{eq:EFT2}) or violate the conservation of quantum numbers that are accidentally conserved in QED (global symmetries), for example the flavour number \emph{of the charged lepton}. The corrections that violate the flavor number of the charged lepton, manifest themselves as weak interactions -- in the form of non-renormalizable four fermion operators -- example of which can be obtained if the tree-level amplitude for the process $\mu^-\rightarrow e^-\bar{\nu}_e\nu_\mu$ is analysed. The latter process is governed by charge current weak interactions described by eq.~\eqref{eq:84}:
\begin{equation}
A = \left(\frac{-ig}{\sqrt{2}}\right)^2(\bar{\nu}_\mu\gamma^\mu P_L\mu)(\bar{e}\gamma^\nu P_L \nu_e)\left(\frac{-ig_{\mu\nu}}{p^2-m_W^2}\right),
\label{eq:SMEFT7}
\end{equation}
where $P_L$ is the left chirality projection and $p^\mu$ is the four-momentum transfer carried by the virtual $W$ propagator. The latter is of order $m_\mu$, hence $p << m_W$. Under such circumstances the propagator can be expanded:
\begin{equation}
\frac{1}{p^2-m_W^2}=-\frac{1}{m_W^2}\left(1+\frac{p^2}{m_W^2}+\frac{p^4}{m_W^4}+\ldots\right).
\label{eq:SMEFT8}
\end{equation}
The effective description emerged -- after expansion, each term in~\eqref{eq:SMEFT8} corresponds to a non-renormalizable operator: the first term contributes to the amplitude as
\begin{equation}
A = \frac{i}{m_W^2}\left(\frac{-ig}{\sqrt{2}}\right)^2 (\bar{\nu}_\mu\gamma^\mu P_L\mu)(\bar{e}\gamma^\nu P_L \nu_e) + O\left(\frac{1}{m_W^4}\right),
\label{eq:SMEFT9}
\end{equation}
and the same amplitude would be produced by a $D=6$ operator of the form:
\begin{equation}
-\frac{g^2}{m_W^2} (\bar{\nu}_\mu\gamma^\mu P_L\mu)(\bar{e}\gamma^\nu P_L \nu_e).
\label{eq:SMEFT10}
\end{equation}
The above current-current interaction constitutes part of the $\mathcal{L}^{eff}_{QED}$. We introduce notation in which $\tilde{\mathcal{L}}_6$ denotes sum of the D=6 operators, $\tilde{\mathcal{L}}_8$ denotes sum of the D=8 operators, etc. Then,
\begin{equation}
\tilde{\mathcal{L}}_6\supset-g^2 (\bar{\nu}_\mu\gamma^\mu P_L\mu)(\bar{e}\gamma^\nu P_L \nu_e), \qquad \mathcal{L}^{eff}_{QED} = \mathcal{L}_{D\leq4} + \frac{\tilde{\mathcal{L}}_{6}}{m_W^2} + \ldots
\label{eq:SMEFT11}
\end{equation}
 The region of validity of $\mathcal{L}_{EFT}$ is for energies below $m_W$, otherwise the expansion~\eqref{eq:SMEFT8} does not make sense. The rest of $\tilde{\mathcal{L}}_6$ is obtained by accounting for the remaining tree-level $W$ and $Z$ boson exchange processes.

Historically the bottom-up approach to weak interactions was first taken to successfully describe the neutron beta decay, long before the SM with its electroweak unification was known~\cite{Fermi:1933jpa}. At the level of the effective QED, the $u\rightarrow d$ transition is governed by the effective operator:
\begin{equation}
-\frac{4G_F}{\sqrt{2}}V_{ud}(\bar{d}\gamma^\mu P_L u)(\bar{e}\gamma^\mu P_L \nu_e).
\label{eq:SMEFT12}
\end{equation} 

Since no new massive particles had been observed in the data at that time, EFT provided an adequate framework to describe the decay. The successful low-energy theory of weak interactions may serve as a proof of principle of efficiency of the bottom-up approach. 

\section{SMEFT}
\label{SMEFT}
We now treat the SM as a low energy effective theory of some unknown more fundamental one, where there exist new degrees of freedom at the scale $\Lambda$ (analogue of $m_W$ in the effective QED example), and their existence is manifested at the $E<< \Lambda$ as higher dimensional operators that supplement the SM. 

The simplest possibility to construct EFT for the SM particle content is to assume that the 125 GeV particle belongs in the deeper theory, as it is in the SM, to the scalar doublet. The doublet transforms linearly under $G_{SM}$. The EW symmetry is then realized linearly on this scalar doublet. 
This approach is called SM Effective Field Theory (SMEFT). 
In principle, the allowed series of higher-dimension operators is infinite:
\begin{equation}
\mathcal{L}_{SMEFT}\equiv \mathcal{L}_{SM} + \Delta\mathcal{L}\equiv \mathcal{L}_{SM}+ \sum_{D>4}\sum_i f^{(D)}_i\mathcal{O}_i^{(D)},
\label{eq:SMEFT1}
\end{equation}
where $\mathcal{O}_i^{(D)}$ denotes the operators of canonical dimension $D$ that are allowed by SM gauge symmetry. The basic building blocks of the operators are the SM fields, including the scalar doublet that subsequently triggers SSB. Hence the SMEFT is constructed for the unbroken phase.
The following identification holds for the effective (Wilson) coefficients $f_i$:
\begin{equation}
f_i = \frac{c^{(D)}_i}{\Lambda^{D-4}},
\label{eq:SMEFT2}
\end{equation}
$c^{(D)}_i$ are dimensionless quantities, to be identified -- up to normalization choices\footnote{The physically well justified normalization in which the constants $c_i$ have the most straightforward interpretation in terms of BSM couplings is discussed in Sec.~\ref{NDA}} for the operators $\mathcal{O}_i$ -- with combinations of couplings present in the underlying deeper QFT that embeds the SM. Usually $\Lambda$ is identified with a mass scale of new particles. For the following discussion it is convenient to combine all operators of dimension $D$ in the $\tilde{\mathcal{L}}_D$:
\begin{eqnarray}
\mathcal{L}_{SMEFT}&= &\mathcal{L}_{SM} +\sum_{D>4}\sum_i \frac{c^{(D)}_i\mathcal{O}_i^{(D)}}{\Lambda^{D-4}}\equiv \mathcal{L}_{SM} + \sum_{D>4}\frac{\tilde{\mathcal{L}}_D}{\Lambda^{D-4}} \nonumber  \\ \mbox{} \\  
&=&  \mathcal{L}_{SM} + \frac{\tilde{\mathcal{L}}_5}{\Lambda} +  \frac{\tilde{\mathcal{L}}_6}{\Lambda^{2}} + \ldots
\label{eq:SMEFT3}
\end{eqnarray}
In particular, the fact that $\mathcal{L}_{SM}$ constitutes very good physics description, together with no new heavy states present at currently available energies suggests, that there is a large enough gap between the EW scale $v$ and some scale $\Lambda$, $v<<\Lambda$, because the larger the gap the larger the suppression of the effects of $D>4$ effective operators; we justify the latter statement below.
 
Let's consider a scattering amplitude $A$ for a binary reaction. On the one hand, the amplitude is dimensionless. On the other, single insertion of an effective operator of dimension $D>4$ gives a contribution to $A$ of order:
\begin{equation}
A\sim \frac{1}{\Lambda^{D-4}}.
\label{eq:SMEFT4}
\end{equation}
By dimensional analysis the remaining dimensions must be produced by some kinetic factors -- the amplitude must remain dimensionless:
\begin{equation}
A\sim \left(\frac{p}{\Lambda}\right)^{D-4},
\label{eq:SMEFT5}
\end{equation}
where $p$ denotes energy scale of the process. Therefore, in general an insertion of a set of $D>4$ operators in a diagram, leads to a contribution to  an amplitude
\begin{equation}
A\sim \left(\frac{p}{\Lambda}\right)^{n},
\label{eq:SMEFT5p}
\end{equation}
where 
\begin{equation}
n = \sum_i (D_i - 4 ); 
\label{eq:SMEFT6}
\end{equation}
the sum over $i$ is over the inserted operators into a given diagram. It holds for any diagram, tree or loop~\cite{Manohar:2018aog}. The power counting formula~\eqref{eq:SMEFT5p} tells how to systematically organize the calculation in QFT. The leading order is governed by the renormalizable part. The $p/\Lambda$ corrections are given by a single insertion of $\tilde{\mathcal{L}}_5$. The $(p/\Lambda)^2$ corrections are given by diagrams with a single insertion of $\tilde{\mathcal{L}}_6$ or two insertions of $\tilde{\mathcal{L}}_5$, etc. Hence in effective QFT $p/\Lambda$ plays the role of an expansion parameter. Obviously it make only sense to use the EFT expansion in $p/\Lambda$ as long as $p<\Lambda$. In fact, the SMEFT is renormalizable order by order in the expansion in $1/\Lambda$: independently
from the number of loops of a given diagram, the quantum corrections calculated
from the Lagrangian truncated at some order of $1/\Lambda$ and calculated to the same order
in $1/\Lambda$ can be renormalised by the couplings present in the effective Lagrangian, i.e. all the necessary counterterms are already present.  The effective Lagrangian is systematically expandable in the canonical dimension $D$, or equivalently in powers of $1/\Lambda$. 

Determination of a complete and non-redundant set (basis) of higher-dimension operators in SMEFT is highly non-trivial. For example, two operators can differ by equations of motion. Such operators are equivalent and only one is to be included in the basis. There are more such constraints, e.g. Fierz identities. The basis of $D\leq6$ operators has been found relatively recently~\cite{Grzadkowski:2010es}. Several bases, more suitable in various contexts, have been also developed~\cite{Falkowski:2015wza}. The number of independent operators in any basis is the same. Below $B$ and $L$ denote baryon and lepton quantum numbers. Assuming only one generation of quark and leptons, there are: 76 operators of $D=6$ that preserve $B$ and $L$ ($\Delta B=\Delta L = 0$), 8 $D=6$ operators that have $\Delta B = \Delta L = \pm 1$, and 2 operators of $D=5$ that have $\Delta L = \pm2$. Assuming 3 generations, the number of operators increases considerably: it is 2499, 546, 12 respectively. Hermitian conjugates are treated as independent operators in the above counting. Hence the counting is equivalent to counting the number of all free real parameters appearing in front of all non-renormalizable operators of $D\leq6$.

\section{HEFT}
\label{HEFT}
We discuss now the EFT based on the non-linear approach to the EW symmetry (HEFT). The construction and possible physical interpretations in the deeper theory will be illustrated on two examples: first, the so-called linear sigma model which will serve as a toy-model and second, the description of interactions between pseudoscalar mesons in QCD (for comprehensive reviews and/or lecture notes see e.g.~\cite{Pich:1998xt,Pich:1995bw,Pich:2018ltt,Scherer:2005ri}; here we highlight the most important aspects). Both examples are motivated by the fact that they feature the same global SSB pattern as in the SM. The latter aspect is discussed in sec~\ref{custodial}. The HEFT basis is discussed afterwards in sec~\ref{heftBasis}. 
\subsection{The linear sigma model} 
\label{linearSigmaModel}
We consider a scalar 4-plet $\Phi^T\equiv (\pi_1,\pi_2,\pi_3,\sigma)\
\equiv (\vec{\pi},\sigma)$ described by a $SO(4)$ symmetric Lagrangian:
\begin{equation}
\mathcal{L}_\sigma = \frac{1}{2}\partial_\mu\Phi^T\partial^\mu\Phi - \frac{\mu^2}{2}\Phi^T\Phi - \frac{\lambda}{4}\left(\Phi^T\Phi\right)^2. 
\label{eq:HEFT1}
\end{equation}
The $\mu^2$ sign choice: $\mu^2<0$ triggers SSB; $-\mu^2=2\lambda v^2$. Adopting the vacuum choice
\begin{equation}
\left<0\right|\sigma\left|0\right>= v,\qquad\left<0\right|\vec{\pi}\left|0\right>= 0
\label{eq:HEFT2p}
\end{equation}
and parametrizing the fluctuations around the vacuum $\hat{\sigma}=\sigma-v$ the Lagrangian reads:
\begin{equation}
\mathcal{L}_\sigma = \frac{1}{4}\left(\partial_\mu\hat{\sigma}\partial^\mu\hat{\sigma}-2\lambda v^2\hat{\sigma}^2+\partial_\mu\vec{\pi}\partial^\mu\vec{\pi}\right) -\lambda v\hat{\sigma}\left(\hat{\sigma}^2+\vec{\pi}^2\right) -\frac{\lambda}{4}\left(\hat{\sigma}^2+\vec{\pi}^2\right)^2.
\label{eq:HEFT2}
\end{equation}
The vacuum is invariant under $SO(3)$ rotations over the $\vec{\pi}$ fields. Hence, the SSB pattern is
\begin{equation}
SO(4)\rightarrow SO(3).
\label{eq:HEFT6p}
\end{equation} The example is very similar to point \textbf{B} in Sec.~\ref{subsub:SSB}, except now the choice is $SO(4)$ instead of $SO(3)$. The fields $\vec{\pi}$ are the three Goldstone bosons associated with the broken generators. 

We shall rewrite the scalar 4-plet in a form of a $2\times2$ matrix
\begin{equation}
\Sigma(x)\equiv \sigma(x) I_2 +i\vec{\tau}\vec{\pi}(x),
\label{eq:HEFT3}
\end{equation}
where $I_2$ is the identity matrix; $\mathcal{L}_\sigma$ can be rewritten as follows:
\begin{equation}
\mathcal{L}_\sigma = \frac{1}{4}\left<\partial_\mu\Sigma^\dagger\partial^\mu\Sigma\right>-\frac{\lambda}{16}\left(\left<\Sigma^\dagger\Sigma\right> - 2v^2\right)^2,
\label{eq:HEFT4}
\end{equation}
where $\left<A\right>$ denotes trace of the matrix $A$. In this notation $\mathcal{L}_\sigma$ is explicitly invariant under global $G\equiv SU(2)_L\times SU(2)_R$ defined as
\begin{equation}
\Sigma\stackrel{G}{\rightarrow}g_L\Sigma g_R^\dagger,\qquad g_{L,R}\in SU(2)_{L,R}.
\label{eq:HEFT5}
\end{equation}
However the vacuum choice is $\left<0\right|\Sigma\left|0\right>=vI_2$. It means the vacuum is invariant only under those transformations in eq.~\eqref{eq:HEFT5} satisfying $g_L=g_R$. This subgroup shall be referred to as the diagonal subgroup and denoted as $H\equiv SU(2)_{L+R}$. Therefore the pattern of symmetry breaking in~\eqref{eq:HEFT4} reads:
\begin{equation}
SU(2)_L\times SU(2)_R\longrightarrow SU(2)_{L+R}.
\label{eq:HEFT6}
\end{equation}
The physics content does not depend on the choice of fields redefinitions (see~\cite{Criado:2018sdb} and references therein). The symmetry breaking patterns in eq.~\eqref{eq:HEFT6} and~\eqref{eq:HEFT6p} must be equivalent, which reflects the well known equialence of the corresponding groups.

The field $\Sigma$ can be rewritten using the polar decomposition:
\begin{equation}
\Sigma = (v+S(x))U(\vec{\phi}),\qquad U(\vec{\phi}) =\exp\left\{i\frac{\vec{\tau}}{v}\vec{\phi}(x)\right\},
\label{eq:HEFT7}
\end{equation}
where $S,\vec{\phi}$ are real fields. The transformations in eq.~\eqref{eq:HEFT5} are realized on these fields as follows:
\begin{equation}
S\stackrel{G}{\longrightarrow} S,\qquad U(\vec{\phi})\stackrel{G}{\longrightarrow}g_L U(\vec{\phi}) g_R^\dagger,\qquad g_{L,R}\in SU(2)_{L,R},
\label{eq:HEFT8}
\end{equation}
i.e. the matrix $U$ inherits the transformations of $\Sigma$. While the matrix $U$ transforms linearly, obviously the fields $\vec{\phi}$ in the exponent transform non-linearly. The Lagrangian takes the form
\begin{equation}
\mathcal{L}_\sigma = \frac{v^2}{4}\left(1+\frac{S}{v}\right)^2\left<\partial_\mu U^\dagger\partial^\mu U\right>+\frac{1}{2}\left(\partial_\mu S \partial^\mu S-m_S^2 S^2\right) - \frac{m_S^2}{2v} S^3 - \frac{m_S^2}{8v^2}S^4.
\label{eq:HEFT9}
\end{equation}
The fields $\vec{\phi}$ interact only through derivatives and are massless. They are the three Goldstone boson fields associated with the three generators broken. The remaining symmetry $SU(2)_{L+R}$ is realized on the $\vec{\phi}$ fields linearly -- after expansion of $U$ fields, each term in $\left<\partial_\mu U^\dagger\partial^\mu U\right>$ is invariant under $SU(2)_{L+R}$ with $\vec{\phi}$ forming an $SU(2)$ triplet. 

The massive field S can be integrated out, leading to an effective theory of Goldstone bosons interactions. The larger the mass $m_S$ the more useful the EFT. At the leading order the low-energy Lagrangian of the $\vec{\phi}$ Goldstones reduces to:
\begin{equation}
\mathcal{L}_2 = \frac{v^2}{4}\left<\partial_\mu U^\dagger\partial^\mu U\right>.
\label{eq:HEFT9p}
\end{equation}
The sub-leading term is generated by tree-level exchange of S. It reads:
\begin{equation}
\mathcal{L}_4 = \frac{v^2}{8m_S^2}\left<\partial_\mu U^\dagger\partial^\mu U\right>^2
\label{eq:HEFT10}
\end{equation}
and is suppressed by a factor $p^2/M^2$ with respect to~\eqref{eq:HEFT9p}; any term in the series generated by integrating out S is solely constructed in terms of the matrices U and its derivatives. 

Remarkably, parametrization of the Goldstone triplet by the matrix U in eq.~\eqref{eq:HEFT7} with its transformation property~\eqref{eq:HEFT8} is a direct consequence of the SSB pattern~\eqref{eq:HEFT6} -- it is universal and applies to any physics scenario that features the SSB pattern~\eqref{eq:HEFT6}~\cite{Coleman:1969sm,Callan:1969sn}. In particular any effective theory of Goldstone interactions, provided the SSB pattern~\eqref{eq:HEFT6}, is constructable on the same matrix U and its derivatives. Of course, depending on the UV completion, differences appear at the level of effective coefficient values and interpretation of the Goldstons. Moreover, the derivatives of the matrix $U$ maybe covariant if the global symmetry is (partially) gauged. In the example above the UV completion is the Lagrangian~\eqref{eq:HEFT1}, which is renormalizable with a linearly realized $SO(4)$ on a scalar 4-plet, or equivalently $SU(2)_L\times SU(2)_R$ symmetry on a bi-doublet $\Sigma$. Nonetheless, the same effective Lagrangian applies to any other scenario with the same SSB pattern. For example it has been applied successfully to the QCD $\pi^0,\pi^\pm$ mesons dynamics, where the SSB is non-perturbative and the UV completion is QCD. The latter example is briefly described in sec~\ref{chiral}.

Hence, the bottom-up description of Goldstone dynamics corresponding to SSB pattern~\eqref{eq:HEFT6} at low-energies requires the most general Lagrangian involving U, invariant under transformations~\eqref{eq:HEFT8}. 
It is straightforwardly generalized to the case of the following symmetry breaking pattern 
\begin{equation}
SU(n_f)_L\times SU(n_f)_R \longrightarrow SU(n_f)_{L+R},
\label{eq:HEFT11p}
\end{equation}
i.e. when $n_f\geq2$. The U field again transforms as in~\eqref{eq:HEFT8}, only now $g_L\in SU(n_f)_L$, $g_R\in SU(n_f)_R$. The  form of the exponent inside $U$ depends on $n_f$. For $n_f=2$ it has already been introduced:
\begin{equation}
U=\exp\left\{i\frac{\tau^a\phi^a}{f}\right\},\qquad (a=1,2,3),
\label{eq:HEFT11b}
\end{equation}
while for $n_f=3$ it reads
\begin{equation}
U=\exp\left\{i\frac{\lambda^a\phi^a}{f}\right\},\qquad (a=1,\ldots,8),
\label{eq:HEFT11c}
\end{equation}
where $f$ denotes the scale of spontaneous breaking of the global symmetry; $\lambda^a$ are the Gell-Mann matrices.
We shall refer to the symmetry breaking pattern~\eqref{eq:HEFT11p} as the chiral symmetry breaking ($\chi SB$). Due to derivative interactions, Goldstones scattering amplitudes vanish in the limit $p\rightarrow 0$, where $p$ represents typical external Goldstone momenta. Correspondingly, the EFT of Goldstones, the so-called chiral EFT, is organized by the number of derivatives, e.g. operator~\eqref{eq:HEFT9p} is LO and~\eqref{eq:HEFT10} is NLO in the chiral expansion. Assuming parity conservation the number of derivatives in each term is even; the chiral EFT reads schematically:
\begin{equation}
\mathcal{L}_{EFT}(U) = \sum_{2n}\mathcal{L}_{2n}
\label{eq:HEFT11}
\end{equation}
where $2n$ denotes the number of derivatives. The zero-derivative term is trivial: $U^\dagger U=I$. Hence the expansion starts with two derivatives. There is only one structure available at LO:
\begin{equation}
\mathcal{L}_2 = \frac{f^2}{4}\left<\partial_\mu U^\dagger \partial^\mu U\right>.
\label{eq:HEFT12p}
\end{equation}
The normalization is chosen such as to recover canonical Goldstone kinetic terms. At NLO, depending of $n_f$ different number of independent terms appear. For $n_f=3$ one has:
\begin{equation}
\mathcal{L}_4^{SU(3)} = c_1 \left<\partial_\mu U^\dagger\partial^\mu U\right>^2 + c_2  \left<\partial_\mu U^\dagger\partial_\nu U\right> \left<\partial^\mu U^\dagger\partial^\nu U\right> + c_3 \left<\partial_\mu U^\dagger\partial^\mu U \partial_\nu U^\dagger\partial^\nu U\right>,
\label{eq:HEFT12}
\end{equation}
while for $n_f=2$ one has:
\begin{equation}
\mathcal{L}_4^{SU(2)} = c_1 \left<\partial_\mu U^\dagger\partial^\mu U\right>^2 + c_2  \left<\partial_\mu U^\dagger\partial_\nu U\right> \left<\partial^\mu U^\dagger\partial^\nu U\right>.
\label{eq:HEFT13}
\end{equation}

Concerning renormalizability: the Goldstones scattering diagram $\Gamma$ scales as $p^{d_\Gamma}$, where 
\begin{equation}
d_\Gamma = 2L+2+\sum_d(d-2)N_d,
\label{eq:HEFT14}
\end{equation}
in which $L$ denotes number of loops and $N_d$ number of vertices of order $O(p^d)$. The property goes under the name of the Weinberg power-counting theorem~\cite{Weinberg:1978kz}. Thus, each loop increases the momentum power suppression by two units. This establishes a crucial power counting that allows to organise the loop expansion as a low-energy expansion in powers of momenta. The leading $O(p^2)$ contributions are obtained with $L = 0$ and $N_{d>2} = 0$. Therefore, at LO one must only consider tree-level diagrams with $\mathcal{L}_2$ insertions. At $O(p^4)$, one must include tree-level contributions with a single insertion of $\mathcal{L}_4$ $(L = 0, N_4 = 1, N_{d>4} = 0)$ plus any number of $\mathcal{L}_2$ vertices, and one-loop graphs with the LO Lagrangian only $(L = 1, N_{d>2} = 0)$. The $O(p^6)$ corrections would involve tree-level diagrams with a single insertion of $\mathcal{L}_6$ $(L = 0, N_4 = 0, N_6 = 1, N_{d>6} = 0)$, one-loop graphs with one insertion of $\mathcal{L}_4$ $(L = 1, N_4 = 1, N_{d>4} = 0)$ and two-loop contributions from $\mathcal{L}_2$ $(L = 2, N_{d>2} = 0)$. 

The ultraviolet loop divergences need to be renormalized. This can be done order by order in the momentum expansion, thanks to the Weinberg's power-counting. Adopting a regularization that preserves the symmetries of the Lagrangian, such as dimensional regularization, the divergences generated by the loops have a symmetric local structure and the needed counterterms necessarily correspond to operators that are already included in the effective Lagrangian, because $\mathcal{L}_{EFT}(U)$ contains by construction all terms permitted by the symmetry. Therefore, the loop divergences can be reabsorbed through a renormalization of the corresponding effective coefficients, appearing at the same order in momentum.

The chiral EFT is similar to the EFT with $1/\Lambda$ expansion, e.g. SMEFT. While the latter is renormalized order by order in $1/\Lambda$ expansion, the former is renormalized order by order in $p$ expansion. The technical difference is that in SMEFT, at any fixed order in $1/\Lambda$, arbitrary loop diagrams (unlimited $D\leq4$ vertices insertions) are renormalizable. On the other hand in the chiral EFT the number of loops is limited at each fixed order in $p$. 
\subsection{Chiral perturbation theory in QCD}
\label{chiral}
In this Section we discuss $\chi SB$ in QCD. The QCD Lagrangian is defined in eq.~\eqref{eq:22}. In the massless limit and collecting $\Psi_f$ in a flavor-space multiplet $q^T=(\Psi_u,\Psi_d,\ldots)$ the Lagrangian reads:
\begin{equation}
\mathcal{L}^0_{QCD}=-\frac{1}{4}G^a_{\mu\nu}G_a^{\mu\nu}+i\bar{q}_L\gamma^\mu D_\mu q_L + i \bar{q}_R\gamma^\mu D_\mu q_R,
\label{eq:CHIRAL1}
\end{equation}
where decomposition into left $L$ and right $R$ chiral components was used. Hence, in the massless limit left and right quarks separate into two different sectors that communicate via gluons exchange. As a consequence $\mathcal{L}^0_{QCD}$ is invariant under global $SU(n_f)_L\times SU(n_f)_R$ transformations of the form
\begin{equation}
q_L\rightarrow g_L q_L,\qquad q_R\rightarrow g_R q_R,\qquad g_{L,R}\in SU(n_f)_{L,R},
\label{eq:CHIRAL2}
\end{equation}
where $n_f$ denotes the number of quark flavors. Since quark mass terms violate explicitly the chiral symmetry, the maximal sensible value of $n_f$ is $n_f=3$, i.e. $q^T=(u,d,s)$. 

The corresponding conserved currents read
\begin{equation}
J^{a\mu}_L = \bar{q}_L\gamma^\mu T^a q_L,\qquad J^{a\mu}_R = \bar{q}_R\gamma^\mu T^a q_R,\qquad (a = 1, \ldots, n^2_f-1), 
\label{eq:CHIRAL3}
\end{equation}
where $T^a$ denote the $SU(n_f)$ generators. Since parity exchanges $L$ with $R$, parity eigen-currents can conveniently be introduced:
\begin{equation}
\begin{array}{l}
J_V^{a\mu}\equiv J_L^{a\mu}+J_R^{a\mu}, \\
J_A^{a\mu}\equiv J_L^{a\mu}-J_R^{a\mu},
\end{array}
\label{eq:CHIRAL4}
\end{equation}
which are parity +1 and parity -1 conserved currents, respectively.

The chiral transformations~\eqref{eq:CHIRAL2} should constitute an approximately good symmetry in the light quark sector $(u,d,s)$. The symmetry would imply existence of degenerate mirror multiplets with opposite chiralities. This in turn implies the existence of mirror multiplets of opposite parity. However, while hadrons form multiplets of $SU(3)_{L+R}$, no mirror hadronic multiplets with parity $p=-1$ exist. Interestingly, the octet of pseudoscalar mesons ($\pi^+,\pi^-,\pi^0,\eta,K^+,K^-,K^0,\bar{K}^0$) is much lighter than the rest of the hadronic states. These empirical facts indicate that the vacuum is not symmetric under the full chiral group. Only those transformations with $g_R = g_L$ remain a symmetry of the physical QCD vacuum. Thus, the $SU(3)_L \times SU(3)_R$
symmetry is broken spontaneously to $SU(3)_{L+R}$. The eight pseudoscalar mesons are to be identified with the Goldstones of spontaneously broken chiral symmetry with $n_f=3$.

The nature of the SSB in QCD is different than in the linear sigma model discussed previously. While in the sigma model the SSB was triggered by a scalar potential, in case of QCD the effect is due to quark-antiquark condensate forming:
\begin{equation}
\left<0\right|\bar{u}u\left|0\right>=\left<0\right|\bar{d}d\left|0\right> = \left<0\right|\bar{s}s\left|0\right>\neq0.
\label{eq:CHIRAL5}
\end{equation}
The effect is non-perturbative. The Goldstones are excitations over the condensate. Hence they are composite objects, bound states of quark-antiquark pairs. Nevertheless as already stated, due to universality of the goldstone matrix $U$, chiral EFT applies to describe the pseudoscalar mesons dynamics. The Goldstones are represented by the matrix~\eqref{eq:HEFT11c}. The explicit field content reads
\begin{equation}
\lambda^a\phi^a \equiv
\left(\begin{array}{ccc}
	\frac{1}{\sqrt{2}}\pi^0+\frac{1}{\sqrt{6}}\eta_8 & \pi^+ & K^+ \\
	\pi^- & -\frac{1}{\sqrt{2}}\pi^0+\frac{1}{\sqrt{6}}\eta_8 & K^0 \\
	K^- & \bar{K}^0 & -\frac{2}{\sqrt{6}} \eta_8
\end{array}\right)
\label{eq:CHIRAL6}
\end{equation}
For $n_f=2$, \eqref{eq:CHIRAL6} reduces to the upper $2\times2$ submatrix with $\pi$ fields only. 

The leading term in the chiral EFT of QCD mesons is~\eqref{eq:HEFT12p}. We shall denote the scale of global symmetry breaking by $F$, i.e. $f\rightarrow F$. Expanding the exponent of $U$, $\mathcal{L}_2$ generates mesons kinetic terms and an infinite tower of interactions with increasing number of fields. For example,  the elastic scattering of $\pi^+\pi^0$ is described by the following tree-level amplitude
\begin{equation}
(\pi^+\pi^0\rightarrow \pi^+\pi^0)= \frac{t}{F},
\label{eq:CHIRAL7}
\end{equation}
where $t=(p_{\pi^+}' - p_{\pi^+})^2$ is the Mandelstam variable. The chiral EFT $\mathcal{L}_2$ allows to relate processes with different number of $\pi$: $\pi\pi\rightarrow 2\pi,4\pi,6\pi,\ldots$ in terms of a single scale $F$. Its value is \mbox{$F=92.2$ MeV}.
 
Up till now, the QCD mesons dynamics was described as Goldstone bosons interactions. Goldstones are massless, while the real mesons are massive, although light. Their masses are generated by accounting for the quarks mass term $-\bar{q}\mathcal{M}q$ in the QCD Lagrangian, which explicitly breaks the chiral symmetry~\eqref{eq:CHIRAL2}. Moreover, the mesons are sensitive to the electroweak interactions, which also breaks the chiral symmetry explicitly. To account for these sources of explicit chiral symmetry breaking, external sources are formally introduced to the chiral EFT. The external sources are the electroweak gauge fields and the quark mass matrix. The external sources can be assigned certain momentum power counting rules consistently. The so-called chiral perturbation theory ($\chi$PT) emerges with a systematic momentum expansion, 
which gives a systematic way of including higher-order corrections -- the $\mathcal{L}_{\chi PT}$ is renormalizable at each order in $O(p)$, as the Chiral EFT. The complete set of operators $\mathcal{L}_4$ and $\mathcal{L}_6$ are known and the renormalizability at the corresponding orders $O(p^4)$ and $O(p^6)$ has been checked explicity~\cite{Gasser:1983yg,Gasser:1984gg,Bijnens:1999sh,Haefeli:2007ty,Bijnens:2001bb,Ebertshauser:2001nj}.
The LO Lagrangian reads:
\begin{equation}
\mathcal{L}_2=\frac{F^2}{4}\left<D_\mu U^\dagger D^\mu U + U^\dagger \chi + \chi^\dagger U\right>,
\label{eq:CHIRAL8}
\end{equation}
where $\chi = 2 B_0 \mathcal{M}$ and the coupling $B_0$ is another effective coupling, like $F$. Its value must be set by measurement; $D_\mu$ is a covariant derivative with the electroweak gauge fields. Nonetheless, the gauge fields are not quantized in the $\chi$PT, i.e. their kinetic terms are absent in the EFT. Again, they are realized as external sources. All in all, at $O(p^2)$ the $\chi$PT Lagrangian is able to describe all QCD pseudomesons Green functions with only two parameters $F$ and $B_0$.

The two terms involving the $\chi$ source in eq.~\eqref{eq:CHIRAL8} generate masses for the meson fields (now pseudo-Goldstones). One of $\chi$PT predictions is that owing to the chiral symmetry, the meson masses squared are proportional to a single power of the quark masses. The proportionality coefficient is $B_0$. This prediction allows in turn for the prediction of quarks mass ratios~\cite{Pich:2018ltt}; explicitly:
\begin{equation}
\begin{array}{c}
\frac{m_d-m_u}{m_d+m_u} = \frac{\left(M_{K^0}^2-M^2_{K^+}\right)-\left(M^2_{\pi^0}-M^2_{\pi^+}\right)}{{M^2_{\pi^0}}}=0.29,\\ \mbox{} \\
\frac{m_s-\frac{1}{2}m_u - \frac{1}{2}m_d}{m_u+m_d} = \frac{M^2_{K^0}-M^2_{\pi^0}}{M^2_{\pi^0}} = 12.6,
\end{array}
\label{eq:CHIRAL8p}
\end{equation}  
which implies 
\begin{equation}
m_u:m_d:m_s = 0.55:1:20.3.
\label{eq:CHIRAL9}
\end{equation}
The $\chi$-dependent terms in~\eqref{eq:CHIRAL8} introduce moreover corrections to the $\pi\pi$ scattering. For example~\eqref{eq:CHIRAL7} is corrected as follows
\begin{equation}
A(\pi^+\pi^0\rightarrow\pi^+\pi^0) = \frac{t-m_\pi^2}{F^2},
\label{eq:CHIRAL10}
\end{equation}
which vanish at $t=m^2_\pi$. Eq.~\eqref{eq:CHIRAL10}, together with~\eqref{eq:CHIRAL9}, are examples of predictions of the $\chi$PT. These successful phenomenological predictions corroborate the pattern of $\chi SB$ in~\eqref{eq:HEFT11p} and the explicit breaking incorporated by the QCD quark masses.

Concerning the range of validity of $\chi$PT: this EFT is an expansion in powers of momenta over some typical hadronic scale $\Lambda_\chi$, which can be expected to be of order of the light-quark resonances. A natural order-of-magnitude estimate of $\Lambda_\chi$ reads~\cite{Gavela:2016bzc}:
\begin{equation}
\Lambda_\chi \lesssim 4\pi F,
\label{eq:CHIRAL11}
\end{equation}
which in case of QCD is $\sim1.2$ GeV. This value sets the upper bound on the cut-off scale and hence determines the maximal $\chi$PT range of validity. 
\subsection{The custodial symmetry}
\label{custodial}
The purpose of this Section is to discuss the SSB pattern in the SM. While the gauge symmetry breaking is~\eqref{eq:eq2}, the global SSB pattern has a richer structure. To see it, it is convenient to rewrite the scalar doublet $\Phi\equiv\left(\begin{array}{c}\phi^+\\ \phi^0\end{array}\right)$ as a $2\times2$ matrix~\cite{Appelquist:1980vg}
\begin{equation}
\Sigma\equiv(\Phi^c,\Phi)\equiv\left(\begin{array}{cc} 
\phi^{0\ast} & \phi^+ \\ -\phi^- & \phi^0
\end{array}\right),
\label{eq:custodialp}
\end{equation}
where $\Phi^c\equiv i\tau_2\Phi^\ast$; $(\cdot)^\ast$ denotes complex conjugation. Then, the pure scalar part of the SM Lagrangian~\eqref{eq:52}, up to an irrelevant constant term, reads
\begin{equation}
\mathcal{L}(\Phi) = \frac{1}{2}\left<(\partial^\mu\Sigma)^\dagger \partial_\mu\Sigma\right>-\frac{\lambda}{16}\left(\left<\Sigma^\dagger\Sigma\right>-v^2\right)^2.
\label{eq:custodial}
\end{equation}
The vacuum choice is $\left<0\right|\phi^0\left|0\right> = v$ or equivalently $\left<0\right|\Sigma\left|0\right>=vI_2$.
Up to normalization factors the Lagrangian is formally identical to the linear sigma model~\eqref{eq:HEFT4}. Hence also the Lagrangian symmetry is formally the same, with the $\Sigma$ transformation as in eq.~\eqref{eq:HEFT5}. It implies that the global SSB pattern in the SM reads:
\begin{equation}
SU(2)_L\times SU(2)_R\longrightarrow SU(2)_{L+R}.
\label{eq:custodial1}
\end{equation}
The symmetry on the right side of eq.~\eqref{eq:custodial1} is called the custodial symmetry, $SU(2)_C$.
Three Goldstone bosons emerge. They can be parametrized with the already familiar $U$ matrix, using the polar decomposition
\begin{equation}
\Sigma(x) = \frac{1}{\sqrt{2}}(v+h(x))U(\vec{\phi}),\qquad U(\vec{\phi}) = \exp\left\{i\vec{\tau}\vec{\phi}/v\right\},
\label{eq:custodial2}
\end{equation}
where again $U$ inherits the transformation properties of $\Sigma$. Not surprisingly, the Goldstone part of $\mathcal{L}_\Phi$ has the form of the universal chiral LO term~\eqref{eq:HEFT12p} with the $f$ scale set to the electroweak $v$:
\begin{equation}
\mathcal{L}_\Phi = \frac{v^2}{4}\left<\partial_\mu U^\dagger \partial^\mu U\right> + O(h/v).
\label{eq:custodial3}
\end{equation}

In the SM however the $SU(2)_L\times U(1)_Y$ subgroup of the global $SU(2)_L\times SU(2)_R$ is gauged and the SM scalar Lagrangian in terms of the $\Sigma$ field reads:
\begin{equation}
\mathcal{L}(\phi) = \frac{1}{2}\left<(D^\mu\Sigma)^\dagger D_\mu\Sigma\right>-\frac{\mu^2}{8}\left<\Sigma^\dagger\Sigma\right>-\frac{\lambda}{16}\left<\Sigma^\dagger\Sigma\right>^2,
\label{eq:custodial1p}
\end{equation}
The $U$ part is adjusted accordingly
\begin{equation}
\mathcal{L}_2 = \frac{v^2}{4}\left<D_\mu U^\dagger D^\mu U\right>,
\label{eq:custodial2p}
\end{equation}
where the covariant derivative reads:
\begin{equation}
D_\mu U(x)\equiv\partial_\mu U(x)+ig W_\mu(x) U(x)-\frac{ig'}{2}B_\mu(x) U(x)\tau_3.
\label{eq:custodial3p}
\end{equation}
Gauging the $U(1)_Y$ explicitly breaks the global $SU(2)_R$ and in turn the $SU(2)_C$ is explicitly broken as well. Another source of explicit custodial symmetry breaking is the Yukawa sector, the largest effect is given by non-vanishing $m_t-m_b$. 
The $U$ term in eq.~\eqref{eq:custodial2p} generates $W$ and $Z$ boson masses $m_W=m_Z\cos\theta_W$.  
It is clearly visible in the unitary gauge where $U(\vec{\phi})=1$, i.e. only the first term in the $U$ expansion survives (as opposed to $\chi$PT in QCD, in the SM the $U(\vec{\phi})$ can be rotated away due to the gauging). In the end, the Goldstone bosons constitute longitudinal polarizations of the weak vector bosons $W$ and $Z$. This successful mass relation is a direct consequence of the pattern of SSB~\eqref{eq:custodial1} combined with the gauging $SU(2)_L\times U(1)_Y$, providing a clear confirmation of the symmetry pattern that is realized at the electroweak scale. Interestingly, a very similar term to~\eqref{eq:custodial2p} was present in eq.~\eqref{eq:CHIRAL8} in QCD $\chi$PT. In fact, the QCD pseudo-Goldstone mesons generate, in addition, a tiny correction $\delta m_W = \delta m_Z \cos\theta_W = Fg/2$.
\subsection{The HEFT Lagrangian}
\label{heftBasis}
Contrary to SMEFT, HEFT  includes the possibility of non-linearly realized $SU(2)_L\times SU(2)_R\rightarrow SU(2)_{L+R}$ where the three Goldstones form a triplet of $SU(2)_{L+R}$ and the $h$ is a singlet. Depending on the UV completion of the SM, the Higgs may or may not form the Higgs doublet. This approach leads to the most general description of the EW and Higgs couplings, satisfying the gauge symmetry of the SM. In specific limits, it may reduce to the SM Lagrangian, or may coincide with the SMEFT. Also contrary to SMEFT, the HEFT is constructed for the broken phase. 

In order to formulate the electroweak effective theory the most general low-energy Lagrangian must be considered, that satisfies the SM gauge symmetries and only contains the known light spectrum: the SM gauge bosons and fermions, the electroweak Nambu-Goldstone modes and the Higgs field $h$. In addition to the SM gauge symmetries, our main assumption will be the pattern of global SSB~\eqref{eq:custodial1}. Hence, in analogy to the $\chi$PT construction for QCD, the Lagrangian shall be organized as an expansion in powers of derivatives and explicit symmetry breaking terms over the $v$ scale. As already illustrated in Sec.~\ref{custodial}, the purely Goldstone terms are formally identical to those present in $\chi$PT with $n_f = 2$. The electroweak effective theory contains, however, a richer variety of ingredients, since included must be the SM gauge symmetries and the fermion sector. In the literature this electroweak effective theory often goes under the name of Higgs Effective Field Theory (HEFT). Sometimes the corresponding Lagrangian is called the Electroweak Chiral Lagrangian (EW$\chi$L); the use of EFTs, in the case of the EW$\chi$L, dates back to the late '80s~\cite{Dobado:1989ax,Dobado:1989ue,Dobado:1989gr,Dobado:1990jy,Dobado:1995qy,Dobado:1999xb,Alboteanu:2008my}.

All the ingredients can be systematically introduced in HEFT in terms of invariants under the global chiral group~\eqref{eq:custodial1} of which $SU(2)_L\times U(1)_Y$ is gauged. In the remaining we shall however limit the discussion to the bosonic part. It is dictated by our phenomenological purpose: to study tree-level BSM effects in vector boson scattering.

The momentum power-counting rules for all the ingredients can be consistently introduced. In particular~\cite{Pich:2016lew}:
\begin{itemize}
	\item as previously, the Goldstone modes, and here also the Higgs $h$, are $O(p^0)=O(1)$; the vev $v$ is also counted as $O(1)$; $U$ is $O(1)$;
	\item all mass parameters $m_h,m_W,\ldots$ are counted as of order $O(p)$;
	\item as a consequence, the gauge couplings $g,g',g_s$ are $O(p)$;
	\item gauge fields $\hat{W}_\mu,\hat{B}_\mu$ defined as
	\begin{equation}
	\hat{W}_\mu\equiv -g\frac{\tau^i}{2}W^i_\mu,\qquad \hat{B}_\mu = -g'\frac{\tau_3}{2}B_\mu,
	\label{eq:heft1}
	\end{equation}
	are $O(p)$, as these constitute part of the (covariant) derivative.
\end{itemize}
Therefore, the fields $W^i_\mu, B_\mu$ are $O(1)$. The above chiral power-counting rules, together with the rules for the fermionic sector, allow for systematic calculations in HEFT, accounting for renormalization of the loop UV divergences order by order in the derivative expansion. The discussion on renormalizability is a straightforward generalization of what is discussed in Sec.~\ref{linearSigmaModel}. 

It is customary to introduce two gauge covariant objects $\V_\mu$ and $\T$, the vector and the scalar chiral fields respectively, that transform in the adjoint of $SU(2)_L$:
\begin{equation}
\V_\mu(x)\equiv\left(D_\mu U(x)\right)U(x)^\dagger,\qquad
\T(x)\equiv U(x)\tau_3 U(x)^\dagger,
\label{eq:heft2}
\end{equation}
where the covariant derivative is the same as in eq.~\eqref{eq:custodial3p}. While $\V_\mu(x)$  is covariant under $SU(2)_C$, $\T(x)$ is not and therefore plays a role of a source of explicit custodial symmetry breaking in the HEFT construction. The counting implies that $\V_\mu$ is $O(p)$ and $\T$ is $O(1)$.

The HEFT Lagrangian can be written as a sum of two terms:
\begin{equation}
\mathcal{L}_{HEFT}\equiv\mathcal{L}_0+\Delta\mathcal{L}.
\label{eq:heft3}
\end{equation}
According to the chiral power-counting rules listed above, the first term is $O(p^2)$, i.e. is LO in chiral derivative expansion. It basically describes the SM part: the kinetic terms for all the particles in the spectrum, the Yukawa couplings and the scalar potential.
The second one starts with terms of order $O(p^4)$ and describes new interactions and deviations from the LO contributions. The LO Lagrangian reads:
\begin{equation}
\begin{split}
\mathcal{L}_0=&
-\dfrac{1}{4}G_{\mu\nu}^a G^{a\,\mu\nu}
-\dfrac{1}{4}W_{\mu\nu}^a W^{a\,\mu\nu}
-\dfrac{1}{4}B_{\mu\nu}B^{\mu\nu}+\\
&
+\dfrac{1}{2}\partial_\mu h\partial^\mu h
-\dfrac{v^2}{4}\left<\V_\mu \V^\mu\right>\mathcal{F}_C(h/v)
-V(h/v)+\text{\tt fermions}\,,
\end{split}
\label{eq:heft4}
\end{equation}
where $\tt fermions$ refers to all the terms involving fermions and these will not be considered here as they do not enter the present analysis. 
The first line describes the kinetic terms for the gauge bosons. The second line contains: the Higgs and the Goldstone bosons kinetic terms and the mass terms for $W$ and $Z$ gauge bosons, and the scalar potential. The electroweak scale $v$ is used to compensate the powers of both the Higgs and the Goldstone fields, because they are expected to have a similar underlying origin. Model-dependent parametric differences, e.g. different physical scales ratios, are implicitly present in  effective coefficients (see below). 

The general form of the Higgs potential in the chiral approach reads:
\begin{equation}
V(h/v) = v^4 \sum_{n=3}c^{(V)}_n \left(\frac{h}{v}\right)^n,
\label{eq:heft4p}
\end{equation}
where $c_n^{(V)}$ are dimensionless effective couplings. The scalar Lagrangian of the SM corresponds to:
\begin{equation}
c_3^{(V)} = \frac{1}{2}m_h^2/v^2,\qquad c_4^{(V)}=\frac{1}{8}m^2_h/v^2,\qquad c^{(V)}_{n>4} = 0.
\label{eq:heft4b}
\end{equation}

The function $\mathcal{F}_C(h/v)$ in the second line of Eq.~\eqref{eq:heft4} is conventionally written as
\begin{equation}
\mathcal{F}_C(h) = 1 + 2 a_C \frac{h}{v}+b_C \frac{h^2}{v^2}+\dots,
\label{eq:heft5}
\end{equation}
where the dots refer to higher powers in $h/v$. In the SM case, the first two coefficients of $\mathcal{F}_C(h/v)$ are exactly equal to $a_C=1=b_C$, while the ones corresponding to higher orders are identically vanishing. In a general analysis $a_C$ and $b_C$ are free, effective parameters.

In general, each chiral-invariant structure should be multiplied with an arbitrary function of $h/v$ because the Higgs field is a singlet under $SU(2)_L \times SU(2)_R$. For instance, the quadratic derivative term of the Higgs should also be multiplied with an arbitrary function $F_h(h/v)$. However, this function can be reabsorbed into a redefinition of the field $h$~\cite{Giudice:2007fh}. Moreover, the kinetic terms for the gauge bosons in the first line of eq.~\eqref{eq:heft4} do not come associated with any $\mathcal{F}(h/v)$. Instead such corrections are shifted to the $\Delta\mathcal{L}$, where they are described by $O(p^4)$ terms:
\begin{equation}
c_B\left<\hat{B}_{\mu\nu}\hat{B}_{\mu\nu}\right>\mathcal{F}_B(h/v),\qquad c_W\left<\hat{W}_{\mu\nu}\hat{W}_{\mu\nu}\right>\mathcal{F}_W(h/v),\qquad c_G\left<\hat{G}_{\mu\nu}\hat{G}_{\mu\nu}\right>\mathcal{F}_G(h/v),
\label{eq:heft6}
\end{equation}
where the stress-tensors of fields~\eqref{eq:heft1} are defined as follows
\begin{equation}
\hat{W}_{\mu\nu}\equiv \partial_\mu\hat{W}_\nu - \partial_\nu\hat{W}_\mu - i [\hat{W}_\mu,\hat{W}_\nu], \qquad \hat{B}_{\mu\nu} \equiv \partial_\mu\hat{B}_\nu - \partial_\nu\hat{B}_\mu - i [\hat{B}_\mu,\hat{B}_\nu]
\label{eq:heft7}
\end{equation}
and similarly for the gluons (this construction has already been introduced in eq.~\eqref{eq:21}). The above illustrates that the $\mathcal{L}_{HEFT}$ splitting into $\mathcal{L}_0$ and $\Delta\mathcal{L}$ in \eqref{eq:heft3} is partially data-driven. 

The basis of $O(p^4)$ HEFT operators has recently been constructed. As in case of SMEFT, the task is highly non-trivial, because redundancies has to be properly accounted for. Moreover, ref.~\cite{Eboli:2016kko} contains the complete classification of operators $O(p^4)$ and $O(p^6)$ that possess quartic interactions among the electroweak gauge bosons and that at the same time do not exhibit triple gauge-boson vertices associated to them. We shall utilize these results in the phenomenological analysis of $WW$ scattering in Sec.~\ref{VVHEFT}.

The HEFT Lagrangian is most suitable to describe low-energy effects of UV physics scenarios where the electroweak symmetry breaking has dynamical, non-perturbative origin, similar to the case of chiral symmetry breaking in the physics of color (QCD). In such a scenario some yet unknown strong dynamics should intervene at a scale $\Lambda_s$, and the characteristic scale of the associated (composite) Goldstone bosons $f$ respects $\Lambda_s \leq 4\pi f$ (compare to eq.~\eqref{eq:CHIRAL11}). In the original formulation, so-called ''technicolor''~\cite{Susskind:1978ms,Dimopoulos:1979es,Dimopoulos:1981xc}, the physical Higgs particle is simply removed from the low-energy spectrum and only the three would-be-Goldstone bosons are retained, in order to give masses to the weak gauge bosons, with $f = v$, where $v = 246$ GeV is the electroweak scale defined via the $W$ mass, $m_W = gv/2$. 
%The smoking gun signature of the technicolor ansatz is the appearance of several resonances at the scale $\lesssim 4\pi v$, i.e. at the TeV scale.

Interestingly, several variants of the strong interacting ansatz exist in the context of electroweak symmetry breaking, with some of them accounting for the existence of a light Higgs resonance in the spectrum. The idea of a light composite Higgs originating in the context of a strongly interacting dynamics was first developed in the 1980s and underwent a recent revival of interest (for a review, see e.g. \cite{Contino:2010rs,Panico:2015jxa}). In this framework, a global symmetry group $G$ is postulated at high energies and broken spontaneously by some strong dynamics mechanism (analogue of the quark-antiquark condensate forming) to a subgroup $H$ at a scale $\Lambda_s$. Among the corresponding Goldstone bosons, three are usually identified with the longitudinal components of the SM gauge bosons and one with the Higgs field, $h$. A scalar potential for the Higgs field is dynamically generated, inducing EWSB and providing a (light) mass to the Higgs particle. Being a pseudo-Goldstone boson arising from the global symmetry breaking, the Higgs boson mass is protected against quantum corrections of the high-energy symmetric theory, providing an elegant solution to the EW hierarchy problem. The EWSB scale, identified with the vacuum expectation value of the Higgs field $\left<h\right>$ does not need to coincide with the EW scale $v$ defined by the EW gauge boson masses. On the other hand, $\left<h\right>$ is typically predicted in any specific Composite Higgs model to obey a constraint linking it to the EW scale $v$ and to the GB scale $f$. 
%In Sec.~\ref{CHmodel} the so-called minimal $SO(5)/SO(4)$ custodial symmetry intrinsically preserving CH model shall be discussed to illustrate the general idea. 
In ref.~\cite{Alonso:2014wta} several concrete CH models were matched to the low-energy HEFT Lagrangian, arguing moreover that any CH model can be matched to HEFT.
\subsection{The primary dimension $d_p$}
\label{primaryDim}
For HEFT, the (chiral) distinction in LO, NLO, etc\ldots, sometimes fails in ordering the impact of the different operators. The latter depends on the structure of the operators and on the energy involved in the observables under consideration. Once the energy is smaller but close to the cut-off, a counting based on the so-called primary dimension $d_p$~\cite{Gavela:2016bzc} is more suitable~\cite{Gavela:2016bzc} (for the details on the argument, see~\cite{Gavela:2016bzc}): it counts the canonical dimension of the leading terms in the expansion of a given object. Indeed, the matrix $U$ and the functions $\F(h)$ hide the dependence on the scale $v$ :
\be
U=1+2i\dfrac{\sigma_a\pi^a}{v}+\ldots\,,\qquad\qquad
\F(h)=1+2a\dfrac{h}{v}+\ldots\,,
\label{GenericUF}
\ee 
and as a consequence $U$, $\F(h)$ and $\T$ (see eq.~\eqref{eq:heft2}) has $d_p=0$, while $\partial_\mu\F(h)$ and $\V_\mu$ (see eq.~\eqref{eq:heft2} together with eq.~\eqref{eq:heft2}) have $d_p=2$.

Any operator in HEFT may be ordered in terms of its $d_p$ and it allows to link the particular structure of an operator to the strength of a physical signal measured by  cross sections~\cite{Gavela:2016bzc}. An interesting application is that operators with the same $d_p$ are expected to have similar impact on  given observables: this information may be used to identify the complete set of operators describing in a similar way the same process, although they belong to different orders in the chiral expansion. In SMEFT the primary dimension is simply understood as the canonical dimension, which implies another application of the primary dimension: if the $d_p$ of an HEFT operator is smaller than the canonical dimension of the SMEFT operator that contributes to  a same observable (this SMEFT operator will be refereed to as ``linear sibling'' of this HEFT operator), then the process described by these operators is expected to have a higher cross section in the HEFT than in the SMEFT: this process may be used to test the linearity of the Higgs sector dynamics ~\cite{Brivio:2013pma,Brivio:2014pfa,Gavela:2014vra,Brivio:2015kia,Brivio:2016fzo,Brivio:2017ije}. 

We shall use the $d_p$ counting in order to justify our choice for HEFT operators in Sec.~\ref{whichNonRen} after our choice of the SMEFT ''models'' will be justified in that Section.

%\subsection{The $SO(5)/SO(4)$ CH model}
%\label{CHmodel}
%xxx

\chapter{The EFT approach to $W^+W^+$ scattering at the LHC}
\label{WWscatEFT}
In this Chapter we describe our strategy for the EFT approach to the same-sign $WW$ scattering data, once available in future and yield significant discrepancies from the SM predictions. We also discuss the method for determining the discovery potential in this EFT approach. Both aspects are applicable in principle to any chosen EFT ''model'', specified by the choice of the effective non-renormalizable operators added to the $\mathcal{L}_{SM}$ and values of the corresponding Wilson coefficients. We shall illustrate our approach by determining discovery regions of certain classes of EFT ''models'', specified below.  We start in Sec.~\ref{preTech} with preliminary technicalities, where we shall: 
\begin{itemize}
	\item define our choice (and justify it) of operators and the normalizations,
	\item argue about the fact that certain qualitative features of $W^+W^+$ scattering in the full reaction can be understood at the level of on-shell $W^+W^+\rightarrow W^+W^+$ amplitudes,
	\item study in detail both on-shell cross sections,
	\item and constraints from perturbative partial wave unitarity bounds in the on-shell scattering.
	\end{itemize}
The EFT approach to the full reaction $pp\rightarrow 2\ jets+ l\nu_l l'\nu_l'$ is discussed in Sec.~\ref{EFTapproachToWW}. Then, in Sec.~\ref{WWinSMEFT} and~\ref{VVHEFT} numerical results for the discovery regions are presented for the chosen EFT ''models'' in SMEFT and HEFT, respectively. 

Moreover, although we do not discuss in this thesis the bounds on the Wilson coefficients obtained from the data analysis when no
statistically significant signal on new physics is observed (such an analysis requires a dedicated discussion), it shall be self-explanatory that it
will be also considerably influenced by the results of this work. 

\section{Preliminary technicalities}
\label{preTech}
\subsection{Which non-renormalizable operators/EFT ''models''?}
\label{whichNonRen}
The same-sign $pp\to W^+W^+ jj$ process probes a number of higher dimension operators.
Among them in the SMEFT expansion are dimension-6 operators which modify only the Higgs-to-gauge coupling:
\begin{equation}
\begin{aligned}
{\cal O}_{\Phi d} = \partial_\mu(\Phi^\dagger\Phi)\partial^\mu(\Phi^\dagger\Phi),\\
{\cal O}_{\Phi W} = (\Phi^\dagger\Phi)\mbox{Tr}[W^{\mu\nu}W_{\mu\nu}],\\
{\cal O}_{\tilde WW}=\Phi^\dagger{\tilde W}_{\mu\nu}W^{\mu\nu}\Phi
\end{aligned}
\label{eq:preTech1}
\end{equation}
(the last one being $\mathcal{CP}$-violating),
dimension-6 operators which induce anomalous triple gauge couplings (aTGC):
\begin{equation}
\begin{aligned}
{\cal O}_{WWW}=\mbox{Tr}[W_{\mu\nu}W^{\nu\rho}W_{\rho}^{\mu}],\\
{\cal O}_W=(D_\mu\Phi)^\dagger W^{\mu\nu}(D_\nu\Phi),\\
{\cal O}_B=(D_\mu\Phi)^\dagger B^{\mu\nu}(D_\nu\Phi),\\
{\cal O}_{\tilde WWW}=\mbox{Tr}[{\tilde W}_{\mu\nu}W^{\nu\rho}W_{\rho}^{\mu}],\\
{\cal O}_{\tilde W}=(D_\mu\Phi)^\dagger {\tilde W}^{\mu\nu}(D_\nu\Phi)
\end{aligned}
\label{eq:preTech2}
\end{equation}
(the last two of which are $\mathcal{CP}$-violating), as well as higher dimension operators.
The field strength tensors are:
\begin{equation}
\begin{aligned}
 W_{\mu\nu} =  \frac{1}{2} \tau^i (\partial_\mu W^i_\nu - \partial_\nu W^i_\mu
       - g \epsilon_{ijk} W^j_\mu W^k_\nu ), 
\\
 B_{\mu \nu}  = \frac{1}{2}  (\partial_\mu B_\nu - \partial_\nu B_\mu),\\
\tilde{W}_{\mu\nu} = \epsilon_{\mu\nu\alpha\beta} W^{\alpha\beta}.
\end{aligned}
\label{eq:fields}
\end{equation}

A common practice in the LHC data analyses in the EFT framework is to derive uncorrelated 
limits on one operator at a time while setting all the remaining Wilson coefficients
to zero.  This in fact means {\it choosing} different EFT ''models'':
such limits are valid only under the assumption that just
one chosen operator dominates BSM effects in the studied process in the available energy
range.  In this work we shall consider only variations of single  at a time operator that modify Quartic Gauge Couplings (QGC) and simultaneously leave intact TGC and Higgs-to-gauge couplings. Such operators are often called genuine QGC (gQGC) operators and in SMEFT these start at dimension-8 ($n=8$). The physical reason for omitting in the phenomenological analysis the non-gQGC operators (in particular $n=6$ operators) is that Higgs and triple gauge couplings can be accessed experimentally via other processes, e.g.  in diboson
production, vector boson fusion, or Higgs production and
decay measurements. If anomalous TGC and/or Higgs couplings are present,
we expect to first probe them in these processes, which are moreover presently known to agree with the SM within a
few per cent~\cite{Butter:2016cvz}. This fact translates into stringent limits on the 
dimension-6 operators (see also~\cite{Falkowski:2016cxu} and references in the introduction therein). On top of that, interestingly recent theoretical development~\cite{Adams:2006sv,deRham:2017avq,deRham:2017zjm,Zhang:2018shp} shows that the numerators $c_i^{(8)}$ of Wilson coefficients of $n=8$ gQGC operators (defined in eq.~\eqref{eq:SMEFT2}) should be somewhat larger than $c_i^{(6)}$ of $n=6$, in a sense that they should dominate the latter in certain linear combinations of $c_i$ so that certain quantities stay positive. These ''positivity bounds'' are derived in the low-energy effective description as consequence of the cherished fundamental principles of QFT such as unitarity, Lorentz invariance, locality, and causality/analyticity of scattering
amplitudes which are assumed to hold in the deeper theory. Hence the ''positivity bounds'' are general, model-independent and suggest dominating effects of $n=8$ over $n=6$ on theoretical grounds. 
%It is worth to mention at this point, that instances have been noticed where the contribution of $n=6$ operators to a given process can be 
%suppressed compared to $n=8$ operators contrary to a naive 
%$(E/\Lambda)$ power counting \cite{Liu:2016idz,Contino:2016jqw,Azatov:2016sqh, 
%Franceschini:2017xkh}. Nonetheless contributions of the $n=6$ SMEFT operators to VBS has been recently investigated. For a discussion on the effects of $n=6$ SMEFT operators in $ZZ$ elastic scattering at LHC, taking into account the bounds on the $n=6$ operators, see\REF.

Therefore we shall take the following point of view, namely that a genuine feature of VBS processes is that they probe BSM effects that manifest as anomalous QGC couplings in the lower energy theory. In turn, concerning the discovery potential we analyse EFT ''models'' that are defined as the SM Lagrangian plus $n=8$ gQGC operators that affect the WWWW vertex, in the SMEFT case. 

In the case of the HEFT Lagrangian, the gQGC operators start at the primary dimension $d_p=8$. Since the $d_p$ counting determines the impact of the non-renormalizable operators on the cross section (Sec.~\ref{primaryDim}), and we want to examine HEFT at the same order as SMEFT, we shall only consider the $d_p=8$ gQGC operators that affect $WWWW$ vertex.

More precisely, we choose our EFT ''models'' as the SM Lagrangian plus a single at a time non-renormalizable operator, for simplicity purposes. Clearly, the assumption about the choice of operators in the truncation in
the EFT series used to analyze a process of our interest introduces a strong model dependent aspect of
that analysis: one is implicitly assuming that there exist a class of UV complete models such that
the chosen truncation is a good approximation. Having said that, the single operator at a time analysis seems quite a severe restriction to UV completions. Nevertheless, as we argue later on based on recent theoretical progress in polarization extraction of VBS scattering from the real data, it is possibly not so severe as naively expected. 

The strategy and the methods we present can be extended to the case of many operators at a time (of arbitrary dimensions),
keeping in mind that varying more than one operator substantially
complicates the analysis.

\subsection{The normalization choice for the non-renormalizable operators}
\label{NDA}

Now that the EFT ''models'' are characterized in terms of precise definition of our choice of effective non-renormalizable operators, we discuss the issue of proper normalization of these effective operators in the EFT ''models'' Lagrangians. Keep in mind that ultimately we want to use the EFT approach to learn something about the couplings and the scale $\Lambda$ of the underlying theory (the bottom-up approach) either in the future data or to determine the discovery potential of EFT. Since the information on the couplings is encoded in $c_i$ of the Wilson coefficients $f_i=c_i/\Lambda$, the proper normalization, i.e. fixing the numerical factors in front of the $f_i\mathcal{O}_i$ terms in our EFT ''models'', is crucial for finding out e.g. whether our discovery regions correspond to strongly or weakly interacting BSM physics. In the next two paragraphs we briefly discuss most important conclusions on the proper normalization of effective operators, that were derived and discussed in detail in~\cite{Manohar:2018aog}, and then we discuss how this formula is useful in the bottom-up approach. %

We first remind the reader that the non-renormalizable operators are to be understood as effective low-energy parametrization of the effects of BSM heavy particles that are integrated out from certain Feynman diagrams in concrete deeper BSM models (think of the example in eq.~\ref{eq:SMEFT7}). Each such BSM model predicts concrete coefficients (Wilson coefficients) that multiply effective operators after the effective Lagrangian is derived. The coefficients include e.g. combination of powers of BSM couplings and/or $1/(4\pi)^2$ loop factors, that occur in Feynman diagrams, powers of $1/\Lambda$ as well as some other, typically $O(1)$, numbers, all these factors specific to the BSM models from which the low-energy effect is derived. 
%Using the bottom-up approach one parametrizes the ignorance about possible completions of the SM by introducing the coefficients $c_i$ or $f_i$. 

Let's now assume we have a generic Feynman diagram with certain set of external legs of a light sector, while the internal propagators are of heavy states. We want to integrate out the heavy particles and derive the effective operators that occur in our EFT Lagrangian for the light sector. In this picture each external leg corresponds to a field (of the light sector) to be present in the effective operators. Interestingly, after a certain input is assumed about the Feynman diagram (number of external legs of each type, number of vertices of each type etc.) one can relate (using topological properties of Feyman diagrams) the sum of powers of couplings of different types (gauge, Yukawa, scalar), the numbers of external legs of each type, with the number of loops in the diagram (equivalently powers of $1/(4\pi)$ in the expressions for the amplitudes). In this way proper normalization of the effective operators generated after decoupling heavy particles in our Feynman diagram is obtained, again as a function of certain input of the diagram (see below). This normalization accounts properly for the $1/(4\pi)$ loop factors. It goes under the name of Naive Dimension Analysis (NDA) Master Formula. The NDA formula reads:
\begin{equation}
\frac{\Lambda^4}{16 \pi^2 } \left[\frac{\partial}{\Lambda}\right]^{N_p}  
\left[\frac{ 4 \pi\,  \phi}{ \Lambda} \right]^{N_\phi}
 \left[\frac{ 4 \pi\,  A}{ \Lambda } \right]^{N_A}  
\left[\frac{ 4 \pi \,  \psi}{\Lambda^{3/2}}\right]^{N_\psi} 
\left[ \frac{g_\ast}{4 \pi }  \right]^{N_g}
\left[\frac{y_\ast}{4 \pi } \right]^{N_y} \left[\frac{\lambda_\ast}{16 \pi^2 } \right]^{N_\lambda}\,,
\label{MasterFormula}
\end{equation}
where $\phi$ represents a scalar: either the Higgs doublet (SMEFT) or the field $h$ (HEFT) in the context of our analysis, $\psi$ a generic fermion, $A$ a generic gauge field, $g_\ast$ the generic gauge coupling, $y_\ast$ the generic Yukawa coupling, $\lambda_\ast$ scalar coupling, while $N_i$ refer to the number of times each type of external line or the coupling appears in the Feynman diagram. 
We emphasis that the sum of $N_g,N_y,N_\lambda$ together with the sum of $N_\phi,N_A,N_\psi$ are related via a single equation with the number of loops $L$ in the Feynman diagram. Obviously, in concrete BSM models there will always be some model specific dimensionless factors in addition to what accounted for in~\eqref{MasterFormula}. But these are naturally expected to be $O(1)$ and are not part of the generic NDA. Below we explain how the formula~\eqref{MasterFormula} is useful in the bottom-up approach, where we do not know Feynman diagrams of the deeper theory.

In the bottom-up approach, we do not need to specify $N_g,N_y,N_\lambda$ or $L$ -- we parametrize our ignorance introducing effective Wilson coefficients $c_i$. Nonetheless, formula~\ref{MasterFormula} is indeed very useful in setting the normalization of effective operators that we add to the SM Lagrangian. First of all, notice that for each operator we know $N_p,N_\phi,N_A$ and $N_\psi$. Then, we can use the first part of formula~\eqref{MasterFormula} formally setting $N_g=N_y=N_\lambda=0$: 
\begin{equation}
\frac{\Lambda^4}{16 \pi^2 } \left[\frac{\partial}{\Lambda}\right]^{N_p}  
\left[\frac{ 4 \pi\,  \phi}{ \Lambda} \right]^{N_\phi}
 \left[\frac{ 4 \pi\,  A}{ \Lambda } \right]^{N_A}  
\left[\frac{ 4 \pi \,  \psi}{\Lambda^{3/2}}\right]^{N_\psi}\,,
\label{MasterFormula1}
\end{equation}
 and introduce the effective coefficient $c_i$ upon~\eqref{MasterFormula1}. From eq.~\eqref{MasterFormula} we know that our effective coefficients $c_i$ are products of $g_\ast/4 \pi$ and/or 
$y_\ast/4 \pi$ and/or  $\lambda_\ast/16 \pi^2$ and some unknown model specific $O(1)$ numbers depending on the completion. In particular, formula~\eqref{MasterFormula} implies that in perturbative completions the  $c_i$ multiplying~\eqref{MasterFormula1} are boundend by 1. Therefore, the NDA normalisation is very useful because it relates the values of the effective Wilson coefficients to the weak or strong interacting phase of the underlying theory: if $c_i$ turns out to be equal of larger than $1$, then the corresponding interactions become strongly coupled; while if it is smaller than $1$, then the interactions are weakly coupled.

%Concerning normalization factors of the non-renormalizable operators specifying our EFT ''models'' we shall make use of the so-called Naive Dimensional Analysis (NDA) master formula, first introduced in Ref.~\cite{Manohar:1983md} and later modified in Refs.~\cite{Brass:2018hfw,Gavela:2016bzc}. Following the notation of Ref.~\cite{Gavela:2016bzc} the formula reads:

In particular, the formula~\eqref{MasterFormula1} implies the following counting rules for $\V_\mu$ and $W_{\mu\nu}$ that will constitute our effective operators: 
\begin{equation}
\left[\frac{\V_\mu}{\Lambda}\right]^{N_{\V_\mu}},\qquad \left[\frac{4\pi}{\Lambda^2}W_{\mu\nu}\right]^{N_{W_{\mu\nu}}}\,.
\label{eq:preTech3}
\end{equation}

Later on, we shall use the formula~\eqref{MasterFormula1} to establish normalizations of the effective operators in our EFT ''models'' for the analysis of discovery regions.
%The NDA normalisation is very useful because relates the values of the Wilson coefficients to the weak or strong interacting phase of the considered theory: if a Wilson coefficients, i.e. the dimensionless coefficient that accompanies the operator upon the normalization according to eq.~\eqref{MasterFormula}, turns out to be equal of larger than $1$, then the corresponding interactions become strongly coupled; while if it is smaller than $1$, the corrections induced by the corresponding interactions are subdominant, a sign of a weakly coupled theory. 

\subsection{Qualitative features on the full $pp$ process from the on-shell vector boson scattering}
In the physical process $pp\to jjll'\nu\nu_{l'}$ the $W$ bosons  are off-shell. Later on in Sec.~\ref{WWinSMEFT} and~\ref{VVHEFT} we shall determine discovery regions of our EFT ''models'' using this process. Here, we argue that 
 qualitative conclusions on the influence of dimension-8 operators  on the full process  can be drawn from 
 the analysis of the on-shell $WW$ scattering. To this end, let us employ the identity for the numerator of the gauge boson propagator:
\begin{equation}
g_{\mu\nu}+\frac{k_\mu k_\nu}{M^2_W} = \sum_{\lambda=1}^{4} \epsilon_\lambda^\mu(k)\left(\epsilon_\lambda^\nu(k)\right)^\ast,
\label{eq:identity}
\end{equation}
where in the frame in which the spatial component of  $k_\mu$ is in the $z$ direction, $k_\mu=(E,0,0,k)$, the explicit form of each polarization vector reads:
\begin{equation}
\begin{array}{llll}
\epsilon^\mu_- &=& \frac{1}{\sqrt{2}}(0,+1-i,0) &\qquad \mathrm{(left),}\\
\epsilon^\mu_+ &=& \frac{1}{\sqrt{2}}(0,-1-i,0) &\qquad \mathrm{(right),}\\
\epsilon^\mu_0 &=& (k,0,0,E)/\sqrt{k^2} &\qquad \mathrm{(longitudinal),}\\
\epsilon^\mu_A &=& (E,0,0,k)/\sqrt{\frac{k^2-m^2_W}{k^2m^2_W}}&\qquad (\mathrm{auxiliary}),
\end{array}
\label{eq:pols}
\end{equation}%
where $k^2\equiv k_\mu k^\mu$. In the on-shell limit $k^2\rightarrow m^2_W$ the auxiliary polarization vanishes and $\epsilon_0$ approaches the exact on-shell form of longitudinal polarization (see eq.~\eqref{eq:VVscattering}). 
With the help of eq.~\eqref{eq:identity} one can then rewrite each of the 4 $W$ propagators in each of the diagram that has VBS topology (see fig~\ref{fig:EWWWprod}), as
\begin{equation}
\frac{-i\sum_{\lambda=1}^4\epsilon_\lambda^\mu\left(\epsilon_\lambda^\nu\right)^\ast}{k^2 - m_W^2}.
\label{eq:identityy}
\end{equation}
Then the parton-level amplitude $qq\rightarrow qqll'v_lv_l'$ with VBS topology can be decomposed as follows
\begin{equation}
M\equiv\frac{\sum_{\lambda_1\lambda_2\lambda_3\lambda_4} M_{\lambda_1}^{q1}M_{\lambda_2}^{q2} M^{WW}_{\lambda_1\lambda_2\lambda_3\lambda_4}M_{\lambda_3}^{l1}M_{\lambda_4}^{l2}}{(k_1^2-m_W^2)(k_2^2-m_W^2)(k_3^2-m_W^2)(k_4^2-m_W^2)}, \qquad \lambda_i\in\{\epsilon_{-},\epsilon_{+},\epsilon_{0},\epsilon_{A}\}.
\label{eq:fullamp}
\end{equation}
The $M_{\lambda_i}^{qi}$ ($M_{\lambda_i}^{li}$)  terms are the trilinear $qqW$ ($llW$) vertices  contracted with  $\epsilon^\ast$ ($\epsilon$) of eq.~\eqref{eq:identityy}, while the  $M^{WW}_{\lambda_1\lambda_2\lambda_3\lambda_4}$ term is  the (off-shell) $WW$ elastic scattering amplitude. 
The sum over $i$ includes necessarily polarization configurations in which  the $W$  polarizations are auxiliary.
Now, the effect of dimension-8 operators grows with the scattering energy ($M_{WW}$) and in the region $M_{WW}>>m_W$ modifies significantly helicity amplitudes so that  deviations from the SM behavior become non-negligible. Since the off-shellness $k^2_i$   are  suppressed dynamically by propagators $1/(k_i^2-m^2_W)$, in this kinematic  limit  the scattered vector bosons must be fast, $|\vec{k_i}|\sim E_i>>m_W$,    Therefore in the high $M_{WW}$ region $\epsilon^\mu_0$ is proportional to $\epsilon^\mu_A$ and both approach the on-shell form of the longitudinal polarization vector. As a result,  the sum in eq.~(\ref{eq:identityy}) runs effectively over $\epsilon_i=\epsilon_0,\, \epsilon_+,\, \epsilon_-$ and the off-shell helicity amplitude can be approximated by the on-shell one, accounting for the factors $\sqrt{(k^2-m^2_W)/(k^2m^2_W)}$ or $1/\sqrt{k^2}$ occurring in~\eqref{eq:pols}. 

Therefore in the following Sections we will discuss in detail the high-energy behavior of the on-shell $W^+W^+$ scattering in the presence of contributions from dimension-8 operators.

%that modify the interactions among electroweak gauge bosons which are relevant for this process
%\begin{equation}
% \begin{aligned}
%   {\cal O}_{WWW}&=\mbox{Tr}[W_{\mu\nu}W^{\nu\rho}W_{\rho}^{\mu}],\\
%{\cal O}_W&=(D_\mu\Phi)^\dagger W^{\mu\nu}(D_\nu\Phi),\\
%{\cal O}_B&=(D_\mu\Phi)^\dagger B^{\mu\nu}(D_\nu\Phi),\\
%{\cal O}_{\tilde WWW}&=\mbox{Tr}[{\tilde W}_{\mu\nu}W^{\nu\rho}W_{\rho}^{\mu}],\\
%{\cal O}_{\tilde W}&=(D_\mu\Phi)^\dagger {\tilde W}^{\mu\nu}(D_\nu\Phi),
%\label{eq:dim6}
%\end{aligned}
%\end{equation}

\subsection{On-shell $W^+W^+$ scattering in SMEFT}
\label{onshellSMEFT}
%operatory
The list of our gQGC dimension-8 operators that contribute to the $WWWW$ vertex reads~\cite{Degrande:2013rea}\footnote{$M6$ is redundant: $\mathcal{O}_{M6}=\frac{1}{2}\mathcal{O}_{M0}$; we usually omit this operator in further analysis.}:
\begin{equation}
\begin{aligned}
  {\cal O}_{S0} &= \left [ \left ( D_\mu \Phi \right)^\dagger
 D_\nu \Phi \right ] \times
\left [ \left ( D^\mu \Phi \right)^\dagger
D^\nu \Phi \right ],
\\
  {\cal O}_{S1} &= \left [ \left ( D_\mu \Phi \right)^\dagger
 D^\mu \Phi  \right ] \times
\left [ \left ( D_\nu \Phi \right)^\dagger
D^\nu \Phi \right ],
\\
 {\cal O}_{M0} &=   \hbox{Tr}\left [ W_{\mu\nu} W^{\mu\nu} \right ]
\times  \left [ \left ( D_\beta \Phi \right)^\dagger
D^\beta \Phi \right ],
\\
 {\cal O}_{M1} &=   \hbox{Tr}\left [ W_{\mu\nu} W^{\nu\beta} \right ]
\times  \left [ \left ( D_\beta \Phi \right)^\dagger
D^\mu \Phi \right ],
\\
  {\cal O}_{M6} &= \left [ \left ( D_\mu \Phi \right)^\dagger W_{\beta\nu}
W^{\beta\nu} D^\mu \Phi  \right ],
\\
  {\cal O}_{M7} &= \left [ \left ( D_\mu \Phi \right)^\dagger W_{\beta\nu}
W^{\beta\mu} D^\nu \Phi  \right ],
\\
 {\cal O}_{T0} &=   \hbox{Tr}\left [ W_{\mu\nu} W^{\mu\nu} \right ]
\times   \hbox{Tr}\left [ W_{\alpha\beta} W^{\alpha\beta} \right ],
\\
 {\cal O}_{T1} &=   \hbox{Tr}\left [ W_{\alpha\nu} W^{\mu\beta} \right ] 
\times   \hbox{Tr}\left [ W_{\mu\beta} W^{\alpha\nu} \right ],
\\
 {\cal O}_{T2} &=   \hbox{Tr}\left [ W_{\alpha\mu} W^{\mu\beta} \right ]
\times   \hbox{Tr}\left [ W_{\beta\nu} W^{\nu\alpha} \right ] .
\end{aligned}
\label{eq:onshellSMEFT1}
\end{equation}
The EFT ''models'' that we study in SMEFT are SM plus a single operator from the list~\eqref{eq:onshellSMEFT1}, each. The on-shell results are presented in the following normalization
\begin{equation}
\Delta\mathcal{L}\supset \sum_{i=S0,S1,M0,M1,M7,T0,T1,T2}f_i\mathcal{O}_i, 
\label{eq:onshellSMEFT1p}
\end{equation}
where the dimensionful $f_i$ are defined via eq.~\eqref{eq:SMEFT2}. The above normalization is not the one predicted by the NDA master formula, but for the on-shell analysis, where we illustrate qualitative features of $W^+W^+$ scattering, a particular choice if normalization is irrelevant. Our choice here is dictated simply by convenience.
%the purpose %

As anticipated, in this Section we give an overview of the behavior of individual helicity amplitudes as a function of energy and their contributions to the total unpolarized cross section, with special attention paid to the partial wave unitarity constraints in the EFT ``models''. We shall illustrate the main points using the operators ${\cal O}_{S0}$ and ${\cal O}_{T1}$ as examples. The qualitative picture remains the same for the other operators as well. The analysis is similar to what already presented in Chapter~\ref{sec:VVinSM} for the SM. The difference is that now, in the presence of non-renormalizable operators, unitarity is necessarily violated above a certain energy scale denoted by $\sqrt{s^U}$, limiting the range of validity of EFT amplitudes to c.o.m. energies $\sqrt{s}$ lower than $\sqrt{s^U}$, i.e. $\sqrt{s}<\sqrt{s^U}$. It is a consequence of the fact that in the presence of non-renormalizable operators amplitudes grow with energy. In the SM the unitarity violation in partial waves did not take place, but occurred in our discussion only hypothetically as a consequence of the $\sim1$ TeV Higgs mass. In Chapter~\ref{sec:VVinSM} we discussed two ways how to regularize the Coulomb singularity, so that the partial waves can be sensibly determined. First, one could consider the custodial limit $g'\rightarrow 0$. An alternative was to subtract the Coulomb pole in the amplitudes and it was the latter that we used to determine the bound on the Higgs mass scale above which fully longitudinal scattering amplitude identically violates unitarity. Here we use another regularization: a $1^{\circ}$ cut in the integration region in eq.~\eqref{eq:VVscattering1p} is applied for partial amplitudes in both forward and backward region. All the computations concerning the unitarity limits, with the emphasis on diagonalization in the helicity space, have been performed using our dedicated \verb-Mathematica- codes. These computations were cross-checked with the \verb-VBFNLO- calculator~\cite{Arnold:2008rz} (where the same $1^{\circ}$ cut is applied) and very good agreement was found ($<5\%$ differences). Concerning cross sections determination, same as in the on-shell analysis in the SM, we apply a $10^{\circ}$ cut in the forward and backward scattering regions. The manually determined Feynman rules were cross checked with the output of the \verb-FeynRules- package~\cite{Christensen:2008py}. The analytic formulas for tree-level scattering amplitudes were obtained semi-automatically with a use of our dedicated \verb-C++- routines that governed Lorentz index contractions etc. These computations were cross checked with the help of the \verb-FeynCalc- package~\cite{Mertig:1990an}. For additional ready to use \verb-Mathematica- codes for tree-level electroweak scattering see~\cite{Romao:2016ien}.

%analytical formulae
Let us consider  the elastic on-shell $W^+W^+\rightarrow W^+W^+$ in the presence of BSM part in a form of a single at a time dimension-8 operator.  The scattering amplitude $iM$ can be written as:
\begin{equation}
iM = A_{SM}+A_{BSM},
\label{eq:pk1}
\end{equation}%
where $A_{SM}$ denotes the SM part and $A_{BSM}$ represents the BSM part that depends on a single Wilson coefficient $f_i$. 
As discussed in Chapter~\ref{sec:VVinSM}, in principle there are 81 independent polarization combinations for massive vector bosons, but we shall apply the reductions in terms of the classes of inequivalent polarization combinations. In particular there are 13 (17) such classes in case of $W^+W^+$ ($W^+W^-$) elastic scattering. 

We start with discussing the scattering energy $\sqrt{s^U}$ at which partial wave unitarity is violated by different helicity amplitudes for the two exemplary operators. We shall apply the tree-level bounds from eq.~\eqref{eq:uni28} which are equivalent to~\eqref{eq:re} that apply when there is no spin-flip  (the tree-level amplitudes $\mathcal{M}$ are real). These bounds are for each helicity combination separately. We also analyse the diagonalized version from eq.~\eqref{eq:uniDiagRe}. The result are shown in Table \ref{tab:unitarityS0text} for the ${\cal O}_{S0}$ operator (positive $f$) and in Table \ref{tab:unitarityT1text} for ${\cal O}_{T1}$ (negative $f$), as a function of the values of $f$, where ''diag.'' denotes the bounds from diagonalization using the $j=0$ partial waves.

\begin{table}[h]%
\begin{center}
\begin{tabular}{c|cccc}
 $ \lambda_1\lambda_2\lambda_1'\lambda_2' $ & $ 0.01 $ & $ 0.1 $ & $ 1. $ & $ 10. $ \\ \hline $
 \text{- - - -} $ & $ \text{x} $ & $ \text{x} $ & $ \text{x} $ & $ \text{x} $  \\ $
 \text{- - - 0} $ & $ \text{x} $ & $ \text{x} $ & $ \text{x} $ & $ \text{x} $ \\ $
 \text{- - - +} $ & $ \text{x} $ & $ \text{x} $ & $ \text{x} $ & $ \text{x} $  \\ $
 \text{- - 0 0} $ & $ 620. $ & $ 200. $ & $ 62. $ & $ 20. $  \\ $
 \text{- - 0 +} $ & $ \text{x} $ & $ \text{x} $ & $ \text{x} $ & $ \text{x} $  \\ $
 \text{- - + +} $ & $ \text{x} $ & $ \text{x} $ & $ \text{x} $ & $ \text{x} $  \\ $
 \text{- 0 - 0} $ & $ \text{x} $ & $ \text{x} $ & $ \text{x} $ & $ \text{x} $  \\ $
 \text{- 0 - +} $ & $ \text{x} $ & $ \text{x} $ & $ \text{x} $ & $ \text{x} $  \\ $
 \text{- 0 0 0} $ & $ \text{x} $ & $ \text{x} $ & $ \text{x} $ & $ \text{x} $  \\ $
 \text{- 0 0 +} $ & $ \text{x} $ & $ \text{x} $ & $ \text{x} $ & $ \text{x} $  \\ $
 \text{- + - +} $ & $ \text{x} $ & $ \text{x} $ & $ \text{x} $ & $ \text{x} $  \\ $
 \text{- + 0 0} $ & $ \text{x} $ & $ \text{x} $ & $ \text{x} $ & $ \text{x} $  \\ $
 \text{0 0 0 0} $ & $ 8.6 $ & $ 4.9 $ & $ 2.7 $ & $ 1.5 $ \\ \hline $
 \text{diag.} $ & $ 8.6 $ & $ 4.9 $ & $ 2.7 $ & $ 1.5 $  \\ \hline
\end{tabular}
\end{center}
\caption{Values of $\sqrt{s^U}$ (in TeV) from the tree-level partial wave unitarity bounds for all elastic on-shell $W^+W^+$ helicity amplitudes for a chosen set of $f_{S0}$ values (first row, in TeV$^{−4}$
); $\lambda_i$ ($\lambda_i'$) denote ingoing (outgoing) W's helicities; ''$x$'' denotes no unitarity violation; ''diag.'' denotes unitarity bounds from diagonalization in the helicity space.
}
\label{tab:unitarityS0text}
\end{table}

\begin{table}[h]%
\begin{center}
\begin{tabular}{c|cccc}
 $\lambda_1\lambda_2\lambda_1'\lambda_2' $ & $ -0.01 $ & $ -0.1 $ & $ -1. $ & $ -10. $ \\ \hline $
 \text{- - - -} $ & $ 6.5 $ & $ 3.7 $ & $ 2.1 $ & $ 1.2 $ \\ $
 \text{- - - 0} $ & $ 2.9\times 10^7 $ & $ 2.9\times 10^6 $ & $ 2.9\times 10^5 $ & $ 2.9\times 10^4 $ \\ $
 \text{- - - +} $  & $ 1.3\times 10^3 $ & $ 410. $ & $ 130. $ & $ 41. $ \\ $
 \text{- - 0 0} $ & $ 410. $ & $ 130. $ & $ 41. $ & $ 13. $ \\ $
 \text{- - 0 +} $ & $ 72. $ & $ 34. $ & $ 16. $ & $ 7.2 $ \\ $
 \text{- - + +} $ & $ 6.6 $ & $ 3.7 $ & $ 2.1 $ & $ 1.2 $ \\ $
 \text{- 0 - 0} $ & $ 1.6\times 10^3 $ & $ 510. $ & $ 160. $ & $ 51. $ \\ $
 \text{- 0 - +} $ & $ 60. $ & $ 28. $ & $ 13. $ & $ 6.0 $ \\ $
 \text{- 0 0 0} $ & $ 1.5\times 10^7 $ & $ 1.5\times 10^6 $ & $ 1.5\times 10^5 $ & $ 1.5\times 10^4 $ \\ $
 \text{- 0 0 +} $ & $ 1.5\times 10^3 $ & $ 480. $ & $ 150. $ & $ 48. $ \\ $
 \text{- + - +} $ & $ 9.9 $ & $ 5.6 $ & $ 3.2 $ & $ 1.8 $ \\ $
 \text{- + 0 0} $ & $ 1.3\times 10^3 $ & $ 410. $ & $ 130. $ & $ 41. $ \\ $
 \text{0 0 0 0} $ & $ \text{x} $ & $ \text{x} $ & $ \text{x} $ & $ \text{x} $ \\ \hline$
 \text{diag.} $ & $ 5.3 $ & $ 3.0 $ & $ 1.7 $ & $ 0.97$ \\ \hline

\end{tabular}
\end{center}
\caption{Values of $\sqrt{s^U}$ (in TeV) from the tree-level partial wave unitarity bounds for all elastic on-shell $W^+W^+$ helicity amplitudes for a chosen set of $f_{T1}$ values (first row, in TeV$^{−4}$
); $\lambda_i$ ($\lambda_i'$) denote ingoing (outgoing) W's helicities; ''$x$'' denotes no unitarity violation; ''diag.'' denotes unitarity bounds from diagonalization in the helicity space.
}
\label{tab:unitarityT1text}
\end{table}

We see that partial wave unitarity is first violated in the $0000$ amplitude for the first operator and in $----$ (very closely followed by $--++$ and somewhat closely followed by $-+-+$) for the second one.  Unitarity is violated at vastly different energies for different helicity amplitudes, depending on the operator considered.  The analytic formulae for the leading energy dependences of the helicity amplitudes $iM$ and the partial waves $\mathcal{T}^{(j=j_{min})}_{\lambda_1'\lambda_2';\lambda_1\lambda_2}$ are shown in Tab.~\ref{tab:analyticS0text} and~\ref{tab:analyticT1text}. The corresponding formulas for the remaining SMEFT operators are listed in Appendix~\ref{app:analiticFormSMEFT}.
\begin{table}[h]%
\begin{center}
\bgroup
\def\arraystretch{1.3}
\begin{tabular}{c|c|c}
 $\lambda_1\lambda_2\lambda_1'\lambda_2' $ &  $i\mathcal{M}_{\lambda_1'\lambda_2';\lambda_1\lambda_2}(\sqrt{s},\theta ,f_{S0}) $ & $ \mathcal{T}_{\lambda_1'\lambda_2';\lambda_1\lambda_2}^{(j=j_{min})} $ \\ \hline $
 \text{- - - -} $ & $ 4 i c_W^4 f_{\text{S0}} m_Z^4-\frac{32 i \csc ^2(\theta ) c_W^2 m_Z^2}{v^2} $ & $ \frac{c_W^4 f_{\text{S0}} m_Z^4}{4 \pi } $ \\ $
 \text{- - - 0} $ & $ \text{pure SM } O(\frac{1}{s^{3/2}}) \text{ term} $ & $ 0 $  \\ $
 \text{- - - +} $ & $ \frac{16 i c_W^4 m_Z^4}{s v^2} $ & $ \frac{c_W^4 m_Z^4}{\sqrt{6} \pi  s v^2} $ \\ $
 \text{- - 0 0} $ & $ 2 i s c_W^2 f_{\text{S0}} m_Z^2 $ & $ \frac{s c_W^2 f_{\text{S0}} m_Z^2}{8 \pi } $ \\ $
 \text{- - 0 +} $ & $ \frac{4 i \sqrt{2} \cot (\theta ) c_W^3 m_Z^3 \left(\left(4 c_W^2+1\right) m_Z^2-m_h^2\right)}{s^{3/2} v^2} $ & $ -\frac{c_W^3 m_Z^3
   \left(\left(4 c_W^2+1\right) m_Z^2-m_h^2\right)}{4 \sqrt{3} \pi  s^{3/2} v^2} $ \\ $
 \text{- - + +} $ & $ 4 i c_W^4 f_{\text{S0}} m_Z^4 $ & $ \frac{c_W^4 f_{\text{S0}} m_Z^4}{4 \pi } $ \\ $
 \text{- 0 - 0} $ & $ \frac{8 i c_W^2 m_Z^2}{v^2 (\cos (\theta )-1)} $ & $ 0 $ \\ $
 \text{- 0 - +} $ & $ -\frac{8 i \sqrt{2} \cot \left(\frac{\theta }{2}\right) c_W^3 m_Z^3}{\sqrt{s} v^2} $ & $ \frac{\sqrt{2} c_W^3 m_Z^3}{3 \pi  \sqrt{s} v^2} $ \\ $
 -000 $ & $ \frac{2 i \sqrt{2} \cot (\theta ) c_W m_Z \left(m_Z-m_h\right) \left(m_h+m_Z\right)}{\sqrt{s} v^2} $ & $ \frac{c_W m_Z \left(m_Z^2-m_h^2\right)}{8
   \sqrt{3} \pi  \sqrt{s} v^2} $ \\ $
 \text{- 0 0 +} $ & $ \frac{4 i c_W^2 m_Z^2 \left(\cos (\theta ) \left(\left(4 c_W^2+1\right) m_Z^2-m_h^2\right)-4 c_W^2 m_Z^2\right)}{s v^2 (\cos (\theta )-1)} $ & $
   \frac{c_W^2 m_Z^2 \left(\left(4 c_W^2+1\right) m_Z^2-m_h^2\right)}{8 \pi  s v^2} $ \\ $
 \text{- + - +} $ & $ -\frac{8 i \cot ^2\left(\frac{\theta }{2}\right) c_W^2 m_Z^2}{v^2} $ & $ \frac{c_W^2 m_Z^2}{6 \pi  v^2} $ \\ $
 \text{- + 0 0} $ & $ -\frac{4 i c_W^2 m_Z^2 \left(2 \left(m_h^2-m_Z^2\right) \csc ^2(\theta )-m_h^2+\left(8 c_W^2+1\right) m_Z^2\right)}{s v^2} $ & $ -\frac{c_W^2
   m_Z^2 \left(m_h^2+\left(4 c_W^2-1\right) m_Z^2\right)}{2 \sqrt{6} \pi  s v^2} $ \\ $
 \text{0 0 0 0} $ & $ i s^2 f_{\text{S0}} $ & $ \frac{s^2 f_{\text{S0}}}{16 \pi }$ \\ \hline

\end{tabular}
\egroup
\caption{Analytic formulas for on-shell $W^+W^+$ elastic scattering helicity amplitudes $iM$ and the minimal $j$ partial waves $\mathcal{T}^{(j=j_{min})}$; $\lambda_i$ ($\lambda_i'$) denote ingoing (outgoing) W's helicities; $c_W\equiv \cos\theta_W$; $\theta$ is the scattering angle; the $\mathcal{O}_{S0}$ operator case; only leading terms in the $\sqrt{s}$  expansion in the limit $s\rightarrow\infty$ are shown.}
\label{tab:analyticS0text}
\end{center}
\end{table}
%%%
\begin{table}[h]%
\begin{center}
\bgroup
\def\arraystretch{1.3}
\begin{tabular}{c|c|c}
 $\lambda_1\lambda_2\lambda_1'\lambda_2' $ &  $i\mathcal{M}_{\lambda_1'\lambda_2';\lambda_1\lambda_2}(\sqrt{s},\theta ,f_{T1}) $ & $ \mathcal{T}_{\lambda_1'\lambda_2';\lambda_1\lambda_2}^{(j=j_{min})} $ \\ \hline $
 \text{- - - -} $ & $ 2 i s^2 f_{\text{T1}} $ & $ \frac{s^2 f_{\text{T1}}}{8 \pi } $ \\ $
 \text{- - - 0} $ & $ -i \sqrt{2} \sqrt{s} c_W^3 f_{\text{T1}} \sin (\theta ) \cos (\theta ) m_Z^3 $ & $ \frac{\sqrt{s} c_W^3
   f_{\text{T1}} m_Z^3}{40 \sqrt{3} \pi } $ \\ $
 \text{- - - +} $ & $ -i s c_W^2 f_{\text{T1}} \sin ^2(\theta ) m_Z^2 $ & $ -\frac{s c_W^2 f_{\text{T1}} m_Z^2}{20 \sqrt{6} \pi } $ \\ $
 \text{- - 0 0} $ & $ -\frac{1}{2} i s c_W^2 f_{\text{T1}} (\cos (2 \theta )-9) m_Z^2 $ & $ \frac{7 s c_W^2 f_{\text{T1}} m_Z^2}{24 \pi
   } $ \\ $
 \text{- - 0 +} $ & $ -\frac{i s^{3/2} c_W f_{\text{T1}} \sin (\theta ) \cos (\theta ) m_Z}{\sqrt{2}} $ & $ \frac{s^{3/2} c_W
   f_{\text{T1}} m_Z}{80 \sqrt{3} \pi } $ \\ $
 \text{- - + +} $ & $ \frac{1}{4} i s^2 f_{\text{T1}} (\cos (2 \theta )+11) $ & $ \frac{s^2 f_{\text{T1}}}{6 \pi } $ \\ $
 \text{- 0 - 0} $ & $ -\frac{1}{2} i s c_W^2 f_{\text{T1}} \sin ^2(\theta ) m_Z^2 $ & $ -\frac{s c_W^2 f_{\text{T1}} m_Z^2}{96 \pi } $ \\ $
 \text{- 0 - +} $ & $ i \sqrt{2} s^{3/2} c_W f_{\text{T1}} \sin \left(\frac{\theta }{2}\right) \cos ^3\left(\frac{\theta
   }{2}\right) m_Z $ & $ -\frac{s^{3/2} c_W f_{\text{T1}} m_Z}{80 \sqrt{2} \pi } $ \\ $
 \text{- 0 0 0} $ & $ -i \sqrt{2} \sqrt{s} c_W^3 f_{\text{T1}} \sin (2 \theta ) m_Z^3 $ & $ -\frac{\sqrt{s} c_W^3 f_{\text{T1}}
   m_Z^3}{20 \sqrt{3} \pi } $ \\ $
 \text{- 0 0 +} $ & $ \frac{1}{4} i s c_W^2 f_{\text{T1}} (4 \cos (\theta )+3 \cos (2 \theta )+1) m_Z^2 $ & $ \frac{s c_W^2
   f_{\text{T1}} m_Z^2}{96 \pi } $ \\ $
 \text{- + - +} $ & $ i s^2 f_{\text{T1}} \cos ^4\left(\frac{\theta }{2}\right) $ & $ \frac{s^2 f_{\text{T1}}}{80 \pi } $ \\ $
 \text{- + 0 0} $ & $ i s c_W^2 f_{\text{T1}} \sin ^2(\theta ) m_Z^2 $ & $ \frac{s c_W^2 f_{\text{T1}} m_Z^2}{20 \sqrt{6} \pi } $ \\ $
 \text{0 0 0 0} $ & $ \frac{2 i \left(2 v^2 c_W^4 f_{\text{T1}} \cos (2 \theta ) m_Z^4+6 v^2 c_W^4 f_{\text{T1}} m_Z^4-m_h^2-4 \csc
   ^2(\theta ) m_Z^2+m_Z^2\right)}{v^2} $ & $ \frac{16 v^4 c_W^4 f_{\text{T1}} m_Z^4+3 v^2 \left(m_Z^2-m_h^2\right)}{24 \pi  v^4} $ \\ \hline

\end{tabular}
\egroup
\caption{For description, see caption of Fig.~\ref{tab:analyticS0text}; the $\mathcal{O}_{T1}$ operator case.}
\label{tab:analyticT1text}
\end{center}
\end{table}

%unitarity
Unitarity bound calculated from diagonalization in the helicity space of $j=0$ is virtually identical to the strongest bound among bounds for helicity partial waves for ${\cal O}_{S0}$, while 
for ${\cal O}_{T1}$ they are about 20\% lower. From now on we denote by $\sqrt{s^U}$ the ''diagonalized'' unitarity bounds unless explicitly stated. 

\begin{figure}[ht]%
\begin{center}
\includegraphics[width=0.7\columnwidth]{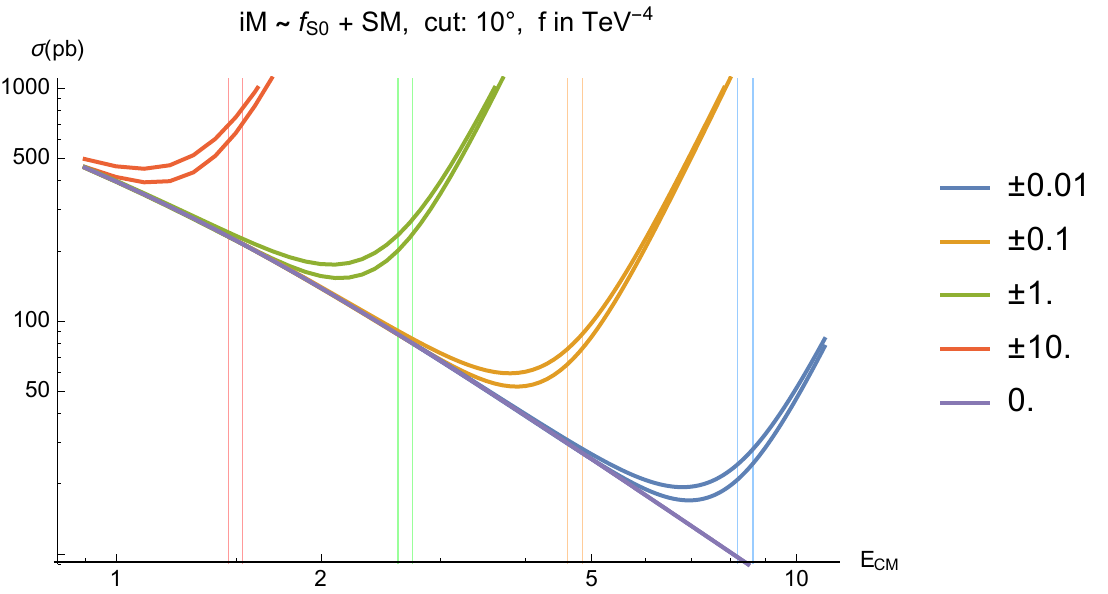}%
\end{center}
\caption{Energy dependence of the total unpolarized elastic on-shell $W^+W^+$ cross sections ($E_{CM} \equiv \sqrt{s}$, in TeV) for a chosen set of $f_i$ values of the SMEFT operators studied.  Vertical lines denote the unitarity bound $\sqrt{s^U}$ (color correspondence).  There is no color distinction between the signs: upper cross section curves correspond to $f<0$; stronger unitarity limits correspond to $f<0$.}
\label{fig:tutalUnpolS0text}
\end{figure}

\begin{figure}[ht]
\begin{center}
 \includegraphics[width=0.9\columnwidth]{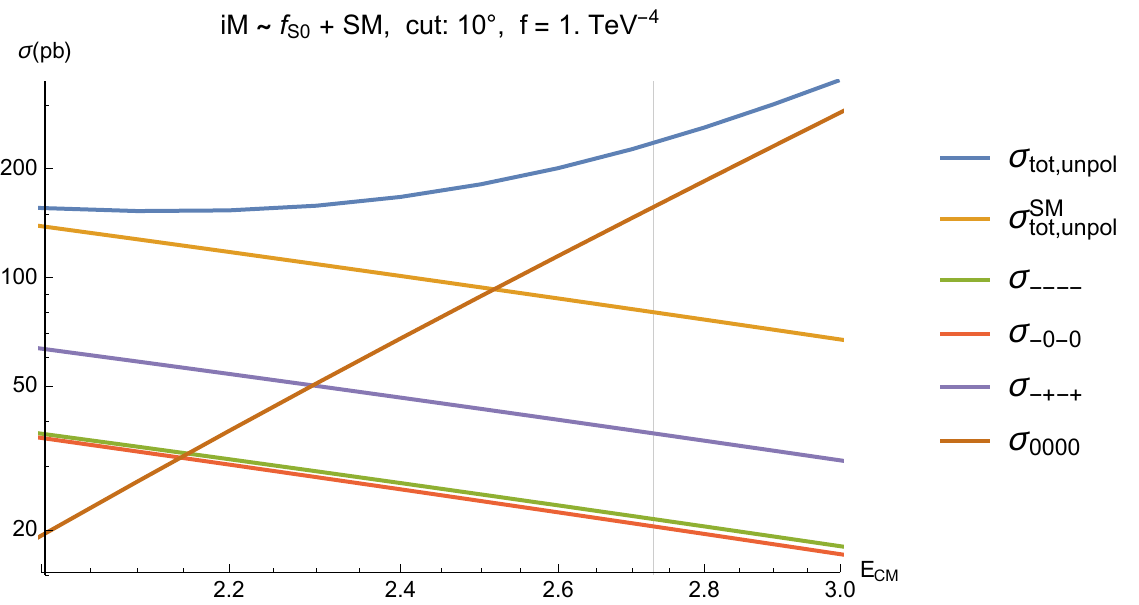}
\end{center}
\caption{Contributions of the polarized cross sections (multiplicity taken into account) as functions of the center-of-mass collision energy
($E_{CM} \equiv \sqrt{s}$, in TeV) for a chosen value of $f_i>0$ for the SMEFT ${\cal O}_{S0}$ operator. The remaining (not shown) polarized contributions are negligibly small. In each plot shown is in addition the total cross section and the total cross section in the SM. }
\label{fig:polsSMEFTS0text}
\end{figure}

\begin{figure}[ht]%
\begin{center}
\includegraphics[width=0.7\columnwidth]{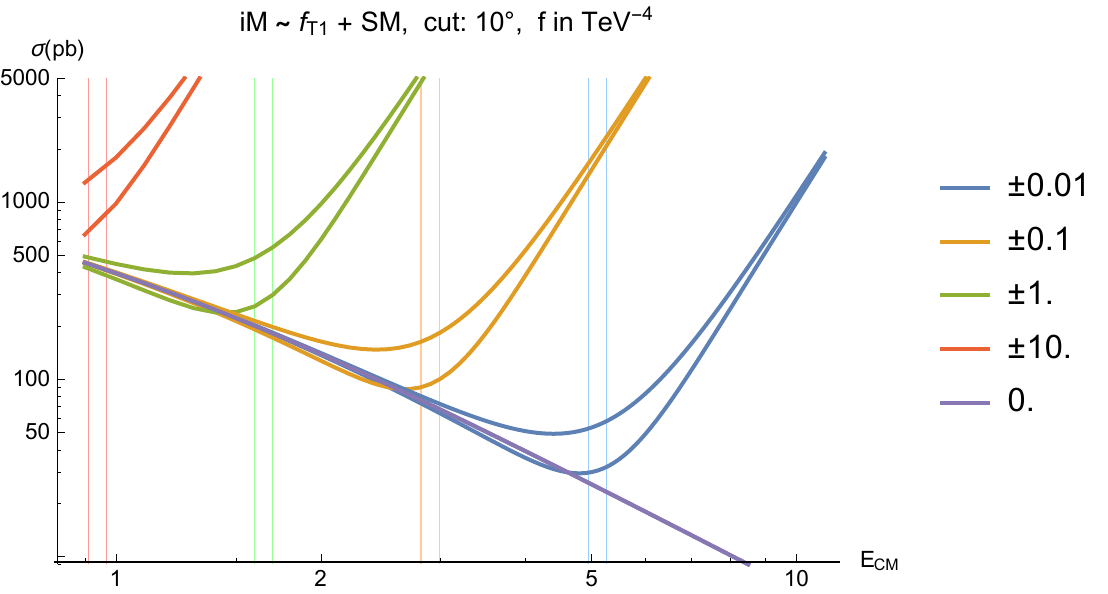}%
\end{center}
\caption{For description see the caption of Fig.~\ref{fig:tutalUnpolS0text}; the ${\cal O}_{T1}$ case.}
\label{fig:tutalUnpolT1text}
\end{figure}

\begin{figure}[ht]
\begin{center}
 \includegraphics[width=0.9\columnwidth]{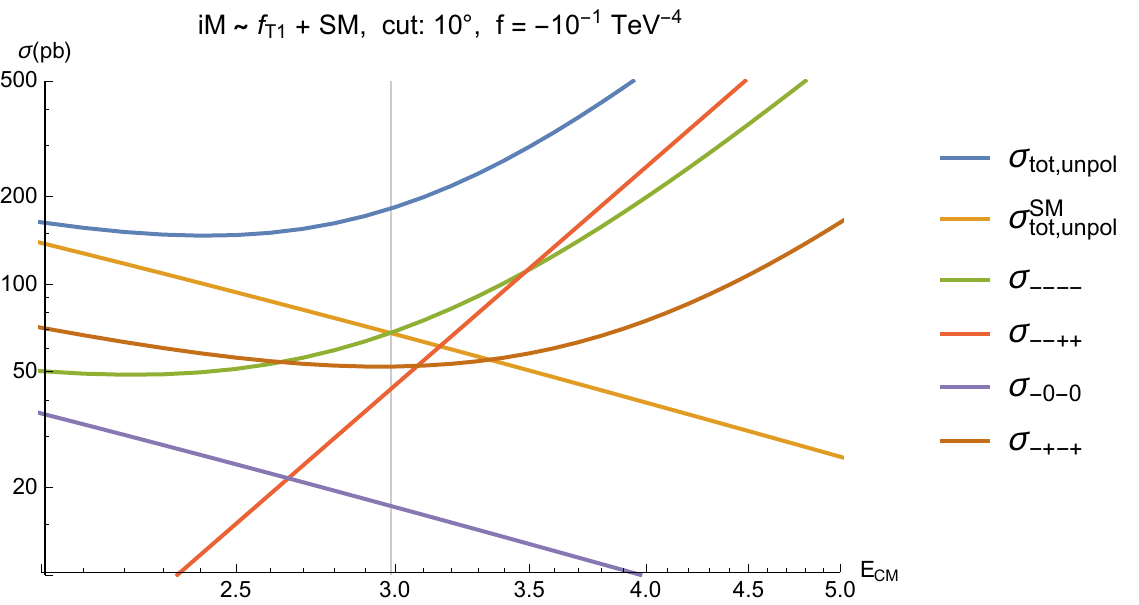}
\end{center}
\caption{For description see the caption of Fig.~\ref{fig:polsSMEFTS0text}; the $f<0$, ${\cal O}_{T1}$ case.}
\label{fig:polsSMEFTT1text}
\end{figure}

%przekroj czynny
We continue with discussion of the on-shell cross sections. 
The total unpolarized on-shell $WW$ cross section can schematically be written as:
\begin{equation}
\sigma\sim\frac{1}{9}\ \sum_{i,j,k,l}\ \ \left|A_{SM}(ij\rightarrow kl)\right|^2 + (A_{SM}(ij\rightarrow kl)A_{BSM}(ij\rightarrow kl)^{\ast} + h.c.) + \left|A_{BSM}(ij\rightarrow kl)\right|^2
\label{eq:pk3}
\end{equation}
Since in the hypothetical on-shell $WW$ scattering (weak bosons are not stable particles) the helicity is an observable, different helicity amplitudes $iM(ij\rightarrow kl)$ (corresponding to helicity configurations $(ijkl)$) do not interfere among themselves, which is explicitly accounted for in~\eqref{eq:pk3}.
Some examples of the c.o.m energy dependence of the cross sections for both the operators are shown in the following figures: for the total unpolarized cross sections with ${\cal O}_{S0}$ in  Fig.~\ref{fig:tutalUnpolS0text}, for polarized cross sections with ${\cal O}_{S0}$ in Fig.~\ref{fig:polsSMEFTS0text}, for unpolarized with ${\cal O}_{T1}$ in Fig.~\ref{fig:tutalUnpolT1text}, and for polarized with ${\cal O}_{T1}$ in Fig.~\ref{fig:polsSMEFTT1text}. Corresponding plots for all the SMEFT operators studied in this work are compiled in figures of Appendix~\ref{app:plotsTotPolAndUnpolSMEFT}.

We observe that both operators show several similar interesting features.  Below $s^U$, sizable deviations from the SM predictions occur only for small energy intervals close to those bounds.  
This is the region where the quadratic term in Eq.~\eqref{eq:pk3} begins to dominate BSM effects
(see Figs.~\ref{fig:tutalUnpolS0text} and \ref{fig:tutalUnpolT1text}).
The contribution of the interference term (the middle term in parenthesis in eq.~\eqref{eq:pk3}) at this point generally depends on
which helicity combinations get affected by a given operator and how much they
contribute to the total cross section in the SM (see Sec.~\ref{interference} for a more detailed discussion).  
%Interference is visible
%for ${\cal O}_{S0}$, ${\cal O}_{S1}$, ${\cal O}_{T0}$, ${\cal O}_{T1}$, ${\cal O}_{T2}$,
%${\cal O}_{M1}$, ${\cal O}_{M7}$, and negligible for ${\cal O}_{M0}$ and ${\cal O}_{M6}$.
%}\
The energy dependence of the unpolarized cross sections around $s^U$ is moreover somewhat weakened by the contribution from the  helicity amplitudes that have not reached the unitarity limit.

Numerical results for contributions of polarized cross sections to the total unpolarized cross section at $s^U$ are shown in the Appendix~\ref{app:analiticFormSMEFT} as a function of $f_i$, for all the SMEFT operators studied.
One can see (also from the polarized contributions plots) that for example for ${\cal O}_{S0}$, the $0000$ cross section (related to the amplitude which violates unitarity first) gives about  65\% of the total cross sections, independently of the value of $f$, for the corresponding values of $s^U$.  For ${\cal O}_{T1}$, it is the $----$ cross section, closely followed by $-+-+$ and $--++$ with an about 90\% combined contribution to the total unpolarized cross section, independently of the value of $f$, for the corresponding values $s^U$.  The rest of the unpolarized cross sections at $s^U$ come (for both operators) from the helicity amplitudes that saturate the cross section in the SM, which either remain constant with energy (although weakly dependent on the value of $f$) or violate perturbative partial 
unitarity at a higher energy.

Importantly, correct assessment of the EFT ''model'' validity range in the $W^+W^+$ scattering process
requires in general also consideration of the $W^+W^-$ scattering amplitudes which by construction
probe the same couplings and are sensitive to exactly the same operators. More precisely, in the end the $\sqrt{s^U}$ should be identified as the stronger unitarity limit between same and opposite $WW$ scattering. For most
higher dimension operators,
this usually significantly reduces their range of validity in $W^+W^+$ analyses. Actually, the choice of $S0$ and $T1$ examples and their Wilson coefficient signs were determined by the requirement that the $s^U$ bounds for $W^+W^+$ are stronger than for $W^+W^-$, in order ease the discussion. Tree-level partial wave unitarity bounds for the full set of operators, for both signs of $f_i$ for both same and opposite sign $WW$ scattering are shown in the Appendix~\ref{app:unitLimitsSMEFT}. The above issue is taken into account in all the vertical lines denoting unitarity limits (e.g. in Fig.~\ref{fig:tutalUnpolS0text} or~\ref{fig:polsSMEFTS0text}). The vertical lines denoting unitarity limits in all the same-sign cross section plots are determined accounting for the opposite $WW$ scattering.

%przekroj - rysunek zbiorczy f<0 i f>0
%The contribution from the saturating helicities to the total unpolarized
%cross-section as a function of c.o.m energy for all the operators is compiled in Fig.~\ref{} (~\ref{}) for $f_i>0$ ($f_i<0$). In Fig.~\ref{} compiled is energy behavior of the total unpolarized cross section for both signs of $f_i$ at once.

In the presence of $n=8$ operators some of the saturating helicities grow with $s$, maximally as $s^2$. 
For each ``EFT model'' there is at least one polarization configuration with the asymptotic $s^2$ energy dependence providing  dominant contribution to the unpolarized cross section at $M_{WW}=\sqrt{s^U}\equiv M^U$. In particular, in the case of ``EFT models'' with scalar operators ($S$)  only  the amplitude with all $W$ bosons polarized longitudinally grows as $s^2$. In the case of transverse  operators ($T$)   some  amplitudes with all $W$ polarized transversally grow as $s^2$, while  for the case of mixed operators ($M$) it happens for amplitudes with two longitudinal and  two  transverse polarizations. It  follows from $D_\mu \Phi$ and $W_{\mu\nu}$ building blocks of BSM operators  which project mostly on the longitudinal and transverse modes, respectively. It is interesting to notice, however,  that for different $S,\, T$  and $M$  distinct  polarization configurations of the outgoing $W$'s  dominate the total cross section at large $M_{WW}$. Measurement of final state $W$ polarizations would give an insight to the dynamics of their interactions. We shall comment on that possibility further in the outlook of Chapter~\ref{conclusions}. 
\subsection{Discussion of the interference effects}
\label{interference}
As already noticed below eq.~\eqref{eq:pk3}, helicity is an observable for the on-shell $WW$ scattering reaction and different helicity configurations do not interfere among themselves.  
The sign dependence of the total unpolarized cross section, most visible for $T0,\, T2$ and also present for $T1,\, M1,\, M7,\, S0,\, S1$, is due to 
the interference terms in eq.~\eqref{eq:pk3}, which are the only non-positive definite terms. The dependence on the sign of $f_i$ is determined by the magnitude of 
SM-BSM terms relative to the BSM$^2$ ones in the region $E\lesssim\Lambda\leq M^U$. While there are always BSM$^2$ 
terms that asymptotically behave as $s^4/\Lambda^8$, the growth as 
$\sim s^2/\Lambda^4$ of the interference terms is not necessarily visible in each of the ``EFT models''.  If the helicity configurations for which the amplitude depends on energy as $s^2$ are not among the saturating helicities of the SM, extra suppression factor(s) of $v/\Lambda<<1$ with respect to the opposite case, will be present in the SM-BSM terms. The latter means suppressed sign dependence of the unpolarized cross section, i.e. suppressed interference. That it is the case for the $M0$ operator, it  can be inferred from the polarized contributions plots in figs.~\ref{tab:polsSMEFT},~\ref{tab:polsSMEFT2},~\ref{tab:polsSMEFTNeg}, and~\ref{tab:polsSMEFTNeg2} (Appendix~\ref{app:plotsTotPolAndUnpolSMEFT}). Practically invisible interference for this operator can be directly observed in the unpolarized cross section plot in Fig.~\ref{tab:tutalUnpol} (Appendix~\ref{app:plotsTotPolAndUnpolSMEFT}). The discussion in this paragraph can be regarded as an additional cross-check of our computations. For general theoretical discussion of SM-BSM interference patterns and some interesting implications to phenomenology associated to $n=6$ operators, see~\cite{Azatov:2016sqh}. %27TeVdraftV5 
\clearpage
\subsection{Practical comments on the $\sqrt{s^U}$ computations}
\label{app:unitarity}
The dominating polarization configurations in the total unpolarized cross section can be read from Fig.~\ref{tab:polsSMEFT},~\ref{tab:polsSMEFT2},~\ref{tab:polsSMEFTNeg}, and~\ref{tab:polsSMEFTNeg2} (Appendix~\ref{app:plotsTotPolAndUnpolSMEFT}). However, the helicity combination that determines the $\sqrt{s^U}$, i.e. that yields strongest unitarity bound,   is not necessarily  among them. The reason is as follows. The partial wave expansion of  helicity amplitude starts with $j_{\mathrm{min}}=\mathrm{max\{|\lambda_{1}-\lambda_{2}|,|\lambda_{3}-\lambda_{4}|\}}$, where $\lambda_{1,2}$ and  $\lambda_{3,4}$ correspond to initial and final $W$ polarizations,  and it is the $j=j_{\mathrm{min}}$ partial wave that yields the strongest unitarity limit. It has been checked that  for the same-sign $WW$ all the  helicity amplitudes that depend on energy as $s^2$ for the case of $M$ operators have $j_{\mathrm{min}}=1$, while for  the $S$ and $T$ operators it is $j_{\mathrm{min}}=0$.  It would imply then that the unitarity limit for the $M$ operators would be weaker than for $S$ and $T$, especially if only the $j=0$ partial waves are considered. 
However, the same operators affect both the same-sign and opposite-sign $WW$ scattering processes, and both processes should be 
considered for the  determination of  the unitarity bounds.  In the case of opposite-sign $WW$, for each  ``EFT model'', including the $M$ ones,  there exists a helicity configuration 
that depends on energy as $s^2$ and has $j_{\mathrm{min}}=0$. As a result, for the $M$-type ``EFT model''  the  unitarity limit is stronger as compared to the limit derived  from same-sign $WW$ partial wave expansion, again especially if only $j=0$ partial waves are considered. This should be kept in mind  
in particular when using the VBFNLO calculator to determine the unitarity bounds that both same- and opposite-sign $WW$ scattering 
processes  are looked at (VBFNLO checks only $j=0$ partial waves).  
The helicity combination yielding the strongest helicity partial wave unitarity limits for each operator are summed up in Table~\ref{tab:pk}.
\begin{table}
\vspace{8mm}
\begin{center}
\begin{tabular}{|c|c|c|c|c|c|c|c|c|}
\hline
 $i=$ & $S0$ & $S1$ & $T0$ & $T1$ & $T2$ & $M0$ & $M1$ & $M7$ \\ \hline
$f_i>0$ & $0000$ & $0000$ & $----$ and $--++$ & $--++$ & $----$ & $--00$ & $--00$ & $--00$ \\
\hline
$f_i<0$ & $0000$ & $0000$ & $--++$ & $--++$ & $--++$ & $--00$ & $--00$ & $--00$ \\ \hline
\end{tabular}
\end{center}
\vspace{3mm}
\caption{The helicity combinations yielding the strongest helicity partial waves unitarity limits for each operator in case of each sign of $f_i$. It is always a j=0 partial wave that yields the strongest  unitarity limits}
\label{tab:pk}
\end{table}
\subsection{On-shell $W^+W^+$ scattering in HEFT}
\label{onshellHEFT}
%operatory
Here we analyze the on-shell amplitudes for HEFT ''models'', i.e.  cross sections and unitarity bounds as was done for the SMEFT case in Sec.~\ref{onshellSMEFT}. The different is, of course, the list of operators that specify the EFT ''models'' in the HEFT case. We distinguish the following gQGC operators contributing to the $WWWW$ vertex:
\begin{equation}
\begin{aligned}
\cP_{6}=& \Tr(\V_\mu\V^\mu)\Tr(\V_\nu\V^\nu)\\
\cP_{11}=&  \Tr(\V_\mu\V_\nu)\Tr(\V^\mu\V^\nu) \\[5mm]
\cT_{42}=&\Tr(\V_\alpha W_{\mu\nu})\Tr(\V^\alpha W^{\mu\nu})\\
\cT_{43}=&\Tr(\V_\alpha W_{\mu\nu})\Tr(\V^\nu W^{\mu\alpha})\\
\cT_{44}=&\Tr(\V^\nu W_{\mu\nu})\Tr(\V_\alpha W^{\mu\alpha})\\
\cT_{61}=&W^a_{\mu\nu}W^{a\mu\nu}\Tr(\V_\alpha\V^\alpha )\\
\cT_{62}=&W^a_{\mu\nu}W^{a\mu\alpha}\Tr(\V_\alpha\V^\nu )\\[5mm]
\cO_{T0}=&W^a_{\mu\nu}W^{a\mu\nu}W^b_{\alpha\beta}W^{b\alpha\beta}\\
\cO_{T1}=&W^a_{\alpha\nu}W^{a\mu\beta}W^b_{\mu\beta}W^{b\alpha\nu}\\
\cO_{T2}=&W^a_{\alpha\mu}W^{a\mu\beta}W^b_{\beta\nu}W^{b\nu\alpha},
\end{aligned}
\label{HEFTOps}
\end{equation}
where as previously, $W_{\mu\nu}\equiv W_{\mu\nu}^a\tau^a/2$. To facilitate the matching between this notation and the one used in previous literature, the operators $\cP_{6}$ and $\cP_{11}$ are also known as $\cL_5$ and $\cL_4$, respectively, of the EW$\chi$L.

In papers where these operators are listed according to the number of derivatives, that is the typical counting of the EW$\chi$L, they belong to three different groups: $\cP_6$ and $\cP_{11}$ are listed among the $\cO(p^4)$ operators; $\cT_{42}$, $\cT_{43}$, $\cT_{44}$, $\cT_{61}$ and $\cT_{62}$ are considered $\cO(p^6)$; finally, $\cO_{T0}$, $\cO_{T1}$ and $\cO_{T2}$ are inserted in the $\cO(p^8)$ group. However, at the phenomenological level, the primary dimension is what matters to establish the impact of an operator (Sec.~\ref{primaryDim}). All these operators have $d_p=8$ and therefore the list~\eqref{HEFTOps} is in accord with the discussion of Sec.~\ref{whichNonRen}.

Ref.~\cite{Eboli:2016kko} presents a much longer list of operators which present a similar structure to the ones listed in Eq.~(\ref{HEFTOps}). However, any operator containing the scalar chiral field $\T$ do not contain interaction between four $W$ and therefore should not be considered in this analysis. The other operators that are not listed above and do not  contain $\T$ have higher primary dimension and therefore are expected to provide subdominant contributions.
%trywialne zwiazki ze smeft

Study of $WWWW$ Lorentz structures given by each operator allows to identify a correlation between the HEFT and SMEFT operators:
\begin{equation}
\begin{aligned}
c_6\cP_6\qquad\Longleftrightarrow&\qquad c^{(8)}_{S1}\cO_{S1}\\
c_{11}\cP_{11}\qquad\Longleftrightarrow&\qquad c^{(8)}_{S0}\cO_{}+c^{(8)}_{S_1}\cO_{S1}\\
c_{61}\cT_{61}\qquad\Longleftrightarrow&\qquad c^{(8)}_{M0}\cO_{M0}\\
c_{62}\cT_{62}\qquad\Longleftrightarrow&\qquad c^{(8)}_{M1}\cO_{M1}\, \end{aligned}
\label{eq:correlations}
\end{equation}
(the operators $\cO_{T0}$, $\cO_{T1}$ and $\cO_{T2}$ belong to both the bases and therefore the correlation is trivial). The SMEFT operator $\cO_{M7}$ contains only part of the interactions described by $\cT_{43}$: the other interactions are described by SMEFT with higher dimensions. Finally, and interestingly, the HEFT operators $\cT_{42}$ and $\cT_{44}$ do not have any correspondence with any of the SMEFT operators of dimension 8. Exemplary results for this pair of operators will be presented in this Section. We work with the HEFT operators normalized according to the NDA formula~\eqref{MasterFormula1}:%tutaj
\begin{equation}
\Delta\mathcal{L}\supset \sum_{i=6,11}\frac{c_i}{16\pi^2}\cP_i\F_i(h)+\sum_{i=42,43,44,61,62}\frac{c_{i}}{\Lambda^2}\cT_i\F_i(h)+\sum_{i=0,1,2}\frac{16\pi^2c_{Ti}}{\Lambda^4}\cO_{Ti}\F_i(h)\,;
\label{DeltaLHEFT}
\end{equation}
where $c_i$ are the dimensionless factors introduced upon the NDA normalization. The $f_i$ dimensionful coefficients are defined via $c_i$ and $\Lambda$ solely, i.e.
\begin{equation}
\begin{array}{l}
	f_i\equiv c_i,\qquad (i=6,11), \\
	f_i \equiv \frac{c_i}{\Lambda^2},\qquad (i=42,43,44,61,62),\\
	f_i \equiv \frac{c_i}{\Lambda^4},\qquad (i=T0,T1,T2).\\
\end{array}
\label{eq:onshellHEFT1}
\end{equation}
(Again, a particular choice of normalization for the purpose of our on-shell analysis is irrelevant).
%przekroje spolaryzowane

\begin{figure}[h]
\begin{center}
 \includegraphics[width=0.7\columnwidth]{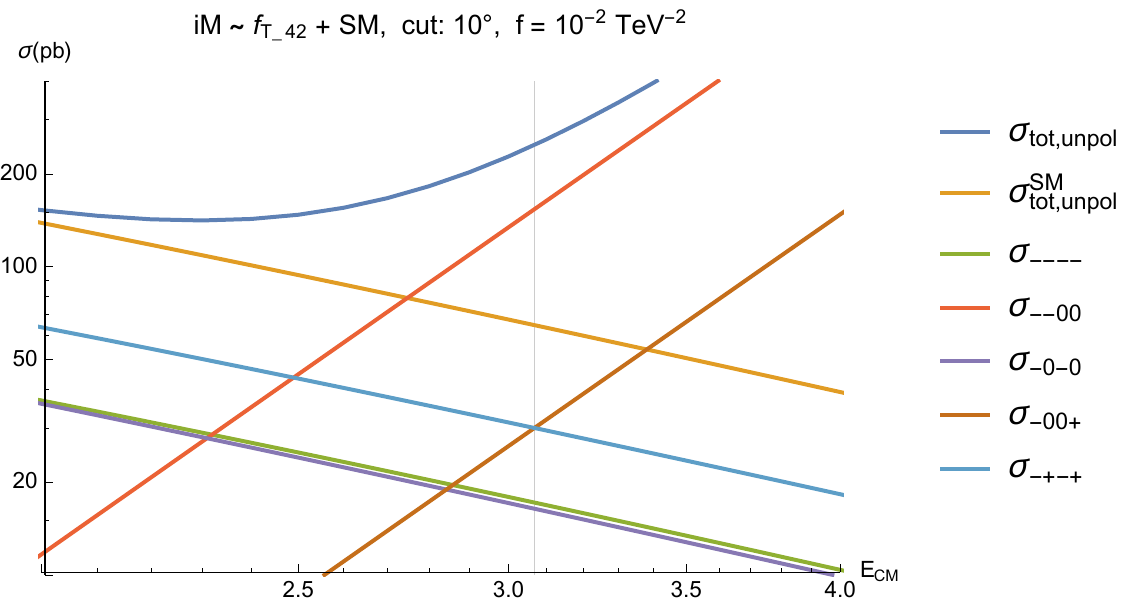} 
\end{center}
\caption{Contributions of the polarized cross sections (multiplicity taken into account) as functions of the center-of-mass collision energy
($E_{CM} \equiv \sqrt{s}$, in TeV) for chosen values of $f_i>0$ of the HEFT $\mathcal{T}_{42}$. The remaining (not shown) polarized contributions are negligibly small. In each plot shown is in addition the total cross section of the EFT ''model'' and the total cross section in the SM.}
\label{fig:polsHEFTT42text}
\end{figure}

\begin{figure}[h]
\begin{center}
 \includegraphics[width=.7\columnwidth]{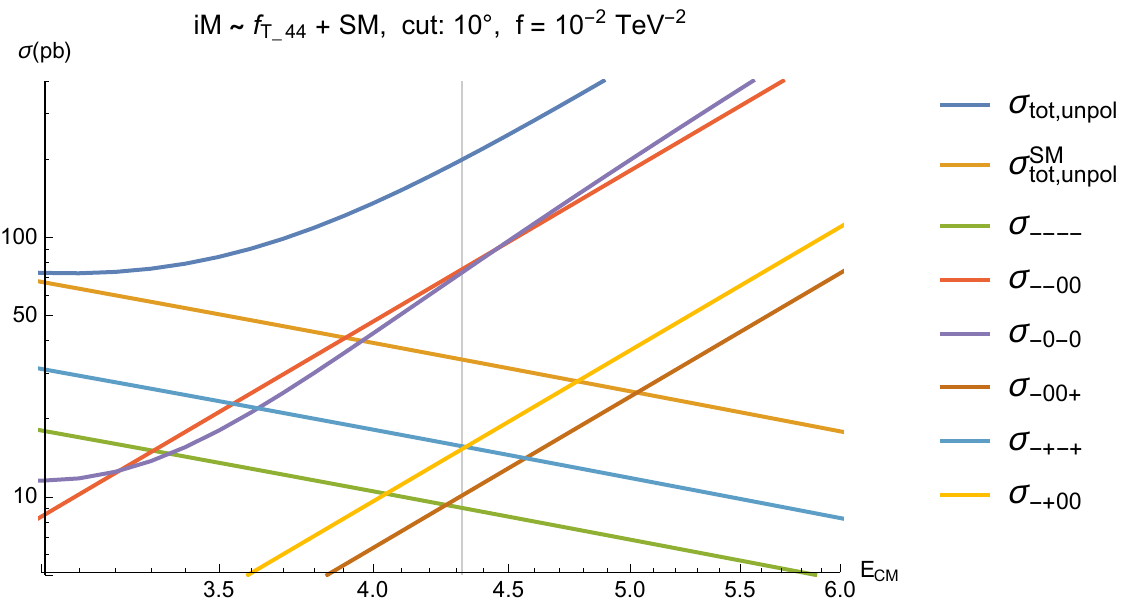} 
\end{center}
\caption{For description see Fig.~\ref{fig:polsHEFTT42text}; the $\mathcal{T}_{44}$.}
\label{fig:polsHEFTT44text}
\end{figure}

Fig.~\ref{fig:polsHEFTT42text} and~\ref{fig:polsHEFTT44text} shows the polarised cross section fractions for $\cT_{42}$ and $\cT_{44}$ and for chosen values of $f_i>0$. The enhancement of $--00$ and/or $-+00$ fractions in the region $M_{WW}\lesssim\Lambda$, visible in these figures, occurs only for the operators $\cT_{42}$, $\cT_{43}$, and $\cT_{44}$ among all the HEFT operators studied in this work. Therefore it is a distinct feature of the non-linear hypothesis, which must be associated to the fact of occurrence in HEFT of distinct Lorentz structures, mentioned above. We shall refer to this interesting feature in the context of phenomenology in Sec.~\ref{VVHEFT}.
%\includegraphics[width=0.52\columnwidth]{plotlistPlot10DegreeParticPolsft44Eq1.pdf} &  \includegraphics[width=0.52\columnwidth]{plotlistPlot10DegreeParticPolsft44Eq1Zoom.pdf} 

%przekroje niespolaryzowane, brak interferencji

%Fig.~\ref{fig:unpol} shows the total unpolarised $WW$ cross sections as functions of $M_{WW}$ for $\cT_{42}$, $\cT_{43}$, and $\cT_{44}$ operators, for both signs of the Wilson coefficients.
%Notice however that no sign dependence is present in $\cT_{42}$, due to the fact that the enhanced polarised cross sections $--00$ and $-+00$ are not among the polarised cross sections that saturate in the SM.

In the Appendix~\ref{app:VVonshellHEFT}, we presented the same set of results as done in the case of SMEFT, for the three distinct HEFT operators; the results consist in: partial wave unitarity limits for chosen values of Wilson coefficients (\ref{app:unitLimitsHEFT}), analytic expressions for both the amplitude $iM$ and the partial waves $\mathcal{T}^{(j)}$, numerical results for the polarized contributions to the total unpolarized cross sections at the scale $\sqrt{s^U}$ (\ref{app:analiticFormHEFT}) and plots of energy dependence of total and polarized contributions cross sections (\ref{app:plotsTotPolAndUnpolHEFT}). In general, the behavior of amplitudes is qualitatively similar to what found in case of SMEFT EFT ''models''.

\section{The EFT approach to $W^+W^+$ scattering at LHC}
\label{EFTapproachToWW}

Searches for deviations from the SM predictions in processes involving 
interactions between known particles are a well established technique to study possible
contributions from Beyond the Standard Model (BSM) physics.
In this Section we address two questions: i) how much we can learn about the scale of new 
physics and its strength using EFT approach 
to $W^+W^+$ scattering if a statistically significant deviation from the SM predictions is observed
in the expected LHC data for the process 
$pp \rightarrow 2 jets + W^+W^+ $, ii)  what is the practical usefulness of
the EFT language to describe the VBS data and whether or not this 
can indeed be the right framework to observe the first hints of new physics at the LHC. In both issues, our specific focus is on the proper use of the EFT in its 
 range of validity.

One should stress that the usefulness of any EFT analysis of a given process relies on the assumption that only few terms in the SMEFT or HEFT expansion give for that process an adequate approximation to the underlying UV theory.  
The necessary condition obviously is that the energy scale
of the considered process, $E<\Lambda$.
%\footnote{\color{red} For $WW$ scattering, we consider  large center of mass energy and large scattering angles, namely large Mandelstam variables $s\sim t\sim u\sim E^2 \gg M^2_W$. }
However, the effective parameters in these expansions 
 are the $f$'s and not the scale $\Lambda$ itself.  Neither 
$\Lambda$ nor the $c_i$'s are known without referring to specific UV complete models. 
As we already argued in Sec.~\ref{whichNonRen}, even for $E<<\Lambda$ a simple counting of powers of $E/\Lambda$ can be misleading as far
as the contribution of various operators to a given process is concerned. The concept of EFT ''models'' defined by the choice of operators ${\cal O}_i$ and the values of $f_i$ is a convenient notion for the remaining discussion.  The questions of this Sections are then: one, about the strategy towards data analysis once available and significant discrepancies with the SM are reported; and two, about the discovery potential at the LHC for BSM physics described by various EFT ''models''.

The crucial question is what the range of validity can be of a given EFT ''model''.  
There is no precise answer to this question unless one starts with
a specific theory and derives $\mathcal{L}_{SMEFT}$ or $\mathcal{L}_{HEFT}$ by decoupling the new degrees of 
freedom.  However, in addition to the obvious constraint that the EFT approach can be 
valid only for the energy scale $E<\Lambda$ (unfortunately with unknown value 
of $\Lambda$), for theoretical consistency the partial wave amplitudes 
should satisfy the perturbative unitarity condition.  The latter
requirement translates into the condition $E^2<\Lambda^2\leq s^U$, where $s^U\equiv s^U(f_i)$ is the 
perturbative partial wave unitarity bound as a function of the chosen operators and the 
values of the coefficients $f_i$'s. 
Thus, the value of $\Lambda^2_{max}=s^U$ gives the upper bound on the validity of the
EFT based ''model''. 
Since the magnitude  of the expected (or observed) experimental effects also depends on
the same $f_i$, one has a frame for a consistent use of the EFT ''model'' to describe
the data once they are available.
For a BSM discovery in the EFT framework, proper usage of the ''model'' is a vital issue.
It makes no physical sense to extend the EFT ''model'' beyond its range of applicability,
set by the condition $E<\Lambda<\sqrt{s^U}$.
We shall illustrate this logic in more detail in the following.

As already discussed in Sec.~\ref{whichNonRen} we shall consider only variations of single dimension-8 operators. However again, the strategy we present in this Section can be applied to the case of many operators at a time, including dimension-6. For a given EFT ''model''
\begin{equation}
d\sigma \propto |i\mathcal{M}|^2=|A_{SM}|^2+(A_{SM}\times A_{BSM}^*+\mathrm{h.c.})+|A_{BSM}|^2,
\label{full1}
\end{equation}
where $d\sigma$ denotes the differential cross sectin for the process
\begin{equation}
pp\rightarrow 2 jets + W^+ W^+ \rightarrow 2 jets + l^+\nu +l'^+\nu^\prime,
\label{process}
\end{equation}
where $l$ and $l'$ stand for any combination of electrons and muons.  The lepton $WW$ decay channel is often referred to as "gold-plated" due to its relatively good signal to background ratio.
The process depends on the $W^+W^+$ scattering amplitude (as illustrated in Fig.~\ref{fig:EWWWprod}).  
The EFT ''models'' can be maximally valid up to certain invariant mass $M=\sqrt{s}$ of the 
$W^+W^+$ system 
\begin{equation}
M<\Lambda\leq M^U(f_i),
\label{cutoff}
\end{equation}
where $M^U(f_i)$ is fixed by the partial wave perturbative unitarity constraint, $(M^U(f_i))^2=s^U(f_i)$ as computed in Sec.~\ref{onshellSMEFT} or~\ref{onshellHEFT}.

The differential cross section
$\frac{d\sigma}{dM}$ reads (actual 
calculations must include also all non-VBS diagrams leading to the same final states):
\begin{equation}
\frac{d\sigma}{dM}\sim \Sigma_{ijkl} \int dx_1 dx_2 q_i(x_1) q_j(x_2) | {\cal M}(ij\to klW^+W^+)|^2 d\Omega\; 
\delta(M-\sqrt{(p_{W^+}+p_{W^+})^2})
\label{cross}
\end{equation}
where $q_i(x)$ is the Parton Distribution Function (PD)F for parton $i$, the sum runs over partons in the initial ($ij$)
and final ($kl$) states and over helicities, the amplitude ${\cal M}$ is for the parton level process $ij\to
klW^+W^+$ and $d\Omega$ denotes the final state phase space integration.  The special 
role of the distribution $\frac{d\sigma}{dM}$ follows from the fact that it is 
straightforward to impose the cutoff $M\leq\Lambda$, Eq.~(\ref{cutoff}), for the $WW$ scattering amplitude. 
The differential cross section, $\frac{d\sigma}{dM}$, is therefore a very sensitive and straightforward test of new physics defined by a given EFT ''model''. Unfortunately, the $W^+W^+$ invariant 
mass in the purely leptonic $W$ decay channel is not directly accessible experimentally 
and one has to investigate various experimental distributions of the charged particles.
The problem here is that the kinematic range of those distributions is not related to the EFT ''model'' validity cutoff $M<\Lambda$ and if $\Lambda < M_{max}$, where $M_{max}$ is the 
kinematic limit accessible at the LHC for the $WW$ system, there is necessarily also a 
contribution to those distributions from the region $\Lambda < M < M_{max}$.  
The question is then: in case a deviation from SM predictions is indeed observed,
how to verify a ''model'' defined by a single higher-dimension operator ${\cal O}_i^{(k)}$
and a given value of $f_i$ by fitting it to a set of experimental distributions $D_i$ 
and in what range of $f_i$ such a fit is really meaningful \cite{Falkowski:2016cxu}. 

Before we address the above question, it is in order to emphasize most important aspects of $WW$ scattering discussed in detail in Sec.~\ref{preTech}:
\begin{enumerate}
\item For a given EFT ''model'', the unitarity bound is very different for partial 
wave of different helicity amplitudes and depends on their individual energy dependence
(some of them remain even constant and never violate unitarity).
Our $M^U$ has to be taken as the $lowest$ unitarity bound, universally for all helicity 
amplitudes, because it is the lowest bound that determines the scale $\Lambda_{max}$.
More precisely, one should take the value obtained from diagonalization of the matrix of the $j=0$ partial waves in the helicity space.
\item Correct assessment of the EFT ''model'' validity range in the $W^+W^+$ scattering process
requires also consideration of the $W^+W^-$ scattering amplitudes which by construction
probe the same couplings and are sensitive to exactly the same operators.  For most
higher dimension operators,
this actually significantly reduces their range of validity in $W^+W^+$ analyses.
Conversely, the $WZ$ and $ZZ$ processes can be assumed to contain uknown
contributions from additional operators which adjust the value of $\Lambda$ consistently (the operators we study may modify $WWZZ$ and/or $ZZZZ$ vertices).
\item 
It is interesting to note that for the $f_i$ values of practical interest
the deviations 
from SM predictions in the total cross sections become sizable only in a 
narrow range of energies just below the value of $M^U$, where the $|A_{BSM}|^2$ term
in Eq.~(\ref{full1}) takes over.  However, for most dimension-8 operators the contribution
of the interference term is not completely negligible.
Even if deviations from the SM are dominated by the helicity combinations that reach the 
unitarity bound first, the total unpolarized cross sections up to
$M=M^{U}$ get important contributions also from amplitudes which are still far from their own
unitarity limits.  
\end{enumerate}
%unitarity in the full reaction %
The following remark is in order concerning determination of the unitarity limits: since we apply the on-shell unitarity limits to the full $pp$ process, where the $W^+W^+$ are scattered off-shell, one could wonder if the energy bound $\sqrt{s^U}$ assessment is correct. Intuitively, the unitarity limits computed this way are valid: for off-shell scattering one can define the $\sqrt{s^U}$ too, and we are interested in the energy scale $\sqrt{s}$ at which the $W^+W^+$ scattering stops making sense. Nonetheless one could think of determining the $\sqrt{s^U}$ directly for the $qq\rightarrow qq W^+W^+$ amplitude. The problem is that the partial wave decomposition for a $2\rightarrow 4$ process is not known and possibly very complicated due to more kinematic variables in the $4$-particle final state. Interestingly, one can use the Argand circle formula~\eqref{eq:uni24} for the elastic scattering $qq\rightarrow qq$. The $qq\rightarrow qq W^+W^+$ can be absorbed into the $R^2_j(s)$, which is itself bounded  (in a similar way the inelastic binary scattering partial waves were bounded in eq.~\eqref{eq:uni27}). Unfortunately, the $\int d\theta$ integration, which is needed at a certain point to project onto the desired $j$ partial wave,  is not straightforward to perform. We shall leave it for future study and use in this work the on-shell $WW$ bounds. \\\mbox{} \\

We now come back to the problem of testing the EFT ''models'' when the $W^+W^+$ invariant mass is not accessible experimentally.
Any BSM signal, $S$, is defined as the deviation from the SM prediction in the differential  distributions $d\sigma/dx_i$
of some observable $x_i$:
\begin{equation}
\frac{dS}{dx_i}=\left(\frac{d\sigma}{dx_i}\right)^{BSM}-\left(\frac{d\sigma}{dx_i}\right)^{SM},
\label{N}
\end{equation}
The first quantitative estimate of the BSM distribution can be written as
\begin{equation}
\left(\frac{d\sigma}{dx_i}\right)^{EFT}=\int^\Lambda_{2M_W}\left(\frac{d^2\sigma}{dx_idM}\right)^{EFT} dM 
+\int_\Lambda^{M_{max}}\left(\frac{d^2\sigma}{dx_idM}\right)^{SM} dM, 
\label{dsigma}
\end{equation}
It defines signal coming uniquely from the operator that defines the ''model'' in its range
of validity and assumes only the SM contribution in the region $M>\Lambda$.
Realistically one expects some BSM contribution also from the region above
$\Lambda$.  While this additional contribution may enhance the signal and thus
our sensitivity to new physics, it may also preclude proper description of the data in the EFT 
language.  Such description in terms of a particular EFT ''model'' makes sense if and only if this contribution is small enough  when compared to the contribution from the region controlled by the EFT ''model''.  The latter depends on the value of $\Lambda$ and $f_i$, and the former on the unknown physics for $M>\Lambda$, which regularizes the scattering amplitudes and makes them consistent with partial wave unitarity.
Ideally, one would conclude that the EFT ''model'' is tested for values of $(\Lambda\leq M^U, f_i)$ 
such that the signals computed from Eq.~(\ref{dsigma}) are statistically consistent 
(say, within 2 standard deviations) with the signals computed when the tail $\Lambda>M^U$ is 
modeled in {\it any} way that preserves unitarity of the amplitudes,
i.e., the contribution from this region is sufficiently suppressed kinematically by
parton distributions.  This requirement is of course impossible to impose in practice, but
for a rough quantitative estimate of the magnitude of this 
contribution, one can assume that all the helicity amplitudes above 
$\Lambda$ remain 
constant at their respective values they reach at $\Lambda$, and that $\Lambda$ is 
common to all the helicity amplitudes.
For $\Lambda = \Lambda_{max}$, this prescription regularizes the helicity amplitudes that violate unitarity at $M^U$ and also properly accounts for the contributions of the helicity amplitudes that remain constant with energy. It gives a reasonable approximation to the total unpolarized cross sections for $M>M^U$, at least after some averaging over $M$.
More elaborated regularization techniques can also be checked  here.
The full contribution to a given distribution $d\sigma/dx_i$ is therefore taken as
\begin{equation}
\left(\frac{d\sigma}{dx_i}\right)^{BSM}=\int^\Lambda_{2M_W}\left(\frac{d^2\sigma}{dx_idM}\right)^{EFT} dM 
+\int_\Lambda^{M_{max}}\left(\frac{d^2\sigma}{dx_idM}\right)^{A=const} dM, 
\label{unitarized}
\end{equation}

Now, BSM observability imposes some minimum value of $f$ to obtain the required
signal statistical significance.
It can be derived based on Eq.~(\ref{unitarized}).  On the other
hand, description in the EFT language imposes some maximum value of $f$ such that signal
estimates computed from Eqs.~(\ref{dsigma}) and (\ref{unitarized}) remain 
statistically consistent.  Large difference between the two computations implies
significant sensitivity to the region above $\Lambda$.  It impedes a meaningful data
description in the EFT language.

\begin{figure}[hbtp]
\centering
%\vspace{-2.5cm}
\includegraphics[width=0.7\linewidth]{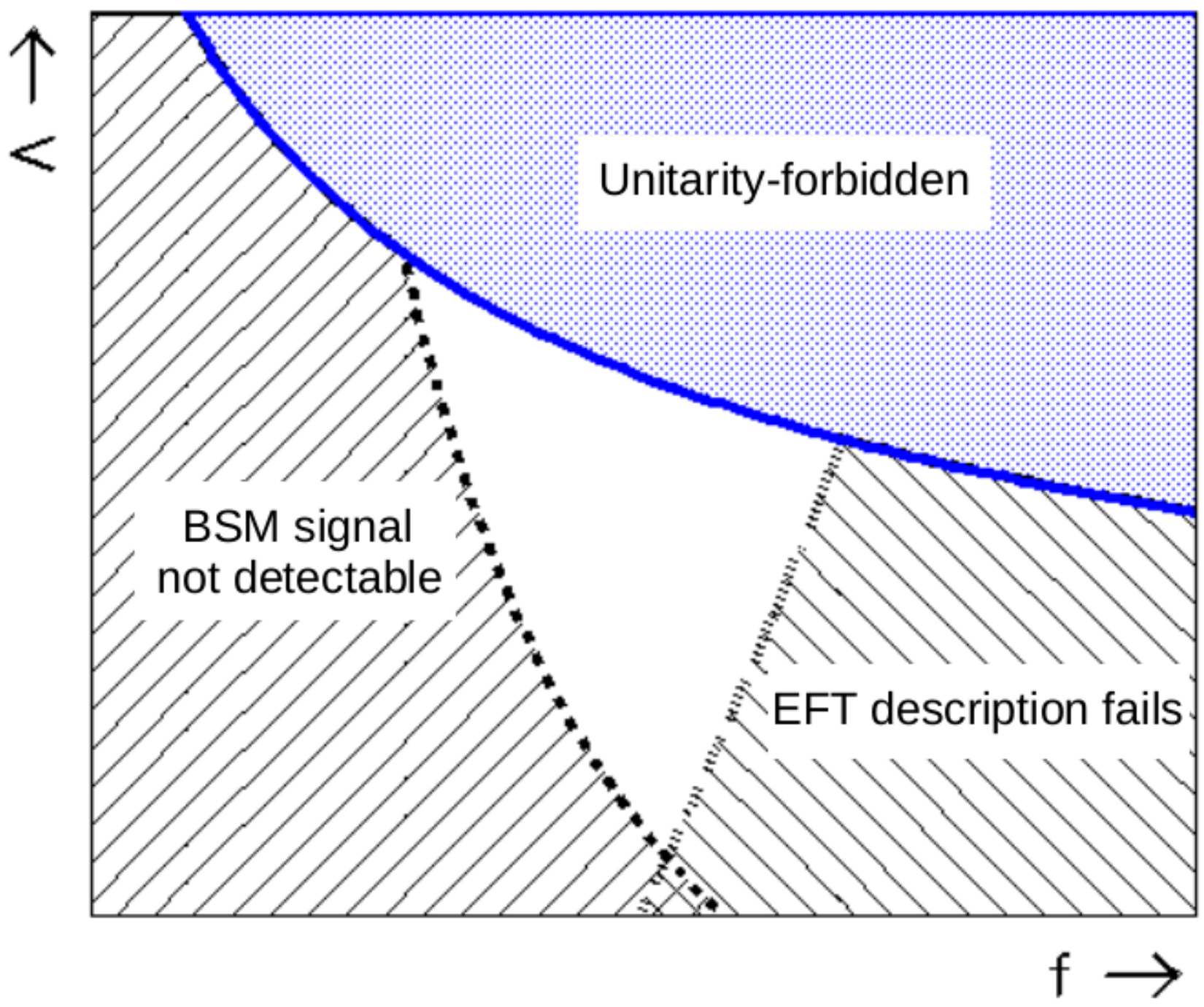}
%\vspace{-2.5cm}
\caption{
Cartoon plot which shows the regions in $f_i$ and $\Lambda$ (for an arbitrary
higher-dimension operator ${\cal O}_i$) in terms of BSM signal observability
and applicability of EFT ''models'' based on the choice of a higher-dimension operator in an analysis of the same-sign VBS process with purely 
leptonic decays.  The central white triangle is the most interesting region
where the underlying BSM physics can be studied within the EFT framework.
}
%\vspace{1cm}
\label{fig:cartoonplot}
\end{figure}

Assuming $\Lambda=\Lambda_{max}$$(=M^U)$, we get a finite interval of possible $f$ values, bounded from two sides,
for which BSM discovery and correct EFT description are both plausible.
In the more general case when\footnote{We remind the reader that $M^U$ is function of $f_i$. The latter has interpretation of ratio of the ''coupling'' $c_i$ over $\Lambda^4$. Hence one has freedom to decrease $\Lambda$ by simultaneously and appropriately decreasing $c_i$ such that $f_i$, hence $M^U$, remains constant. It simply means considering different BSM physics (characterized by couplings and $\Lambda$).}
$\Lambda \leq M^U$, i.e., new physics states may appear before our EFT ''model'' reaches its 
unitarity limit, respective limits on $f$  depend on the actual value of $\Lambda$.
We thus obtain a 2-dimensional region in the plane $(\Lambda, f_i)$, which is
shown in the cartoon plot in Fig.~\ref{fig:cartoonplot}.  This region is bounded from 
above by the unitarity bound $M^U(f_i)$ (solid blue curve), from the left by the
signal significance criterion (dashed black curve)
and from the right by the EFT consistency criterion (dotted black curve).
The EFT could be the right framework to search for BSM physics as long as these
three criteria do not mutually exclude each other, i.e., graphically, the ``triangle"
shown in our cartoon plot is not empty.  In next Sections we will verify whether such 
``triangles" indeed exist in SMEFT and HEFT.

Thus, our preferred strategy for data analysis is as follows:

\begin{enumerate}
\item From collected data measure a distribution $d\sigma/dx_i$ (possibly in more than one
dimension) that
offers the highest sensitivity to the studied operator(s)\footnote{The sensitivity is known from simulations.},
\item If deviations from the SM are indeed observed, fit particular values of 
$(\Lambda\leq M^U, f_i)$ based on EFT simulated templates in which
the contribution from the region $M > \Lambda$ is taken into account according to 
Eq.~(\ref{unitarized}) or using some more elaborated regularization methods,
\item Fixing $f_i$ and $\Lambda$ to the fit values, recalculate the simulated
$d\sigma/dx_i$ template
so that the region $M > \Lambda$ is populated only by the SM contribution
(Eq.~(\ref{dsigma})),
\item Check statistical consistency between the original simulated $d\sigma/dx_i$ template
and the one based on Eq.~(\ref{dsigma}),
\item Physics conclusions from the obtained $(\Lambda, f_i)$ values can only be drawn
if such consistency is found.  
In addition, stability of the result against different
regularization methods provides a measure of uncertainty of the procedure - too much
sensitivity to the region above $\Lambda$ means the procedure is destined to fail
and so the physical conclusion is that data cannot be described with the studied operator.
\end{enumerate}

Our goal in Sec.~\ref{WWinSMEFT} and~\ref{VVHEFT} is to find, numerically, the discovery potential at the HL-LHC of the
BSM physics effectively described by EFT ''models'' constructed within the SMEFT and HEFT hypotheses respectively, with single (primary) dimension-8 operators at a time,
with proper attention paid to the regions of validity of such models, as described in this Section.

\subsection{Discovery potential of SMEFT}
\label{WWinSMEFT}

For the following analysis dedicated event samples of the process 
$pp \to jj\mu^+\mu^+\nu\nu$ at 14 TeV
were generated at LO using the MadGraph5\_aMC@NLO v5.2.2.3 generator~\cite{Alwall:2014hca}, 
with the appropriate UFO~\cite{Degrande:2011ua} files 
containing additional vertices involving the desired dimension-8 operators.
For each dimension-8 operator a sample of at least 500,000 events within a phase space 
consistent with a VBS-like topology (defined below) was generated.  A preselected arbitrary value of 
the relevant $f$ coefficient (from now on, $f \equiv f_i$ with $i = S0, S1, T0, T1, T2,
M0, M1, M6, M7$) was assumed at each generation;
different $f$ values were obtained by applying weights to generated events, using
the \verb+reweight+ command in MadGraph.  The value $f$=0 represents the Standard 
Model predictions for each study.
The Pythia package v6.4.1.9~\cite{Sjostrand:2006za} was used for hadronization as well as initial and final state 
radiation processes.  No detector was simulated.  Cross sections at the
output of MadGraph were multiplied by a factor 4 to
account for all the lepton (electron and/or muon) combinations in the final state.

The results of this Section are based on our ref.~[a] where we used the MadGraph normalization (already described by eq.~\ref{eq:onshellSMEFT1} and~\ref{eq:onshellSMEFT1p}) (which is not equivalent to the NDA normalization prescription). The choice was dictated by consistency with ref.~\cite{Degrande:2013rea}. We shall exploit the NDA normalization for SMEFT in Sec.~\ref{discReg27TeV}, when we estimate and compare ranges of BSM couplings that correspond to the discovery regions found in the 27 and 14 TeV cases. Nevertheless conclusions on the ranges of the $\Lambda$ scale probed by the EFT approach in $W^+W^+$ scattering are independent of the differences in normalizations and therefore can be read off\footnote{Of course, the existence (or not) of a non-empty discovery region is also a normalization independent statement.} from the plots showing the discovery regions in this Section~(\ref{WWinSMEFT}).%

In this analysis, the Standard Model process $pp \to jjl^+l^+\nu\nu$ is treated
as the irreducible background, while signal is defined as the enhancement
(which may be positive or negative in particular cases) of the event
yield in the presence of a given dimension-8 operator relative to the
Standard Model prediction.  No reducible backgrounds were simulated, as they
are known to be strongly detector dependent.  For this reason, results presented
here should be treated mainly as a demonstration of our strategy rather than as a precise determination
of numerical values.
For more realistic results this analysis should be repeated with full
detector simulation for each of the LHC experiments separately.

The final analysis is performed by applying standard VBS-like event
selection criteria, similar to those applied in data analyses carried
by ATLAS and
CMS.  These were: $M_{jj} >$ 500 GeV, $\Delta\eta_{jj} >$ 2.5, $p_T^{~j} >$ 30 GeV,
$|\eta_j| <$ 5, $p_T^{~l} >$25 GeV, $|\eta_l| <$ 2.5.
As already discussed, signal is calculated in two ways.
First, using Eq.~(\ref{dsigma}), where $\Lambda$ can vary in principle between $2m_W$ and
the appropriate unitarity limit for each chosen value of $f$.
The $M_{WW} > \Lambda$ tail of the distribution is then assumed identical as
in the Standard Model case.
Second, using Eq.~(\ref{unitarized}) which accounts for an additional BSM contribution
coming from the region $M_{WW} > \Lambda$.  The latter is estimated under the
assumption that helicity amplitudes remain constant above this limit, as already discussed.  For the case when $\Lambda$ is equal to the
unitarity limit, this corresponds to unitarity saturation.

\begin{figure}[hbtp]
\centering
\includegraphics[width=1.05\linewidth]{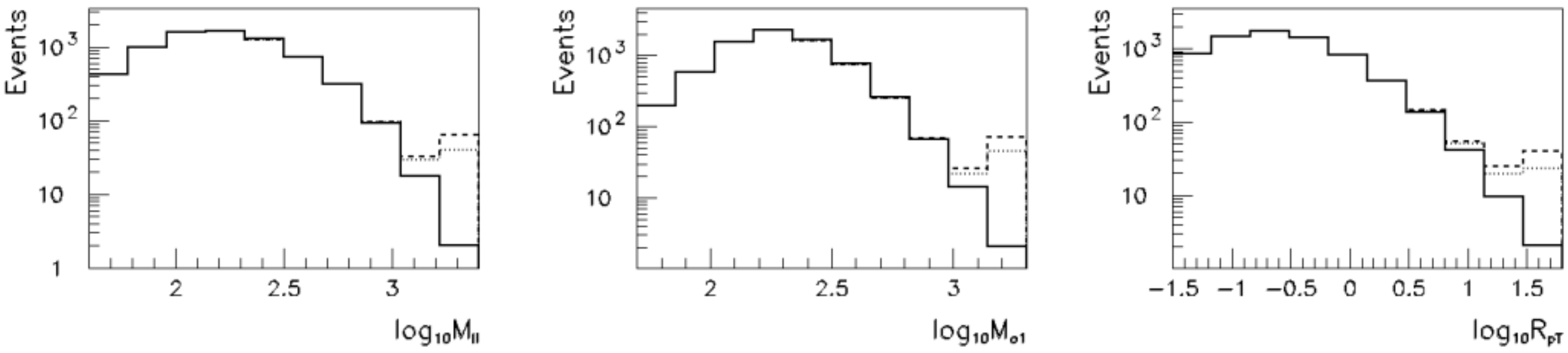}
%\vspace{-8.5cm}
\caption{
Typical examples of kinematic distributions used for the assessment of BSM signal significances.
Shown are the distributions of $M_{ll}$, $M_{o1}$ and $R_{pT}$ (in log scale): in the
Standard Model (solid lines), with $f_{T1}$=0.1/TeV$^{-4}$ and the high-$M_{WW}$ tail
treatment according to Eq.~(\ref{unitarized}) (dashed lines), and with
$f_{T1}$=0.1/TeV$^{-4}$ and the high-$M_{WW}$ tail
treatment according to Eq.~(\ref{dsigma}) (dotted lines).
Assumed is $\sqrt{s}$ = 14 TeV and an integrated luminosity of 3 ab$^{-1}$.
}
\label{fig:Dim8Dist}
\end{figure}

For each $f$ value of every dimension-8 operator, signal significance is
assessed by studying the distributions of a large number of kinematic variables.  
We only considered one-dimensional distributions
of single variables.  Each distribution was divided into 10 bins,
arranged so that the Standard Model prediction in each bin is never lower than
2 events.  Overflows were always included in the respective highest bins.
Ultimately, each distribution had the form of 10 numbers, that represent the expected
event yields normalized to a total integrated luminosity of 3 ab$^{-1}$, each calculated in three
different versions: $N^{SM}_i$ for the Standard Model case, $N^{EFT}_i$ from
applying Eq.~(\ref{dsigma}), and $N^{BSM}_i$ from applying Eq.~(\ref{unitarized})
(here subscript $i$ runs over the bins).
In this analysis, Eq.~(\ref{unitarized}) was implemented by applying additional
weights to events above $M_{WW} = \Lambda$ in the original non-regularized
samples generated by MadGraph.  For the dimension-8 operators, this weight
was equal to $(\Lambda/M_{WW})^4$.
The choice of the power in the exponent takes into account that the non-regularized
total cross section for $WW$ scattering grows less steeply around $M_{WW} = \Lambda$
than its asymptotic
behavior $\sim s^3$, which is valid\footnote{We remind the reader that $\sigma$ carries $1/s$ kinematic factor, i.e. $iM$ independent.} in the limit $M_{WW} \to \infty$.
This follows from the observation that unitarity is first violated much before
the cross section gets dominated by its $\sim s^3$ term, as shown in Sec.~\ref{preTech}.
The applied procedure is supposed to ensure that the total $WW$ scattering
cross section after regularization behaves like $1/s$ for $M_{WW} > \Lambda$,
and so it approximates the
principle of constant amplitude, at least after some averaging over the
individual helicity combinations.
Examples of simulated distributions are shown in Fig.~\ref{fig:Dim8Dist}.

Signal significance expressed in standard deviations ($\sigma$)
is defined as the square root
of a $\chi^2$ resulting from comparing the bin-by-bin event yields:

\begin{equation}
\chi^2 = \sum_i (N^{BSM}_i - N^{SM}_i)^2 / N^{SM}_i.
\label{chi}
\end{equation}

\begin{figure}[hbtp]
\centering
\includegraphics[width=0.95\linewidth]{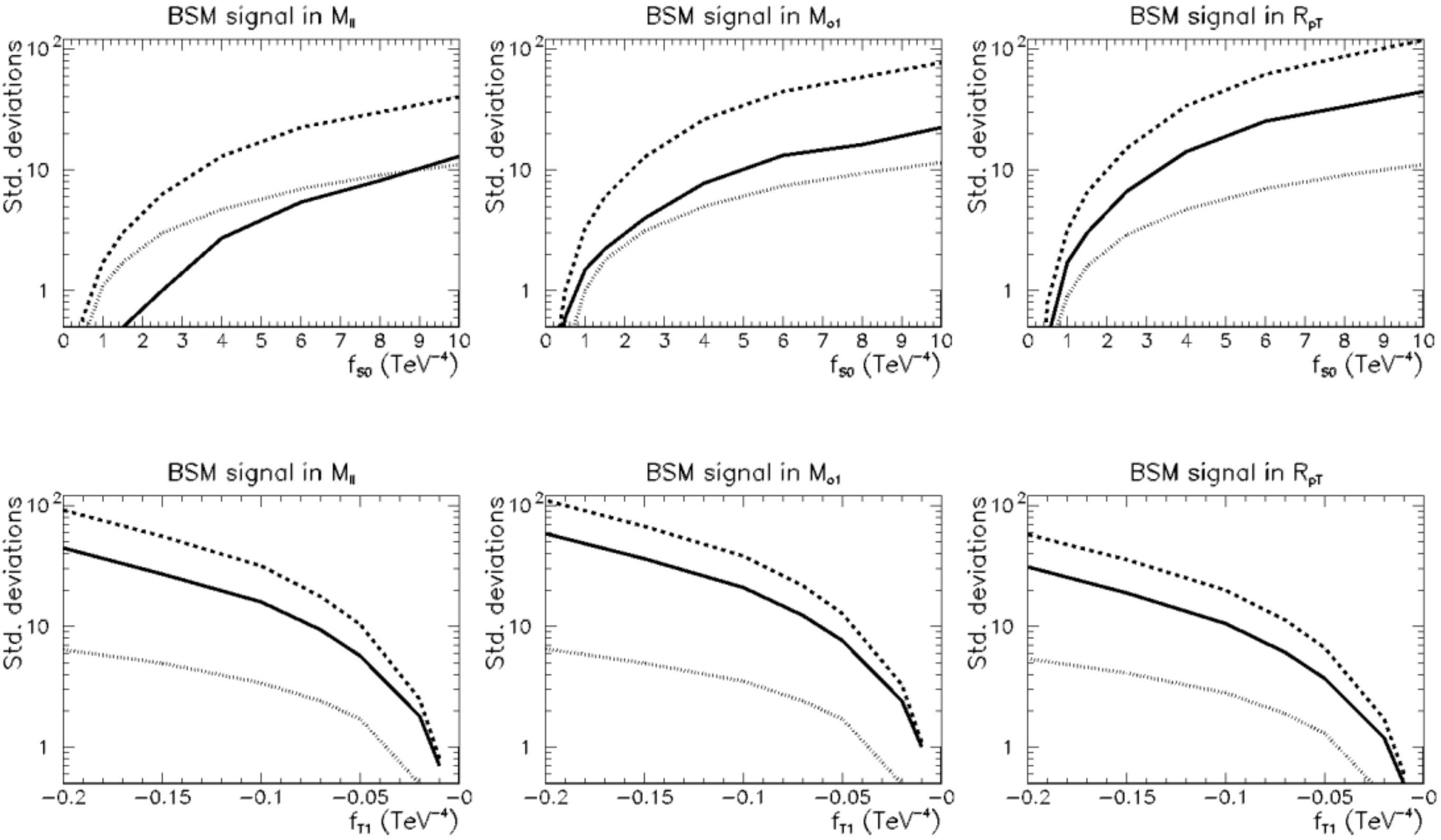}
\caption{
Typical examples of BSM signal significances computed as a function of $f_{S0}$ (upper row)
and $f_{T1}$ (lower row) based on different kinematic distributions.  Here the $\Lambda$
cutoff is assumed equal to the unitarity limit.  Shown are predictions obtained by using
Eq.~(\ref{dsigma}) (solid lines) and Eq.~(\ref{unitarized}) (dashed lines).
The dotted lines show the difference in standard deviations between the two respective 
calculations.
Assumed is $\sqrt{s}$ = 14 TeV and an integrated luminosity of 3 ab$^{-1}$.
}
\label{fig:Sigmas}
\end{figure}

Lower observation limits on each operator are defined by the requirement of
signal significance being above the 5$\sigma$ level.
Small differences between the respective signal
predictions obtained using Eqs.~(\ref{dsigma}) and (\ref{unitarized}), as well
as using other regularization techniques, will be manifest
as slightly different observation limits
and should be understood as the uncertainty margin
arising from the unknown physics above $\Lambda$, no longer described in terms of 
the EFT.  Examples of signal significances 
as a function of $f$ (with $\Lambda=\Lambda_{max}$) are shown in
Fig.~\ref{fig:Sigmas} with dashed curves.
Consistency of the EFT description is determined by requiring
a small difference (at most 2$\sigma$) between the respective predictions from 
Eqs.~(\ref{dsigma}) and (\ref{unitarized}).  To this end $\chi_{add}^2$ is introduced as follows:
\begin{equation}
\chi^2_{add} = \sum_i (N^{EFT}_i - N^{BSM}_i)^2 / N^{BSM}_i.
\label{chiadd}
\end{equation}
In this analysis
we allowed differences amounting to up to 2$\sigma$ in the most sensitive
kinematic distribution, as computed from~\eqref{chiadd}.  This difference as a function of $f$ is shown in
Fig.~\ref{fig:Sigmas} as dotted curves.
These considerations consequently translate into effective upper limits
on the value of $f$ for each operator.

%This number quantifies our experimental
%sensitivity to the unknown region $M_{WW} > \Lambda$.

For each dimension-8 operator we took the distribution that produced the
highest $\chi^2$ (eq.~\eqref{chi}) among the considered variables.  The most sensitive
variables we found to be:
\begin{equation}
R_{p_T} \equiv p_T^{~l1}p_T^{~l2}/(p_T^{~j1}p_T^{~j2})
\label{eq:}
\end{equation} 
(introduced in~\cite{Doroba:2012pd})
for ${\cal O}_{S0}$ and ${\cal O}_{S1}$, and 
\begin{equation}
M_{o1} \equiv \sqrt{(|\vec{p}_T^{~l1}|+|\vec{p}_T^{~l2}|
+|\vec{p}_T^{~miss}|)^2 - (\vec{p}_T^{~l1}+\vec{p}_T^{~l2}+\vec{p}_T^{~miss})^2},
\end{equation}
\cite{bib:M01},
for the remaining operators.

Here, the unitarity limits were taken from the VBFNLO calculator v1.3.0,
after applying appropriate conversion factors to the input values of the Wilson
coefficients, so to make it suitable to the MadGraph 5 normalization.
We used the respective values from \verb+T-matrix diagonalization+\footnote{which are the limits that we referred to as ''diag.'' e.g. in Tab.~\ref{tab:unitarityT1text} in Sec.~\ref{onshellSMEFT}.}, considering both $W^+W^+$
and $W^+W^-$ channels, and taking always the lower value of the two.
For the operators we consider here, unitarity limits are lower for $W^+W^-$ than 
for $W^+W^+$ except for
$f_{S0}$ (both positive and negative) and negative $f_{T1}$.


Assuming $\Lambda$ is equal to the respective unitarity bounds,
the lower and upper limits for the values of $f$ for each dimension-8 operator, for positive
and negative $f$ values, estimated for the HL-LHC with an integrated luminosity of 3 ab$^{-1}$,
are read out directly from graphs such as Fig.~\ref{fig:Sigmas} and
listed below in Table 1.  These limits define the (continuous) sets of testable 
EFT ''models'' based on the choice of single dimension-8 operators.

%\begin{tabular}{c|ccc|c|ccc}
%Operator & Lower limit & Lower limit & Upper limit & Operator & Lower limit & Lower limit & Upper limit \\
%         &  from (\ref{unitarized}) &  from (\ref{dsigma}) & & & from (\ref{unitarized}) &  from (\ref{dsigma}) & \\
%\hline
%$f_{S0}$  & 1.3  & 2.2  & 2.0 & $-f_{S0}$ & 1.2  & 1.8  & 2.0 \\
%$f_{S1}$  & 8.0  & 12   & 6.5 & $-f_{S1}$ & 5.5  & 11   & 6.0 \\ 
%$f_{T0}$  & 0.08 & 0.20 & 0.13 & $-f_{T0}$ & 0.05 & 0.09 & 0.12 \\
%$f_{T1}$  & 0.03 & 0.04 & 0.06 & $-f_{T1}$ & 0.03 & 0.04 & 0.06 \\
%$f_{T2}$  & 0.20 & 0.25 & 0.25 & $-f_{T2}$ & 0.10 & 0.12 & 0.20 \\
%$f_{M0}$  & 1.0  & 1.5  & 1.2 & $-f_{M0}$ & 1.0  & 1.5  & 1.2 \\ 
%$f_{M1}$  & 1.0  & 2.0  & 1.9 & $-f_{M1}$ & 0.9  & 1.2  & 1.8 \\ 
%$f_{M6}$  & 2.0  & 3.0  & 2.4 & $-f_{M6}$ & 2.0  & 3.0  & 2.4 \\
%$f_{M7}$  & 1.1  & 1.4  & 2.8 & $-f_{M7}$ & 1.3  & 2.0  & 2.8 \\
%\end{tabular}
\begin{table}[h]
\vspace{8mm}
\begin{center}
\begin{tabular}{|c|cc|c|cc|}
\hline
Coeff.   & Lower limit & Upper limit & Coeff.   & Lower limit & Upper limit \\
         &  (TeV$^{-4}$) &  (TeV$^{-4}$) & & (TeV$^{-4}$) &  (TeV$^{-4}$) \\
\hline
$f_{S0}$  & 1.3  & 2.0 & $-f_{S0}$ & 1.2  & 2.0 \\
$f_{S1}$  & 8.0  & 6.5 & $-f_{S1}$ & 5.5  & 6.0 \\
$f_{T0}$  & 0.08 & 0.13 & $-f_{T0}$ & 0.05 & 0.12 \\
$f_{T1}$  & 0.03 & 0.06 & $-f_{T1}$ & 0.03 & 0.06 \\
$f_{T2}$  & 0.20 & 0.25 & $-f_{T2}$ & 0.10 & 0.20 \\
$f_{M0}$  & 1.0  & 1.2 & $-f_{M0}$ & 1.0  & 1.2 \\
$f_{M1}$  & 1.0  & 1.9 & $-f_{M1}$ & 0.9  & 1.8 \\
$f_{M6}$  & 2.0  & 2.4 & $-f_{M6}$ & 2.0  & 2.4 \\
$f_{M7}$  & 1.1  & 2.8 & $-f_{M7}$ & 1.3  & 2.8 \\
\hline
\end{tabular}
\end{center}
\vspace{3mm}
\caption{
Estimated lower limits for BSM signal significance and upper limits for
EFT consistency for each dimension-8 operator
(positive and negative $f$ values), for the case when $\Lambda$ is equal
to the unitarity bound,
in the $W^+W^+$ scattering process at the LHC with 3 ab$^{-1}$.}
\end{table}

The fact that the obtained lower limits are more optimistic than those from
several earlier studies (see, e.g., Ref.~\cite{Degrande:2013yda})
reflects our lack of detector simulation and reducible
background treatment, but may be partly due to the use of
the most sensitive kinematic variables.  It must be stressed, nonetheless, that
both these factors affect all lower and upper limits likewise, so their relative
positions with respect to each other are unlikely to change much.

As can be seen, the ranges are rather narrow, but in most cases non-empty.
Rather wide regions where BSM signal significance does not preclude consistent 
EFT description
can be identified for $f_{T1}$ and $f_{M7}$ regardless of sign, as well as
somewhat smaller regions for $f_{T0}$, $f_{T2}$ and $f_{M1}$.
Prospects for $f_{M0}$, $f_{M6}$ and $f_{S0}$ may depend on the accuracy
of the high-$M_{WW}$ tail modelling and
a narrow window is also likely to open up unless measured
signal turns out very close to its most conservative prediction.
Only for positive values of $f_{S1}$, the resulting upper limit for consistent
EFT description remains entirely below the lower limit for signal significance.

\begin{figure}[hbtp]
\centering
\vspace{5mm}
\includegraphics[width=1.02\linewidth]{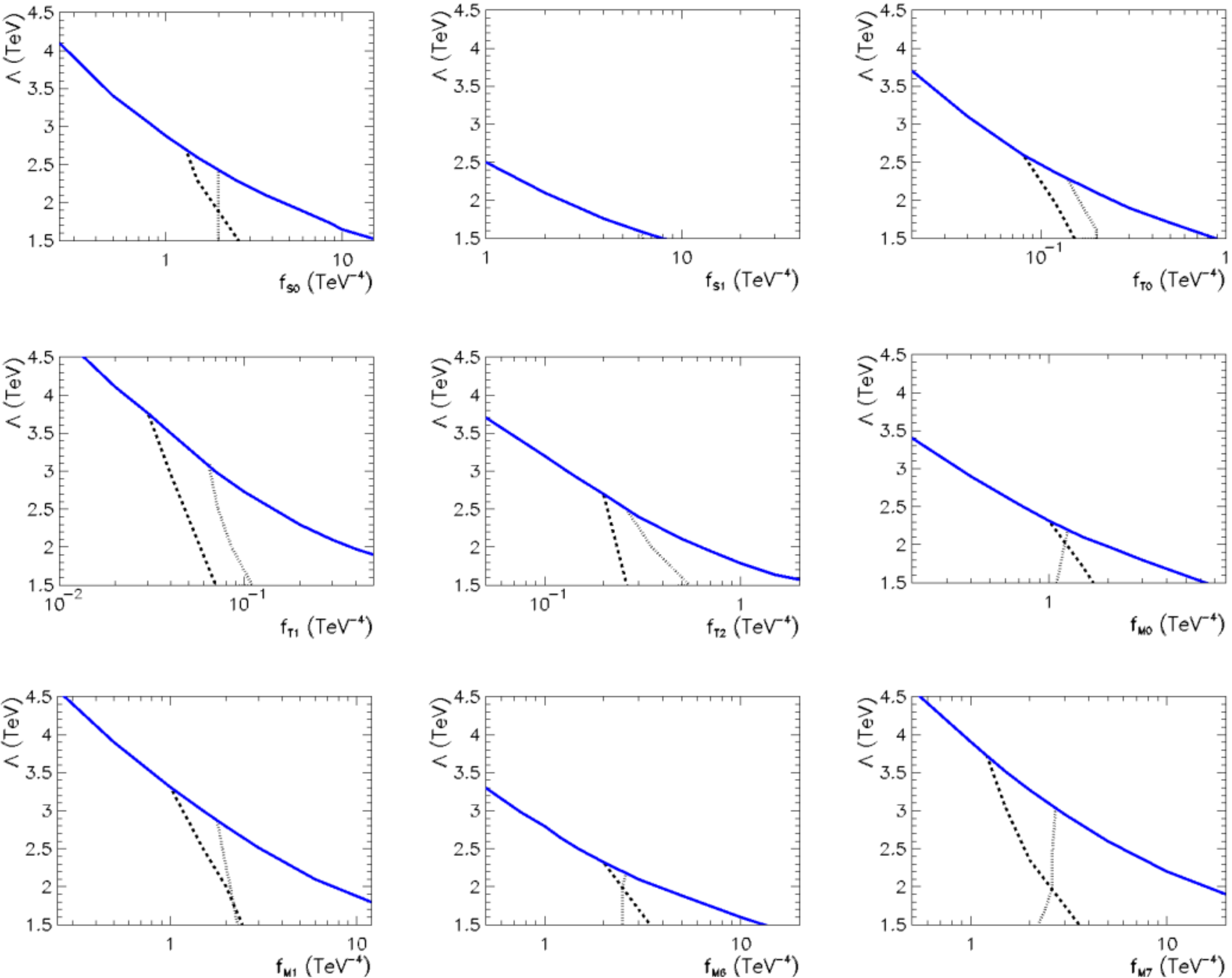}
\vspace{5mm}
\caption{
Regions in the $\Lambda$ vs $f$ (positive $f$ values) space for dimension-8 operators 
in which a 5$\sigma$ BSM signal can be observed and the EFT is applicable.  
The unitarity limit is shown in blue.  Also shown are the lower limits for a 
5$\sigma$ signal
significance from Eq.~(\ref{unitarized}) (dashed lines) and the upper limit on $2\sigma$ EFT
consistency (dotted lines).
Assumed is $\sqrt{s}$ = 14 TeV and an integrated luminosity of 3 ab$^{-1}$.   
}
\label{fig:Dim8Limits3}
\end{figure}

\begin{figure}[hbt]
\centering
\includegraphics[width=1.02\linewidth]{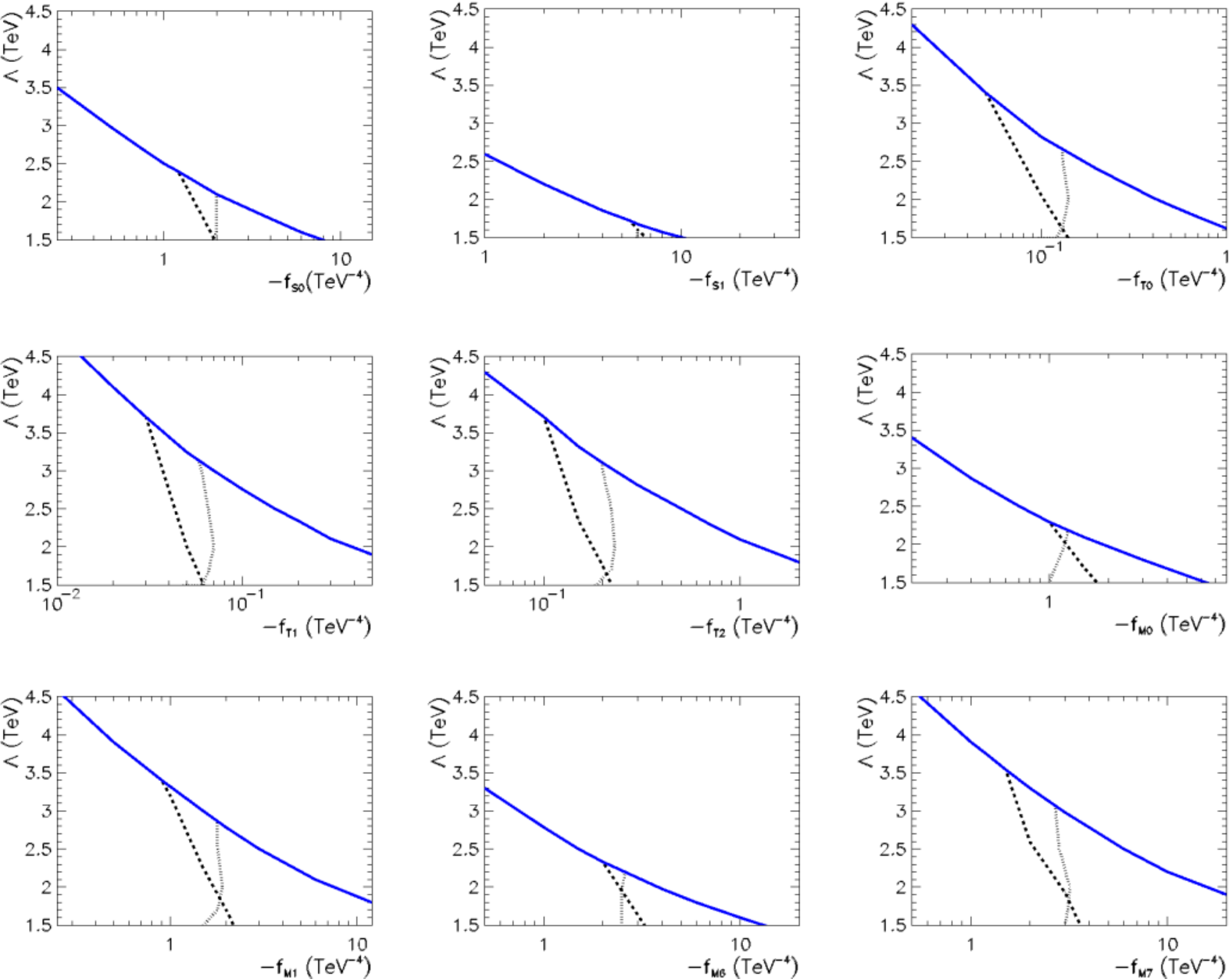}
\vspace{1cm}
\caption{
Regions in the $\Lambda$ vs $f$ (negative $f$ values) space for dimension-8 operators in which a 5$\sigma$ BSM signal can be observed and the EFT is applicable. For the meaning of curves see caption of Fig.~\ref{fig:Dim8Limits3}.
Assumed is $\sqrt{s}$ = 14 TeV and an integrated luminosity of 3 ab$^{-1}$.        
}
\label{fig:Dim8Limits4}
\end{figure}

Allowing that the scale of new physics $\Lambda$ may be lower than the actual
unitarity bound results in 2-dimensional limits in the $(f,\Lambda)$ plane.
Usually this means further reduction of the allowed $f$ ranges for lower $\Lambda$ 
values and the resulting regions take the form of an irregular triangle.
Respective results for all the dimension-8 operators are depicted in 
Figs.~\ref{fig:Dim8Limits3} and \ref{fig:Dim8Limits4}.  It is interesting to note
that in many cases this puts an effective lower limit on $\Lambda$ itself, in
addition to the upper limit derived from the unitarity condition.
In particular, the adopted criteria bound the value of $\Lambda$ to being above
$\sim$2 TeV for the ${\cal O}_M$ operators as well as for ${\cal O}_{S0}$.  
The ${\cal O}_T$ operators still allow a wider range of $\Lambda$.
Unfortunately, there is little we can learn from fitting $f_{S1}$, since
signal observability requires very low $\Lambda$ values, for which the new physics
could probably be detected directly.

\begin{figure}[hbt]
\vspace{-2cm}
\centering
\includegraphics[width=1.0\linewidth]{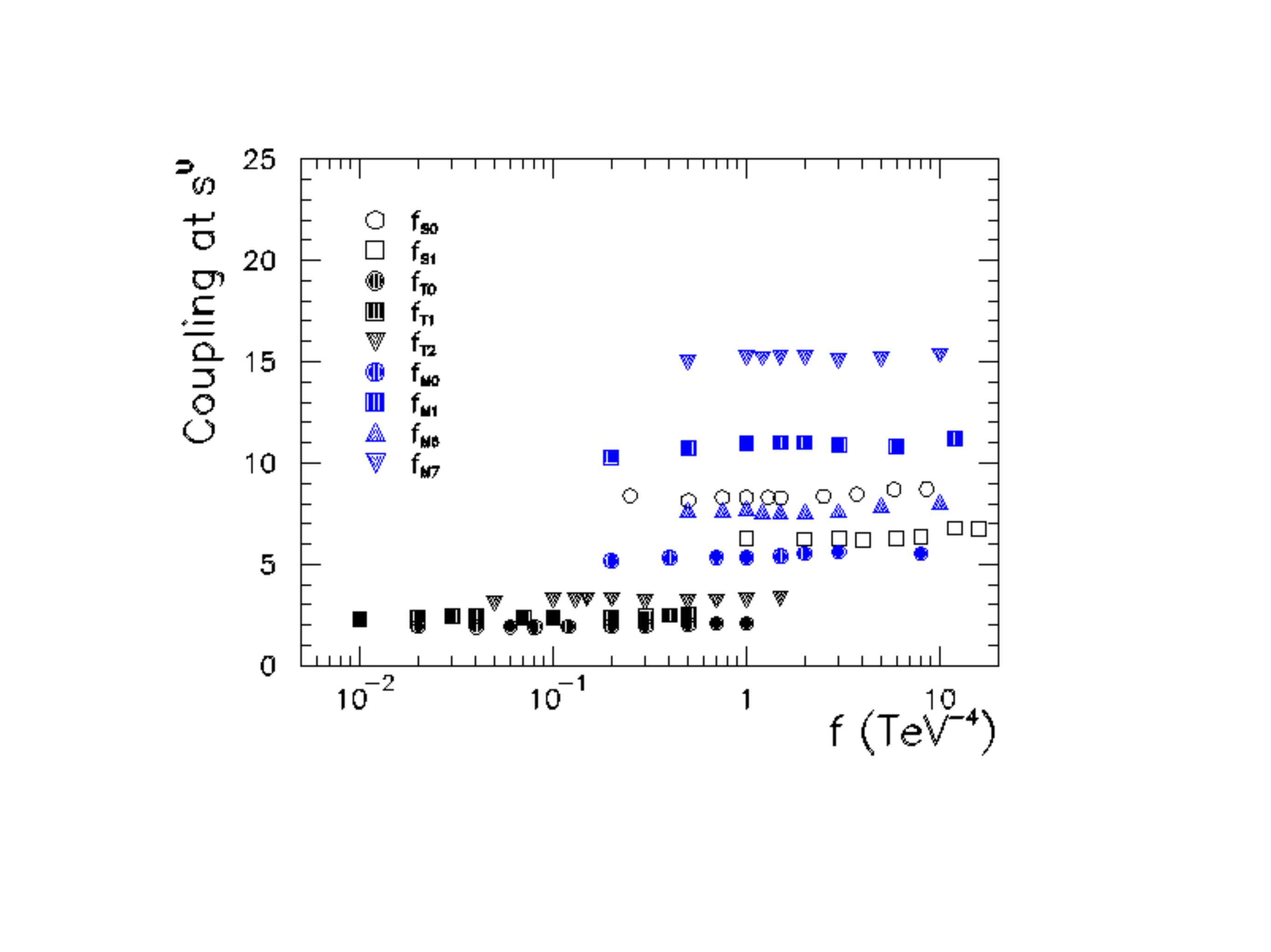}
\vspace{-3cm}
\caption{
Maximum value $\sqrt{c_i^{max}}$ of the coupling constants related to individual
dimension-8 operators, calculated at the energy where the unitarity limit
is reached, as a function of the relevant $f$ value.
}
\label{fig:slaweksplot}
\end{figure}

It is interesting to plot the values of the couplings $\sqrt{c_i}$ 
in Eq.~\eqref{eq:SMEFT2} as a function of $f_i$  assuming $\Lambda_{max}=M^U$ i.e.,
$c^{max}=f \times (M^{U})^{k-4}$, where $k$ is the dimensionality of the operator that defines the EFT ''model''. 
In models with one BSM scale and one BSM coupling constant $\sqrt{c_i}$ has roughly\footnote{The well justified matching is obtained with the use of the NDA normalization. This, together with a more detailed discussion, is done for SMEFT in Sec.~\ref{discReg27TeV}. The purpose in this Section is to give an idea of such matching and present qualitative features of the couplings in the context of our discovery regions.} the interpretation of the coupling constant \cite{Giudice:2007fh}.
The values of $c_i^{max}$ are to a good approximation
independent of $f$ (see Fig.~\ref{fig:slaweksplot}) and, being generally
in the range ($\sqrt{4\pi}, 4\pi$), reflect the 
approach to a strongly interacting regime (which is to be identified with $4\pi$ for the $\sqrt{c_i}$) in an underlying (unknown) UV complete theory.
The EFT discovery regions depicted in Figs.~\ref{fig:Dim8Limits3} and
\ref{fig:Dim8Limits4} have further interesting 
implications for the couplings $c_i$.  For a fixed $f$, the unitarity bound $\Lambda^2<s^U$ implies that 
$c_i<c_i^{max}=f(M^U)^4$, whereas
the lower bound on $\Lambda$ that comes from the combination of the
signal significance and EFT consistency criteria gives us
$c_i>\Lambda_{min}^ 4f$.  Thus, a given range $(\Lambda_{min}, \Lambda_{max})$ 
corresponds to a range of values of the couplings $c_i$, so that we could not only discover
an indirect sign of BSM physics, but also learn something about the nature of the 
complete theory, whether it is strongly or weakly interacting. 
In particular, for the following operators: ${\cal O}_{S0}$, ${\cal O}_{M0}$, ${\cal O}_{M1}$, ${\cal O}_{M6}$ and
${\cal O}_{M7}$, only models with $c_i$ being close to the strong interaction limit will be
experimentally testable, while a wider range of $c_i$ may be testable for ${\cal O}_{T0}$,
${\cal O}_{T1}$ and ${\cal O}_{T2}$.

\clearpage
\subsection{Discovery potential of HEFT}
\label{VVHEFT}

This Section is devoted to presentation of the results on the discovery regions when our EFT ''models'' are defined within the HEFT ansatz. More specifically, the EFT Lagrangian considered in the analysis is defined as the sum of the SM Lagrangian, by taking $\mathcal{L}_0$ from~\eqref{eq:heft4} with the choice~\eqref{eq:heft4b} and fixing $a_C=1=b_C$ in Eq.~\eqref{eq:heft5} and neglecting any higher order contribution in powers of $h/v$, plus $\Delta\mathcal{L}$ introduced in Eq.~\eqref{DeltaLHEFT}. The SM predictions are then recovered selecting $c_i=0$ for all $i$. The analysis including BSM effects is performed considering only one operator of $\Delta\mathcal{L}$ at a time, i.e. switching off all the other operator coefficients. %

The approach to numerical analysis is almost a carbon copy of what presented in Sec.~\ref{WWinSMEFT}. The minor differences correspond to somewhat updated or different software tools used: this time the samples events are generated in MadGraph5\_aMC@NLO
%~\cite{Alwall:2014hca} 
v2.6.2. The HEFT Lagrangian implementation is obtained with a private UFO 
%~\cite{Degrande:2011ua} 
file, created using FeynRules v2.3.32. Hadronization is performed with Pythia v8.2~\cite{Sjostrand:2006za}, run within MadGraph. Event files at the reconstructed level are generated with the help of the MadAnalysis5~\cite{Conte:2012fm} v1.6.33 package (available within MadGraph). Within the latter the FastJet v3.3.0 package~\cite{Cacciari:2011ma} is used with the jet clustering \verb+antikt+ algorithm with \verb+radius=0.35+ and \verb+ptmin=20+; the detector efficiencies is set to 100\%. 

In this Section (\ref{VVHEFT}) we work in the readly interpretable NDA normalization. Its explicit form for our choice of the HEFT operators was already introduced in Sec.~\eqref{onshellHEFT} (equations~\eqref{DeltaLHEFT} and~\eqref{eq:onshellHEFT1}).

\begin{figure}[h!] 
\begin{tabular}{cc}
\includegraphics[width=0.5\linewidth]{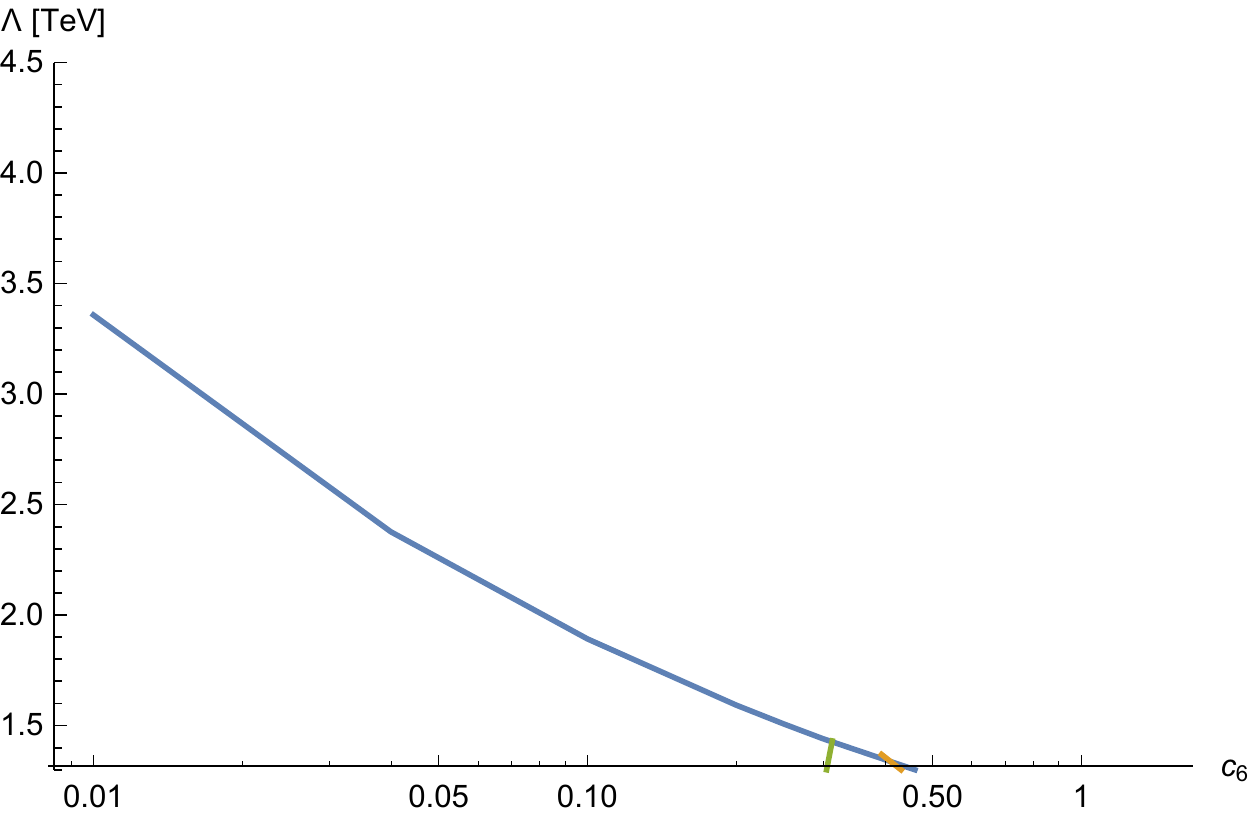} 
& \includegraphics[width=0.5\linewidth]{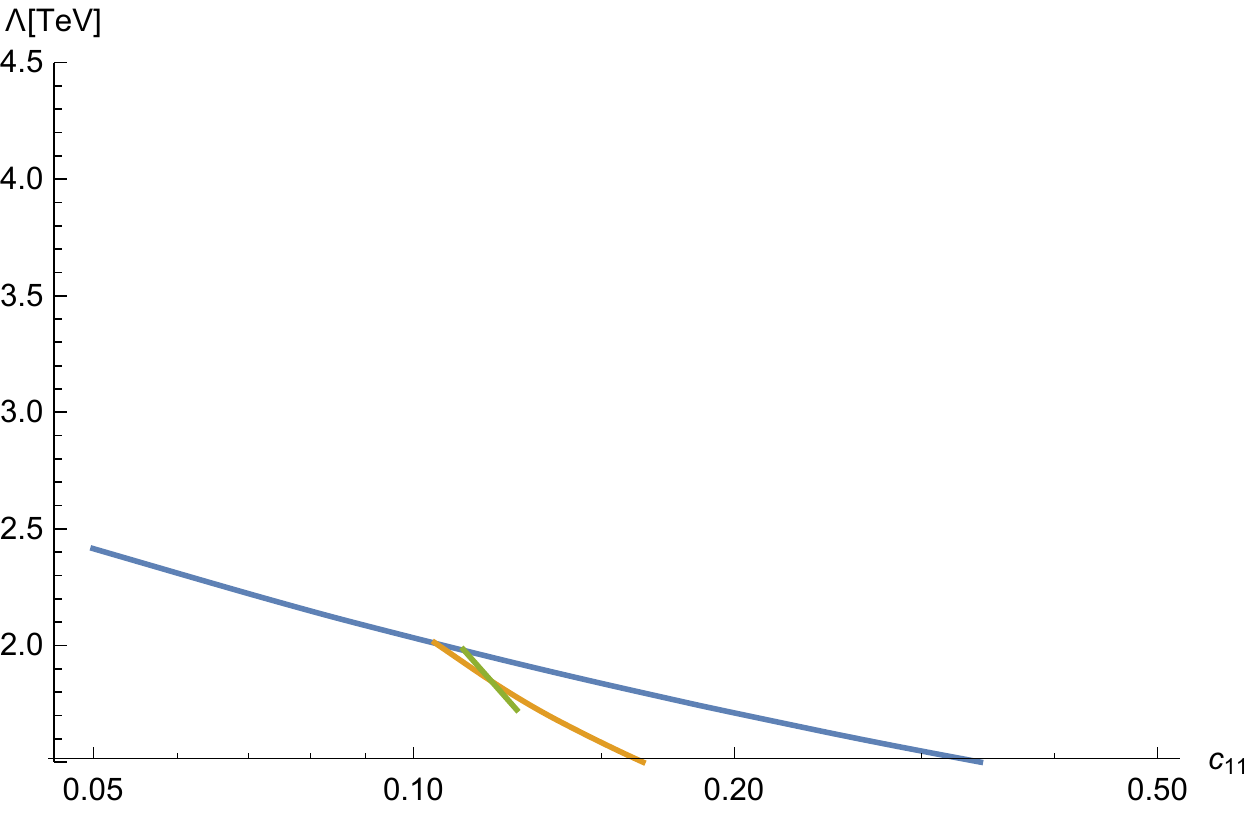} \\
\includegraphics[width=0.5\linewidth]{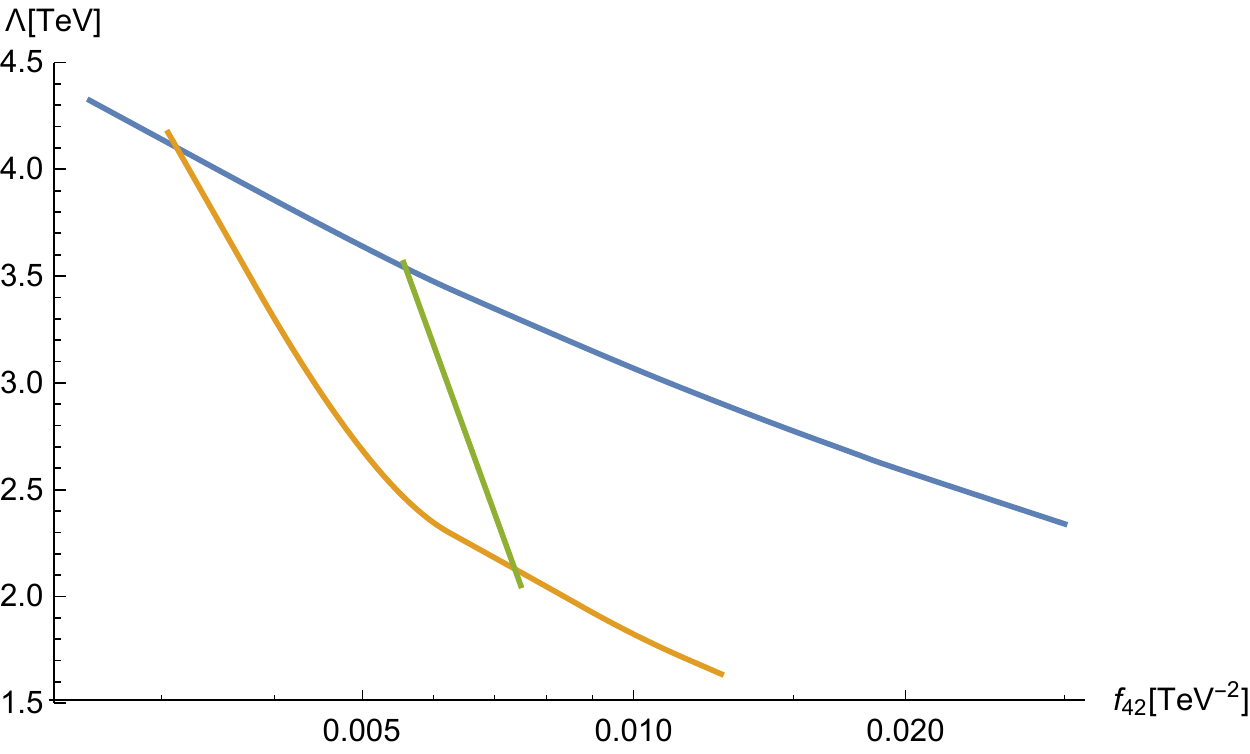} 
& \includegraphics[width=0.5\linewidth]{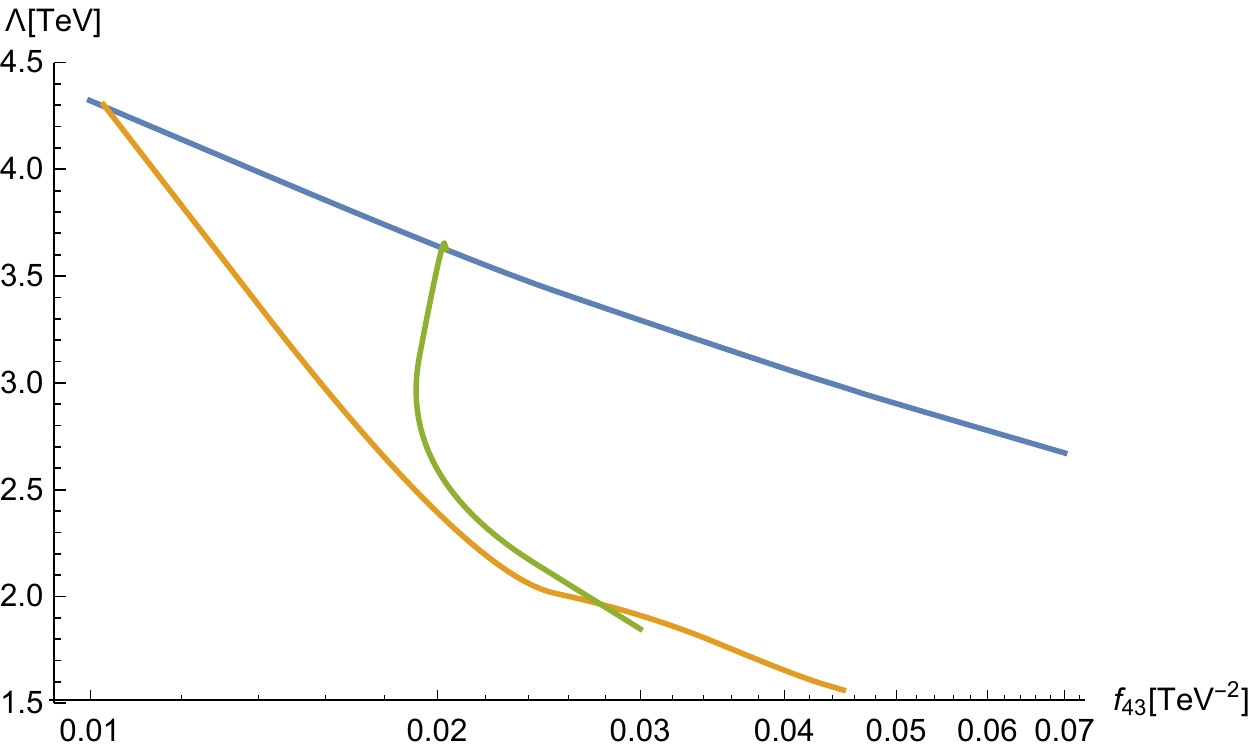} \\
\includegraphics[width=0.5\linewidth]{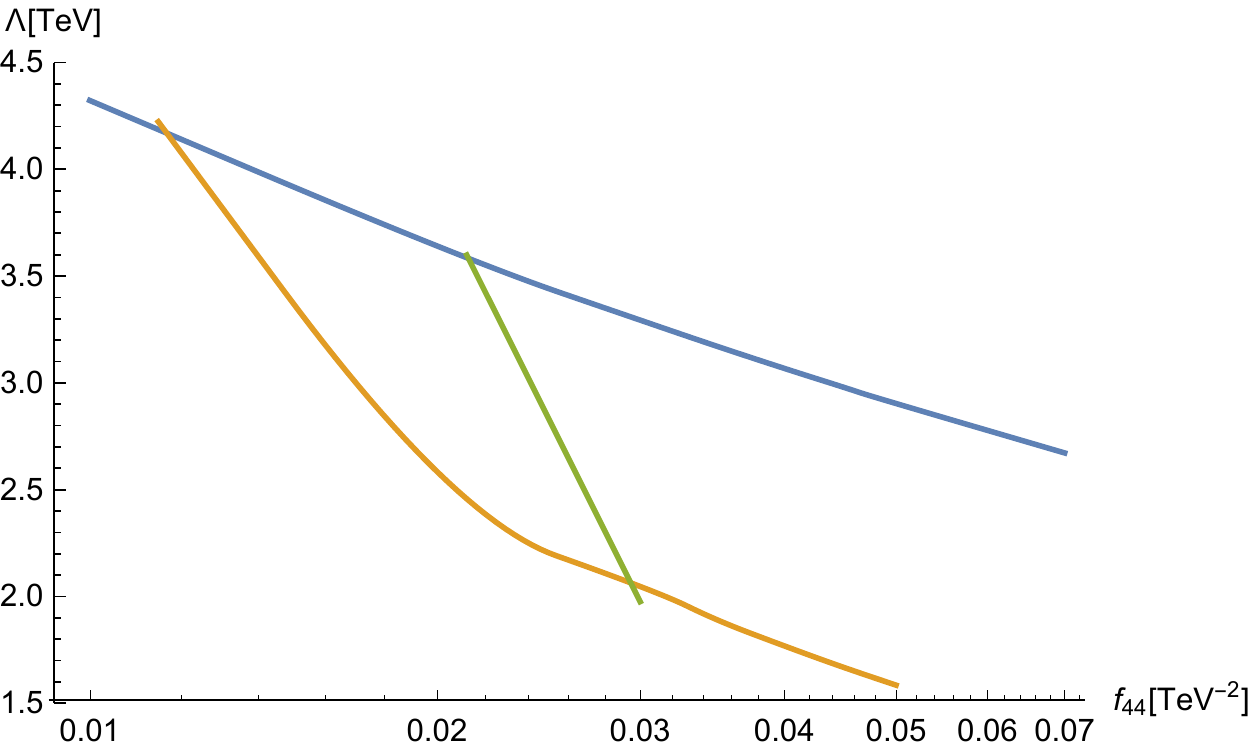} 
& \includegraphics[width=0.5\linewidth]{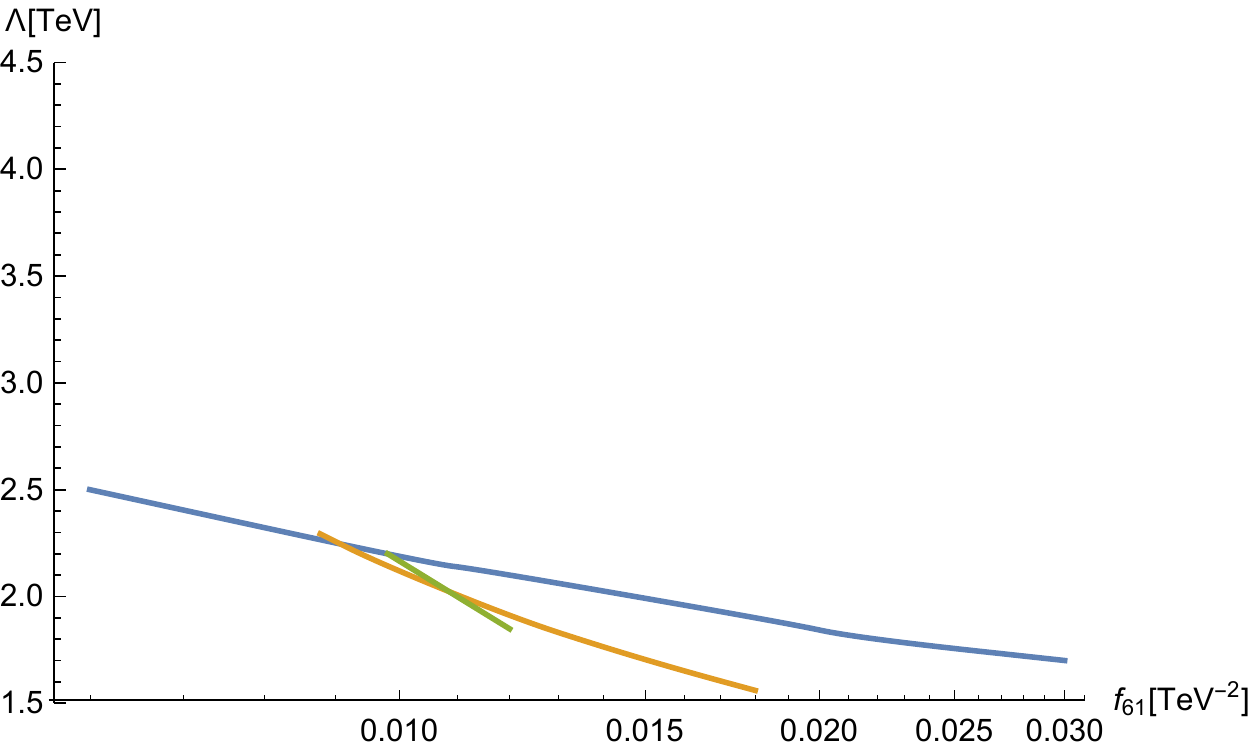} 
\end{tabular}
\caption{Regions in the $\Lambda$ {\it vs.} $c_i$ or $f_i$ (with $c_i,\,f_i>0$) space for $d_p=8$ HEFT operators in which a $5\sigma$ BSM signal can be observed and the EFT is applicable. The unitarity limit is shown in blue, the lower limits for a 5$\sigma$ signal significance from Eq.~(\ref{unitarized}) is in yellow, the upper limit on $2\sigma$ EFT consistency in green. $\sqrt{s} = 14 \TeV$ and an integrated luminosity of 3 $ab^{-1}$ are assumed. From top to bottom and from left to right, the operators considered are $\cP_6$, $\cP_{11}$, $\cT_{42}$, $\cT_{43}$, $\cT_{44}$ and $\cT_{61}$.}
\label{fig:trianglesPositivePart1}
\end{figure}

\begin{figure}[h!] 
\begin{tabular}{cc}
\includegraphics[width=0.5\linewidth]{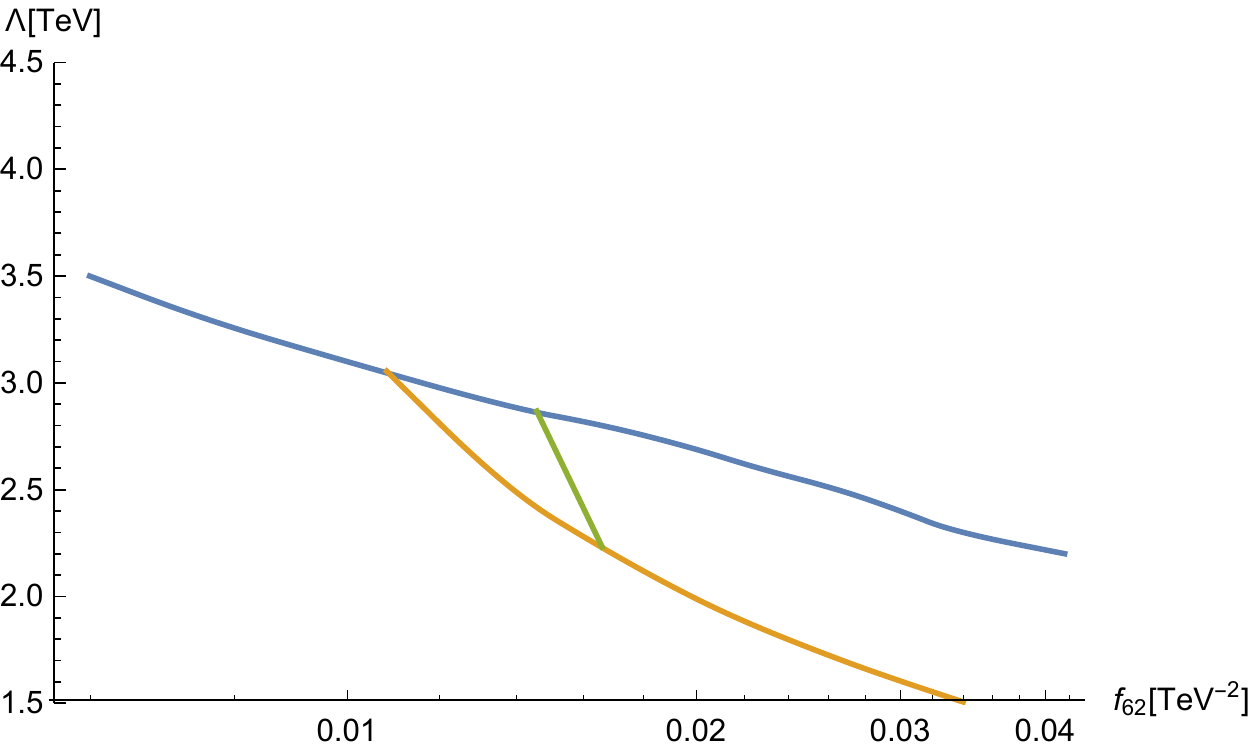}  
&\includegraphics[width=0.5\linewidth]{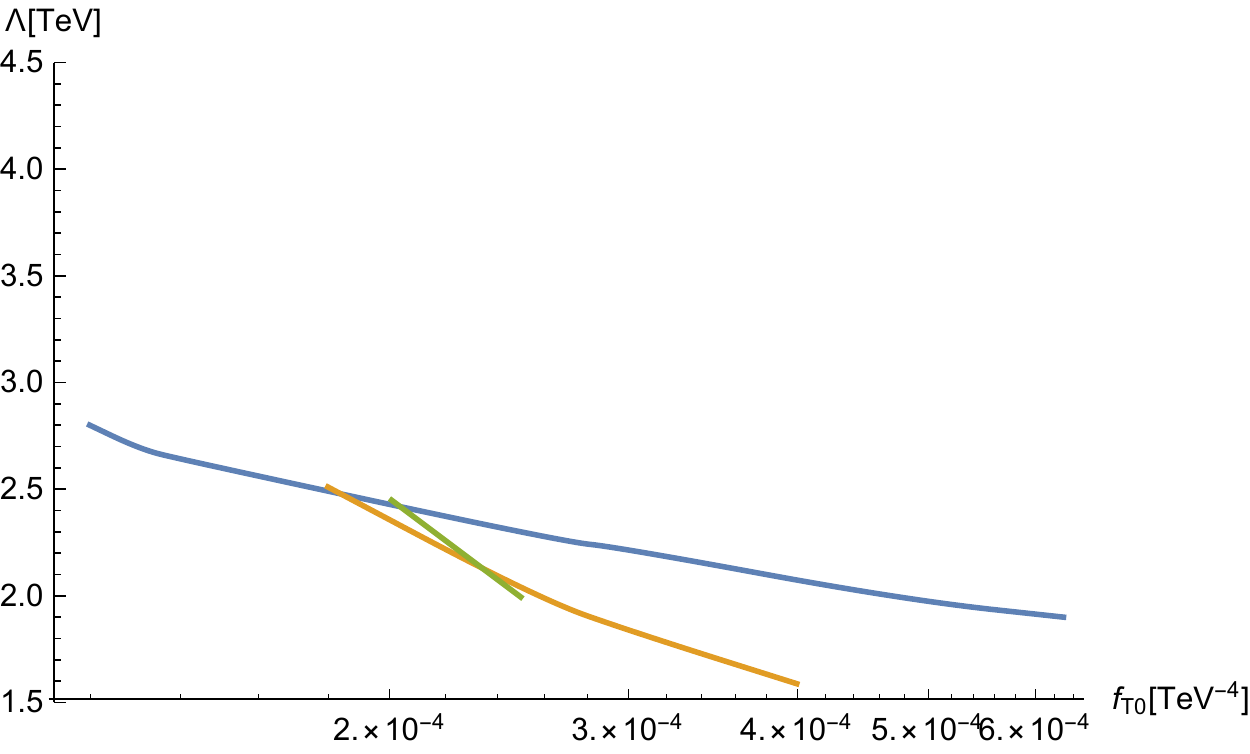}\\ 
\includegraphics[width=0.5\linewidth]{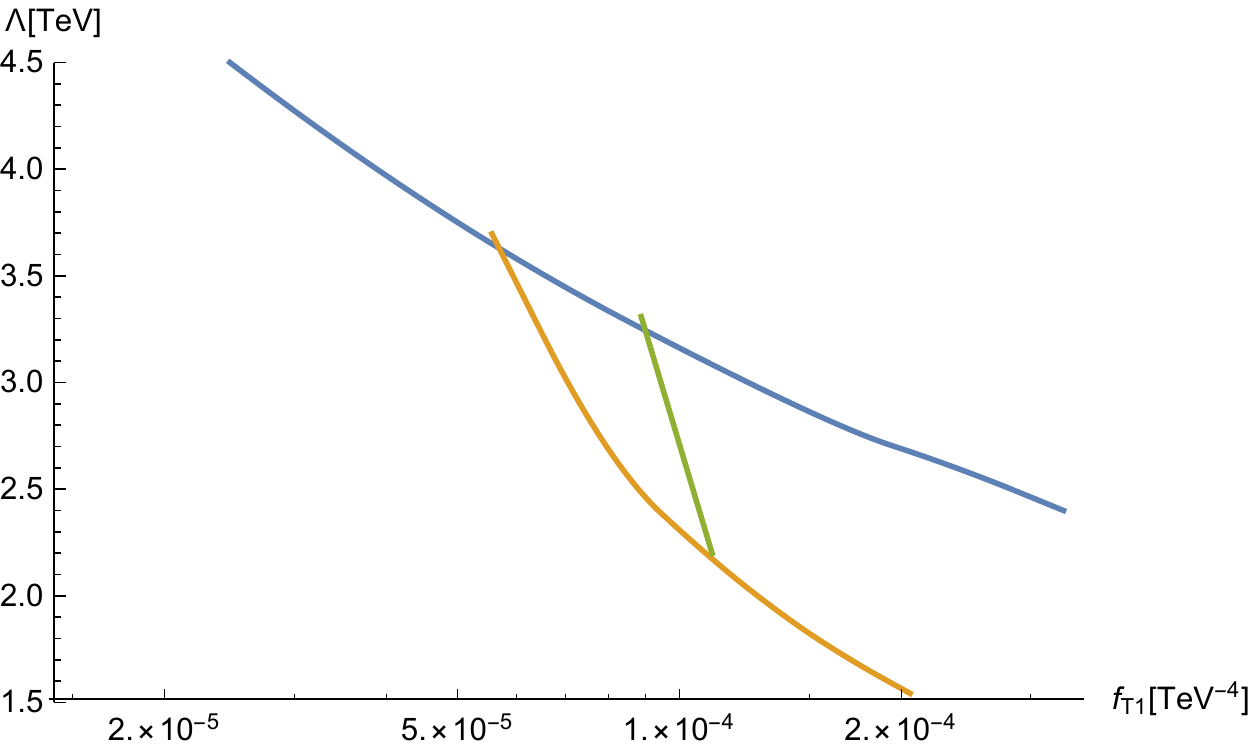} 
&\includegraphics[width=0.5\linewidth]{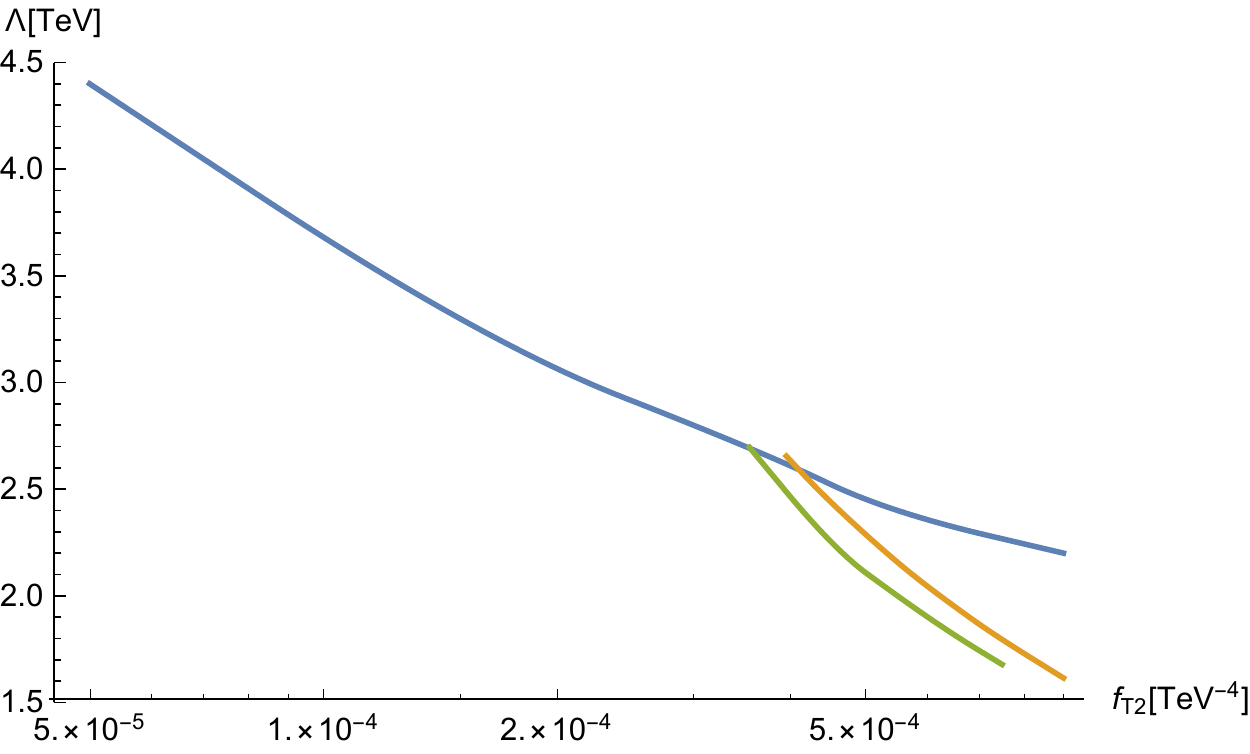}
\end{tabular}
\caption{Same description as in Fig.~\ref{fig:trianglesPositivePart1}. From top to bottom and from left to right, the operators considered are $\cT_{62}$, $\cO_{T1}$, $\cO_{T2}$ and $\cO_{T2}$.}
\label{fig:trianglesPositivePart2}
\end{figure}

\begin{figure}[h!] 
\begin{tabular}{cc}
\includegraphics[width=0.5\linewidth]{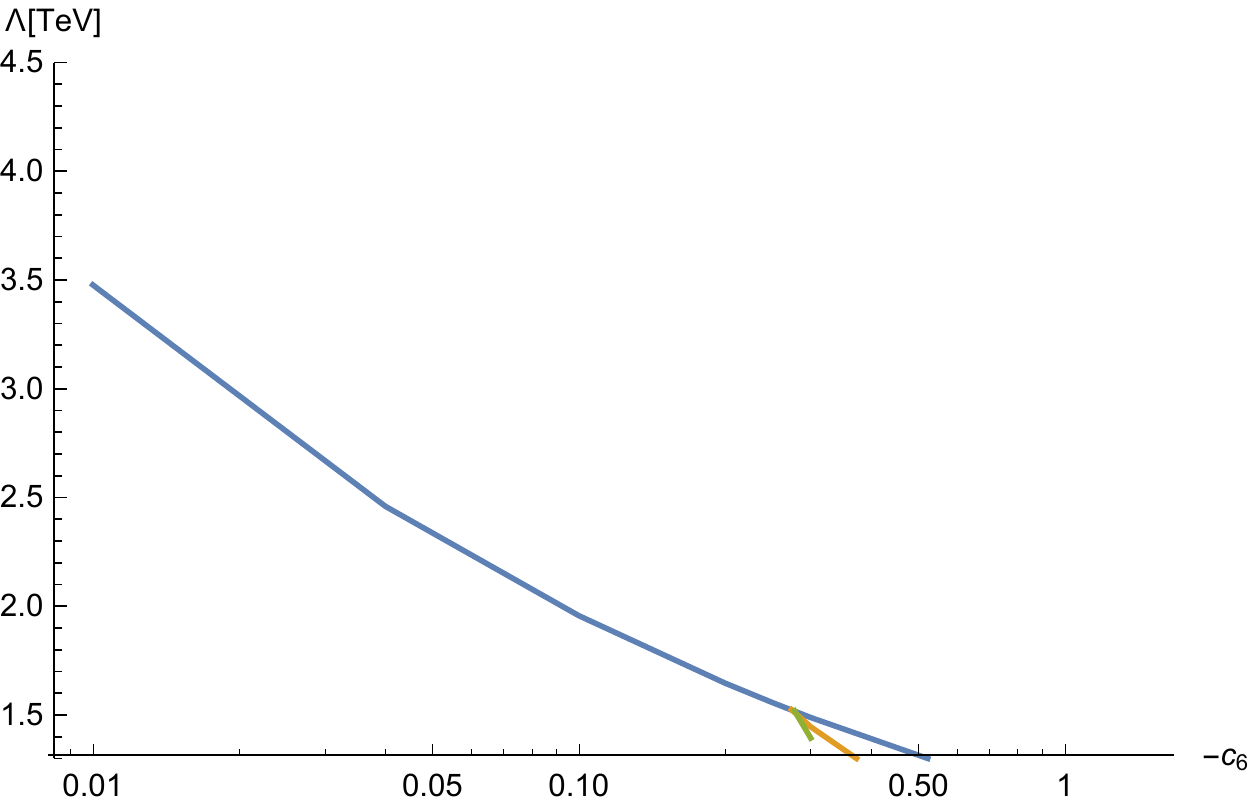} 
& \includegraphics[width=0.5\linewidth]{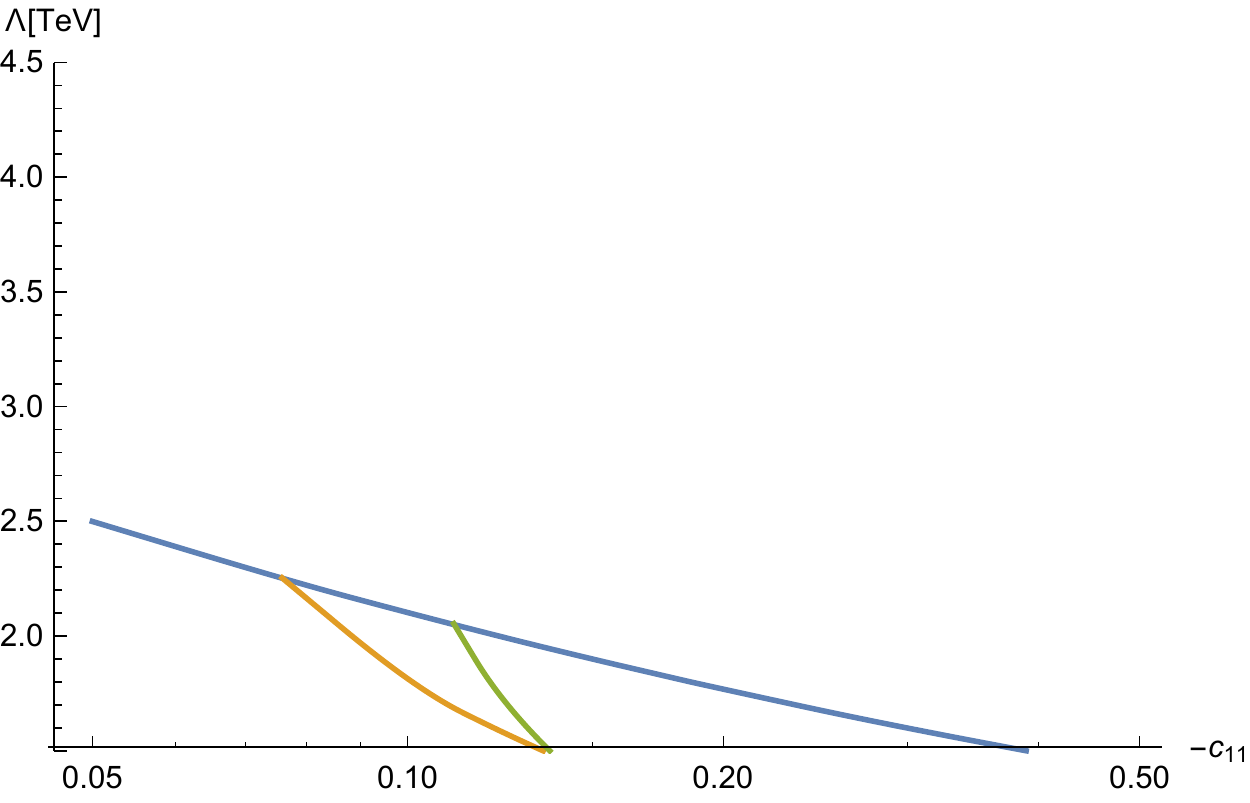} \\
\includegraphics[width=0.5\linewidth]{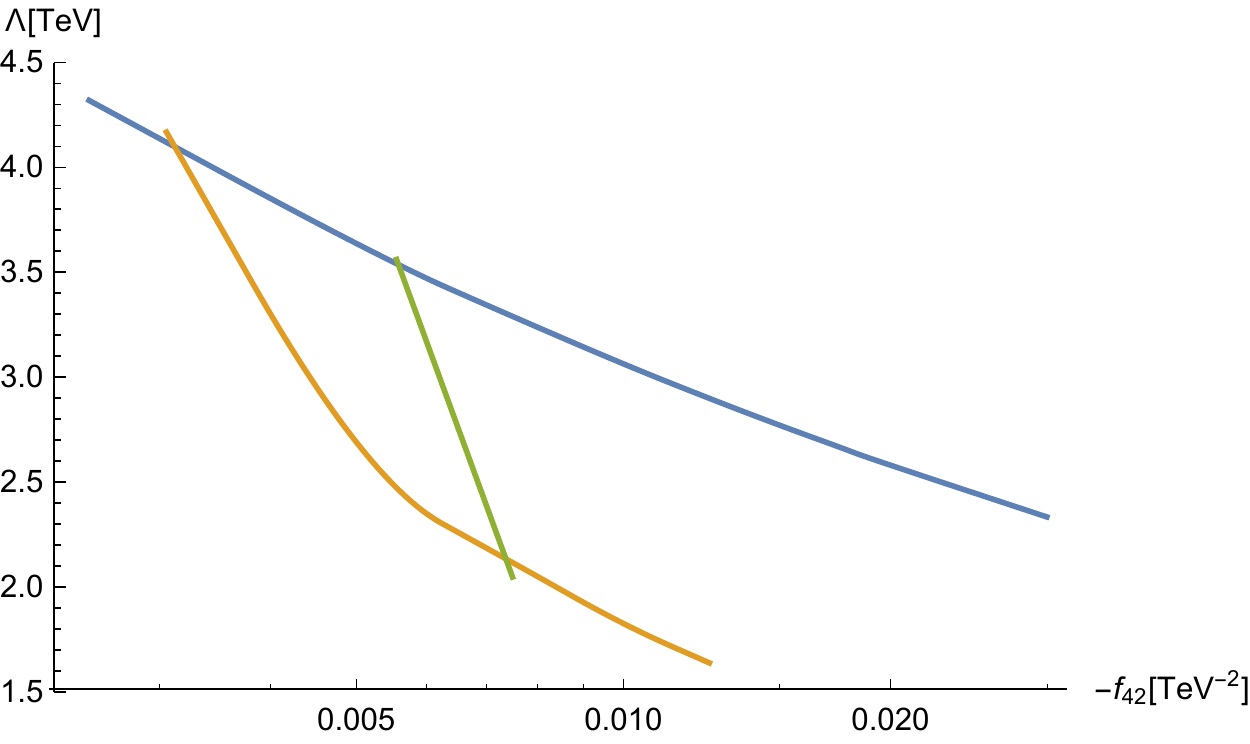} 
& \includegraphics[width=0.5\linewidth]{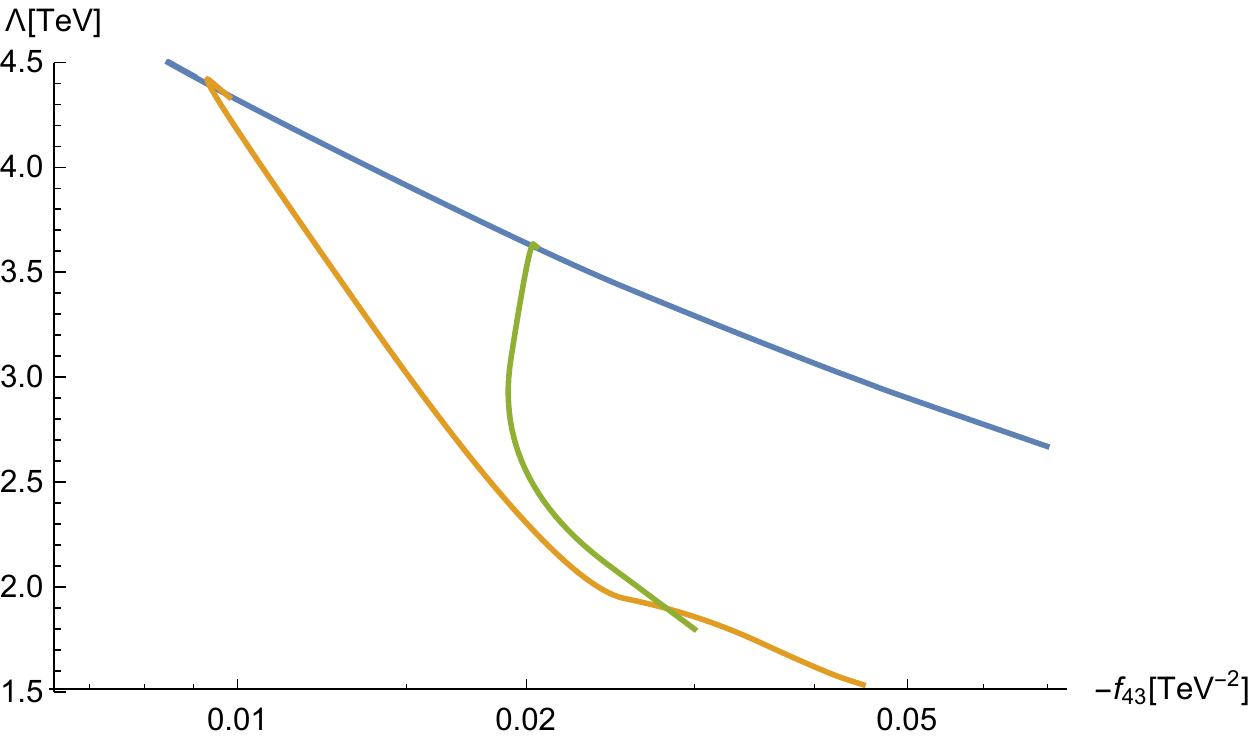} \\
\includegraphics[width=0.5\linewidth]{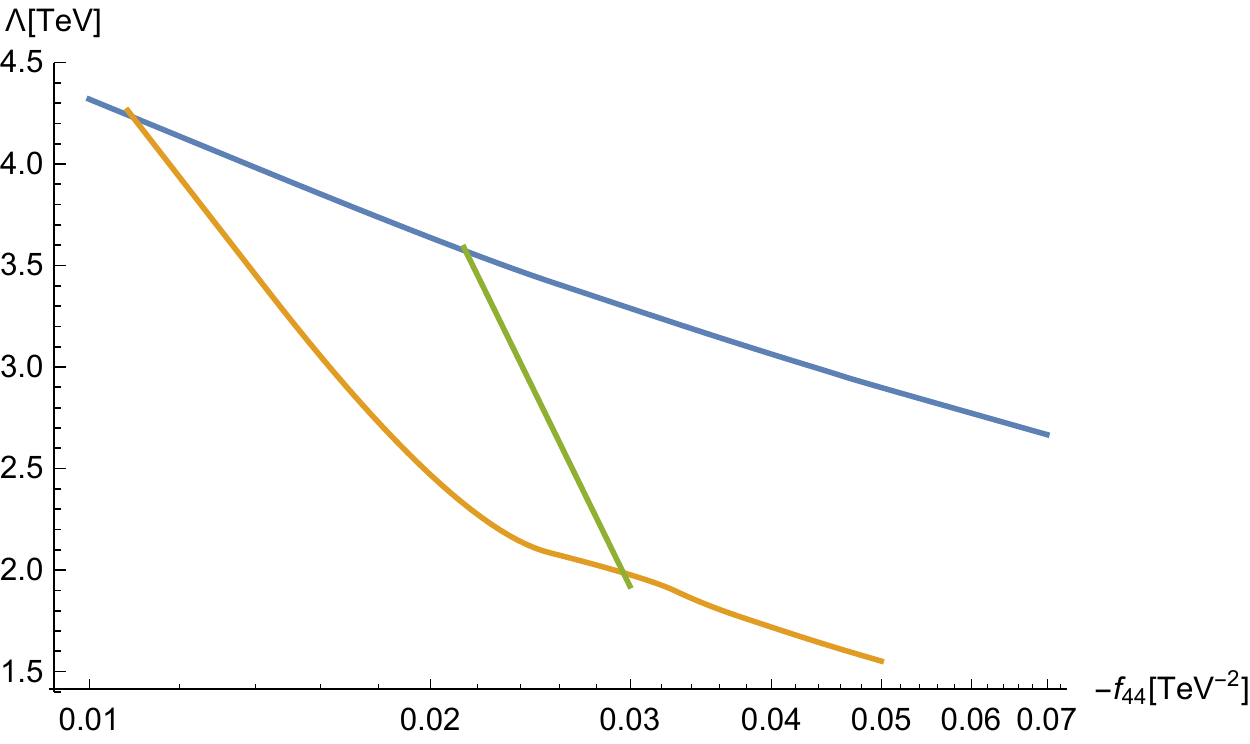} 
&\includegraphics[width=0.5\linewidth]{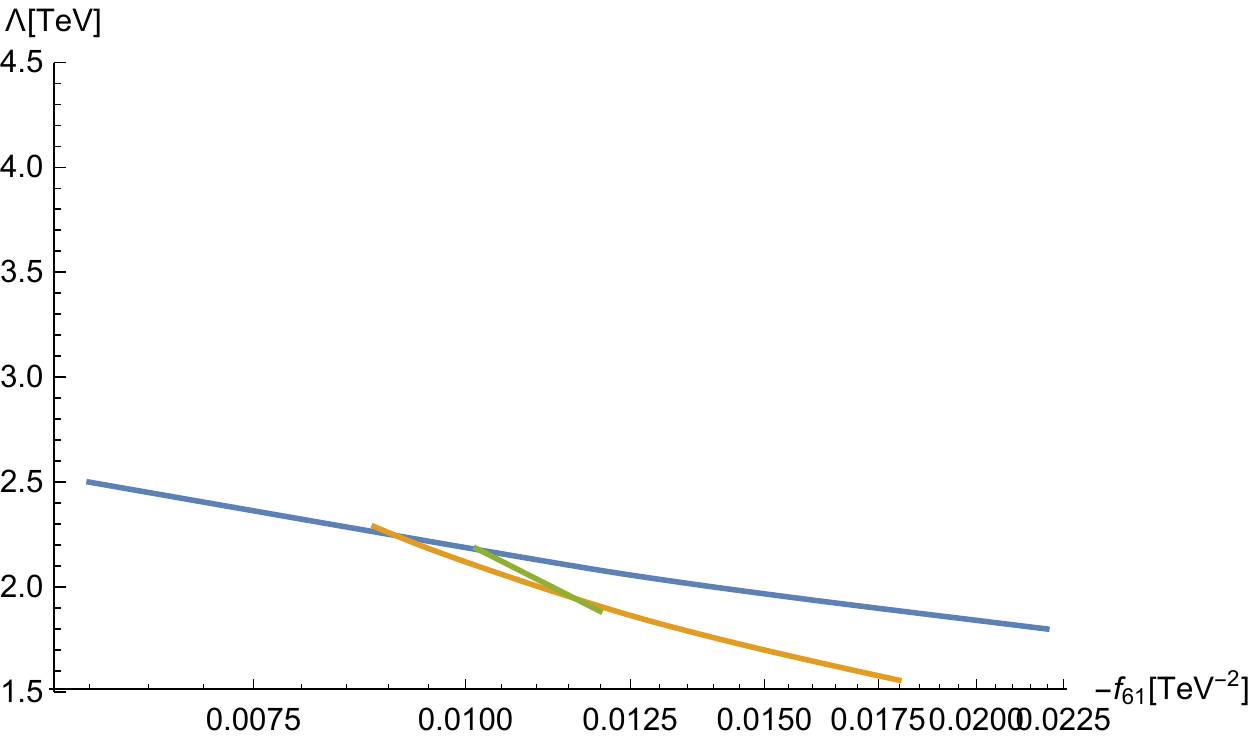} 
\end{tabular}
\caption{Same description and operators as in Fig.~\ref{fig:trianglesPositivePart1}. Negative values for $c_i$ and $f_i$ are considered.}
\label{fig:trianglesNegativePart1}
\end{figure}

\begin{figure}[h!] 
\begin{tabular}{cc}
\includegraphics[width=0.5\linewidth]{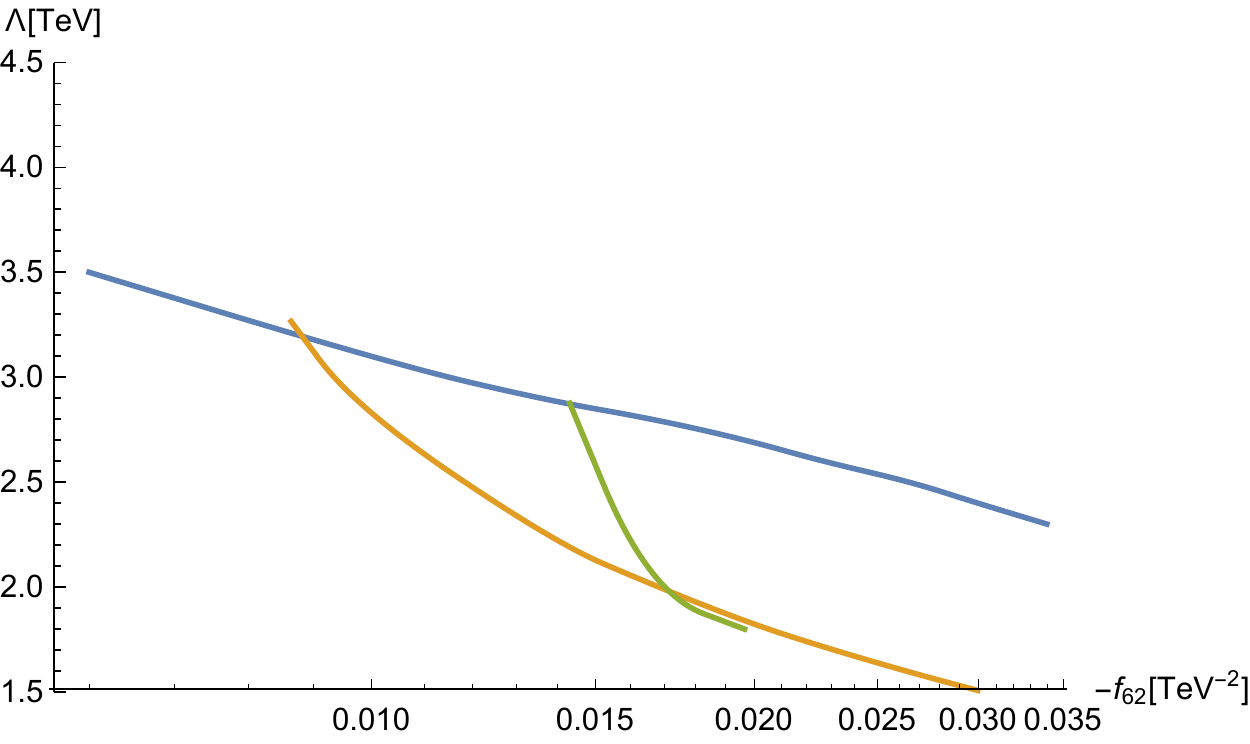}  
& \includegraphics[width=0.5\linewidth]{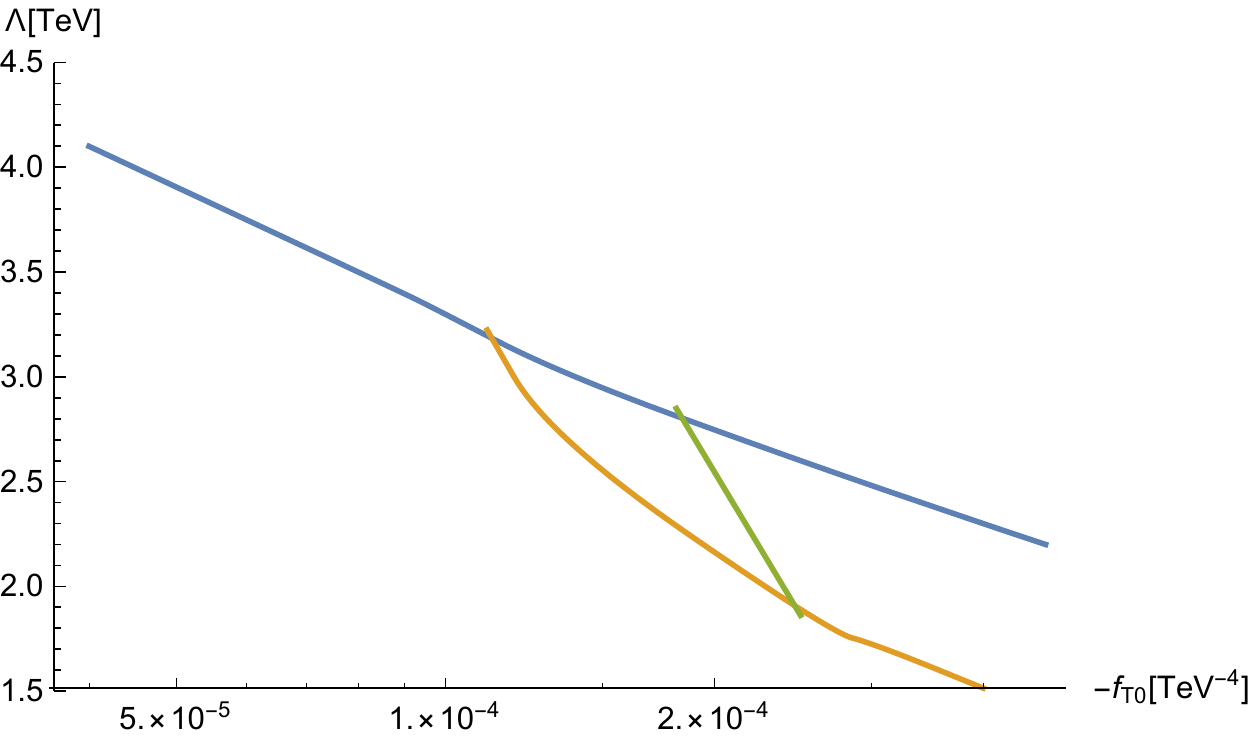}\\
\includegraphics[width=0.5\linewidth]{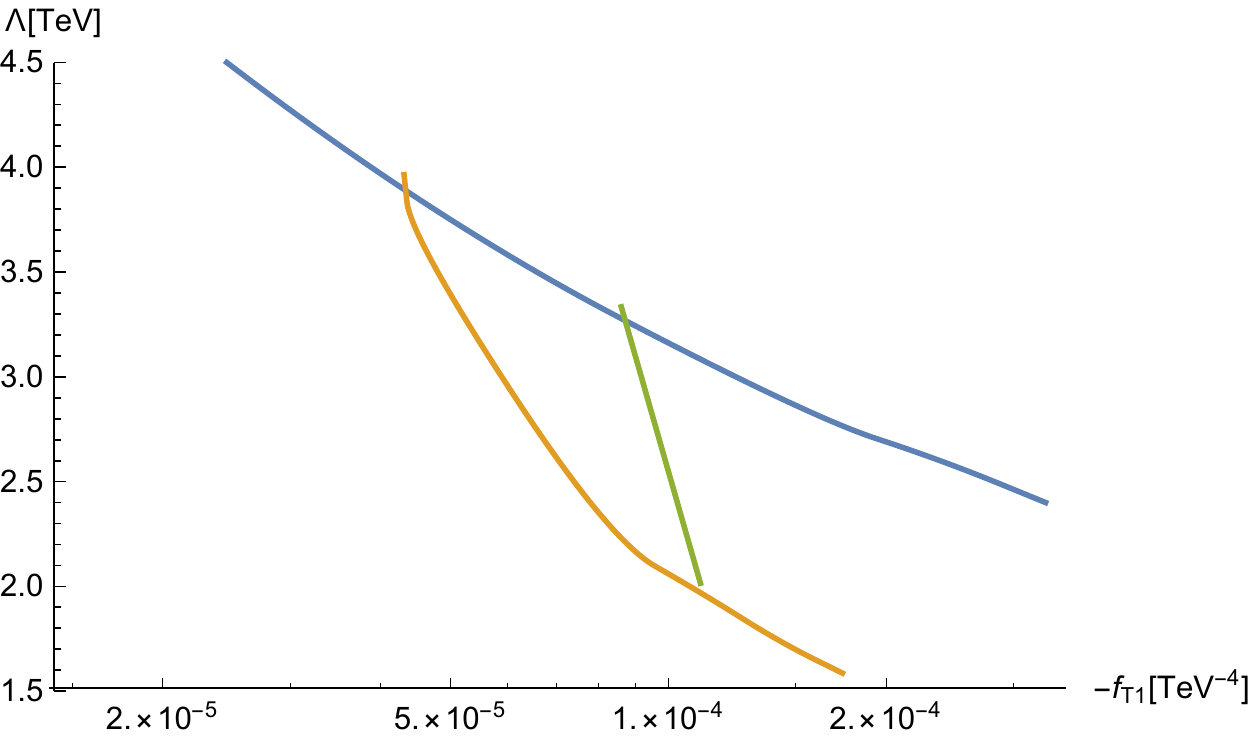} 
& \includegraphics[width=0.5\linewidth]{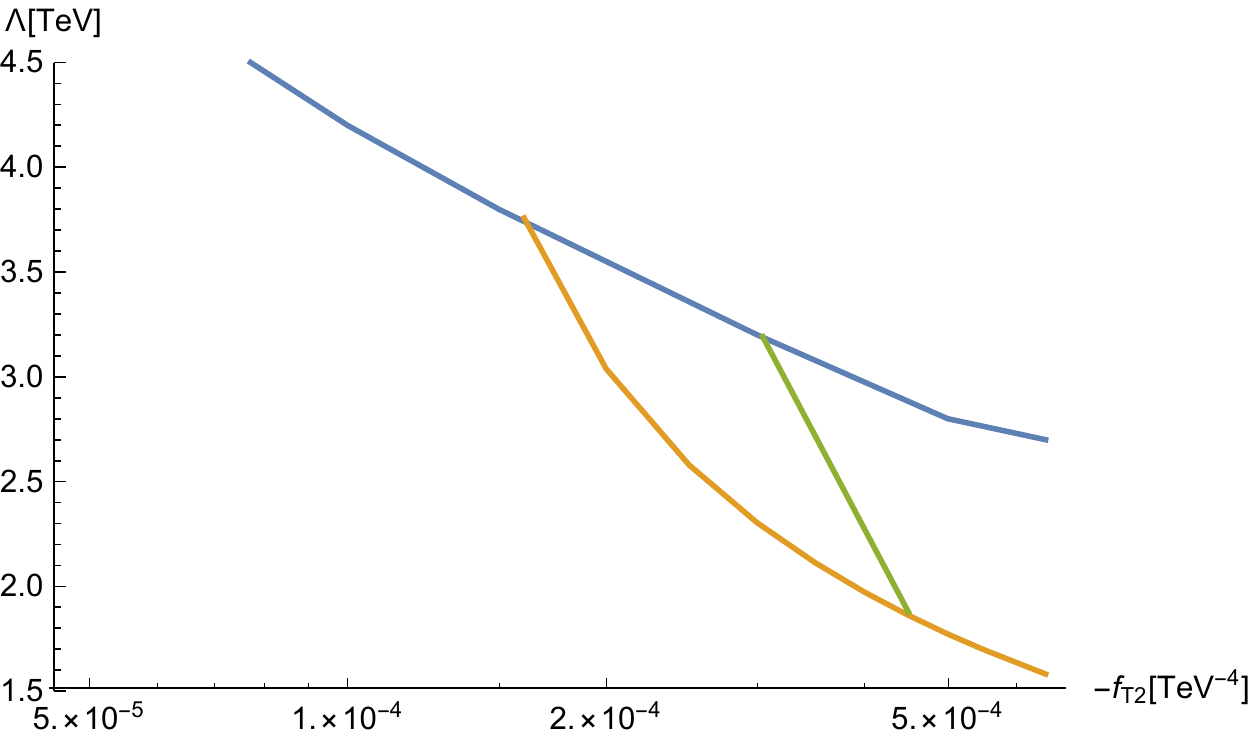}
\end{tabular}
\caption{Same description and operators as in Fig.~\ref{fig:trianglesPositivePart2}. Negative values for $c_i$ and $f_i$ are considered.}
\label{fig:trianglesNegativePart2}
\end{figure}

All in all, Figs.~\ref{fig:trianglesPositivePart1} and \ref{fig:trianglesPositivePart2} show the results on the discovery regions for the individual operators $\cP_{6}, \cP_{11},\cT_{42},\cT_{43},\cT_{44}, \cT_{61}, \cT_{62}, \cO_{T0},\cO_{T1}, \cO_{T2}$ for positive $f_i$. The case of $f_i<0$ is shown in Figs.~\ref{fig:trianglesNegativePart1} and ~\ref{fig:trianglesNegativePart2}. The obtained regions resemble triangles in which the left side (yellow line) is bounded by the $5\sigma$ discoverability, the upper side (blue line) by the unitarity violation limit, while the right side (green line) by the 2$\sigma$-EFT consistency. The comment on the most sensitive variables is identical to what described in the SMEFT analysis (due to trivial correspondence of operators) and the $M_{o1}$ was found most sensitive for $\cT_{42},\cT_{43},\cT_{44}$.

%For the $\cP_6$ and $\cP_{11}$ plots, the vertical axis represents the cut-off scale $\Lambda$ (Eq.~\eqref{eq:pk}) bounded by the energy $M^U$ at which unitarity is violated. 

The discovery region for $\cP_{6}$, in the specific case of a negative Wilson coefficient (Fig.~\ref{fig:trianglesNegativePart1} up-left) is tiny, pointing our a very narrow range of values for $c_6$, centered in $c_6\approx-0.3$, for which a signal could be seen. For a positive Wilson coefficient (Fig.~\ref{fig:trianglesPositivePart1} up-left), there is not any available region, as the ''discoverability line'' is on the right of the $2\sigma$-consistency line: in the region where data may point out a signal, the EFT description breaks down. For this second case, with positive $c_6$, no effects are expected considering the operator $\cP_6$. 

The situation for $\cP_{11}$ (Figs.~\ref{fig:trianglesPositivePart1} and \ref{fig:trianglesNegativePart1} up-right) is different. While for positive Wilson coefficient, the discoverability region is tiny, with $c_{11}\sim0.11$, for a negative Wilson coefficient this region is much larger, with $c_{11}\in-[0.076,\,0.14]$.

These results for $\cP_6$ and $\cP_{11}$ are interesting because they show the impact of the $2\sigma$ EFT consistency requirement considered here for the first time in the HEFT context: in the literature, the coefficients $c_6$ and $c_{11}$ (typically labelled $a_5$ and $a_4$, respectively) may vary within a large range of values providing hypothetically visible signals at colliders; here, instead, chances to find a signal of NP in the $W_LW_L\to W_LW_L$ scattering, described in a consistent HEFT framework by the operators $\cP_6$ and $\cP_{11}$, are present essentially only for negative $c_{11}$, or extremely tuned values of $c_6$ and positive $c_{11}$.

Other cases where the discoverability region is tiny are for $\cT_{61}$ for both signs of the corresponding Wilson coefficients and for $\cO_{T0}$ for $c_{T0}>0$, as can be seen in Figs.~\ref{fig:trianglesPositivePart1} and \ref{fig:trianglesNegativePart1} lower-right and Fig.~\ref{fig:trianglesPositivePart2} upper-right, respectively. These regions are centered in $[f_{61}\sim\pm0.01\TeV^{-2},\,\Lambda\sim2.2\TeV]$ for $\cT_{61}$ and $[f_{T0}\sim2\times 10^{-4}\TeV^{-4},\,\Lambda\sim2.4\TeV]$ for $\cO_{T0}$, corresponding to $c_{61}\sim\pm0.046$ and $c_{T0}\sim0.006$. On the other side, another case where no discoverability region is found is for $\cO_{T2}$ and for positive Wilson coefficient.

\begin{table}[h!]
\hspace{-0.6cm}
\begin{tabular}{|c|c|c|c|c|c|}
\hline
&&&&&\\[-4mm]
   		& $\cP_6$ 	& $\cP_{11}$ 		& $\cT_{42}$ 		& $\cT_{43}$ 		& $\cT_{44}$ 		\\[1mm] 
\hline
&&&&&\\[-3mm]
$c_i>0$ 	& - 		& $0.11$ 			& $[0.033,\,0.007]$ 	& $[0.11,\,0.27]$ 	& $[0.13,\,0.27]$ 	\\[1mm] 
$c_i<0$ 	& $0.3$	& $-[0.076,\,0.14]$ 	& $-[0.034,\,0.070]$	& $-[0.11,\,0.27]$	& $-[0.11,\,0.28]$	\\[1mm]
\hline
\hline
&&&&&\\[-4mm]
  		& $\cT_{61}$ 		& $\cT_{62}$ 		& $\cO_{T0}$ 			& $\cO_{T1}$ 			& $\cO_{T2}$ \\[1mm] 
\hline
&&&&&\\[-3mm]
$c_i>0$  	& $[0.045,\,0.047]$  	& $[0.083,\,0.120]$	& $[0.0051,\,0.0072]$ 	& $[0.0026,\,0.0110]$	& - \\[1mm] 
$c_i<0$  	& $-[0.044,\,0.048]$ 	& $-[0.072,\,0.12]$ 	& $-[0.003,\,0.012]$ 		& $-[0.0018,\,0.0110]$ 	& $-[0.0052,\,0.032]$ \\[1mm] 
\hline
\end{tabular}
\caption{Ranges of values for the dimensionless $c_i$ operator coefficients corresponding to the discovery regions in Figs.~\ref{fig:trianglesPositivePart1}-\ref{fig:trianglesNegativePart2}. The normalisation is defined in Eq.~\eqref{DeltaLHEFT}. ``-'' denotes no available discovery region.}
\label{tab:cminCmaxHEFT}
\end{table}

For all the other cases, there are relatively large discovery regions where the corresponding operators enhance the signal with respect to the SM prediction, with a possibility for this enhancement to be measured at colliders and with a consistent EFT description. The Tab.~\ref{tab:cminCmaxHEFT} shows the ranges of values for the dimensionless coefficients $c_i$, with the normalisation as in Eq.~\eqref{DeltaLHEFT}, for all the discovery regions in Figs.~\ref{fig:trianglesPositivePart1}-\ref{fig:trianglesNegativePart2}. As can be seen, all the coefficients are smaller than 1 and, considering that the NDA normalisation has been adopted in Eq.~\eqref{DeltaLHEFT}, this would lead to the conclusion that the discovery regions correspond to (some) weakly coupled BSM theories. While this statement is true for $\mathcal{P}_6,\mathcal{P}_{11}$, the cases where $W_{\mu\nu}$ in the operators as building blocks, is more subtle and requires a separate discussion. See Sec.~\ref{discReg27TeV} where it is discussed.

As pointed out in Sect.~\ref{onshellHEFT}, a few HEFT operators considered here can find a sibling in the SMEFT at $n=8$, that have been the focus of the study in Sec.~\ref{WWinSMEFT}. The discoverability regions show in 
Figs.~\ref{fig:trianglesPositivePart1}-\ref{fig:trianglesNegativePart2} for these operators indeed match with the results presented in Sec.~\ref{WWinSMEFT}, after the rescaling in order to account for the normalisations. This can be easily checked for the operators $\cO_{T1}$ that are the same in both the bases and the operator $\cP_6$ that is equivalent to $\cO_{S1}$; it is slightly more difficult for the operator $\cP_{11}$ that is equivalent to a combination between $\cO_{S0}$ and $\cO_{S1}$, as explicitly shown in Eq.~\eqref{eq:correlations}. The small differences occur for the regions of smaller $\Lambda$, where there is less available discoverability region, see e.g. for $\cO_{T0}$, below $\Lambda\sim 2.5\TeV$. They are due to different (updated) software tools and different analysis algorithms within these tools. On the other side, these differences may represent the uncertainties on the discovery region determination.

The last three operators, $\cT_{42}$, $\cT_{43}$ and $\cT_{44}$, do no have a sibling in the SMEFT Lagrangian at $n\leq8$. The interactions described by $\cT_{42}$, $\cT_{43}$ and $\cT_{44}$ will be governed by $n>8$ SMEFT operators, and therefore the strength of their signal is expected to be much more suppressed in the linear anzatz. Moreover, the helicity amplitudes that are considerably enhanced or even dominating the cross section at $M_{WW}\lesssim\Lambda$ are different than those found for the SMEFT operators (for the total unpolarized cross sections and its polarised fractions in on-shell $WW$ scattering with $\cT_{42}$, $\cT_{43}$ and $\cT_{44}$ insertions see Sec.~\ref{onshellHEFT}; see also the Appendix). This suggests that the same-sign $WW$ scattering could be a sensitive channel to disentangle between the SMEFT and HEFT descriptions of the EWSB sector, but a more dedicated analysis would be necessary to investigate this possibility.%
 
The results presented so far for the HEFT Lagrangian can be translated in terms of more fundamental theories whose dynamics takes place at a much higher energy scale. We provide below two examples. The first example is the minimal $SO(5)/SO(4)$ CH model which was matched to the HEFT Lagrangian in Refs.~\cite{Alonso:2014wta,Hierro:2015nna}. In this CH model the initial Lagrangian is written invariant under the global $SO(5)$ symmetry; after the spontaneous breaking down to $SO(4)$, the SM GBs and the physical Higgs arise as  a bi-doublet of $SU(2)_L\times SU(2)_R\equiv SO(4)$ Goldstone bosons, described altogether by a single unitary matrix. After this breaking, the Lagrangian describing the low-energy model matches the HEFT one, where the Wilson coefficients $c_i$ are written in terms of the high-energy coefficients, $\tilde c_i$, in the notation of Refs.~\cite{Alonso:2014wta,Hierro:2015nna}. 

The analysis in Refs.~\cite{Alonso:2014wta,Hierro:2015nna} considered up to four-derivative operators and therefore, once focusing only on the genuine quartic operators, the results of Tab.~\ref{tab:cminCmaxHEFT} in this paper can be used to constrain the operator coefficients $\tilde c_4$, $\tilde c_5$ and $\tilde c_6$ that appear in Tab.~1 in Ref.~\cite{Alonso:2014wta}. Adopting the NDA normalisation, the $c_6$ and $c_{11}$ coefficients can be written as
\be
c_6=-8\,\pi^2\,\xi\,\tilde c_6+16\,\pi^2\,\xi^2\,\tilde c_4\,,\qquad\qquad
c_{11}=16\,\pi^2\,\xi^2\,\tilde c_5\,.
\ee
An explicit value for $\xi$ can be extracted once taking the values in Tab.~\ref{tab:cminCmaxHEFT}: assuming, for example, the specific values for $\tilde c_i$ coefficients indicated inside the brackets,
\be
\begin{aligned}
c_6(\tilde c_4=-1,\,\tilde c_6=0)&=-0.3\quad&&\Longrightarrow\quad \xi=0.04\\
c_6(\tilde c_4=0,\,\tilde c_6=1)&=-0.3\quad&&\Longrightarrow\quad \xi=0.004\\
c_{11}(\tilde c_5=1)&=0.11\quad&&\Longrightarrow\quad \xi=0.026\\
c_{11}(\tilde c_5=-1)&=-[0.076,\,0.14]\quad&&\Longrightarrow\quad \xi=[0.026,\,0.03]\,.
\end{aligned}
\label{XiValuesfromCoefficients}
\ee

As a second example, one can consider the so-called Minimal Linear $\sigma$ Model~\cite{Feruglio:2016zvt}, that is a renormalizable model which may represent the UV completion of the Minimal CH model. Also in this case, only up to four-derivative operators have been considered in the low-energy limit; while the Wilson coefficient $c_{11}$ does not receive any contribution, $c_6$ turns out to be dependent only from $\tilde c_4=1/64$. The parameter $\xi$ is fixed to be
\be
\xi=0.35\,.
\label{xiFromP6}
\ee
This result should be interpreted as follows: in case of a NP discovery in the $W^+W^+$ scattering with a significance of more than $5\sigma$, in the Lorentz configurations described by the $\cP_6$ operator, and assuming that this NP corresponds to the Minimal Linear $\sigma$ Model, the parameter $\xi$ would take the value in Eq.~\eqref{xiFromP6}. However, such a large value would be already excluded considering EW precision observable constraints~\cite{Feruglio:2016zvt} (for a review see Ref.~\cite{Panico:2015jxa}). It follows that the Minimal Linear $\sigma$ Model cannot explain such a NP signal on $W^+W^+$ scattering.

An alternative possibility to extract bounds on $\xi$ is to consider the traditional relation between the scale $f$ and the cut-off $\Lambda$ present in the CH scenario~\cite{Kaplan:1983fs},
\be
f<\Lambda<4\pi f\,.
\ee
After a simple manipulation, this expression can be written in terms of the parameter $\xi$ as
\be
\dfrac{v^2}{\Lambda^2}<\xi<16\pi^2\dfrac{v^2}{\Lambda^2}\,.
\ee
This relation provides model independent bounds on $\xi$ that can be derived looking at the discoverability plots in Figs.~\ref{fig:trianglesPositivePart1}-\ref{fig:trianglesNegativePart2}. Considering again, as an example, the $\cP_{11}$ operator, 
\be
\begin{cases}
c_{11}>0\quad\rightarrow\quad\Lambda\approx2\TeV\quad&\Longrightarrow\quad 0.015<\xi<2.4\\
c_{11}<0\quad\rightarrow\quad1.5\TeV\lesssim\Lambda\lesssim2.3\TeV\quad&\Longrightarrow\quad 0.011<\xi<4.2\\
\end{cases}
\ee
compatible with the results in Eqs.~\eqref{XiValuesfromCoefficients} and \eqref{xiFromP6}. Similar ranges of values can be found for the other operators.

\clearpage

\subsection{The SMEFT discovery regions in HE-LHC phase}
\label{discReg27TeV}
In this Section we report the results on the discovery regions calculated for a 27 TeV machine and compare them with
the discovery regions obtained at 14 TeV (HL-LHC). In the HL-LHC case, for all the individual dimension-8 operators that 
affect genuinely the $WWWW$ quartic coupling  
the triangles were found to be rather narrow or even entirely empty (for
${\cal O}_{S1}$).
This Section is devoted to the verification if
an increased beam energy and integrated luminosity will translate into larger EFT triangles. The resuls are presented in the Madgraph convention (same as was the case of 14 TeV SMEFT analysis). Moreover, we interpret both the HL-LHC and HE-LHC SMEFT discovery regions, i.e. we check what values of the BSM couplings the discovery regions that we found, correspond to. For the latter goal we shall subsequently translate our results to be presented first in the Madgraph normalization, to the physically well justified NDA normalization. %tutaj

The methodology is a copy of the one presented in~\ref{WWinSMEFT}. The only difference is the updated MadGraph version (v5.2.6.2). We disregard here the fact that an increased beam energy may lead to a
re-optimization of the detector geometries and of selection criteria in order to accommodate
for the slightly different event topologies.
\begin{figure}[h] 
  \begin{tabular}{cc}
      \includegraphics[width=0.5\linewidth]{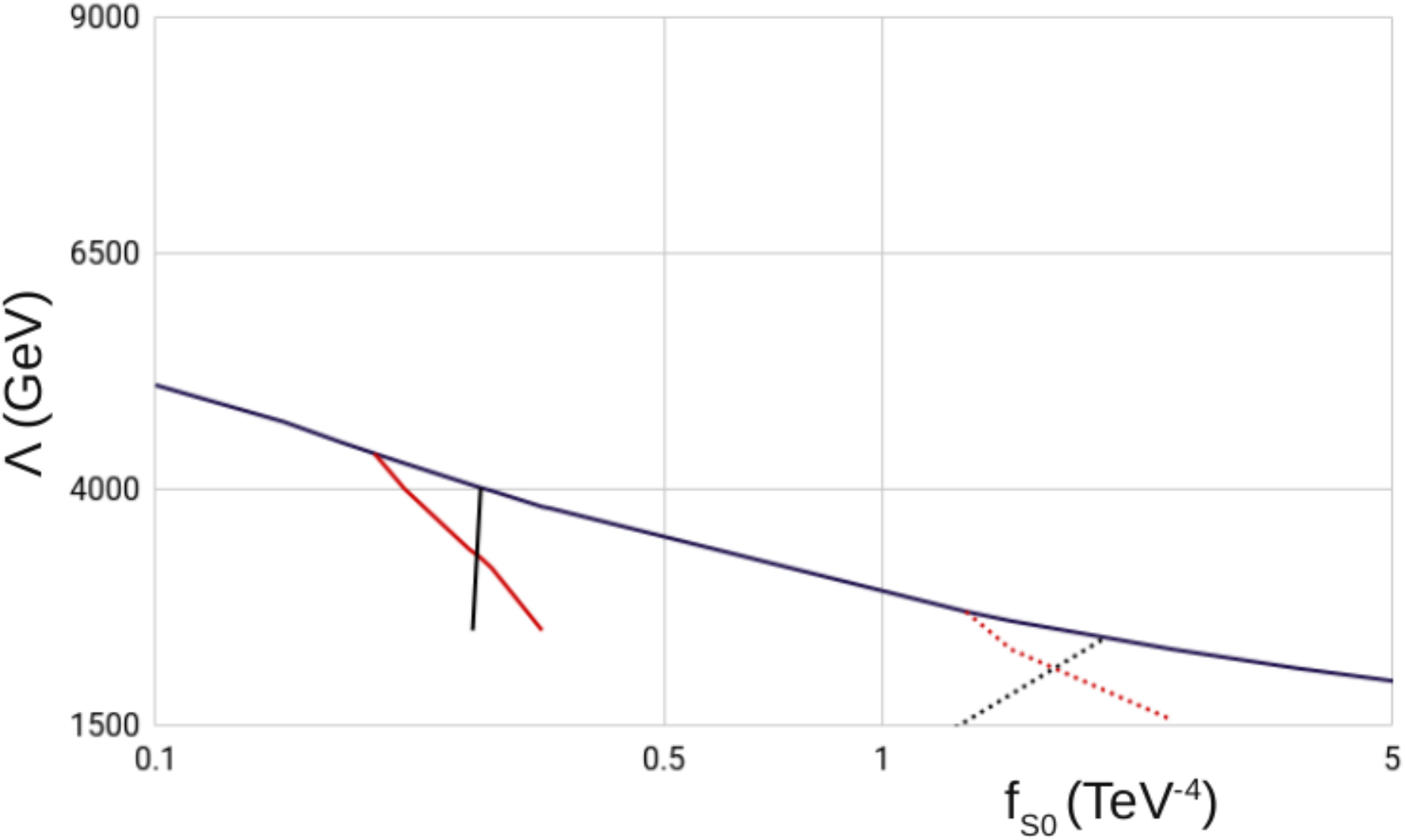} 
& \includegraphics[width=0.5\linewidth]{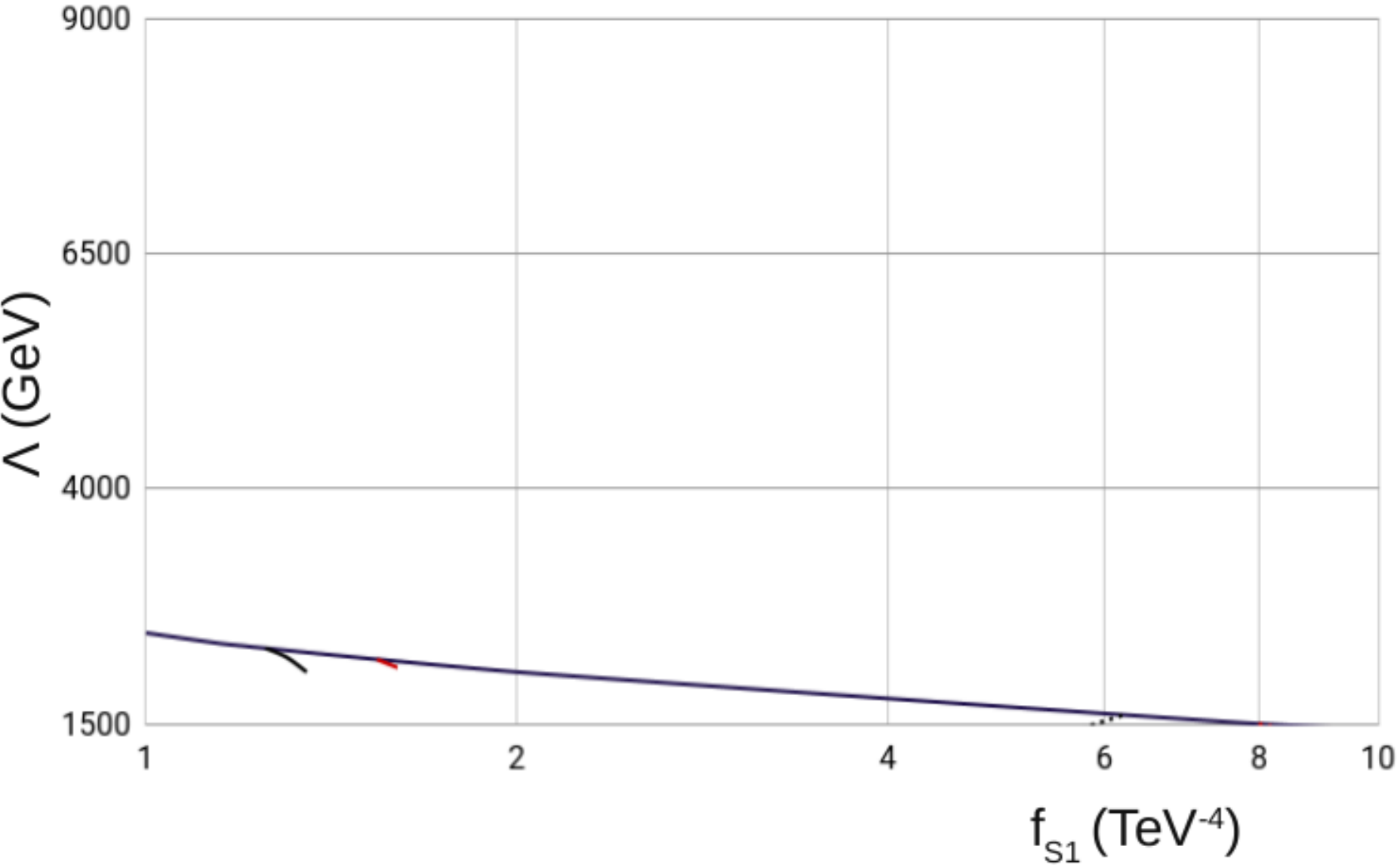} \\
\includegraphics[width=0.5\linewidth]{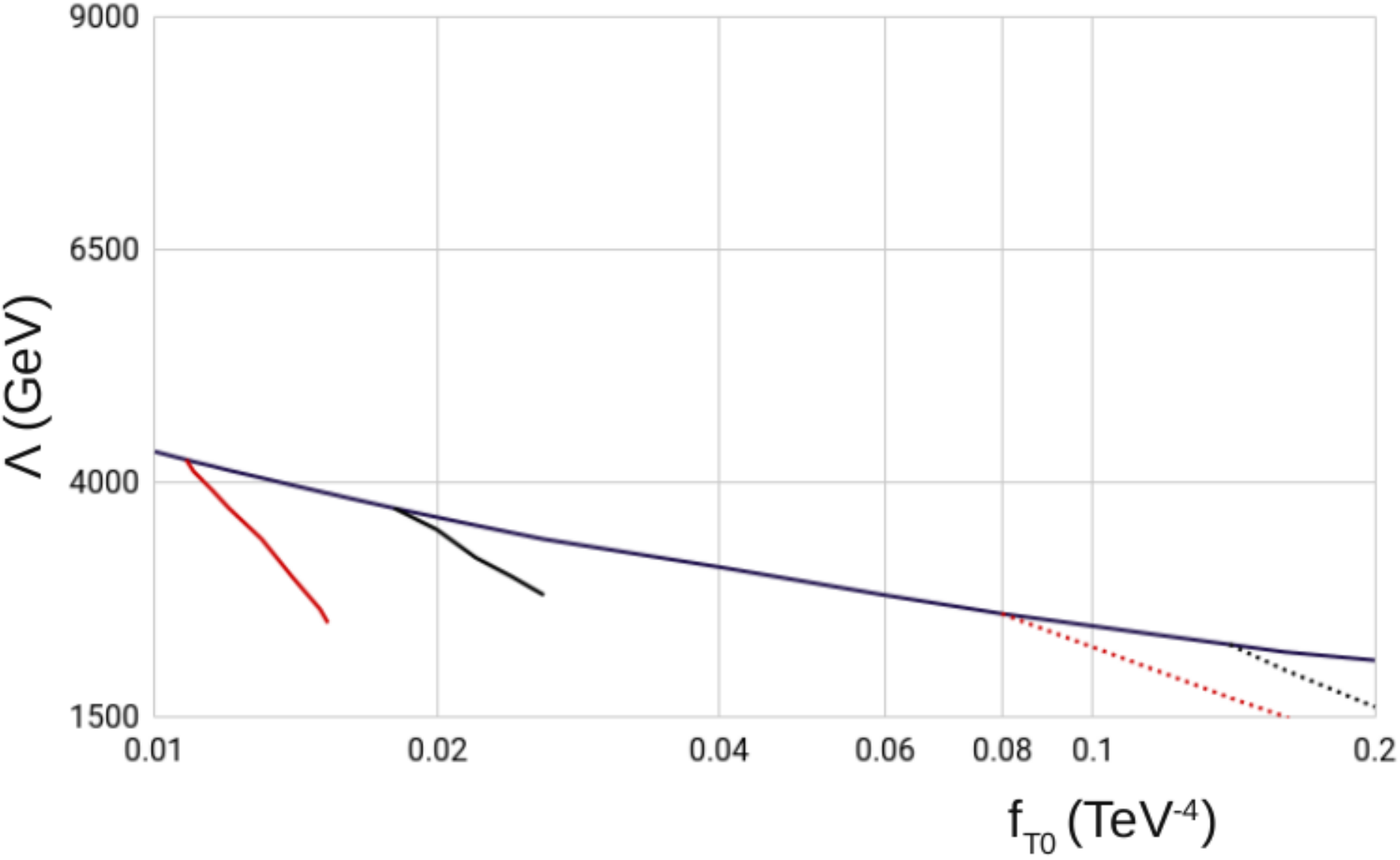} 
& \includegraphics[width=0.5\linewidth]{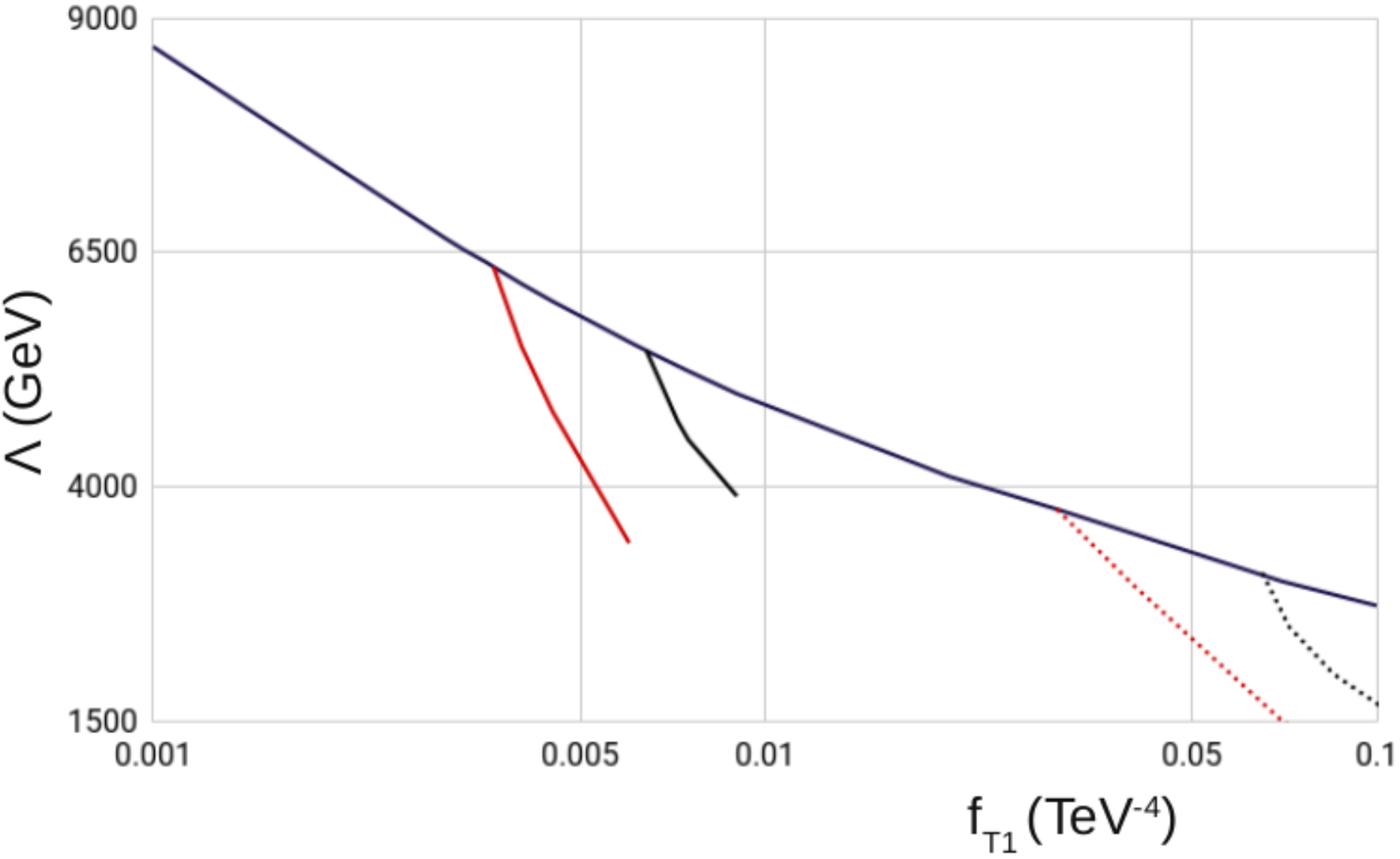} \\
\includegraphics[width=0.5\linewidth]{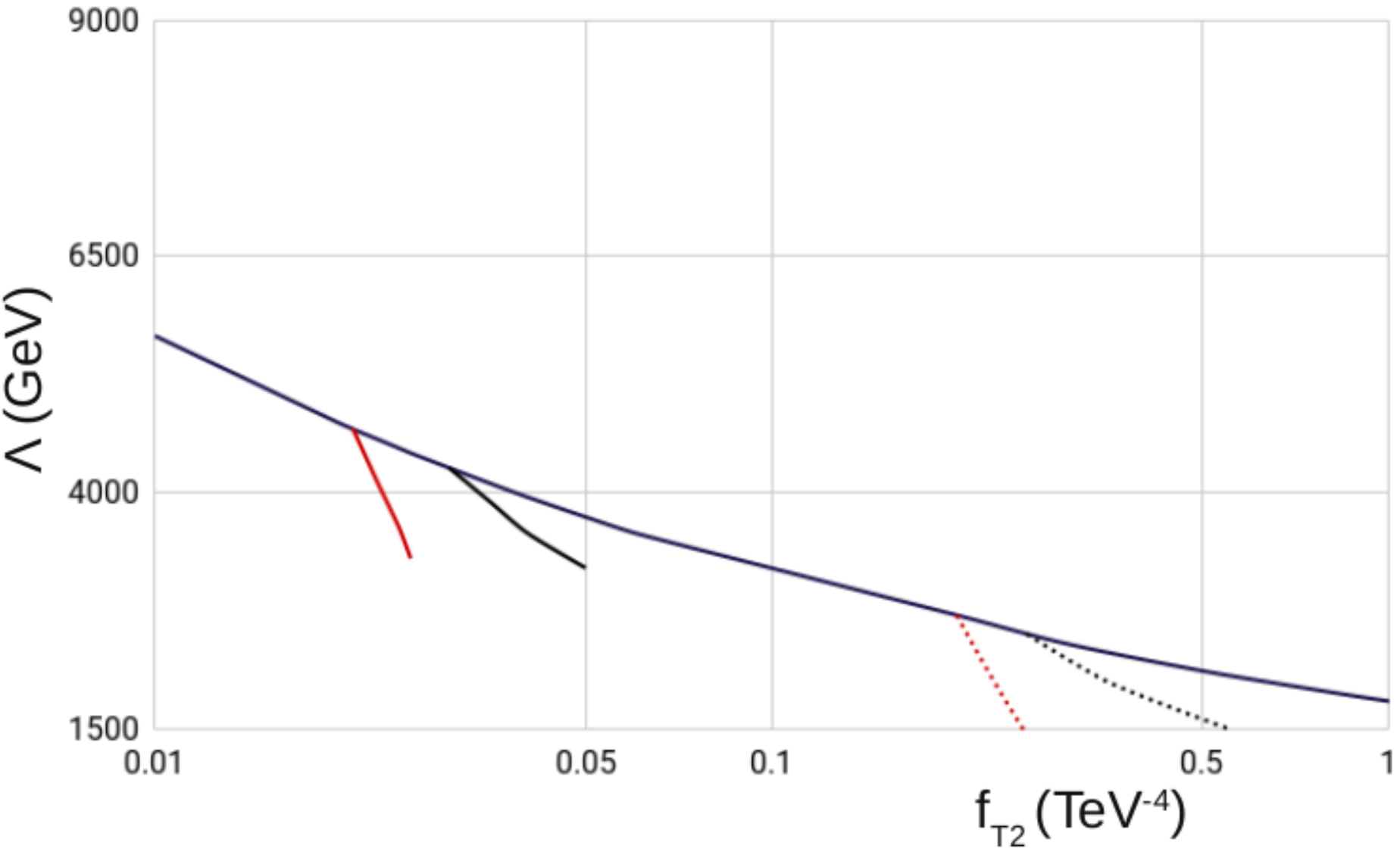} 
& \includegraphics[width=0.5\linewidth]{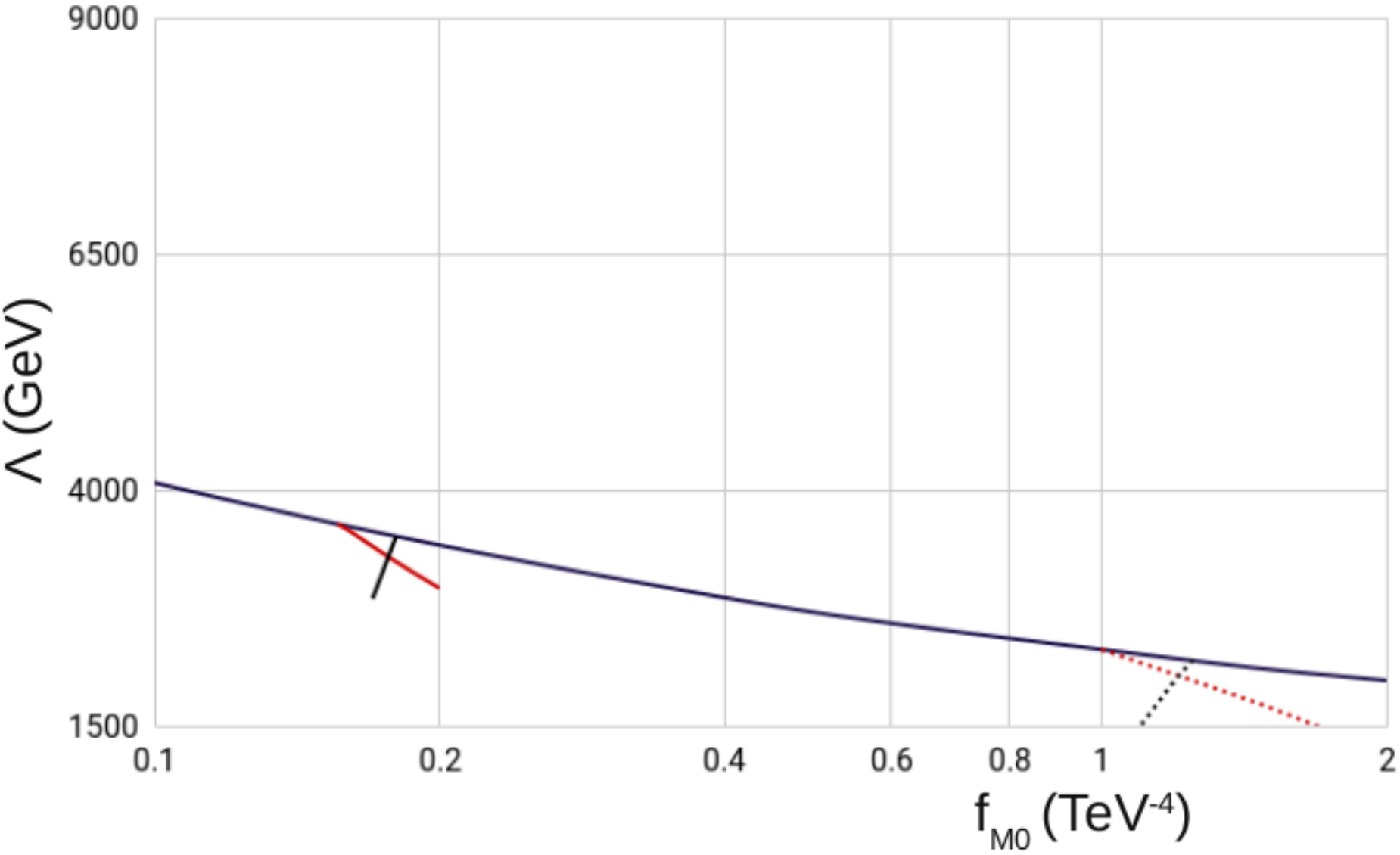} \\
\includegraphics[width=0.5\linewidth]{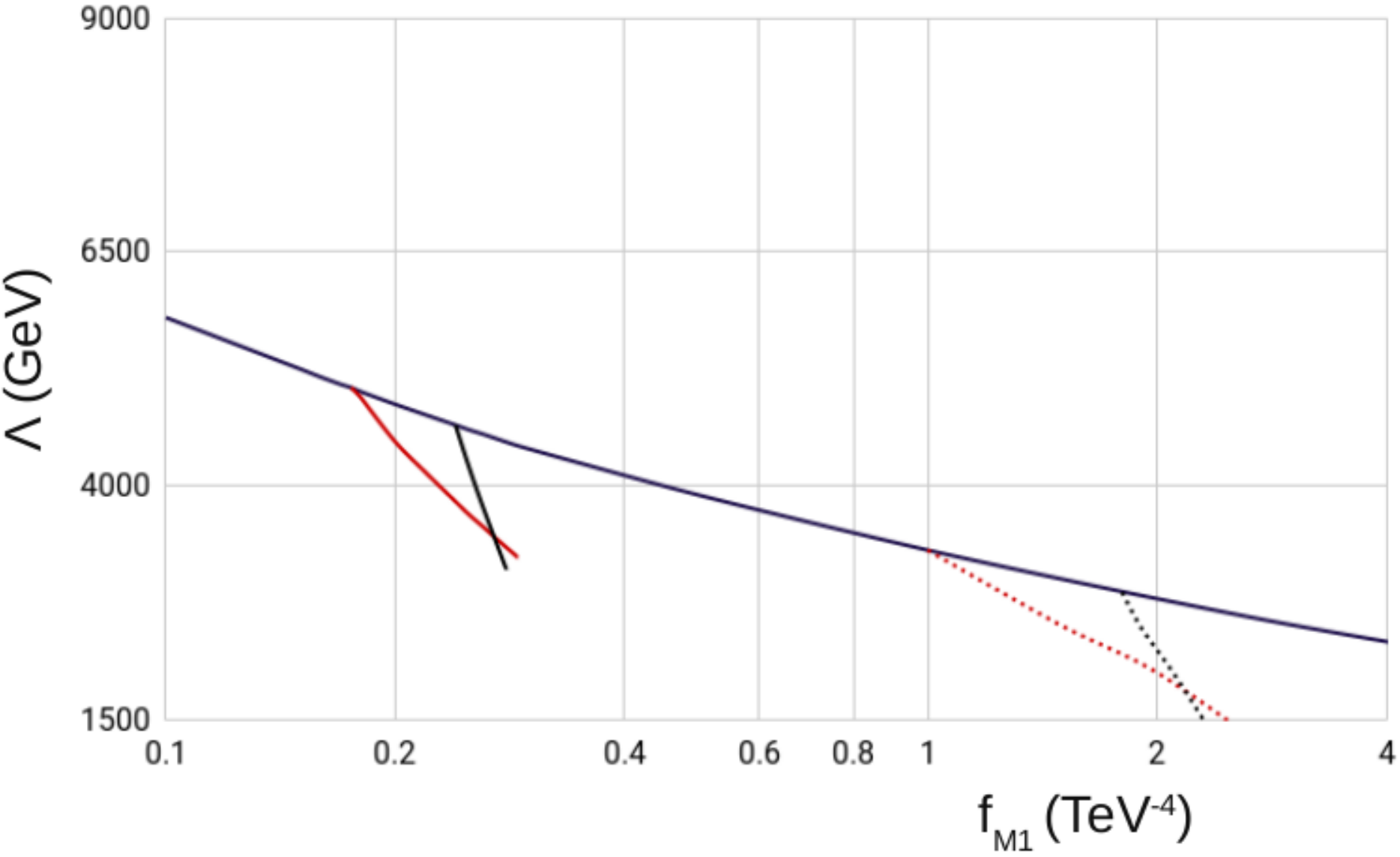} 
& \includegraphics[width=0.5\linewidth]{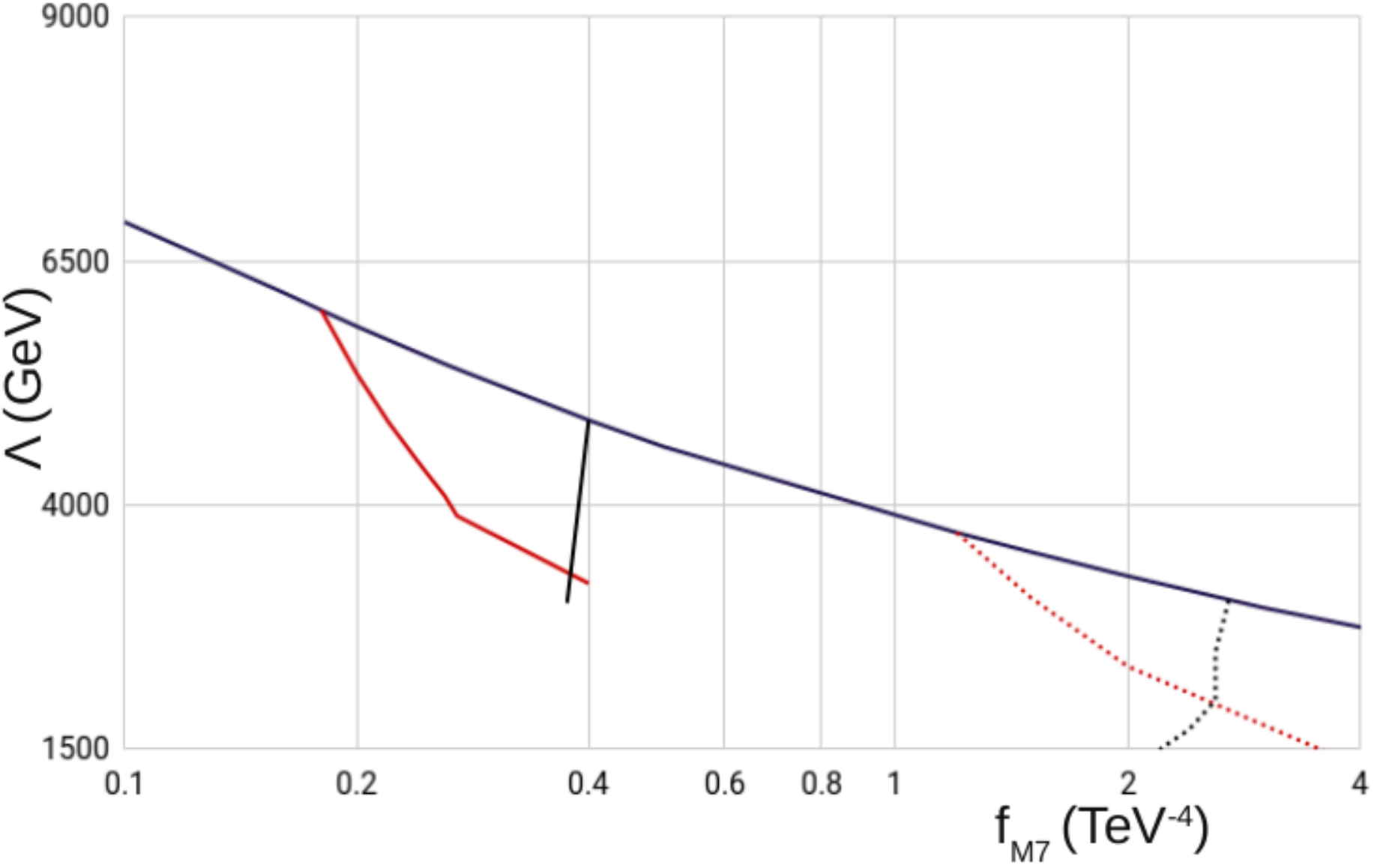} 
  \end{tabular}
\caption{Regions in the $\Lambda$ vs $f$ (positive $f$ values) space for dimension-8 operators in which a $5\sigma$
BSM signal can be observed and the EFT is applicable. The unitarity limit is shown in blue; the lower limits for a $5\sigma$ signal significance from Eq.~\eqref{unitarized} (red) and the upper
limit on $2\sigma$ EFT consistency (black). The solid (dotted) lines correspond to $\sqrt{s} = 27\ (14)$ TeV. 
Assumed is the integrated luminosity of 3 ab$^{-1}$.}
\label{fig:compPos}
\end{figure}%

Fig.~\ref{fig:compPos} shows the results for the individual operators $S0,\, S1,\, T0,\, T1,\, T2,\, M0,\, M1$
and $M7$, in comparison with results at 14 TeV (for positive $f$ values).
Not unexpectedly, all the triangles are shifted to lower $f$ values compared
to 14 TeV, the shift being as large as almost an order of magnitude.  However,
the total area of the triangles does not get significantly larger as we increase
the energy.  This is because the EFT consistency criterion pushes the effective
upper limits on $f$ in a similar manner as does the BSM observability criterion
for the lower limits.  Overall, the shapes and sizes of all the EFT triangles
are remarkably similar for 27 TeV as for 14 TeV, only their respective positions
differ.  
\begin{figure}[h] 
  \begin{tabular}{cc}
\includegraphics[width=0.5\linewidth]{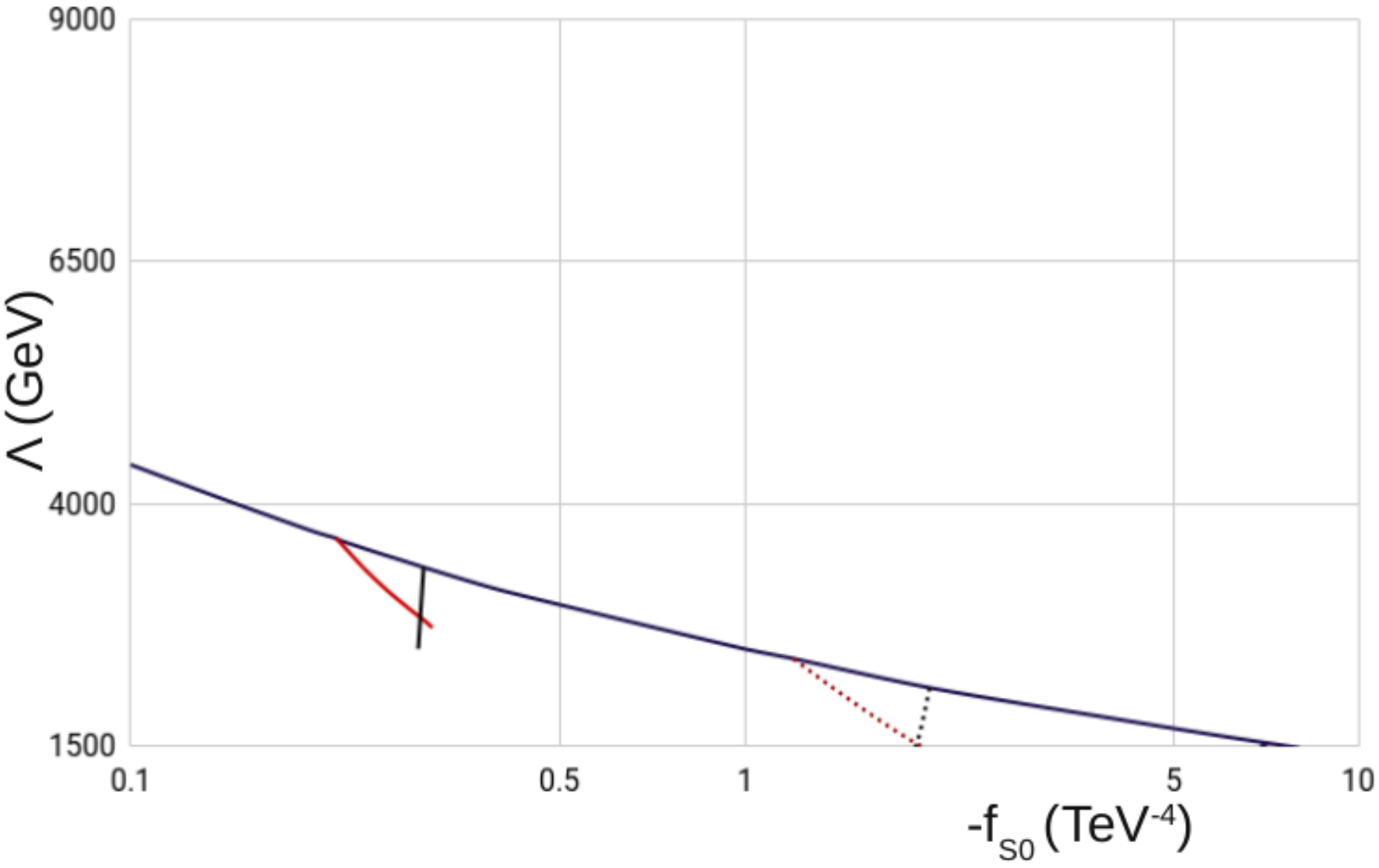} 
& \includegraphics[width=0.5\linewidth]{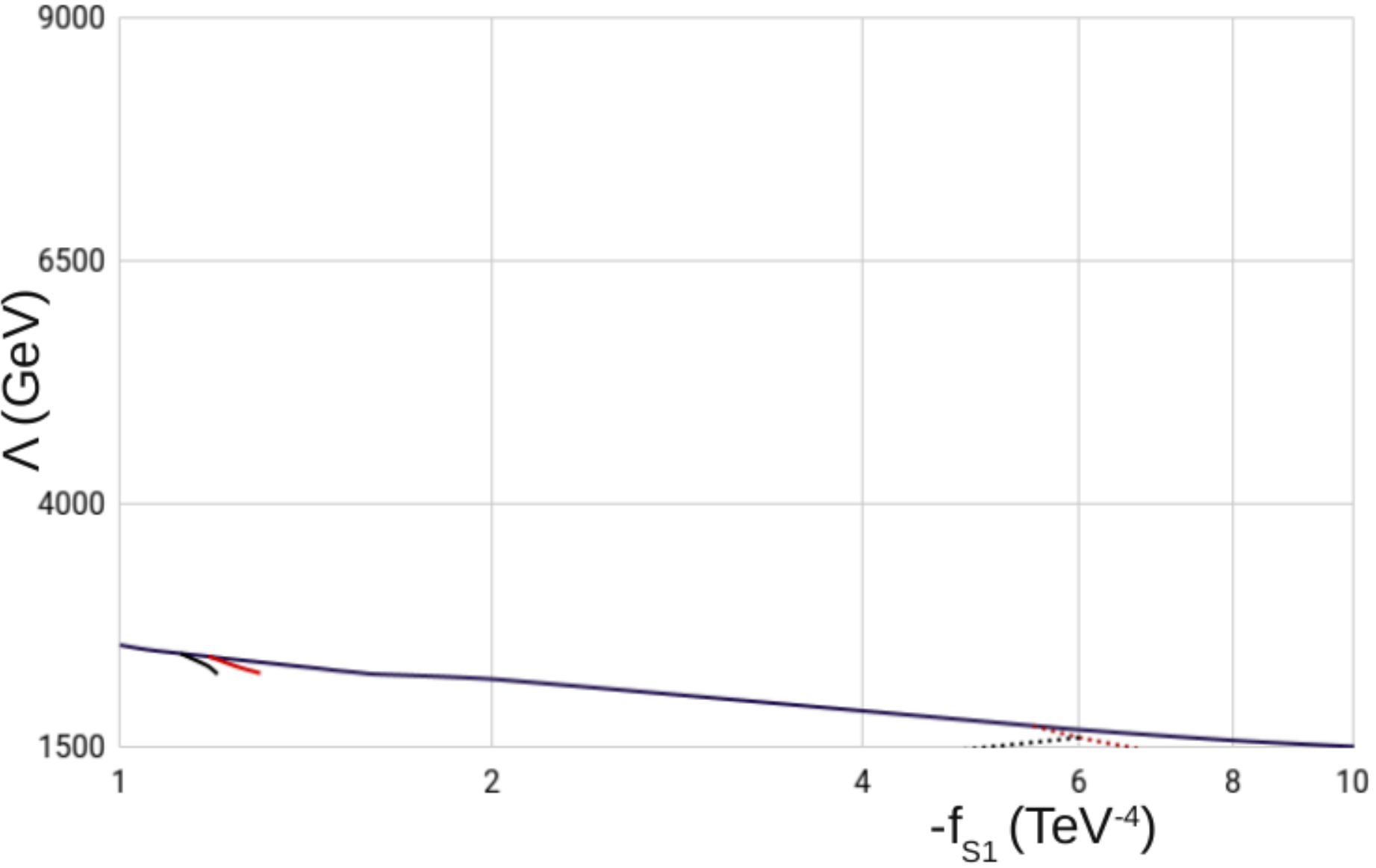} \\     
\includegraphics[width=0.5\linewidth]{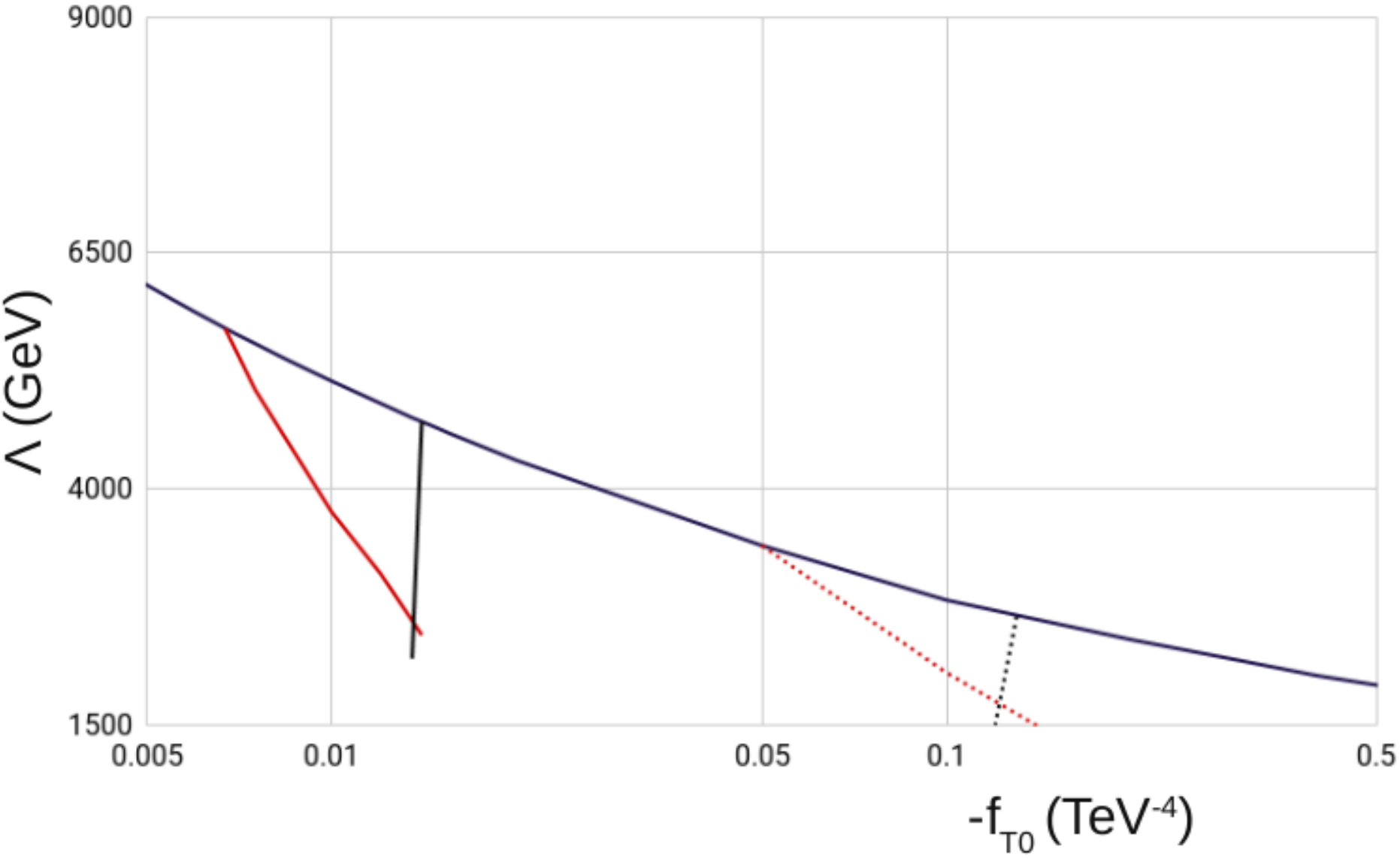} 
& \includegraphics[width=0.5\linewidth]{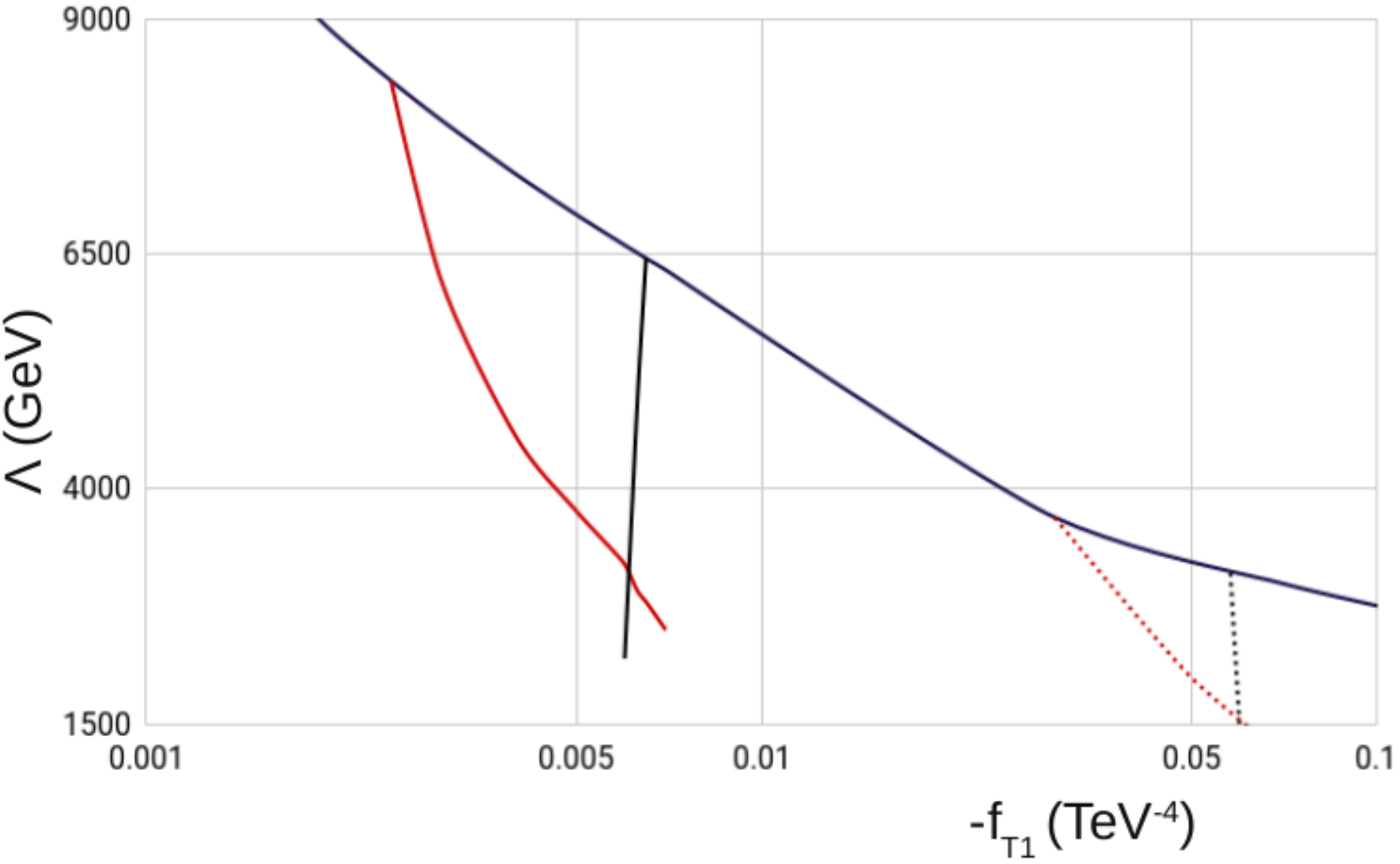} \\
\includegraphics[width=0.5\linewidth]{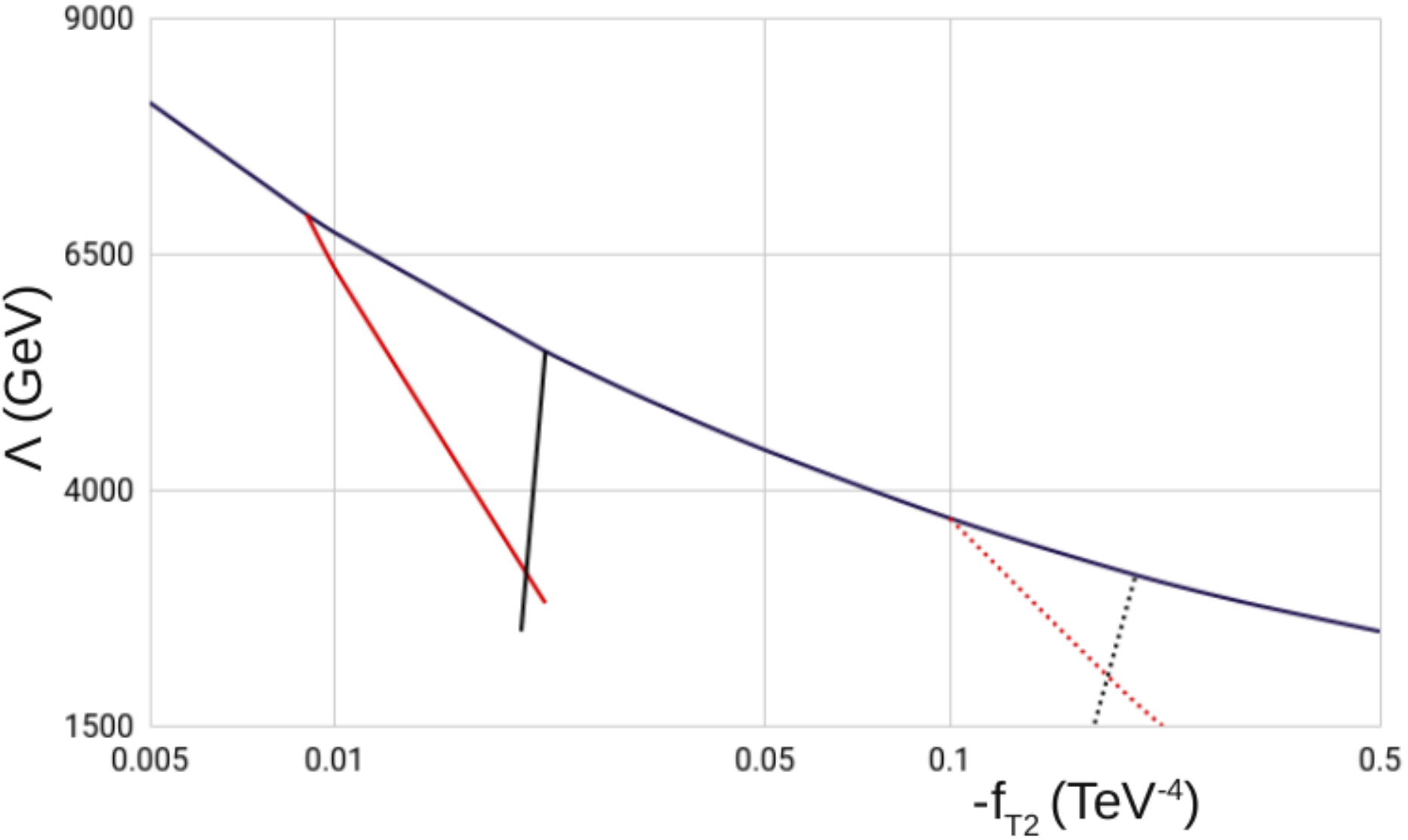} 
& \includegraphics[width=0.5\linewidth]{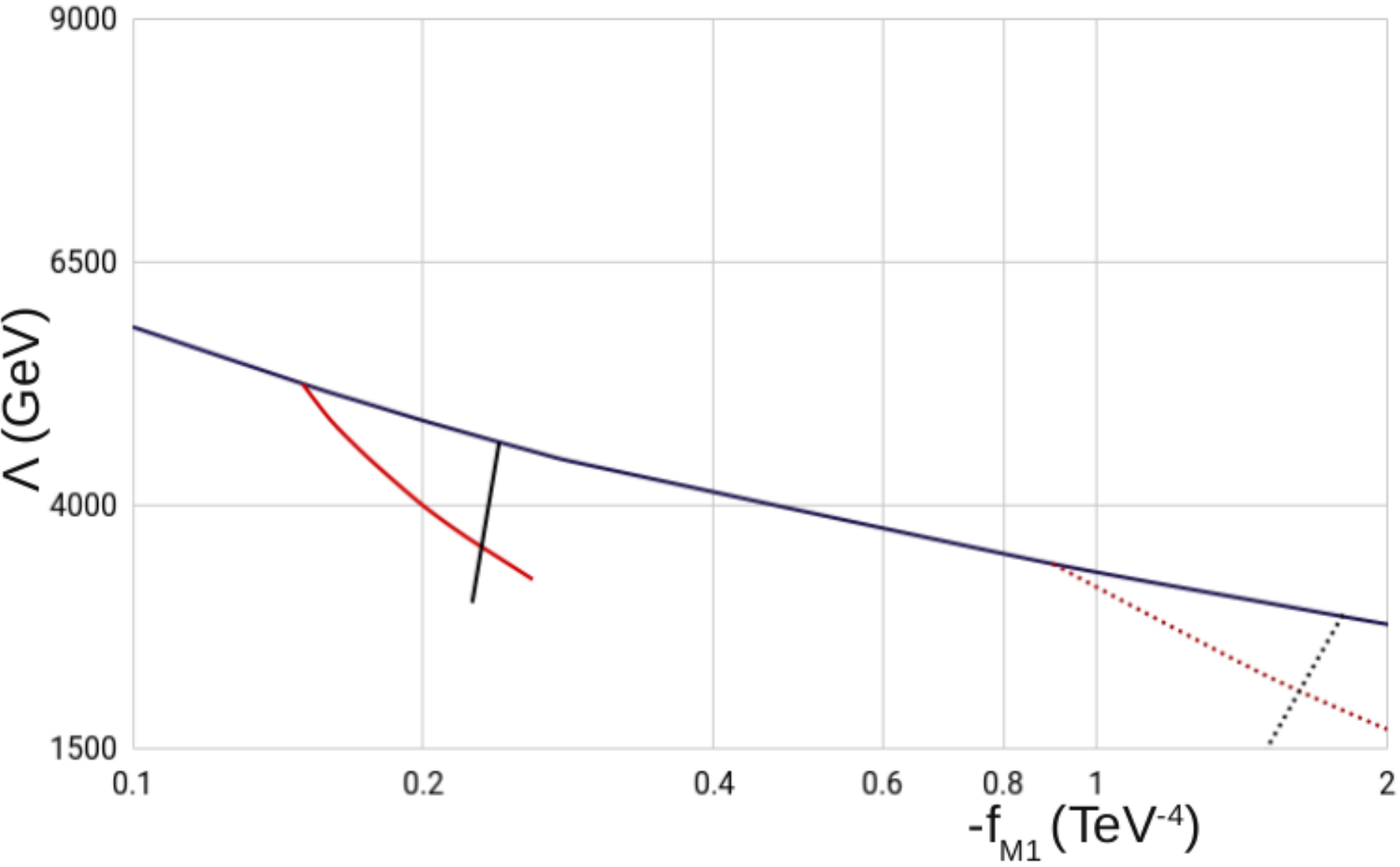} \\
\includegraphics[width=0.5\linewidth]{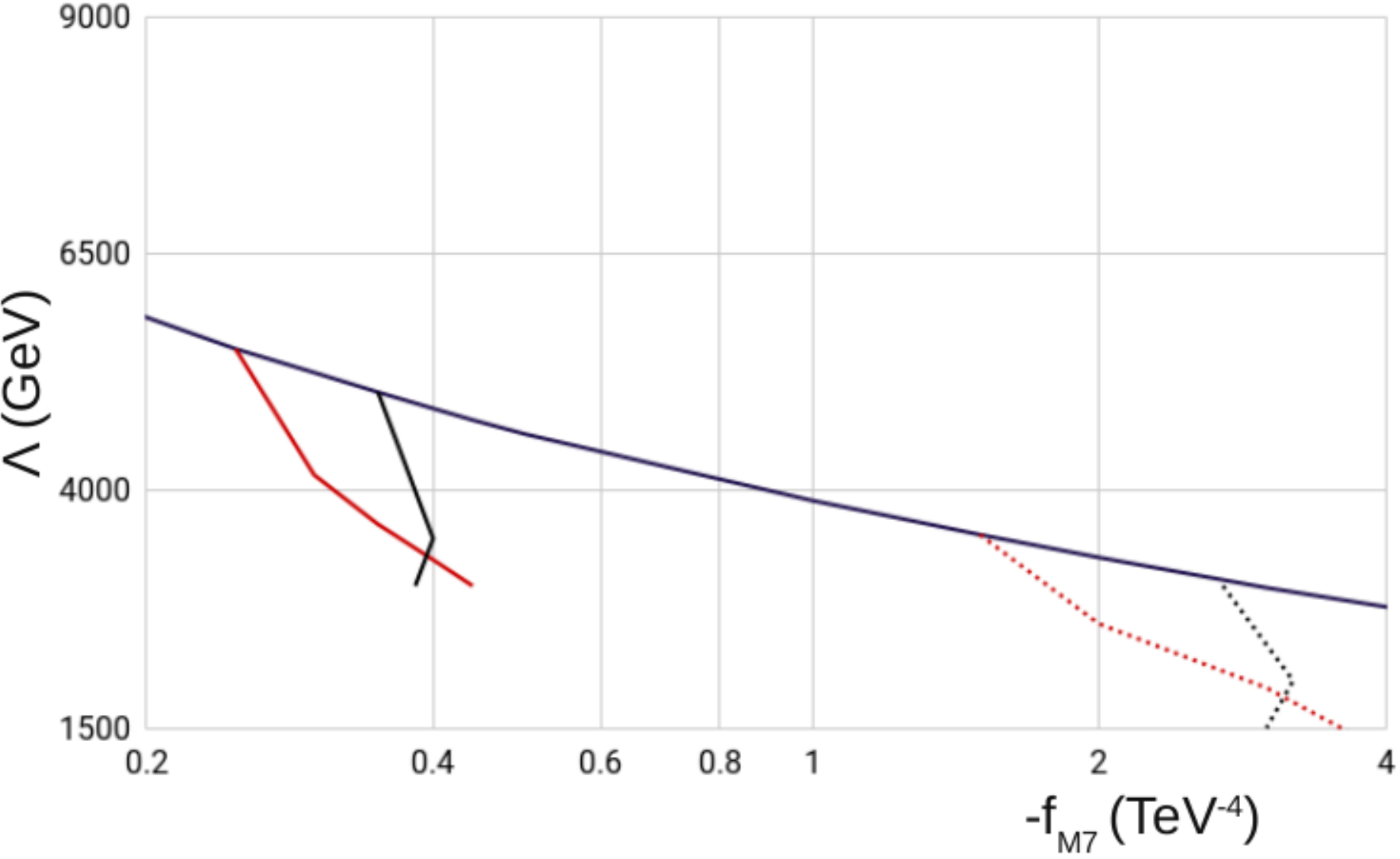} 
  \end{tabular}
\caption{Regions in the $\Lambda$ vs $f$ (negative $f$ values) space for dimension-8 operators in which a $5\sigma$
BSM signal can be observed and the EFT is applicable. The unitarity limit is shown in blue; the lower limits for a $5\sigma$ signal significance from Eq.~\eqref{unitarized} (red) and the upper
limit on $2\sigma$ EFT consistency (black). The solid (dotted) lines correspond to $\sqrt{s} = 27\ (14)$ TeV. 
Assumed is the integrated luminosity of 3 ab$^{-1}$.}
\label{fig:compNeg}
\end{figure}
Fig.~\ref{fig:compNeg} shows the respective results for negative $f$ values
of $S0,\, S1,\, T0,\, T1,\, T2,\, M1$ and $M7$.  
Here exactly the same observations can be made again.
The negative $f$ values of $M0$  look virtually identical to their
positive counterparts, since for these operators the SM-BSM interference term in the
total amplitude calculation is practically negligible (see the discussion of Sec.~\ref{interference}), and so we do not show them here.
There is no triangle at all for $S1$, for which the overall lower limit for
BSM observability is about 1.2~TeV$^{-4}$ and the upper limit for EFT 
consistency is 1.4 ~TeV$^{-4}$.
Here as well we observe a similar behavior as for 14 TeV.
Such as for 14 TeV, full detector simulation with reducible backgrounds included
can only make the picture worse.

\begin{figure}[htb] 
\begin{center}
\includegraphics[width=0.7\linewidth]{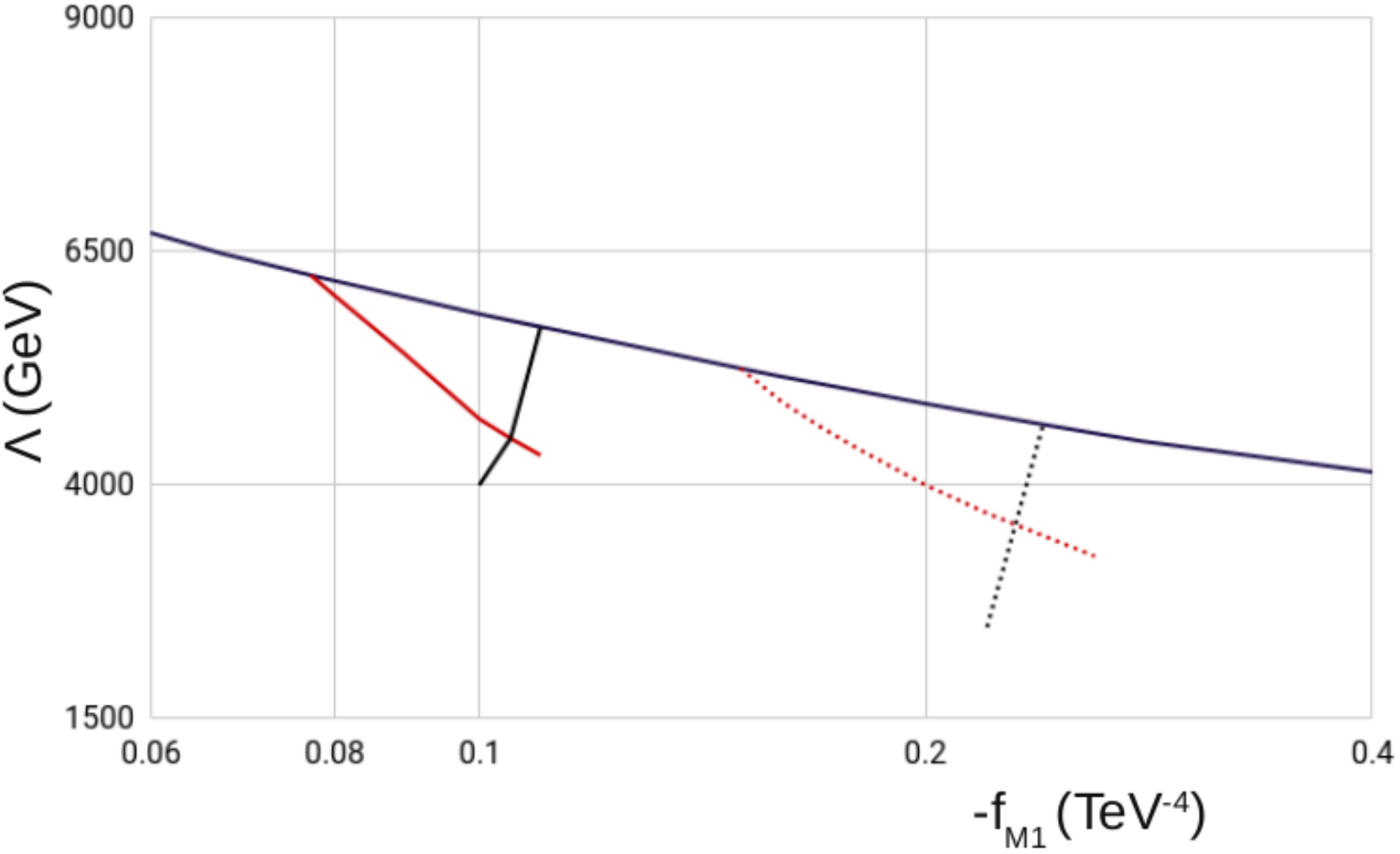} 
\end{center}
\caption{Regions in the $\Lambda$ vs $f$ (negative $f$ values) space for $M1$ operator in which a $5\sigma$
BSM signal can be observed and the EFT is applicable. The unitarity limit is shown in blue; the lower limits for a $5\sigma$ signal significance from Eq.~\eqref{unitarized} (red) and the upper
limit on $2\sigma$ EFT consistency (black). The solid (dotted) lines correspond to 15 ab$^{-1}$ (3 ab$^{-1}$ ). 
Assumed is $\sqrt{s} = 27$ TeV.}
\label{fig:compLumis}
\end{figure}

For the sake of a convenient comparison between the respective results
at two different $pp$ beam energies, in the bulk of this study
we have always assumed the same integrated
luminosity of 3 ab$^{-1}$ for both cases.  This number is appropriate for the
HL-LHC stage, but underestimates the expected statistical power of the
HE-LHC.  However, it is trivial to recalculate all the results to 15/ab
in order to get the true expected discovery reach of the HE-LHC, taking 
into account
its actual expected luminosity.  An increase of statistics by a factor 5
will lead to a further shift of all the EFT triangles by a factor close 
to $\sqrt{5}$,
both in the 5$\sigma$ discovery and the 2$\sigma$ consistency curves
(in fact, somewhat less than that because of non-linear dependence of 
the BSM signal on the value of the individual Wilson coefficients).  It 
will not significantly change either the shape nor the size of triangles.  
A comparison of results calculated for the same $pp$ beam energy of
27 TeV and two different integrated luminosities is exemplified for 
the $M1$ operator in Fig.~\ref{fig:compLumis}.

Our simple procedure to suppress the high-mass tails
by applying a { $(\Lambda/M_{WW})^4$ }weight to events generated above the scale
of $\Lambda$ works reasonably well in the vicinity of the unitarity limit.
In this region 
it produces a tail falling approximately like $1/M_{WW}^2$, which is the expected
asymptotic (i.e. for $M_{WW}>>\Lambda$) behavior of the total cross section after regularization.
It nonetheless becomes too strong as we go to $\Lambda << M^U$, where the
total cross section is still dominated by the SM contribution which does
not require any further suppression.
Moreover, for low values of $\Lambda$ the tail itself becomes large, leading
to large uncertainties due to the details of its modelling.
We have discontinued the curves on Figs.~\ref{fig:compPos} and~\ref{fig:compNeg} below the values
at which we find the method lead to the unphysical result of
signal being suppressed below the SM level itself.
For this reason the EFT triangles for $T0, T1$ and $T2$ do not close.
For the remaining cases, however, they
are completely contained in the region where our simple method is 
still viable.

%-------------------------------------------------------------------%

%%%%%%%%%%%%%%%%%%%%%%%%%%%%%%%%%%%%%%%%%%%%%%%%%%%%%%%%%%%%%%%%%%
Assuming that  the departure form the SM predictions is indeed observed at the HL/HE-LHC we turn to the question what can be said about the couplings corresponding to the discovery regions found for each ``EFT model''.  While the probed $\Lambda$ scale can be read off directly from figs.~\ref{fig:compPos},~\ref{fig:compNeg}, some matching between fundamental parameters of a deeper BSM physics and the Wilson coefficients $c_i$ of the low energy approximation is needed to extract the information about couplings.

Let us start with operators that contain the stress tensor $W_{\mu\nu}$. Since $W_{\mu}^i$ is an elementary gauge field that couples to particles
in the high-energy theory only via electroweak interactions, and the EFT is given by integrating out heavy particles, then every gauge field comes along with a factor of the SM electroweak gauge coupling $g$. Therefore, the counting rule for $W_{\mu\nu}^i$ as known from eq.~\eqref{eq:preTech3} should be replaced by the following, more specific one:  
\begin{equation}
\left[\frac{g}{\Lambda^2}W_{\mu\nu}\right]^{N_{W_{\mu\nu}}}.
\label{eq:preTech4}
\vspace{.7cm}
\end{equation}
which is implied by eq.~\eqref{MasterFormula}. The Madgraph normalization (left hand side) is related to normalization predicted by NDA (right hand side) as follows:
\begin{equation}
\begin{array}{ll}
	\mathcal{L}_{SMEFT}\supset & f_{i}\mathcal{O}_{i}\equiv c_{i}\cdot 2\frac{g^2}{\Lambda^4}\mathcal{O}_{i}, \qquad i=M0,M1\\ \mbox{}\\
	&f_{i}\mathcal{O}_{i}\equiv c_{i}\cdot 2^2\frac{g^4}{16\pi^2\Lambda^4}\mathcal{O}_{i}, \qquad i=T0,T1,T2\\ \mbox{}\\
		&f_{i}\mathcal{O}_{i}\equiv c_{i}\cdot 2^2\frac{g^2}{\Lambda^4}\mathcal{O}_{i}, \qquad i=M6,M7
\end{array}
\label{eq:NDA}
\end{equation}
The factor  2   follows from the relation 
$\mathrm{Tr}[ W_{\alpha\beta}W_{\mu\nu}]=\frac{1}{2}W^i_{\alpha\beta}W^i_{\mu\nu}$  since in NDA the stress tensor  $W^i_{\mu\nu}$ is used for counting purposes rather than the matrix form $W_{\mu\nu}$.  Extra factor 2  in case of $M6,\, M7$ operators is due to differences in the SU(2) structure, as can be seen from the relation $\mathcal{O}_{M7}=\frac{1}{2}\mathcal{O}_{M1} + \ldots,$ while $\frac{1}{16\pi^2}$ in front of $T$ operators is a single loop suppression factor suggested by the 4-th power of the electroweak coupling factored out.
We remind the reader that the remaining dimensionless factors $c_i$ in eq.~\eqref{eq:NDA}
are naturally expected to be bounded by $O(1)$ numbers if the underlying BSM theory is weakly-coupled  (i.e. is in the perturbative regime). 

Using~\eqref{eq:NDA}, it is straightforward to find the range of $c_i$ corresponding to the discovery regions shown in Fig.~\ref{fig:compPos},~\ref{fig:compNeg}. 
The numerical values, presented in Table~\ref{tab:cminCmax} and~\ref{tab:cminCmax14} for 27 and 14 TeV case, respectively, are found to be roughly consistent. 
\begin{table}[ht]
\vspace{8mm}
\begin{center}
\begin{tabular}{|c|c|c|c|c|c|c|}
\hline
  $f_i>0$ & $T0$ & $T1$ & $T2$ & $M0$ & $M1$ & $M7$ \\ \hline
$c^{min}_i$--$c^{max}_i$ & 130.--770.  & 120.--1300. & 670.--2200. & 23.--32. & 45.--133. & 33.--140. \\ \hline
%$g_\ast^{min}$--$g_\ast^{max}$  &--& --& --& --& --& --\\ \hline%2.2--3.4 & 2.2--3.9 & 3.3--4.5 & 6.3--6.8 & 7.4--9.7 & 6.9--9.8 \\ \hline
\end{tabular}
%& g*: 8.4--10. c_i: 0.2--0.46 dla f_i=s0>0 & 27 
\begin{tabular}{|c|c|c|c|c|c|c|c|}
\hline
  $f_i<0$ & $T0$ & $T1$ & $T2$ & $M1$ & $M7$ \\ \hline
$c^{min}_i$--$c^{max}_i$  & 110.--1500.  & 140.--2600. & 410.--4500.  & 48.--130. & 54.--270. \\ \hline
%$g_\ast^{min}$--$g_\ast^{max}$  & -- & -- & -- & -- & -- \\ \hline%& 2.2--4.7 & 2.9--5.3  & 7.6--9.7 & 6.5--9.7 \\ \hline
\end{tabular}
\end{center}
\vspace{3mm}
\caption{For each ``EFT model'' characterized by a choice of a single $n=8$ operator shown are the ranges of the overall coefficients $c_i$ in eq.~\ref{eq:NDA} that correspond to the discovery regions found in the 27 TeV study.}
\label{tab:cminCmax}
\end{table}
\begin{table}[ht]
\vspace{8mm}
\begin{center}
\begin{tabular}{|c|c|c|c|c|c|c|}
\hline
  $f_i>0$ & $T0$ & $T1$ & $T2$ & $M0$ & $M1$ & $M7$ \\ \hline
$c^{min}_i$--$c^{max}_i$  & 137.--790.  & 76.--1300. & 280.--2200. & 23.-33. & 38.-140. & 24.-130. \\ \hline%& 137.--790.  & 76.--1300. & 280.--2200. & 23.-33. & 38.-140. & 24.-130. \\ \hline
%$g_\ast^{min}$--$g_\ast^{max}$  & -- & -- & -- & -- & -- & -- 
%& 2.4--3.5 & 1.9--3.9 & 2.7--4.5 & 6.2--6.9 & 7.1-9.9 & 6.4--9.8 \\ \hline
\end{tabular}
%& c_i=s0 0.21--0.44; g* 8.5--10.
\begin{tabular}{|c|c|c|c|c|c|c|c|}
\hline
  $f_i<0$ & $T0$ & $T1$ & $T2$ & $M0$ & $M1$ & $M7$\\ \hline
$c^{min}_i$--$c^{max}_i$  & 510.--1400.  & 170.--1200. & 700.--4100. & 23.-33. & 45.-140. & 24.-140. \\ \hline
%$g_\ast^{min}$--$g_\ast^{max}$  & -- & -- & -- & -- & -- & -- 
%& 2.6--4.0 & 2.4--3.9 & 3.3--5.2 & 6.2--6.9 & 7.4-9.9 & 6.3--9.8 \\ \hline
\end{tabular}
%& c_i 0.061--0.25; g* 6.2--8.9
\end{center}
\vspace{3mm}
\caption{See Table~\ref{tab:cminCmax} for description; 14 TeV case.}
\label{tab:cminCmax14} 	
\end{table}
However, instead of being of order 1 they are much larger. It suggests that in case of linearly realized spontaneous breaking of $SU(2)\times U(1)$ symmetry our method of probing BSM physics  is sensitive only to strong dynamics.

Since the character of the gauge field $W_{\mu}^i$ is the same in both SMEFT and HEFT hypotheses, the above conclusion holds for HEFT discovery regions as well\footnote{We want to emphasis to the reader that the conclusions on matching to concrete CH models presented in Sec.~\ref{VVHEFT} are left intact by current considerations -- the operators $\mathcal{P}_6,\mathcal{P}_{11}$ are not built out of $W_{\mu\nu}$ tensors.}, as far as the $\mathcal{O}_{T0},\mathcal{O}_{T1},\mathcal{O}_{T2}$ operators are considered (these operators are common to both bases).   

We turn now to the discussion of the $S0$ and $S1$ operators. We assume these are generated at loop level in the BSM. Otherwise they would come associated with $n=6$ operators, which are neglected in our analysis. Then, the NDA suggests 
\begin{equation}
	\mathcal{L}\supset f_{i}\mathcal{O}_{i}\equiv c_i\cdot\frac{g_\ast^4}{16\pi^2\Lambda^4}\mathcal{O}_{i},\qquad i=S0,S1 \\ \mbox{}
\label{eq:NDA2}
\end{equation}
Again, $c_i$ are some combinations of BSM couplings and naturally expected to be bounded by 1. 
If we set $c_i=1$ in eq.~\eqref{eq:NDA2}, then for $f_{S0}>0$ we find that that $g_\ast\in(8.5;10.)$ and $g_\ast\in(8.4;10.)$ in the 14 and 27 TeV case respectively. For $f_{S0}<0$  we find $g_\ast\in(6.2;8.9)$ and $g_\ast\in(7.3;8.8)$ for the 14 and 27 TeV case,  respectively. The coupling is large, but  interestingly it satisfies $g_\ast<4\pi$.
For $f_{S1}$ the discovery regions are empty for both 14 and 27 TeV cases  for both signs.

\clearpage
\chapter{Conclusions and Outlook}
\label{conclusions}
The Standard Model is most probably only an effective low energy description of particles and their interactions -- there are many observational facts (e.g. the existence of Dark Matter, matter-antimatter asymmetry in the Universe) as well as some theoretical puzzles of the SM itself  that show that the SM needs  to be embedded into a deeper theory. In particular the spontaneous symmetry breaking triggered by the single scalar doublet $\Phi$ can be regarded as a simple parametrization of the SSB to be completed with some larger sector. The process that is particularly sensitive to  effects of new particles interacting electroweakly and probing the Higgs mechanism is scattering of the $W$ and $Z$ vector bosons. It is quite possible that measurements of the vector boson scattering processes at the LHC
will reveal some deviations from the  Standard Model predictions, but no new particles will be
observed directly. 

Broadly speaking, the goal of this thesis has been, first of all, to estimate  in the Effective Field Theory framework the discovery potential for the physics beyond the Standard Model in the vector boson  scattering process. The second question that has been addressed is what can we learn about new physics by analysing the data for the vector boson scattering in the framework of the EFT once some deviations from the Standard Model predictions are indeed  observed. For both issues, the special attention and emphasis has been given to the use of the EFT in its range of validity, the point overlooked in the earlier literature.

Two different effective field theories for the SM physical degrees of freedom have been used, to describe the linear and the non-linear realisation of the Higgs sector dynamics.
First, there is the Standard Model Effective Field Theory (SMEFT) where the $SU(2)_L\times SU(2)_R$ symmetry in the Higgs sector is realised linearly,  with the Higgs field being a $SU(2)_L\times SU(2)_R$ bi-doublet. 
%The perturbative expansion is characterised by inverse powers of the cut-off scale $\Lambda$. 
The other one is the Higgs Effective Field Theory (HEFT) in which the $SU(2)_L\times SU(2)_R$ symmetry is  realized non-linearly, on the three Goldstone bosons eaten up by the gauge fields.
The physical Higgs field may be either a singlet of the diagonal, the so-called custodial, symmetry $SU(2)_C$ (if in an UV completion  the $SU(2)_L\times SU(2)_R$ symmetry is realised linearly at a certain level) or an additional field never forming the Higgs doublet with the three Goldstone bosons. This Lagrangian is the most general description of the EW and Higgs couplings, satisfying the gauge symmetry of the SM: in specific limits, it may reduce to the SM Lagrangian, or may coincide with the SMEFT one, or may match the description of Composite Higgs (CH) models.
%o tym ze need for new physics i że ww jest ok; SMEFT & HEFT

In this work the prospects have been analyzed for discovering physics beyond the SM at the HL-LHC and HE-LHC in the EFT framework applied to the VBS amplitudes in the process $pp\rightarrow W^+W^+jj$.  We have introduced the concept of EFT ''models'' defined by the choice of non-renormalizable operators and values of the Wilson coefficients and analyzed ''models'' based on a single operator at a time added to the SM Lagrangian. The effective parameters in the EFT approach are the Wilson coefficients $f_i$ and not the cut-off scale $\Lambda$ itself. The latter is  unknown without referring to specific UV complete models. In fact $\Lambda$ can be varied in a certain range after the EFT ''model'' is specified: $\Lambda <\sqrt{s^U}$, where $\sqrt{s^U}$ is the partial wave unitarity bound, which is a  function of $f_i$. Clearly, the unitarity bound plays a crucial role in the context of proper use of the EFT description. The region of validity of the EFT description of the $WW$ scattering is
determined by the condition $M_{WW}<\Lambda < \sqrt{s^U}$  where $M_{WW}$ is the $WW$ invariant mass. This aspect had not been properly accounted for in the literature on VBS at the LHC. In this work this criterion has been applied to the binary VBS amplitude, which is an approximation
in the context of the process $pp\rightarrow 2 jets + 2 \mathrm{lepton\ pair}$. In the thesis we have also developed a formalism for a determination of partial wave
unitarity constraints directly in the process $pp\rightarrow 2 jets + WW$  but  left its application to VBS described by the EFT ''models'' studied in this thesis for future study.  

We have determined the discovery regions in the parameters $(\Lambda, f_i)$ for certain class of EFT ''models'' in their range of validity. Thus, if the BSM physics would be well approximated by those EFT
"models"  with the values of the parameters in the obtained discovery regions, it should manifest itself at the LHC. These positive results
obtained on the basis of the simple EFT "models" strongly suggest that similar studies should evolve
in the direction of multidimensional analysis with
many higher dimension operators included at a time.  This in turn may require
global simultaneous analysis of many processes (including $ WZ,\, ZZ$ and semi-leptonic
$WV$, if not other processes) to help to disentangle the correlations between
signals originating from different operators.

With the choice of the same sign $WW$ scattering process for our study, often considered as  a "gold-plated" process due to its relatively good signal to background ratio, we have been facing at the same time  a strong technical challenge  due to the lack of direct experimental access to the $WW$  invariant mass.
The invariant mass $M_{WW}$ is directly related
to the intrinsic range of validity of the EFT approach
%, $M_{WW} < \Lambda\leq s^U$,
and it is important to tackle this issue correctly  in the estimation of the discovery potential and in the future
data analysis in order to
study the underlying BSM physics.  While this is relatively simple (in principle)
for final states where $M_{WW}$ can be determined on an event-by-event
basis, the value of $M_{WW}$ is unfortunately not available in leptonic $W$ decays.
We argue that use of EFT ''models'' in the analysis of purely
leptonic $W$ decay channels requires bounding the possible contribution from
the region $M_{WW} > \Lambda$ (the tail), no longer described by the ''model'',
and ensuring that it does not significantly distort the measured distributions 
compared to what they would have looked from the region of EFT validity alone. In particular the tail of the $M_{WW}$ distribution with $M_{WW}>\Lambda$ should not have any significant impact on the fit of an EFT ''model'' to the data. We have proposed a solution  to this problem.

Considering first the SMEFT case, our choice for the  higher-dimension operators was motivated by the fact that VBS can be regarded as a genuine process for probing the dimension-8 operators that modify the quartic WWWW coupling but do not affect triple electroweak couplings.  
We have found that with a possible exception  of one operator (denoted as ${\cal O}_{S1}$) all dimension-8 operators which affect the
$WWWW$ quartic coupling have regions of  the "model" validity where
a 5$\sigma$ BSM signal can be observed at HL-LHC with 3 ab$^{-1}$ of data  and the contribution from  the region $M_{WW} > \Lambda$ is small.

Going now the HEFT case, we have used the Naive Dimension Analysis considerations to justify  the choice of the HEFT operators -- we  have considered primary dimension-8 operators since these are expected to have similar impact on the phenomenology as the dimension-8 SMEFT operators already chosen. Interestingly among these operators there are three, denoted by $\mathcal{T}_{42},\mathcal{T}_{43},\mathcal{T}_{44}$,
that do not have siblings in SMEFT at dimension-8. They correspond to distinct interactions (Lorentz structures of the $WWWW$ vertex). Their discovery regions are relatively large with the following bound on the probed BSM scale: $2\mathrm{TeV}\leq\Lambda\leq3.5\mathrm{TeV}$. These findings are to be regarded as a first step in a well motivated pursue for establishing a global analysis framework aiming at distinguishing between the SMEFT and HEFT hypotheses in the real forthcoming data, once the latter exhibits significant deviations from the SM predictions.

%%%
The obtained results for the discovery potential of the  HEFT ''models'' can be translated into bounds on the parameters of UV theories that project at low-energy on the HEFT operators. The case of the Composite Higgs model has been discussed, considering first a generic $SO(5)/SO(4)$ CH description and then the so-called Minimal Linear $\sigma$ Model. In both cases, interesting bounds have been found on the values that can be tested of the parameter $\xi$, that measures the level of non-linearity of the Higgs sector and its fine-tuning. These bounds are independent from those extracted from  EW precision observables or the absence of any composite resonance.

When comparing the results presented here for the HEFT operators with the earlier literature, it is worth mentioning that in Ref.~\cite{Delgado:2013hxa} the value of 0.8  for the coefficients $c_{11}$ and $c_6$ has been considered (in the NDA normalisation adopted in this paper) . Therefore it is outside the discoverability regions identified for the operators $\cP_{6}$ and $\cP_{11}$. The reason for this discrepancy is the additional constraint, namely that the tail, i.e. the $M_{WW}>\Lambda$ region, must be negligible in the signal estimate in the EFT framework. This exemplifies the relevance of the procedure illustrated here to interpret data on same-sign $WW$ scattering in terms of the HEFT Lagrangian. 

%27TeV results
Although an increase of the LHC energy vastly improves the sensitivity to
new physics effects in VBS processes, the question of EFT applicability is
a different one and, as a result of our investigation, we conclude that it cannot be solved by changing the energy; we examined the HE-LHC case (27 TeV c.o.m $pp$ collision energy) and compared it with the HL-LHC (14 TeV c.o.m) case. Rephrasing it, the discovery regions do not get significantly larger in the HE-LHC study compared to the HL-LHC; they only shift to lower values of $f_i$ and larger scales of $\Lambda$. 
This is because of the lack of experimental access to the $WW$ invariant mass and the need to control the tail of the distribution  for describing the data in terms of the EFT. 
This conclusion holds regardless of the actual proton beam energy.

We  turn now to the second question. Discovering experimentally some statistically significant deviations from  the SM predictions in the VBS would clearly indicate the presence of BSM physics, even if no new particles are directly observed.  Obviously, fitting them to EFT "models" makes sense only if one takes into account the region of validity of such `'models". Therefore, the results of this thesis will be very relevant for such fits. Moreover, such fits will be useful if a successful fit in  in an EFT framework would hint to (select) some classes of UV complete BSM theories.

Indeed, we have shown that from such analysis it may be possible to learn something about
the underlying UV completion of the full theory.  Successful determination of a
given $f$ value, using a procedure that respects the EFT restricted range of 
applicability,
will put non-trivial bounds on the value of $\Lambda$ and consequently, the BSM
coupling $c_i$.  These bounds are rather weak for ${\cal O}_{T0}$, ${\cal O}_{T1}$ and ${\cal O}_{T2}$
operators, but potentially stronger for ${\cal O}_{M0}$, ${\cal O}_{M1}$, ${\cal O}_{M6}$ and ${\cal O}_{M7}$.
In particular, applicability of the EFT in terms of these operators already requires
$\Lambda \ge$ 2 TeV, while stringent upper limits arise from the unitarity condition.
Because of relatively low sensitivity to $f_{S0}$ and $f_{S1}$, it will unfortunately
be hard to learn much about $W_LW_L$ physics using the 
EFT approach with dimension-8 operators. Our results are important for both the SMEFT and HEFT fits and for future multidimensional analyses mentioned earlier.

We end this thesis with a short discussion  of future directions and prospects.
Consideration of other VBS processes and $W$ decay channels may significantly 
improve
the situation with regard to our both questions.  In particular, the semileptonic decays, where one $W^+$ decays
leptonically and the other $W^+$ into hadrons, have never been studied in VBS
analyses because of their more complicated jet combinatorics and consequently much
higher background.  Progress in the implementation of $W$-jet tagging
techniques based on jet substructure algorithms may render these channels interesting
again.
However, they are presently faced with two other experimental challenges.  One is
the precision of the $M_{WW}$ determination which relies on the missing-$E_T$
measurement resolution.  The other one is poor control over the sign over the
hadronic $W$.
The advantages would however be substantial.
If $M_{WW}$ can be reconstructed with reasonable accuracy,
it is straightforward to fit $f$ and $\Lambda$ to the measured
distribution in an EFT-consistent way even for arbitrarily large $f$.  Existence
of a high-$M_{WW}$ tail above $\Lambda$ is then not a problem,
but a bonus, as it may give us additional hints about the BSM physics.
Finally, because of the invariant mass issue, the $ZZ$ scattering channel,
despite its lowest cross section, may ultimately prove to be the process from which we
can learn the most about BSM in case the LHC fails to discover new physics directly.

%we argue about infering from on-shell; we study in detail...

Last but not least, we have argued that qualitative conclusions on the influence of non-renormalizable operators on
the full process can be drawn from the analysis of the polarized on-shell $WW$ scattering amplitudes. It may have interesting phenomenological consequences: the polarizations of the outgoing W bosons that dominate the unpolarized cross section at large $M_{WW}$ are different for SMEFT ''models'' defined by operators belonging to different S, T, M subsets. The latter fact can be inferred from our detailed on-shell $W^+W^+$ scattering analysis bulk of which is presented in the Appendices. There are new theoretical ideas how to learn from experimental data about those polarizations~\cite{Ballestrero:2017bxn}.  If indeed possible,  one could  discriminate between contributions from different EFT models  defined by a single (or at most two) operator. This would be a  test  for classes of
UV completions containing it in its low energy approximation such an operator,  independently of the  presence  of other  operators as well.

In the above, there are already several important directions pointed out, in which the study can be extended. A few more interesting examples we mention below:

First, apart from restrictions from unitarity, there is possibly another set of bounds on the EFT ''models'', the so-called ''positivity bounds''. Since these are inferred from fundamental features of QFT, e.g. analyticity and unitarity, they are model independent and therefore are valid in the context of the EFT approach. For a single operator at a time analysis these limitations  set bound on the sign of Wilson coefficient $f_i$. For more non-zero Wilson coefficients considered simultaneously, these bounds have form of certain bounded regions in the $f_i$ space. Interestingly, the framework of how to solve for these ''positivity bounds'' in the general case has been established~\cite{Bi:2019phv}. Application of such bounds would be relevant in the context of global analyses where more than single $f_i$ at a time is considered.

Second, the analysis of discovery regions can be straightforwardly extended to $\mathcal{CP}$ odd operators contributing to same-sign $WW$
scattering. The contribution of the $\mathcal{CP}$-odd operators to the $WW$ scattering has not yet been discussed in
the literature. The study of $\mathcal{CP}$ odd operators is of particular interest as $\mathcal{C}, \mathcal{CP}$ violation by BSM is a
necessary condition for explaining the observed matter-antimatter asymmetry in the universe (which
the SM fails to explain). 

Third, dimension-6 operators could be discussed in the context of $WW$ scattering,
with inclusion of the bounds, consistently obtained with our EFT strategy, from independent measurements in the Higgs
sector and other precision electroweak data.  For a recent discussion on the effects of $n=6$ SMEFT operators in $ZZ$ elastic scattering at LHC, taking into account the bounds on the $n=6$ operators but not the unitarity issue, see~\cite{Gomez-Ambrosio:2018pnl}. The EFT approach to the VBS  is a subject under active investigations and those are just some examples  of possible directions.

\clearpage
\appendix
\chapter{On-shell gauge boson scattering in the SM: unitary cancelations}
\label{app:VVonshellSM}
This appendix is a supplement to the content of Chapter~\ref{sec:VVinSM}. Shown are explicit formulas that illustrate cancellations among tree-level vector bosons scattering diagrams, leading to a perturbative partial wave unitarity bounds fulfilment. For details on the discussion see that Chapter.
\newgeometry{tmargin=2cm, bmargin=3cm, lmargin=1.5cm, rmargin=1.5cm} 
\begin{table}
%\left(
\begin{center}
\begin{tabular}{c|c|c|c|c||c}
 helicity & $iM^{Z+A}$ & $\sim s^n$& $iM^{Z+A+Contact}$ & $\sim s^n$ & $\sim s^n$\\ \mbox{} &\mbox{}&$n=$ &\mbox{}&  $n=$ & $n=$ \\ \hline{} &&&&& \\
$ \text{- - - - } $ & $  -\frac{i (-12 \cos (2 \theta )+\cos (4 \theta )+139) \csc ^2(\theta ) c_W^2 m_Z^2}{4 v^2} $ & $  0 $ & $  -\frac{32 i \csc ^2(\theta ) c_W^2
   m_Z^2}{v^2} $ & $  0 $ & $  0 $  \\ &&&&&  \\
	
 $\text{- - - 0 } $ & $  -\frac{i \sqrt{2} \sqrt{s} \cos (\theta ) \sin (\theta ) c_W m_Z}{v^2} $ & $  \frac{1}{2} $ & $  \frac{4 i \sqrt{2} \cot (\theta ) c_W^3
   m_Z^3}{\sqrt{s} v^2} $ & $  -\frac{1}{2} $ & $  -\frac{3}{2} $  \\ &&&&& \\	
	
 $\text{- - - + } $ & $  -\frac{2 i \sin ^2(\theta ) c_W^2 m_Z^2}{v^2} $ & $  0 $ & $  \frac{20 i c_W^4 m_Z^4}{s v^2} $ & $  -1 $ & $  -1 $  \\ &&&&& \\

 $\text{- - 0 0 } $ & $  -\frac{i s (\cos (2 \theta )+7)}{2 v^2} $ & $  1 $ & $  -\frac{2 i c_W^2 m_Z^2}{v^2} $ & $  0 $ & $  -1 $  \\ &&&&& \\

 $\text{- - 0 + } $ & $  -\frac{i \sqrt{2} \sqrt{s} \cos (\theta ) \sin (\theta ) c_W m_Z}{v^2} $ & $  \frac{1}{2} $ & $  \frac{4 i \sqrt{2} \cot (\theta ) c_W^3 \left(4
   c_W^2+1\right) m_Z^5}{s^{3/2} v^2} $ & $  -\frac{3}{2} $ & $  -\frac{3}{2} $  \\ &&&&& \\
	
 $\text{- - + + } $ & $  \frac{i (\cos (2 \theta )-5) c_W^2 m_Z^2}{v^2} $ & $  0 $ & $  -\frac{4 i c_W^4 m_Z^4}{s v^2} $ & $  -1 $ & $  -2 $  \\ &&&&& \\

 $\text{- 0 - 0 } $ & $  -\frac{i s \sin ^2(\theta )}{v^2} $ & $  1 $ & $  \frac{2 i (\cos (\theta )+3) c_W^2 m_Z^2}{v^2 (\cos (\theta )-1)} $ & $  0 $ & $  0 $  \\ &&&&& \\

 $\text{- 0 - + } $ & $  \frac{i \sqrt{s} \csc ^2\left(\frac{\theta }{2}\right) \sin ^3(\theta ) c_W m_Z}{\sqrt{2} v^2} $ & $  \frac{1}{2} $ & $  -\frac{10 i \sqrt{2} \cot
   \left(\frac{\theta }{2}\right) c_W^3 m_Z^3}{\sqrt{s} v^2} $ & $  -\frac{1}{2} $ & $  -\frac{1}{2} $  \\ &&&&& \\
	
 $\text{- 0 0 0 } $ & $  -\frac{i s^{3/2} \cos (\theta ) \sin (\theta )}{\sqrt{2} v^2 c_W m_Z} $ & $  \frac{3}{2} $ & $  \frac{2 i \sqrt{2} \cot (\theta ) c_W \left(2
   c_W^2+1\right) m_Z^3}{\sqrt{s} v^2} $ & $  -\frac{1}{2} $ & $  -\frac{1}{2} $  \\ &&&&& \\
	
 $\text{- 0 0 + } $ & $  \frac{2 i s \cos ^2\left(\frac{\theta }{2}\right) \cos (\theta )}{v^2} $ & $  1 $ & $  \frac{4 i c_W^2 \left(5 \cos (\theta ) c_W^2-5 c_W^2+\cos
   (\theta )\right) m_Z^4}{s v^2 (\cos (\theta )-1)} $ & $  -1 $ & $  -1 $  \\ &&&&& \\
	
 $\text{- + - + } $ & $  -\frac{i (\cos (2 \theta )+7) \cot ^2\left(\frac{\theta }{2}\right) c_W^2 m_Z^2}{v^2} $ & $  0 $ & $  -\frac{8 i \cot ^2\left(\frac{\theta
   }{2}\right) c_W^2 m_Z^2}{v^2} $ & $  0 $ & $  0 $  \\ &&&&& \\
	
 $\text{- + 0 0 } $ & $  \frac{i s \sin ^2(\theta )}{v^2} $ & $  1 $ & $  -\frac{2 i c_W^2 m_Z^2}{v^2} $ & $  0 $ & $  -1 $  \\ &&&&& \\

 $\text{0 0 0 0 } $ & $  \frac{i s^2 (\cos (2 \theta )-5)}{4 v^2 c_W^2 m_Z^2} $ & $  2 $ & $  -\frac{i s}{v^2} $ & $  1 $ & $  0 $ \\ \hline
\end{tabular}
\end{center}
%\right)
\caption{Shown are cancellations between various diagram subsets leading to unitarity fulfilment. Analyzed are tree-level helicity amplitudes for $W^+W^+\rightarrow W^+W^+$ process. $iM^{Z+A}$ denotes sum over $Z_\mu$ and $A_\mu$ exchange diagrams; in $iM^{Z+A+Contact}$ the contact term is added; shown are only the leading terms of the helicity amplitudes in the asymptotic limit $s\rightarrow\infty$; column 3. and 5. show the leading asymptotic $s$ dependence in the amplitudes $iM^{Z+A}$ and $iM^{Z+A+Contact}$ respectively; the last column corresponds to the full tree-level amplitude.}
\label{tab:WWonshell1}
\end{table}
\begin{table}%
\begin{center}
\begin{tabular}{c|c|c}
$ \text{helicity} $ & $iM(s,\theta ,f)$ & $ \sigma _{\text{ijkl}} $ \\ \hline \mbox{} && \\ 
 $\text{- - - -} $ & $ -\frac{32 i c_W^2 \csc ^2(\theta ) m_Z^2}{v^2} $ & $ 100. $ \\ \mbox{} && \\
 $\text{- - - 0} $ & $ -\frac{2 i \sqrt{2} c_W^3 \cot (\theta ) \csc ^2(\theta ) m_Z^3 \left(\cos (2 \theta ) \left(\left(4 c_W^2+1\right)
   m_Z^2-m_h^2\right)+\left(28 c_W^2-25\right) m_Z^2-7 m_h^2\right)}{s^{3/2} v^2} $ & $ 2.6\times 10^{-3} $ \\ \mbox{}&& \\
 $\text{- - - +} $ & $ \frac{16 i c_W^4 m_Z^4}{s v^2} $ & $ 6.1\times 10^{-4} $ \\ \mbox{} &&\\
 $\text{- - 0 0} $ & $ \frac{2 i c_W^2 \csc ^2(\theta ) m_Z^2 \left(\cos (2 \theta ) \left(m_Z^2-m_h^2\right)-3 m_h^2-5 m_Z^2\right)}{s v^2} $ & $ 4.8\times 10^{-3}
   $ \\ \mbox{} &&\\
 $\text{- - 0 +} $ & $ \frac{4 i \sqrt{2} c_W^3 \cot (\theta ) m_Z^3 \left(\left(4 c_W^2+1\right) m_Z^2-m_h^2\right)}{s^{3/2} v^2} $ & $ 5.1\times 10^{-6} $ \\ \mbox{} &&\\
 $\text{- - + +} $ & $ \frac{8 i c_W^4 m_Z^4 \left(\left(4 c_W^2+1\right) m_Z^2-m_h^2-4 \csc ^2(\theta ) m_Z^2\right)}{s^2 v^2} $ & $ 2.2\times 10^{-7} $ \\ \mbox{} &&\\
 $\text{- 0 - 0} $ & $ \frac{8 i c_W^2 m_Z^2}{v^2 (\cos (\theta )-1)} $ & $ 110. $ \\ \mbox{} &&\\
 $\text{- 0 - +} $ & $ -\frac{8 i \sqrt{2} c_W^3 \cot \left(\frac{\theta }{2}\right) m_Z^3}{\sqrt{s} v^2} $ & $ 0.42 $ \\ \mbox{} &&\\
 $\text{- 0 0 0} $ & $ \frac{2 i \sqrt{2} c_W \cot (\theta ) m_Z \left(m_Z-m_h\right) \left(m_h+m_Z\right)}{\sqrt{s} v^2} $ & $ 1.7\times 10^{-3} $ \\ \mbox{} &&\\
 $\text{- 0 0 +} $ & $ \frac{4 i c_W^2 m_Z^2 \left(\cos (\theta ) \left(\left(4 c_W^2+1\right) m_Z^2-m_h^2\right)-4 c_W^2 m_Z^2\right)}{s v^2 (\cos (\theta
   )-1)} $ & $ 1.4\times 10^{-3} $ \\ \mbox{} &&\\
 $\text{- + - +} $ & $ -\frac{8 i c_W^2 \cot ^2\left(\frac{\theta }{2}\right) m_Z^2}{v^2} $ & $ 170. $ \\ \mbox{} &&\\
 $\text{- + 0 0} $ & $ -\frac{4 i c_W^2 m_Z^2 \left(\left(8 c_W^2+1\right) m_Z^2+2 \csc ^2(\theta ) \left(m_h^2-m_Z^2\right)-m_h^2\right)}{s v^2} $ & $ 2.4\times
   10^{-3} $ \\ \mbox{} &&\\
 $\text{0 0 0 0} $ & $ -\frac{2 i \left(m_h^2+4 \csc ^2(\theta ) m_Z^2-m_Z^2\right)}{v^2} $ & $ 7.8 $  \\ \hline
\end{tabular}
\end{center}
\caption{The leading terms in the asymptotic limit $s\rightarrow\infty$ of the tree-level helicity amplitudes for $W^+W^+\rightarrow W^+W^+$ process (column 2.); The leading terms are at most constant in energy -- unitarity is preserved in the SM; polarized fractions contributions to the total unpolarized cross sections in $pb$ (column 3.)}
\label{tab:WWonshell2}
\end{table}
\begin{table}%
\begin{center}
\begin{tabular}{c|c|c|c|c||c}
helicity & $iM^{Z+A}$ & $\sim s^n$& $iM^{Z+A+Quartic}$ & $\sim s^n$ & $\sim s^n$\\ \mbox{} &\mbox{}&$n=$ &\mbox{}&  $n=$ & $n=$ \\ \hline{} &&&&& \\
$\text{- - - - } $ & $  \frac{i (15 \cos (\theta )-14 \cos (2 \theta )+\cos (3 \theta )+62) \csc ^2\left(\frac{\theta }{2}\right) c_W^2 m_Z^2}{8 v^2} $ & $  0 $ & $  \frac{8 i \csc ^2\left(\frac{\theta }{2}\right) c_W^2 m_Z^2}{v^2} $ & $  0 $ & $  0 $ \\ &&&&& \\
 $\text{- - - 0 } $ & $  \frac{i \sqrt{s} (\cos (\theta )-3) \sin (\theta ) c_W m_Z}{\sqrt{2} v^2} $ & $  \frac{1}{2} $ & $  \frac{i \sqrt{2} (\cos (\theta )-3) \cot \left(\frac{\theta }{2}\right) c_W^3 m_Z^3}{\sqrt{s} v^2} $ & $  -\frac{1}{2} $ & $  -\frac{1}{2} $ \\ &&&&& \\
 $\text{- - - + } $ & $  \frac{i \sin ^2(\theta ) c_W^2 m_Z^2}{v^2} $ & $  0 $ & $  -\frac{10 i (\cos (\theta )+1) c_W^4 m_Z^4}{s v^2} $ & $  -1 $ & $  -1 $ \\ &&&&& \\
 $\text{- - 0 0 } $ & $  \frac{i s (\cos (2 \theta )+7)}{4 v^2} $ & $  1 $ & $  -\frac{i (\cos (\theta )-1) c_W^2 m_Z^2}{v^2} $ & $  0 $ & $  -1 $ \\ &&&&& \\
 $\text{- - 0 + } $ & $  \frac{i \sqrt{s} (\cos (\theta )+3) \sin (\theta ) c_W m_Z}{\sqrt{2} v^2} $ & $  \frac{1}{2} $ & $  -\frac{i \sqrt{2} \sin (\theta ) c_W^3 m_Z^3}{\sqrt{s} v^2} $ & $  -\frac{1}{2} $ & $  -\frac{3}{2} $ \\ &&&&& \\
 $\text{- - + + } $ & $  -\frac{i (12 \cos (\theta )+\cos (2 \theta )-5) c_W^2 m_Z^2}{2 v^2} $ & $  0 $ & $  \frac{2 i (\cos (\theta )+1) c_W^4 m_Z^4}{s v^2} $ & $  -1 $ & $  -2 $ \\ &&&&& \\
 $\text{- 0 - 0 } $ & $  \frac{i s \sin ^2(\theta )}{2 v^2} $ & $  1 $ & $  -\frac{i (\cos (\theta )-5) \cot ^2\left(\frac{\theta }{2}\right) c_W^2 m_Z^2}{v^2} $ & $  0 $ & $  0 $ \\ &&&&& \\
 $\text{- 0 - + } $ & $  -\frac{i \sqrt{s} (\cos (\theta )+1) \sin (\theta ) c_W m_Z}{\sqrt{2} v^2} $ & $  \frac{1}{2} $ & $  \frac{10 i \sqrt{2} \cos ^2\left(\frac{\theta }{2}\right) \cot \left(\frac{\theta }{2}\right) c_W^3 m_Z^3}{\sqrt{s} v^2} $ & $  -\frac{1}{2} $ & $  -\frac{1}{2} $ \\ &&&&& \\
 $\text{- 0 0 - } $ & $  \frac{i s \sin ^2(\theta )}{2 v^2} $ & $  1 $ & $  \frac{i (\cos (\theta )+1) c_W^2 m_Z^2}{v^2} $ & $  0 $ & $  -1 $ \\ &&&&& \\
 $\text{- 0 0 0 } $ & $  \frac{i s^{3/2} (\cos (\theta )+3) \sin (\theta )}{2 \sqrt{2} v^2 c_W m_Z} $ & $  \frac{3}{2} $ & $  -\frac{i \sqrt{s} \sin (\theta ) c_W m_Z}{\sqrt{2} v^2} $ & $  \frac{1}{2} $ & $  -\frac{1}{2} $ \\ &&&&& \\
 $\text{- 0 0 + } $ & $  -\frac{i s \cos ^2\left(\frac{\theta }{2}\right) (\cos (\theta )+3)}{v^2} $ & $  1 $ & $  \frac{i (\cos (\theta )+1) c_W^2 m_Z^2}{v^2} $ & $  0 $ & $  -1 $ \\ &&&&& \\
 $\text{- 0 + - } $ & $  \frac{i \sqrt{2} \sqrt{s} \sin ^2\left(\frac{\theta }{2}\right) \sin (\theta ) c_W m_Z}{v^2} $ & $  \frac{1}{2} $ & $  -\frac{5 i \sqrt{2} \sin (\theta ) c_W^3 m_Z^3}{\sqrt{s} v^2} $ & $  -\frac{1}{2} $ & $  -\frac{1}{2} $ \\ &&&&& \\
 $\text{- 0 + 0 } $ & $  \frac{i s (\cos (\theta )+3) \sin ^2\left(\frac{\theta }{2}\right)}{v^2} $ & $  1 $ & $  \frac{i (\cos (\theta )-1) c_W^2 m_Z^2}{v^2} $ & $  0 $ & $  -1 $ \\ &&&&& \\
 $\text{- + - + } $ & $  \frac{2 i \cos ^2\left(\frac{\theta }{2}\right) (\cos (\theta )+3) \cot ^2\left(\frac{\theta }{2}\right) c_W^2 m_Z^2}{v^2} $ & $  0 $ & $  \frac{i \csc ^6\left(\frac{\theta }{2}\right) \sin ^4(\theta ) c_W^2 m_Z^2}{2 v^2} $ & $  0 $ & $  0 $ \\ &&&&& \\
 $\text{- + 0 0 } $ & $  -\frac{i s \sin ^2(\theta )}{2 v^2} $ & $  1 $ & $  \frac{i (\cos (\theta )+1) c_W^2 m_Z^2}{v^2} $ & $  0 $ & $  0 $ \\ &&&&& \\
 $\text{- + + - } $ & $  \frac{2 i (\cos (\theta )+3) \sin ^2\left(\frac{\theta }{2}\right) c_W^2 m_Z^2}{v^2} $ & $  0 $ & $  -\frac{4 i (\cos (\theta )-1) c_W^2 m_Z^2}{v^2} $ & $  0 $ & $  0 $ \\ &&&&& \\
 $\text{0 0 0 0 } $ & $  -\frac{i s^2 (12 \cos (\theta )+\cos (2 \theta )-5)}{8 v^2 c_W^2 m_Z^2} $ & $  2 $ & $  \frac{i s (\cos (\theta )+1)}{2 v^2} $ & $  1 $ & $  0 $ \\ \hline
\end{tabular}
\end{center}
\caption{Shown are cancellations between various diagram subsets leading to unitarity fulfilment. Analyzed are tree-level helicity amplitudes for $W^+W^-\rightarrow W^+W^-$ process. $iM^{Z+A}$ denotes sum over $Z_\mu$ and $A_\mu$ exchange diagrams; in $iM^{Z+A+Contact}$ the contact term is added; shown are only the leading terms of the helicity amplitudes in the asymptotic limit $s\rightarrow\infty$; column 3. and 5. show the leading asymptotic $s$ dependence in the amplitudes $iM^{Z+A}$ and $iM^{Z+A+Contact}$ respectively; the last column corresponds to the full tree-level amplitude.}
\label{tab:WWonshell3}
\end{table}
\begin{table}%
\begin{center}
\begin{tabular}{c|c|c}
helicity & $iM^{full}$ & $\sim s^n$, $n=$ \\\hline{} && \\
 $ \text{- - - - } $ & $ \frac{8 i \csc ^2\left(\frac{\theta }{2}\right) c_W^2 m_Z^2}{v^2} $ & $ 0 $ \\  && \\
 $ \text{- - - 0 } $ & $ -\frac{4 i \sqrt{2} \cot \left(\frac{\theta }{2}\right) c_W^3 m_Z^3}{\sqrt{s} v^2} $ & $ -\frac{1}{2} $ \\  && \\
 $ \text{- - - + } $ & $ -\frac{12 i (\cos (\theta )+1) c_W^4 m_Z^4}{s v^2} $ & $ -1 $ \\  && \\
 $ \text{- - 0 0 } $ & $ \frac{2 i c_W^2 m_Z^2 \left(\left(m_Z-m_h\right) \left(m_h+m_Z\right) \csc ^2\left(\frac{\theta
   }{2}\right)+\left(2 \cos (\theta ) c_W^2+6 c_W^2+\cos (\theta )+1\right) m_Z^2\right)}{s v^2} $ & $ -1 $ \\  && \\
 $ \text{- - 0 + } $ & $ -\frac{2 i \sqrt{2} \cot \left(\frac{\theta }{2}\right) c_W^3 m_Z^3 \left(m_h^2-4 c_W^2 m_Z^2+\cos (\theta )
   \left(8 c_W^2+1\right) m_Z^2\right)}{s^{3/2} v^2} $ & $ -\frac{3}{2} $ \\  && \\
 $ \text{- - + + } $ & $ -\frac{4 i c_W^4 m_Z^4 \left(2 m_h^2+\frac{(\cos (\theta )+3) m_Z^2}{\cos (\theta )-1}+\cos (\theta ) \left(8
   c_W^2+1\right) m_Z^2\right)}{s^2 v^2} $ & $ -2 $ \\  && \\
 $ \text{- 0 - 0 } $ & $ \frac{4 i \cot ^2\left(\frac{\theta }{2}\right) c_W^2 m_Z^2}{v^2} $ & $ 0 $ \\  && \\
 $ \text{- 0 - + } $ & $ \frac{3 i \csc ^4\left(\frac{\theta }{2}\right) \sin ^3(\theta ) c_W^3 m_Z^3}{\sqrt{2} \sqrt{s} v^2} $ & $
   -\frac{1}{2} $ \\  && \\
 $ \text{- 0 0 - } $ & $ \frac{2 i \cot ^2\left(\frac{\theta }{2}\right) c_W^2 m_Z^2 \left(m_h^2-2 \left(4 c_W^2+1\right) m_Z^2+\cos
   (\theta ) \left(8 c_W^2+1\right) m_Z^2\right)}{s v^2} $ & $ -1 $ \\  && \\
 $ \text{- 0 0 0 } $ & $ -\frac{i \sqrt{2} \cot \left(\frac{\theta }{2}\right) c_W m_Z \left(m_h^2+2 c_W^2 m_Z^2+\cos (\theta )
   \left(2 c_W^2+1\right) m_Z^2\right)}{\sqrt{s} v^2} $ & $ -\frac{1}{2} $ \\  && \\
 $ \text{- 0 0 + } $ & $ \frac{2 i \cot ^2\left(\frac{\theta }{2}\right) c_W^2 m_Z^2 \left(m_h^2-8 c_W^2 m_Z^2+\cos (\theta ) \left(8
   c_W^2+1\right) m_Z^2\right)}{s v^2} $ & $ -1 $ \\  && \\
 $ \text{- 0 + - } $ & $ -\frac{6 i \sqrt{2} \sin (\theta ) c_W^3 m_Z^3}{\sqrt{s} v^2} $ & $ -\frac{1}{2} $ \\  && \\
 $ \text{- 0 + 0 } $ & $ -\frac{2 i c_W^2 m_Z^2 \left(m_h^2+4 c_W^2 m_Z^2+\cos (\theta ) \left(8 c_W^2+1\right) m_Z^2\right)}{s v^2} $ & $
   -1 $ \\  && \\
 $ \text{- + - + } $ & $ -\frac{4 i \sin ^4(\theta ) c_W^2 m_Z^2}{v^2 (\cos (\theta )-1)^3} $ & $ 0 $ \\  && \\
 $ \text{- + 0 0 } $ & $ \frac{2 i (\cos (\theta )+1) c_W^2 m_Z^2}{v^2} $ & $ 0 $ \\  && \\
 $ \text{- + + - } $ & $ -\frac{4 i (\cos (\theta )-1) c_W^2 m_Z^2}{v^2} $ & $ 0 $ \\  && \\
 $ \text{0 0 0 0 } $ & $ \frac{i \csc ^2\left(\frac{\theta }{2}\right) \left(4 \cos (\theta ) m_h^2-4 m_h^2+\cos (2 \theta ) m_Z^2+7
   m_Z^2\right)}{4 v^2} $ & $ 0 $ \\  \hline
\end{tabular}
\end{center}
\caption{The leading terms in the asymptotic limit $s\rightarrow\infty$ of the tree-level helicity amplitudes for $W^+W^-\rightarrow W^+W^-$ process (column 2.); the exponent of the leading behavior (column 3.)}
\label{tab:WWonshell4}
\end{table}

%%%%%%%%%%%%%%%%%%%%%%% WW ZZ %%%%%%%%%%%%%%%%%%%%%%%%%%%%%%%%%%%%%%%

\begin{table}%
\begin{center}
\bgroup
\def\arraystretch{1.6}
\begin{tabular}{c|c|c|c|c||c}
helicity & $iM^{A}$ & $\sim s^n$& $iM^{A+Quartic}$ & $\sim s^n$ & $\sim s^n$\\ \mbox{} &\mbox{}&$n=$ &\mbox{}&  $n=$ & $n=$ \\ \hline{} &&&&& \\
$\text{- - - - } $ & $ \frac{i (-12 \cos (2 \theta )+\cos (4 \theta )+139) \csc ^2(\theta ) c_W^4 m_Z^2}{4 v^2} $ & $ 0 $ & $ \frac{32 i \csc ^2(\theta ) c_W^4 m_Z^2}{v^2} $ & $ 0 $ & $ 0 $ \\
 $\text{- - - 0 } $ & $ \frac{i \sqrt{2} \sqrt{s} \cos (\theta ) \sin (\theta ) c_W^4 m_Z}{v^2} $ & $ \frac{1}{2} $ & $ -\frac{4 i \sqrt{2} \cot (\theta ) c_W^2 m_Z^3}{\sqrt{s} v^2} $ & $ -\frac{1}{2} $ & $ -\frac{1}{2} $ \\
 $\text{- - - + } $ & $ \frac{2 i \sin ^2(\theta ) c_W^4 m_Z^2}{v^2} $ & $ 0 $ & $ -\frac{4 i c_W^2 \left(8 c_W^4-4 c_W^2+1\right) m_Z^4}{s v^2} $ & $ -1 $ & $ -1 $ \\
 $\text{- - 0 0 } $ & $ \frac{i s (\cos (2 \theta )+7) c_W^4}{2 v^2} $ & $ 1 $ & $ \frac{2 i c_W^2 m_Z^2}{v^2} $ & $ 0 $ & $ -1 $ \\
 $\text{- - 0 + } $ & $ \frac{i \sqrt{2} \sqrt{s} \cos (\theta ) \sin (\theta ) c_W^4 m_Z}{v^2} $ & $ \frac{1}{2} $ & $ -\frac{4 i \sqrt{2} \cot (\theta ) c_W^4 \left(8 c_W^2-3\right) m_Z^5}{s^{3/2} v^2} $ & $ -\frac{3}{2} $ & $ -\frac{3}{2} $ \\
 $\text{- - + + } $ & $ -\frac{i (\cos (2 \theta )-5) c_W^4 m_Z^2}{v^2} $ & $ 0 $ & $ \frac{4 i c_W^2 m_Z^4}{s v^2} $ & $ -1 $ & $ -2 $ \\
 $\text{- 0 - - } $ & $ -\frac{i \sqrt{2} \sqrt{s} \cos (\theta ) \sin (\theta ) c_W^3 m_Z}{v^2} $ & $ \frac{1}{2} $ & $ \frac{4 i \sqrt{2} \cot (\theta ) c_W m_Z^3}{\sqrt{s} v^2} $ & $ -\frac{1}{2} $ & $ -\frac{1}{2} $ \\
 $\text{- 0 - 0 } $ & $ \frac{i s \sin ^2(\theta ) c_W^3}{v^2} $ & $ 1 $ & $ \frac{i \csc ^2\left(\frac{\theta }{2}\right) c_W \left(4 c_W^2+\cos (\theta )-1\right) m_Z^2}{v^2} $ & $ 0 $ & $ 0 $ \\
 $\text{- 0 - + } $ & $ -\frac{i \sqrt{s} \csc ^2\left(\frac{\theta }{2}\right) \sin ^3(\theta ) c_W^3 m_Z}{\sqrt{2} v^2} $ & $ \frac{1}{2} $ & $ \frac{2 i \sqrt{2} \cot \left(\frac{\theta }{2}\right) c_W \left(8 c_W^4-4 c_W^2+1\right) m_Z^3}{\sqrt{s} v^2} $ & $ -\frac{1}{2} $ & $ -\frac{1}{2} $ \\
 $\text{- 0 0 0 } $ & $ \frac{i s^{3/2} \cos (\theta ) \sin (\theta ) c_W^3}{\sqrt{2} v^2 m_Z} $ & $ \frac{3}{2} $ & $ -\frac{2 i \sqrt{2} \cot (\theta ) c_W \left(2 c_W^2+1\right) m_Z^3}{\sqrt{s} v^2} $ & $ -\frac{1}{2} $ & $ -\frac{1}{2} $ \\
 $\text{- 0 0 + } $ & $ -\frac{2 i s \cos ^2\left(\frac{\theta }{2}\right) \cos (\theta ) c_W^3}{v^2} $ & $ 1 $ & $ {\scriptstyle \frac{1}{s v^2} i \csc ^2\left(\frac{\theta }{2}\right) c_W \left(-12 c_W^4+3 c_W^2+\right.}$ & $ -1 $ & $ -1 $ \\

&&& $ {\scriptstyle \left.\cos (\theta ) \left(16 c_W^4-5 c_W^2+1\right)-1\right) m_Z^4 }$ &&  \\
 $\text{- 0 + + } $ & $ -\frac{i \sqrt{2} \sqrt{s} \cos (\theta ) \sin (\theta ) c_W^3 m_Z}{v^2} $ & $ \frac{1}{2} $ & $ \frac{4 i \sqrt{2} \cot (\theta ) c_W^3 \left(8 c_W^2-3\right) m_Z^5}{s^{3/2} v^2} $ & $ -\frac{3}{2} $ & $ -\frac{3}{2} $ \\
 $\text{- + - - } $ & $ \frac{2 i \sin ^2(\theta ) c_W^4 m_Z^2}{v^2} $ & $ 0 $ & $ -\frac{4 i c_W^2 \left(4 c_W^2+1\right) m_Z^4}{s v^2} $ & $ -1 $ & $ -1 $ \\
 $\text{- + - 0 } $ & $ \frac{i \sqrt{s} \csc ^2\left(\frac{\theta }{2}\right) \sin ^3(\theta ) c_W^4 m_Z}{\sqrt{2} v^2} $ & $ \frac{1}{2} $ & $ -\frac{2 i \sqrt{2} \cot \left(\frac{\theta }{2}\right) c_W^2 \left(4 c_W^2+1\right) m_Z^3}{\sqrt{s} v^2} $ & $ -\frac{1}{2} $ & $ -\frac{1}{2} $ \\
 $\text{- + - + } $ & $ \frac{i (\cos (2 \theta )+7) \cot ^2\left(\frac{\theta }{2}\right) c_W^4 m_Z^2}{v^2} $ & $ 0 $ & $ \frac{8 i \cot ^2\left(\frac{\theta }{2}\right) c_W^4 m_Z^2}{v^2} $ & $ 0 $ & $ 0 $ \\
 $\text{- + 0 0 } $ & $ -\frac{i s \sin ^2(\theta ) c_W^4}{v^2} $ & $ 1 $ & $ \frac{2 i c_W^2 m_Z^2}{v^2} $ & $ 0 $ & $ 0 $ \\
 $\text{0 0 - - } $ & $ \frac{i s (\cos (2 \theta )+7) c_W^2}{2 v^2} $ & $ 1 $ & $ \frac{2 i m_Z^2}{v^2} $ & $ 0 $ & $ -1 $ \\
 $\text{0 0 - 0 } $ & $ -\frac{i s^{3/2} \cos (\theta ) \sin (\theta ) c_W^2}{\sqrt{2} v^2 m_Z} $ & $ \frac{3}{2} $ & $ \frac{2 i \sqrt{2} \cot (\theta ) \left(4 c_W^4-2 c_W^2+1\right) m_Z^3}{\sqrt{s} v^2} $ & $ -\frac{1}{2} $ & $ -\frac{1}{2} $ \\
 $\text{0 0 - + } $ & $ -\frac{i s \sin ^2(\theta ) c_W^2}{v^2} $ & $ 1 $ & $ \frac{2 i \left(1-2 c_W^2\right){}^2 m_Z^2}{v^2} $ & $ 0 $ & $ 0 $ \\
 $\text{0 0 0 0 } $ & $ -\frac{i s^2 (\cos (2 \theta )-5) c_W^2}{4 v^2 m_Z^2} $ & $ 2 $ & $ \frac{i s}{v^2} $ & $ 1 $ & $ 0 $ \\ \hline
\end{tabular}
\egroup
\end{center}
\caption{Shown are cancellations between various diagram subsets leading to unitarity fulfilment. Analysed are tree-level helicity amplitudes for $W^+W^-\rightarrow ZZ$ process. $iM^{A}$ denotes sum over $A_\mu$ exchange diagrams; in $iM^{A+Contact}$ the contact term is added; shown are only the leading terms of the helicity amplitudes in the asymptotic limit $s\rightarrow\infty$; column 3. and 5. show the leading asymptotic $s$ dependence in the amplitudes $iM^{A}$ and $iM^{A+Contact}$ respectively; the last column corresponds to the full tree-level amplitude.}
\label{tab:WWonshell7}
\end{table}
\begin{table}%
\begin{center}
\bgroup
\def\arraystretch{1.6}
\begin{tabular}{c|c|c}
helicity & $iM^{full}$ & $\sim s^n$, $n=$ \\\hline{} && \\
 $\text{- - - - } $ & $ \frac{32 i \csc ^2(\theta ) c_W^4 m_Z^2}{v^2} $ & $ 0 $ \\
 $\text{- - - 0 } $ & $ -\frac{4 i \sqrt{2} \cot (\theta ) c_W^2 m_Z^3}{\sqrt{s} v^2} $ & $ -\frac{1}{2} $ \\
 $\text{- - - + } $ & $ -\frac{4 i c_W^2 \left(8 c_W^4-4 c_W^2+1\right) m_Z^4}{s v^2} $ & $ -1 $ \\
 $\text{- - 0 0 } $ & $ \frac{2 i c_W^2 m_Z^2 \left(-m_h^2+4 \csc ^2(\theta ) c_W^2 m_Z^2+4 \left(c_W^2+1\right) m_Z^2\right)}{s v^2} $ & $ -1 $ \\
 $\text{- - 0 + } $ & $ -\frac{4 i \sqrt{2} \cot (\theta ) c_W^4 \left(8 c_W^2-3\right) m_Z^5}{s^{3/2} v^2} $ & $ -\frac{3}{2} $ \\
 $\text{- - + + } $ & $ \frac{4 i c_W^2 m_Z^4 \left(8 \csc ^2(\theta ) m_Z^2 c_W^6-m_h^2+2 \left(-8 c_W^4+4 c_W^2+1\right) m_Z^2\right)}{s^2 v^2} $ & $ -2 $ \\
 $\text{- 0 - - } $ & $ \frac{4 i \sqrt{2} \cot (\theta ) c_W m_Z^3}{\sqrt{s} v^2} $ & $ -\frac{1}{2} $ \\
 $\text{- 0 - 0 } $ & $ \frac{i \csc ^2\left(\frac{\theta }{2}\right) c_W \left(4 c_W^2+\cos (\theta )-1\right) m_Z^2}{v^2} $ & $ 0 $ \\
 $\text{- 0 - + } $ & $ \frac{2 i \sqrt{2} \cot \left(\frac{\theta }{2}\right) c_W \left(8 c_W^4-4 c_W^2+1\right) m_Z^3}{\sqrt{s} v^2} $ & $ -\frac{1}{2} $ \\
 $\text{- 0 0 0 } $ & $ -\frac{2 i \sqrt{2} \cot (\theta ) c_W \left(2 c_W^2+1\right) m_Z^3}{\sqrt{s} v^2} $ & $ -\frac{1}{2} $ \\
 $\text{- 0 0 + } $ & $ \frac{i \csc ^2\left(\frac{\theta }{2}\right) c_W \left(-12 c_W^4+3 c_W^2+\cos (\theta ) \left(16 c_W^4-5 c_W^2+1\right)-1\right) m_Z^4}{s v^2} $ & $ -1 $ \\
 $\text{- 0 + + } $ & $ \frac{4 i \sqrt{2} \cot (\theta ) c_W^3 \left(8 c_W^2-3\right) m_Z^5}{s^{3/2} v^2} $ & $ -\frac{3}{2} $ \\
 $\text{- + - - } $ & $ -\frac{4 i c_W^2 \left(4 c_W^2+1\right) m_Z^4}{s v^2} $ & $ -1 $ \\
 $\text{- + - 0 } $ & $ -\frac{2 i \sqrt{2} \cot \left(\frac{\theta }{2}\right) c_W^2 \left(4 c_W^2+1\right) m_Z^3}{\sqrt{s} v^2} $ & $ -\frac{1}{2} $ \\
 $\text{- + - + } $ & $ \frac{8 i \cot ^2\left(\frac{\theta }{2}\right) c_W^4 m_Z^2}{v^2} $ & $ 0 $ \\
 $\text{- + 0 0 } $ & $ \frac{2 i c_W^2 m_Z^2}{v^2} $ & $ 0 $ \\
 $\text{0 0 - - } $ & $ \frac{8 i \csc ^2(\theta ) \left(2 c_W^3-c_W\right){}^2 m_Z^4+4 i \left(4 c_W^4-c_W^2+1\right) m_Z^4-2 i m_h^2 m_Z^2}{s v^2} $ & $ -1 $ \\
 $\text{0 0 - 0 } $ & $ \frac{2 i \sqrt{2} \cot (\theta ) \left(4 c_W^4-2 c_W^2+1\right) m_Z^3}{\sqrt{s} v^2} $ & $ -\frac{1}{2} $ \\
 $\text{0 0 - + } $ & $ \frac{2 i \left(1-2 c_W^2\right){}^2 m_Z^2}{v^2} $ & $ 0 $ \\
 $\text{0 0 0 0 } $ & $ -\frac{i \left(m_h^2-8 \csc ^2(\theta ) c_W^2 m_Z^2+2 c_W^2 m_Z^2\right)}{v^2} $ & $ 0 $ \\ \hline
\end{tabular}
\egroup
\end{center}
\caption{The leading terms in the asymptotic limit $s\rightarrow\infty$ of the tree-level helicity amplitudes for $W^+W^-\rightarrow ZZ$ process (column 2.); the exponent of the leading behavior (column 3.)}
\label{tab:WWonshell8}
\end{table}

%\begin{table}%
%\begin{center}
%\begin{tabular}{cccccccccc}
 %\text{- - - - } & \text{- - - 0 } & \text{- - - + } & \text{- - 0 0 } & \text{- - 0 + } & \text{- - + + } & \text{- 0 - - } & \text{- 0 - 0 } & \text{- 0 - + } & \text{- 0 0 0 } \\ 
%\text{- 0 0 + } & \text{- 0 + + } & \text{- + - - } & \text{- + - 0 } & \text{- + - + } & \text{- + 0 0 } & \text{0 0 - - } & \text{0 0 - 0 } & \text{0 0 - + } & \text{0 0 0 0 }
%\end{tabular}
%\end{center}
%%\caption{}
%%\label{}
%%\end{table}
%The corresponding multiplicities read:
%%\begin{table}%
%\begin{center}\begin{tabular}{cccccccccc}
%2 & 4 & 4 & 2 & 4 & 2 & 4 & 8 & 8 & 4 \\ 
%8 & 4 & 4 & 8 & 4 & 2 & 2 & 4 & 2 & 1 \\
%\end{tabular}
%\end{center}

%\left(
%\begin{array}{cccccccccccccccccccc}
 %\text{- - - - } & \text{- - - 0 } & \text{- - - + } & \text{- - 0 0 } & \text{- - 0 + } & \text{- - + + } & \text{- 0 - - } & \text{- 0 - 0 } & \text{- 0 - + } & \text{- 0 0 0 } & \text{- 0 0 + } & \text{- 0 + + } & \text{- + - - } & \text{- + - 0 } & \text{- + - + } & \text{- +
   %0 0 } & \text{0 0 - - } & \text{0 0 - 0 } & \text{0 0 - + } & \text{0 0 0 0 } \\
 %2 & 4 & 4 & 2 & 4 & 2 & 4 & 8 & 8 & 4 & 8 & 4 & 4 & 8 & 4 & 2 & 2 & 4 & 2 & 1 \\
%\end{array}
%\right)

%%%%%%%%%%%%%%%%%%%%%%%%%%%%%%%%%%%%%%%%%%%%%%%%%%%%%%%%%%%%%%

%%%%%%%%%%%%%%%%%%%%%%% WZ WZ %%%%%%%%%%%%%%%%%%%%%%%%%%%%%%%%%%%%%%%

\begin{table}%
\begin{center}
\bgroup
\def\arraystretch{1.5}
\begin{tabular}{c|c|c|c|c||c}
helicity & $iM^{A}$ & $\sim s^n$& $iM^{A+Quartic}$ & $\sim s^n$ & $\sim s^n$\\ \mbox{} &\mbox{}&$n=$ &\mbox{}&  $n=$ & $n=$ \\ \hline{} &&&&& \\
 $\text{- - - - } $ & $ \frac{i (15 \cos (\theta )+14 \cos (2 \theta )+\cos (3 \theta )-62) c_W^4 m_Z^2}{4 v^2 (\cos (\theta )+1)} $ & $ 0 $ & $ -\frac{8 i \sec ^2\left(\frac{\theta }{2}\right) c_W^4 m_Z^2}{v^2} $ & $ 0 $ & $ 0 $ \\
 $\text{- - - 0 } $ & $ -\frac{i \sqrt{s} (\cos (\theta )+3) \sin (\theta ) c_W^4 m_Z}{\sqrt{2} v^2} $ & $ \frac{1}{2} $ & $ -\frac{i \sqrt{2} c_W^2 \left(4 c_W^2+\cos (\theta )-1\right) m_Z^3 \tan \left(\frac{\theta }{2}\right)}{\sqrt{s} v^2} $ & $ -\frac{1}{2} $ & $ -\frac{1}{2} $ \\
 $\text{- - - + } $ & $ -\frac{i \sin ^2(\theta ) c_W^4 m_Z^2}{v^2} $ & $ 0 $ & $ -\frac{2 i (\cos (\theta )-1) c_W^2 \left(2 c_W^4+4 c_W^2-1\right) m_Z^4}{s v^2} $ & $ -1 $ & $ -1 $ \\
 $\text{- - 0 - } $ & $ -\frac{i \sqrt{s} (\cos (\theta )+3) \sin (\theta ) c_W^3 m_Z}{\sqrt{2} v^2} $ & $ \frac{1}{2} $ & $ -\frac{i \sqrt{2} c_W \left(-8 c_W^4+12 c_W^2+\cos (\theta )-1\right) m_Z^3 \tan \left(\frac{\theta }{2}\right)}{\sqrt{s} v^2} $ & $ -\frac{1}{2} $ & $ -\frac{1}{2} $ \\
 $\text{- - 0 0 } $ & $ -\frac{i s (\cos (2 \theta )+7) c_W^3}{4 v^2} $ & $ 1 $ & $ -\frac{i (\cos (\theta )+1) c_W m_Z^2}{v^2} $ & $ 0 $ & $ -1 $ \\
 $\text{- - 0 + } $ & $ -\frac{i \sqrt{s} (\cos (\theta )-3) \sin (\theta ) c_W^3 m_Z}{\sqrt{2} v^2} $ & $ \frac{1}{2} $ & $ -\frac{i \sqrt{2} \sin (\theta ) c_W m_Z^3}{\sqrt{s} v^2} $ & $ -\frac{1}{2} $ & $ -\frac{3}{2} $ \\
 $\text{- - + - } $ & $ -\frac{i \sin ^2(\theta ) c_W^4 m_Z^2}{v^2} $ & $ 0 $ & $ -\frac{2 i (\cos (\theta )-1) c_W^2 \left(2 c_W^4+4 c_W^2-1\right) m_Z^4}{s v^2} $ & $ -1 $ & $ -1 $ \\
 $\text{- - + 0 } $ & $ -\frac{i \sqrt{s} (\cos (\theta )-3) \sin (\theta ) c_W^4 m_Z}{\sqrt{2} v^2} $ & $ \frac{1}{2} $ & $ -\frac{i \sqrt{2} \sin (\theta ) c_W^2 m_Z^3}{\sqrt{s} v^2} $ & $ -\frac{1}{2} $ & $ -\frac{3}{2} $ \\
 $\text{- - + + } $ & $ \frac{i (-12 \cos (\theta )+\cos (2 \theta )-5) c_W^4 m_Z^2}{2 v^2} $ & $ 0 $ & $ \frac{2 i (\cos (\theta )-1) c_W^2 m_Z^4}{s v^2} $ & $ -1 $ & $ -2 $ \\
 $\text{- 0 - 0 } $ & $ -\frac{i s \sin ^2(\theta ) c_W^4}{2 v^2} $ & $ 1 $ & $ \frac{i (\cos (\theta )-1) c_W^2 m_Z^2}{v^2} $ & $ 0 $ & $ 0 $ \\
 $\text{- 0 - + } $ & $ \frac{i \sqrt{s} (\cos (\theta )+1) \sin (\theta ) c_W^4 m_Z}{\sqrt{2} v^2} $ & $ \frac{1}{2} $ & $ \frac{i \sqrt{2} \sin (\theta ) c_W^2 \left(1-6 c_W^2\right) m_Z^3}{\sqrt{s} v^2} $ & $ -\frac{1}{2} $ & $ -\frac{1}{2} $ \\
 $\text{- 0 0 - } $ & $ -\frac{i s \sin ^2(\theta ) c_W^3}{2 v^2} $ & $ 1 $ & $ -\frac{i c_W \left(4 c_W^2+\cos (\theta )+1\right) m_Z^2 \tan ^2\left(\frac{\theta }{2}\right)}{v^2} $ & $ 0 $ & $ 0 $ \\
 $\text{- 0 0 0 } $ & $ -\frac{i s^{3/2} (\cos (\theta )-3) \sin (\theta ) c_W^3}{2 \sqrt{2} v^2 m_Z} $ & $ \frac{3}{2} $ & $ -\frac{i \sqrt{s} \sin (\theta ) c_W m_Z}{\sqrt{2} v^2} $ & $ \frac{1}{2} $ & $ -\frac{1}{2} $ \\
 $\text{- 0 0 + } $ & $ \frac{i s \cos ^2\left(\frac{\theta }{2}\right) (\cos (\theta )-3) c_W^3}{v^2} $ & $ 1 $ & $ \frac{i (\cos (\theta )+1) c_W m_Z^2}{v^2} $ & $ 0 $ & $ -1 $ \\
 $\text{- 0 + - } $ & $ -\frac{i \sqrt{2} \sqrt{s} \sin ^2\left(\frac{\theta }{2}\right) \sin (\theta ) c_W^4 m_Z}{v^2} $ & $ \frac{1}{2} $ & $ \frac{4 i \sqrt{2} \csc (\theta ) \sin ^4\left(\frac{\theta }{2}\right) c_W^2 \left(6 c_W^2-1\right) m_Z^3}{\sqrt{s} v^2} $ & $ -\frac{1}{2} $ & $
   -\frac{1}{2} $ \\
 $\text{- 0 + 0 } $ & $ -\frac{i s (\cos (\theta )-3) \sin ^2\left(\frac{\theta }{2}\right) c_W^4}{v^2} $ & $ 1 $ & $ \frac{i (\cos (\theta )-1) c_W^2 m_Z^2}{v^2} $ & $ 0 $ & $ -1 $ \\
 $\text{- + - + } $ & $ \frac{2 i \cos ^2\left(\frac{\theta }{2}\right) (\cos (\theta )-3) c_W^4 m_Z^2}{v^2} $ & $ 0 $ & $ -\frac{4 i (\cos (\theta )+1) c_W^4 m_Z^2}{v^2} $ & $ 0 $ & $ 0 $ \\
 $\text{- + 0 - } $ & $ \frac{i \sqrt{2} \sqrt{s} \sin ^2\left(\frac{\theta }{2}\right) \sin (\theta ) c_W^3 m_Z}{v^2} $ & $ \frac{1}{2} $ & $ -\frac{4 i \sqrt{2} \csc (\theta ) \sin ^4\left(\frac{\theta }{2}\right) c_W \left(4 c_W^4+2 c_W^2-1\right) m_Z^3}{\sqrt{s} v^2} $ & $ -\frac{1}{2} $ & $
   -\frac{1}{2} $ \\
 $\text{- + 0 0 } $ & $ \frac{i s \sin ^2(\theta ) c_W^3}{2 v^2} $ & $ 1 $ & $ \frac{i (\cos (\theta )-1) c_W \left(2 c_W^2-1\right) m_Z^2}{v^2} $ & $ 0 $ & $ 0 $ \\
 $\text{- + 0 + } $ & $ -\frac{i \sqrt{s} (\cos (\theta )+1) \sin (\theta ) c_W^3 m_Z}{\sqrt{2} v^2} $ & $ \frac{1}{2} $ & $ \frac{i \sqrt{2} \sin (\theta ) c_W \left(4 c_W^4+2 c_W^2-1\right) m_Z^3}{\sqrt{s} v^2} $ & $ -\frac{1}{2} $ & $ -\frac{1}{2} $ \\
 $\text{- + + - } $ & $ \frac{4 i (\cos (\theta )-3) \sin ^4\left(\frac{\theta }{2}\right) c_W^4 m_Z^2}{v^2 (\cos (\theta )+1)} $ & $ 0 $ & $ -\frac{32 i \csc ^2(\theta ) \sin ^6\left(\frac{\theta }{2}\right) c_W^4 m_Z^2}{v^2} $ & $ 0 $ & $ 0 $ \\
 $\text{0 - 0 - } $ & $ -\frac{i s \sin ^2(\theta ) c_W^2}{2 v^2} $ & $ 1 $ & $ \frac{i \left(-8 c_W^4+8 c_W^2+\cos (\theta )-1\right) m_Z^2}{v^2} $ & $ 0 $ & $ 0 $ \\
 $\text{0 - 0 0 } $ & $ -\frac{i s^{3/2} (\cos (\theta )-3) \sin (\theta ) c_W^2}{2 \sqrt{2} v^2 m_Z} $ & $ \frac{3}{2} $ & $ -\frac{i \sqrt{s} \sin (\theta ) m_Z}{\sqrt{2} v^2} $ & $ \frac{1}{2} $ & $ -\frac{1}{2} $ \\
 $\text{0 - 0 + } $ & $ -\frac{i s (\cos (\theta )-3) \sin ^2\left(\frac{\theta }{2}\right) c_W^2}{v^2} $ & $ 1 $ & $ \frac{i (\cos (\theta )-1) m_Z^2}{v^2} $ & $ 0 $ & $ -1 $ \\
 $\text{0 0 0 0 } $ & $ \frac{i s^2 (-12 \cos (\theta )+\cos (2 \theta )-5) c_W^2}{8 v^2 m_Z^2} $ & $ 2 $ & $ \frac{i s (\cos (\theta )-1)}{2 v^2} $ & $ 1 $ & $ 0 $ \\ \hline
\end{tabular}
\egroup
\end{center}
\caption{Shown are cancellations between various diagram subsets leading to unitarity fulfilment. Analyzed are tree-level helicity amplitudes for $W^+Z\rightarrow W^+Z$ process. $iM^{A}$ denotes sum over $A_\mu$ exchange diagrams; in $iM^{A+Contact}$ the contact term is added; shown are only the leading terms of the helicity amplitudes in the asymptotic limit $s\rightarrow\infty$; column 3. and 5. show the leading asymptotic $s$ dependence in the amplitudes $iM^{A}$ and $iM^{A+Contact}$ respectively; the last column corresponds to the full tree-level amplitude.}
\label{tab:WWonshell9}
\end{table}
\begin{table}%
\begin{center}
\bgroup
\def\arraystretch{1.6}
\begin{tabular}{c|c|c}
helicity & $iM^{full}$ & $\sim s^n$, $n=$ \\\hline{} && \\
 $\text{- - - - } $ & $ -\frac{8 i \sec ^2\left(\frac{\theta }{2}\right) c_W^4 m_Z^2}{v^2} $ & $ 0 $ \\
 $\text{- - - 0 } $ & $ \frac{4 i \sqrt{2} \csc (\theta ) c_W^2 \left((\cos (\theta )-1) c_W^2-\cos (\theta )\right) m_Z^3}{\sqrt{s} v^2} $ & $ -\frac{1}{2} $ \\
 $\text{- - - + } $ & $ -\frac{4 i c_W^2 \left((\cos (\theta )-1) \left(c_W^2+2\right) c_W^2+1\right) m_Z^4}{s v^2} $ & $ -1 $ \\
 $\text{- - 0 - } $ & $ -\frac{4 i \sqrt{2} \csc (\theta ) c_W \left((\cos (\theta )-1) \left(2 c_W^2-3\right) c_W^2+\cos (\theta )\right) m_Z^3}{\sqrt{s} v^2} $ & $ -\frac{1}{2} $ \\
 $\text{- - 0 0 } $ & $ {\scriptstyle-\frac{1}{s v^2}2 i \csc ^2(\theta ) c_W m_Z^2 \left(4 \cos ^4\left(\frac{\theta }{2}\right) m_h^2-\sin ^2\left(\frac{\theta }{2}\right) \left((4 \cos (\theta )+\cos (2 \theta )-5) c_W^4+\right.\right. }$ & \\
&  ${\scriptstyle \left.\left.+2 (-3 \cos (\theta )+\cos (2 \theta )-2) c_W^2+6 (\cos (\theta )+1)\right)
   m_Z^2\right)} $ & $ -1 $ \\
 $\text{- - 0 + } $ & $ \frac{2 i \sqrt{2} c_W m_Z^3 \left(c_W^2 \left(2 (\cos (\theta )-1) c_W^4+(5 \cos (\theta )+3) c_W^2+2 \cos (\theta )+3\right) m_Z^2-\cot ^2\left(\frac{\theta }{2}\right) m_h^2\right) \tan \left(\frac{\theta }{2}\right)}{s^{3/2} v^2} $ & $ -\frac{3}{2} $ \\
 $\text{- - + - } $ & $ -\frac{4 i c_W^2 \left((\cos (\theta )-1) \left(c_W^2+2\right) c_W^2+1\right) m_Z^4}{s v^2} $ & $ -1 $ \\
 $\text{- - + 0 } $ & $ -\frac{2 i \sqrt{2} c_W^2 m_Z^3 \left(\cot \left(\frac{\theta }{2}\right) m_h^2-c_W^2 \left(3 (\cos (\theta )-1) c_W^2+6 \cos (\theta )+7\right) m_Z^2 \tan \left(\frac{\theta }{2}\right)\right)}{s^{3/2} v^2} $ & $ -\frac{3}{2} $ \\
 $\text{- - + + } $ & $ -\frac{4 i c_W^2 m_Z^4 \left(m_h^2+\left(\frac{\left((\cos (\theta )-1) c_W^3+2 (\cos (\theta )+1) c_W\right){}^2}{\cos (\theta )+1}-2\right) m_Z^2\right)}{s^2 v^2} $ & $ -2 $ \\
 $\text{- 0 - 0 } $ & $ -\frac{2 i c_W^2 m_Z^2}{v^2} $ & $ 0 $ \\
 $\text{- 0 - + } $ & $ \frac{2 i \sqrt{2} \cot \left(\frac{\theta }{2}\right) c_W^2 \left(3 (\cos (\theta )-1) c_W^2+1\right) m_Z^3}{\sqrt{s} v^2} $ & $ -\frac{1}{2} $ \\
 $\text{- 0 0 - } $ & $ \frac{2 i c_W \left(2 (\cos (\theta )-1) c_W^2-\cos (\theta )-1\right) m_Z^2}{v^2 (\cos (\theta )+1)} $ & $ 0 $ \\
 $\text{- 0 0 0 } $ & $ -\frac{i \csc (\theta ) c_W m_Z \left(2 (\cos (\theta )+1) m_h^2+\left((-8 \cos (\theta )+3 \cos (2 \theta )+5) c_W^2+6 \cos (\theta )+2\right) m_Z^2\right)}{\sqrt{2} \sqrt{s} v^2} $ & $ -\frac{1}{2} $ \\
 $\text{- 0 0 + } $ & $ \frac{8 i \cos ^2\left(\frac{\theta }{2}\right) \csc ^2(\theta ) c_W m_Z^2 \left(\frac{1}{2} (\cos (\theta )+1) m_h^2-\sin ^2\left(\frac{\theta }{2}\right) \left(6 (\cos (\theta )-1) c_W^4+3 (\cos (\theta )+1) c_W^2+1\right) m_Z^2\right)}{s v^2} $ & $ -1 $ \\
 $\text{- 0 + - } $ & $ -\frac{2 i \sqrt{2} c_W^2 \left(3 (\cos (\theta )-1) c_W^2+1\right) m_Z^3 \tan \left(\frac{\theta }{2}\right)}{\sqrt{s} v^2} $ & $ -\frac{1}{2} $ \\
 $\text{- 0 + 0 } $ & $ -\frac{2 i \csc ^2(\theta ) \sin ^2\left(\frac{\theta }{2}\right) c_W^2 m_Z^2 \left(2 (\cos (\theta )+1) m_h^2+\left((-2 \cos (\theta )+9 \cos (2 \theta )-7) c_W^2+4 (\cos (\theta )+1)\right) m_Z^2\right)}{s v^2} $ & $ -1 $ \\
 $\text{- + - + } $ & $ -\frac{4 i (\cos (\theta )+1) c_W^4 m_Z^2}{v^2} $ & $ 0 $ \\
 $\text{- + 0 - } $ & $ \frac{2 i \sqrt{2} c_W \left((\cos (\theta )-1) \left(2 c_W^2+1\right) c_W^2+1\right) m_Z^3 \tan \left(\frac{\theta }{2}\right)}{\sqrt{s} v^2} $ & $ -\frac{1}{2} $ \\
 $\text{- + 0 0 } $ & $ \frac{2 i c_W \left((\cos (\theta )-1) c_W^2+1\right) m_Z^2}{v^2} $ & $ 0 $ \\
 $\text{- + 0 + } $ & $ \frac{2 i \sqrt{2} c_W \left(2 \sin (\theta ) c_W^4+\sin (\theta ) c_W^2-\cot \left(\frac{\theta }{2}\right)\right) m_Z^3}{\sqrt{s} v^2} $ & $ -\frac{1}{2} $ \\
 $\text{- + + - } $ & $ -\frac{32 i \csc ^2(\theta ) \sin ^6\left(\frac{\theta }{2}\right) c_W^4 m_Z^2}{v^2} $ & $ 0 $ \\
 $\text{0 - 0 - } $ & $ -\frac{2 i \left(1-2 c_W^2\right){}^2 m_Z^2}{v^2} $ & $ 0 $ \\
 $\text{0 - 0 0 } $ & $ -\frac{i \csc (\theta ) m_Z \left(2 (\cos (\theta )+1) m_h^2+\left(-4 \sin ^2(\theta ) c_W^4+(-4 \cos (\theta )+\cos (2 \theta )+11) c_W^2+2 (\cos (\theta )-1)\right) m_Z^2\right)}{\sqrt{2} \sqrt{s} v^2} $ & $ -\frac{1}{2} $ \\
 $\text{0 - 0 + } $ & $ -\frac{2 i m_Z^2 \left(m_h^2+\left(-8 c_W^6+\sec ^2\left(\frac{\theta }{2}\right) \left(1-2 c_W^2\right){}^2 c_W^2+\cos (\theta ) \left(2 c_W^3+c_W\right){}^2\right) m_Z^2\right)}{s v^2} $ & $ -1 $ \\
 $\text{0 0 0 0 } $ & $ -\frac{i \sec ^2\left(\frac{\theta }{2}\right) \left(2 (\cos (\theta )+1) m_h^2+(\cos (2 \theta )+7) c_W^2 m_Z^2\right)}{4 v^2} $ & $ 0 $ \\ \hline
\end{tabular}
\egroup
\end{center}
\caption{The leading terms in the asymptotic limit $s\rightarrow\infty$ of the tree-level helicity amplitudes for $W^+Z\rightarrow W^+Z$ process (column 2.); the exponent of the leading behavior (column 3.)}
\label{tab:WWonshell10}
\end{table}
%%%%%%%%%%%%%%%%%%%%%%%%%%%%%%%%%%%%%%%%%%%%%%%%%%%%%%%%%%%%%%

%%%% SM partial waves %%%%%%%%%%%%%%%%%%%%%%%%%%%%%%%%%%%%%%%%

\begin{table}%
\begin{center}
\begin{tabular}{c|c|c}
$\lambda_1\lambda_2\lambda_1'\lambda_2'$ & $\mathcal{T}^{(j=j_{min})}_{\lambda_1'\lambda_2';\lambda_1\lambda_2} $ & $j_{min} = $ \\ \hline && \\
 $ \text{- - - - } $ & $ 0 $ & $ 0 $ \\ && \\
 $ \text{- - - 0 } $ & $ 0 $ & $ 2 $ \\ && \\
 $ \text{- - - + } $ & $ \frac{c_W^4 m_Z^4}{\sqrt{6} \pi  s v^2} $ & $ 2 $ \\ && \\
 $ \text{- - 0 0 } $ & $ \frac{c_W^2 m_Z^2 \left(m_h^2-m_Z^2\right)}{4 \pi  s v^2} $ & $ 0 $ \\ && \\
 $ \text{- - 0 + } $ & $ -\frac{c_W^3 m_Z^3 \left(\left(4 c_W^2+1\right) m_Z^2-m_h^2\right)}{4 \sqrt{3} \pi  s^{3/2} v^2} $ & $ 2 $ \\ && \\
 $ \text{- - + + } $ & $ \frac{c_W^4 m_Z^4 \left(\left(4 c_W^2+1\right) m_Z^2-m_h^2\right)}{2 \pi  s^2 v^2} $ & $ 0 $ \\ && \\
 $ \text{- 0 - 0 } $ & $ 0 $ & $ 1 $ \\ && \\
 $ \text{- 0 - + } $ & $ \frac{\sqrt{2} c_W^3 m_Z^3}{3 \pi  \sqrt{s} v^2} $ & $ 2 $ \\ && \\
 $ \text{- 0 0 0 } $ & $ \frac{c_W m_Z \left(m_Z^2-m_h^2\right)}{8 \sqrt{3} \pi  \sqrt{s} v^2} $ & $ 2 $ \\ && \\
 $ \text{- 0 0 + } $ & $ \frac{c_W^2 m_Z^2 \left(\left(4 c_W^2+1\right) m_Z^2-m_h^2\right)}{8 \pi  s v^2} $ & $ 1 $ \\ && \\
 $ \text{- + - + } $ & $ \frac{c_W^2 m_Z^2}{6 \pi  v^2} $ & $ 2 $ \\ && \\
 $ \text{- + 0 0 } $ & $ -\frac{c_W^2 m_Z^2 \left(m_h^2+\left(4 c_W^2-1\right) m_Z^2\right)}{2 \sqrt{6} \pi  s v^2} $ & $ 2 $ \\ && \\
 $ \text{0 0 0 0 } $ & $ -\frac{m_h^2-m_Z^2}{8 \pi  v^2} $ & $ 0 $ \\ \hline 
\end{tabular}
\end{center}
\caption{The leading terms in the asymptotic limit $s\rightarrow\infty$ of the tree-level partial wave amplitudes (column 2.) for $W^+W^+\rightarrow W^+W^+$ process}
\label{tab:WWonshell11}
\end{table}

\begin{table}
\begin{center}
\begin{tabular}{c|c|c}
$\lambda_1\lambda_2\lambda_1'\lambda_2'$ & $\mathcal{T}^{(j=j_{min})}_{\lambda_1'\lambda_2';\lambda_1\lambda_2} $ & $j_{min} = $ \\ \hline && \\
 $ \text{- - - - } $ & $ 0 $ & $ 0 $ \\ && \\
 $ \text{- - - 0 } $ & ${  \frac{c_W^3 m_Z^3}{4 \sqrt{3} \pi  \sqrt{s} v^2} }$ & $ 2 $ \\ && \\
 $ \text{- - - + } $ & ${  -\frac{\sqrt{\frac{3}{2}} c_W^4 m_Z^4}{4 \pi  s v^2} }$ & $ 2 $ \\ && \\
 $ \text{- - 0 0 } $ & ${  \frac{c_W^2 \left(6 c_W^2+1\right) m_Z^4}{8 \pi  s v^2} }$ & $ 0 $ \\ && \\
 $ \text{- - 0 + } $ & ${  \frac{c_W^3 m_Z^3 \left(m_h^2+\left(4 c_W^2+1\right) m_Z^2\right)}{8 \sqrt{3} \pi  s^{3/2} v^2} }$ & $ 2 $ \\ && \\
 $ \text{- - + + } $ & ${  -\frac{c_W^4 m_Z^4 \left(2 m_h^2+m_Z^2\right)}{4 \pi  s^2 v^2} }$ & $ 0 $ \\ && \\
 $ \text{- 0 - 0 } $ & ${  -\frac{c_W^2 m_Z^2}{8 \pi  v^2} }$ & $ 1 $ \\ && \\
 $ \text{- 0 - + } $ & ${  -\frac{3 c_W^3 m_Z^3}{4 \sqrt{2} \pi  \sqrt{s} v^2} }$ & $ 2 $ \\ && \\
 $ \text{- 0 0 - } $ & ${  \frac{c_W^2 m_Z^2 \left(m_h^2-\frac{1}{3} \left(16 c_W^2+5\right) m_Z^2\right)}{16 \pi  s v^2} }$ & $ 1 $ \\ && \\
 $ \text{- 0 0 0 } $ & ${  -\frac{c_W m_Z \left(m_h^2+\left(4 c_W^2+1\right) m_Z^2\right)}{16 \sqrt{3} \pi  \sqrt{s} v^2} }$ & $ 2 $ \\ && \\
 $ \text{- 0 0 + } $ & ${  -\frac{c_W^2 m_Z^2 \left(3 m_h^2+\left(32 c_W^2+7\right) m_Z^2\right)}{48 \pi  s v^2} }$ & $ 1 $ \\ && \\
 $ \text{- 0 + - } $ & ${  -\frac{c_W^3 m_Z^3}{4 \sqrt{2} \pi  \sqrt{s} v^2} }$ & $ 2 $ \\ && \\
 $ \text{- 0 + 0 } $ & ${  -\frac{c_W^2 m_Z^2 \left(m_h^2+\frac{1}{3} \left(4 c_W^2-1\right) m_Z^2\right)}{16 \pi  s v^2} }$ & $ 1 $ \\ && \\
 $ \text{- + - + } $ & ${  -\frac{7 c_W^2 m_Z^2}{24 \pi  v^2} }$ & $ 2 $ \\ && \\
 $ \text{- + 0 0 } $ & ${  \frac{c_W^2 m_Z^2}{8 \sqrt{6} \pi  v^2} }$ & $ 2 $ \\ && \\
 $ \text{- + + - } $ & ${  \frac{c_W^2 m_Z^2}{8 \pi  v^2} }$ & $ 2 $ \\ && \\
 $ \text{0 0 0 0 } $ & ${  -\frac{2 m_h^2+m_Z^2}{16 \pi  v^2} }$ & $ 0 $ \\ \hline
\end{tabular}
\end{center}
\caption{The leading terms in the asymptotic limit $s\rightarrow\infty$ of the tree-level partial wave amplitudes (column 2.) for $W^+W^-\rightarrow W^+W^-$ process}
\label{tab:WWonshell12}
\end{table}

\begin{table}
\begin{center}
\begin{tabular}{c|c|c}
$\lambda_1\lambda_2\lambda_1'\lambda_2'$ & $\mathcal{T}^{(j=j_{min})}_{\lambda_1'\lambda_2';\lambda_1\lambda_2} $ & $j_{min} = $ \\ \hline && \\
\text{- - - - } & 0 & 0 \\
 $ \text{- - - 0 } $ & $ \frac{c_W^2 m_Z^3}{4 \sqrt{3} \pi  \sqrt{s} v^2} $ & $ 2 $ \\ && \\
 $ \text{- - - + } $ & $ -\frac{c_W^2 \left(8 c_W^4-4 c_W^2+1\right) m_Z^4}{4 \sqrt{6} \pi  s v^2} $ & $ 2 $ \\ && \\
 $ \text{- - 0 0 } $ & $ \frac{c_W^2 m_Z^2 \left(4 \left(c_W^2+1\right) m_Z^2-m_h^2\right)}{8 \pi  s v^2} $ & $ 0 $ \\ && \\
 $ \text{- - 0 + } $ & $ \frac{c_W^4 \left(8 c_W^2-3\right) m_Z^5}{4 \sqrt{3} \pi  s^{3/2} v^2} $ & $ 2 $ \\ && \\
 $ \text{- - + + } $ & $ -\frac{c_W^2 m_Z^4 \left(m_h^2+2 \left(8 c_W^4-4 c_W^2-1\right) m_Z^2\right)}{4 \pi  s^2 v^2} $ & $ 0 $ \\ && \\
 $ \text{- 0 - - } $ & $ \frac{c_W m_Z^3}{4 \sqrt{3} \pi  \sqrt{s} v^2} $ & $ 2 $ \\ && \\
 $ \text{- 0 - 0 } $ & $ -\frac{c_W m_Z^2}{16 \pi  v^2} $ & $ 1 $ \\ && \\
 $ \text{- 0 - + } $ & $ -\frac{c_W \left(8 c_W^4-4 c_W^2+1\right) m_Z^3}{6 \sqrt{2} \pi  \sqrt{s} v^2} $ & $ 2 $ \\ && \\
 $ \text{- 0 0 0 } $ & $ -\frac{c_W \left(2 c_W^2+1\right) m_Z^3}{8 \sqrt{3} \pi  \sqrt{s} v^2} $ & $ 2 $ \\ && \\
 $ \text{- 0 0 + } $ & $ -\frac{c_W \left(16 c_W^4-5 c_W^2+1\right) m_Z^4}{16 \pi  s v^2} $ & $ 1 $ \\ && \\
 $ \text{- 0 + + } $ & $ \frac{c_W^3 \left(8 c_W^2-3\right) m_Z^5}{4 \sqrt{3} \pi  s^{3/2} v^2} $ & $ 2 $ \\ && \\
 $ \text{- + - - } $ & $ -\frac{c_W^2 \left(4 c_W^2+1\right) m_Z^4}{4 \sqrt{6} \pi  s v^2} $ & $ 2 $ \\ && \\
 $ \text{- + - 0 } $ & $ -\frac{c_W^2 \left(4 c_W^2+1\right) m_Z^3}{6 \sqrt{2} \pi  \sqrt{s} v^2} $ & $ 2 $ \\ && \\
 $ \text{- + - + } $ & $ -\frac{c_W^4 m_Z^2}{6 \pi  v^2} $ & $ 2 $ \\ && \\
 $ \text{- + 0 0 } $ & $ \frac{c_W^2 m_Z^2}{8 \sqrt{6} \pi  v^2} $ & $ 2 $ \\ && \\
 $ \text{0 0 - - } $ & $ -\frac{m_Z^2 \left(m_h^2+2 \left(-4 c_W^4+c_W^2-1\right) m_Z^2\right)}{8 \pi  s v^2} $ & $ 0 $ \\ && \\
 $ \text{0 0 - 0 } $ & $ -\frac{\left(4 c_W^4-2 c_W^2+1\right) m_Z^3}{8 \sqrt{3} \pi  \sqrt{s} v^2} $ & $ 2 $ \\ && \\
 $ \text{0 0 - + } $ & $ \frac{\left(1-2 c_W^2\right){}^2 m_Z^2}{8 \sqrt{6} \pi  v^2} $ & $ 2 $ \\ && \\
 $ \text{0 0 0 0 } $ & $ -\frac{m_h^2+2 c_W^2 m_Z^2}{16 \pi  v^2} $ & $ 0 $ \\ \hline
\end{tabular}
\end{center}
\caption{The leading terms in the asymptotic limit $s\rightarrow\infty$ of the tree-level partial wave amplitudes (column 2.) for $W^+W^-\rightarrow ZZ$ process}
\label{tab:WWonshell13}
\end{table}

\begin{table}
\begin{center}
\bgroup
\def\arraystretch{1.7}
\begin{tabular}{c|c|c}
$\lambda_1\lambda_2\lambda_1'\lambda_2'$ & $\mathcal{T}^{(j=j_{min})}_{\lambda_1'\lambda_2';\lambda_1\lambda_2} $ & $j_{min} = $ \\ \hline && \\
 $ \text{- - - - } $ & $ 0 $ & $ 0 $ \\
 $ \text{- - - 0 } $ & $ -\frac{c_W^2 \left(c_W^2-1\right) m_Z^3}{4 \sqrt{3} \pi  \sqrt{s} v^2} $ & $ 2 $ \\
 $ \text{- - - + } $ & $ \frac{c_W^2 \left(c_W^4+2 c_W^2-1\right) m_Z^4}{4 \sqrt{6} \pi  s v^2} $ & $ 2 $ \\
 $ \text{- - 0 - } $ & $ -\frac{c_W \left(2 c_W^4-3 c_W^2+1\right) m_Z^3}{4 \sqrt{3} \pi  \sqrt{s} v^2} $ & $ 2 $ \\
 $ \text{- - 0 0 } $ & $ \frac{c_W m_Z^2 \left(m_h^2+\left(c_W^4-5 c_W^2+3\right) m_Z^2\right)}{8 \pi  s v^2} $ & $ 0 $ \\
 $ \text{- - 0 + } $ & $ \frac{c_W m_Z^3 \left(m_h^2+c_W^2 \left(-4 c_W^4-2 c_W^2+1\right) m_Z^2\right)}{8 \sqrt{3} \pi  s^{3/2} v^2} $ & $ 2 $ \\
 $ \text{- - + - } $ & $ \frac{c_W^2 \left(c_W^4+2 c_W^2-1\right) m_Z^4}{4 \sqrt{6} \pi  s v^2} $ & $ 2 $ \\
 $ \text{- - + 0 } $ & $ -\frac{c_W^2 m_Z^3 \left(m_h^2+c_W^2 \left(1-6 c_W^2\right) m_Z^2\right)}{8 \sqrt{3} \pi  s^{3/2} v^2} $ & $ 2 $ \\
 $ \text{- - + + } $ & $ -\frac{c_W^2 m_Z^4 \left(m_h^2-\left(3 c_W^6+4 c_W^4-4 c_W^2+2\right) m_Z^2\right)}{4 \pi  s^2 v^2} $ & $ 0 $ \\
 $ \text{- 0 - 0 } $ & $ -\frac{c_W^2 m_Z^2}{16 \pi  v^2} $ & $ 1 $ \\
 $ \text{- 0 - + } $ & $ \frac{c_W^2 \left(3 c_W^2-2\right) m_Z^3}{12 \sqrt{2} \pi  \sqrt{s} v^2} $ & $ 2 $ \\
 $ \text{- 0 0 - } $ & $ \frac{c_W \left(2 c_W^2-1\right) m_Z^2}{16 \pi  v^2} $ & $ 1 $ \\
 $ \text{- 0 0 0 } $ & $ -\frac{c_W m_Z \left(m_h^2+\left(3-4 c_W^2\right) m_Z^2\right)}{16 \sqrt{3} \pi  \sqrt{s} v^2} $ & $ 2 $ \\
 $ \text{- 0 0 + } $ & $ -\frac{c_W m_Z^2 \left(m_h^2+\left(-4 c_W^4+4 c_W^2+1\right) m_Z^2\right)}{16 \pi  s v^2} $ & $ 1 $ \\
 $ \text{- 0 + - } $ & $ \frac{c_W^2 \left(9 c_W^2-2\right) m_Z^3}{12 \sqrt{2} \pi  \sqrt{s} v^2} $ & $ 2 $ \\
 $ \text{- 0 + 0 } $ & $ -\frac{c_W^2 m_Z^2 \left(m_h^2+\left(2-13 c_W^2\right) m_Z^2\right)}{16 \pi  s v^2} $ & $ 1 $ \\
 $ \text{- + - + } $ & $ -\frac{c_W^4 m_Z^2}{8 \pi  v^2} $ & $ 2 $ \\
 $ \text{- + 0 - } $ & $ -\frac{c_W \left(6 c_W^4+3 c_W^2-2\right) m_Z^3}{12 \sqrt{2} \pi  \sqrt{s} v^2} $ & $ 2 $ \\
 $ \text{- + 0 0 } $ & $ -\frac{c_W \left(c_W^2-1\right) m_Z^2}{8 \sqrt{6} \pi  v^2} $ & $ 2 $ \\
 $ \text{- + 0 + } $ & $ \frac{c_W \left(2 c_W^4+c_W^2-2\right) m_Z^3}{12 \sqrt{2} \pi  \sqrt{s} v^2} $ & $ 2 $ \\
 $ \text{- + + - } $ & $ \frac{7 c_W^4 m_Z^2}{24 \pi  v^2} $ & $ 2 $ \\
 $ \text{0 - 0 - } $ & $ -\frac{\left(1-2 c_W^2\right){}^2 m_Z^2}{16 \pi  v^2} $ & $ 1 $ \\
 $ \text{0 - 0 0 } $ & $ \frac{m_Z \left(m_h^2+\left(1-2 c_W^2\right) m_Z^2\right)}{16 \sqrt{3} \pi  \sqrt{s} v^2} $ & $ 2 $ \\
 $ \text{0 - 0 + } $ & $ -\frac{m_Z^2 \left(m_h^2-\frac{1}{3} c_W^2 \left(28 c_W^4+4 c_W^2+1\right) m_Z^2\right)}{16 \pi  s v^2} $ & $ 1 $ \\
 $ \text{0 0 0 0 } $ & $ \frac{c_W^2 m_Z^2-m_h^2}{16 \pi  v^2} $ & $ 0 $\\ \hline
\end{tabular}
\egroup
\end{center}
\caption{The leading terms in the asymptotic limit $s\rightarrow\infty$ of the tree-level partial wave amplitudes (column 2.) for $W^+Z\rightarrow W^+Z$ process}
\label{tab:WWonshell14}
\end{table}
\clearpage
%%%%%%%%%%%%%%%%%%%%%%%%%%%%%%%%%%%%%%%%%%%%%%%%%%%%%%%%%%%%%%
\chapter{On-shell $WW$ scattering in the SMEFT: full results}
\label{app:VVonshellSMEFT}%tutaj
In Sec.~\ref{onshellSMEFT} certain aspects of the on-shell $W^+W^+$ scattering were presented for the SMEFT Lagrangian, based on $\mathcal{O}_{S0}$ and $\mathcal{O}_{T1}$ operators as examples. Here we present the results for all the EFT ''models''. See that Section for explicit reference to these results. 
\newgeometry{tmargin=2cm, bmargin=3cm, lmargin=1.5cm, rmargin=1.5cm}
\section{Unitarity limits: the numerical results}
\thispagestyle{empty} 
\label{app:unitLimitsSMEFT}

%%%%%%%%%%%%%%%%% unitairty on-shell WW scattering %%%%%%%%%%%%%%%%%%%%%%%%
\begin{table}[h]%
\begin{center}
%\resizebox{0.8\columnwidth}{!}{
\begin{tabular}{c|cccc||cccc}
 $ \lambda_1\lambda_2\lambda_1'\lambda_2' $ & $ 0.01 $ & $ 0.1 $ & $ 1. $ & $ 10. $ & $ -0.01 $ & $ -0.1 $ & $ -1. $ & $ -10. $ \\ \hline $
 \text{- - - -} $ & $ \text{x} $ & $ \text{x} $ & $ \text{x} $ & $ \text{x} $ & $ \text{x} $ & $ \text{x} $ & $ \text{x} $ & $ \text{x} $ \\ $
 \text{- - - 0} $ & $ \text{x} $ & $ \text{x} $ & $ \text{x} $ & $ \text{x} $ & $ \text{x} $ & $ \text{x} $ & $ \text{x} $ & $ \text{x} $ \\ $
 \text{- - - +} $ & $ \text{x} $ & $ \text{x} $ & $ \text{x} $ & $ \text{x} $ & $ \text{x} $ & $ \text{x} $ & $ \text{x} $ & $ \text{x} $ \\ $
 \text{- - 0 0} $ & $ 620. $ & $ 200. $ & $ 62. $ & $ 20. $ & $ 620. $ & $ 200. $ & $ 62. $ & $ 20. $ \\ $
 \text{- - 0 +} $ & $ \text{x} $ & $ \text{x} $ & $ \text{x} $ & $ \text{x} $ & $ \text{x} $ & $ \text{x} $ & $ \text{x} $ & $ \text{x} $ \\ $
 \text{- - + +} $ & $ \text{x} $ & $ \text{x} $ & $ \text{x} $ & $ \text{x} $ & $ \text{x} $ & $ \text{x} $ & $ \text{x} $ & $ \text{x} $ \\ $
 \text{- 0 - 0} $ & $ \text{x} $ & $ \text{x} $ & $ \text{x} $ & $ \text{x} $ & $ \text{x} $ & $ \text{x} $ & $ \text{x} $ & $ \text{x} $ \\ $
 \text{- 0 - +} $ & $ \text{x} $ & $ \text{x} $ & $ \text{x} $ & $ \text{x} $ & $ \text{x} $ & $ \text{x} $ & $ \text{x} $ & $ \text{x} $ \\ $
 \text{- 0 0 0} $ & $ \text{x} $ & $ \text{x} $ & $ \text{x} $ & $ \text{x} $ & $ \text{x} $ & $ \text{x} $ & $ \text{x} $ & $ \text{x} $ \\ $
 \text{- 0 0 +} $ & $ \text{x} $ & $ \text{x} $ & $ \text{x} $ & $ \text{x} $ & $ \text{x} $ & $ \text{x} $ & $ \text{x} $ & $ \text{x} $ \\ $
 \text{- + - +} $ & $ \text{x} $ & $ \text{x} $ & $ \text{x} $ & $ \text{x} $ & $ \text{x} $ & $ \text{x} $ & $ \text{x} $ & $ \text{x} $ \\ $
 \text{- + 0 0} $ & $ \text{x} $ & $ \text{x} $ & $ \text{x} $ & $ \text{x} $ & $ \text{x} $ & $ \text{x} $ & $ \text{x} $ & $ \text{x} $ \\ $
 \text{0 0 0 0} $ & $ 8.6 $ & $ 4.9 $ & $ 2.7 $ & $ 1.5 $ & $ 8.2 $ & $ 4.6 $ & $ 2.6 $ & $ 1.5 $ \\ \hline $
 \text{diag.} $ & $ 8.6 $ & $ 4.9 $ & $ 2.7 $ & $ 1.5 $ & $ 8.2 $ & $ 4.6 $ & $ 2.6 $ & $ 1.5 $ \\ \hline
\end{tabular}
%}
\end{center}
\caption{Values of $\sqrt{s^U}$ (in TeV) from the tree-level partial wave unitarity bounds for all elastic on-shell $W^+W^+$ helicity amplitudes for a chosen set of $f_{S0}$ values (first row, in TeV$^{−4}$
); $\lambda_i$ ($\lambda_i'$) denote ingoing (outgoing) W's helicities; ''$x$'' denotes no unitarity violation; ''diag.'' denotes unitarity bounds from diagonalization in the helicity space.
}
\label{tab:unitarityS0}
\end{table}

\begin{table}[h]%
\begin{center}
\resizebox{0.8\columnwidth}{!}{
\begin{tabular}{c|cccc||cccc}
 $ \lambda_1\lambda_2\lambda_1'\lambda_2' $ & $ 0.01 $ & $ 0.1 $ & $ 1. $ & $ 10. $ & $ -0.01 $ & $ -0.1 $ & $ -1. $ & $ -10. $ \\ \hline $
 \text{- - - - } $ & $ \text{x} $ & $ \text{x} $ & $ \text{x} $ & $ \text{x} $ & $ \text{x} $ & $ \text{x} $ & $ \text{x} $ & $ \text{x} $ \\ $
 \text{- - - 0 } $ & $ 4.2\times 10^7 $ & $ 4.2\times 10^6 $ & $ 4.2\times 10^5 $ & $ 4.2\times 10^4 $ & $ 4.2\times 10^7 $ & $ 4.2\times 10^6 $ & $ 4.2\times 10^5 $ & $ 4.2\times 10^4 $ \\ $
 \text{- - - + } $ & $ \text{x} $ & $ \text{x} $ & $ \text{x} $ & $ \text{x} $ & $ \text{x} $ & $ \text{x} $ & $ \text{x} $ & $ \text{x} $ \\ $
 \text{- - 0 0 } $ & $ 1.1\times 10^3 $ & $ 340. $ & $ 110. $ & $ 34. $ & $ 1.1\times 10^3 $ & $ 340. $ & $ 110. $ & $ 34. $ \\ $
 \text{- - 0 + } $ & $ 4.2\times 10^7 $ & $ 4.2\times 10^6 $ & $ 4.2\times 10^5 $ & $ 4.2\times 10^4 $ & $ 4.2\times 10^7 $ & $ 4.2\times 10^6 $ & $ 4.2\times 10^5 $ & $ 4.2\times 10^4 $ \\ $
 \text{- - + + } $ & $ \text{x} $ & $ \text{x} $ & $ \text{x} $ & $ \text{x} $ & $ \text{x} $ & $ \text{x} $ & $ \text{x} $ & $ \text{x} $ \\ $
 \text{- 0 - 0 } $ & $ 1.6\times 10^3 $ & $ 510. $ & $ 160. $ & $ 51. $ & $ 1.4\times 10^3 $ & $ 450. $ & $ 140. $ & $ 45. $ \\ $
 \text{- 0 - + } $ & $ 1.7\times 10^7 $ & $ 1.7\times 10^6 $ & $ 1.7\times 10^5 $ & $ 1.7\times 10^4 $ & $ 1.7\times 10^7 $ & $ 1.7\times 10^6 $ & $ 1.7\times 10^5 $ & $ 1.7\times 10^4 $ \\ $
 \text{- 0 0 - } $ & $ 1.5\times 10^3 $ & $ 480. $ & $ 150. $ & $ 48. $ & $ 1.5\times 10^3 $ & $ 480. $ & $ 150. $ & $ 48. $ \\ $
 \text{- 0 0 0 } $ & $ 82. $ & $ 38. $ & $ 18. $ & $ 8.2 $ & $ 82. $ & $ 38. $ & $ 18. $ & $ 8.2 $ \\ $
 \text{- 0 0 + } $ & $ 880. $ & $ 280. $ & $ 88. $ & $ 28. $ & $ 880. $ & $ 280. $ & $ 88. $ & $ 28. $ \\ $
 \text{- 0 + - } $ & $ 1.7\times 10^7 $ & $ 1.7\times 10^6 $ & $ 1.7\times 10^5 $ & $ 1.7\times 10^4 $ & $ 1.7\times 10^7 $ & $ 1.7\times 10^6 $ & $ 1.7\times 10^5 $ & $ 1.7\times 10^4 $ \\ $
 \text{- 0 + 0 } $ & $ 1.5\times 10^3 $ & $ 480. $ & $ 150. $ & $ 48. $ & $ 1.5\times 10^3 $ & $ 480. $ & $ 150. $ & $ 48. $ \\ $
 \text{- + - + } $ & $ \text{x} $ & $ \text{x} $ & $ \text{x} $ & $ \text{x} $ & $ \text{x} $ & $ \text{x} $ & $ \text{x} $ & $ \text{x} $ \\ $
 \text{- + 0 0 } $ & $ 1.5\times 10^3 $ & $ 490. $ & $ 150. $ & $ 49. $ & $ 1.5\times 10^3 $ & $ 490. $ & $ 150. $ & $ 49. $ \\ $
 \text{- + + - } $ & $ \text{x} $ & $ \text{x} $ & $ \text{x} $ & $ \text{x} $ & $ \text{x} $ & $ \text{x} $ & $ \text{x} $ & $ \text{x} $ \\ $
 \text{0 0 0 0 } $ & $ 9.1 $ & $ 5.1 $ & $ 2.9 $ & $ 1.6 $ & $ 9.5 $ & $ 5.3 $ & $ 3.0 $ & $ 1.7 $ \\ \hline $
 \text{diag.} $ & $ 9.1 $ & $ 5.1 $ & $ 2.9 $ & $ 1.6 $ & $ 9.5 $ & $ 5.3 $ & $ 3.0 $ & $ 1.7 $\\ \hline

\end{tabular}
}
\end{center}
\caption{See description of Tab.~\ref{tab:unitarityS0}; reaction $W^+W^-\rightarrow W^+W^-$.
%Values of $\sqrt{s^U}$ (in TeV) from the tree-level partial wave unitarity bounds for all elastic on-shell $W^+W^-$ helicity amplitudes in the reaction $W^+W^-\rightarrow W^+W^-$ for a chosen set of $f_{S0}$ values (first row, in TeV$^{−4}$
%); $\lambda_i$ ($\lambda_i'$) denote ingoing (outgoing) W's helicities; ''$x$'' denotes no unitarity violation; ''diag.'' denotes unitarity bounds from diagonalization in the helicity space.
}
\label{tab:unitarityS0ww}
\end{table}
\clearpage
%--------------------------------------------------------------------%
\begin{table}%
\begin{center}
\begin{tabular}{c|cccc||cccc}
 $\lambda_1\lambda_2\lambda_1'\lambda_2' $ & $ 0.01 $ & $ 0.1 $ & $ 1. $ & $ 10. $ & $ -0.01 $ & $ -0.1 $ & $ -1. $ & $ -10. $ \\ \hline $
 \text{- - - -} $ & $ \text{x} $ & $ \text{x} $ & $ \text{x} $ & $ \text{x} $ & $ \text{x} $ & $ \text{x} $ & $ \text{x} $ & $ \text{x} $ \\ $
 \text{- - - 0} $ & $ 5.9\times 10^7 $ & $ 5.9\times 10^6 $ & $ 5.9\times 10^5 $ & $ 5.9\times 10^4 $ & $ 5.9\times 10^7 $ & $ 5.9\times 10^6 $ & $ 5.9\times 10^5 $ & $ 5.9\times
   10^4 $ \\ $
 \text{- - - +} $ & $ \text{x} $ & $ \text{x} $ & $ \text{x} $ & $ \text{x} $ & $ \text{x} $ & $ \text{x} $ & $ \text{x} $ & $ \text{x} $ \\ $
 \text{- - 0 0} $ & $ 1.5\times 10^3 $ & $ 480. $ & $ 150. $ & $ 48. $ & $ 1.5\times 10^3 $ & $ 480. $ & $ 150. $ & $ 48. $ \\ $
 \text{- - 0 +} $ & $ 5.9\times 10^7 $ & $ 5.9\times 10^6 $ & $ 5.9\times 10^5 $ & $ 5.9\times 10^4 $ & $ 5.9\times 10^7 $ & $ 5.9\times 10^6 $ & $ 5.9\times 10^5 $ & $ 5.9\times
   10^4 $ \\ $
 \text{- - + +} $ & $ \text{x} $ & $ \text{x} $ & $ \text{x} $ & $ \text{x} $ & $ \text{x} $ & $ \text{x} $ & $ \text{x} $ & $ \text{x} $ \\ $
 \text{- 0 - 0} $ & $ 1.4\times 10^3 $ & $ 450. $ & $ 140. $ & $ 45. $ & $ 1.6\times 10^3 $ & $ 520. $ & $ 160. $ & $ 52. $ \\ $
 \text{- 0 - +} $ & $ 1.7\times 10^7 $ & $ 1.7\times 10^6 $ & $ 1.7\times 10^5 $ & $ 1.7\times 10^4 $ & $ 1.7\times 10^7 $ & $ 1.7\times 10^6 $ & $ 1.7\times 10^5 $ & $ 1.7\times
   10^4 $ \\ $
 \text{- 0 0 0} $ & $ 91. $ & $ 42. $ & $ 20. $ & $ 9.1 $ & $ 91. $ & $ 42. $ & $ 20. $ & $ 9.1 $ \\ $
 \text{- 0 0 +} $ & $ 1.5\times 10^3 $ & $ 480. $ & $ 150. $ & $ 48. $ & $ 1.5\times 10^3 $ & $ 480. $ & $ 150. $ & $ 48. $ \\ $
 \text{- + - +} $ & $ \text{x} $ & $ \text{x} $ & $ \text{x} $ & $ \text{x} $ & $ \text{x} $ & $ \text{x} $ & $ \text{x} $ & $ \text{x} $ \\ $
 \text{- + 0 0} $ & $ 1.8\times 10^3 $ & $ 580. $ & $ 180. $ & $ 58. $ & $ 1.8\times 10^3 $ & $ 580. $ & $ 180. $ & $ 58. $ \\ $
 \text{0 0 0 0} $ & $ 11. $ & $ 6.4 $ & $ 3.6 $ & $ 2.0 $ & $ 11. $ & $ 6.1 $ & $ 3.4 $ & $ 1.9 $ \\ \hline $
 \text{diag.} $ & $ 11. $ & $ 6.4 $ & $ 3.6 $ & $ 2.0 $ & $ 11. $ & $ 6.1 $ & $ 3.4 $ & $ 1.9$ \\ \hline

\end{tabular}
\end{center}
\caption{Values of $\sqrt{s^U}$ (in TeV) from the tree-level partial wave unitarity bounds for all elastic on-shell $W^+W^+$ helicity amplitudes for a chosen set of $f_{S1}$ values (first row, in TeV$^{−4}$
); $\lambda_i$ ($\lambda_i'$) denote ingoing (outgoing) W's helicities; ''$x$'' denotes no unitarity violation; ''diag.'' denotes unitarity bounds from diagonalization in the helicity space.
}
\label{tab:unitarityS1}
\end{table}

\begin{table}%
\begin{center}
\begin{tabular}{c|cccc||cccc}
 $\lambda_1\lambda_2\lambda_1'\lambda_2' $ & $ 0.01 $ & $ 0.1 $ & $ 1. $ & $ 10. $ & $ -0.01 $ & $ -0.1 $ & $ -1. $ & $ -10. $ \\ \hline $
 \text{- - - - } $ & $ \text{x} $ & $ \text{x} $ & $ \text{x} $ & $ \text{x} $ & $ \text{x} $ & $ \text{x} $ & $ \text{x} $ & $ \text{x} $ \\ $
 \text{- - - 0 } $ & $ 8.4\times 10^7 $ & $ 8.4\times 10^6 $ & $ 8.4\times 10^5 $ & $ 8.4\times 10^4 $ & $ 8.4\times 10^7 $ & $ 8.4\times 10^6 $ & $ 8.4\times 10^5 $ & $ 8.4\times 10^4 $ \\ $
 \text{- - - + } $ & $ \text{x} $ & $ \text{x} $ & $ \text{x} $ & $ \text{x} $ & $ \text{x} $ & $ \text{x} $ & $ \text{x} $ & $ \text{x} $ \\ $
 \text{- - 0 0 } $ & $ 580. $ & $ 180. $ & $ 58. $ & $ 18. $ & $ 580. $ & $ 180. $ & $ 58. $ & $ 18. $ \\ $
 \text{- - 0 + } $ & $ 8.4\times 10^7 $ & $ 8.4\times 10^6 $ & $ 8.4\times 10^5 $ & $ 8.4\times 10^4 $ & $ 8.4\times 10^7 $ & $ 8.4\times 10^6 $ & $ 8.4\times 10^5 $ & $ 8.4\times 10^4 $ \\ $
 \text{- - + + } $ & $ \text{x} $ & $ \text{x} $ & $ \text{x} $ & $ \text{x} $ & $ \text{x} $ & $ \text{x} $ & $ \text{x} $ & $ \text{x} $ \\ $
 \text{- 0 - 0 } $ & $ 2.3\times 10^3 $ & $ 730. $ & $ 230. $ & $ 73. $ & $ 2.0\times 10^3 $ & $ 640. $ & $ 200. $ & $ 64. $ \\ $
 \text{- 0 - + } $ & $ 3.4\times 10^7 $ & $ 3.4\times 10^6 $ & $ 3.4\times 10^5 $ & $ 3.4\times 10^4 $ & $ 3.4\times 10^7 $ & $ 3.4\times 10^6 $ & $ 3.4\times 10^5 $ & $ 3.4\times 10^4 $ \\ $
 \text{- 0 0 - } $ & $ 2.2\times 10^3 $ & $ 680. $ & $ 220. $ & $ 68. $ & $ 2.2\times 10^3 $ & $ 680. $ & $ 220. $ & $ 68. $ \\ $
 \text{- 0 0 0 } $ & $ 100. $ & $ 48. $ & $ 22. $ & $ 10. $ & $ 100. $ & $ 48. $ & $ 22. $ & $ 10. $ \\ $
 \text{- 0 0 + } $ & $ 2.2\times 10^3 $ & $ 680. $ & $ 220. $ & $ 68. $ & $ 2.2\times 10^3 $ & $ 680. $ & $ 220. $ & $ 68. $ \\ $
 \text{- 0 + - } $ & $ 3.4\times 10^7 $ & $ 3.4\times 10^6 $ & $ 3.4\times 10^5 $ & $ 3.4\times 10^4 $ & $ 3.4\times 10^7 $ & $ 3.4\times 10^6 $ & $ 3.4\times 10^5 $ & $ 3.4\times 10^4 $ \\ $
 \text{- 0 + 0 } $ & $ 1.2\times 10^3 $ & $ 390. $ & $ 120. $ & $ 39. $ & $ 1.2\times 10^3 $ & $ 390. $ & $ 120. $ & $ 39. $ \\ $
 \text{- + - + } $ & $ \text{x} $ & $ \text{x} $ & $ \text{x} $ & $ \text{x} $ & $ \text{x} $ & $ \text{x} $ & $ \text{x} $ & $ \text{x} $ \\ $
 \text{- + 0 0 } $ & $ 2.2\times 10^3 $ & $ 690. $ & $ 220. $ & $ 69. $ & $ 2.2\times 10^3 $ & $ 690. $ & $ 220. $ & $ 69. $ \\ $
 \text{- + + - } $ & $ \text{x} $ & $ \text{x} $ & $ \text{x} $ & $ \text{x} $ & $ \text{x} $ & $ \text{x} $ & $ \text{x} $ & $ \text{x} $ \\ $
 \text{0 0 0 0 } $ & $ 7.7 $ & $ 4.3 $ & $ 2.4 $ & $ 1.4 $ & $ 8.0 $ & $ 4.5 $ & $ 2.5 $ & $ 1.4 $ \\ \hline $
 \text{diag.} $ & $ 7.7 $ & $ 4.3 $ & $ 2.4 $ & $ 1.4 $ & $ 8.0 $ & $ 4.5 $ & $ 2.5 $ & $ 1.4 $\\ \hline
\end{tabular}
\end{center}
\caption{See description of Tab.~\ref{tab:unitarityS1}; reaction $W^+W^-\rightarrow W^+W^-$.
%Values of $\sqrt{s^U}$ (in TeV) from the tree-level partial wave unitarity bounds for all elastic on-shell $W^+W^-$ helicity amplitudes for a chosen set of $f_{S1}$ values (first row, in TeV$^{−4}$
%); $\lambda_i$ ($\lambda_i'$) denote ingoing (outgoing) W's helicities; ''$x$'' denotes no unitarity violation.
}
\label{tab:unitarityS1ww}
\end{table}

\clearpage

%--------------------------------------------------------------------%

\begin{table}%
\begin{center}
\begin{tabular}{c|cccc||cccc}
 $\lambda_1\lambda_2\lambda_1'\lambda_2' $ & $ 0.01 $ & $ 0.1 $ & $ 1. $ & $ 10. $ & $ -0.01 $ & $ -0.1 $ & $ -1. $ & $ -10. $ \\ \hline $
 \text{- - - -} $ & $ \text{x} $ & $ \text{x} $ & $ \text{x} $ & $ \text{x} $ & $ \text{x} $ & $ \text{x} $ & $ \text{x} $ & $ \text{x} $ \\ $
 \text{- - - 0} $ & $ 1.5\times 10^7 $ & $ 1.5\times 10^6 $ & $ 1.5\times 10^5 $ & $ 1.5\times 10^4 $ & $ 1.5\times 10^7 $ & $ 1.5\times 10^6 $ & $
   1.5\times 10^5 $ & $ 1.5\times 10^4 $ \\ $
 \text{- - - +} $ & $ 910. $ & $ 290. $ & $ 91. $ & $ 29. $ & $ 910. $ & $ 290. $ & $ 91. $ & $ 29. $ \\ $
 \text{- - 0 0} $ & $ 760. $ & $ 240. $ & $ 76. $ & $ 24. $ & $ 760. $ & $ 240. $ & $ 76. $ & $ 24. $ \\ $
 \text{- - 0 +} $ & $ 57. $ & $ 27. $ & $ 12. $ & $ 5.7 $ & $ 57. $ & $ 27. $ & $ 12. $ & $ 5.7 $ \\ $
 \text{- - + +} $ & $ 7.8 $ & $ 4.4 $ & $ 2.5 $ & $ 1.4 $ & $ 7.8 $ & $ 4.4 $ & $ 2.5 $ & $ 1.4 $ \\ $
 \text{- 0 - 0} $ & $ 990. $ & $ 310. $ & $ 99. $ & $ 31. $ & $ 1.1\times 10^3 $ & $ 360. $ & $ 110. $ & $ 36. $ \\ $
 \text{- 0 - +} $ & $ 48. $ & $ 22. $ & $ 10. $ & $ 4.8 $ & $ 48. $ & $ 22. $ & $ 10. $ & $ 4.8 $ \\ $
 \text{- 0 0 0} $ & $ 7.3\times 10^6 $ & $ 7.3\times 10^5 $ & $ 7.3\times 10^4 $ & $ 7.3\times 10^3 $ & $ 7.3\times 10^6 $ & $ 7.3\times 10^5 $ & $
   7.3\times 10^4 $ & $ 7.3\times 10^3 $ \\ $
 \text{- 0 0 +} $ & $ 1.1\times 10^3 $ & $ 340. $ & $ 110. $ & $ 34. $ & $ 1.1\times 10^3 $ & $ 340. $ & $ 110. $ & $ 34. $ \\ $
 \text{- + - +} $ & $ 9.4 $ & $ 5.2 $ & $ 2.9 $ & $ 1.6 $ & $ 8.3 $ & $ 4.7 $ & $ 2.7 $ & $ 1.5 $ \\ $
 \text{- + 0 0} $ & $ 910. $ & $ 290. $ & $ 91. $ & $ 29. $ & $ 910. $ & $ 290. $ & $ 91. $ & $ 29. $ \\ $
 \text{0 0 0 0} $ & $ \text{x} $ & $ \text{x} $ & $ \text{x} $ & $ \text{x} $ & $ \text{x} $ & $ \text{x} $ & $ \text{x} $ & $ \text{x} $ \\ \hline $
 \text{diag.} $ & $ 7.2 $ & $ 4.1 $ & $ 2.3 $ & $ 1.3 $ & $ 7.2 $ & $ 4.1 $ & $ 2.3 $ & $ 1.3 $\\ \hline

\end{tabular}
\end{center}
\caption{Values of $\sqrt{s^U}$ (in TeV) from the tree-level partial wave unitarity bounds for all elastic on-shell $W^+W^+$ helicity amplitudes for a chosen set of $f_{T0}$ values (first row, in TeV$^{−4}$
); $\lambda_i$ ($\lambda_i'$) denote ingoing (outgoing) W's helicities; ''$x$'' denotes no unitarity violation; ''diag.'' denotes unitarity bounds from diagonalization in the helicity space.
}
\label{tab:unitarityT0}
\end{table}

\begin{table}%
\begin{center}
\begin{tabular}{c|cccc||cccc}
 $\lambda_1\lambda_2\lambda_1'\lambda_2' $ & $ 0.01 $ & $ 0.1 $ & $ 1. $ & $ 10. $ & $ -0.01 $ & $ -0.1 $ & $ -1. $ & $ -10. $ \\ \hline $
  \text{- - - - } $ & $ 5.5 $ & $ 3.1 $ & $ 1.8 $ & $ 1.0 $ & $ 6.3 $ & $ 3.5 $ & $ 2.0 $ & $ 1.1 $ \\ \hline $
 \text{- - - 0 } $ & $ 2.1\times 10^7 $ & $ 2.1\times 10^6 $ & $ 2.1\times 10^5 $ & $ 2.1\times 10^4 $ & $ 2.1\times 10^7 $ & $ 2.1\times 10^6 $ & $ 2.1\times 10^5 $ & $ 2.1\times 10^4 $ \\ $
 \text{- - - + } $ & $ 1.1\times 10^3 $ & $ 340. $ & $ 110. $ & $ 34. $ & $ 1.1\times 10^3 $ & $ 340. $ & $ 110. $ & $ 34. $ \\ $
 \text{- - 0 0 } $ & $ 290. $ & $ 91. $ & $ 29. $ & $ 9.1 $ & $ 290. $ & $ 91. $ & $ 29. $ & $ 9.1 $ \\ $
 \text{- - 0 + } $ & $ 64. $ & $ 30. $ & $ 14. $ & $ 6.4 $ & $ 64. $ & $ 30. $ & $ 14. $ & $ 6.4 $ \\ $
 \text{- - + + } $ & $ 5.5 $ & $ 3.1 $ & $ 1.8 $ & $ 0.99 $ & $ 5.5 $ & $ 3.1 $ & $ 1.8 $ & $ 0.99 $ \\ $
 \text{- 0 - 0 } $ & $ \text{x} $ & $ \text{x} $ & $ \text{x} $ & $ \text{x} $ & $ \text{x} $ & $ \text{x} $ & $ \text{x} $ & $ \text{x} $ \\ $
 \text{- 0 - + } $ & $ 8.4\times 10^6 $ & $ 8.4\times 10^5 $ & $ 8.4\times 10^4 $ & $ 8.4\times 10^3 $ & $ 8.4\times 10^6 $ & $ 8.4\times 10^5 $ & $ 8.4\times 10^4 $ & $ 8.4\times 10^3 $ \\ $
 \text{- 0 0 - } $ & $ 1.1\times 10^3 $ & $ 340. $ & $ 110. $ & $ 34. $ & $ 1.1\times 10^3 $ & $ 340. $ & $ 110. $ & $ 34. $ \\ $
 \text{- 0 0 0 } $ & $ 1.0\times 10^7 $ & $ 1.0\times 10^6 $ & $ 1.0\times 10^5 $ & $ 1.0\times 10^4 $ & $ 1.0\times 10^7 $ & $ 1.0\times 10^6 $ & $ 1.0\times 10^5 $ & $ 1.0\times 10^4 $ \\ $
 \text{- 0 0 + } $ & $ 1.1\times 10^3 $ & $ 340. $ & $ 110. $ & $ 34. $ & $ 1.1\times 10^3 $ & $ 340. $ & $ 110. $ & $ 34. $ \\ $
 \text{- 0 + - } $ & $ 48. $ & $ 22. $ & $ 10. $ & $ 4.8 $ & $ 48. $ & $ 22. $ & $ 10. $ & $ 4.8 $ \\ $
 \text{- 0 + 0 } $ & $ 760. $ & $ 240. $ & $ 76. $ & $ 24. $ & $ 760. $ & $ 240. $ & $ 76. $ & $ 24. $ \\ $
 \text{- + - + } $ & $ \text{x} $ & $ \text{x} $ & $ \text{x} $ & $ \text{x} $ & $ \text{x} $ & $ \text{x} $ & $ \text{x} $ & $ \text{x} $ \\ $
 \text{- + 0 0 } $ & $ 1.1\times 10^3 $ & $ 340. $ & $ 110. $ & $ 34. $ & $ 1.1\times 10^3 $ & $ 340. $ & $ 110. $ & $ 34. $ \\ $
 \text{- + + - } $ & $ 8.9 $ & $ 5.0 $ & $ 2.8 $ & $ 1.6 $ & $ 8.9 $ & $ 5.0 $ & $ 2.8 $ & $ 1.6 $ \\ $
 \text{0 0 0 0 } $ & $ \text{x} $ & $ \text{x} $ & $ \text{x} $ & $ \text{x} $ & $ \text{x} $ & $ \text{x} $ & $ \text{x} $ & $ \text{x} $ \\ \hline $
 \text{diag.} $ & $ 4.4 $ & $ 2.5 $ & $ 1.4 $ & $ 0.82 $ & $ 5.1 $ & $ 2.9 $ & $ 1.6 $ & $ 0.90 $\\ \hline

\end{tabular}
\end{center}
\caption{See description of Tab.~\ref{tab:unitarityT0}; reaction $W^+W^-\rightarrow W^+W^-$.
%Values of $\sqrt{s^U}$ (in TeV) from the tree-level partial wave unitarity bounds for all elastic on-shell $W^+W^-$ helicity amplitudes for a chosen set of $f_{T0}$ values (first row, in TeV$^{−4}$
%); $\lambda_i$ ($\lambda_i'$) denote ingoing (outgoing) W's helicities; ''$x$'' denotes no unitarity violation.
}
\label{tab:unitarityT0ww}
\end{table}
\clearpage
%--------------------------------------------------------------------%

\begin{table}%
\begin{center}
\begin{tabular}{c|cccc||cccc}
 $\lambda_1\lambda_2\lambda_1'\lambda_2' $ & $ 0.01 $ & $ 0.1 $ & $ 1. $ & $ 10. $ & $ -0.01 $ & $ -0.1 $ & $ -1. $ & $ -10. $ \\ \hline $
 \text{- - - -} $ & $ 7.5 $ & $ 4.2 $ & $ 2.4 $ & $ 1.3 $ & $ 6.5 $ & $ 3.7 $ & $ 2.1 $ & $ 1.2 $ \\ $
 \text{- - - 0} $ & $ 2.9\times 10^7 $ & $ 2.9\times 10^6 $ & $ 2.9\times 10^5 $ & $ 2.9\times 10^4 $ & $ 2.9\times 10^7 $ & $ 2.9\times 10^6 $ & $ 2.9\times 10^5 $ & $ 2.9\times 10^4 $ \\ $
 \text{- - - +} $ & $ 1.3\times 10^3 $ & $ 410. $ & $ 130. $ & $ 41. $ & $ 1.3\times 10^3 $ & $ 410. $ & $ 130. $ & $ 41. $ \\ $
 \text{- - 0 0} $ & $ 410. $ & $ 130. $ & $ 41. $ & $ 13. $ & $ 410. $ & $ 130. $ & $ 41. $ & $ 13. $ \\ $
 \text{- - 0 +} $ & $ 72. $ & $ 34. $ & $ 16. $ & $ 7.2 $ & $ 72. $ & $ 34. $ & $ 16. $ & $ 7.2 $ \\ $
 \text{- - + +} $ & $ 6.6 $ & $ 3.7 $ & $ 2.1 $ & $ 1.2 $ & $ 6.6 $ & $ 3.7 $ & $ 2.1 $ & $ 1.2 $ \\ $
 \text{- 0 - 0} $ & $ 1.4\times 10^3 $ & $ 440. $ & $ 140. $ & $ 44. $ & $ 1.6\times 10^3 $ & $ 510. $ & $ 160. $ & $ 51. $ \\ $
 \text{- 0 - +} $ & $ 60. $ & $ 28. $ & $ 13. $ & $ 6.0 $ & $ 60. $ & $ 28. $ & $ 13. $ & $ 6.0 $ \\ $
 \text{- 0 0 0} $ & $ 1.5\times 10^7 $ & $ 1.5\times 10^6 $ & $ 1.5\times 10^5 $ & $ 1.5\times 10^4 $ & $ 1.5\times 10^7 $ & $ 1.5\times 10^6 $ & $ 1.5\times 10^5 $ & $ 1.5\times 10^4 $ \\ $
 \text{- 0 0 +} $ & $ 1.5\times 10^3 $ & $ 480. $ & $ 150. $ & $ 48. $ & $ 1.5\times 10^3 $ & $ 480. $ & $ 150. $ & $ 48. $ \\ $
 \text{- + - +} $ & $ 11. $ & $ 6.2 $ & $ 3.5 $ & $ 2.0 $ & $ 9.9 $ & $ 5.6 $ & $ 3.2 $ & $ 1.8 $ \\ $
 \text{- + 0 0} $ & $ 1.3\times 10^3 $ & $ 410. $ & $ 130. $ & $ 41. $ & $ 1.3\times 10^3 $ & $ 410. $ & $ 130. $ & $ 41. $ \\ $
 \text{0 0 0 0} $ & $ \text{x} $ & $ \text{x} $ & $ \text{x} $ & $ \text{x} $ & $ \text{x} $ & $ \text{x} $ & $ \text{x} $ & $ \text{x} $ \\ \hline$
 \text{diag.} $ & $ 6.1 $ & $ 3.4 $ & $ 1.9 $ & $ 1.1 $ & $ 5.3 $ & $ 3.0 $ & $ 1.7 $ & $ 0.97$ \\ \hline

\end{tabular}
\end{center}
\caption{Values of $\sqrt{s^U}$ (in TeV) from the tree-level partial wave unitarity bounds for all elastic on-shell $W^+W^+$ helicity amplitudes for a chosen set of $f_{T1}$ values (first row, in TeV$^{−4}$
); $\lambda_i$ ($\lambda_i'$) denote ingoing (outgoing) W's helicities; ''$x$'' denotes no unitarity violation; ''diag.'' denotes unitarity bounds from diagonalization in the helicity space.
}
\label{tab:unitarityT1}
\end{table}

\begin{table}%
\begin{center}
\begin{tabular}{c|cccc||cccc}
 $\lambda_1\lambda_2\lambda_1'\lambda_2' $ & $ 0.01 $ & $ 0.1 $ & $ 1. $ & $ 10. $ & $ -0.01 $ & $ -0.1 $ & $ -1. $ & $ -10. $ \\ \hline $
 \text{- - - - } $ & $ 6.5 $ & $ 3.7 $ & $ 2.1 $ & $ 1.2 $ & $ 7.5 $ & $ 4.2 $ & $ 2.4 $ & $ 1.3 $ \\ $
 \text{- - - 0 } $ & $ 1.4\times 10^7 $ & $ 1.4\times 10^6 $ & $ 1.4\times 10^5 $ & $ 1.4\times 10^4 $ & $ 1.4\times 10^7 $ & $ 1.4\times 10^6 $ & $ 1.4\times 10^5 $ & $ 1.4\times 10^4 $ \\ $
 \text{- - - + } $ & $ 890. $ & $ 280. $ & $ 89. $ & $ 28. $ & $ 890. $ & $ 280. $ & $ 89. $ & $ 28. $ \\ $
 \text{- - 0 0 } $ & $ 360. $ & $ 110. $ & $ 36. $ & $ 11. $ & $ 360. $ & $ 110. $ & $ 36. $ & $ 11. $ \\ $
 \text{- - 0 + } $ & $ 56. $ & $ 26. $ & $ 12. $ & $ 5.6 $ & $ 56. $ & $ 26. $ & $ 12. $ & $ 5.6 $ \\ $
 \text{- - + + } $ & $ 5.9 $ & $ 3.3 $ & $ 1.9 $ & $ 1.1 $ & $ 5.9 $ & $ 3.3 $ & $ 1.9 $ & $ 1.1 $ \\ $
 \text{- 0 - 0 } $ & $ 1.1\times 10^3 $ & $ 360. $ & $ 110. $ & $ 36. $ & $ 1000. $ & $ 320. $ & $ 100. $ & $ 32. $ \\ $
 \text{- 0 - + } $ & $ 48. $ & $ 22. $ & $ 10. $ & $ 4.8 $ & $ 48. $ & $ 22. $ & $ 10. $ & $ 4.8 $ \\ $
 \text{- 0 0 - } $ & $ 1.5\times 10^3 $ & $ 480. $ & $ 150. $ & $ 48. $ & $ 1.5\times 10^3 $ & $ 480. $ & $ 150. $ & $ 48. $ \\ $
 \text{- 0 0 0 } $ & $ 6.9\times 10^6 $ & $ 6.9\times 10^5 $ & $ 6.9\times 10^4 $ & $ 6.9\times 10^3 $ & $ 6.9\times 10^6 $ & $ 6.9\times 10^5 $ & $ 6.9\times 10^4 $ & $ 6.9\times 10^3 $ \\ $
 \text{- 0 0 + } $ & $ 880. $ & $ 280. $ & $ 88. $ & $ 28. $ & $ 880. $ & $ 280. $ & $ 88. $ & $ 28. $ \\ $
 \text{- 0 + - } $ & $ 60. $ & $ 28. $ & $ 13. $ & $ 6.0 $ & $ 60. $ & $ 28. $ & $ 13. $ & $ 6.0 $ \\ $
 \text{- 0 + 0 } $ & $ \text{x} $ & $ \text{x} $ & $ \text{x} $ & $ \text{x} $ & $ \text{x} $ & $ \text{x} $ & $ \text{x} $ & $ \text{x} $ \\ $
 \text{- + - + } $ & $ 8.3 $ & $ 4.7 $ & $ 2.7 $ & $ 1.5 $ & $ 9.3 $ & $ 5.2 $ & $ 2.9 $ & $ 1.6 $ \\ $
 \text{- + 0 0 } $ & $ 880. $ & $ 280. $ & $ 88. $ & $ 28. $ & $ 890. $ & $ 280. $ & $ 89. $ & $ 28. $ \\ $
 \text{- + + - } $ & $ 11. $ & $ 5.9 $ & $ 3.3 $ & $ 1.9 $ & $ 11. $ & $ 5.9 $ & $ 3.3 $ & $ 1.9 $ \\ $
 \text{0 0 0 0 } $ & $ \text{x} $ & $ \text{x} $ & $ \text{x} $ & $ \text{x} $ & $ \text{x} $ & $ \text{x} $ & $ \text{x} $ & $ \text{x} $ \\ \hline $
 \text{diag.} $ & $ 4.9 $ & $ 2.8 $ & $ 1.6 $ & $ 0.91 $ & $ 5.7 $ & $ 3.2 $ & $ 1.8 $ & $ 1.0 $\\ \hline 

\end{tabular}
\end{center}
\caption{See description of Tab.~\ref{tab:unitarityT1}; reaction $W^+W^-\rightarrow W^+W^-$.
%Values of $\sqrt{s^U}$ (in TeV) from the tree-level partial wave unitarity bounds for all elastic on-shell $W^+W^-$ helicity amplitudes for a chosen set of $f_{T1}$ values (first row, in TeV$^{−4}$
%); $\lambda_i$ ($\lambda_i'$) denote ingoing (outgoing) W's helicities; ''$x$'' denotes no unitarity violation.
}
\label{tab:unitarityT1ww}
\end{table}
\clearpage
%--------------------------------------------------------------------%

\begin{table}%
\begin{center}
\begin{tabular}{c|cccc||cccc}
 $\lambda_1\lambda_2\lambda_1'\lambda_2' $ & $ 0.01 $ & $ 0.1 $ & $ 1. $ & $ 10. $ & $ -0.01 $ & $ -0.1 $ & $ -1. $ & $ -10. $ \\ \hline $
 \text{- - - -} $ & $ 9.0 $ & $ 5.0 $ & $ 2.8 $ & $ 1.6 $ & $ 7.7 $ & $ 4.4 $ & $ 2.5 $ & $ 1.4 $ \\ $
 \text{- - - 0} $ & $ 3.9\times 10^7 $ & $ 3.9\times 10^6 $ & $ 3.9\times 10^5 $ & $ 3.9\times 10^4 $ & $ 3.9\times 10^7 $ & $ 3.9\times 10^6 $ & $
   3.9\times 10^5 $ & $ 3.9\times 10^4 $ \\ $
 \text{- - - +} $ & $ 1.5\times 10^3 $ & $ 470. $ & $ 150. $ & $ 47. $ & $ 1.5\times 10^3 $ & $ 470. $ & $ 150. $ & $ 47. $ \\ $
 \text{- - 0 0} $ & $ 760. $ & $ 240. $ & $ 76. $ & $ 24. $ & $ 760. $ & $ 240. $ & $ 76. $ & $ 24. $ \\ $
 \text{- - 0 +} $ & $ 91. $ & $ 42. $ & $ 20. $ & $ 9.1 $ & $ 91. $ & $ 42. $ & $ 20. $ & $ 9.1 $ \\ $
 \text{- - + +} $ & $ 11. $ & $ 6.2 $ & $ 3.5 $ & $ 2.0 $ & $ 11. $ & $ 6.2 $ & $ 3.5 $ & $ 2.0 $ \\ $
 \text{- 0 - 0} $ & $ 1.4\times 10^3 $ & $ 440. $ & $ 140. $ & $ 44. $ & $ 1.6\times 10^3 $ & $ 510. $ & $ 160. $ & $ 51. $ \\ $
 \text{- 0 - +} $ & $ 60. $ & $ 28. $ & $ 13. $ & $ 6.0 $ & $ 60. $ & $ 28. $ & $ 13. $ & $ 6.0 $ \\ $
 \text{- 0 0 0} $ & $ 2.0\times 10^7 $ & $ 2.0\times 10^6 $ & $ 2.0\times 10^5 $ & $ 2.0\times 10^4 $ & $ 2.0\times 10^7 $ & $ 2.0\times 10^6 $ & $
   2.0\times 10^5 $ & $ 2.0\times 10^4 $ \\ $
 \text{- 0 0 +} $ & $ 1.5\times 10^3 $ & $ 480. $ & $ 150. $ & $ 48. $ & $ 1.5\times 10^3 $ & $ 480. $ & $ 150. $ & $ 48. $ \\ $
 \text{- + - +} $ & $ 11. $ & $ 6.2 $ & $ 3.5 $ & $ 2.0 $ & $ 9.9 $ & $ 5.6 $ & $ 3.2 $ & $ 1.8 $ \\ $
 \text{- + 0 0} $ & $ 1.3\times 10^3 $ & $ 410. $ & $ 130. $ & $ 41. $ & $ 1.3\times 10^3 $ & $ 410. $ & $ 130. $ & $ 41. $ \\ $
 \text{0 0 0 0} $ & $ \text{x} $ & $ \text{x} $ & $ \text{x} $ & $ \text{x} $ & $ \text{x} $ & $ \text{x} $ & $ \text{x} $ & $ \text{x} $ \\ \hline $
 \text{diag.} $ & $ 8.3 $ & $ 4.7 $ & $ 2.6 $ & $ 1.5 $ & $ 7.2 $ & $ 4.1 $ & $ 2.3 $ & $ 1.3$ \\ \hline

\end{tabular}
\end{center}
\caption{Values of $\sqrt{s^U}$ (in TeV) from the tree-level partial wave unitarity bounds for all elastic on-shell $W^+W^+$ helicity amplitudes for a chosen set of $f_{T2}$ values (first row, in TeV$^{−4}$
); $\lambda_i$ ($\lambda_i'$) denote ingoing (outgoing) W's helicities; ''$x$'' denotes no unitarity violation; ''diag.'' denotes unitarity bounds from diagonalization in the helicity space.
}
\label{tab:unitarityT2}
\end{table}

\begin{table}%
\begin{center}
\begin{tabular}{c|cccc||cccc}
 $\lambda_1\lambda_2\lambda_1'\lambda_2' $ & $ 0.01 $ & $ 0.1 $ & $ 1. $ & $ 10. $ & $ -0.01 $ & $ -0.1 $ & $ -1. $ & $ -10. $ \\ \hline $
  \text{- - - - } $ & $ 6.5 $ & $ 3.7 $ & $ 2.1 $ & $ 1.2 $ & $ 7.5 $ & $ 4.2 $ & $ 2.4 $ & $ 1.3 $ \\ $
 \text{- - - 0 } $ & $ 3.3\times 10^7 $ & $ 3.3\times 10^6 $ & $ 3.3\times 10^5 $ & $ 3.3\times 10^4 $ & $ 3.3\times 10^7 $ & $ 3.3\times 10^6 $ & $ 3.3\times 10^5 $ & $ 3.3\times 10^4 $ \\ $
 \text{- - - + } $ & $ 1.4\times 10^3 $ & $ 430. $ & $ 140. $ & $ 43. $ & $ 1.4\times 10^3 $ & $ 430. $ & $ 140. $ & $ 43. $ \\ $
 \text{- - 0 0 } $ & $ 480. $ & $ 150. $ & $ 48. $ & $ 15. $ & $ 480. $ & $ 150. $ & $ 48. $ & $ 15. $ \\ $
 \text{- - 0 + } $ & $ 100. $ & $ 47. $ & $ 22. $ & $ 10. $ & $ 100. $ & $ 47. $ & $ 22. $ & $ 10. $ \\ $
 \text{- - + + } $ & $ 7.8 $ & $ 4.4 $ & $ 2.5 $ & $ 1.4 $ & $ 7.8 $ & $ 4.4 $ & $ 2.5 $ & $ 1.4 $ \\ $
 \text{- 0 - 0 } $ & $ 1.6\times 10^3 $ & $ 510. $ & $ 160. $ & $ 51. $ & $ 1.4\times 10^3 $ & $ 450. $ & $ 140. $ & $ 45. $ \\ $
 \text{- 0 - + } $ & $ 60. $ & $ 28. $ & $ 13. $ & $ 6.0 $ & $ 60. $ & $ 28. $ & $ 13. $ & $ 6.0 $ \\ $
 \text{- 0 0 - } $ & $ 1.5\times 10^3 $ & $ 480. $ & $ 150. $ & $ 48. $ & $ 1.5\times 10^3 $ & $ 480. $ & $ 150. $ & $ 48. $ \\ $
 \text{- 0 0 0 } $ & $ 1.7\times 10^7 $ & $ 1.7\times 10^6 $ & $ 1.7\times 10^5 $ & $ 1.7\times 10^4 $ & $ 1.7\times 10^7 $ & $ 1.7\times 10^6 $ & $ 1.7\times 10^5 $ & $ 1.7\times 10^4 $ \\ $
 \text{- 0 0 + } $ & $ 2.1\times 10^3 $ & $ 680. $ & $ 210. $ & $ 68. $ & $ 2.1\times 10^3 $ & $ 680. $ & $ 210. $ & $ 68. $ \\ $
 \text{- 0 + - } $ & $ 60. $ & $ 28. $ & $ 13. $ & $ 6.0 $ & $ 60. $ & $ 28. $ & $ 13. $ & $ 6.0 $ \\ $
 \text{- 0 + 0 } $ & $ 1.2\times 10^3 $ & $ 390. $ & $ 120. $ & $ 39. $ & $ 1.2\times 10^3 $ & $ 390. $ & $ 120. $ & $ 39. $ \\ $
 \text{- + - + } $ & $ 9.9 $ & $ 5.6 $ & $ 3.2 $ & $ 1.8 $ & $ 11. $ & $ 6.2 $ & $ 3.5 $ & $ 2.0 $ \\ $
 \text{- + 0 0 } $ & $ 1.1\times 10^3 $ & $ 340. $ & $ 110. $ & $ 34. $ & $ 1.1\times 10^3 $ & $ 340. $ & $ 110. $ & $ 34. $ \\ $
 \text{- + + - } $ & $ 11. $ & $ 5.9 $ & $ 3.3 $ & $ 1.9 $ & $ 11. $ & $ 5.9 $ & $ 3.3 $ & $ 1.9 $ \\ $
 \text{0 0 0 0 } $ & $ \text{x} $ & $ \text{x} $ & $ \text{x} $ & $ \text{x} $ & $ \text{x} $ & $ \text{x} $ & $ \text{x} $ & $ \text{x} $ \\ \hline$
 \text{diag.} $ & $ 5.7 $ & $ 3.2 $ & $ 1.8 $ & $ 1.1 $ & $ 6.6 $ & $ 3.7 $ & $ 2.1 $ & $ 1.2 $\\ \hline

\end{tabular}
\end{center}
\caption{See description of Tab.~\ref{tab:unitarityT2}; reaction $W^+W^-\rightarrow W^+W^-$.
%Values of $\sqrt{s^U}$ (in TeV) from the tree-level partial wave unitarity bounds for all elastic on-shell $W^+W^-$ helicity amplitudes for a chosen set of $f_{T2}$ values (first row, in TeV$^{−4}$
%); $\lambda_i$ ($\lambda_i'$) denote ingoing (outgoing) W's helicities; ''$x$'' denotes no unitarity violation.
}
\label{tab:unitarityT2ww}
\end{table}

\clearpage
%--------------------------------------------------------------------%

\begin{table}%
\begin{center}
\begin{tabular}{c|cccc||cccc}
 $\lambda_1\lambda_2\lambda_1'\lambda_2' $ & $ 0.01 $ & $ 0.1 $ & $ 1. $ & $ 10. $ & $ -0.01 $ & $ -0.1 $ & $ -1. $ & $ -10. $ \\ \hline$
 \text{- - - -} $ & $ \text{x} $ & $ \text{x} $ & $ \text{x} $ & $ \text{x} $ & $ \text{x} $ & $ \text{x} $ & $ \text{x} $ & $ \text{x} $ \\ $
 \text{- - - 0} $ & $ 2.9\times 10^7 $ & $ 2.9\times 10^6 $ & $ 2.9\times 10^5 $ & $ 2.9\times 10^4 $ & $ 2.9\times 10^7 $ & $ 2.9\times 10^6 $ & $ 2.9\times 10^5 $ & $ 2.9\times 10^4 $ \\ $
 \text{- - - +} $ & $ 1.8\times 10^3 $ & $ 580. $ & $ 180. $ & $ 58. $ & $ 1.8\times 10^3 $ & $ 580. $ & $ 180. $ & $ 58. $ \\ $
 \text{- - 0 0} $ & $ 1.1\times 10^3 $ & $ 340. $ & $ 110. $ & $ 34. $ & $ 1.1\times 10^3 $ & $ 340. $ & $ 110. $ & $ 34. $ \\ $
 \text{- - 0 +} $ & $ 91. $ & $ 42. $ & $ 20. $ & $ 9.1 $ & $ 91. $ & $ 42. $ & $ 20. $ & $ 9.1 $ \\ $
 \text{- - + +} $ & $ 760. $ & $ 240. $ & $ 76. $ & $ 24. $ & $ 760. $ & $ 240. $ & $ 76. $ & $ 24. $ \\ $
 \text{- 0 - 0} $ & $ 1.2\times 10^3 $ & $ 360. $ & $ 120. $ & $ 36. $ & $ 1.3\times 10^3 $ & $ 420. $ & $ 130. $ & $ 42. $ \\ $
 \text{- 0 - +} $ & $ 76. $ & $ 35. $ & $ 16. $ & $ 7.6 $ & $ 76. $ & $ 35. $ & $ 16. $ & $ 7.6 $ \\ $
 \text{- 0 0 0} $ & $ 91. $ & $ 42. $ & $ 20. $ & $ 9.1 $ & $ 91. $ & $ 42. $ & $ 20. $ & $ 9.1 $ \\ $
 \text{- 0 0 +} $ & $ 12. $ & $ 6.7 $ & $ 3.8 $ & $ 2.1 $ & $ 12. $ & $ 6.7 $ & $ 3.8 $ & $ 2.1 $ \\ $
 \text{- + - +} $ & $ 1.1\times 10^3 $ & $ 350. $ & $ 110. $ & $ 35. $ & $ 850. $ & $ 270. $ & $ 85. $ & $ 27. $ \\ $
 \text{- + 0 0} $ & $ 1.3\times 10^3 $ & $ 410. $ & $ 130. $ & $ 41. $ & $ 1.3\times 10^3 $ & $ 410. $ & $ 130. $ & $ 41. $ \\ $
 \text{0 0 0 0} $ & $ 1.1\times 10^3 $ & $ 360. $ & $ 110. $ & $ 36. $ & $ 1.0\times 10^3 $ & $ 320. $ & $ 100. $ & $ 32. $ \\\hline $
 \text{diag.} $ & $ 630. $ & $ 200. $ & $ 63. $ & $ 20. $ & $ 530. $ & $ 170. $ & $ 53. $ & $ 17. $ \\ \hline

\end{tabular}
\end{center}
\caption{Values of $\sqrt{s^U}$ (in TeV) from the tree-level partial wave unitarity bounds for all elastic on-shell $W^+W^+$ helicity amplitudes for a chosen set of $f_{M0}$ values (first row, in TeV$^{−4}$
); $\lambda_i$ ($\lambda_i'$) denote ingoing (outgoing) W's helicities; ''$x$'' denotes no unitarity violation; ''diag.'' denotes unitarity bounds from diagonalization in the helicity space.
}
\label{tab:unitarityM0}
\end{table}

\begin{table}%
\begin{center}
\begin{tabular}{c|cccc||cccc}
 $\lambda_1\lambda_2\lambda_1'\lambda_2' $ & $ 0.01 $ & $ 0.1 $ & $ 1. $ & $ 10. $ & $ -0.01 $ & $ -0.1 $ & $ -1. $ & $ -10. $ \\ \hline$
 \text{- - - - } $ & $ 510. $ & $ 160. $ & $ 51. $ & $ 16. $ & $ 360. $ & $ 110. $ & $ 36. $ & $ 12. $ \\ $
 \text{- - - 0 } $ & $ 4.2\times 10^7 $ & $ 4.2\times 10^6 $ & $ 4.2\times 10^5 $ & $ 4.2\times 10^4 $ & $ 4.2\times 10^7 $ & $ 4.2\times 10^6 $ & $ 4.2\times 10^5 $ & $ 4.2\times 10^4 $ \\ $
 \text{- - - + } $ & $ 2.2\times 10^3 $ & $ 690. $ & $ 220. $ & $ 69. $ & $ 2.2\times 10^3 $ & $ 690. $ & $ 220. $ & $ 69. $ \\ $
 \text{- - 0 0 } $ & $ 8.4 $ & $ 4.7 $ & $ 2.7 $ & $ 1.5 $ & $ 8.4 $ & $ 4.7 $ & $ 2.7 $ & $ 1.5 $ \\ $
 \text{- - 0 + } $ & $ 100. $ & $ 48. $ & $ 22. $ & $ 10. $ & $ 100. $ & $ 48. $ & $ 22. $ & $ 10. $ \\ $
 \text{- - + + } $ & $ 540. $ & $ 170. $ & $ 54. $ & $ 17. $ & $ 540. $ & $ 170. $ & $ 54. $ & $ 17. $ \\ $
 \text{- 0 - 0 } $ & $ 2.3\times 10^3 $ & $ 730. $ & $ 230. $ & $ 73. $ & $ 2.0\times 10^3 $ & $ 630. $ & $ 200. $ & $ 63. $ \\ $
 \text{- 0 - + } $ & $ 1.7\times 10^7 $ & $ 1.7\times 10^6 $ & $ 1.7\times 10^5 $ & $ 1.7\times 10^4 $ & $ 1.7\times 10^7 $ & $ 1.7\times 10^6 $ & $ 1.7\times 10^5 $ & $ 1.7\times 10^4 $ \\ $
 \text{- 0 0 - } $ & $ 1.5\times 10^3 $ & $ 480. $ & $ 150. $ & $ 48. $ & $ 1.5\times 10^3 $ & $ 480. $ & $ 150. $ & $ 48. $ \\ $
 \text{- 0 0 0 } $ & $ 100. $ & $ 48. $ & $ 22. $ & $ 10. $ & $ 100. $ & $ 48. $ & $ 22. $ & $ 10. $ \\ $
 \text{- 0 0 + } $ & $ 1.5\times 10^3 $ & $ 480. $ & $ 150. $ & $ 48. $ & $ 1.5\times 10^3 $ & $ 480. $ & $ 150. $ & $ 48. $ \\ $
 \text{- 0 + - } $ & $ 76. $ & $ 35. $ & $ 16. $ & $ 7.6 $ & $ 76. $ & $ 35. $ & $ 16. $ & $ 7.6 $ \\ $
 \text{- 0 + 0 } $ & $ 12. $ & $ 6.7 $ & $ 3.8 $ & $ 2.1 $ & $ 12. $ & $ 6.7 $ & $ 3.8 $ & $ 2.1 $ \\ $
 \text{- + - + } $ & $ \text{x} $ & $ \text{x} $ & $ \text{x} $ & $ \text{x} $ & $ \text{x} $ & $ \text{x} $ & $ \text{x} $ & $ \text{x} $ \\ $
 \text{- + 0 0 } $ & $ 1.5\times 10^3 $ & $ 490. $ & $ 150. $ & $ 49. $ & $ 1.5\times 10^3 $ & $ 490. $ & $ 150. $ & $ 49. $ \\ $
 \text{- + + - } $ & $ 980. $ & $ 310. $ & $ 98. $ & $ 31. $ & $ 990. $ & $ 310. $ & $ 99. $ & $ 31. $ \\ $
 \text{0 0 0 0 } $ & $ 500. $ & $ 160. $ & $ 50. $ & $ 16. $ & $ 460. $ & $ 150. $ & $ 46. $ & $ 15. $ \\ \hline$
 \text{diag.} $ & $ 7.3 $ & $ 4.1 $ & $ 2.3 $ & $ 1.3 $ & $ 7.3 $ & $ 4.1 $ & $ 2.3 $ & $ 1.3 $ \\ \hline

\end{tabular}
\end{center}
\caption{See description of Tab.~\ref{tab:unitarityM0}; reaction $W^+W^-\rightarrow W^+W^-$.
%Values of $\sqrt{s^U}$ (in TeV) from the tree-level partial wave unitarity bounds for all elastic on-shell $W^+W^-$ helicity amplitudes for a chosen set of $f_{M0}$ values (first row, in TeV$^{−4}$
%); $\lambda_i$ ($\lambda_i'$) denote ingoing (outgoing) W's helicities; ''$x$'' denotes no unitarity violation.
}
\label{tab:unitarityM0ww}
\end{table}
\clearpage
%--------------------------------------------------------------------%

\begin{table}%
\begin{center}
\begin{tabular}{c|cccc||cccc}
 $\lambda_1\lambda_2\lambda_1'\lambda_2' $ & $ 0.01 $ & $ 0.1 $ & $ 1. $ & $ 10. $ & $ -0.01 $ & $ -0.1 $ & $ -1. $ & $ -10. $ \\ \hline$
 \text{- - - -} $ & $ 720. $ & $ 230. $ & $ 72. $ & $ 23. $ & $ 1.0\times 10^3 $ & $ 320. $ & $ 100. $ & $ 32. $ \\ $
 \text{- - - 0} $ & $ 1.2\times 10^8 $ & $ 1.2\times 10^7 $ & $ 1.2\times 10^6 $ & $ 1.2\times 10^5 $ & $ 1.2\times 10^8 $ & $ 1.2\times 10^7 $ & $ 1.2\times 10^6 $ & $ 1.2\times 10^5 $ \\ $
 \text{- - - +} $ & $ 3.7\times 10^3 $ & $ 1.2\times 10^3 $ & $ 370. $ & $ 120. $ & $ 3.7\times 10^3 $ & $ 1.2\times 10^3 $ & $ 370. $ & $ 120. $ \\ $
 \text{- - 0 0} $ & $ 1.1\times 10^3 $ & $ 340. $ & $ 110. $ & $ 34. $ & $ 1.1\times 10^3 $ & $ 340. $ & $ 110. $ & $ 34. $ \\ $
 \text{- - 0 +} $ & $ 140. $ & $ 67. $ & $ 31. $ & $ 14. $ & $ 140. $ & $ 67. $ & $ 31. $ & $ 14. $ \\ $
 \text{- - + +} $ & $ 1.5\times 10^3 $ & $ 480. $ & $ 150. $ & $ 48. $ & $ 1.5\times 10^3 $ & $ 480. $ & $ 150. $ & $ 48. $ \\ $
 \text{- 0 - 0} $ & $ 14. $ & $ 7.6 $ & $ 4.3 $ & $ 2.4 $ & $ 13. $ & $ 7.1 $ & $ 4.0 $ & $ 2.3 $ \\ $
 \text{- 0 - +} $ & $ 120. $ & $ 56. $ & $ 26. $ & $ 12. $ & $ 120. $ & $ 56. $ & $ 26. $ & $ 12. $ \\ $
 \text{- 0 0 0} $ & $ 140. $ & $ 67. $ & $ 31. $ & $ 14. $ & $ 140. $ & $ 67. $ & $ 31. $ & $ 14. $ \\ $
 \text{- 0 0 +} $ & $ 17. $ & $ 9.5 $ & $ 5.3 $ & $ 3.0 $ & $ 17. $ & $ 9.5 $ & $ 5.3 $ & $ 3.0 $ \\ $
 \text{- + - +} $ & $ 1.7\times 10^3 $ & $ 530. $ & $ 170. $ & $ 54. $ & $ 2.2\times 10^3 $ & $ 700. $ & $ 220. $ & $ 70. $ \\ $
 \text{- + 0 0} $ & $ 2.6\times 10^3 $ & $ 820. $ & $ 260. $ & $ 82. $ & $ 2.6\times 10^3 $ & $ 820. $ & $ 260. $ & $ 82. $ \\ $
 \text{0 0 0 0} $ & $ 1.0\times 10^3 $ & $ 320. $ & $ 100. $ & $ 32. $ & $ 1.1\times 10^3 $ & $ 360. $ & $ 110. $ & $ 36. $ \\ \hline $
 \text{diag.} $ & $ 530. $ & $ 170. $ & $ 53. $ & $ 17. $ & $ 690. $ & $ 220. $ & $ 69. $ & $ 22.$ \\ \hline

\end{tabular}
\end{center}
\caption{Values of $\sqrt{s^U}$ (in TeV) from the tree-level partial wave unitarity bounds for all elastic on-shell $W^+W^+$ helicity amplitudes for a chosen set of $f_{M1}$ values (first row, in TeV$^{−4}$
); $\lambda_i$ ($\lambda_i'$) denote ingoing (outgoing) W's helicities; ''$x$'' denotes no unitarity violation; ''diag.'' denotes unitarity bounds from diagonalization in the helicity space.
}
\label{tab:unitarityM1}
\end{table}

\begin{table}%
\begin{center}
\begin{tabular}{c|cccc||cccc}
 $\lambda_1\lambda_2\lambda_1'\lambda_2' $ & $ 0.01 $ & $ 0.1 $ & $ 1. $ & $ 10. $ & $ -0.01 $ & $ -0.1 $ & $ -1. $ & $ -10. $ \\ \hline$
  \text{- - - - } $ & $ \text{x} $ & $ \text{x} $ & $ \text{x} $ & $ \text{x} $ & $ \text{x} $ & $ \text{x} $ & $ \text{x} $ & $ \text{x} $ \\ $
 \text{- - - 0 } $ & $ 3.2\times 10^5 $ & $ 1.9\times 10^5 $ & $ 8.3\times 10^4 $ & $ 3.6\times 10^4 $ & $ 3.2\times 10^5 $ & $ 1.9\times 10^5 $ & $ 8.3\times 10^4 $ & $ 3.6\times 10^4 $ \\ $
 \text{- - - + } $ & $ 4.4\times 10^3 $ & $ 1.4\times 10^3 $ & $ 440. $ & $ 140. $ & $ 4.4\times 10^3 $ & $ 1.4\times 10^3 $ & $ 440. $ & $ 140. $ \\ $
 \text{- - 0 0 } $ & $ 12. $ & $ 6.7 $ & $ 3.8 $ & $ 2.1 $ & $ 12. $ & $ 6.7 $ & $ 3.8 $ & $ 2.1 $ \\ $
 \text{- - 0 + } $ & $ 160. $ & $ 76. $ & $ 35. $ & $ 16. $ & $ 160. $ & $ 76. $ & $ 35. $ & $ 16. $ \\ $
 \text{- - + + } $ & $ 1.1\times 10^3 $ & $ 340. $ & $ 110. $ & $ 34. $ & $ 1.1\times 10^3 $ & $ 340. $ & $ 110. $ & $ 34. $ \\ $
 \text{- 0 - 0 } $ & $ 13. $ & $ 7.2 $ & $ 4.0 $ & $ 2.3 $ & $ 14. $ & $ 7.6 $ & $ 4.3 $ & $ 2.4 $ \\ $
 \text{- 0 - + } $ & $ 96. $ & $ 44. $ & $ 21. $ & $ 9.6 $ & $ 96. $ & $ 44. $ & $ 21. $ & $ 9.6 $ \\ $
 \text{- 0 0 - } $ & $ 3.0\times 10^3 $ & $ 960. $ & $ 300. $ & $ 96. $ & $ 3.0\times 10^3 $ & $ 960. $ & $ 300. $ & $ 96. $ \\ $
 \text{- 0 0 0 } $ & $ 160. $ & $ 76. $ & $ 35. $ & $ 16. $ & $ 160. $ & $ 76. $ & $ 35. $ & $ 16. $ \\ $
 \text{- 0 0 + } $ & $ 1.4\times 10^3 $ & $ 430. $ & $ 140. $ & $ 43. $ & $ 1.4\times 10^3 $ & $ 430. $ & $ 140. $ & $ 43. $ \\ $
 \text{- 0 + - } $ & $ 120. $ & $ 56. $ & $ 26. $ & $ 12. $ & $ 120. $ & $ 56. $ & $ 26. $ & $ 12. $ \\ $
 \text{- 0 + 0 } $ & $ 17. $ & $ 9.5 $ & $ 5.3 $ & $ 3.0 $ & $ 17. $ & $ 9.5 $ & $ 5.3 $ & $ 3.0 $ \\ $
 \text{- + - + } $ & $ 1.2\times 10^3 $ & $ 380. $ & $ 120. $ & $ 38. $ & $ 1.6\times 10^3 $ & $ 490. $ & $ 160. $ & $ 49. $ \\ $
 \text{- + 0 0 } $ & $ 19. $ & $ 11. $ & $ 5.9 $ & $ 3.3 $ & $ 19. $ & $ 11. $ & $ 5.9 $ & $ 3.3 $ \\ $
 \text{- + + - } $ & $ 2.0\times 10^3 $ & $ 620. $ & $ 200. $ & $ 62. $ & $ 2.0\times 10^3 $ & $ 620. $ & $ 200. $ & $ 62. $ \\ $
 \text{0 0 0 0 } $ & $ 1.0\times 10^3 $ & $ 330. $ & $ 100. $ & $ 33. $ & $ 1.1\times 10^3 $ & $ 350. $ & $ 110. $ & $ 35. $ \\ \hline$
 \text{diag.} $ & $ 10. $ & $ 5.8 $ & $ 3.3 $ & $ 1.9 $ & $ 10. $ & $ 5.8 $ & $ 3.3 $ & $ 1.9 $\\ \hline

\end{tabular}
\end{center}
\caption{See description of Tab.~\ref{tab:unitarityM1}; reaction $W^+W^-\rightarrow W^+W^-$.
%Values of $\sqrt{s^U}$ (in TeV) from the tree-level partial wave unitarity bounds for all elastic on-shell $W^+W^-$ helicity amplitudes for a chosen set of $f_{M1}$ values (first row, in TeV$^{−4}$
%); $\lambda_i$ ($\lambda_i'$) denote ingoing (outgoing) W's helicities; ''$x$'' denotes no unitarity violation.
}
\label{tab:unitarityM1ww}
\end{table}
\clearpage

%--------------------------------------------------------------------%

\begin{table}%
\begin{center}
\begin{tabular}{c|cccc||cccc}
 $\lambda_1\lambda_2\lambda_1'\lambda_2' $ & $ 0.01 $ & $ 0.1 $ & $ 1. $ & $ 10. $ & $ -0.01 $ & $ -0.1 $ & $ -1. $ & $ -10. $ \\ \hline$
 \text{- - - -} $ & $ 1.4\times 10^3 $ & $ 450. $ & $ 140. $ & $ 45. $ & $ 1.0\times 10^3 $ & $ 320. $ & $ 100. $ & $ 32. $ \\ $
 \text{- - - 0} $ & $ \text{x} $ & $ \text{x} $ & $ \text{x} $ & $ \text{x} $ & $ \text{x} $ & $ \text{x} $ & $ \text{x} $ & $ \text{x} $ \\ $
 \text{- - - +} $ & $ \text{x} $ & $ \text{x} $ & $ \text{x} $ & $ \text{x} $ & $ \text{x} $ & $ \text{x} $ & $ \text{x} $ & $ \text{x} $ \\ $
 \text{- - 0 0} $ & $ 1.2\times 10^3 $ & $ 390. $ & $ 120. $ & $ 39. $ & $ 1.2\times 10^3 $ & $ 390. $ & $ 120. $ & $ 39. $ \\ $
 \text{- - 0 +} $ & $ \text{x} $ & $ \text{x} $ & $ \text{x} $ & $ \text{x} $ & $ \text{x} $ & $ \text{x} $ & $ \text{x} $ & $ \text{x} $ \\ $
 \text{- - + +} $ & $ 1.2\times 10^3 $ & $ 390. $ & $ 120. $ & $ 39. $ & $ 1.2\times 10^3 $ & $ 390. $ & $ 120. $ & $ 39. $ \\ $
 \text{- 0 - 0} $ & $ 15. $ & $ 8.5 $ & $ 4.8 $ & $ 2.7 $ & $ 16. $ & $ 9.1 $ & $ 5.1 $ & $ 2.9 $ \\ $
 \text{- 0 - +} $ & $ \text{x} $ & $ \text{x} $ & $ \text{x} $ & $ \text{x} $ & $ \text{x} $ & $ \text{x} $ & $ \text{x} $ & $ \text{x} $ \\ $
 \text{- 0 0 0} $ & $ \text{x} $ & $ \text{x} $ & $ \text{x} $ & $ \text{x} $ & $ \text{x} $ & $ \text{x} $ & $ \text{x} $ & $ \text{x} $ \\ $
 \text{- 0 0 +} $ & $ 16. $ & $ 8.8 $ & $ 4.9 $ & $ 2.8 $ & $ 16. $ & $ 8.8 $ & $ 4.9 $ & $ 2.8 $ \\ $
 \text{- + - +} $ & $ \text{x} $ & $ \text{x} $ & $ \text{x} $ & $ \text{x} $ & $ \text{x} $ & $ \text{x} $ & $ \text{x} $ & $ \text{x} $ \\ $
 \text{- + 0 0} $ & $ \text{x} $ & $ \text{x} $ & $ \text{x} $ & $ \text{x} $ & $ \text{x} $ & $ \text{x} $ & $ \text{x} $ & $ \text{x} $ \\ $
 \text{0 0 0 0} $ & $ 1.3\times 10^3 $ & $ 410. $ & $ 130. $ & $ 41. $ & $ 1.2\times 10^3 $ & $ 370. $ & $ 120. $ & $ 37. $ \\ \hline $
 \text{diag.} $ & $ 800. $ & $ 250. $ & $ 80. $ & $ 25. $ & $ 620. $ & $ 190. $ & $ 62. $ & $ 20.$ \\ \hline

\end{tabular}
\end{center}
\caption{Values of $\sqrt{s^U}$ (in TeV) from the tree-level partial wave unitarity bounds for all elastic on-shell $W^+W^+$ helicity amplitudes for a chosen set of $f_{M7}$ values (first row, in TeV$^{−4}$
); $\lambda_i$ ($\lambda_i'$) denote ingoing (outgoing) W's helicities; ''$x$'' denotes no unitarity violation; ''diag.'' denotes unitarity bounds from diagonalization in the helicity space.
}
\label{tab:unitarityM7}
\end{table}

\begin{table}%
\begin{center}
\begin{tabular}{c|cccc||cccc}
 $\lambda_1\lambda_2\lambda_1'\lambda_2' $ & $ 0.01 $ & $ 0.1 $ & $ 1. $ & $ 10. $ & $ -0.01 $ & $ -0.1 $ & $ -1. $ & $ -10. $ \\ \hline$
 \text{- - - - } $ & $ \text{x} $ & $ \text{x} $ & $ \text{x} $ & $ \text{x} $ & $ \text{x} $ & $ \text{x} $ & $ \text{x} $ & $ \text{x} $ \\ $
 \text{- - - 0 } $ & $ \text{x} $ & $ \text{x} $ & $ \text{x} $ & $ \text{x} $ & $ \text{x} $ & $ \text{x} $ & $ \text{x} $ & $ \text{x} $ \\ $
 \text{- - - + } $ & $ \text{x} $ & $ \text{x} $ & $ \text{x} $ & $ \text{x} $ & $ \text{x} $ & $ \text{x} $ & $ \text{x} $ & $ \text{x} $ \\ $
 \text{- - 0 0 } $ & $ 14. $ & $ 8.0 $ & $ 4.5 $ & $ 2.5 $ & $ 14. $ & $ 8.0 $ & $ 4.5 $ & $ 2.5 $ \\ $
 \text{- - 0 + } $ & $ 160. $ & $ 76. $ & $ 35. $ & $ 16. $ & $ 160. $ & $ 76. $ & $ 35. $ & $ 16. $ \\ $
 \text{- - + + } $ & $ 1.1\times 10^3 $ & $ 340. $ & $ 110. $ & $ 34. $ & $ 1.1\times 10^3 $ & $ 340. $ & $ 110. $ & $ 34. $ \\ $
 \text{- 0 - 0 } $ & $ 16. $ & $ 9.1 $ & $ 5.1 $ & $ 2.9 $ & $ 15. $ & $ 8.5 $ & $ 4.8 $ & $ 2.7 $ \\ $
 \text{- 0 - + } $ & $ 120. $ & $ 56. $ & $ 26. $ & $ 12. $ & $ 120. $ & $ 56. $ & $ 26. $ & $ 12. $ \\ $
 \text{- 0 0 - } $ & $ \text{x} $ & $ \text{x} $ & $ \text{x} $ & $ \text{x} $ & $ \text{x} $ & $ \text{x} $ & $ \text{x} $ & $ \text{x} $ \\ $
 \text{- 0 0 0 } $ & $ \text{x} $ & $ \text{x} $ & $ \text{x} $ & $ \text{x} $ & $ \text{x} $ & $ \text{x} $ & $ \text{x} $ & $ \text{x} $ \\ $
 \text{- 0 0 + } $ & $ 1.5\times 10^3 $ & $ 480. $ & $ 150. $ & $ 48. $ & $ 1.5\times 10^3 $ & $ 480. $ & $ 150. $ & $ 48. $ \\ $
 \text{- 0 + - } $ & $ \text{x} $ & $ \text{x} $ & $ \text{x} $ & $ \text{x} $ & $ \text{x} $ & $ \text{x} $ & $ \text{x} $ & $ \text{x} $ \\ $
 \text{- 0 + 0 } $ & $ 22. $ & $ 12. $ & $ 7.0 $ & $ 3.9 $ & $ 22. $ & $ 12. $ & $ 7.0 $ & $ 3.9 $ \\ $
 \text{- + - + } $ & $ 2.2\times 10^3 $ & $ 700. $ & $ 220. $ & $ 70. $ & $ 1.7\times 10^3 $ & $ 540. $ & $ 170. $ & $ 54. $ \\ $
 \text{- + 0 0 } $ & $ 22. $ & $ 13. $ & $ 7.0 $ & $ 4.0 $ & $ 22. $ & $ 12. $ & $ 7.0 $ & $ 4.0 $ \\ $
 \text{- + + - } $ & $ \text{x} $ & $ \text{x} $ & $ \text{x} $ & $ \text{x} $ & $ \text{x} $ & $ \text{x} $ & $ \text{x} $ & $ \text{x} $ \\ $
 \text{0 0 0 0 } $ & $ 1.3\times 10^3 $ & $ 410. $ & $ 130. $ & $ 41. $ & $ 1.2\times 10^3 $ & $ 380. $ & $ 120. $ & $ 38. $ \\ \hline$
 \text{diag.} $ & $ 12. $ & $ 6.9 $ & $ 3.9 $ & $ 2.2 $ & $ 12. $ & $ 6.9 $ & $ 3.9 $ & $ 2.2 $ \\ \hline

\end{tabular}
\end{center}
\caption{See description of Tab.~\ref{tab:unitarityM7}; reaction $W^+W^-\rightarrow W^+W^-$.
%Values of $\sqrt{s^U}$ (in TeV) from the tree-level partial wave unitarity bounds for all elastic on-shell $W^+W^-$ helicity amplitudes for a chosen set of $f_{M7}$ values (first row, in TeV$^{−4}$
%); $\lambda_i$ ($\lambda_i'$) denote ingoing (outgoing) W's helicities; ''$x$'' denotes no unitarity violation.
}
\label{tab:unitarityM7ww}
\end{table}
\clearpage

\section{Analytic formulae for both $iM$ and $\mathcal{T}^{(j=j_{min})}$ and contributions of $\sigma_{pol}$ to $\sigma^{tot}_{unpol}$ at $\sqrt{s^U}$ (numerical result)} 
\thispagestyle{empty} 
\label{app:analiticFormSMEFT}
%%%%%%%%%%%%%%%%%%%% cross-sections at s^U %%%%%%%%%%%%%%%%%%%%%%%%%%%%%%%%%%%%%%%%%%%%%%
\begin{table}[h]%
\begin{center}
\bgroup
\def\arraystretch{1.3}
\begin{tabular}{c|c|c}
 $\lambda_1\lambda_2\lambda_1'\lambda_2' $ &  $i\mathcal{M}_{\lambda_1'\lambda_2';\lambda_1\lambda_2}(\sqrt{s},\theta ,f_{S0}) $ & $ \mathcal{T}_{\lambda_1'\lambda_2';\lambda_1\lambda_2}^{(j=j_{min})} $ \\ \hline $
 \text{- - - -} $ & $ 4 i c_W^4 f_{\text{S0}} m_Z^4-\frac{32 i \csc ^2(\theta ) c_W^2 m_Z^2}{v^2} $ & $ \frac{c_W^4 f_{\text{S0}} m_Z^4}{4 \pi } $ \\ $
 \text{- - - 0} $ & $ -\frac{2 i \sqrt{2} \cot (\theta ) \csc ^2(\theta ) c_W^3 m_Z^3 \left(-7 m_h^2+\left(28 c_W^2-25\right) m_Z^2+\cos (2 \theta ) \left(\left(4
   c_W^2+1\right) m_Z^2-m_h^2\right)\right)}{s^{3/2} v^2} $ & $ 0 $ \\ $
 \text{- - - +} $ & $ \frac{16 i c_W^4 m_Z^4}{s v^2} $ & $ \frac{c_W^4 m_Z^4}{\sqrt{6} \pi  s v^2} $ \\ $
 \text{- - 0 0} $ & $ 2 i s c_W^2 f_{\text{S0}} m_Z^2 $ & $ \frac{s c_W^2 f_{\text{S0}} m_Z^2}{8 \pi } $ \\ $
 \text{- - 0 +} $ & $ \frac{4 i \sqrt{2} \cot (\theta ) c_W^3 m_Z^3 \left(\left(4 c_W^2+1\right) m_Z^2-m_h^2\right)}{s^{3/2} v^2} $ & $ -\frac{c_W^3 m_Z^3
   \left(\left(4 c_W^2+1\right) m_Z^2-m_h^2\right)}{4 \sqrt{3} \pi  s^{3/2} v^2} $ \\ $
 \text{- - + +} $ & $ 4 i c_W^4 f_{\text{S0}} m_Z^4 $ & $ \frac{c_W^4 f_{\text{S0}} m_Z^4}{4 \pi } $ \\ $
 \text{- 0 - 0} $ & $ \frac{8 i c_W^2 m_Z^2}{v^2 (\cos (\theta )-1)} $ & $ 0 $ \\ $
 \text{- 0 - +} $ & $ -\frac{8 i \sqrt{2} \cot \left(\frac{\theta }{2}\right) c_W^3 m_Z^3}{\sqrt{s} v^2} $ & $ \frac{\sqrt{2} c_W^3 m_Z^3}{3 \pi  \sqrt{s} v^2} $ \\ $
 -000 $ & $ \frac{2 i \sqrt{2} \cot (\theta ) c_W m_Z \left(m_Z-m_h\right) \left(m_h+m_Z\right)}{\sqrt{s} v^2} $ & $ \frac{c_W m_Z \left(m_Z^2-m_h^2\right)}{8
   \sqrt{3} \pi  \sqrt{s} v^2} $ \\ $
 \text{- 0 0 +} $ & $ \frac{4 i c_W^2 m_Z^2 \left(\cos (\theta ) \left(\left(4 c_W^2+1\right) m_Z^2-m_h^2\right)-4 c_W^2 m_Z^2\right)}{s v^2 (\cos (\theta )-1)} $ & $
   \frac{c_W^2 m_Z^2 \left(\left(4 c_W^2+1\right) m_Z^2-m_h^2\right)}{8 \pi  s v^2} $ \\ $
 \text{- + - +} $ & $ -\frac{8 i \cot ^2\left(\frac{\theta }{2}\right) c_W^2 m_Z^2}{v^2} $ & $ \frac{c_W^2 m_Z^2}{6 \pi  v^2} $ \\ $
 \text{- + 0 0} $ & $ -\frac{4 i c_W^2 m_Z^2 \left(2 \left(m_h^2-m_Z^2\right) \csc ^2(\theta )-m_h^2+\left(8 c_W^2+1\right) m_Z^2\right)}{s v^2} $ & $ -\frac{c_W^2
   m_Z^2 \left(m_h^2+\left(4 c_W^2-1\right) m_Z^2\right)}{2 \sqrt{6} \pi  s v^2} $ \\ $
 \text{0 0 0 0} $ & $ i s^2 f_{\text{S0}} $ & $ \frac{s^2 f_{\text{S0}}}{16 \pi }$ \\ \hline

\end{tabular}
\egroup
\caption{Analytic formulas for on-shell $W^+W^+$ elastic scattering helicity amplitudes $iM$ and the minimal $j$ partial waves $\mathcal{T}^{(j=j_{min})}$; $\lambda_i$ ($\lambda_i'$) denote ingoing (outgoing) W's helicities; $c_W\equiv \cos\theta_W$; $\theta$ is the scattering angle; the $\mathcal{O}_{S0}$ operator case; only leading terms in the $\sqrt{s}$  expansion in the limit $s\rightarrow\infty$ are shown.}
\label{tab:analyticS0}
\end{center}
\end{table}
\begin{table}[h]
\begin{center}
\resizebox{0.8\columnwidth}{!}{
\begin{tabular}{c|cccc||cccc}
 $\lambda_1\lambda_2\lambda_1'\lambda_2' $ & $ 0.01 $ & $ 0.1 $ & $ 1. $ & $ 10. $ & $ -0.01 $ & $ -0.1 $ & $ -1. $ & $ -10. $ \\ \hline $
 \text{- - - -} $ & $ 2.3 $ & $ 7.2 $ & $ 21. $ & $ 57. $ & $ 2.6 $ & $ 8.0 $ & $ 23. $ & $ 61. $ \\ $
 \text{- - - 0} $ & $ 3.7\times 10^{-10} $ & $ 3.4\times 10^{-8} $ & $ 2.8\times 10^{-6} $ & $ 1.6\times 10^{-4} $ & $ 5.6\times 10^{-10} $ & $ 5.1\times 10^{-8} $ & $ 4.0\times 10^{-6}
   $ & $ 2.3\times 10^{-4} $ \\ $
 \text{- - - +} $ & $ 1.5\times 10^{-9} $ & $ 4.7\times 10^{-8} $ & $ 1.5\times 10^{-6} $ & $ 4.7\times 10^{-5} $ & $ 2.1\times 10^{-9} $ & $ 6.4\times 10^{-8} $ & $ 2.0\times 10^{-6} $ & $
   6.2\times 10^{-5} $ \\ $
 \text{- - 0 0} $ & $ 1.9\times 10^{-6} $ & $ 5.9\times 10^{-5} $ & $ 1.9\times 10^{-3} $ & $ 0.059 $ & $ 2.2\times 10^{-6} $ & $ 7.1\times 10^{-5} $ & $ 2.2\times 10^{-3} $ & $ 0.069 $ \\ $
 \text{- - 0 +} $ & $ 2.0\times 10^{-13} $ & $ 2.0\times 10^{-11} $ & $ 1.9\times 10^{-9} $ & $ 1.8\times 10^{-7} $ & $ 3.0\times 10^{-13} $ & $ 3.0\times 10^{-11} $ & $ 2.9\times
   10^{-9} $ & $ 2.6\times 10^{-7} $ \\ $
 \text{- - + +} $ & $ 3.0\times 10^{-14} $ & $ 9.3\times 10^{-12} $ & $ 3.0\times 10^{-9} $ & $ 9.4\times 10^{-7} $ & $ 3.9\times 10^{-14} $ & $ 1.2\times 10^{-11} $ & $ 3.8\times
   10^{-9} $ & $ 1.2\times 10^{-6} $ \\ $
 \text{- 0 - 0} $ & $ 2.2 $ & $ 6.8 $ & $ 20. $ & $ 57. $ & $ 2.4 $ & $ 7.5 $ & $ 22. $ & $ 61. $ \\ $
 \text{- 0 - +} $ & $ 9.1\times 10^{-5} $ & $ 9.0\times 10^{-4} $ & $ 8.8\times 10^{-3} $ & $ 0.083 $ & $ 1.1\times 10^{-4} $ & $ 1.1\times 10^{-3} $ & $ 0.011 $ & $ 0.099 $ \\ $
 \text{- 0 0 0} $ & $ 9.7\times 10^{-7} $ & $ 8.6\times 10^{-6} $ & $ 6.1\times 10^{-5} $ & $ 3.0\times 10^{-4} $ & $ 1.2\times 10^{-6} $ & $ 1.0\times 10^{-5} $ & $ 7.1\times 10^{-5} $ & $
   3.4\times 10^{-4} $ \\ $
 \text{- 0 0 +} $ & $ 8.1\times 10^{-9} $ & $ 2.4\times 10^{-7} $ & $ 6.6\times 10^{-6} $ & $ 1.5\times 10^{-4} $ & $ 1.1\times 10^{-8} $ & $ 3.3\times 10^{-7} $ & $ 8.6\times 10^{-6} $ & $
   1.9\times 10^{-4} $ \\ $
 \text{- + - +} $ & $ 4.1 $ & $ 13. $ & $ 37. $ & $ 97. $ & $ 4.5 $ & $ 14. $ & $ 40. $ & $ 100. $ \\ $
 \text{- + 0 0} $ & $ 9.9\times 10^{-9} $ & $ 3.0\times 10^{-7} $ & $ 8.6\times 10^{-6} $ & $ 2.2\times 10^{-4} $ & $ 1.4\times 10^{-8} $ & $ 4.1\times 10^{-7} $ & $ 1.1\times 10^{-5} $ & $
   2.8\times 10^{-4} $ \\ $
 \text{0 0 0 0} $ & $ 16. $ & $ 50. $ & $ 160. $ & $ 490. $ & $ 15. $ & $ 46. $ & $ 150. $ & $ 460.$ \\ \hline
\end{tabular}
}
\end{center}
\caption{Contribution of polarized elastic on-shell $W^+W^+$ scattering cross sections (in $pb$) to the total unpolarized cross sections at $\sqrt{s^U}$ for a chosen set of $f_{S0}$ values (first row, in TeV$^{−4}$).}
\label{tab:polarizedContributionsS0}
\end{table}
\clearpage

%----------------------------------------------------------------%
\begin{table}%
\begin{center}
\bgroup
\def\arraystretch{1.5}
\begin{tabular}{c|c|c}
 $\lambda_1\lambda_2\lambda_1'\lambda_2' $ &  $i\mathcal{M}_{\lambda_1'\lambda_2';\lambda_1\lambda_2}(\sqrt{s},\theta ,f_{S1}) $ & $ \mathcal{T}_{\lambda_1'\lambda_2';\lambda_1\lambda_2}^{(j=j_{min})} $ \\ \hline $
 \text{- - - -} $ & $ \frac{1}{2} i c_W^2 m_Z^2 \left(c_W^2 f_{\text{S1}} \cos (2 \theta ) m_Z^2+3 c_W^2 f_{\text{S1}} m_Z^2-\frac{64 \csc ^2(\theta )}{v^2}\right) $ & $ \frac{c_W^4 f_{\text{S1}} m_Z^4}{12 \pi } $ \\ $
 \text{- - - 0} $ & $ -\frac{i \sqrt{s} c_W^3 f_{\text{S1}} \sin (\theta ) \cos (\theta ) m_Z^3}{\sqrt{2}} $ & $ \frac{\sqrt{s} c_W^3 f_{\text{S1}} m_Z^3}{80 \sqrt{3} \pi } $ \\ $
 \text{- - - +} $ & $ -i c_W^4 f_{\text{S1}} \sin ^2(\theta ) m_Z^4 $ & $ -\frac{c_W^4 f_{\text{S1}} m_Z^4}{20 \sqrt{6} \pi } $ \\ $
 \text{- - 0 0} $ & $ \frac{1}{2} i s c_W^2 f_{\text{S1}} \sin ^2(\theta ) m_Z^2 $ & $ \frac{s c_W^2 f_{\text{S1}} m_Z^2}{48 \pi } $ \\ $
 \text{- - 0 +} $ & $ -\frac{i \sqrt{s} c_W^3 f_{\text{S1}} \sin (\theta ) \cos (\theta ) m_Z^3}{\sqrt{2}} $ & $ \frac{\sqrt{s} c_W^3 f_{\text{S1}} m_Z^3}{80 \sqrt{3} \pi } $ \\ $
 \text{- - + +} $ & $ \frac{1}{2} i c_W^4 f_{\text{S1}} (\cos (2 \theta )+3) m_Z^4 $ & $ \frac{c_W^4 f_{\text{S1}} m_Z^4}{12 \pi } $ \\ $
 \text{- 0 - 0} $ & $ -\frac{1}{2} i s c_W^2 f_{\text{S1}} \sin ^2(\theta ) m_Z^2 $ & $ -\frac{s c_W^2 f_{\text{S1}} m_Z^2}{96 \pi } $ \\ $
 \text{- 0 - +} $ & $ \frac{i \sqrt{s} c_W^3 f_{\text{S1}} \sin (\theta ) (\cos (\theta )+1) m_Z^3}{\sqrt{2}} $ & $ -\frac{\sqrt{s} c_W^3 f_{\text{S1}} m_Z^3}{40 \sqrt{2} \pi } $ \\ $
 \text{- 0 0 0} $ & $ -\frac{i s^{3/2} c_W f_{\text{S1}} \sin (2 \theta ) m_Z}{4 \sqrt{2}} $ & $ -\frac{s^{3/2} c_W f_{\text{S1}} m_Z}{160 \sqrt{3} \pi } $ \\ $
 \text{- 0 0 +} $ & $ i s c_W^2 f_{\text{S1}} \cos ^2\left(\frac{\theta }{2}\right) \cos (\theta ) m_Z^2 $ & $ \frac{s c_W^2 f_{\text{S1}} m_Z^2}{96 \pi } $ \\ $
 \text{- + - +} $ & $ -\frac{i c_W^2 \cot ^2\left(\frac{\theta }{2}\right) m_Z^2 \left(v^2 c_W^2 f_{\text{S1}} \cos (2 \theta ) m_Z^2-v^2 c_W^2 f_{\text{S1}} m_Z^2+16\right)}{2 v^2} $ & $ \frac{c_W^2 m_Z^2 \left(3 v^2 c_W^2 f_{\text{S1}} m_Z^2+10\right)}{60 \pi  v^2} $ \\ $
 \text{- + 0 0} $ & $ \frac{1}{2} i s c_W^2 f_{\text{S1}} \sin ^2(\theta ) m_Z^2 $ & $ \frac{s c_W^2 f_{\text{S1}} m_Z^2}{40 \sqrt{6} \pi } $ \\ $
 \text{0 0 0 0} $ & $ \frac{1}{8} i s^2 f_{\text{S1}} (\cos (2 \theta )+3) $ & $ \frac{s^2 f_{\text{S1}}}{48 \pi }$ \\ \hline
\end{tabular}
\egroup
\caption{Analytic formulas for on-shell $W^+W^+$ elastic scattering helicity amplitudes $iM$ and the minimal $j$ partial waves $\mathcal{T}^{(j=j_{min})}$; $\lambda_i$ ($\lambda_i'$) denote ingoing (outgoing) W's helicities; $c_W\equiv \cos\theta_W$; $\theta$ is the scattering angle; the $\mathcal{O}_{S1}$ operator case; only leading terms in the $\sqrt{s}$  expansion in the limit $s\rightarrow\infty$ are shown.}
\label{tab:analyticS1}
\end{center}
\end{table}

\begin{table}%
\begin{center}
\resizebox{1.05\columnwidth}{!}{
\begin{tabular}{c|cccc||cccc}
 $\lambda_1\lambda_2\lambda_1'\lambda_2' $ & $ 0.01 $ & $ 0.1 $ & $ 1. $ & $ 10. $ & $ -0.01 $ & $ -0.1 $ & $ -1. $ & $ -10. $ \\\hline  $
 \text{- - - -} $ & $ 2.9 $ & $ 9.0 $ & $ 26. $ & $ 67. $ & $ 2.7 $ & $ 8.4 $ & $ 25. $ & $ 64. $ \\ $ 
 \text{- - - 0} $ & $ 8.8\times 10^{-10} $ & $ 7.9\times 10^{-8} $ & $ 6.0\times 10^{-6} $ & $ 3.1\times 10^{-4} $ & $ 7.3\times 10^{-10} $ & $ 6.8\times 10^{-8} $ & $
   5.4\times 10^{-6} $ & $ 3.2\times 10^{-4} $ \\ $
 \text{- - - +} $ & $ 3.0\times 10^{-9} $ & $ 9.3\times 10^{-8} $ & $ 2.9\times 10^{-6} $ & $ 8.2\times 10^{-5} $ & $ 2.4\times 10^{-9} $ & $ 7.7\times 10^{-8} $ & $ 2.5\times
   10^{-6} $ & $ 8.3\times 10^{-5} $ \\ $
 \text{- - 0 0} $ & $ 8.0\times 10^{-8} $ & $ 2.4\times 10^{-6} $ & $ 6.9\times 10^{-5} $ & $ 1.8\times 10^{-3} $ & $ 1.4\times 10^{-7} $ & $ 4.3\times 10^{-6} $ & $ 1.3\times
   10^{-4} $ & $ 3.7\times 10^{-3} $ \\ $
 \text{- - 0 +} $ & $ 6.7\times 10^{-12} $ & $ 6.7\times 10^{-10} $ & $ 6.7\times 10^{-8} $ & $ 6.7\times 10^{-6} $ & $ 1.2\times 10^{-11} $ & $ 1.2\times 10^{-9} $ & $
   1.2\times 10^{-7} $ & $ 1.2\times 10^{-5} $ \\ $
 \text{- - + +} $ & $ 3.2\times 10^{-15} $ & $ 1.0\times 10^{-12} $ & $ 3.2\times 10^{-10} $ & $ 1.0\times 10^{-7} $ & $ 6.3\times 10^{-15} $ & $ 2.0\times 10^{-12} $ & $
   6.2\times 10^{-10} $ & $ 1.9\times 10^{-7} $ \\ $
 \text{- 0 - 0} $ & $ 2.7 $ & $ 8.5 $ & $ 25. $ & $ 68. $ & $ 2.6 $ & $ 7.9 $ & $ 24. $ & $ 64. $ \\ $
 \text{- 0 - +} $ & $ 1.4\times 10^{-4} $ & $ 1.4\times 10^{-3} $ & $ 0.014 $ & $ 0.12 $ & $ 1.2\times 10^{-4} $ & $ 1.2\times 10^{-3} $ & $ 0.012 $ & $ 0.12 $ \\ $
 \text{- 0 0 0} $ & $ 2.2\times 10^{-4} $ & $ 2.2\times 10^{-3} $ & $ 0.022 $ & $ 0.21 $ & $ 2.0\times 10^{-4} $ & $ 2.0\times 10^{-3} $ & $ 0.020 $ & $ 0.21 $ \\ $
 \text{- 0 0 +} $ & $ 3.2\times 10^{-7} $ & $ 9.9\times 10^{-6} $ & $ 3.0\times 10^{-4} $ & $ 9.0\times 10^{-3} $ & $ 1.6\times 10^{-7} $ & $ 5.2\times 10^{-6} $ & $ 1.7\times
   10^{-4} $ & $ 5.5\times 10^{-3} $ \\ $
 \text{- + - +} $ & $ 5.1 $ & $ 16. $ & $ 45. $ & $ 110. $ & $ 4.7 $ & $ 15. $ & $ 43. $ & $ 110. $ \\ $
 \text{- + 0 0} $ & $ 5.2\times 10^{-8} $ & $ 1.6\times 10^{-6} $ & $ 4.9\times 10^{-5} $ & $ 1.5\times 10^{-3} $ & $ 1.6\times 10^{-7} $ & $ 4.9\times 10^{-6} $ & $ 1.5\times
   10^{-4} $ & $ 4.8\times 10^{-3} $ \\ $
 \text{0 0 0 0} $ & $ 0.58 $ & $ 1.9 $ & $ 5.9 $ & $ 19. $ & $ 2.0 $ & $ 6.4 $ & $ 20. $ & $ 62. $\\ \hline
\end{tabular}
}
\end{center}
\caption{Contribution of polarized elastic on-shell $W^+W^+$ scattering cross sections (in $pb$) to the total unpolarized cross sections at $\sqrt{s^U}$ for a chosen set of $f_{S1}$ values (first row, in TeV$^{−4}$).}
\label{tab:polarizedContributionsS1}
\end{table}
\clearpage

%----------------------------------------------------------------%

\begin{table}%
\begin{center}
\bgroup
\def\arraystretch{1.5}
\begin{tabular}{c|c|c}
 $\lambda_1\lambda_2\lambda_1'\lambda_2' $ &  $i\mathcal{M}_{\lambda_1'\lambda_2';\lambda_1\lambda_2}(\sqrt{s},\theta ,f_{T0}) $ & $ \mathcal{T}_{\lambda_1'\lambda_2';\lambda_1\lambda_2}^{(j=j_{min})} $ \\ \hline $
  \text{- - - -} $ & $ 2 i c_W^2 m_Z^2 \left(c_W^2 f_{\text{T0}} \cos (2 \theta ) m_Z^2+3 c_W^2 f_{\text{T0}} m_Z^2-\frac{16 \csc
   ^2(\theta )}{v^2}\right) $ & $ \frac{c_W^4 f_{\text{T0}} m_Z^4}{3 \pi } $ \\ $
 \text{- - - 0} $ & $ -i \sqrt{2} \sqrt{s} c_W^3 f_{\text{T0}} \sin (2 \theta ) m_Z^3 $ & $ \frac{\sqrt{s} c_W^3 f_{\text{T0}} m_Z^3}{20
   \sqrt{3} \pi } $ \\ $
 \text{- - - +} $ & $ -2 i s c_W^2 f_{\text{T0}} \sin ^2(\theta ) m_Z^2 $ & $ -\frac{s c_W^2 f_{\text{T0}} m_Z^2}{10 \sqrt{6} \pi } $ \\ $
 \text{- - 0 0} $ & $ 2 i s c_W^2 f_{\text{T0}} \sin ^2(\theta ) m_Z^2 $ & $ \frac{s c_W^2 f_{\text{T0}} m_Z^2}{12 \pi } $ \\ $
 \text{- - 0 +} $ & $ -i \sqrt{2} s^{3/2} c_W f_{\text{T0}} \sin (\theta ) \cos (\theta ) m_Z $ & $ \frac{s^{3/2} c_W f_{\text{T0}}
   m_Z}{40 \sqrt{3} \pi } $ \\ $
 \text{- - + +} $ & $ \frac{1}{2} i s^2 f_{\text{T0}} (\cos (2 \theta )+3) $ & $ \frac{s^2 f_{\text{T0}}}{12 \pi } $ \\ $
 \text{- 0 - 0} $ & $ -i s c_W^2 f_{\text{T0}} \sin ^2(\theta ) m_Z^2 $ & $ -\frac{s c_W^2 f_{\text{T0}} m_Z^2}{48 \pi } $ \\ $
 \text{- 0 - +} $ & $ 2 i \sqrt{2} s^{3/2} c_W f_{\text{T0}} \sin \left(\frac{\theta }{2}\right) \cos ^3\left(\frac{\theta
   }{2}\right) m_Z $ & $ -\frac{s^{3/2} c_W f_{\text{T0}} m_Z}{40 \sqrt{2} \pi } $ \\ $
 \text{- 0 0 0} $ & $ -2 i \sqrt{2} \sqrt{s} c_W^3 f_{\text{T0}} \sin (2 \theta ) m_Z^3 $ & $ -\frac{\sqrt{s} c_W^3 f_{\text{T0}}
   m_Z^3}{10 \sqrt{3} \pi } $ \\ $
 \text{- 0 0 +} $ & $ \frac{1}{2} i s c_W^2 f_{\text{T0}} (4 \cos (\theta )+3 \cos (2 \theta )+1) m_Z^2 $ & $ \frac{s c_W^2
   f_{\text{T0}} m_Z^2}{48 \pi } $ \\ $
 \text{- + - +} $ & $ 2 i s^2 f_{\text{T0}} \cos ^4\left(\frac{\theta }{2}\right) $ & $ \frac{s^2 f_{\text{T0}}}{40 \pi } $ \\ $
 \text{- + 0 0} $ & $ 2 i s c_W^2 f_{\text{T0}} \sin ^2(\theta ) m_Z^2 $ & $ \frac{s c_W^2 f_{\text{T0}} m_Z^2}{10 \sqrt{6} \pi } $ \\ $
 \text{0 0 0 0} $ & $ \frac{2 i \left(4 v^2 c_W^4 f_{\text{T0}} \cos (2 \theta ) m_Z^4+4 v^2 c_W^4 f_{\text{T0}} m_Z^4-m_h^2-4 \csc
   ^2(\theta ) m_Z^2+m_Z^2\right)}{v^2} $ & $ \frac{8 v^4 c_W^4 f_{\text{T0}} m_Z^4+3 v^2 \left(m_Z^2-m_h^2\right)}{24 \pi  v^4}$ \\ \hline
\end{tabular}
\egroup
\caption{Analytic formulas for on-shell $W^+W^+$ elastic scattering helicity amplitudes $iM$ and the minimal $j$ partial waves $\mathcal{T}^{(j=j_{min})}$; $\lambda_i$ ($\lambda_i'$) denote ingoing (outgoing) W's helicities; $c_W\equiv \cos\theta_W$; $\theta$ is the scattering angle; the $\mathcal{O}_{T0}$ operator case; only leading terms in the $\sqrt{s}$  expansion in the limit $s\rightarrow\infty$ are shown.}
\label{tab:analyticT0}
\end{center}
\end{table}

\begin{table}%
\begin{center}
\resizebox{1.05\columnwidth}{!}{
\begin{tabular}{c|cccc||cccc}
 $\lambda_1\lambda_2\lambda_1'\lambda_2' $ & $ 0.01 $ & $ 0.1 $ & $ 1. $ & $ 10. $ & $ -0.01 $ & $ -0.1 $ & $ -1. $ & $ -10. $ \\ \hline $
 \text{- - - -} $ & $ 8.6 $ & $ 25. $ & $ 63. $ & $ 130. $ & $ 6.6 $ & $ 20. $ & $ 53. $ & $ 120. $ \\ $
 \text{- - - 0} $ & $ 6.7\times 10^{-8} $ & $ 4.8\times 10^{-6} $ & $ 2.5\times 10^{-4} $ & $ 8.0\times 10^{-3} $ & $ 2.4\times 10^{-8} $ & $ 2.1\times
   10^{-6} $ & $ 1.3\times 10^{-4} $ & $ 5.9\times 10^{-3} $ \\ $
 \text{- - - +} $ & $ 5.0\times 10^{-7} $ & $ 1.7\times 10^{-5} $ & $ 5.4\times 10^{-4} $ & $ 0.018 $ & $ 1.6\times 10^{-6} $ & $ 5.2\times 10^{-5} $ & $
   1.6\times 10^{-3} $ & $ 0.050 $ \\ $
 \text{- - 0 0} $ & $ 1.4\times 10^{-6} $ & $ 3.5\times 10^{-5} $ & $ 7.3\times 10^{-4} $ & $ 0.012 $ & $ 1.4\times 10^{-6} $ & $ 4.1\times 10^{-5} $ & $
   1.2\times 10^{-3} $ & $ 0.030 $ \\ $
 \text{- - 0 +} $ & $ 3.5\times 10^{-4} $ & $ 3.6\times 10^{-3} $ & $ 0.038 $ & $ 0.39 $ & $ 6.2\times 10^{-4} $ & $ 6.1\times 10^{-3} $ & $ 0.059 $ & $ 0.58
   $ \\ $
 \text{- - + +} $ & $ 1.2 $ & $ 4.0 $ & $ 13. $ & $ 44. $ & $ 2.8 $ & $ 8.7 $ & $ 26. $ & $ 79. $ \\ $
 \text{- 0 - 0} $ & $ 8.1 $ & $ 24. $ & $ 63. $ & $ 150. $ & $ 6.1 $ & $ 19. $ & $ 53. $ & $ 130. $ \\ $
 \text{- 0 - +} $ & $ 7.0\times 10^{-4} $ & $ 6.8\times 10^{-3} $ & $ 0.063 $ & $ 0.56 $ & $ 5.2\times 10^{-3} $ & $ 0.052 $ & $ 0.51 $ & $ 5.0 $ \\ $
 \text{- 0 0 0} $ & $ 1.2\times 10^{-5} $ & $ 8.2\times 10^{-5} $ & $ 4.6\times 10^{-4} $ & $ 6.8\times 10^{-3} $ & $ 7.0\times 10^{-6} $ & $ 5.0\times
   10^{-5} $ & $ 2.0\times 10^{-4} $ & $ 3.0\times 10^{-3} $ \\ $
 \text{- 0 0 +} $ & $ 2.2\times 10^{-6} $ & $ 6.6\times 10^{-5} $ & $ 1.8\times 10^{-3} $ & $ 0.049 $ & $ 1.2\times 10^{-6} $ & $ 3.9\times 10^{-5} $ & $
   1.3\times 10^{-3} $ & $ 0.043 $ \\ $
 \text{- + - +} $ & $ 9.9 $ & $ 27. $ & $ 63. $ & $ 110. $ & $ 26. $ & $ 78. $ & $ 220. $ & $ 570. $ \\ $
 \text{- + 0 0} $ & $ 3.3\times 10^{-7} $ & $ 8.6\times 10^{-6} $ & $ 2.0\times 10^{-4} $ & $ 4.6\times 10^{-3} $ & $ 1.3\times 10^{-6} $ & $ 4.1\times
   10^{-5} $ & $ 1.3\times 10^{-3} $ & $ 0.039 $ \\ $
 \text{0 0 0 0} $ & $ 0.48 $ & $ 1.5 $ & $ 4.1 $ & $ 11. $ & $ 0.37 $ & $ 1.1 $ & $ 3.4 $ & $ 9.3$ \\ \hline

\end{tabular}
}
\end{center}
\caption{Contribution of polarized elastic on-shell $W^+W^+$ scattering cross sections (in $pb$) to the total unpolarized cross sections at $\sqrt{s^U}$ for a chosen set of $f_{T0}$ values (first row, in TeV$^{−4}$).}
\label{tab:polarizedContributionsT0}
\end{table}
\clearpage
%----------------------------------------------------------------%

\begin{table}%
\begin{center}
\bgroup
\def\arraystretch{1.5}
\begin{tabular}{c|c|c}
 $\lambda_1\lambda_2\lambda_1'\lambda_2' $ &  $i\mathcal{M}_{\lambda_1'\lambda_2';\lambda_1\lambda_2}(\sqrt{s},\theta ,f_{T1}) $ & $ \mathcal{T}_{\lambda_1'\lambda_2';\lambda_1\lambda_2}^{(j=j_{min})} $ \\ \hline $
 \text{- - - -} $ & $ 2 i s^2 f_{\text{T1}} $ & $ \frac{s^2 f_{\text{T1}}}{8 \pi } $ \\ $
 \text{- - - 0} $ & $ -i \sqrt{2} \sqrt{s} c_W^3 f_{\text{T1}} \sin (\theta ) \cos (\theta ) m_Z^3 $ & $ \frac{\sqrt{s} c_W^3
   f_{\text{T1}} m_Z^3}{40 \sqrt{3} \pi } $ \\ $
 \text{- - - +} $ & $ -i s c_W^2 f_{\text{T1}} \sin ^2(\theta ) m_Z^2 $ & $ -\frac{s c_W^2 f_{\text{T1}} m_Z^2}{20 \sqrt{6} \pi } $ \\ $
 \text{- - 0 0} $ & $ -\frac{1}{2} i s c_W^2 f_{\text{T1}} (\cos (2 \theta )-9) m_Z^2 $ & $ \frac{7 s c_W^2 f_{\text{T1}} m_Z^2}{24 \pi
   } $ \\ $
 \text{- - 0 +} $ & $ -\frac{i s^{3/2} c_W f_{\text{T1}} \sin (\theta ) \cos (\theta ) m_Z}{\sqrt{2}} $ & $ \frac{s^{3/2} c_W
   f_{\text{T1}} m_Z}{80 \sqrt{3} \pi } $ \\ $
 \text{- - + +} $ & $ \frac{1}{4} i s^2 f_{\text{T1}} (\cos (2 \theta )+11) $ & $ \frac{s^2 f_{\text{T1}}}{6 \pi } $ \\ $
 \text{- 0 - 0} $ & $ -\frac{1}{2} i s c_W^2 f_{\text{T1}} \sin ^2(\theta ) m_Z^2 $ & $ -\frac{s c_W^2 f_{\text{T1}} m_Z^2}{96 \pi } $ \\ $
 \text{- 0 - +} $ & $ i \sqrt{2} s^{3/2} c_W f_{\text{T1}} \sin \left(\frac{\theta }{2}\right) \cos ^3\left(\frac{\theta
   }{2}\right) m_Z $ & $ -\frac{s^{3/2} c_W f_{\text{T1}} m_Z}{80 \sqrt{2} \pi } $ \\ $
 \text{- 0 0 0} $ & $ -i \sqrt{2} \sqrt{s} c_W^3 f_{\text{T1}} \sin (2 \theta ) m_Z^3 $ & $ -\frac{\sqrt{s} c_W^3 f_{\text{T1}}
   m_Z^3}{20 \sqrt{3} \pi } $ \\ $
 \text{- 0 0 +} $ & $ \frac{1}{4} i s c_W^2 f_{\text{T1}} (4 \cos (\theta )+3 \cos (2 \theta )+1) m_Z^2 $ & $ \frac{s c_W^2
   f_{\text{T1}} m_Z^2}{96 \pi } $ \\ $
 \text{- + - +} $ & $ i s^2 f_{\text{T1}} \cos ^4\left(\frac{\theta }{2}\right) $ & $ \frac{s^2 f_{\text{T1}}}{80 \pi } $ \\ $
 \text{- + 0 0} $ & $ i s c_W^2 f_{\text{T1}} \sin ^2(\theta ) m_Z^2 $ & $ \frac{s c_W^2 f_{\text{T1}} m_Z^2}{20 \sqrt{6} \pi } $ \\ $
 \text{0 0 0 0} $ & $ \frac{2 i \left(2 v^2 c_W^4 f_{\text{T1}} \cos (2 \theta ) m_Z^4+6 v^2 c_W^4 f_{\text{T1}} m_Z^4-m_h^2-4 \csc
   ^2(\theta ) m_Z^2+m_Z^2\right)}{v^2} $ & $ \frac{16 v^4 c_W^4 f_{\text{T1}} m_Z^4+3 v^2 \left(m_Z^2-m_h^2\right)}{24 \pi  v^4} $ \\ \hline

\end{tabular}
\egroup
\caption{Analytic formulas for on-shell $W^+W^+$ elastic scattering helicity amplitudes $iM$ and the minimal $j$ partial waves $\mathcal{T}^{(j=j_{min})}$; $\lambda_i$ ($\lambda_i'$) denote ingoing (outgoing) W's helicities; $c_W\equiv \cos\theta_W$; $\theta$ is the scattering angle; the $\mathcal{O}_{T1}$ operator case; only leading terms in the $\sqrt{s}$  expansion in the limit $s\rightarrow\infty$ are shown.}
\label{tab:analyticT1}
\end{center}
\end{table}

\begin{table}%
\begin{center}
\resizebox{1.05\columnwidth}{!}{
\begin{tabular}{c|cccc||cccc}
 $\lambda_1\lambda_2\lambda_1'\lambda_2' $ & $ 0.01 $ & $ 0.1 $ & $ 1. $ & $ 10. $ & $ -0.01 $ & $ -0.1 $ & $ -1. $ & $ -10. $ \\ \hline$
 \text{- - - -} $ & $ 5.0 $ & $ 15. $ & $ 41. $ & $ 100. $ & $ 22. $ & $ 68. $ & $ 210. $ & $ 610. $ \\ $
 \text{- - - 0} $ & $ 2.9\times 10^{-8} $ & $ 2.2\times 10^{-6} $ & $ 1.2\times 10^{-4} $ & $ 4.3\times 10^{-3} $ & $ 1.9\times 10^{-8} $ & $ 1.4\times 10^{-6} $ & $ 8.7\times 10^{-5} $ & $ 3.5\times 10^{-3} $ \\ $
 \text{- - - +} $ & $ 1.3\times 10^{-7} $ & $ 4.2\times 10^{-6} $ & $ 1.4\times 10^{-4} $ & $ 4.6\times 10^{-3} $ & $ 5.3\times 10^{-7} $ & $ 1.7\times 10^{-5} $ & $ 5.3\times 10^{-4} $ & $ 0.017 $ \\ $
 \text{- - 0 0} $ & $ 2.8\times 10^{-6} $ & $ 8.9\times 10^{-5} $ & $ 2.8\times 10^{-3} $ & $ 0.092 $ & $ 6.6\times 10^{-6} $ & $ 2.1\times 10^{-4} $ & $ 6.3\times 10^{-3} $ & $ 0.19 $ \\ $
 \text{- - 0 +} $ & $ 1.4\times 10^{-4} $ & $ 1.4\times 10^{-3} $ & $ 0.015 $ & $ 0.15 $ & $ 1.7\times 10^{-4} $ & $ 1.8\times 10^{-3} $ & $ 0.019 $ & $ 0.20 $ \\ $
 \text{- - + +} $ & $ 9.0 $ & $ 30. $ & $ 100. $ & $ 330. $ & $ 13. $ & $ 44. $ & $ 150. $ & $ 480. $ \\ $
 \text{- 0 - 0} $ & $ 6.5 $ & $ 19. $ & $ 53. $ & $ 130. $ & $ 5.8 $ & $ 17. $ & $ 48. $ & $ 120. $ \\ $
 \text{- 0 - +} $ & $ 3.8\times 10^{-4} $ & $ 3.5\times 10^{-3} $ & $ 0.031 $ & $ 0.24 $ & $ 2.4\times 10^{-3} $ & $ 0.023 $ & $ 0.23 $ & $ 2.3 $ \\ $
 \text{- 0 0 0} $ & $ 8.1\times 10^{-6} $ & $ 5.8\times 10^{-5} $ & $ 3.1\times 10^{-4} $ & $ 3.5\times 10^{-3} $ & $ 6.3\times 10^{-6} $ & $ 4.5\times 10^{-5} $ & $ 2.0\times 10^{-4} $ & $ 1.7\times 10^{-3} $ \\ $
 \text{- 0 0 +} $ & $ 8.4\times 10^{-7} $ & $ 2.5\times 10^{-5} $ & $ 6.7\times 10^{-4} $ & $ 0.017 $ & $ 3.3\times 10^{-7} $ & $ 1.1\times 10^{-5} $ & $ 3.5\times 10^{-4} $ & $ 0.012 $ \\ $
 \text{- + - +} $ & $ 8.7 $ & $ 24. $ & $ 60. $ & $ 110. $ & $ 17. $ & $ 52. $ & $ 150. $ & $ 370. $ \\ $
 \text{- + 0 0} $ & $ 1.5\times 10^{-7} $ & $ 3.9\times 10^{-6} $ & $ 8.1\times 10^{-5} $ & $ 1.4\times 10^{-3} $ & $ 5.8\times 10^{-7} $ & $ 1.7\times 10^{-5} $ & $ 5.1\times 10^{-4} $ & $ 0.015 $ \\ $
 \text{0 0 0 0} $ & $ 0.39 $ & $ 1.2 $ & $ 3.4 $ & $ 9.1 $ & $ 0.35 $ & $ 1.0 $ & $ 3.0 $ & $ 8.2$ \\ \hline

\end{tabular}
}
\end{center}
\caption{Contribution of polarized elastic on-shell $W^+W^+$ scattering cross sections (in $pb$) to the total unpolarized cross sections at $\sqrt{s^U}$ for a chosen set of $f_{T1}$ values (first row, in TeV$^{−4}$).}
\label{tab:polarizedContributionsT1}
\end{table}
\clearpage
%----------------------------------------------------------------%

\begin{table}%
\begin{center}
\bgroup
\def\arraystretch{1.5}
\begin{tabular}{c|c|c}
 $\lambda_1\lambda_2\lambda_1'\lambda_2' $ &  $i\mathcal{M}_{\lambda_1'\lambda_2';\lambda_1\lambda_2}(\sqrt{s},\theta ,f_{T2}) $ & $ \mathcal{T}_{\lambda_1'\lambda_2';\lambda_1\lambda_2}^{(j=j_{min})} $ \\ \hline $
  \text{- - - -} $ & $ i s^2 f_{\text{T2}} $ & $ \frac{s^2 f_{\text{T2}}}{16 \pi } $ \\ $
 \text{- - - 0} $ & $ -\frac{3 i \sqrt{s} c_W^3 f_{\text{T2}} \sin (2 \theta ) m_Z^3}{4 \sqrt{2}} $ & $ \frac{\sqrt{3} \sqrt{s} c_W^3
   f_{\text{T2}} m_Z^3}{160 \pi } $ \\ $
 \text{- - - +} $ & $ -\frac{3}{4} i s c_W^2 f_{\text{T2}} \sin ^2(\theta ) m_Z^2 $ & $ -\frac{\sqrt{\frac{3}{2}} s c_W^2 f_{\text{T2}}
   m_Z^2}{80 \pi } $ \\ $
 \text{- - 0 0} $ & $ -\frac{1}{4} i s c_W^2 f_{\text{T2}} (\cos (2 \theta )-5) m_Z^2 $ & $ \frac{s c_W^2 f_{\text{T2}} m_Z^2}{12 \pi }
   $ \\ $
 \text{- - 0 +} $ & $ -\frac{i s^{3/2} c_W f_{\text{T2}} \sin (2 \theta ) m_Z}{4 \sqrt{2}} $ & $ \frac{s^{3/2} c_W f_{\text{T2}}
   m_Z}{160 \sqrt{3} \pi } $ \\ $
 \text{- - + +} $ & $ \frac{1}{8} i s^2 f_{\text{T2}} (\cos (2 \theta )+3) $ & $ \frac{s^2 f_{\text{T2}}}{48 \pi } $ \\ $
 \text{- 0 - 0} $ & $ -\frac{1}{2} i s c_W^2 f_{\text{T2}} \sin ^2(\theta ) m_Z^2 $ & $ -\frac{s c_W^2 f_{\text{T2}} m_Z^2}{96 \pi } $ \\ $
 \text{- 0 - +} $ & $ i \sqrt{2} s^{3/2} c_W f_{\text{T2}} \sin \left(\frac{\theta }{2}\right) \cos ^3\left(\frac{\theta }{2}\right)
   m_Z $ & $ -\frac{s^{3/2} c_W f_{\text{T2}} m_Z}{80 \sqrt{2} \pi } $ \\ $
 \text{- 0 0 0} $ & $ -\frac{3 i \sqrt{s} c_W^3 f_{\text{T2}} \sin (\theta ) \cos (\theta ) m_Z^3}{\sqrt{2}} $ & $ -\frac{\sqrt{3}
   \sqrt{s} c_W^3 f_{\text{T2}} m_Z^3}{80 \pi } $ \\ $
 \text{- 0 0 +} $ & $ \frac{1}{4} i s c_W^2 f_{\text{T2}} (3 \cos (\theta )+2 \cos (2 \theta )+1) m_Z^2 $ & $ \frac{s c_W^2
   f_{\text{T2}} m_Z^2}{96 \pi } $ \\ $
 \text{- + - +} $ & $ i s^2 f_{\text{T2}} \cos ^4\left(\frac{\theta }{2}\right) $ & $ \frac{s^2 f_{\text{T2}}}{80 \pi } $ \\ $
 \text{- + 0 0} $ & $ i s c_W^2 f_{\text{T2}} \sin ^2(\theta ) m_Z^2 $ & $ \frac{s c_W^2 f_{\text{T2}} m_Z^2}{20 \sqrt{6} \pi } $ \\ $
 \text{0 0 0 0} $ & $ \frac{i \left(3 v^2 c_W^4 f_{\text{T2}} \cos (2 \theta ) m_Z^4+5 v^2 c_W^4 f_{\text{T2}} m_Z^4-2 m_h^2-8 \csc
   ^2(\theta ) m_Z^2+2 m_Z^2\right)}{v^2} $ & $ \frac{2 v^4 c_W^4 f_{\text{T2}} m_Z^4+v^2 \left(m_Z^2-m_h^2\right)}{8 \pi  v^4} $
   \\ \hline

\end{tabular}
\egroup
\caption{Analytic formulas for on-shell $W^+W^+$ elastic scattering helicity amplitudes $iM$ and the minimal $j$ partial waves $\mathcal{T}^{(j=j_{min})}$; $\lambda_i$ ($\lambda_i'$) denote ingoing (outgoing) W's helicities; $c_W\equiv \cos\theta_W$; $\theta$ is the scattering angle; the $\mathcal{O}_{T2}$ operator case; only leading terms in the $\sqrt{s}$  expansion in the limit $s\rightarrow\infty$ are shown.}
\label{tab:analyticT2}
\end{center}
\end{table}

\begin{table}%
\begin{center}
\resizebox{1.05\columnwidth}{!}{
\begin{tabular}{c|cccc||cccc}
 $\lambda_1\lambda_2\lambda_1'\lambda_2' $ & $ 0.01 $ & $ 0.1 $ & $ 1. $ & $ 10. $ & $ -0.01 $ & $ -0.1 $ & $ -1. $ & $ -10. $ \\ \hline $
 \text{- - - -} $ & $ 3.6 $ & $ 11. $ & $ 29. $ & $ 69. $ & $ 18. $ & $ 54. $ & $ 160. $ & $ 470. $ \\ $
 \text{- - - 0} $ & $ 9.2\times 10^{-9} $ & $ 7.3\times 10^{-7} $ & $ 4.5\times 10^{-5} $ & $ 1.8\times 10^{-3} $ & $ 3.1\times 10^{-9} $ & $ 2.9\times
   10^{-7} $ & $ 2.2\times 10^{-5} $ & $ 1.1\times 10^{-3} $ \\ $
 \text{- - - +} $ & $ 1.2\times 10^{-7} $ & $ 4.0\times 10^{-6} $ & $ 1.3\times 10^{-4} $ & $ 4.2\times 10^{-3} $ & $ 3.8\times 10^{-7} $ & $ 1.2\times
   10^{-5} $ & $ 3.8\times 10^{-4} $ & $ 0.012 $ \\ $
 \text{- - 0 0} $ & $ 3.8\times 10^{-7} $ & $ 1.1\times 10^{-5} $ & $ 3.1\times 10^{-4} $ & $ 8.6\times 10^{-3} $ & $ 1.0\times 10^{-6} $ & $ 3.1\times
   10^{-5} $ & $ 9.5\times 10^{-4} $ & $ 0.028 $ \\ $
 \text{- - 0 +} $ & $ 6.1\times 10^{-5} $ & $ 6.3\times 10^{-4} $ & $ 6.5\times 10^{-3} $ & $ 0.068 $ & $ 1.1\times 10^{-4} $ & $ 1.1\times 10^{-3} $ & $
   0.011 $ & $ 0.10 $ \\ $
 \text{- - + +} $ & $ 0.35 $ & $ 1.2 $ & $ 3.9 $ & $ 13. $ & $ 0.84 $ & $ 2.6 $ & $ 7.8 $ & $ 24. $ \\ $
 \text{- 0 - 0} $ & $ 4.9 $ & $ 15. $ & $ 41. $ & $ 100. $ & $ 3.7 $ & $ 11. $ & $ 34. $ & $ 88. $ \\ $
 \text{- 0 - +} $ & $ 4.4\times 10^{-4} $ & $ 4.5\times 10^{-3} $ & $ 0.046 $ & $ 0.47 $ & $ 3.0\times 10^{-3} $ & $ 0.030 $ & $ 0.29 $ & $ 2.9 $ \\ $
 \text{- 0 0 0} $ & $ 4.7\times 10^{-6} $ & $ 3.6\times 10^{-5} $ & $ 2.1\times 10^{-4} $ & $ 1.7\times 10^{-3} $ & $ 2.7\times 10^{-6} $ & $ 2.2\times
   10^{-5} $ & $ 1.2\times 10^{-4} $ & $ 5.1\times 10^{-4} $ \\ $
 \text{- 0 0 +} $ & $ 5.3\times 10^{-7} $ & $ 1.6\times 10^{-5} $ & $ 4.6\times 10^{-4} $ & $ 0.012 $ & $ 2.5\times 10^{-7} $ & $ 7.9\times 10^{-6} $ & $
   2.5\times 10^{-4} $ & $ 8.5\times 10^{-3} $ \\ $
 \text{- + - +} $ & $ 5.6 $ & $ 16. $ & $ 41. $ & $ 82. $ & $ 21. $ & $ 63. $ & $ 190. $ & $ 520. $ \\ $
 \text{- + 0 0} $ & $ 1.1\times 10^{-7} $ & $ 3.4\times 10^{-6} $ & $ 9.4\times 10^{-5} $ & $ 2.6\times 10^{-3} $ & $ 4.5\times 10^{-7} $ & $ 1.4\times
   10^{-5} $ & $ 4.4\times 10^{-4} $ & $ 0.013 $ \\ $
 \text{0 0 0 0} $ & $ 0.29 $ & $ 0.89 $ & $ 2.6 $ & $ 7.1 $ & $ 0.22 $ & $ 0.69 $ & $ 2.1 $ & $ 6.0$ \\ \hline

\end{tabular}
}
\end{center}
\caption{Contribution of polarized elastic on-shell $W^+W^+$ scattering cross sections (in $pb$) to the total unpolarized cross sections at $\sqrt{s^U}$ for a chosen set of $f_{T2}$ values (first row, in TeV$^{−4}$).}
\label{tab:polarizedContributionsT2}
\end{table}
\clearpage
%----------------------------------------------------------------%
\begin{table}%
\begin{center}
\bgroup
\def\arraystretch{1.5}
\begin{tabular}{c|c|c}
 $\lambda_1\lambda_2\lambda_1'\lambda_2' $ &  $i\mathcal{M}_{\lambda_1'\lambda_2';\lambda_1\lambda_2}(\sqrt{s},\theta ,f_{M0}) $ & $ \mathcal{T}_{\lambda_1'\lambda_2';\lambda_1\lambda_2}^{(j=j_{min})} $ \\ \hline $
\text{- - - -} $ & $ i c_W^2 m_Z^2 \left(c_W^2 f_{\text{M0}} \cos (2 \theta ) m_Z^2+3 c_W^2 f_{\text{M0}} m_Z^2-\frac{32 \csc
   ^2(\theta )}{v^2}\right) $ & $ \frac{c_W^4 f_{\text{M0}} m_Z^4}{6 \pi } $ \\ $
 \text{- - - 0} $ & $ -i \sqrt{2} \sqrt{s} c_W^3 f_{\text{M0}} \sin (\theta ) \cos (\theta ) m_Z^3 $ & $ \frac{\sqrt{s} c_W^3
   f_{\text{M0}} m_Z^3}{40 \sqrt{3} \pi } $ \\ $
 \text{- - - +} $ & $ -\frac{1}{2} i s c_W^2 f_{\text{M0}} \sin ^2(\theta ) m_Z^2 $ & $ -\frac{s c_W^2 f_{\text{M0}} m_Z^2}{40 \sqrt{6}
   \pi } $ \\ $
 \text{- - 0 0} $ & $ i s c_W^2 f_{\text{M0}} \sin ^2(\theta ) m_Z^2 $ & $ \frac{s c_W^2 f_{\text{M0}} m_Z^2}{24 \pi } $ \\ $
 \text{- - 0 +} $ & $ -\frac{i s^{3/2} c_W f_{\text{M0}} \sin (\theta ) \cos (\theta ) m_Z}{2 \sqrt{2}} $ & $ \frac{s^{3/2} c_W
   f_{\text{M0}} m_Z}{160 \sqrt{3} \pi } $ \\ $
 \text{- - + +} $ & $ \frac{1}{2} i s c_W^2 f_{\text{M0}} (\cos (2 \theta )+3) m_Z^2 $ & $ \frac{s c_W^2 f_{\text{M0}} m_Z^2}{12 \pi }
   $ \\ $
 \text{- 0 - 0} $ & $ -\frac{3}{4} i s c_W^2 f_{\text{M0}} \sin ^2(\theta ) m_Z^2 $ & $ -\frac{s c_W^2 f_{\text{M0}} m_Z^2}{64 \pi } $ \\ $
 \text{- 0 - +} $ & $ \frac{i s^{3/2} c_W f_{\text{M0}} \sin (\theta ) (\cos (\theta )+1) m_Z}{4 \sqrt{2}} $ & $ -\frac{s^{3/2} c_W
   f_{\text{M0}} m_Z}{160 \sqrt{2} \pi } $ \\ $
 \text{- 0 0 0} $ & $ -\frac{i s^{3/2} c_W f_{\text{M0}} \sin (\theta ) \cos (\theta ) m_Z}{2 \sqrt{2}} $ & $ -\frac{s^{3/2} c_W
   f_{\text{M0}} m_Z}{160 \sqrt{3} \pi } $ \\ $
 \text{- 0 0 +} $ & $ \frac{1}{2} i s^2 f_{\text{M0}} \cos ^4\left(\frac{\theta }{2}\right) $ & $ \frac{s^2 f_{\text{M0}}}{128 \pi } $ \\ $
 \text{- + - +} $ & $ 2 i s c_W^2 f_{\text{M0}} \cos ^4\left(\frac{\theta }{2}\right) m_Z^2 $ & $ \frac{s c_W^2 f_{\text{M0}} m_Z^2}{40
   \pi } $ \\ $
 \text{- + 0 0} $ & $ i s c_W^2 f_{\text{M0}} \sin ^2(\theta ) m_Z^2 $ & $ \frac{s c_W^2 f_{\text{M0}} m_Z^2}{20 \sqrt{6} \pi } $ \\ $
 \text{0 0 0 0} $ & $ 2 i s c_W^2 f_{\text{M0}} \cos ^2(\theta ) m_Z^2 $ & $ \frac{s c_W^2 f_{\text{M0}} m_Z^2}{24 \pi }$ \\ \hline

\end{tabular}
\egroup
\caption{Analytic formulas for on-shell $W^+W^+$ elastic scattering helicity amplitudes $iM$ and the minimal $j$ partial waves $\mathcal{T}^{(j=j_{min})}$; $\lambda_i$ ($\lambda_i'$) denote ingoing (outgoing) W's helicities; $c_W\equiv \cos\theta_W$; $\theta$ is the scattering angle; the $\mathcal{O}_{M0}$ operator case; only leading terms in the $\sqrt{s}$  expansion in the limit $s\rightarrow\infty$ are shown.}
\label{tab:analyticM0}
\end{center}
\end{table}

\begin{table}%
\begin{center}
\resizebox{1.05\columnwidth}{!}{
\begin{tabular}{c|cccc||cccc}
 $\lambda_1\lambda_2\lambda_1'\lambda_2' $ & $ 0.01 $ & $ 0.1 $ & $ 1. $ & $ 10. $ & $ -0.01 $ & $ -0.1 $ & $ -1. $ & $ -10. $ \\ \hline $
 \text{- - - -} $ & $ 3.2 $ & $ 9.9 $ & $ 28. $ & $ 70. $ & $ 3.2 $ & $ 9.9 $ & $ 28. $ & $ 71. $ \\ $
 \text{- - - 0} $ & $ 1.3\times 10^{-9} $ & $ 1.1\times 10^{-7} $ & $ 7.9\times 10^{-6} $ & $ 3.8\times 10^{-4} $ & $ 1.5\times 10^{-9} $ & $ 1.3\times 10^{-7} $ & $ 1.0\times 10^{-5} $ & $ 5.7\times 10^{-4} $ \\ $
 \text{- - - +} $ & $ 1.1\times 10^{-7} $ & $ 3.7\times 10^{-6} $ & $ 1.2\times 10^{-4} $ & $ 3.9\times 10^{-3} $ & $ 2.1\times 10^{-7} $ & $ 6.5\times 10^{-6} $ & $ 2.1\times 10^{-4} $ & $ 6.6\times 10^{-3} $ \\ $
 \text{- - 0 0} $ & $ 2.1\times 10^{-7} $ & $ 6.5\times 10^{-6} $ & $ 2.0\times 10^{-4} $ & $ 5.8\times 10^{-3} $ & $ 3.6\times 10^{-7} $ & $ 1.1\times 10^{-5} $ & $ 3.4\times 10^{-4} $ & $ 0.010 $ \\ $
 \text{- - 0 +} $ & $ 1.6\times 10^{-4} $ & $ 1.6\times 10^{-3} $ & $ 0.017 $ & $ 0.18 $ & $ 1.6\times 10^{-4} $ & $ 1.6\times 10^{-3} $ & $ 0.017 $ & $ 0.17 $ \\ $
 \text{- - + +} $ & $ 3.5\times 10^{-7} $ & $ 1.1\times 10^{-5} $ & $ 3.6\times 10^{-4} $ & $ 0.011 $ & $ 3.5\times 10^{-7} $ & $ 1.1\times 10^{-5} $ & $ 3.6\times 10^{-4} $ & $ 0.011 $ \\ $
 \text{- 0 - 0} $ & $ 3.0 $ & $ 9.3 $ & $ 27. $ & $ 72. $ & $ 3.0 $ & $ 9.3 $ & $ 27. $ & $ 72. $ \\ $
 \text{- 0 - +} $ & $ 3.2\times 10^{-4} $ & $ 3.3\times 10^{-3} $ & $ 0.034 $ & $ 0.37 $ & $ 1.3\times 10^{-3} $ & $ 0.013 $ & $ 0.13 $ & $ 1.4 $ \\ $
 \text{- 0 0 0} $ & $ 1.9\times 10^{-4} $ & $ 1.9\times 10^{-3} $ & $ 0.019 $ & $ 0.20 $ & $ 1.4\times 10^{-4} $ & $ 1.4\times 10^{-3} $ & $ 0.015 $ & $ 0.16 $ \\ $
 \text{- 0 0 +} $ & $ 5.1 $ & $ 17. $ & $ 54. $ & $ 180. $ & $ 5.1 $ & $ 17. $ & $ 54. $ & $ 170. $ \\ $
 \text{- + - +} $ & $ 5.6 $ & $ 17. $ & $ 48. $ & $ 120. $ & $ 5.6 $ & $ 17. $ & $ 49. $ & $ 120. $ \\ $
 \text{- + 0 0} $ & $ 2.2\times 10^{-7} $ & $ 7.0\times 10^{-6} $ & $ 2.2\times 10^{-4} $ & $ 7.3\times 10^{-3} $ & $ 4.6\times 10^{-7} $ & $ 1.4\times 10^{-5} $ & $ 4.5\times 10^{-4} $ & $ 0.014 $ \\ $
 \text{0 0 0 0} $ & $ 0.18 $ & $ 0.55 $ & $ 1.7 $ & $ 4.6 $ & $ 0.18 $ & $ 0.56 $ & $ 1.7 $ & $ 5.0 $ \\ \hline

\end{tabular}
}
\end{center}
\caption{Contribution of polarized elastic on-shell $W^+W^+$ scattering cross sections (in $pb$) to the total unpolarized cross sections at $\sqrt{s^U}$ for a chosen set of $f_{M0}$ values (first row, in TeV$^{−4}$).}
\label{tab:polarizedContributionsM0}
\end{table}
\clearpage
%----------------------------------------------------------------%

\begin{table}%
\begin{center}
\bgroup
\def\arraystretch{1.5}
\begin{tabular}{c|c|c}
$\lambda_1\lambda_2\lambda_1'\lambda_2' $ &  $i\mathcal{M}_{\lambda_1'\lambda_2';\lambda_1\lambda_2}(\sqrt{s},\theta ,f_{M1}) $ & $ \mathcal{T}_{\lambda_1'\lambda_2';\lambda_1\lambda_2}^{(j=j_{min})} $ \\ \hline $
 \text{- - - -} $ & $ -i s c_W^2 f_{\text{M1}} m_Z^2 $ & $ -\frac{s c_W^2 f_{\text{M1}} m_Z^2}{16 \pi } $ \\ $
 \text{- - - 0} $ & $ \frac{i \sqrt{s} c_W^3 f_{\text{M1}} \sin (\theta ) \cos (\theta ) m_Z^3}{2 \sqrt{2}} $ & $ -\frac{\sqrt{s} c_W^3
   f_{\text{M1}} m_Z^3}{160 \sqrt{3} \pi } $ \\ $
 \text{- - - +} $ & $ \frac{1}{8} i s c_W^2 f_{\text{M1}} \sin ^2(\theta ) m_Z^2 $ & $ \frac{s c_W^2 f_{\text{M1}} m_Z^2}{160 \sqrt{6}
   \pi } $ \\ $
 \text{- - 0 0} $ & $ \frac{1}{8} i s c_W^2 f_{\text{M1}} (\cos (2 \theta )-5) m_Z^2 $ & $ -\frac{s c_W^2 f_{\text{M1}} m_Z^2}{24 \pi }
   $ \\ $
 \text{- - 0 +} $ & $ \frac{i s^{3/2} c_W f_{\text{M1}} \sin (\theta ) \cos (\theta ) m_Z}{8 \sqrt{2}} $ & $ -\frac{s^{3/2} c_W
   f_{\text{M1}} m_Z}{640 \sqrt{3} \pi } $ \\ $
 \text{- - + +} $ & $ -\frac{1}{8} i s c_W^2 f_{\text{M1}} (\cos (2 \theta )+3) m_Z^2 $ & $ -\frac{s c_W^2 f_{\text{M1}} m_Z^2}{48 \pi }
   $ \\ $
 \text{- 0 - 0} $ & $ \frac{1}{8} i s^2 f_{\text{M1}} (\cos (\theta )+1) $ & $ \frac{s^2 f_{\text{M1}}}{192 \pi } $ \\ $
 \text{- 0 - +} $ & $ -\frac{i s^{3/2} c_W f_{\text{M1}} \sin (\theta ) (\cos (\theta )+1) m_Z}{16 \sqrt{2}} $ & $ \frac{s^{3/2} c_W
   f_{\text{M1}} m_Z}{640 \sqrt{2} \pi } $ \\ $
 \text{- 0 0 0} $ & $ \frac{i s^{3/2} c_W f_{\text{M1}} \sin (\theta ) \cos (\theta ) m_Z}{8 \sqrt{2}} $ & $ \frac{s^{3/2} c_W
   f_{\text{M1}} m_Z}{640 \sqrt{3} \pi } $ \\ $
 \text{- 0 0 +} $ & $ -\frac{1}{8} i s^2 f_{\text{M1}} \cos ^4\left(\frac{\theta }{2}\right) $ & $ -\frac{s^2 f_{\text{M1}}}{512 \pi }
   $ \\ $
 \text{- + - +} $ & $ -\frac{1}{2} i s c_W^2 f_{\text{M1}} \cos ^4\left(\frac{\theta }{2}\right) m_Z^2 $ & $ -\frac{s c_W^2
   f_{\text{M1}} m_Z^2}{160 \pi } $ \\ $
 \text{- + 0 0} $ & $ -\frac{1}{4} i s c_W^2 f_{\text{M1}} \sin ^2(\theta ) m_Z^2 $ & $ -\frac{s c_W^2 f_{\text{M1}} m_Z^2}{80 \sqrt{6}
   \pi } $ \\ $
 \text{0 0 0 0} $ & $ -\frac{1}{4} i s c_W^2 f_{\text{M1}} (\cos (2 \theta )+3) m_Z^2 $ & $ -\frac{s c_W^2 f_{\text{M1}} m_Z^2}{24 \pi} $ \\
\hline
\end{tabular}
\egroup
\caption{Analytic formulas for on-shell $W^+W^+$ elastic scattering helicity amplitudes $iM$ and the minimal $j$ partial waves $\mathcal{T}^{(j=j_{min})}$; $\lambda_i$ ($\lambda_i'$) denote ingoing (outgoing) W's helicities; $c_W\equiv \cos\theta_W$; $\theta$ is the scattering angle; the $\mathcal{O}_{M1}$ operator case; only leading terms in the $\sqrt{s}$  expansion in the limit $s\rightarrow\infty$ are shown.}
\label{tab:analyticM1}
\end{center}
\end{table}

\begin{table}%
\begin{center}
\resizebox{1.05\columnwidth}{!}{
\begin{tabular}{c|cccc||cccc}
 $\lambda_1\lambda_2\lambda_1'\lambda_2' $ & $ 0.01 $ & $ 0.1 $ & $ 1. $ & $ 10. $ & $ -0.01 $ & $ -0.1 $ & $ -1. $ & $ -10. $ \\\hline $
 \text{- - - -} $ & $ 1.6 $ & $ 5.1 $ & $ 15. $ & $ 42. $ & $ 1.6 $ & $ 5.1 $ & $ 15. $ & $ 41. $ \\ $
 \text{- - - 0} $ & $ 9.8\times 10^{-11} $ & $ 9.1\times 10^{-9} $ & $ 7.7\times 10^{-7} $ & $ 5.2\times 10^{-5} $ & $ 8.4\times 10^{-11} $ & $ 7.7\times 10^{-9} $ & $ 6.3\times 10^{-7} $ & $ 3.9\times 10^{-5} $ \\ $
 \text{- - - +} $ & $ 2.6\times 10^{-8} $ & $ 8.2\times 10^{-7} $ & $ 2.6\times 10^{-5} $ & $ 8.2\times 10^{-4} $ & $ 1.4\times 10^{-8} $ & $ 4.5\times 10^{-7} $ & $ 1.5\times 10^{-5} $ & $ 4.7\times 10^{-4} $ \\ $
 \text{- - 0 0} $ & $ 4.1\times 10^{-7} $ & $ 1.3\times 10^{-5} $ & $ 4.2\times 10^{-4} $ & $ 0.013 $ & $ 3.0\times 10^{-7} $ & $ 9.5\times 10^{-6} $ & $ 3.0\times 10^{-4} $ & $ 9.8\times 10^{-3} $ \\ $
 \text{- - 0 +} $ & $ 4.0\times 10^{-5} $ & $ 4.0\times 10^{-4} $ & $ 4.1\times 10^{-3} $ & $ 0.042 $ & $ 4.0\times 10^{-5} $ & $ 4.0\times 10^{-4} $ & $ 4.1\times 10^{-3} $ & $ 0.043 $ \\ $
 \text{- - + +} $ & $ 4.4\times 10^{-8} $ & $ 1.4\times 10^{-6} $ & $ 4.5\times 10^{-5} $ & $ 1.4\times 10^{-3} $ & $ 4.4\times 10^{-8} $ & $ 1.4\times 10^{-6} $ & $ 4.5\times 10^{-5} $ & $ 1.5\times 10^{-3} $ \\ $
 \text{- 0 - 0} $ & $ 13. $ & $ 41. $ & $ 130. $ & $ 430. $ & $ 25. $ & $ 80. $ & $ 260. $ & $ 820. $ \\ $
 \text{- 0 - +} $ & $ 3.3\times 10^{-4} $ & $ 3.3\times 10^{-3} $ & $ 0.033 $ & $ 0.34 $ & $ 7.9\times 10^{-5} $ & $ 8.1\times 10^{-4} $ & $ 8.4\times 10^{-3} $ & $ 0.088 $ \\ $
 \text{- 0 0 0} $ & $ 3.4\times 10^{-5} $ & $ 3.5\times 10^{-4} $ & $ 3.6\times 10^{-3} $ & $ 0.038 $ & $ 4.7\times 10^{-5} $ & $ 4.7\times 10^{-4} $ & $ 4.8\times 10^{-3} $ & $ 0.048 $ \\ $
 \text{- 0 0 +} $ & $ 2.5 $ & $ 8.2 $ & $ 27. $ & $ 88. $ & $ 2.5 $ & $ 8.2 $ & $ 27. $ & $ 89. $ \\ $
 \text{- + - +} $ & $ 2.9 $ & $ 8.8 $ & $ 26. $ & $ 72. $ & $ 2.9 $ & $ 8.8 $ & $ 26. $ & $ 71. $ \\ $
 \text{- + 0 0} $ & $ 5.7\times 10^{-8} $ & $ 1.8\times 10^{-6} $ & $ 5.7\times 10^{-5} $ & $ 1.8\times 10^{-3} $ & $ 2.7\times 10^{-8} $ & $ 8.7\times 10^{-7} $ & $ 2.8\times 10^{-5} $ & $ 8.9\times 10^{-4} $ \\ $
 \text{0 0 0 0} $ & $ 0.091 $ & $ 0.28 $ & $ 0.88 $ & $ 2.7 $ & $ 0.091 $ & $ 0.28 $ & $ 0.85 $ & $ 2.4$ \\ \hline

\end{tabular}
}
\end{center}
\caption{Contribution of polarized elastic on-shell $W^+W^+$ scattering cross sections (in $pb$) to the total unpolarized cross sections at $\sqrt{s^U}$ for a chosen set of $f_{M1}$ values (first row, in TeV$^{−4}$).}
\label{tab:polarizedContributionsM1}
\end{table}
\clearpage

%----------------------------------------------------------------%

\begin{table}%
\begin{center}
\bgroup
\def\arraystretch{1.5}
\begin{tabular}{c|c|c}
$\lambda_1\lambda_2\lambda_1'\lambda_2' $ &  $i\mathcal{M}_{\lambda_1'\lambda_2';\lambda_1\lambda_2}(\sqrt{s},\theta ,f_{M7}) $ & $ \mathcal{T}_{\lambda_1'\lambda_2';\lambda_1\lambda_2}^{(j=j_{min})} $ \\ \hline $
 \text{- - - -} $ & $ \frac{1}{2} i s c_W^2 f_{\text{M7}} m_Z^2 $ & $ \frac{s c_W^2 f_{\text{M7}} m_Z^2}{32 \pi } $ \\ $
 \text{- - - 0} $ & $ -\frac{2 i \sqrt{2} c_W^3 \cot (\theta ) \csc ^2(\theta ) m_Z^3 \left(\cos (2 \theta ) \left(\left(4
   c_W^2+1\right) m_Z^2-m_h^2\right)+\left(28 c_W^2-25\right) m_Z^2-7 m_h^2\right)}{s^{3/2} v^2} $ & $ 0 $ \\ $
 \text{- - - +} $ & $ \frac{16 i c_W^4 m_Z^4}{s v^2} $ & $ \frac{c_W^4 m_Z^4}{\sqrt{6} \pi  s v^2} $ \\ $
 \text{- - 0 0} $ & $ \frac{1}{2} i s c_W^2 f_{\text{M7}} m_Z^2 $ & $ \frac{s c_W^2 f_{\text{M7}} m_Z^2}{32 \pi } $ \\ $
 \text{- - 0 +} $ & $ \frac{4 i \sqrt{2} c_W^3 \cot (\theta ) m_Z^3 \left(\left(4 c_W^2+1\right) m_Z^2-m_h^2\right)}{s^{3/2} v^2} $ & $
   -\frac{c_W^3 m_Z^3 \left(\left(4 c_W^2+1\right) m_Z^2-m_h^2\right)}{4 \sqrt{3} \pi  s^{3/2} v^2} $ \\ $
 \text{- - + +} $ & $ \frac{1}{2} i s c_W^2 f_{\text{M7}} m_Z^2 $ & $ \frac{s c_W^2 f_{\text{M7}} m_Z^2}{32 \pi } $ \\ $
 \text{- 0 - 0} $ & $ -\frac{1}{16} i s^2 f_{\text{M7}} (\cos (\theta )+1) $ & $ -\frac{s^2 f_{\text{M7}}}{384 \pi } $ \\ $
 \text{- 0 - +} $ & $ -\frac{8 i \sqrt{2} c_W^3 \cot \left(\frac{\theta }{2}\right) m_Z^3}{\sqrt{s} v^2} $ & $ \frac{\sqrt{2} c_W^3
   m_Z^3}{3 \pi  \sqrt{s} v^2} $ \\ $
 \text{- 0 0 0} $ & $ \frac{2 i \sqrt{2} c_W \cot (\theta ) m_Z \left(m_Z-m_h\right) \left(m_h+m_Z\right)}{\sqrt{s} v^2} $ & $ \frac{c_W
   m_Z \left(m_Z^2-m_h^2\right)}{8 \sqrt{3} \pi  \sqrt{s} v^2} $ \\ $
 \text{- 0 0 +} $ & $ \frac{1}{16} i s^2 f_{\text{M7}} (\cos (\theta )+1) $ & $ \frac{s^2 f_{\text{M7}}}{384 \pi } $ \\ $
 \text{- + - +} $ & $ -\frac{8 i c_W^2 \cot ^2\left(\frac{\theta }{2}\right) m_Z^2}{v^2} $ & $ \frac{c_W^2 m_Z^2}{6 \pi  v^2} $ \\ $
 \text{- + 0 0} $ & $ -\frac{4 i c_W^2 m_Z^2 \left(\left(8 c_W^2+1\right) m_Z^2+2 \csc ^2(\theta )
   \left(m_h^2-m_Z^2\right)-m_h^2\right)}{s v^2} $ & $ -\frac{c_W^2 m_Z^2 \left(\left(4 c_W^2-1\right) m_Z^2+m_h^2\right)}{2
   \sqrt{6} \pi  s v^2} $ \\ $
 \text{0 0 0 0} $ & $ \frac{1}{2} i s c_W^2 f_{\text{M7}} m_Z^2 $ & $ \frac{s c_W^2 f_{\text{M7}} m_Z^2}{32 \pi }$ \\ \hline

\end{tabular}
\egroup
\caption{Analytic formulas for on-shell $W^+W^+$ elastic scattering helicity amplitudes $iM$ and the minimal $j$ partial waves $\mathcal{T}^{(j=j_{min})}$; $\lambda_i$ ($\lambda_i'$) denote ingoing (outgoing) W's helicities; $c_W\equiv \cos\theta_W$; $\theta$ is the scattering angle; the $\mathcal{O}_{M7}$ operator case; only leading terms in the $\sqrt{s}$  expansion in the limit $s\rightarrow\infty$ are shown.}
\label{tab:analyticM7}
\end{center}
\end{table}

\begin{table}%
\begin{center}
\resizebox{1.05\columnwidth}{!}{
\begin{tabular}{c|cccc||cccc}
 $\lambda_1\lambda_2\lambda_1'\lambda_2' $ & $ 0.01 $ & $ 0.1 $ & $ 1. $ & $ 10. $ & $ -0.01 $ & $ -0.1 $ & $ -1. $ & $ -10. $ \\\hline $
 \text{- - - -} $ & $ 1.2 $ & $ 3.6 $ & $ 11. $ & $ 31. $ & $ 1.2 $ & $ 3.6 $ & $ 11. $ & $ 32. $ \\ $
 \text{- - - 0} $ & $ 2.2\times 10^{-11} $ & $ 2.1\times 10^{-9} $ & $ 1.8\times 10^{-7} $ & $ 1.3\times 10^{-5} $ & $ 2.2\times 10^{-11} $ & $
   2.1\times 10^{-9} $ & $ 1.8\times 10^{-7} $ & $ 1.3\times 10^{-5} $ \\ $
 \text{- - - +} $ & $ 1.8\times 10^{-10} $ & $ 5.6\times 10^{-9} $ & $ 1.7\times 10^{-7} $ & $ 5.2\times 10^{-6} $ & $ 1.8\times 10^{-10} $ & $
   5.6\times 10^{-9} $ & $ 1.7\times 10^{-7} $ & $ 5.3\times 10^{-6} $ \\ $
 \text{- - 0 0} $ & $ 2.4\times 10^{-7} $ & $ 7.6\times 10^{-6} $ & $ 2.4\times 10^{-4} $ & $ 7.9\times 10^{-3} $ & $ 3.1\times 10^{-7} $ & $ 9.8\times
   10^{-6} $ & $ 3.1\times 10^{-4} $ & $ 9.8\times 10^{-3} $ \\ $
 \text{- - 0 +} $ & $ 1.2\times 10^{-14} $ & $ 1.2\times 10^{-12} $ & $ 1.1\times 10^{-10} $ & $ 1.0\times 10^{-8} $ & $ 1.2\times 10^{-14} $ & $
   1.2\times 10^{-12} $ & $ 1.1\times 10^{-10} $ & $ 1.0\times 10^{-8} $ \\ $
 \text{- - + +} $ & $ 1.3\times 10^{-7} $ & $ 4.3\times 10^{-6} $ & $ 1.4\times 10^{-4} $ & $ 4.4\times 10^{-3} $ & $ 1.3\times 10^{-7} $ & $ 4.3\times
   10^{-6} $ & $ 1.4\times 10^{-4} $ & $ 4.4\times 10^{-3} $ \\ $
 \text{- 0 - 0} $ & $ 17. $ & $ 56. $ & $ 180. $ & $ 580. $ & $ 8.8 $ & $ 29. $ & $ 93. $ & $ 300. $ \\ $
 \text{- 0 - +} $ & $ 2.2\times 10^{-5} $ & $ 2.2\times 10^{-4} $ & $ 2.1\times 10^{-3} $ & $ 0.020 $ & $ 2.2\times 10^{-5} $ & $ 2.2\times 10^{-4} $ & $
   2.1\times 10^{-3} $ & $ 0.020 $ \\ $
 \text{- 0 0 0} $ & $ 2.4\times 10^{-7} $ & $ 2.3\times 10^{-6} $ & $ 1.9\times 10^{-5} $ & $ 1.1\times 10^{-4} $ & $ 2.4\times 10^{-7} $ & $ 2.3\times
   10^{-6} $ & $ 1.9\times 10^{-5} $ & $ 1.1\times 10^{-4} $ \\ $
 \text{- 0 0 +} $ & $ 12. $ & $ 39. $ & $ 130. $ & $ 410. $ & $ 12. $ & $ 39. $ & $ 130. $ & $ 410. $ \\ $
 \text{- + - +} $ & $ 2.0 $ & $ 6.3 $ & $ 19. $ & $ 53. $ & $ 2.0 $ & $ 6.3 $ & $ 19. $ & $ 54. $ \\ $
 \text{- + 0 0} $ & $ 1.2\times 10^{-9} $ & $ 3.7\times 10^{-8} $ & $ 1.1\times 10^{-6} $ & $ 2.8\times 10^{-5} $ & $ 1.2\times 10^{-9} $ & $ 3.7\times
   10^{-8} $ & $ 1.1\times 10^{-6} $ & $ 2.9\times 10^{-5} $ \\ $
 \text{0 0 0 0} $ & $ 0.064 $ & $ 0.20 $ & $ 0.61 $ & $ 1.8 $ & $ 0.064 $ & $ 0.20 $ & $ 0.63 $ & $ 1.9 $\\ \hline

\end{tabular}
}
\end{center}
\caption{Contribution of polarized elastic on-shell $W^+W^+$ scattering cross sections (in $pb$) to the total unpolarized cross sections at $\sqrt{s^U}$ for a chosen set of $f_{M7}$ values (first row, in TeV$^{−4}$).}
\label{tab:polarizedContributionsM7}
\end{table}
\clearpage

\newgeometry{tmargin=2cm, bmargin=1cm, lmargin=2cm, rmargin=2cm} 
\section{Plots of $\sigma_{pol}(s)$ and $\sigma^{tot}_{unpol}(s)$}
\label{app:plotsTotPolAndUnpolSMEFT}
\thispagestyle{empty} 
\begin{figure}[h]
\begin{center}
\resizebox{1.\columnwidth}{!}{
\begin{tabular}{cc}
 \includegraphics[width=0.52\columnwidth]{plotSigmaTotCs0GeqLeqGridZoom.pdf} &\includegraphics[width=0.52\columnwidth]{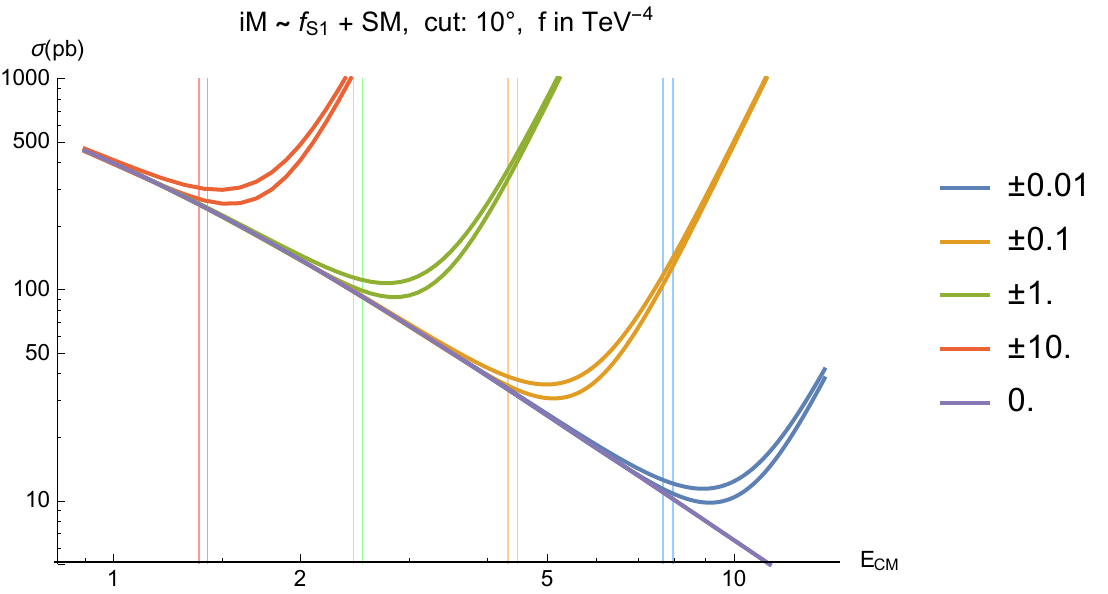} \\
\includegraphics[width=0.52\columnwidth]{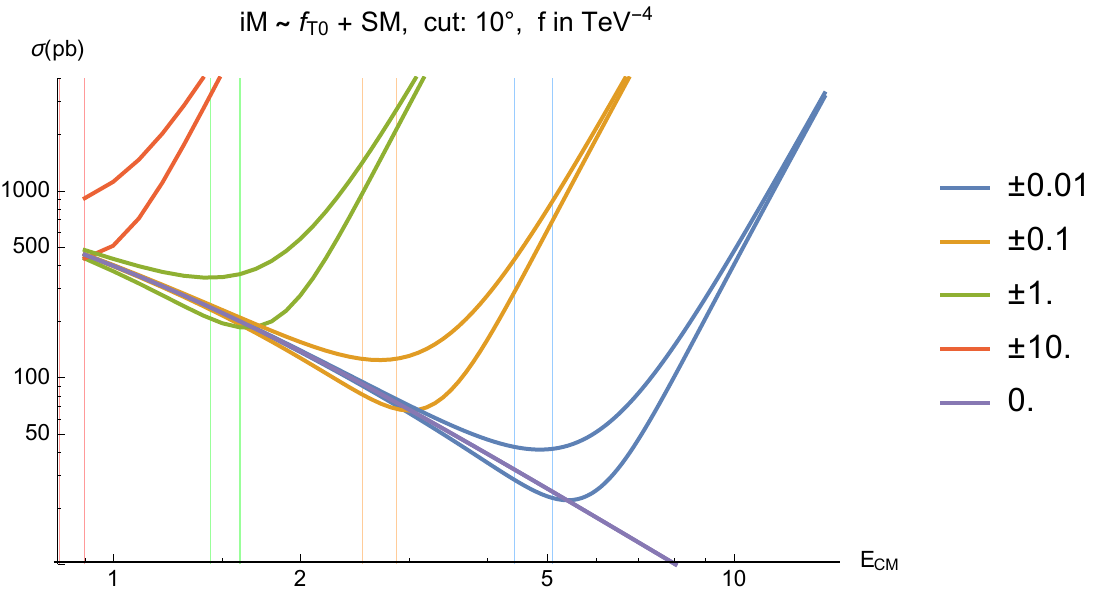} & \includegraphics[width=0.52\columnwidth]{plotSigmaTotCt1GeqLeqGridZoom.pdf}  \\
\includegraphics[width=0.52\columnwidth]{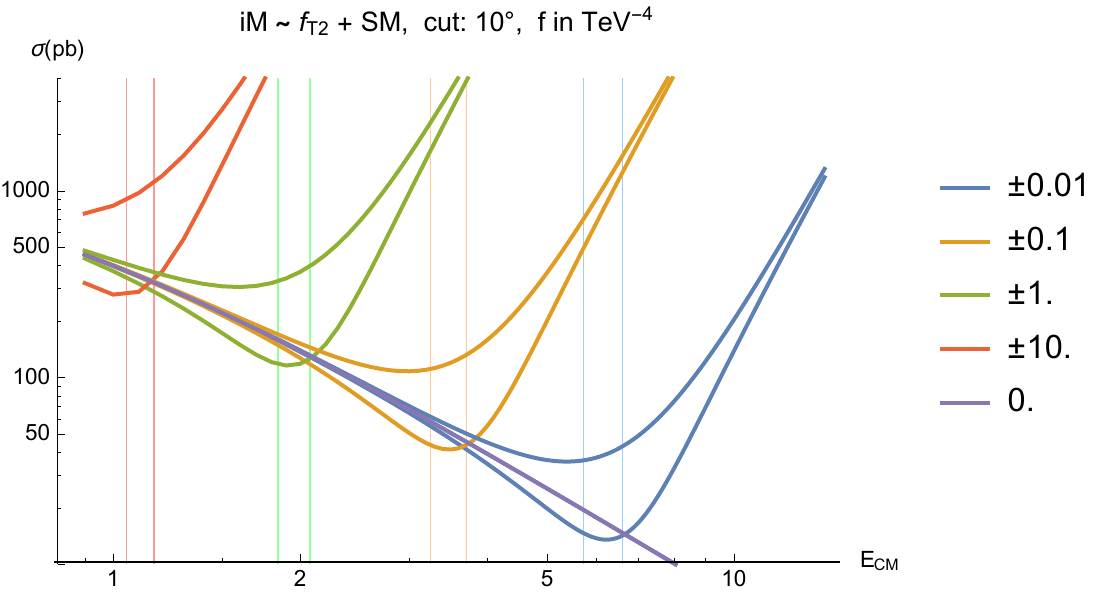} &  \includegraphics[width=0.52\columnwidth]{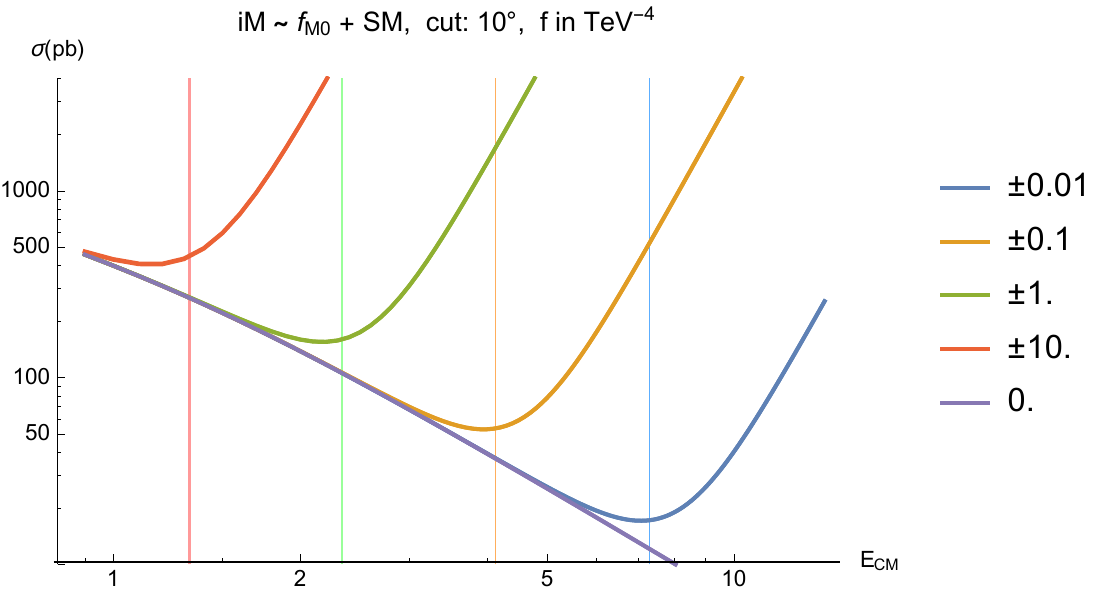} \\
 \includegraphics[width=0.52\columnwidth]{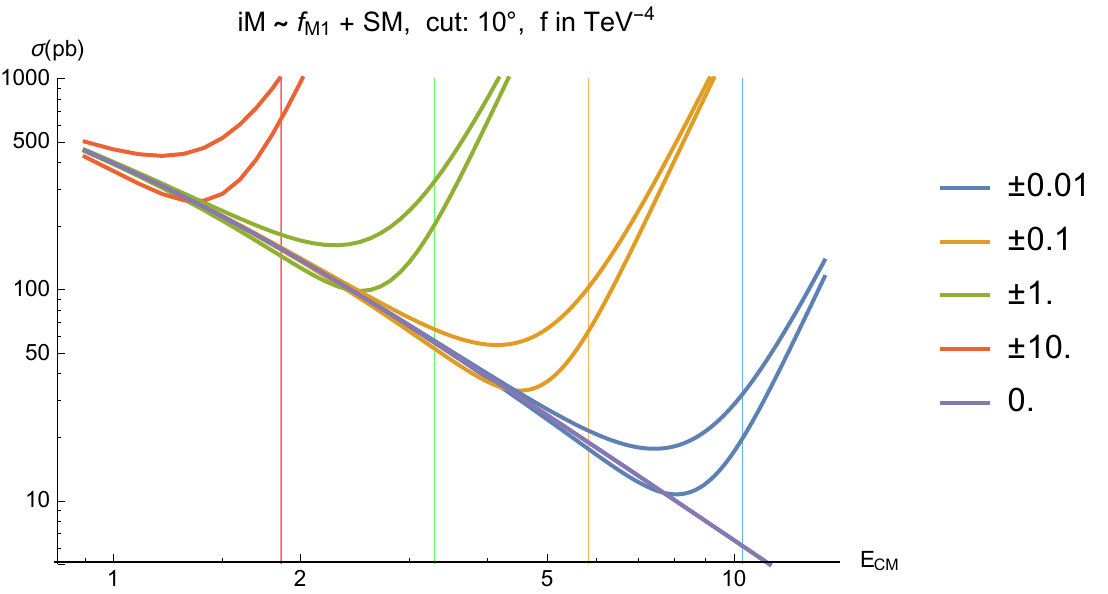} & \includegraphics[width=0.52\columnwidth]{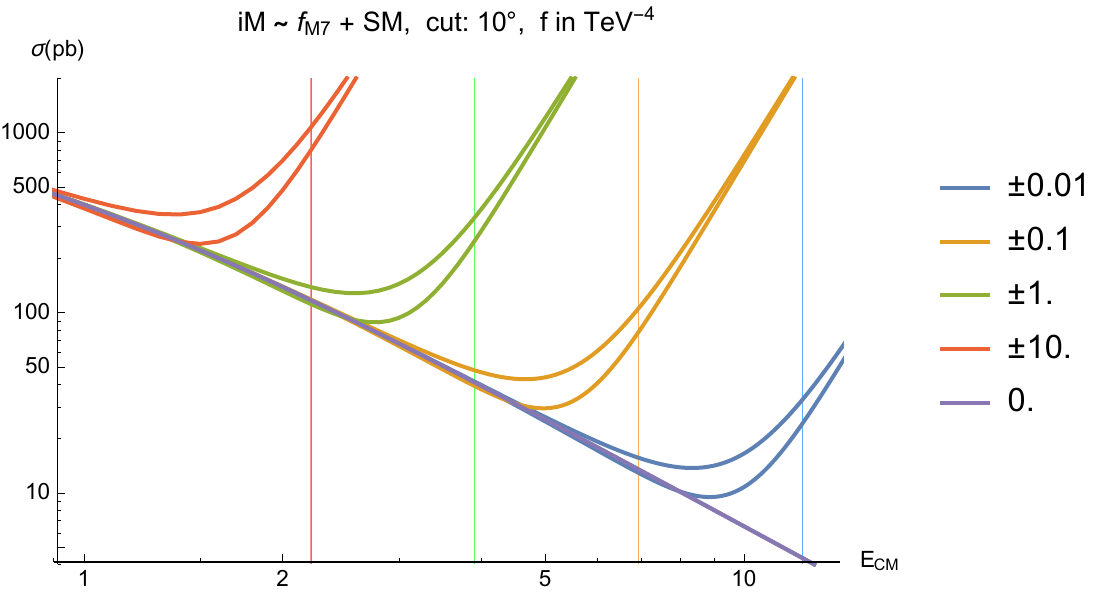}
\end{tabular}
}
\end{center}
\caption{Energy dependence of the total unpolarized elastic on-shell $W^+W^+$ cross sections ($E_{CM} \equiv \sqrt{s}$, in TeV) for a chosen set of $f_i$ values of the SMEFT operators studied.  Vertical lines denote the unitarity bound $\sqrt{s^U}$ (color correspondence).  There is no color distinction between the signs: except for M1 and M7, upper cross section curves correspond to $f<0$; in S0,S1 (T0,T2) stronger unitarity limits correspond to $f<0$ ($f>0$).}
\label{tab:tutalUnpol}
\end{figure}
%\clearpage
%\section{Contributions of polarized cross sections to the total unpolarized cross section at $\sqrt{s^U}$: numerical results}
%\thispagestyle{empty} 
%\label{app:contribPolarizedSMEFT}
%
%\newpage
\clearpage
\newgeometry{tmargin=2cm, bmargin=3cm, lmargin=2cm, rmargin=2cm}
\begin{figure}
\begin{center}
\begin{tabular}{cc}
 \includegraphics[width=0.5\columnwidth]{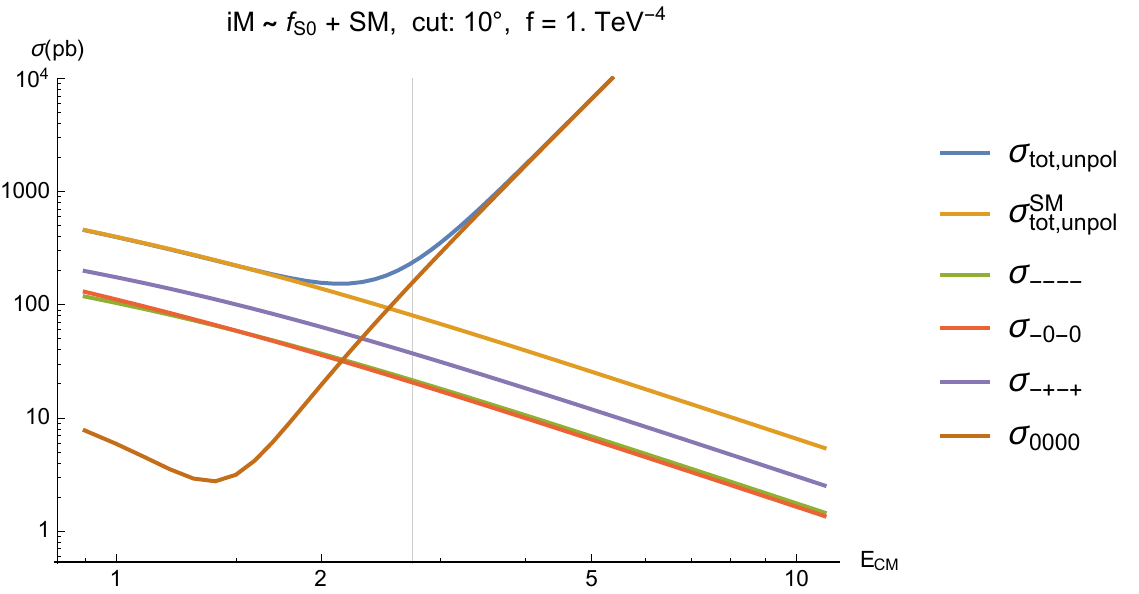}& \includegraphics[width=0.5\columnwidth]{plotlistPlot10DegreeParticPolsfs0Eq1Zoom.pdf} \\

\includegraphics[width=0.50\columnwidth]{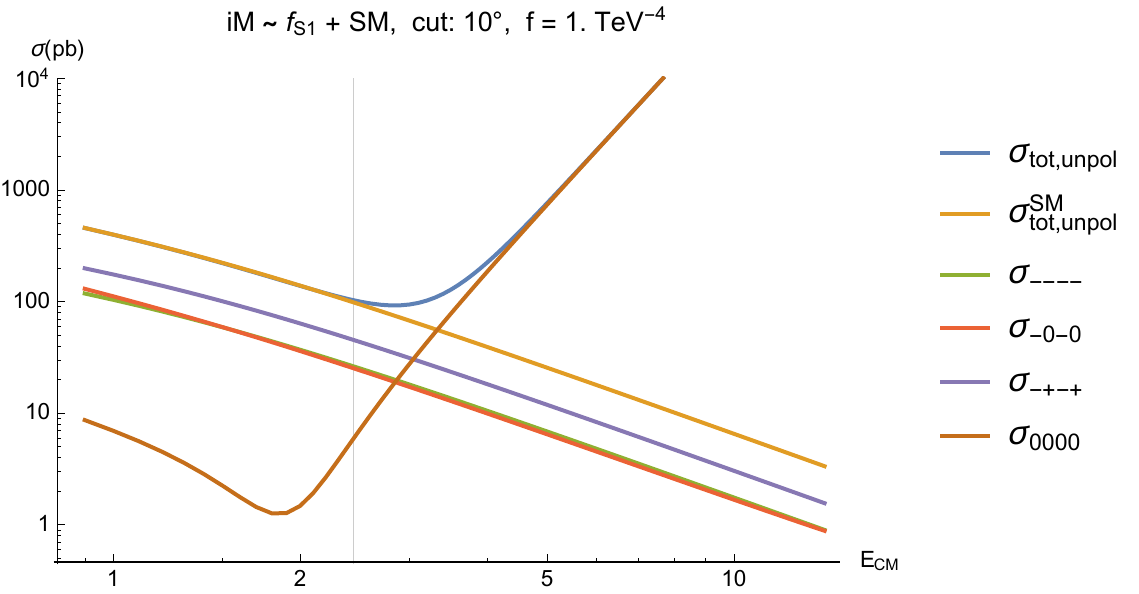}& \includegraphics[width=0.5\columnwidth]{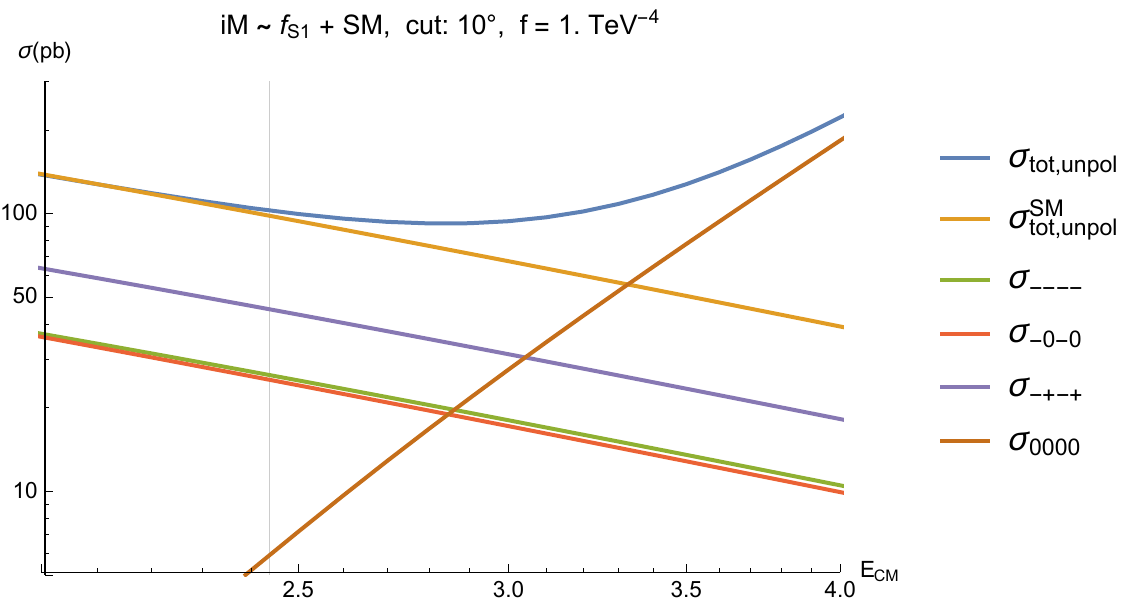} \\

\includegraphics[width=0.5\columnwidth]{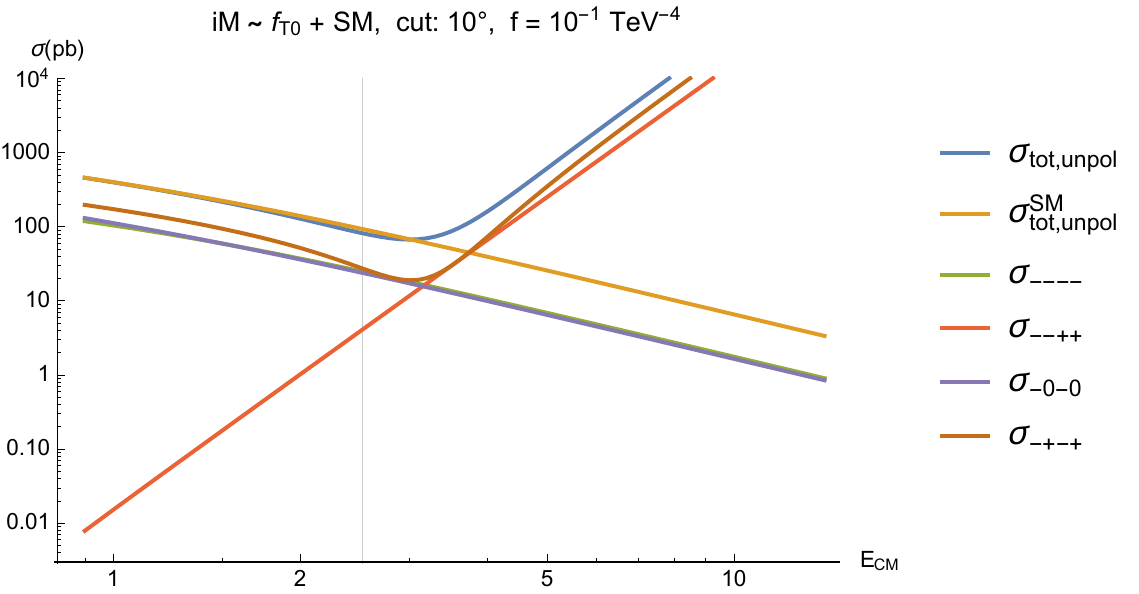}& \includegraphics[width=0.5\columnwidth]{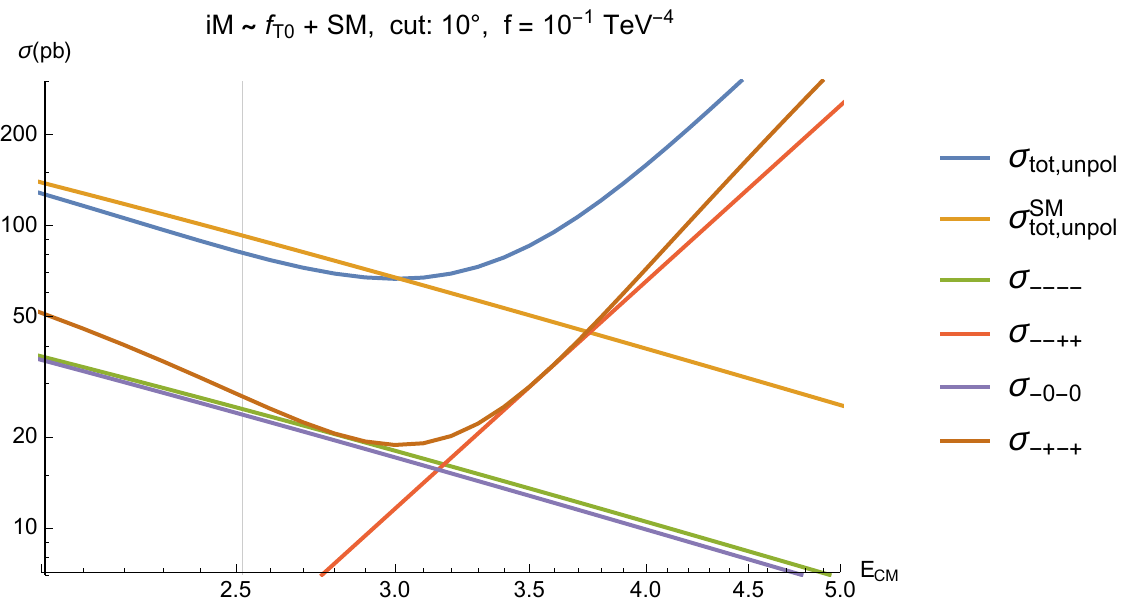} \\

\includegraphics[width=0.5\columnwidth]{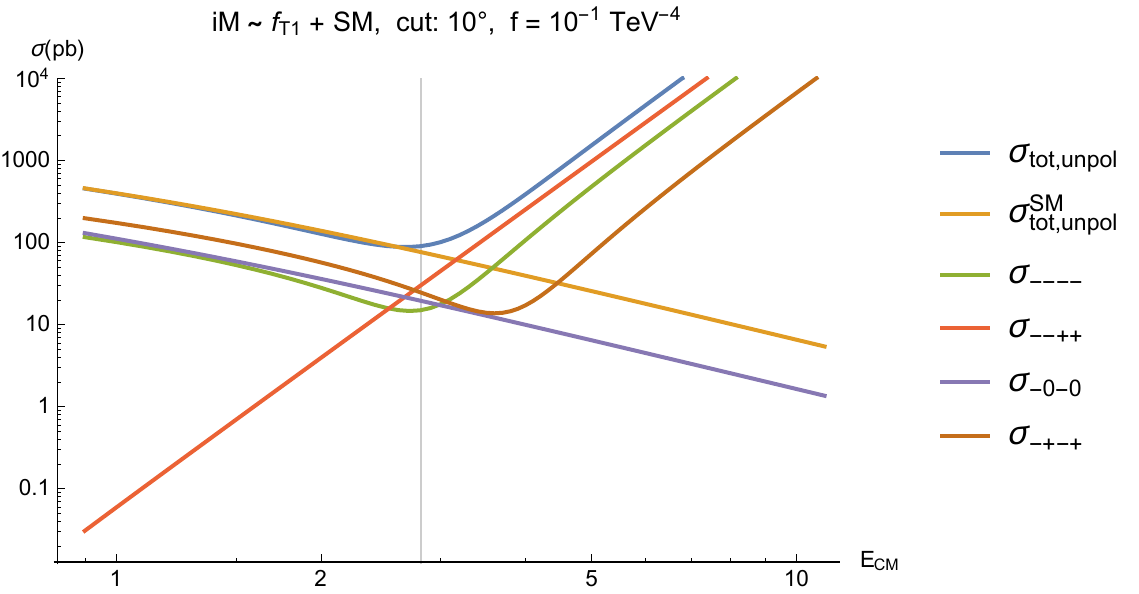}& \includegraphics[width=0.5\columnwidth]{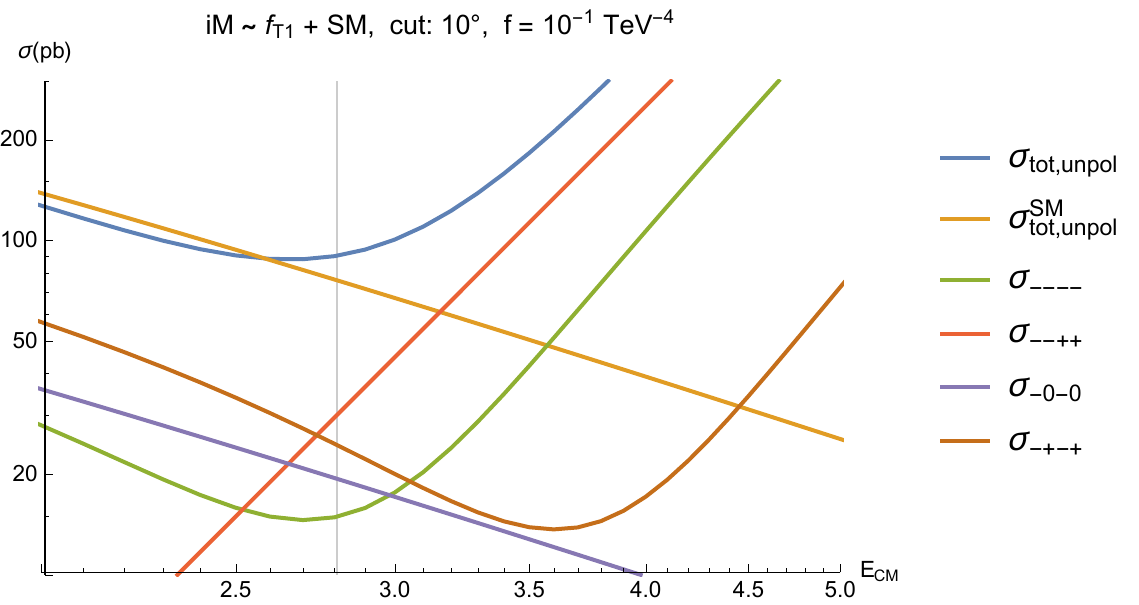} \\
\end{tabular}
\end{center}
\caption{Contributions of the polarized cross sections (multiplicity taken into account) as functions of the center-of-mass collision energy
($E_{CM} \equiv \sqrt{s}$, in TeV) for chosen values of $f_i>0$ of the SMEFT operators studied. The remaining (not shown) polarized contributions are negligibly small. In each plot shown is in addition the total cross section of a EFT ''model'' and the total cross section in the SM. In each row right plot is zoom in of the left one; part 1.}
\label{tab:polsSMEFT}
\end{figure}

\begin{figure}
\begin{center}
\begin{tabular}{cc}
 \includegraphics[width=0.5\columnwidth]{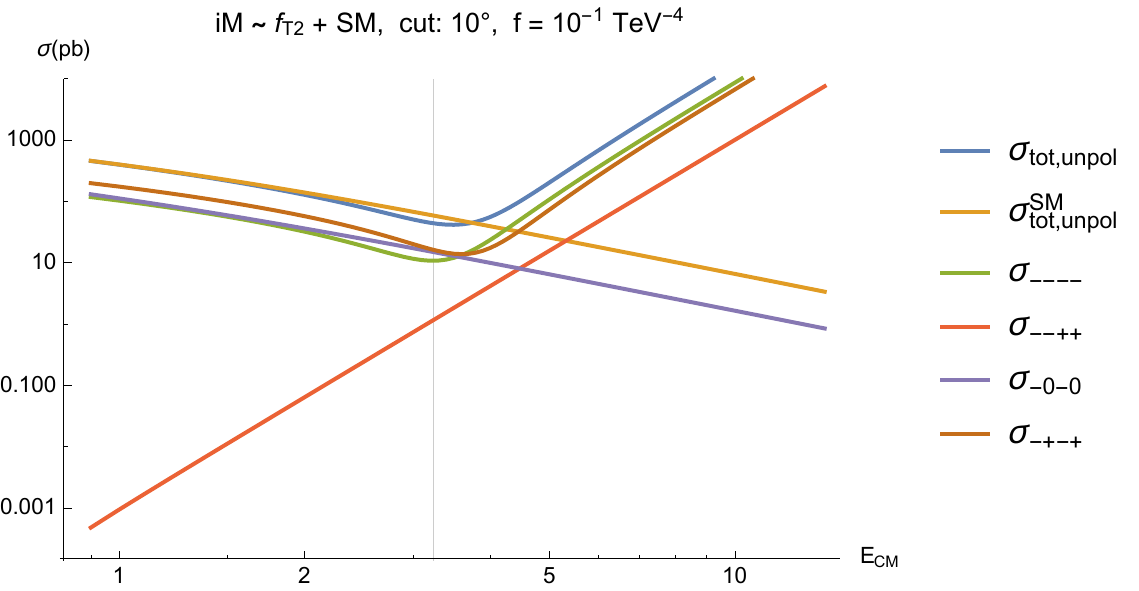}&  \includegraphics[width=0.52\columnwidth]{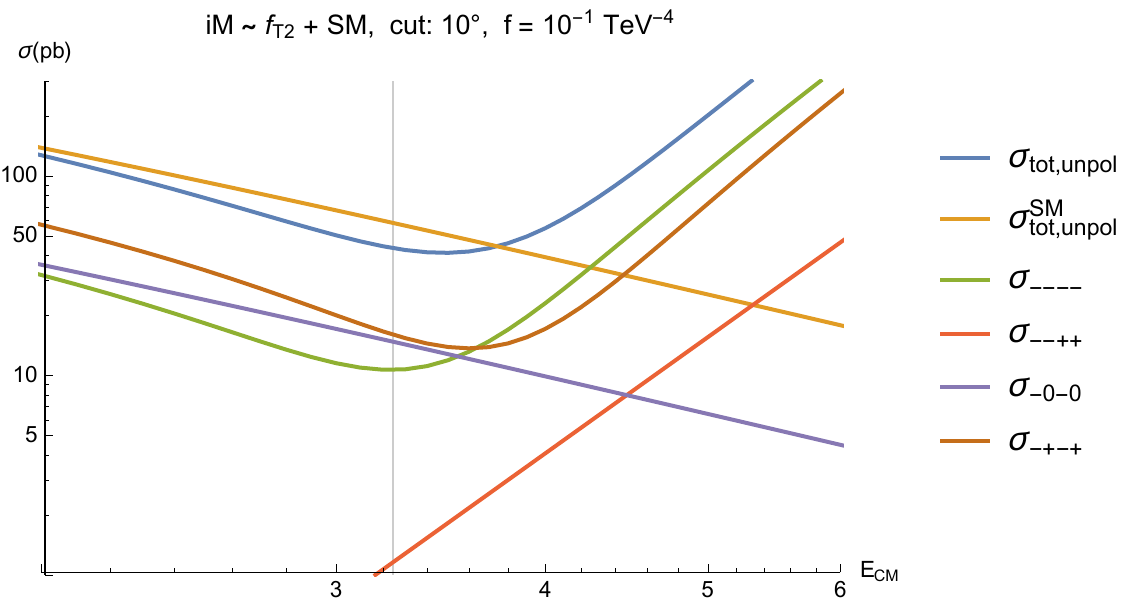} 
\\

\includegraphics[width=0.5\columnwidth]{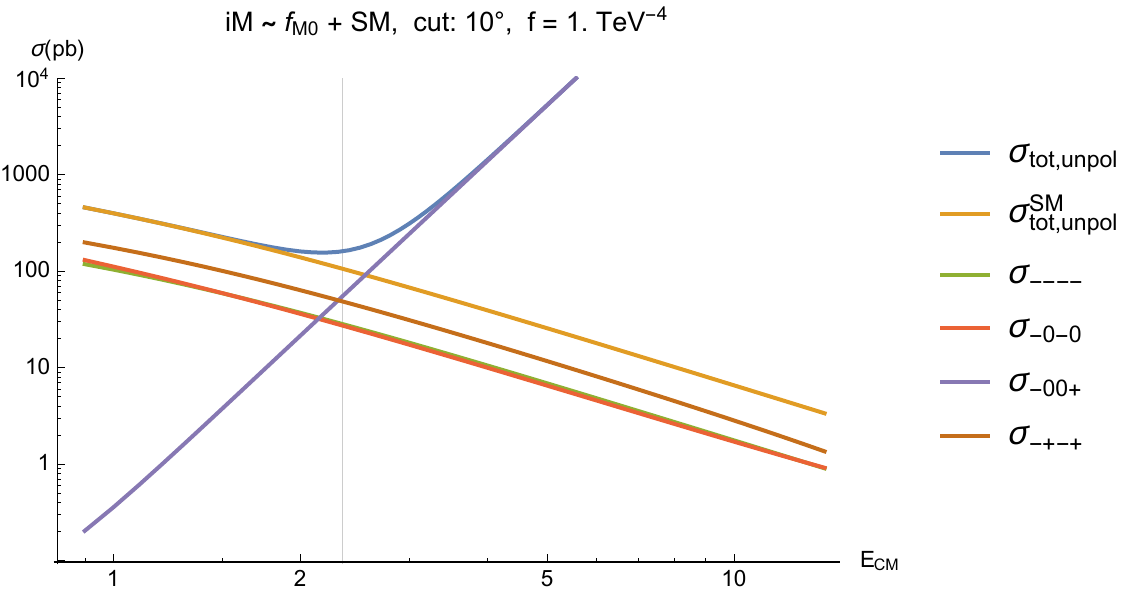}
&\includegraphics[width=0.5\columnwidth]{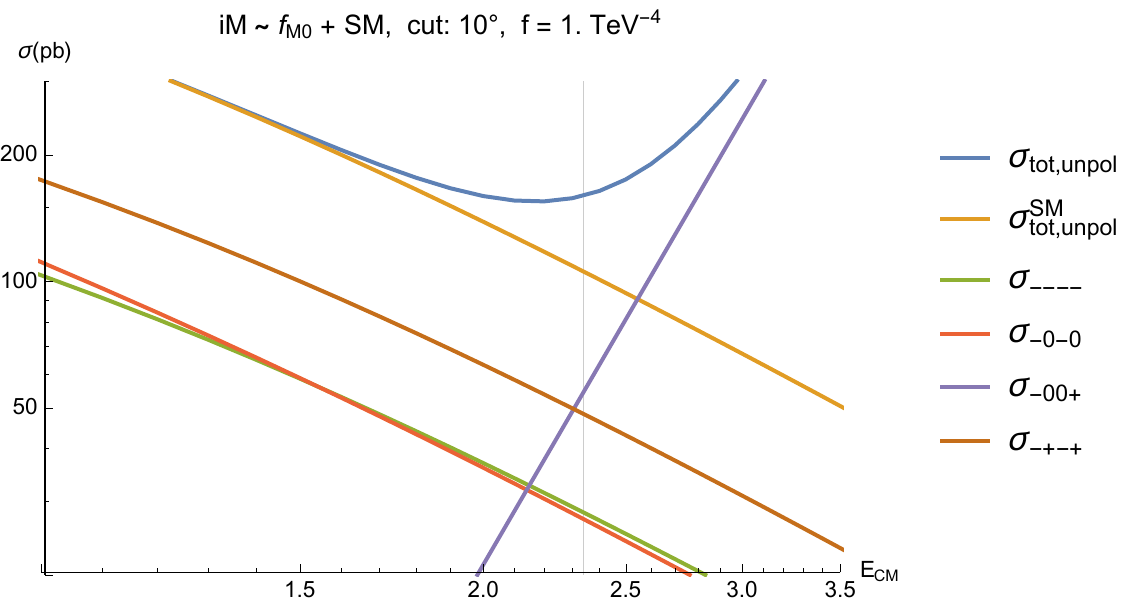} \\

\includegraphics[width=0.5\columnwidth]{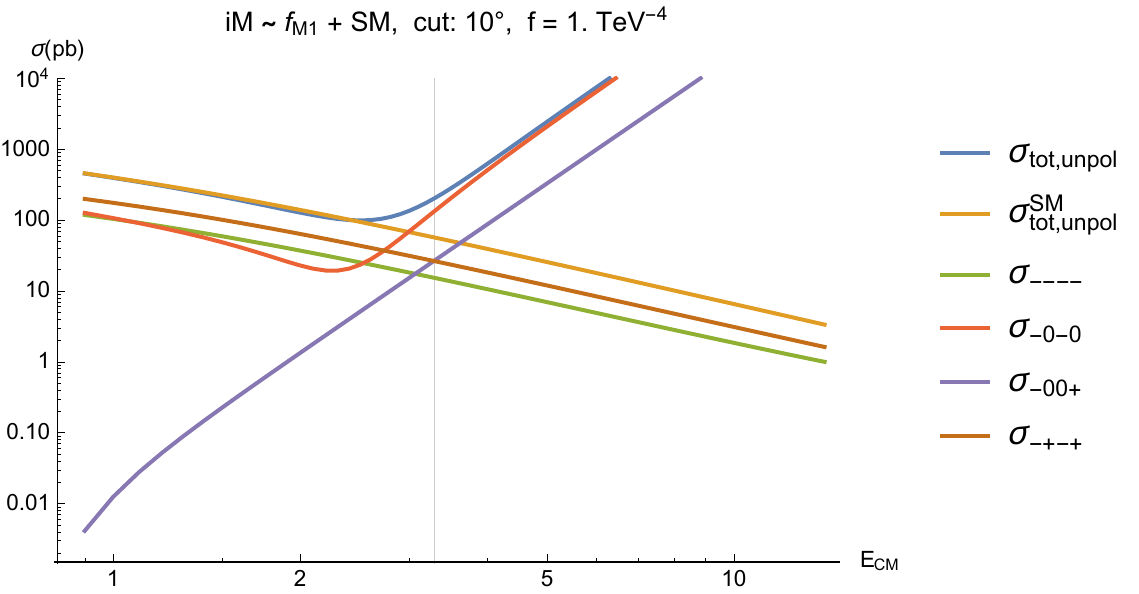}& \includegraphics[width=0.5\columnwidth]{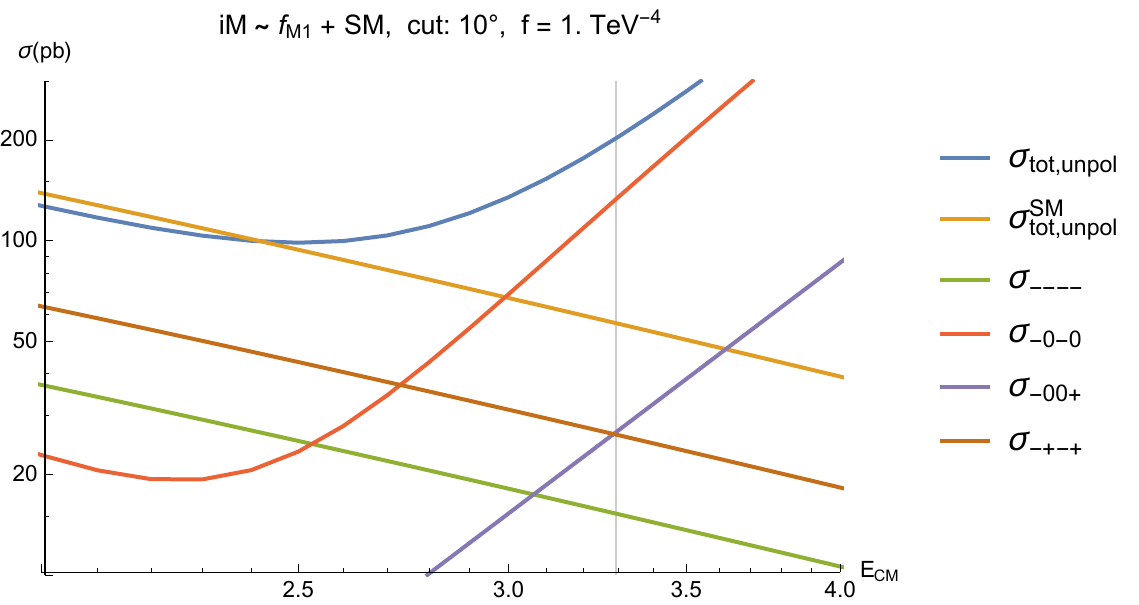} \\

\includegraphics[width=0.5\columnwidth]{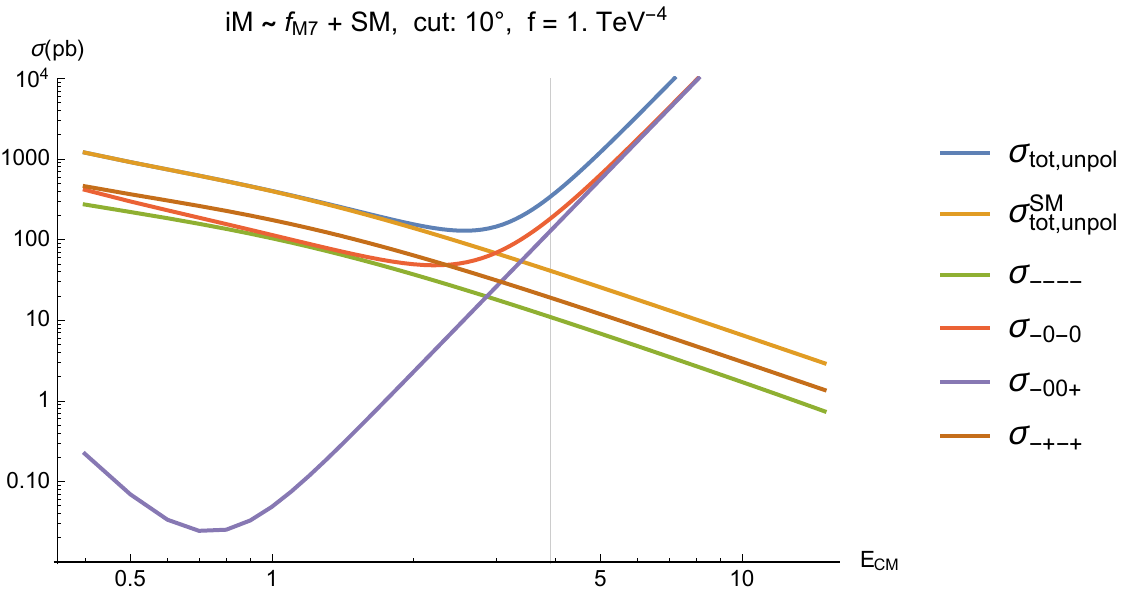}& \includegraphics[width=0.5\columnwidth]{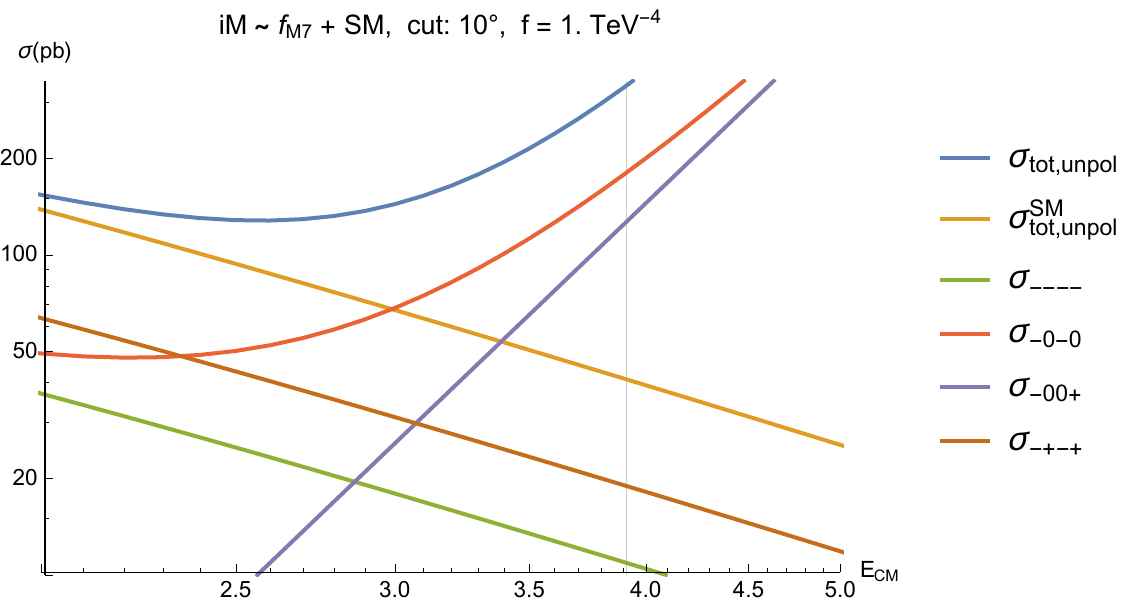} \\
\end{tabular}
\end{center}
\caption{See description of Fig.~\ref{tab:polsSMEFT}; part 2.}
\label{tab:polsSMEFT2}
\end{figure}
\clearpage

%-------- polarized f< 0 ------------------------------%

\begin{figure}
\begin{center}
\begin{tabular}{cc}
 \includegraphics[width=0.5\columnwidth]{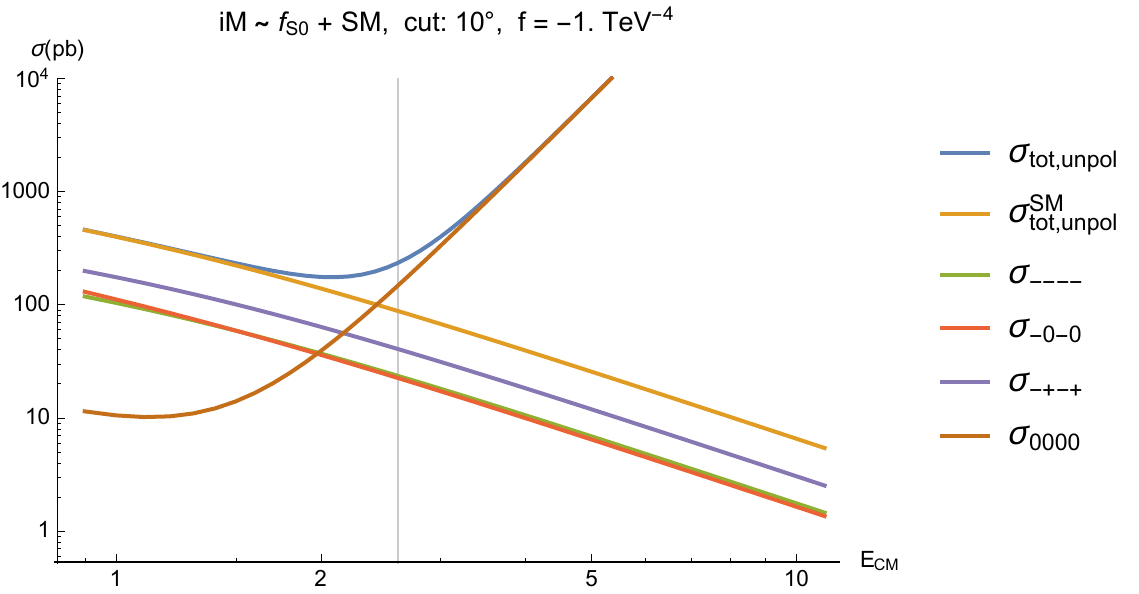}& \includegraphics[width=0.5\columnwidth]{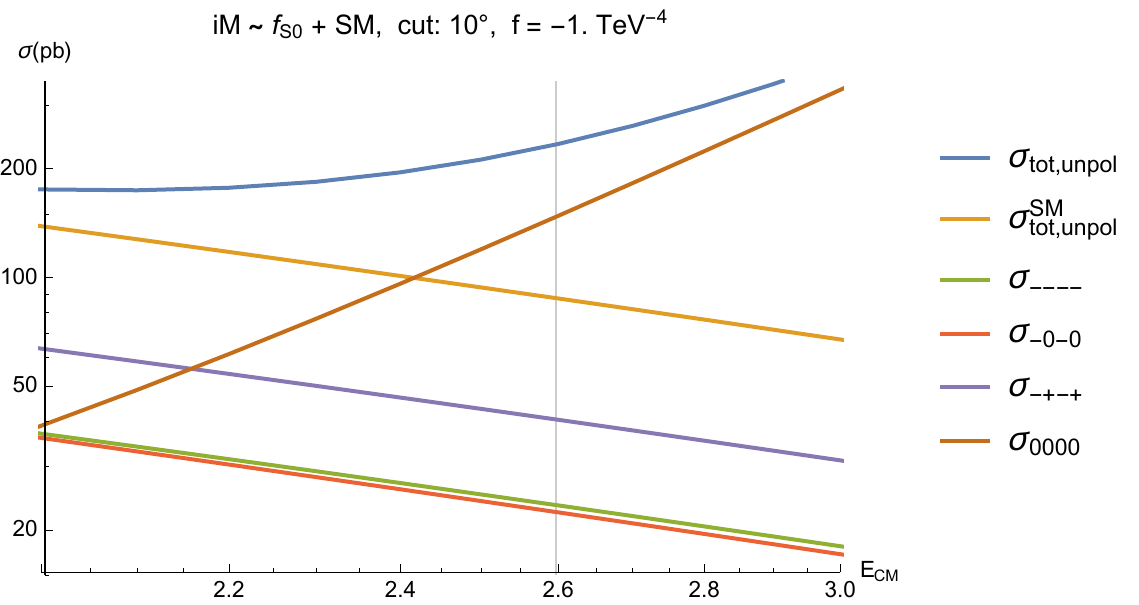} \\

\includegraphics[width=0.5\columnwidth]{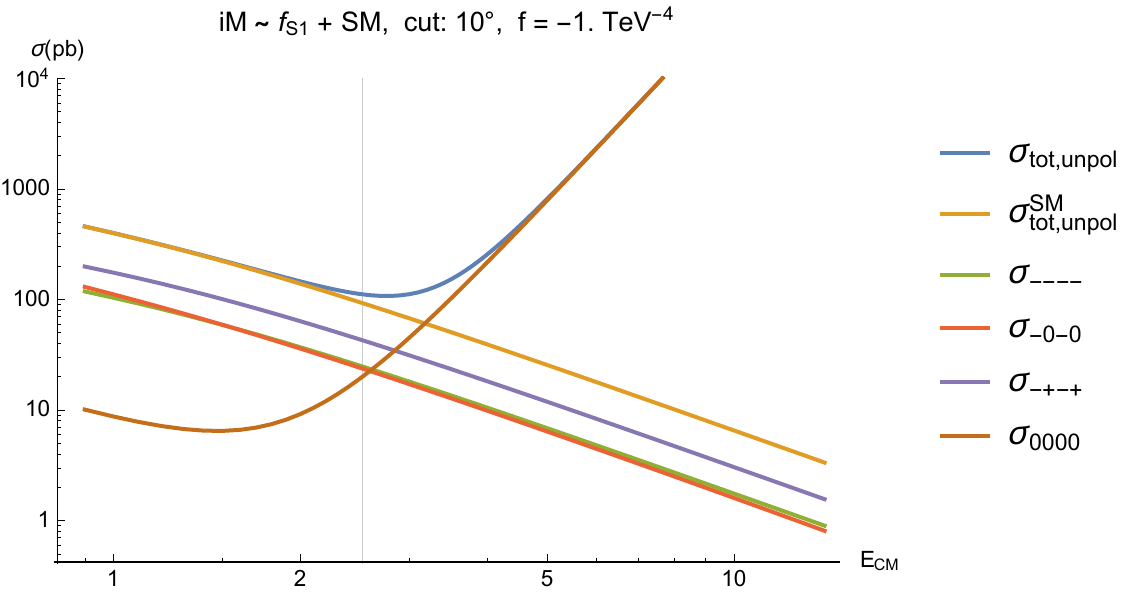}& \includegraphics[width=0.5\columnwidth]{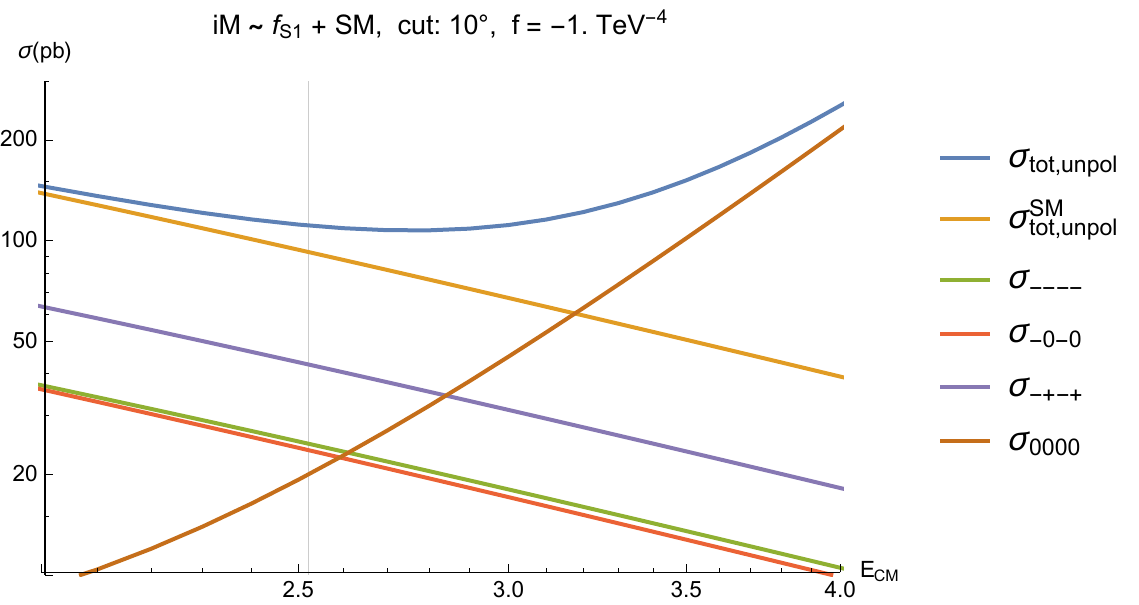} \\

\includegraphics[width=0.5\columnwidth]{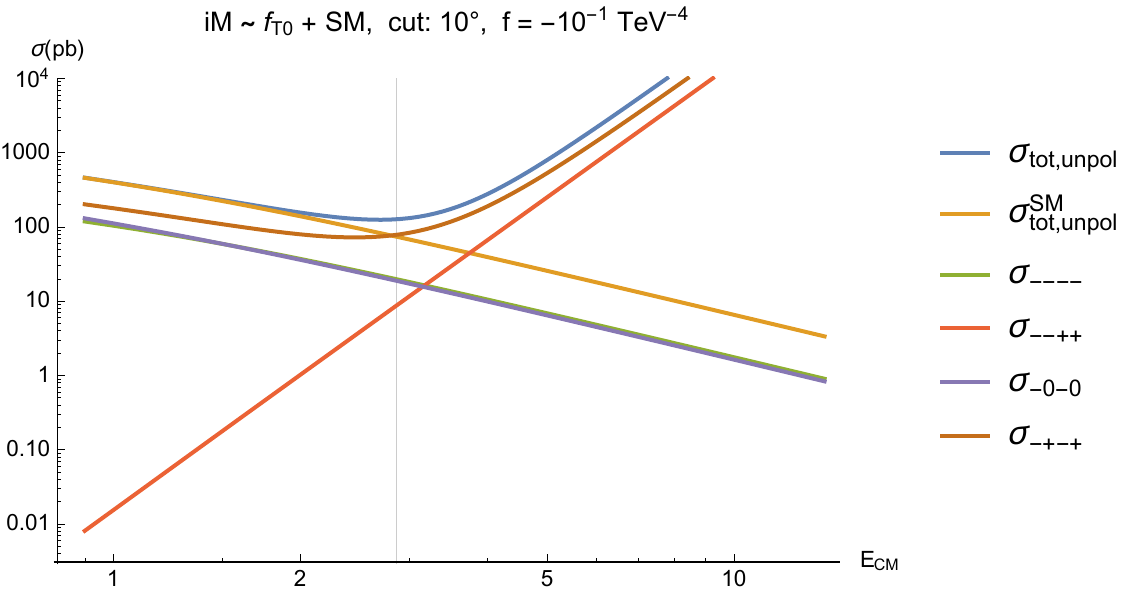}& \includegraphics[width=0.5\columnwidth]{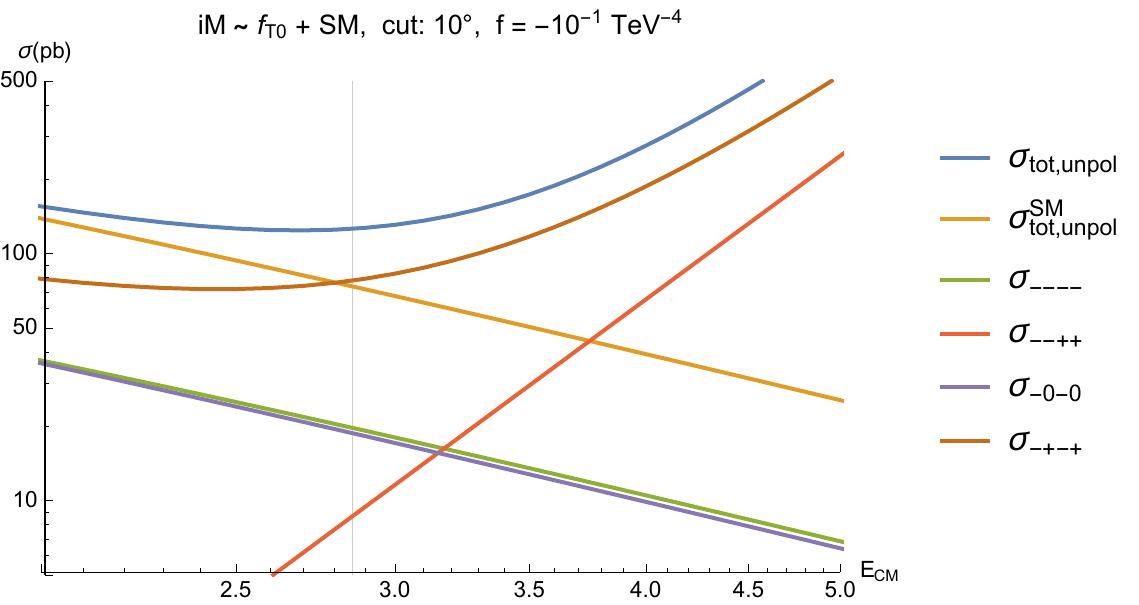} \\

\includegraphics[width=0.5\columnwidth]{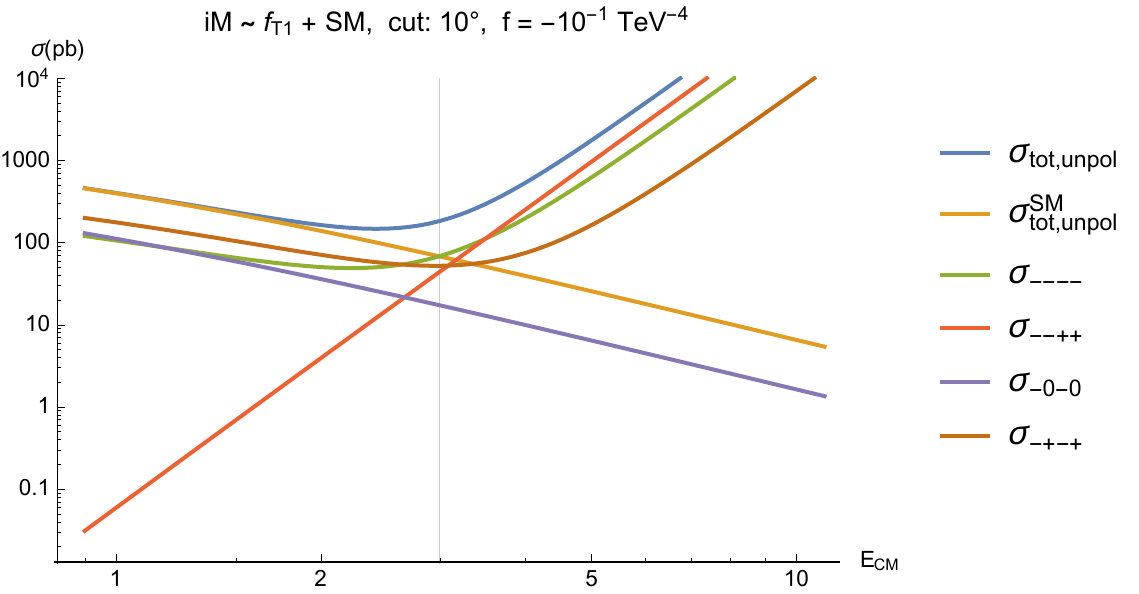}& \includegraphics[width=0.5\columnwidth]{plotlistPlot10DegreeParticPolsft1Eq01LeqZoom.pdf} \\
\end{tabular}
\end{center}
\caption{Contributions of the polarized cross sections (multiplicity taken into account) as functions of the center-of-mass collision energy
($E_{CM} \equiv \sqrt{s}$, in TeV) for chosen values of $f_i<0$ of the SMEFT operators studied. The remaining (not shown) polarized contributions are negligibly small. In each plot shown is in addition the total cross section of a EFT ''model'' and the total cross section in the SM. In each row right plot is zoom in of the left one; part 1.}
\label{tab:polsSMEFTNeg}
\end{figure}

\begin{figure}
\begin{center}
\begin{tabular}{cc}
 \includegraphics[width=0.5\columnwidth]{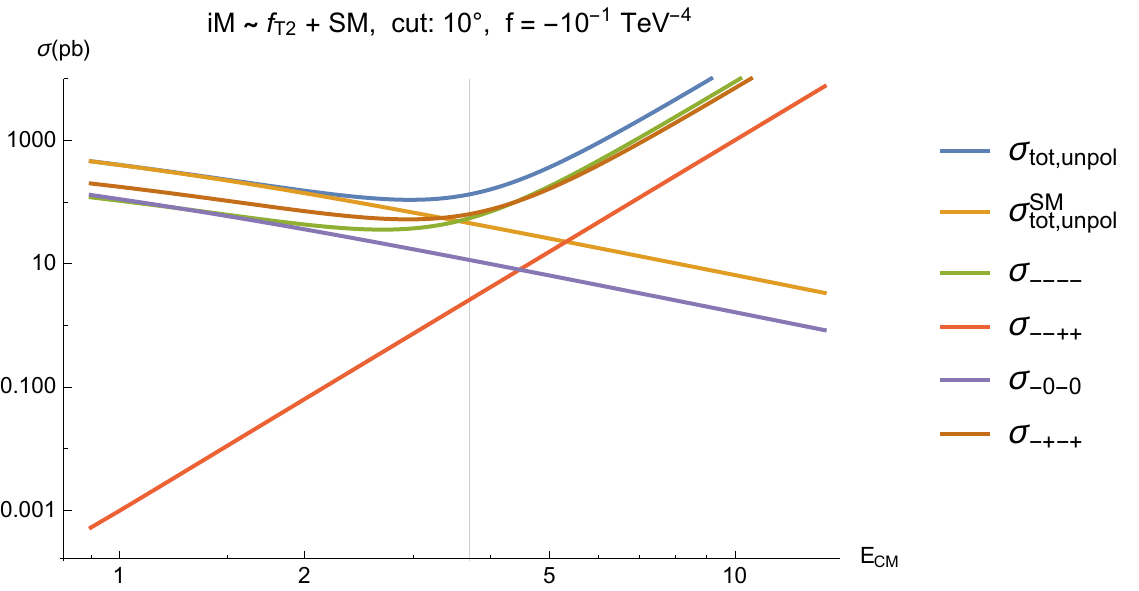}&  \includegraphics[width=0.5\columnwidth]{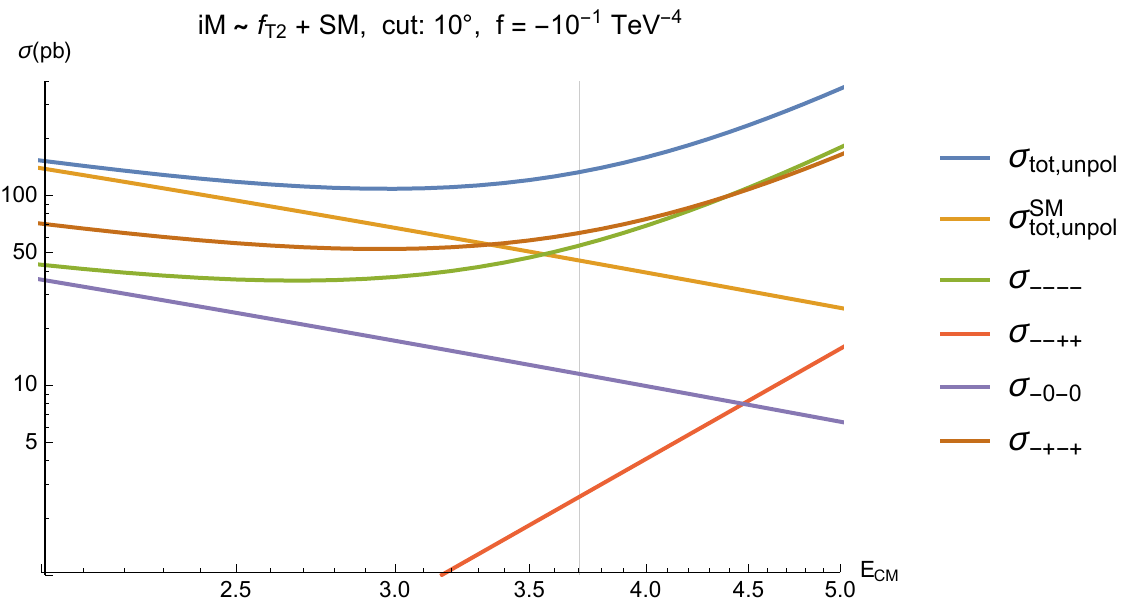} 
\\

\includegraphics[width=0.5\columnwidth]{plotlistPlot10DegreeParticPolsfm0Eq1.pdf}
&\includegraphics[width=0.5\columnwidth]{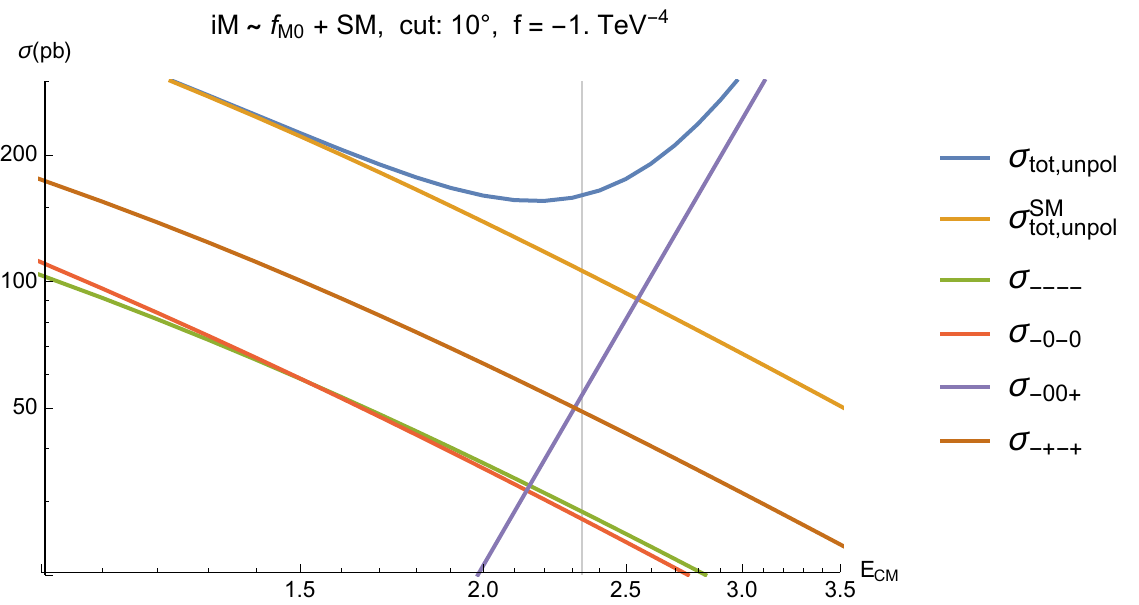} \\

\includegraphics[width=0.5\columnwidth]{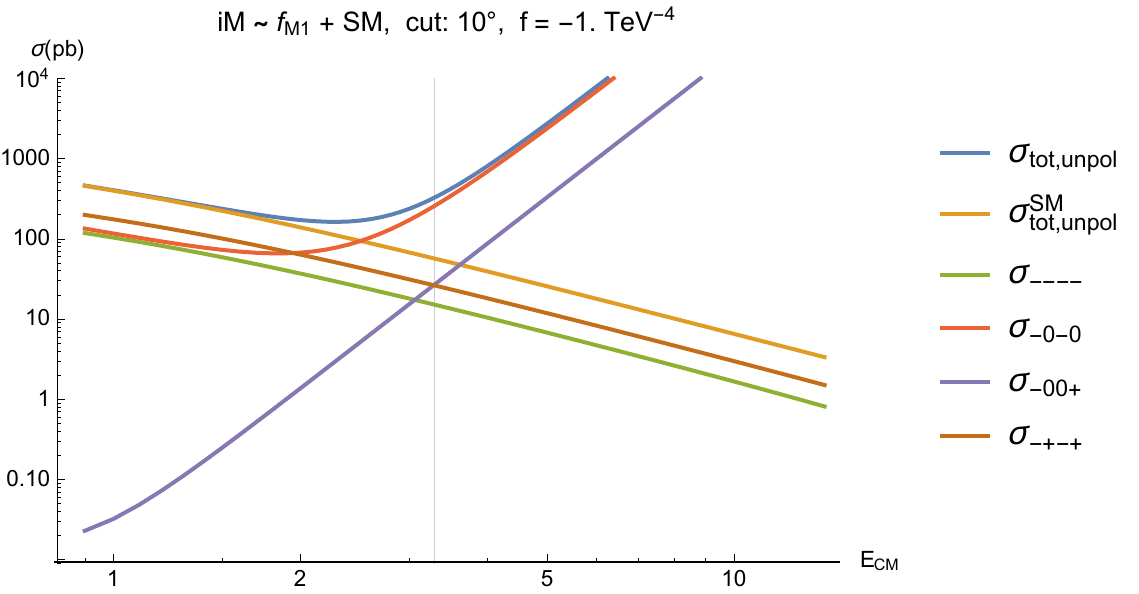}
& \includegraphics[width=0.5\columnwidth]{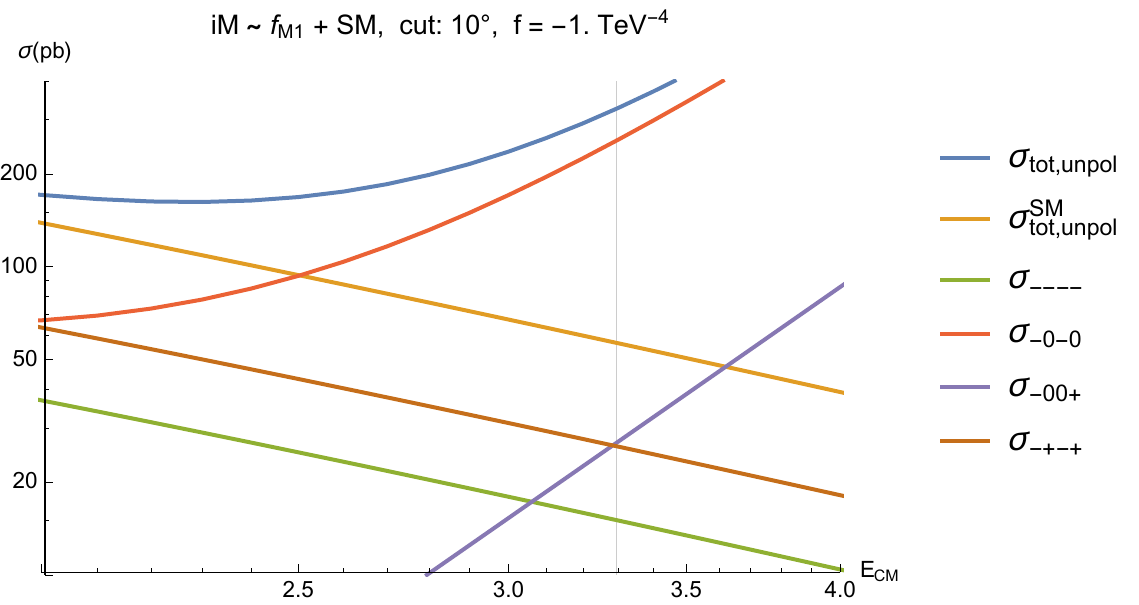} \\

\includegraphics[width=0.5\columnwidth]{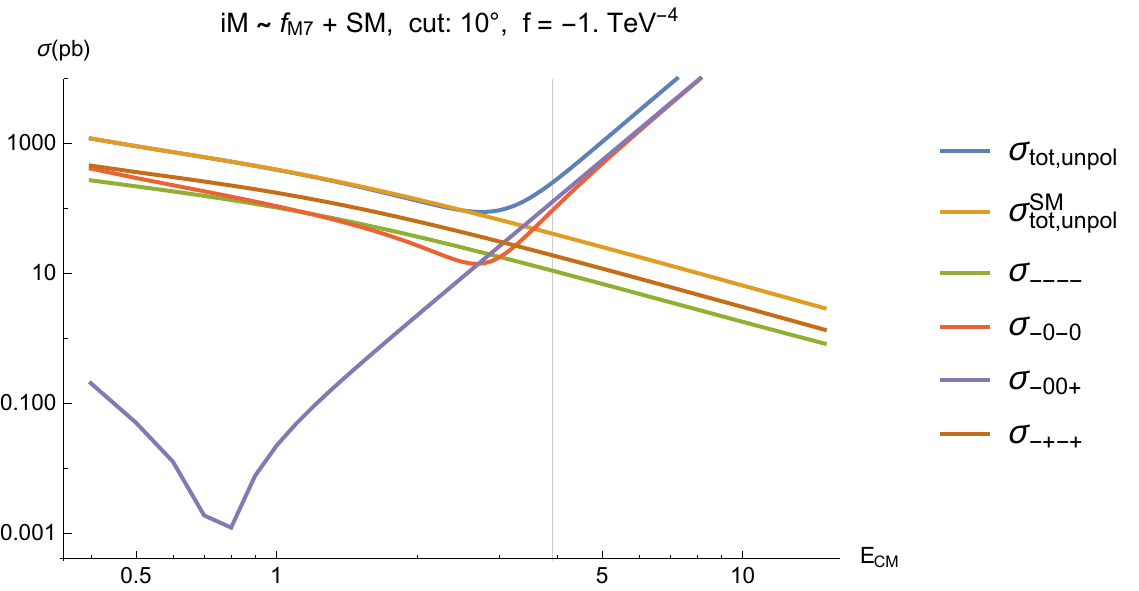}& \includegraphics[width=0.5\columnwidth]{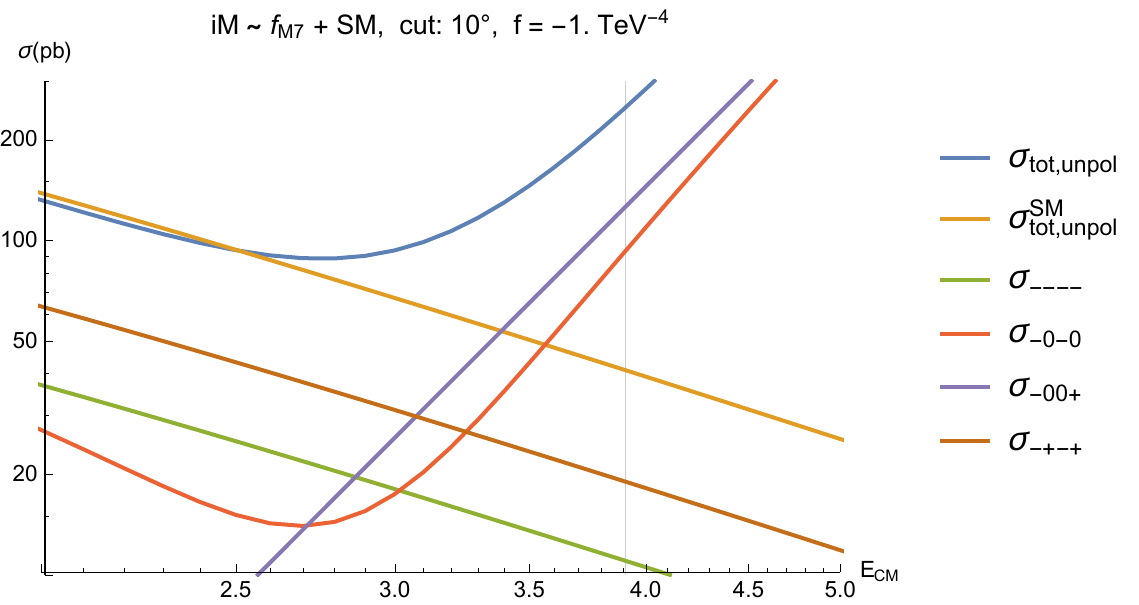} \\
\end{tabular}
\end{center}
\caption{See description of Fig.~\ref{tab:polsSMEFTNeg}; part 2.}
\label{tab:polsSMEFTNeg2}
\end{figure}
\clearpage

\chapter{On-shell gauge boson scattering in the HEFT: more results}
\label{app:VVonshellHEFT}
In Sec.~\ref{onshellHEFT} polarization contributions to the total unpolarized cross sections in the on-shell $W^+W^+$ scattering in HEFT were presented for $\mathcal{T}_{42}$ and $\mathcal{T}_{44}$ operators as examples of features of $W^+W^+$ scattering in HEFT. Here we present the complete set of results for all the three EFT ''models'' distinguished in that Section (see therein for discussion). 
\newgeometry{tmargin=2cm, bmargin=3cm, lmargin=2cm, rmargin=2cm}
\section{Unitarity limits: the numerical results}
\thispagestyle{empty} 
\label{app:unitLimitsHEFT}

\begin{table}[h]%
\begin{center}
\resizebox{0.9\columnwidth}{!}{
\begin{tabular}{c|cccc||cccc}
 $\lambda_1\lambda_2\lambda_1'\lambda_2' $ & $ 0.0001 $ & $ 0.001 $ & $ 0.01 $ & $ 0.1 $ & $ -0.0001 $ & $ -0.001 $ & $ -0.01 $ & $ -0.1 $ \\ \hline $
 \text{- - - -} $ & $ 880. $ & $ 280. $ & $ 88. $ & $ 28. $ & $ 630. $ & $ 200. $ & $ 63. $ & $ 20. $ \\ $
 \text{- - - 0} $ & $ 9.0\times 10^7 $ & $ 9.0\times 10^6 $ & $ 9.0\times 10^5 $ & $ 9.0\times 10^4 $ & $ 9.0\times 10^7 $ & $ 9.0\times 10^6 $ & $ 9.0\times 10^5 $ & $ 9.0\times 10^4 $ \\ $
 \text{- - - +} $ & $ 3.2\times 10^3 $ & $ 1.0\times 10^3 $ & $ 320. $ & $ 100. $ & $ 3.2\times 10^3 $ & $ 1.0\times 10^3 $ & $ 320. $ & $ 100. $ \\ $
 \text{- - 0 0} $ & $ 11. $ & $ 6.2 $ & $ 3.5 $ & $ 2.0 $ & $ 11. $ & $ 6.2 $ & $ 3.5 $ & $ 2.0 $ \\ $
 \text{- - 0 +} $ & $ 130. $ & $ 61. $ & $ 29. $ & $ 13. $ & $ 130. $ & $ 61. $ & $ 29. $ & $ 13. $ \\ $
 \text{- - + +} $ & $ 940. $ & $ 300. $ & $ 94. $ & $ 30. $ & $ 940. $ & $ 300. $ & $ 94. $ & $ 30. $ \\ $
 \text{- 0 - 0} $ & $ 2.3\times 10^3 $ & $ 730. $ & $ 230. $ & $ 73. $ & $ 2.0\times 10^3 $ & $ 630. $ & $ 200. $ & $ 63. $ \\ $
 \text{- 0 - +} $ & $ 110. $ & $ 51. $ & $ 24. $ & $ 11. $ & $ 110. $ & $ 51. $ & $ 24. $ & $ 11. $ \\ $
 \text{- 0 0 0} $ & $ 130. $ & $ 61. $ & $ 29. $ & $ 13. $ & $ 130. $ & $ 61. $ & $ 29. $ & $ 13. $ \\ $
 \text{- 0 0 +} $ & $ 16. $ & $ 8.8 $ & $ 5.0 $ & $ 2.8 $ & $ 16. $ & $ 8.8 $ & $ 5.0 $ & $ 2.8 $ \\ $
 \text{- + - +} $ & $ 1.5\times 10^3 $ & $ 470. $ & $ 150. $ & $ 47. $ & $ 1.9\times 10^3 $ & $ 610. $ & $ 190. $ & $ 61. $ \\ $
 \text{- + 0 0} $ & $ 2.3\times 10^3 $ & $ 710. $ & $ 230. $ & $ 71. $ & $ 2.3\times 10^3 $ & $ 710. $ & $ 230. $ & $ 71. $ \\ $
 \text{0 0 0 0} $ & $ 880. $ & $ 280. $ & $ 88. $ & $ 28. $ & $ 790. $ & $ 250. $ & $ 79. $ & $ 25. $ \\ \hline $
 \text{diag.} $ & $ 9.6 $ & $ 5.4 $ & $ 3.1 $ & $ 1.7 $ & $ 9.6 $ & $ 5.4 $ & $ 3.1 $ & $ 1.7 $ \\ \hline

\end{tabular}
}
\end{center}
\caption{Values of $\sqrt{s^U}$ (in TeV) from the tree-level partial wave unitarity bounds for all elastic on-shell $W^+W^+$ helicity amplitudes for a chosen set of $f_{T42}$ values (first row, in TeV$^{−4}$
); $\lambda_i$ ($\lambda_i'$) denote ingoing (outgoing) W's helicities; ''$x$'' denotes no unitarity violation; ''diag.'' denotes unitarity bounds from diagonalization in the helicity space.
}
\label{tab:unitarityT42}
\end{table}
\begin{table}[h]%
\begin{center}
\resizebox{0.9\columnwidth}{!}{
\begin{tabular}{c|cccc||cccc}
 $\lambda_1\lambda_2\lambda_1'\lambda_2' $ & $ 0.0001 $ & $ 0.001 $ & $ 0.01 $ & $ 0.1 $ & $ -0.0001 $ & $ -0.001 $ & $ -0.01 $ & $ -0.1 $ \\ \hline $
 \text{- - - - } $ & $ 630. $ & $ 200. $ & $ 63. $ & $ 20. $ & $ 880. $ & $ 280. $ & $ 88. $ & $ 28. $ \\ $
 \text{- - - 0 } $ & $ 4.2\times 10^7 $ & $ 4.2\times 10^6 $ & $ 4.2\times 10^5 $ & $ 4.2\times 10^4 $ & $ 4.2\times 10^7 $ & $ 4.2\times 10^6 $ & $ 4.2\times 10^5 $ & $ 4.2\times 10^4 $ \\ $
 \text{- - - + } $ & $ 2.2\times 10^3 $ & $ 690. $ & $ 220. $ & $ 69. $ & $ 2.2\times 10^3 $ & $ 690. $ & $ 220. $ & $ 69. $ \\ $
 \text{- - 0 0 } $ & $ 11. $ & $ 6.2 $ & $ 3.5 $ & $ 2.0 $ & $ 11. $ & $ 6.2 $ & $ 3.5 $ & $ 2.0 $ \\ $
 \text{- - 0 + } $ & $ 100. $ & $ 48. $ & $ 22. $ & $ 10. $ & $ 100. $ & $ 48. $ & $ 22. $ & $ 10. $ \\ $
 \text{- - + + } $ & $ \text{x} $ & $ \text{x} $ & $ \text{x} $ & $ \text{x} $ & $ \text{x} $ & $ \text{x} $ & $ \text{x} $ & $ \text{x} $ \\ $
 \text{- 0 - 0 } $ & $ 1.6\times 10^3 $ & $ 490. $ & $ 160. $ & $ 49. $ & $ 1.8\times 10^3 $ & $ 570. $ & $ 180. $ & $ 57. $ \\ $
 \text{- 0 - + } $ & $ 87. $ & $ 41. $ & $ 19. $ & $ 8.7 $ & $ 87. $ & $ 41. $ & $ 19. $ & $ 8.7 $ \\ $
 \text{- 0 0 - } $ & $ 1.9\times 10^3 $ & $ 590. $ & $ 190. $ & $ 59. $ & $ 1.9\times 10^3 $ & $ 590. $ & $ 190. $ & $ 59. $ \\ $
 \text{- 0 0 0 } $ & $ 100. $ & $ 48. $ & $ 22. $ & $ 10. $ & $ 100. $ & $ 48. $ & $ 22. $ & $ 10. $ \\ $
 \text{- 0 0 + } $ & $ 13. $ & $ 7.4 $ & $ 4.2 $ & $ 2.4 $ & $ 13. $ & $ 7.4 $ & $ 4.2 $ & $ 2.4 $ \\ $
 \text{- 0 + - } $ & $ 110. $ & $ 51. $ & $ 24. $ & $ 11. $ & $ 110. $ & $ 51. $ & $ 24. $ & $ 11. $ \\ $
 \text{- 0 + 0 } $ & $ 16. $ & $ 8.8 $ & $ 5.0 $ & $ 2.8 $ & $ 16. $ & $ 8.8 $ & $ 5.0 $ & $ 2.8 $ \\ $
 \text{- + - + } $ & $ 1.4\times 10^3 $ & $ 430. $ & $ 140. $ & $ 43. $ & $ 1.0\times 10^3 $ & $ 330. $ & $ 100. $ & $ 33. $ \\ $
 \text{- + 0 0 } $ & $ 1.6\times 10^3 $ & $ 490. $ & $ 160. $ & $ 49. $ & $ 1.5\times 10^3 $ & $ 490. $ & $ 150. $ & $ 49. $ \\ $
 \text{- + + - } $ & $ 1.7\times 10^3 $ & $ 540. $ & $ 170. $ & $ 54. $ & $ 1.7\times 10^3 $ & $ 540. $ & $ 170. $ & $ 54. $ \\ $
 \text{0 0 0 0 } $ & $ 1.0\times 10^3 $ & $ 330. $ & $ 100. $ & $ 33. $ & $ 1.1\times 10^3 $ & $ 360. $ & $ 110. $ & $ 36. $ \\ \hline $
 \text{diag.} $ & $ 9.6 $ & $ 5.4 $ & $ 3.1 $ & $ 1.7 $ & $ 9.6 $ & $ 5.4 $ & $ 3.1 $ & $ 1.7 $\\ \hline

\end{tabular}
}
\end{center}
\caption{See description of Tab.~\ref{tab:unitarityT42}; reaction $W^+W^-\rightarrow W^+W^-$.
%Values of $\sqrt{s^U}$ (in TeV) from the tree-level partial wave unitarity bounds for all elastic on-shell $W^+W^-$ helicity amplitudes for a chosen set of $f_{T42}$ values (first row, in TeV$^{−4}$
%); $\lambda_i$ ($\lambda_i'$) denote ingoing (outgoing) W's helicities; ''$x$'' denotes no unitarity violation.
}
\label{tab:unitarityT42ww}
\end{table}

\clearpage
%--------------------------------------------------------------------%
\begin{table}%
\begin{center}
\begin{tabular}{c|cccc||cccc}
 $\lambda_1\lambda_2\lambda_1'\lambda_2' $ & $ 0.0001 $ & $ 0.001 $ & $ 0.01 $ & $ 0.1 $ & $ -0.0001 $ & $ -0.001 $ & $ -0.01 $ & $ -0.1 $ \\ \hline $
 \text{- - - -} $ & $ \text{x} $ & $ \text{x} $ & $ \text{x} $ & $ \text{x} $ & $ \text{x} $ & $ \text{x} $ & $ \text{x} $ & $ \text{x} $ \\ $
 \text{- - - 0} $ & $ 1.8\times 10^8 $ & $ 1.8\times 10^7 $ & $ 1.8\times 10^6 $ & $ 1.8\times 10^5 $ & $ 1.8\times 10^8 $ & $ 1.8\times 10^7 $ & $ 1.8\times 10^6 $ & $ 1.8\times 10^5 $ \\ $
 \text{- - - +} $ & $ 4.5\times 10^3 $ & $ 1.4\times 10^3 $ & $ 450. $ & $ 140. $ & $ 4.5\times 10^3 $ & $ 1.4\times 10^3 $ & $ 450. $ & $ 140. $ \\ $
 \text{- - 0 0} $ & $ 16. $ & $ 8.8 $ & $ 5.0 $ & $ 2.8 $ & $ 16. $ & $ 8.8 $ & $ 5.0 $ & $ 2.8 $ \\ $
 \text{- - 0 +} $ & $ 1.8\times 10^8 $ & $ 1.8\times 10^7 $ & $ 1.8\times 10^6 $ & $ 1.8\times 10^5 $ & $ 1.8\times 10^8 $ & $ 1.8\times 10^7 $ & $ 1.8\times 10^6 $ & $ 1.8\times 10^5 $ \\ $
 \text{- - + +} $ & $ \text{x} $ & $ \text{x} $ & $ \text{x} $ & $ \text{x} $ & $ \text{x} $ & $ \text{x} $ & $ \text{x} $ & $ \text{x} $ \\ $
 \text{- 0 - 0} $ & $ 18. $ & $ 10. $ & $ 5.7 $ & $ 3.2 $ & $ 17. $ & $ 9.4 $ & $ 5.3 $ & $ 3.0 $ \\ $
 \text{- 0 - +} $ & $ 110. $ & $ 51. $ & $ 24. $ & $ 11. $ & $ 110. $ & $ 51. $ & $ 24. $ & $ 11. $ \\ $
 \text{- 0 0 0} $ & $ 170. $ & $ 77. $ & $ 36. $ & $ 17. $ & $ 170. $ & $ 77. $ & $ 36. $ & $ 17. $ \\ $
 \text{- 0 0 +} $ & $ 17. $ & $ 9.8 $ & $ 5.5 $ & $ 3.1 $ & $ 17. $ & $ 9.8 $ & $ 5.5 $ & $ 3.1 $ \\ $
 \text{- + - +} $ & $ 1.9\times 10^3 $ & $ 610. $ & $ 190. $ & $ 61. $ & $ 1.5\times 10^3 $ & $ 470. $ & $ 150. $ & $ 47. $ \\ $
 \text{- + 0 0} $ & $ 23. $ & $ 13. $ & $ 7.2 $ & $ 4.0 $ & $ 23. $ & $ 13. $ & $ 7.2 $ & $ 4.0 $ \\ $
 \text{0 0 0 0} $ & $ 2.8\times 10^3 $ & $ 880. $ & $ 280. $ & $ 88. $ & $ 2.5\times 10^3 $ & $ 790. $ & $ 250. $ & $ 79. $ \\ \hline$
 \text{diag.} $ & $ 14. $ & $ 7.6 $ & $ 4.3 $ & $ 2.4 $ & $ 14. $ & $ 7.6 $ & $ 4.3 $ & $ 2.4 $\\ \hline

\end{tabular}
\end{center}
\caption{Values of $\sqrt{s^U}$ (in TeV) from the tree-level partial wave unitarity bounds for all elastic on-shell $W^+W^+$ helicity amplitudes for a chosen set of $f_{T43}$ values (first row, in TeV$^{−4}$
); $\lambda_i$ ($\lambda_i'$) denote ingoing (outgoing) W's helicities; ''$x$'' denotes no unitarity violation; ''diag.'' denotes unitarity bounds from diagonalization in the helicity space.
}
\label{tab:unitarityT43}
\end{table}

\begin{table}%
\begin{center}
\begin{tabular}{c|cccc||cccc}
 $\lambda_1\lambda_2\lambda_1'\lambda_2' $ & $ 0.0001 $ & $ 0.001 $ & $ 0.01 $ & $ 0.1 $ & $ -0.0001 $ & $ -0.001 $ & $ -0.01 $ & $ -0.1 $ \\ \hline $
 \text{- - - - } $ & $ 1.2\times 10^3 $ & $ 390. $ & $ 120. $ & $ 39. $ & $ 890. $ & $ 280. $ & $ 89. $ & $ 28. $ \\ $
 \text{- - - 0 } $ & $ 7.0\times 10^5 $ & $ 2.8\times 10^5 $ & $ 1.4\times 10^5 $ & $ 6.4\times 10^4 $ & $ 7.0\times 10^5 $ & $ 2.8\times 10^5 $ & $ 1.4\times 10^5 $ & $ 6.4\times 10^4 $ \\ $
 \text{- - - + } $ & $ 5.4\times 10^3 $ & $ 1.7\times 10^3 $ & $ 540. $ & $ 170. $ & $ 5.4\times 10^3 $ & $ 1.7\times 10^3 $ & $ 540. $ & $ 170. $ \\ $
 \text{- - 0 0 } $ & $ 16. $ & $ 8.8 $ & $ 5.0 $ & $ 2.8 $ & $ 16. $ & $ 8.8 $ & $ 5.0 $ & $ 2.8 $ \\ $
 \text{- - 0 + } $ & $ 150. $ & $ 69. $ & $ 32. $ & $ 15. $ & $ 150. $ & $ 69. $ & $ 32. $ & $ 15. $ \\ $
 \text{- - + + } $ & $ 1.9\times 10^3 $ & $ 590. $ & $ 190. $ & $ 59. $ & $ 1.9\times 10^3 $ & $ 590. $ & $ 190. $ & $ 59. $ \\ $
 \text{- 0 - 0 } $ & $ 17. $ & $ 9.4 $ & $ 5.3 $ & $ 3.0 $ & $ 18. $ & $ 10. $ & $ 5.7 $ & $ 3.2 $ \\ $
 \text{- 0 - + } $ & $ 1.0\times 10^8 $ & $ 1.0\times 10^7 $ & $ 1.0\times 10^6 $ & $ 1.0\times 10^5 $ & $ 1.0\times 10^8 $ & $ 1.0\times 10^7 $ & $ 1.0\times 10^6 $ & $ 1.0\times 10^5 $ \\ $
 \text{- 0 0 - } $ & $ 15. $ & $ 8.2 $ & $ 4.6 $ & $ 2.6 $ & $ 15. $ & $ 8.2 $ & $ 4.6 $ & $ 2.6 $ \\ $
 \text{- 0 0 0 } $ & $ 190. $ & $ 87. $ & $ 40. $ & $ 19. $ & $ 190. $ & $ 87. $ & $ 40. $ & $ 19. $ \\ $
 \text{- 0 0 + } $ & $ 19. $ & $ 11. $ & $ 5.9 $ & $ 3.3 $ & $ 19. $ & $ 11. $ & $ 5.9 $ & $ 3.3 $ \\ $
 \text{- 0 + - } $ & $ 1.0\times 10^8 $ & $ 1.0\times 10^7 $ & $ 1.0\times 10^6 $ & $ 1.0\times 10^5 $ & $ 1.0\times 10^8 $ & $ 1.0\times 10^7 $ & $ 1.0\times 10^6 $ & $ 1.0\times 10^5 $ \\ $
 \text{- 0 + 0 } $ & $ 25. $ & $ 14. $ & $ 7.8 $ & $ 4.4 $ & $ 25. $ & $ 14. $ & $ 7.8 $ & $ 4.4 $ \\ $
 \text{- + - + } $ & $ \text{x} $ & $ \text{x} $ & $ \text{x} $ & $ \text{x} $ & $ \text{x} $ & $ \text{x} $ & $ \text{x} $ & $ \text{x} $ \\ $
 \text{- + 0 0 } $ & $ 25. $ & $ 14. $ & $ 7.8 $ & $ 4.4 $ & $ 25. $ & $ 14. $ & $ 7.8 $ & $ 4.4 $ \\ $
 \text{- + + - } $ & $ \text{x} $ & $ \text{x} $ & $ \text{x} $ & $ \text{x} $ & $ \text{x} $ & $ \text{x} $ & $ \text{x} $ & $ \text{x} $ \\ $
 \text{0 0 0 0 } $ & $ 2.8\times 10^3 $ & $ 870. $ & $ 280. $ & $ 87. $ & $ 2.6\times 10^3 $ & $ 810. $ & $ 260. $ & $ 81. $ \\\hline $
 \text{diag.} $ & $ 14. $ & $ 7.7 $ & $ 4.3 $ & $ 2.5 $ & $ 14. $ & $ 7.7 $ & $ 4.3 $ & $ 2.5 $\\ \hline

\end{tabular}
\end{center}
\caption{See description of Tab.~\ref{tab:unitarityT43}; reaction $W^+W^-\rightarrow W^+W^-$.
%Values of $\sqrt{s^U}$ (in TeV) from the tree-level partial wave unitarity bounds for all elastic on-shell $W^+W^-$ helicity amplitudes for a chosen set of $f_{T43}$ values (first row, in TeV$^{−4}$
%); $\lambda_i$ ($\lambda_i'$) denote ingoing (outgoing) W's helicities; ''$x$'' denotes no unitarity violation.
}
\label{tab:unitarityT43ww}
\end{table}
\clearpage
%--------------------------------------------------------------------%
\begin{table}%
\begin{center}
\begin{tabular}{c|cccc||cccc}
 $\lambda_1\lambda_2\lambda_1'\lambda_2' $ & $ 0.0001 $ & $ 0.001 $ & $ 0.01 $ & $ 0.1 $ & $ -0.0001 $ & $ -0.001 $ & $ -0.01 $ & $ -0.1 $ \\ \hline$
 \text{- - - -} $ & $ \text{x} $ & $ \text{x} $ & $ \text{x} $ & $ \text{x} $ & $ \text{x} $ & $ \text{x} $ & $ \text{x} $ & $ \text{x} $ \\ $
 \text{- - - 0} $ & $ \text{x} $ & $ \text{x} $ & $ \text{x} $ & $ \text{x} $ & $ \text{x} $ & $ \text{x} $ & $ \text{x} $ & $ \text{x} $ \\ $
 \text{- - - +} $ & $ \text{x} $ & $ \text{x} $ & $ \text{x} $ & $ \text{x} $ & $ \text{x} $ & $ \text{x} $ & $ \text{x} $ & $ \text{x} $ \\ $
 \text{- - 0 0} $ & $ 16. $ & $ 8.8 $ & $ 5.0 $ & $ 2.8 $ & $ 16. $ & $ 8.8 $ & $ 5.0 $ & $ 2.8 $ \\ $
 \text{- - 0 +} $ & $ 170. $ & $ 77. $ & $ 36. $ & $ 17. $ & $ 170. $ & $ 77. $ & $ 36. $ & $ 17. $ \\ $
 \text{- - + +} $ & $ 1.3\times 10^3 $ & $ 420. $ & $ 130. $ & $ 42. $ & $ 1.3\times 10^3 $ & $ 420. $ & $ 130. $ & $ 42. $ \\ $
 \text{- 0 - 0} $ & $ 18. $ & $ 10. $ & $ 5.7 $ & $ 3.2 $ & $ 17. $ & $ 9.4 $ & $ 5.3 $ & $ 3.0 $ \\ $
 \text{- 0 - +} $ & $ 140. $ & $ 64. $ & $ 30. $ & $ 14. $ & $ 140. $ & $ 64. $ & $ 30. $ & $ 14. $ \\ $
 \text{- 0 0 0} $ & $ \text{x} $ & $ \text{x} $ & $ \text{x} $ & $ \text{x} $ & $ \text{x} $ & $ \text{x} $ & $ \text{x} $ & $ \text{x} $ \\ $
 \text{- 0 0 +} $ & $ 25. $ & $ 14. $ & $ 7.8 $ & $ 4.4 $ & $ 25. $ & $ 14. $ & $ 7.8 $ & $ 4.4 $ \\ $
 \text{- + - +} $ & $ 2.7\times 10^3 $ & $ 860. $ & $ 270. $ & $ 86. $ & $ 2.1\times 10^3 $ & $ 660. $ & $ 210. $ & $ 66. $ \\ $
 \text{- + 0 0} $ & $ 23. $ & $ 13. $ & $ 7.2 $ & $ 4.0 $ & $ 23. $ & $ 13. $ & $ 7.2 $ & $ 4.0 $ \\ $
 \text{0 0 0 0} $ & $ 1.6\times 10^3 $ & $ 510. $ & $ 160. $ & $ 51. $ & $ 1.4\times 10^3 $ & $ 460. $ & $ 140. $ & $ 46. $ \\ \hline$
 \text{diag.} $ & $ 14. $ & $ 7.6 $ & $ 4.3 $ & $ 2.4 $ & $ 14. $ & $ 7.6 $ & $ 4.3 $ & $ 2.4 $ \\ \hline

\end{tabular}
\end{center}
\caption{Values of $\sqrt{s^U}$ (in TeV) from the tree-level partial wave unitarity bounds for all elastic on-shell $W^+W^+$ helicity amplitudes for a chosen set of $f_{T44}$ values (first row, in TeV$^{−4}$
); $\lambda_i$ ($\lambda_i'$) denote ingoing (outgoing) W's helicities; ''$x$'' denotes no unitarity violation; ''diag.'' denotes unitarity bounds from diagonalization in the helicity space.
}
\label{tab:unitarityT44}
\end{table}

\begin{table}%
\begin{center}
\begin{tabular}{c|cccc||cccc}
 $\lambda_1\lambda_2\lambda_1'\lambda_2' $ & $ 0.0001 $ & $ 0.001 $ & $ 0.01 $ & $ 0.1 $ & $ -0.0001 $ & $ -0.001 $ & $ -0.01 $ & $ -0.1 $ \\ \hline$
 \text{- - - - } $ & $ 1.2\times 10^3 $ & $ 390. $ & $ 120. $ & $ 39. $ & $ 890. $ & $ 280. $ & $ 89. $ & $ 28. $ \\ $
 \text{- - - 0 } $ & $ \text{x} $ & $ \text{x} $ & $ \text{x} $ & $ \text{x} $ & $ \text{x} $ & $ \text{x} $ & $ \text{x} $ & $ \text{x} $ \\ $
 \text{- - - + } $ & $ \text{x} $ & $ \text{x} $ & $ \text{x} $ & $ \text{x} $ & $ \text{x} $ & $ \text{x} $ & $ \text{x} $ & $ \text{x} $ \\ $
 \text{- - 0 0 } $ & $ 16. $ & $ 8.8 $ & $ 5.0 $ & $ 2.8 $ & $ 16. $ & $ 8.8 $ & $ 5.0 $ & $ 2.8 $ \\ $
 \text{- - 0 + } $ & $ 190. $ & $ 87. $ & $ 40. $ & $ 19. $ & $ 190. $ & $ 87. $ & $ 40. $ & $ 19. $ \\ $
 \text{- - + + } $ & $ 1.9\times 10^3 $ & $ 590. $ & $ 190. $ & $ 59. $ & $ 1.9\times 10^3 $ & $ 590. $ & $ 190. $ & $ 59. $ \\ $
 \text{- 0 - 0 } $ & $ 17. $ & $ 9.4 $ & $ 5.3 $ & $ 3.0 $ & $ 18. $ & $ 10. $ & $ 5.7 $ & $ 3.2 $ \\ $
 \text{- 0 - + } $ & $ \text{x} $ & $ \text{x} $ & $ \text{x} $ & $ \text{x} $ & $ \text{x} $ & $ \text{x} $ & $ \text{x} $ & $ \text{x} $ \\ $
 \text{- 0 0 - } $ & $ 15. $ & $ 8.2 $ & $ 4.6 $ & $ 2.6 $ & $ 15. $ & $ 8.2 $ & $ 4.6 $ & $ 2.6 $ \\ $
 \text{- 0 0 0 } $ & $ \text{x} $ & $ \text{x} $ & $ \text{x} $ & $ \text{x} $ & $ \text{x} $ & $ \text{x} $ & $ \text{x} $ & $ \text{x} $ \\ $
 \text{- 0 0 + } $ & $ 19. $ & $ 11. $ & $ 5.9 $ & $ 3.3 $ & $ 19. $ & $ 11. $ & $ 5.9 $ & $ 3.3 $ \\ $
 \text{- 0 + - } $ & $ 140. $ & $ 64. $ & $ 30. $ & $ 14. $ & $ 140. $ & $ 64. $ & $ 30. $ & $ 14. $ \\ $
 \text{- 0 + 0 } $ & $ 17. $ & $ 9.8 $ & $ 5.5 $ & $ 3.1 $ & $ 17. $ & $ 9.8 $ & $ 5.5 $ & $ 3.1 $ \\ $
 \text{- + - + } $ & $ \text{x} $ & $ \text{x} $ & $ \text{x} $ & $ \text{x} $ & $ \text{x} $ & $ \text{x} $ & $ \text{x} $ & $ \text{x} $ \\ $
 \text{- + 0 0 } $ & $ 25. $ & $ 14. $ & $ 7.8 $ & $ 4.4 $ & $ 25. $ & $ 14. $ & $ 7.8 $ & $ 4.4 $ \\ $
 \text{- + + - } $ & $ 2.4\times 10^3 $ & $ 760. $ & $ 240. $ & $ 76. $ & $ 2.4\times 10^3 $ & $ 770. $ & $ 240. $ & $ 77. $ \\ $
 \text{0 0 0 0 } $ & $ 1.6\times 10^3 $ & $ 500. $ & $ 160. $ & $ 50. $ & $ 1.5\times 10^3 $ & $ 470. $ & $ 150. $ & $ 47. $ \\ \hline $
 \text{diag.} $ & $ 14. $ & $ 7.7 $ & $ 4.3 $ & $ 2.5 $ & $ 14. $ & $ 7.7 $ & $ 4.3 $ & $ 2.5$ \\ \hline

\end{tabular}
\end{center}
\caption{See description of Tab.~\ref{tab:unitarityT44}; reaction $W^+W^-\rightarrow W^+W^-$.
%Values of $\sqrt{s^U}$ (in TeV) from the tree-level partial wave unitarity bounds for all elastic on-shell $W^+W^-$ helicity amplitudes for a chosen set of $f_{T44}$ values (first row, in TeV$^{−4}$
%); $\lambda_i$ ($\lambda_i'$) denote ingoing (outgoing) W's helicities; ''$x$'' denotes no unitarity violation.
}
\label{tab:unitarityT44ww}
\end{table}
\clearpage

\section{Analytic formulae for both $iM$ and $\mathcal{T}^{(j=j_{min})}$ and contributions of $\sigma_{pol}$ to $\sigma^{tot}_{unpol}$ at $\sqrt{s^U}$ (numerical result)} 
\thispagestyle{empty} 
\label{app:analiticFormHEFT}

\begin{table}[h]%
\begin{center}
\bgroup
\def\arraystretch{1.1}
\begin{tabular}{c|c|c}
 $\lambda_1\lambda_2\lambda_1'\lambda_2' $ &  $i\mathcal{M}_{\lambda_1'\lambda_2';\lambda_1\lambda_2}(\sqrt{s},\theta ,f_{T42}) $ & $ \mathcal{T}_{\lambda_1'\lambda_2';\lambda_1\lambda_2}^{(j=j_{min})} $ \\ \hline $ 
 \text{- - - -} $ & $ \frac{8 i s c_W^2 f_{\text{T42}} m_Z^2}{v^2} $ & $ \frac{s c_W^2 f_{\text{T42}} m_Z^2}{2 \pi  v^2} $ \\ $
 \text{- - - 0} $ & $ \frac{i \sqrt{2} \sqrt{s} c_W^3 f_{\text{T42}} \sin (2 \theta ) m_Z^3}{v^2} $ & $ -\frac{\sqrt{s} c_W^3 f_{\text{T42}} m_Z^3}{20 \sqrt{3} \pi  v^2} $ \\ $
 \text{- - - +} $ & $ \frac{i s c_W^2 f_{\text{T42}} \sin ^2(\theta ) m_Z^2}{v^2} $ & $ \frac{s c_W^2 f_{\text{T42}} m_Z^2}{20 \sqrt{6} \pi  v^2} $ \\ $
 \text{- - 0 0} $ & $ \frac{2 i s^2 f_{\text{T42}}}{v^2} $ & $ \frac{s^2 f_{\text{T42}}}{8 \pi  v^2} $ \\ $
 \text{- - 0 +} $ & $ \frac{i s^{3/2} c_W f_{\text{T42}} \sin (\theta ) \cos (\theta ) m_Z}{\sqrt{2} v^2} $ & $ -\frac{s^{3/2} c_W f_{\text{T42}} m_Z}{80 \sqrt{3} \pi  v^2} $ \\ $
 \text{- - + +} $ & $ -\frac{i s c_W^2 f_{\text{T42}} (\cos (2 \theta )-5) m_Z^2}{v^2} $ & $ \frac{s c_W^2 f_{\text{T42}} m_Z^2}{3 \pi  v^2} $ \\ $
 \text{- 0 - 0} $ & $ \frac{3 i s c_W^2 f_{\text{T42}} \sin ^2(\theta ) m_Z^2}{2 v^2} $ & $ \frac{s c_W^2 f_{\text{T42}} m_Z^2}{32 \pi  v^2} $ \\ $
 \text{- 0 - +} $ & $ -\frac{i s^{3/2} c_W f_{\text{T42}} \sin (\theta ) (\cos (\theta )+1) m_Z}{2 \sqrt{2} v^2} $ & $ \frac{s^{3/2} c_W f_{\text{T42}} m_Z}{80 \sqrt{2} \pi  v^2} $ \\ $
 \text{- 0 0 0} $ & $ \frac{i s^{3/2} c_W f_{\text{T42}} \sin (\theta ) \cos (\theta ) m_Z}{\sqrt{2} v^2} $ & $ \frac{s^{3/2} c_W f_{\text{T42}} m_Z}{80 \sqrt{3} \pi  v^2} $ \\ $
 \text{- 0 0 +} $ & $ -\frac{i s^2 f_{\text{T42}} \cos ^4\left(\frac{\theta }{2}\right)}{v^2} $ & $ -\frac{s^2 f_{\text{T42}}}{64 \pi  v^2} $ \\ $
 \text{- + - +} $ & $ -\frac{4 i s c_W^2 f_{\text{T42}} \cos ^4\left(\frac{\theta }{2}\right) m_Z^2}{v^2} $ & $ -\frac{s c_W^2 f_{\text{T42}} m_Z^2}{20 \pi  v^2} $ \\ $
 \text{- + 0 0} $ & $ -\frac{2 i s c_W^2 f_{\text{T42}} \sin ^2(\theta ) m_Z^2}{v^2} $ & $ -\frac{s c_W^2 f_{\text{T42}} m_Z^2}{10 \sqrt{6} \pi  v^2} $ \\ $
 \text{0 0 0 0} $ & $ -\frac{2 i s c_W^2 f_{\text{T42}} (\cos (2 \theta )-3) m_Z^2}{v^2} $ & $ \frac{5 s c_W^2 f_{\text{T42}} m_Z^2}{12 \pi  v^2}$ \\ \hline

\end{tabular}
\egroup
\caption{Analytic formulas for on-shell $W^+W^+$ elastic scattering helicity amplitudes $iM$ and the minimal $j$ partial waves $\mathcal{T}^{(j=j_{min})}$; $\lambda_i$ ($\lambda_i'$) denote ingoing (outgoing) W's helicities; $c_W\equiv \cos\theta_W$; $\theta$ is the scattering angle; the $\mathcal{T}_{42}$ operator case; only leading terms in the $\sqrt{s}$  expansion in the limit $s\rightarrow\infty$ are shown.}
\label{tab:analyticT42}
\end{center}
\end{table}
\begin{table}[h]%
\begin{center}
\resizebox{0.9\columnwidth}{!}{
\begin{tabular}{c|cccc||cccc}
 $\lambda_1\lambda_2\lambda_1'\lambda_2' $ & $ 0.0001 $ & $ 0.001 $ & $ 0.01 $ & $ 0.1 $ & $ -0.0001 $ & $ -0.001 $ & $ -0.01 $ & $ -0.1 $ \\ \hline $
 \text{- - - -} $ & $ 1.9 $ & $ 5.8 $ & $ 17. $ & $ 45. $ & $ 1.9 $ & $ 5.9 $ & $ 17. $ & $ 48. $ \\ $
 \text{- - - 0} $ & $ 1.7\times 10^{-10} $ & $ 1.6\times 10^{-8} $ & $ 1.3\times 10^{-6} $ & $ 8.5\times 10^{-5} $ & $ 1.5\times 10^{-10} $ & $ 1.3\times 10^{-8} $ & $ 1.1\times 10^{-6} $ & $ 6.6\times 10^{-5} $ \\ $
 \text{- - - +} $ & $ 3.9\times 10^{-8} $ & $ 1.2\times 10^{-6} $ & $ 3.9\times 10^{-5} $ & $ 1.2\times 10^{-3} $ & $ 2.1\times 10^{-8} $ & $ 6.7\times 10^{-7} $ & $ 2.2\times 10^{-5} $ & $ 6.9\times 10^{-4} $ \\ $
 \text{- - 0 0} $ & $ 15. $ & $ 47. $ & $ 150. $ & $ 500. $ & $ 15. $ & $ 47. $ & $ 150. $ & $ 490. $ \\ $
 \text{- - 0 +} $ & $ 5.2\times 10^{-5} $ & $ 5.2\times 10^{-4} $ & $ 5.4\times 10^{-3} $ & $ 0.056 $ & $ 5.2\times 10^{-5} $ & $ 5.2\times 10^{-4} $ & $ 5.3\times 10^{-3} $ & $ 0.055 $ \\ $
 \text{- - + +} $ & $ 2.6\times 10^{-7} $ & $ 8.2\times 10^{-6} $ & $ 2.6\times 10^{-4} $ & $ 8.4\times 10^{-3} $ & $ 2.6\times 10^{-7} $ & $ 8.2\times 10^{-6} $ & $ 2.6\times 10^{-4} $ & $ 8.3\times 10^{-3} $ \\ $
 \text{- 0 - 0} $ & $ 1.8 $ & $ 5.5 $ & $ 16. $ & $ 46. $ & $ 1.8 $ & $ 5.5 $ & $ 16. $ & $ 46. $ \\ $
 \text{- 0 - +} $ & $ 4.3\times 10^{-4} $ & $ 4.3\times 10^{-3} $ & $ 0.044 $ & $ 0.44 $ & $ 1.0\times 10^{-4} $ & $ 1.0\times 10^{-3} $ & $ 0.011 $ & $ 0.11 $ \\ $
 \text{- 0 0 0} $ & $ 4.4\times 10^{-5} $ & $ 4.5\times 10^{-4} $ & $ 4.7\times 10^{-3} $ & $ 0.050 $ & $ 6.1\times 10^{-5} $ & $ 6.1\times 10^{-4} $ & $ 6.2\times 10^{-3} $ & $ 0.062 $ \\ $
 \text{- 0 0 +} $ & $ 2.8 $ & $ 9.2 $ & $ 30. $ & $ 98. $ & $ 2.8 $ & $ 9.2 $ & $ 30. $ & $ 96. $ \\ $
 \text{- + - +} $ & $ 3.3 $ & $ 10. $ & $ 30. $ & $ 80. $ & $ 3.3 $ & $ 10. $ & $ 30. $ & $ 80. $ \\ $
 \text{- + 0 0} $ & $ 8.6\times 10^{-8} $ & $ 2.7\times 10^{-6} $ & $ 8.5\times 10^{-5} $ & $ 2.7\times 10^{-3} $ & $ 4.1\times 10^{-8} $ & $ 1.3\times 10^{-6} $ & $ 4.1\times 10^{-5} $ & $ 1.3\times 10^{-3} $ \\ $
 \text{0 0 0 0} $ & $ 0.10 $ & $ 0.32 $ & $ 0.98 $ & $ 2.8 $ & $ 0.11 $ & $ 0.33 $ & $ 1.0 $ & $ 3.1$ \\ \hline

\end{tabular}
}
\end{center}
\caption{Contribution of polarized elastic on-shell $W^+W^+$ scattering cross sections (in $pb$) to the total unpolarized cross sections at $\sqrt{s^U}$ for a chosen set of $f_{T42}$ values (first row, in TeV$^{−4}$).}
\label{tab:polarizedContributionsT42}
\end{table}
\clearpage

\begin{table}%
\begin{center}
\bgroup
\def\arraystretch{1.5}
\begin{tabular}{c|c|c}
 $\lambda_1\lambda_2\lambda_1'\lambda_2' $ &  $i\mathcal{M}_{\lambda_1'\lambda_2';\lambda_1\lambda_2}(\sqrt{s},\theta ,f_{T43}) $ & $ \mathcal{T}_{\lambda_1'\lambda_2';\lambda_1\lambda_2}^{(j=j_{min})} $ \\ \hline $ 
 \text{- - - -} $ & $ \frac{i c_W^2 m_Z^2 \left(c_W^2 f_{\text{T43}} \cos (2 \theta ) m_Z^2-5 c_W^2 f_{\text{T43}} m_Z^2-32 \csc ^2(\theta )\right)}{v^2} $ & $ -\frac{c_W^4 f_{\text{T43}} m_Z^4}{3 \pi  v^2} $ \\ $
 \text{- - - 0} $ & $ -\frac{i \sqrt{2} \sqrt{s} c_W^3 f_{\text{T43}} \sin (\theta ) \cos (\theta ) m_Z^3}{v^2} $ & $ \frac{\sqrt{s} c_W^3 f_{\text{T43}} m_Z^3}{40 \sqrt{3} \pi  v^2} $ \\ $
 \text{- - - +} $ & $ -\frac{i s c_W^2 f_{\text{T43}} \sin ^2(\theta ) m_Z^2}{2 v^2} $ & $ -\frac{s c_W^2 f_{\text{T43}} m_Z^2}{40 \sqrt{6} \pi  v^2} $ \\ $
 \text{- - 0 0} $ & $ \frac{i s^2 f_{\text{T43}}}{2 v^2} $ & $ \frac{s^2 f_{\text{T43}}}{32 \pi  v^2} $ \\ $
 \text{- - 0 +} $ & $ -\frac{i \sqrt{2} \sqrt{s} c_W^3 f_{\text{T43}} \sin (\theta ) \cos (\theta ) m_Z^3}{v^2} $ & $ \frac{\sqrt{s} c_W^3 f_{\text{T43}} m_Z^3}{40 \sqrt{3} \pi  v^2} $ \\ $
 \text{- - + +} $ & $ \frac{i c_W^4 f_{\text{T43}} (\cos (2 \theta )-5) m_Z^4}{v^2} $ & $ -\frac{c_W^4 f_{\text{T43}} m_Z^4}{3 \pi  v^2} $ \\ $
 \text{- 0 - 0} $ & $ \frac{i s^2 f_{\text{T43}} (\cos (\theta )+1)}{4 v^2} $ & $ \frac{s^2 f_{\text{T43}}}{96 \pi  v^2} $ \\ $
 \text{- 0 - +} $ & $ \frac{i s^{3/2} c_W f_{\text{T43}} \sin (\theta ) (\cos (\theta )+1) m_Z}{2 \sqrt{2} v^2} $ & $ -\frac{s^{3/2} c_W f_{\text{T43}} m_Z}{80 \sqrt{2} \pi  v^2} $ \\ $
 \text{- 0 0 0} $ & $ -\frac{i s^{3/2} c_W f_{\text{T43}} \sin (2 \theta ) m_Z}{4 \sqrt{2} v^2} $ & $ -\frac{s^{3/2} c_W f_{\text{T43}} m_Z}{160 \sqrt{3} \pi  v^2} $ \\ $
 \text{- 0 0 +} $ & $ -\frac{i s^2 f_{\text{T43}} (\cos (\theta )+1)}{4 v^2} $ & $ -\frac{s^2 f_{\text{T43}}}{96 \pi  v^2} $ \\ $
 \text{- + - +} $ & $ \frac{4 i s c_W^2 f_{\text{T43}} \cos ^4\left(\frac{\theta }{2}\right) m_Z^2}{v^2} $ & $ \frac{s c_W^2 f_{\text{T43}} m_Z^2}{20 \pi  v^2} $ \\ $
 \text{- + 0 0} $ & $ \frac{i s^2 f_{\text{T43}} \sin ^2(\theta )}{4 v^2} $ & $ \frac{s^2 f_{\text{T43}}}{80 \sqrt{6} \pi  v^2} $ \\ $
 \text{0 0 0 0} $ & $ \frac{2 i s c_W^2 f_{\text{T43}} \cos ^2(\theta ) m_Z^2}{v^2} $ & $ \frac{s c_W^2 f_{\text{T43}} m_Z^2}{24 \pi  v^2}$ \\ \hline

\end{tabular}
\egroup
\caption{Analytic formulas for on-shell $W^+W^+$ elastic scattering helicity amplitudes $iM$ and the minimal $j$ partial waves $\mathcal{T}^{(j=j_{min})}$; $\lambda_i$ ($\lambda_i'$) denote ingoing (outgoing) W's helicities; $c_W\equiv \cos\theta_W$; $\theta$ is the scattering angle; the $\mathcal{T}_{43}$ operator case; only leading terms in the $\sqrt{s}$  expansion in the limit $s\rightarrow\infty$ are shown.}
\label{tab:analyticT43}
\end{center}
\end{table}

\begin{table}%
\begin{center}
\resizebox{1.05\columnwidth}{!}{
\begin{tabular}{c|cccc||cccc}
$ \lambda_1\lambda_2\lambda_1'\lambda_2' $ & $ 0.0001 $ & $ 0.001 $ & $ 0.01 $ & $ 0.1 $ & $ -0.0001 $ & $ -0.001 $ & $ -0.01 $ & $ -0.1 $ \\ \hline$
 \text{- - - -} $ & $ 0.96 $ & $ 3.0 $ & $ 9.1 $ & $ 26. $ & $ 0.96 $ & $ 3.0 $ & $ 9.1 $ & $ 26. $ \\ $
 \text{- - - 0} $ & $ 9.5\times 10^{-12} $ & $ 8.9\times 10^{-10} $ & $ 7.8\times 10^{-8} $ & $ 5.8\times 10^{-6} $ & $ 1.3\times 10^{-11} $ & $ 1.2\times 10^{-9} $ & $ 1.1\times 10^{-7} $ & $ 8.6\times 10^{-6} $ \\ $
 \text{- - - +} $ & $ 1.2\times 10^{-8} $ & $ 3.9\times 10^{-7} $ & $ 1.3\times 10^{-5} $ & $ 4.0\times 10^{-4} $ & $ 1.7\times 10^{-8} $ & $ 5.3\times 10^{-7} $ & $ 1.7\times 10^{-5} $ & $ 5.4\times 10^{-4} $ \\ $
 \text{- - 0 0} $ & $ 7.2 $ & $ 23. $ & $ 75. $ & $ 240. $ & $ 7.2 $ & $ 23. $ & $ 75. $ & $ 240. $ \\ $
 \text{- - 0 +} $ & $ 9.1\times 10^{-13} $ & $ 9.1\times 10^{-11} $ & $ 9.1\times 10^{-9} $ & $ 9.2\times 10^{-7} $ & $ 1.1\times 10^{-12} $ & $ 1.1\times 10^{-10} $ & $ 1.1\times 10^{-8} $ & $ 1.1\times 10^{-6} $ \\ $
 \text{- - + +} $ & $ 6.7\times 10^{-16} $ & $ 2.1\times 10^{-13} $ & $ 6.6\times 10^{-11} $ & $ 2.0\times 10^{-8} $ & $ 6.1\times 10^{-16} $ & $ 1.9\times 10^{-13} $ & $ 6.0\times 10^{-11} $ & $ 1.9\times 10^{-8} $ \\ $
 \text{- 0 - 0} $ & $ 6.9 $ & $ 22. $ & $ 73. $ & $ 240. $ & $ 14. $ & $ 44. $ & $ 140. $ & $ 460. $ \\ $
 \text{- 0 - +} $ & $ 6.7\times 10^{-4} $ & $ 6.8\times 10^{-3} $ & $ 0.070 $ & $ 0.72 $ & $ 1.0\times 10^{-3} $ & $ 0.010 $ & $ 0.10 $ & $ 1.0 $ \\ $
 \text{- 0 0 0} $ & $ 5.6\times 10^{-5} $ & $ 5.6\times 10^{-4} $ & $ 5.7\times 10^{-3} $ & $ 0.059 $ & $ 4.7\times 10^{-5} $ & $ 4.8\times 10^{-4} $ & $ 4.9\times 10^{-3} $ & $ 0.051 $ \\ $
 \text{- 0 0 +} $ & $ 9.5 $ & $ 31. $ & $ 100. $ & $ 330. $ & $ 9.5 $ & $ 31. $ & $ 100. $ & $ 330. $ \\ $
 \text{- + - +} $ & $ 1.7 $ & $ 5.2 $ & $ 16. $ & $ 44. $ & $ 1.7 $ & $ 5.2 $ & $ 16. $ & $ 46. $ \\ $
 \text{- + 0 0} $ & $ 1.5 $ & $ 4.7 $ & $ 15. $ & $ 50. $ & $ 1.5 $ & $ 4.7 $ & $ 15. $ & $ 50. $ \\ $
 \text{0 0 0 0} $ & $ 0.053 $ & $ 0.16 $ & $ 0.51 $ & $ 1.5 $ & $ 0.053 $ & $ 0.17 $ & $ 0.51 $ & $ 1.6$ \\ \hline

\end{tabular}
}
\end{center}
\caption{Contribution of polarized elastic on-shell $W^+W^+$ scattering cross sections (in $pb$) to the total unpolarized cross sections at $\sqrt{s^U}$ for a chosen set of $f_{T43}$ values (first row, in TeV$^{−4}$).}
\label{tab:polarizedContributionsT43}
\end{table}
\clearpage

\begin{table}%
\begin{center}
\bgroup
\def\arraystretch{1.5}
\begin{tabular}{c|c|c}
 $\lambda_1\lambda_2\lambda_1'\lambda_2' $ &  $i\mathcal{M}_{\lambda_1'\lambda_2';\lambda_1\lambda_2}(\sqrt{s},\theta ,f_{T44}) $ & $ \mathcal{T}_{\lambda_1'\lambda_2';\lambda_1\lambda_2}^{(j=j_{min})} $ \\ \hline $ 
 \text{- - - -} $ & $ -\frac{32 i c_W^2 \csc ^2(\theta ) m_Z^2}{v^2} $ & $ 0 $ \\ $
 \text{- - - 0} $ & $ -\frac{2 i \sqrt{2} c_W^3 \cot (\theta ) \csc ^2(\theta ) m_Z^3 \left(\cos (2 \theta ) \left(\left(4 c_W^2+1\right) m_Z^2-m_h^2\right)+\left(28 c_W^2-25\right) m_Z^2-7 m_h^2\right)}{s^{3/2} v^2} $ & $ 0 $ \\ $
 \text{- - - +} $ & $ \frac{16 i c_W^4 m_Z^4}{s v^2} $ & $ \frac{c_W^4 m_Z^4}{\sqrt{6} \pi  s v^2} $ \\ $
 \text{- - 0 0} $ & $ \frac{i s^2 f_{\text{T44}}}{2 v^2} $ & $ \frac{s^2 f_{\text{T44}}}{32 \pi  v^2} $ \\ $
 \text{- - 0 +} $ & $ \frac{i s^{3/2} c_W f_{\text{T44}} \sin (2 \theta ) m_Z}{4 \sqrt{2} v^2} $ & $ -\frac{s^{3/2} c_W f_{\text{T44}} m_Z}{160 \sqrt{3} \pi  v^2} $ \\ $
 \text{- - + +} $ & $ -\frac{i s c_W^2 f_{\text{T44}} (\cos (2 \theta )-5) m_Z^2}{2 v^2} $ & $ \frac{s c_W^2 f_{\text{T44}} m_Z^2}{6 \pi  v^2} $ \\ $
 \text{- 0 - 0} $ & $ \frac{i s^2 f_{\text{T44}} (\cos (\theta )+1)}{4 v^2} $ & $ \frac{s^2 f_{\text{T44}}}{96 \pi  v^2} $ \\ $
 \text{- 0 - +} $ & $ \frac{i s^{3/2} c_W f_{\text{T44}} \sin (\theta ) (\cos (\theta )+1) m_Z}{4 \sqrt{2} v^2} $ & $ -\frac{s^{3/2} c_W f_{\text{T44}} m_Z}{160 \sqrt{2} \pi  v^2} $ \\ $
 \text{- 0 0 0} $ & $ \frac{2 i \sqrt{2} c_W \cot (\theta ) m_Z \left(m_Z-m_h\right) \left(m_h+m_Z\right)}{\sqrt{s} v^2} $ & $ \frac{c_W m_Z \left(m_Z^2-m_h^2\right)}{8 \sqrt{3} \pi  \sqrt{s} v^2} $ \\ $
 \text{- 0 0 +} $ & $ \frac{i s^2 f_{\text{T44}} \sin ^2(\theta )}{8 v^2} $ & $ \frac{s^2 f_{\text{T44}}}{384 \pi  v^2} $ \\ $
 \text{- + - +} $ & $ \frac{2 i s c_W^2 f_{\text{T44}} \cos ^4\left(\frac{\theta }{2}\right) m_Z^2}{v^2} $ & $ \frac{s c_W^2 f_{\text{T44}} m_Z^2}{40 \pi  v^2} $ \\ $
 \text{- + 0 0} $ & $ \frac{i s^2 f_{\text{T44}} \sin ^2(\theta )}{4 v^2} $ & $ \frac{s^2 f_{\text{T44}}}{80 \sqrt{6} \pi  v^2} $ \\ $
 \text{0 0 0 0} $ & $ \frac{2 i s c_W^2 f_{\text{T44}} m_Z^2}{v^2} $ & $ \frac{s c_W^2 f_{\text{T44}} m_Z^2}{8 \pi  v^2} $ \\ \hline

\end{tabular}
\egroup
\caption{Analytic formulas for on-shell $W^+W^+$ elastic scattering helicity amplitudes $iM$ and the minimal $j$ partial waves $\mathcal{T}^{(j=j_{min})}$; $\lambda_i$ ($\lambda_i'$) denote ingoing (outgoing) W's helicities; $c_W\equiv \cos\theta_W$; $\theta$ is the scattering angle; the $\mathcal{T}_{44}$ operator case; only leading terms in the $\sqrt{s}$  expansion in the limit $s\rightarrow\infty$ are shown.}
\label{tab:analyticT44}
\end{center}
\end{table}

\begin{table}%
\begin{center}
\resizebox{1.05\columnwidth}{!}{
\begin{tabular}{c|cccc||cccc}
 $\lambda_1\lambda_2\lambda_1'\lambda_2' $ & $ 0.0001 $ & $ 0.001 $ & $ 0.01 $ & $ 0.1 $ & $ -0.0001 $ & $ -0.001 $ & $ -0.01 $ & $ -0.1 $ \\ \hline $
 \text{- - - -} $ & $ 0.96 $ & $ 3.0 $ & $ 9.1 $ & $ 26. $ & $ 0.96 $ & $ 3.0 $ & $ 9.1 $ & $ 26. $ \\ $
 \text{- - - 0} $ & $ 1.0\times 10^{-11} $ & $ 9.5\times 10^{-10} $ & $ 8.4\times 10^{-8} $ & $ 6.2\times 10^{-6} $ & $ 1.0\times 10^{-11} $ & $ 9.6\times 10^{-10} $ & $ 8.4\times 10^{-8} $ & $ 6.3\times 10^{-6} $ \\ $
 \text{- - - +} $ & $ 1.0\times 10^{-10} $ & $ 3.1\times 10^{-9} $ & $ 9.5\times 10^{-8} $ & $ 2.9\times 10^{-6} $ & $ 1.0\times 10^{-10} $ & $ 3.1\times 10^{-9} $ & $ 9.5\times 10^{-8} $ & $ 2.9\times 10^{-6} $ \\ $
 \text{- - 0 0} $ & $ 7.2 $ & $ 23. $ & $ 75. $ & $ 250. $ & $ 7.2 $ & $ 23. $ & $ 75. $ & $ 240. $ \\ $
 \text{- - 0 +} $ & $ 5.1\times 10^{-5} $ & $ 5.2\times 10^{-4} $ & $ 5.3\times 10^{-3} $ & $ 0.054 $ & $ 5.1\times 10^{-5} $ & $ 5.2\times 10^{-4} $ & $ 5.3\times 10^{-3} $ & $ 0.053 $ \\ $
 \text{- - + +} $ & $ 1.3\times 10^{-7} $ & $ 4.1\times 10^{-6} $ & $ 1.3\times 10^{-4} $ & $ 4.2\times 10^{-3} $ & $ 1.3\times 10^{-7} $ & $ 4.1\times 10^{-6} $ & $ 1.3\times 10^{-4} $ & $ 4.1\times 10^{-3} $ \\ $
 \text{- 0 - 0} $ & $ 6.9 $ & $ 22. $ & $ 73. $ & $ 240. $ & $ 14. $ & $ 44. $ & $ 140. $ & $ 460. $ \\ $
 \text{- 0 - +} $ & $ 1.4\times 10^{-4} $ & $ 1.4\times 10^{-3} $ & $ 0.014 $ & $ 0.15 $ & $ 3.0\times 10^{-4} $ & $ 3.0\times 10^{-3} $ & $ 0.031 $ & $ 0.31 $ \\ $
 \text{- 0 0 0} $ & $ 1.6\times 10^{-7} $ & $ 1.5\times 10^{-6} $ & $ 1.3\times 10^{-5} $ & $ 8.5\times 10^{-5} $ & $ 1.6\times 10^{-7} $ & $ 1.5\times 10^{-6} $ & $ 1.3\times 10^{-5} $ & $ 8.6\times 10^{-5} $ \\ $
 \text{- 0 0 +} $ & $ 0.97 $ & $ 3.1 $ & $ 10. $ & $ 32. $ & $ 0.97 $ & $ 3.1 $ & $ 10. $ & $ 32. $ \\ $
 \text{- + - +} $ & $ 1.7 $ & $ 5.2 $ & $ 16. $ & $ 45. $ & $ 1.7 $ & $ 5.2 $ & $ 16. $ & $ 45. $ \\ $
 \text{- + 0 0} $ & $ 1.5 $ & $ 4.7 $ & $ 15. $ & $ 50. $ & $ 1.5 $ & $ 4.7 $ & $ 15. $ & $ 49. $ \\ $
 \text{0 0 0 0} $ & $ 0.053 $ & $ 0.16 $ & $ 0.50 $ & $ 1.5 $ & $ 0.053 $ & $ 0.17 $ & $ 0.52 $ & $ 1.6 $\\ \hline

\end{tabular}
}
\end{center}
\caption{Contribution of polarized elastic on-shell $W^+W^+$ scattering cross sections (in $pb$) to the total unpolarized cross sections at $\sqrt{s^U}$ for a chosen set of $f_{T44}$ values (first row, in TeV$^{−4}$).}
\label{tab:polarizedContributionsT44}
\end{table}
\clearpage

\section{Plots of $\sigma_{pol}(s)$ and $\sigma^{tot}_{unpol}(s)$}
\label{app:plotsTotPolAndUnpolHEFT}
\thispagestyle{empty} 
\begin{figure}[h]
\begin{center}
\begin{tabular}{cc}
 \includegraphics[width=0.5\columnwidth]{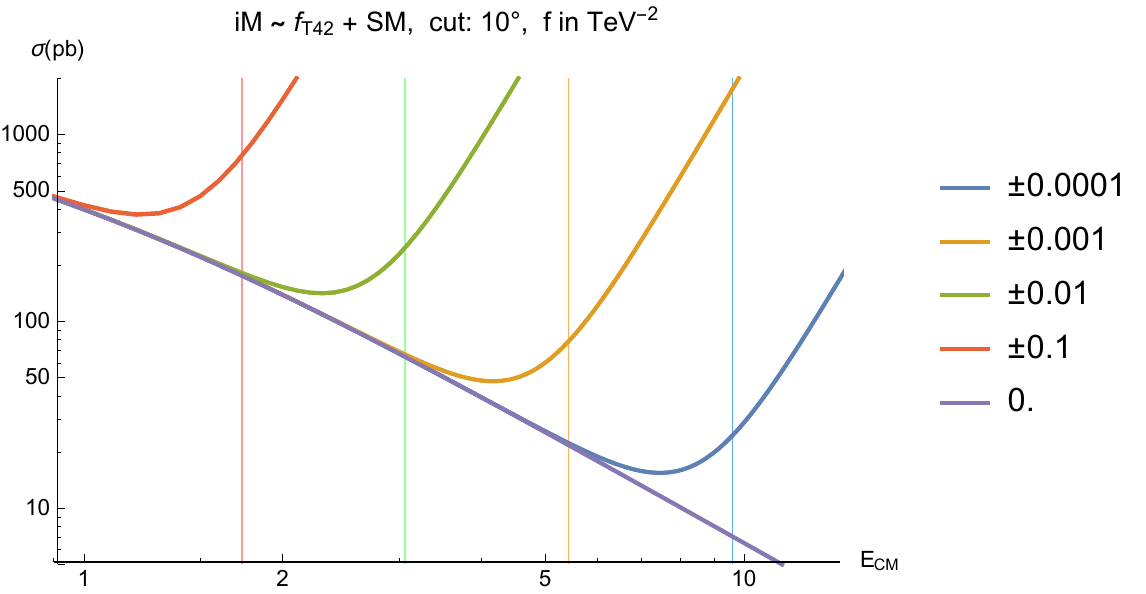}&  \includegraphics[width=0.5\columnwidth]{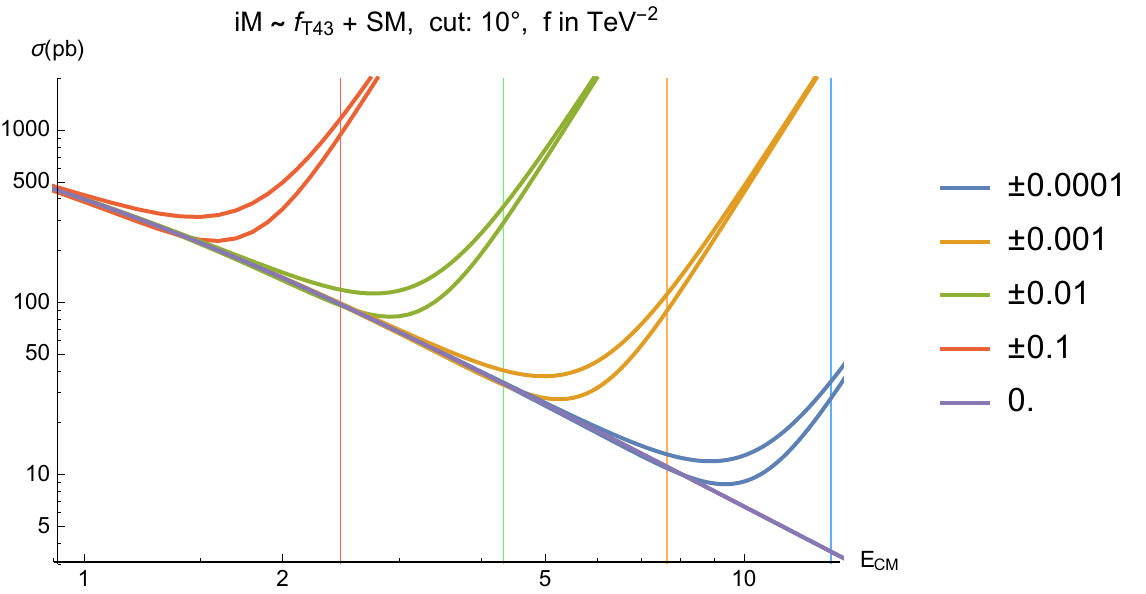} \\
\includegraphics[width=0.5\columnwidth]{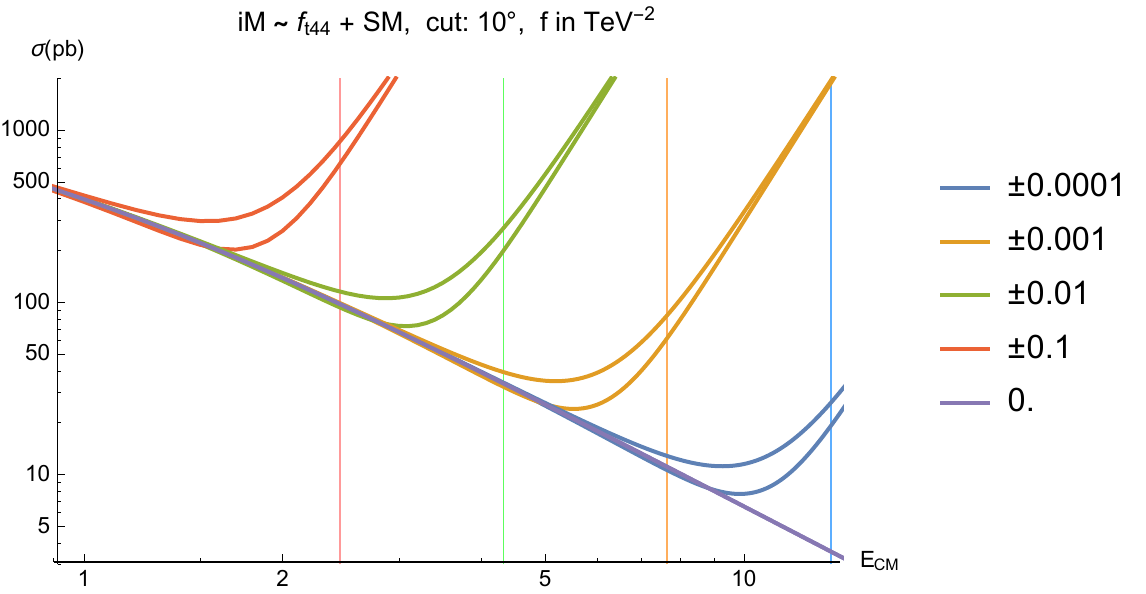}
\end{tabular}
\end{center}
\caption{Energy dependence of the total unpolarized elastic on-shell $W^+W^+$ cross sections ($E_{CM} \equiv \sqrt{s}$, in TeV) for a chosen set of $f_i$ values of the HEFT $\mathcal{T}_{42}$,  $\mathcal{T}_{43}$, $\mathcal{T}_{44}$ operators.  Vertical lines denote the unitarity bound $\sqrt{s^U}$ (color correspondence).  There is no color distinction between the signs: 
both for $\mathcal{T}_{43}$ and $\mathcal{T}_{44}$ upper cross section curves correspond to $f_i<0$; there is practically no sign dependence of unitarity bounds.
}
\label{tab:tutalUnpolHEFT}
\end{figure}
%-------- polarized f> 0 HEFT ------------------------------%

\begin{figure}[h]
\begin{center}
\begin{tabular}{cc}
 \includegraphics[width=0.5\columnwidth]{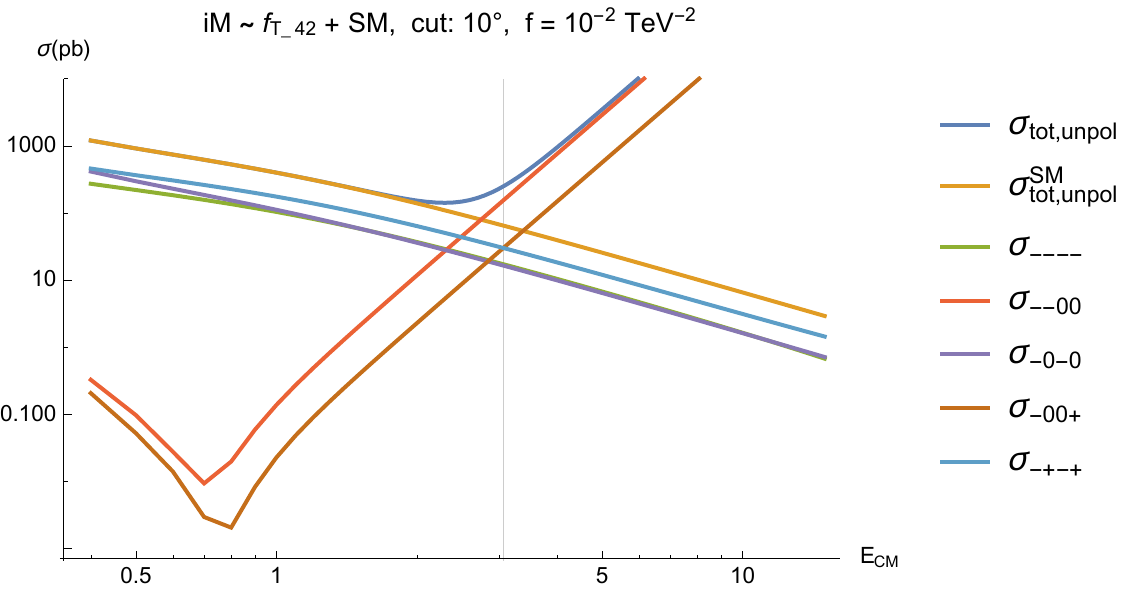} &  \includegraphics[width=0.5\columnwidth]{plotlistPlot10DegreeParticPolsft42Eq1Zoom.pdf} 
\\

\includegraphics[width=0.5\columnwidth]{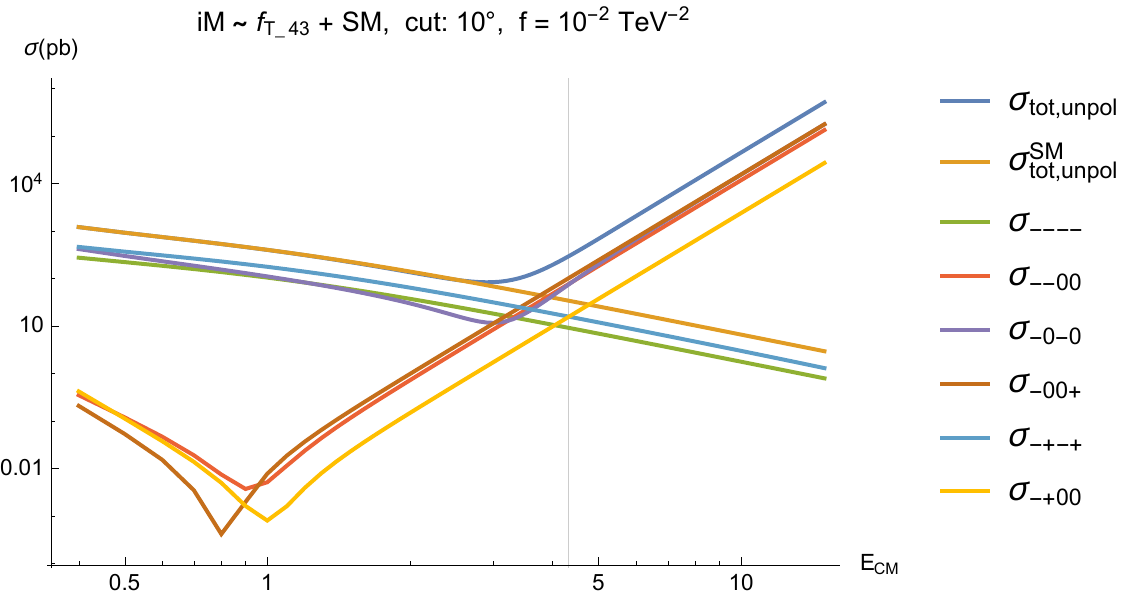} &  \includegraphics[width=0.5\columnwidth]{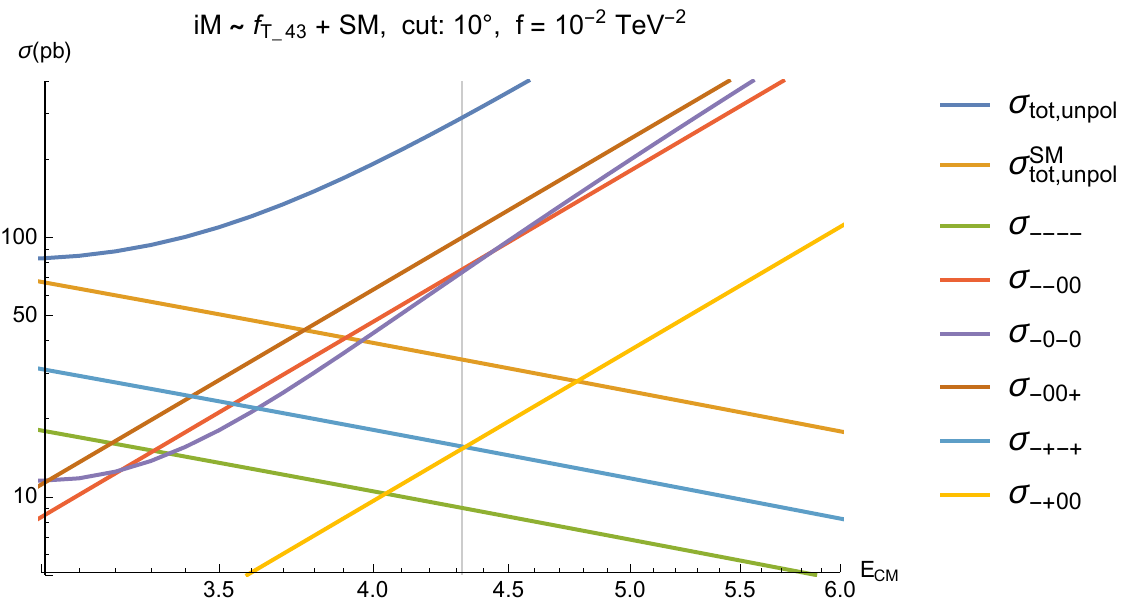} 
\\

\includegraphics[width=0.5\columnwidth]{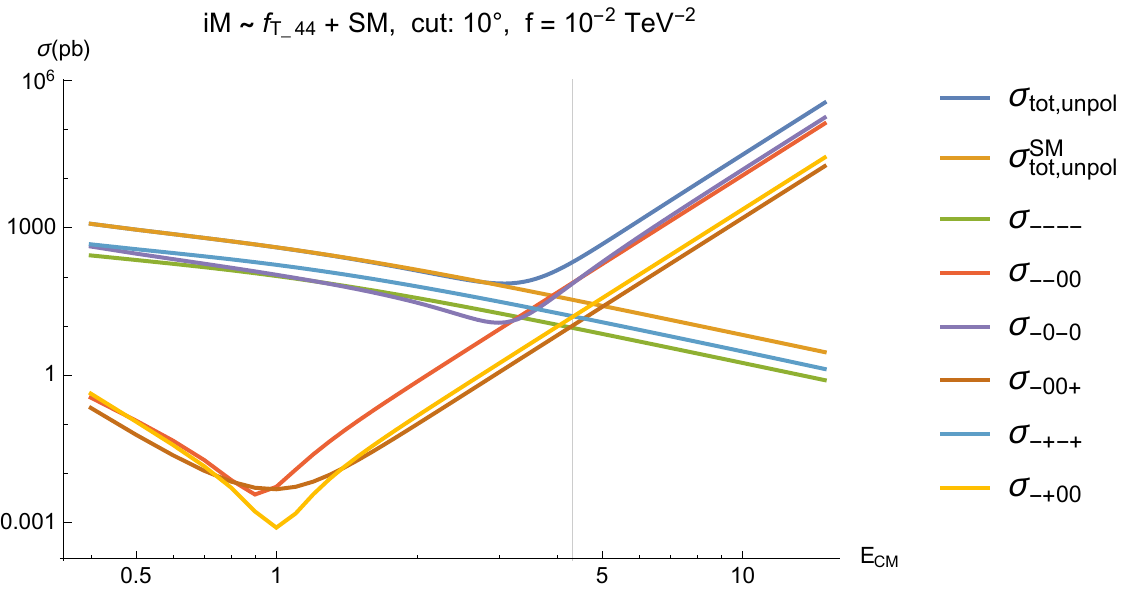} &  \includegraphics[width=0.5\columnwidth]{plotlistPlot10DegreeParticPolsft44Eq1Zoom.pdf} 

\end{tabular}
\end{center}
\caption{Contributions of the polarized cross sections (multiplicity taken into account) as functions of the center-of-mass collision energy
($E_{CM} \equiv \sqrt{s}$, in TeV) for chosen values of $f_i>0$ of the HEFT $\mathcal{T}_{42}$, $\mathcal{T}_{43}$, $\mathcal{T}_{44}$ operators. The remaining (not shown) polarized contributions are negligibly small. In each plot shown is in addition the total cross section of a EFT ''model'' and the total cross section in the SM. In each row right plot is zoom in of the left one.}
\label{tab:polsHEFT}
\end{figure}

%-------- polarized f< 0 HEFT ------------------------------%

\begin{figure}[h]
\begin{center}
\begin{tabular}{cc}
 \includegraphics[width=0.5\columnwidth]{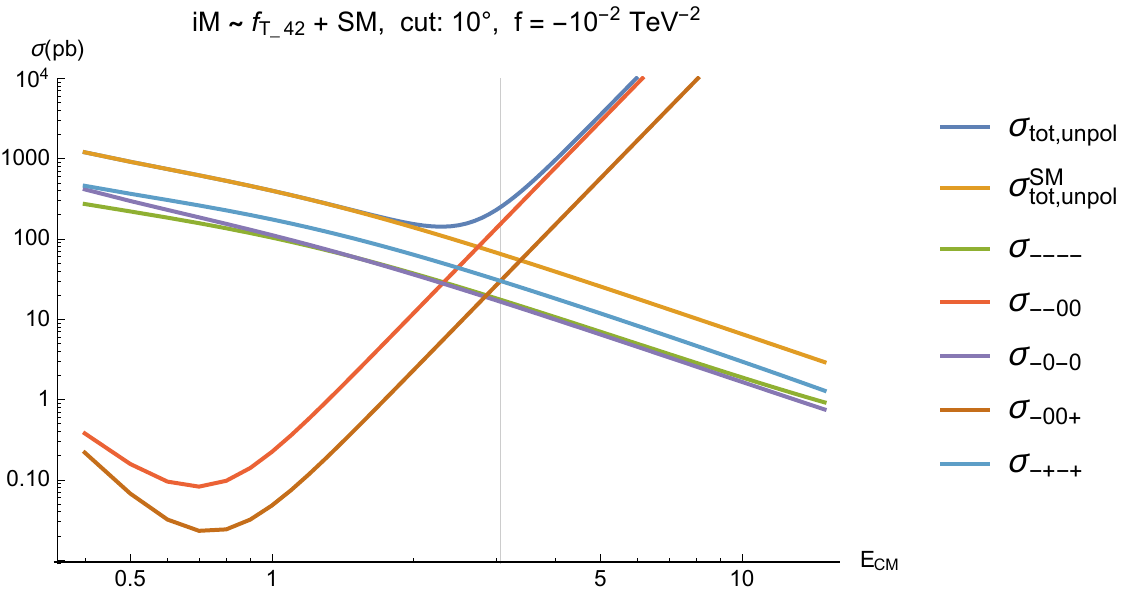} &  \includegraphics[width=0.5\columnwidth]{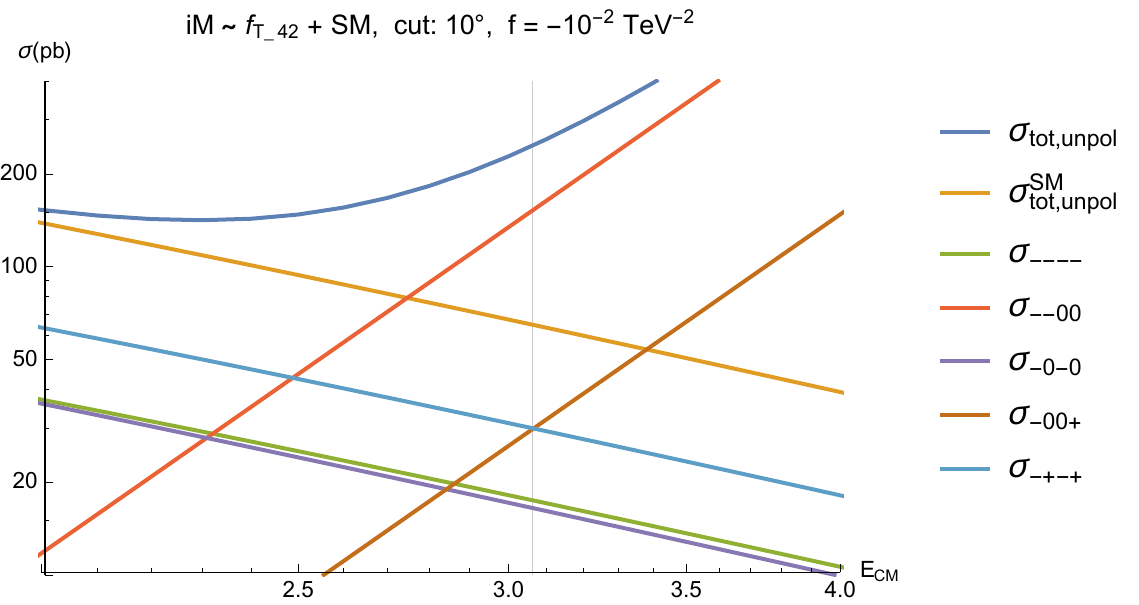} 
\\

 \includegraphics[width=0.5\columnwidth]{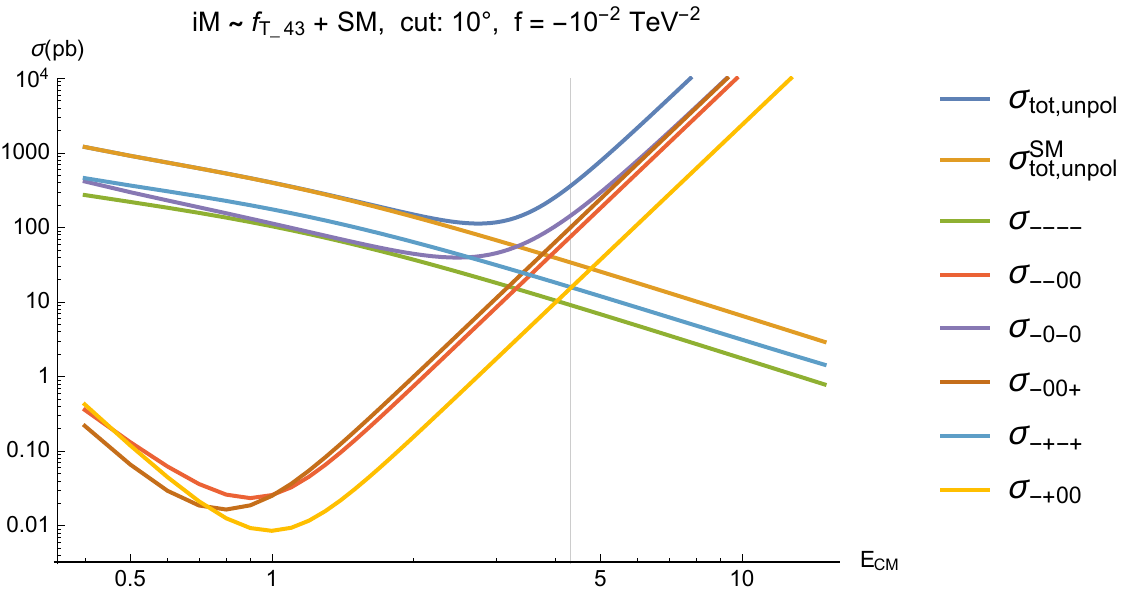} &  \includegraphics[width=0.5\columnwidth]{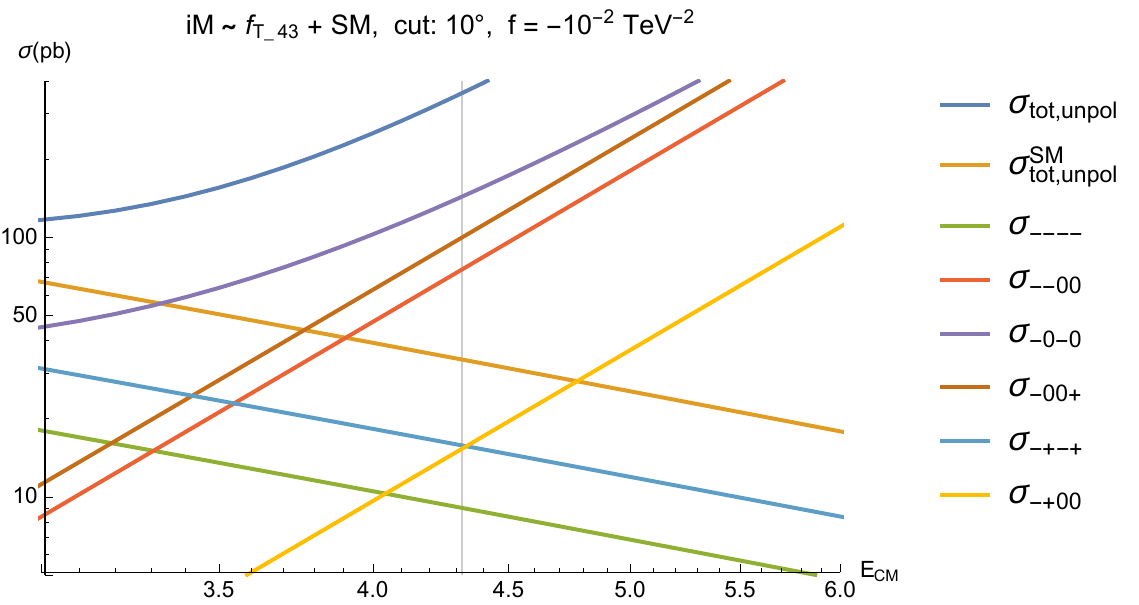} 
\\

 \includegraphics[width=0.5\columnwidth]{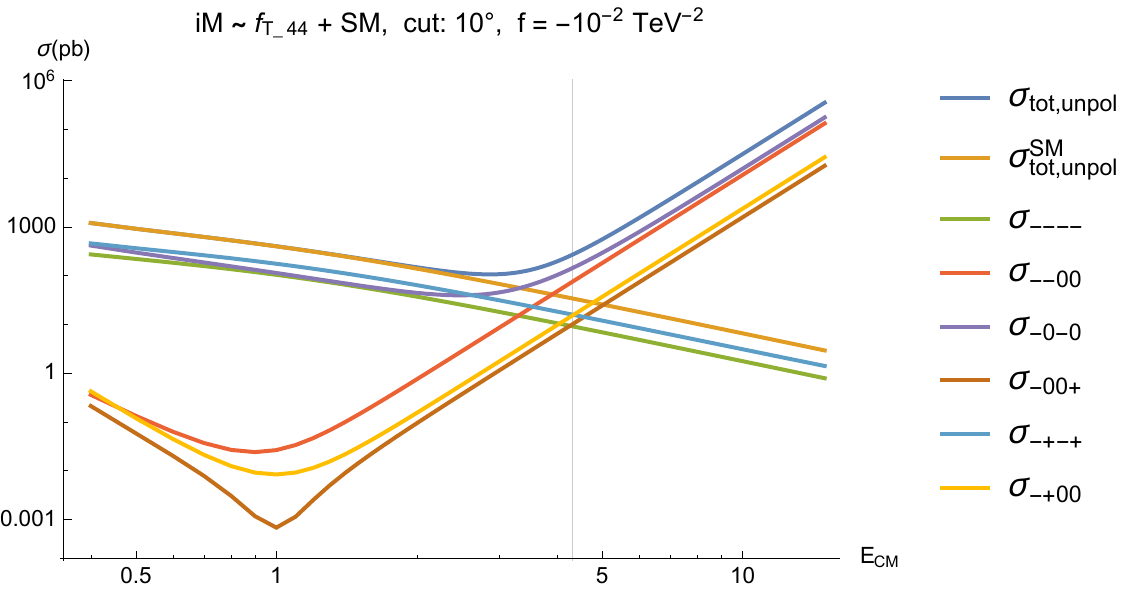} &  \includegraphics[width=0.5\columnwidth]{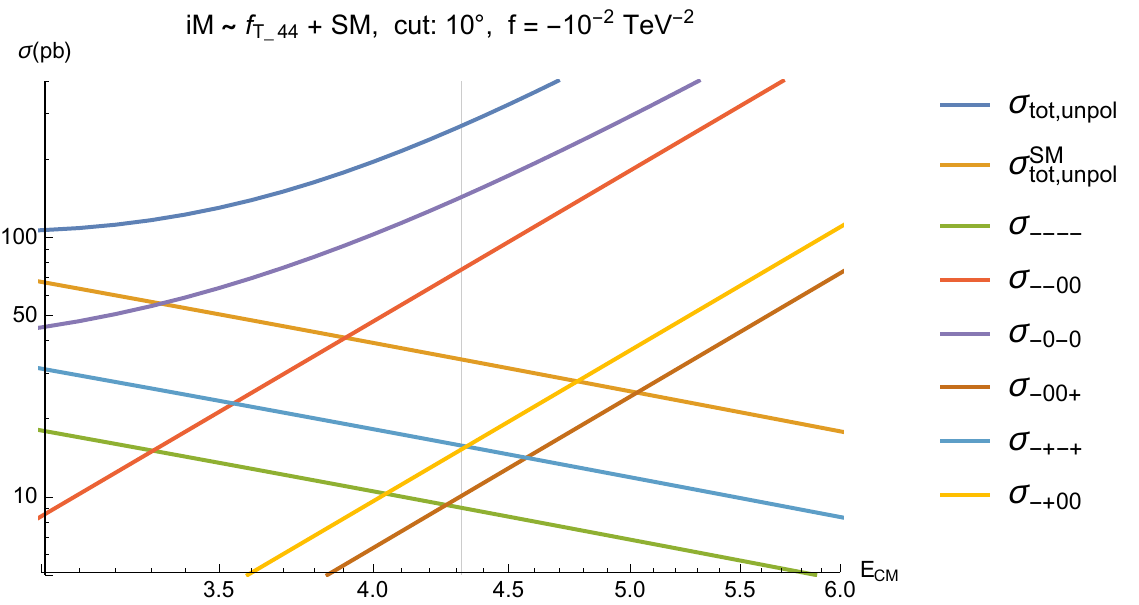} 
\end{tabular}
\end{center}
\caption{See description of Fig.~\ref{tab:polsHEFT}; $f_i<0$ case.}
\label{tab:polsHEFTNeg}
\end{figure}

\newgeometry{tmargin=2cm, bmargin=3cm, lmargin=3cm, rmargin=3cm} 
%
%%%%%
%%%%%%%%%%%%%%%%%%%%%%%%%  Bibliography    %%%%%%%%%%%%%%%%%%%%%%%%
%%%%%
\pagebreak
\newpage
%\footnotesize

%\bibliography{biblio}{}
%\bibliographystyle{BiblioStyle}

\providecommand{\href}[2]{#2}\begingroup\raggedright\endgroup
%
%\bibliography{biblio}{}
%\bibliographystyle{BiblioStyle}

%%%%%%%%%%%%%%%%%%%%%%%%%%%%%%%%%%%%%%%%%%%%%%%%%%%%%%%%%%%%%%%%%%%%%%%%%%%%%%%%%%%%%%%%%%%%%%%%%%%%%%%%%%
% BIBLIOGRAPHY
% REMEMBER TO WRITE THE LINE bibtex article IN THE TERMINAL TO GENERATE THE CORRESPONDING FILES
%\bibliography{}

\begin{thebibliography}{10}

%the SM-------------------------------------

\bibitem{Glashow:1961tr} 
  S.~L.~Glashow,
  %``Partial Symmetries of Weak Interactions,''
  Nucl.\ Phys.\  {\bf 22}, 579 (1961).
  doi:10.1016/0029-5582(61)90469-2
  %%CITATION = doi:10.1016/0029-5582(61)90469-2;%%
  %7443 citations counted in INSPIRE as of 06 Jul 2019
	
	\bibitem{Weinberg:1967tq} 
  S.~Weinberg,
  %``A Model of Leptons,''
  Phys.\ Rev.\ Lett.\  {\bf 19}, 1264 (1967).
  doi:10.1103/PhysRevLett.19.1264
  %%CITATION = doi:10.1103/PhysRevLett.19.1264;%%
  %11511 citations counted in INSPIRE as of 06 Jul 2019
	
	\bibitem{Salam:1968rm} 
  A.~Salam,
  %``Weak and Electromagnetic Interactions,''
  Conf.\ Proc.\ C {\bf 680519}, 367 (1968).
  %%CITATION = CONFP,C680519,367;%%
  %4247 citations counted in INSPIRE as of 06 Jul 2019
	
	\bibitem{tHooft:1972tcz} 
  G.~'t Hooft and M.~J.~G.~Veltman,
  %``Regularization and Renormalization of Gauge Fields,''
  Nucl.\ Phys.\ B {\bf 44}, 189 (1972).
  doi:10.1016/0550-3213(72)90279-9
  %%CITATION = doi:10.1016/0550-3213(72)90279-9;%%
  %4126 citations counted in INSPIRE as of 06 Jul 2019

	
%----------Higgs mechanism------------------------------------------

\bibitem{Englert:1964et} 
  F.~Englert and R.~Brout,
  %``Broken Symmetry and the Mass of Gauge Vector Mesons,''
  Phys.\ Rev.\ Lett.\  {\bf 13}, 321 (1964).
  doi:10.1103/PhysRevLett.13.321
  %%CITATION = doi:10.1103/PhysRevLett.13.321;%%
  %4686 citations counted in INSPIRE as of 06 Jul 2019
	
	\bibitem{Higgs:1964ia} 
  P.~W.~Higgs,
  %``Broken symmetries, massless particles and gauge fields,''
  Phys.\ Lett.\  {\bf 12}, 132 (1964).
  doi:10.1016/0031-9163(64)91136-9
  %%CITATION = doi:10.1016/0031-9163(64)91136-9;%%
  %4315 citations counted in INSPIRE as of 06 Jul 2019
	
	\bibitem{Higgs:1964pj} 
  P.~W.~Higgs,
  %``Broken Symmetries and the Masses of Gauge Bosons,''
  Phys.\ Rev.\ Lett.\  {\bf 13}, 508 (1964).
  doi:10.1103/PhysRevLett.13.508
  %%CITATION = doi:10.1103/PhysRevLett.13.508;%%
  %5085 citations counted in INSPIRE as of 06 Jul 2019
	
	
	\bibitem{Guralnik:1964eu} 
  G.~S.~Guralnik, C.~R.~Hagen and T.~W.~B.~Kibble,
  %``Global Conservation Laws and Massless Particles,''
  Phys.\ Rev.\ Lett.\  {\bf 13}, 585 (1964).
  doi:10.1103/PhysRevLett.13.585
  %%CITATION = doi:10.1103/PhysRevLett.13.585;%%
  %3505 citations counted in INSPIRE as of 06 Jul 2019
	
	
	\bibitem{Higgs:1966ev} 
  P.~W.~Higgs,
  %``Spontaneous Symmetry Breakdown without Massless Bosons,''
  Phys.\ Rev.\  {\bf 145}, 1156 (1966).
  doi:10.1103/PhysRev.145.1156
  %%CITATION = doi:10.1103/PhysRev.145.1156;%%
  %2927 citations counted in INSPIRE as of 06 Jul 2019
	
	\bibitem{Kibble:1967sv} 
  T.~W.~B.~Kibble,
  %``Symmetry breaking in nonAbelian gauge theories,''
  Phys.\ Rev.\  {\bf 155}, 1554 (1967).
  doi:10.1103/PhysRev.155.1554
  %%CITATION = doi:10.1103/PhysRev.155.1554;%%
  %1614 citations counted in INSPIRE as of 06 Jul 2019


	
%-------EFT overview---------------------------------------
\bibitem{Manohar:2018aog} 
  A.~V.~Manohar,
  %``Introduction to Effective Field Theories,''
  arXiv:1804.05863 [hep-ph].
  %%CITATION = ARXIV:1804.05863;%%
  %15 citations counted in INSPIRE as of 06 Jul 2019
	
\bibitem{Pich:2015lkh} 
  A.~Pich,
  %``Electroweak Symmetry Breaking and the Higgs Boson,''
  Acta Phys.\ Polon.\ B {\bf 47}, 151 (2016)
  doi:10.5506/APhysPolB.47.151
  [arXiv:1512.08749 [hep-ph]].
  %%CITATION = doi:10.5506/APhysPolB.47.151;%%
  %11 citations counted in INSPIRE as of 07 Jul 2019

%------------SMEFT -----------------------------------

\bibitem{Buchmuller:1985jz} 
  W.~Buchmuller and D.~Wyler,
  %``Effective Lagrangian Analysis of New Interactions and Flavor Conservation,''
  Nucl.\ Phys.\ B {\bf 268}, 621 (1986).
  doi:10.1016/0550-3213(86)90262-2
  %%CITATION = doi:10.1016/0550-3213(86)90262-2;%%
  %1555 citations counted in INSPIRE as of 06 Jul 2019

\bibitem{Grzadkowski:2010es} 
  B.~Grzadkowski, M.~Iskrzynski, M.~Misiak and J.~Rosiek,
  %``Dimension-Six Terms in the Standard Model Lagrangian,''
  JHEP {\bf 1010}, 085 (2010)
  doi:10.1007/JHEP10(2010)085
  [arXiv:1008.4884 [hep-ph]].
  %%CITATION = doi:10.1007/JHEP10(2010)085;%%
  %907 citations counted in INSPIRE as of 06 Jul 2019

%---------------HEFT--------------------------------------

\bibitem{Feruglio:1992wf}
F.~Feruglio,  
Int.  J. Mod. Phys. {\bf A8} (1993) 4937--4972,
  [hep-ph/9301281].

\bibitem{Grinstein:2007iv}
B.~Grinstein and M.~Trott,  Phys. Rev. {\bf D76} (2007) 073002,
  [arXiv:0704.1505].

\bibitem{Contino:2010mh}
R.~Contino, C.~Grojean, M.~Moretti, F.~Piccinini, and R.~Rattazzi,  JHEP {\bf 05} (2010) 089,
  [arXiv:1002.1011].

\bibitem{Alonso:2012px}
R.~Alonso, M.~B. Gavela, L.~Merlo, S.~Rigolin, and J.~Yepes,  Phys.
  Lett. {\bf B722} (2013) 330--335,
  [arXiv:1212.3305], [Erratum:
  Phys. Lett.B726,926(2013)].

\bibitem{Alonso:2012pz}
R.~Alonso, M.~B. Gavela, L.~Merlo, S.~Rigolin, and J.~Yepes,  Phys. Rev. {\bf D87} (2013), no.~5
  055019, [arXiv:1212.3307].

\bibitem{Buchalla:2013rka}
G.~Buchalla, O.~Cata, and C.~Krause,  Nucl. Phys. {\bf B880} (2014)
  552--573, [arXiv:1307.5017],
  [Erratum: Nucl. Phys.B913,475(2016)].

\bibitem{Brivio:2013pma}
I.~Brivio, T.~Corbett, O.~J.~P. Eboli, M.~B. Gavela, J.~Gonzalez-Fraile, M.~C.
  Gonzalez-Garcia, L.~Merlo, and S.~Rigolin,  JHEP {\bf 03} (2014) 024,
   [arXiv:1311.1823].

\bibitem{Brivio:2014pfa}
I.~Brivio, O.~J.~P. Eboli, M.~B. Gavela, M.~C. Gonzalez-Garcia, L.~Merlo, and
  S.~Rigolin,  JHEP {\bf 12} (2014) 004,
  [arXiv:1405.5412].

\bibitem{Gavela:2014vra}
M.~B. Gavela, J.~Gonzalez-Fraile, M.~C. Gonzalez-Garcia, L.~Merlo, S.~Rigolin,
  and J.~Yepes,  JHEP {\bf 10}
  (2014) 044, [arXiv:1406.6367].

\bibitem{Gavela:2014uta}
M.~B. Gavela, K.~Kanshin, P.~A.~N. Machado, and S.~Saa,  JHEP
  {\bf 03} (2015) 043, [arXiv:1409.1571].

\bibitem{Eboli:2016kko}
O.~J.~P. Eboli and M.~C. Gonzalez-Garcia,  Phys. Rev. {\bf D93} (2016), no.~9 093013,
  [arXiv:1604.03555].

\bibitem{Brivio:2016fzo}
I.~Brivio, J.~Gonzalez-Fraile, M.~C. Gonzalez-Garcia, and L.~Merlo,  Eur. Phys. J. {\bf C76}
  (2016), no.~7 416, [arXiv:1604.06801].

\bibitem{deFlorian:2016spz}
{\bf LHC Higgs Cross Section Working Group} Collaboration, D.~de~Florian {\em
  et.~al.},  [arXiv:1610.07922].

\bibitem{Merlo:2016prs}
L.~Merlo, S.~Saa, and M.~Sacristan-Barbero,  Eur. Phys. J. {\bf C77} (2017),
  no.~3 185, [arXiv:1612.04832].

\bibitem{Buchalla:2017jlu}
G.~Buchalla, O.~Cata, A.~Celis, M.~Knecht, and C.~Krause,  Nucl.
  Phys. {\bf B928} (2018) 93--106,
  [arXiv:1710.06412].

\bibitem{Alonso:2017tdy}
R.~Alonso, K.~Kanshin, and S.~Saa,  Phys. Rev. {\bf D97} (2018), no.~3 035010,
  [arXiv:1710.06848].

\bibitem{Pich:2015kwa} 
  A.~Pich, I.~Rosell, J.~Santos and J.~J.~Sanz-Cillero,
  %``Low-energy signals of strongly-coupled electroweak symmetry-breaking scenarios,''
  Phys.\ Rev.\ D {\bf 93}, no. 5, 055041 (2016)
  doi:10.1103/PhysRevD.93.055041
  [arXiv:1510.03114 [hep-ph]].
  %%CITATION = doi:10.1103/PhysRevD.93.055041;%%
  %16 citations counted in INSPIRE as of 07 Jul 2019

\bibitem{Pich:2016lew} 
  A.~Pich, I.~Rosell, J.~Santos and J.~J.~Sanz-Cillero,
  %``Fingerprints of heavy scales in electroweak effective Lagrangians,''
  JHEP {\bf 1704}, 012 (2017)
  doi:10.1007/JHEP04(2017)012
  [arXiv:1609.06659 [hep-ph]].
  %%CITATION = doi:10.1007/JHEP04(2017)012;%%
  %13 citations counted in INSPIRE as of 07 Jul 2019
%------------Spinors----------------------------------------
\bibitem{Dreiner:2008tw} 
  H.~K.~Dreiner, H.~E.~Haber and S.~P.~Martin,
  %``Two-component spinor techniques and Feynman rules for quantum field theory and supersymmetry,''
  Phys.\ Rept.\  {\bf 494}, 1 (2010)
  doi:10.1016/j.physrep.2010.05.002
  [arXiv:0812.1594 [hep-ph]].
  %%CITATION = doi:10.1016/j.physrep.2010.05.002;%%
  %308 citations counted in INSPIRE as of 06 Jul 2019
	
%---- GB theorem----------------------------------------------

\bibitem{Goldstone:1961eq} 
  J.~Goldstone,
  %``Field Theories with Superconductor Solutions,''
  Nuovo Cim.\  {\bf 19}, 154 (1961).
  doi:10.1007/BF02812722
  %%CITATION = doi:10.1007/BF02812722;%%
  %1772 citations counted in INSPIRE as of 08 Jul 2019
 Y.~Nambu,
  %``Quasiparticles and Gauge Invariance in the Theory of Superconductivity,''
  Phys.\ Rev.\  {\bf 117}, 648 (1960).
  doi:10.1103/PhysRev.117.648
  %%CITATION = doi:10.1103/PhysRev.117.648;%%
  %568 citations counted in INSPIRE as of 08 Jul 2019
Y.~Nambu and G.~Jona-Lasinio,
  %``Dynamical Model of Elementary Particles Based on an Analogy with Superconductivity. 1.,''
  Phys.\ Rev.\  {\bf 122}, 345 (1961).
  doi:10.1103/PhysRev.122.345
  %%CITATION = doi:10.1103/PhysRev.122.345;%%
  %5162 citations counted in INSPIRE as of 08 Jul 2019
 Y.~Nambu and G.~Jona-Lasinio,
  %``Dynamical Model Of Elementary Particles Based On An Analogy With Superconductivity. Ii,''
  Phys.\ Rev.\  {\bf 124}, 246 (1961).
  doi:10.1103/PhysRev.124.246
  %%CITATION = doi:10.1103/PhysRev.124.246;%%
  %2562 citations counted in INSPIRE as of 08 Jul 2019
	
%-------------PDG------------------------------------------------

\bibitem{Nakamura:2010zzi} 
  K.~Nakamura {\it et al.} [Particle Data Group],
  %``Review of particle physics,''
  J.\ Phys.\ G {\bf 37}, 075021 (2010).
  doi:10.1088/0954-3899/37/7A/075021
  %%CITATION = doi:10.1088/0954-3899/37/7A/075021;%%
  %6233 citations counted in INSPIRE as of 06 Jul 2019
	
%---------------SM review---------------------------------------------

\bibitem{Nir} 
 (Lecture notes) Y. Nir,
''The Standard Model and Flavor Physics'', available at http://indico.ictp.it/event/7968/ (under ''The SM and Flavor'')

\bibitem{Grossman:2017thq} 
  Y.~Grossman and P.~Tanedo,
  %``Just a Taste: Lectures on Flavor Physics,''
  doi:10.1142/9789813233348\_0004
  arXiv:1711.03624 [hep-ph].
  %%CITATION = doi:10.1142/9789813233348_0004;%%
  %5 citations counted in INSPIRE as of 06 Jul 2019
	
\bibitem{Pich:2012sx} 
  A.~Pich,
  %``The Standard Model of Electroweak Interactions,''
  arXiv:1201.0537 [hep-ph].
  %%CITATION = ARXIV:1201.0537;%%
  %36 citations counted in INSPIRE as of 06 Jul 2019
	
%\bibitem{Pich:2015lkh} 
  %A.~Pich,
  %%``Electroweak Symmetry Breaking and the Higgs Boson,''
  %Acta Phys.\ Polon.\ B {\bf 47}, 151 (2016)
  %doi:10.5506/APhysPolB.47.151
  %[arXiv:1512.08749 [hep-ph]].
  %%%CITATION = doi:10.5506/APhysPolB.47.151;%%
  %%11 citations counted in INSPIRE as of 07 Jul 2019
	
%----------electron anomalous mag moment ---------------------------------------------

\bibitem{Hanneke:2008tm} 
  D.~Hanneke, S.~Fogwell and G.~Gabrielse,
  %``New Measurement of the Electron Magnetic Moment and the Fine Structure Constant,''
  Phys.\ Rev.\ Lett.\  {\bf 100}, 120801 (2008)
  doi:10.1103/PhysRevLett.100.120801
  [arXiv:0801.1134 [physics.atom-ph]].
  %%CITATION = doi:10.1103/PhysRevLett.100.120801;%%
  %462 citations counted in INSPIRE as of 06 Jul 2019	
%----------alfaQCD-------------------------------------------

\bibitem{Bethke:2011tr} 
  S.~Bethke {\it et al.},
  %``Workshop on Precision Measurements of alphas,''
  arXiv:1110.0016 [hep-ph].
  %%CITATION = ARXIV:1110.0016;%%
  %104 citations counted in INSPIRE as of 06 Jul 2019
%---------aleph-------------------------------------------------

\bibitem{ALEPH:2010aa} 
  LEP Electroweak Working Group [ALEPH and CDF and D0 and DELPHI and L3 and OPAL and SLD Collaborations and LEP Electroweak Working Group and Tevatron Electroweak Working Group and SLD Electroweak and Heavy Flavour Groups],
  %``Precision Electroweak Measurements and Constraints on the Standard Model,''
  arXiv:1012.2367 [hep-ex]; http://www.cern.ch/LEPEWWG/.
  %%CITATION = ARXIV:1012.2367;%%
  %397 citations counted in INSPIRE as of 06 Jul 2019
%-------tevatron,top discovery-----------------------------------

\bibitem{Lancaster:2011wr} 
  [CDF and D0 Collaborations and Tevatron Electroweak Working Group],
  %``Combination of CDF and D0 results on the mass of the top quark using up to 5.8~fb-1 of data,''
  arXiv:1107.5255 [hep-ex].
  %%CITATION = ARXIV:1107.5255;%%
  %309 citations counted in INSPIRE as of 06 Jul 2019
	
%---------indirect upper bound on higgs mass-------------------------------------------------

\bibitem{Baak:2011ze} 
  M.~Baak, M.~Goebel, J.~Haller, A.~Hoecker, D.~Ludwig, K.~Moenig, M.~Schott and J.~Stelzer,
  %``Updated Status of the Global Electroweak Fit and Constraints on New Physics,''
  Eur.\ Phys.\ J.\ C {\bf 72}, 2003 (2012)
  doi:10.1140/epjc/s10052-012-2003-4
  [arXiv:1107.0975 [hep-ph]].
  %%CITATION = doi:10.1140/epjc/s10052-012-2003-4;%%
  %320 citations counted in INSPIRE as of 06 Jul 2019
	
%----------Higgs discovery---------------------------------------

\bibitem{Aad:2012tfa} 
  G.~Aad {\it et al.} [ATLAS Collaboration],
  %``Observation of a new particle in the search for the Standard Model Higgs boson with the ATLAS detector at the LHC,''
  Phys.\ Lett.\ B {\bf 716}, 1 (2012)
  doi:10.1016/j.physletb.2012.08.020
  [arXiv:1207.7214 [hep-ex]].
  %%CITATION = doi:10.1016/j.physletb.2012.08.020;%%
  %9523 citations counted in INSPIRE as of 06 Jul 2019

\bibitem{Chatrchyan:2012xdj} 
  S.~Chatrchyan {\it et al.} [CMS Collaboration],
  %``Observation of a New Boson at a Mass of 125 GeV with the CMS Experiment at the LHC,''
  Phys.\ Lett.\ B {\bf 716}, 30 (2012)
  doi:10.1016/j.physletb.2012.08.021
  [arXiv:1207.7235 [hep-ex]].
  %%CITATION = doi:10.1016/j.physletb.2012.08.021;%%
  %9313 citations counted in INSPIRE as of 06 Jul 2019
		
\bibitem{Aad:2008zzm} 
  G.~Aad {\it et al.} [ATLAS Collaboration],
  %``The ATLAS Experiment at the CERN Large Hadron Collider,''
  JINST {\bf 3}, S08003 (2008).
  doi:10.1088/1748-0221/3/08/S08003
  %%CITATION = doi:10.1088/1748-0221/3/08/S08003;%%
  %7235 citations counted in INSPIRE as of 06 Jul 2019

\bibitem{Chatrchyan:2008aa} 
  S.~Chatrchyan {\it et al.} [CMS Collaboration],
  %``The CMS Experiment at the CERN LHC,''
  JINST {\bf 3}, S08004 (2008).
  doi:10.1088/1748-0221/3/08/S08004
  %%CITATION = doi:10.1088/1748-0221/3/08/S08004;%%
  %5989 citations counted in INSPIRE as of 06 Jul 2019
	
\bibitem{Aad:2015zhl} 
  G.~Aad {\it et al.} [ATLAS and CMS Collaborations],
  %``Combined Measurement of the Higgs Boson Mass in $pp$ Collisions at $\sqrt{s}=7$ and 8 TeV with the ATLAS and CMS Experiments,''
  Phys.\ Rev.\ Lett.\  {\bf 114}, 191803 (2015)
  doi:10.1103/PhysRevLett.114.191803
  [arXiv:1503.07589 [hep-ex]].
  %%CITATION = doi:10.1103/PhysRevLett.114.191803;%%
  %1361 citations counted in INSPIRE as of 06 Jul 2019

\bibitem{Aad:2015mxa} 
  G.~Aad {\it et al.} [ATLAS Collaboration],
  %``Study of the spin and parity of the Higgs boson in diboson decays with the ATLAS detector,''
  Eur.\ Phys.\ J.\ C {\bf 75}, no. 10, 476 (2015)
  Erratum: [Eur.\ Phys.\ J.\ C {\bf 76}, no. 3, 152 (2016)]
  doi:10.1140/epjc/s10052-015-3685-1, 10.1140/epjc/s10052-016-3934-y
  [arXiv:1506.05669 [hep-ex]].
  %%CITATION = doi:10.1140/epjc/s10052-015-3685-1, 10.1140/epjc/s10052-016-3934-y;%%
  %221 citations counted in INSPIRE as of 06 Jul 2019
	
\bibitem{Khachatryan:2014kca} 
  V.~Khachatryan {\it et al.} [CMS Collaboration],
  %``Constraints on the spin-parity and anomalous HVV couplings of the Higgs boson in proton collisions at 7 and 8 TeV,''
  Phys.\ Rev.\ D {\bf 92}, no. 1, 012004 (2015)
  doi:10.1103/PhysRevD.92.012004
  [arXiv:1411.3441 [hep-ex]].
  %%CITATION = doi:10.1103/PhysRevD.92.012004;%%
  %370 citations counted in INSPIRE as of 06 Jul 2019
	
\bibitem{Khachatryan:2016vau} 
  G.~Aad {\it et al.} [ATLAS and CMS Collaborations],
  %``Measurements of the Higgs boson production and decay rates and constraints on its couplings from a combined ATLAS and CMS analysis of the LHC pp collision data at $ \sqrt{s}=7 $ and 8 TeV,''
  JHEP {\bf 1608}, 045 (2016)
  doi:10.1007/JHEP08(2016)045
  [arXiv:1606.02266 [hep-ex]].
  %%CITATION = doi:10.1007/JHEP08(2016)045;%%
  %962 citations counted in INSPIRE as of 06 Jul 2019
	
	
\bibitem{Aad:2015pja} 
  G.~Aad {\it et al.} [ATLAS Collaboration],
  %``Search for flavour-changing neutral current top quark decays $t\to Hq$ in $pp$ collisions at $\sqrt{s}=8$ TeV with the ATLAS detector,''
  JHEP {\bf 1512}, 061 (2015)
  doi:10.1007/JHEP12(2015)061
  [arXiv:1509.06047 [hep-ex]]; V.~Khachatryan {\it et al.} [CMS Collaboration],
  %``Search for top quark decays via Higgs-boson-mediated flavor-changing neutral currents in pp collisions at $ \sqrt{s}=8 $ TeV,''
  JHEP {\bf 1702}, 079 (2017)
  doi:10.1007/JHEP02(2017)079
  [arXiv:1610.04857 [hep-ex]].
  %%CITATION = doi:10.1007/JHEP02(2017)079;%%
  %44 citations counted in INSPIRE as of 06 Jul 2019
  %%CITATION = doi:10.1007/JHEP12(2015)061;%%
  %78 citations counted in INSPIRE as of 06 Jul 2019



%-----------------------------------------------------------------
	
\bibitem{Aad:2016blu} 
  G.~Aad {\it et al.} [ATLAS Collaboration],
  %``Search for lepton-flavour-violating decays of the Higgs and $Z$ bosons with the ATLAS detector,''
  Eur.\ Phys.\ J.\ C {\bf 77}, no. 2, 70 (2017)
  doi:10.1140/epjc/s10052-017-4624-0
  [arXiv:1604.07730 [hep-ex]]; V.~Khachatryan {\it et al.} [CMS Collaboration],
  %``Search for lepton flavour violating decays of the Higgs boson to $e \tau$ and $e \mu$ in proton-proton collisions at $\sqrt s=$ 8 TeV,''
  Phys.\ Lett.\ B {\bf 763}, 472 (2016)
  doi:10.1016/j.physletb.2016.09.062
  [arXiv:1607.03561 [hep-ex]]; CMS Collaboration, CMS-PAS-HIG-17-001.
  %%CITATION = doi:10.1016/j.physletb.2016.09.062;%%
  %47 citations counted in INSPIRE as of 06 Jul 2019
  %%CITATION = doi:10.1140/epjc/s10052-017-4624-0;%%
  %104 citations counted in INSPIRE as of 06 Jul 2019
	
%----why beyond----------------------------------------------------------

\bibitem{Pokorski:2005fb} 
  S.~Pokorski,
  %``Phenomenological guide to physics beyond the standard model,''
  doi:10.1007/1-4020-3733-3\_9
  hep-ph/0502132.
  %%CITATION = doi:10.1007/1-4020-3733-3_9;%%
  %10 citations counted in INSPIRE as of 06 Jul 2019	
	
%-----------EFT in flavor and higgs--------------------------------

\bibitem{Fajfer:2012vx} 
  For example: S.~Fajfer, J.~F.~Kamenik and I.~Nisandzic,
  %``On the $B \to D^* \tau \bar \nu_{\tau}$ Sensitivity to New Physics,''
  Phys.\ Rev.\ D {\bf 85}, 094025 (2012)
  doi:10.1103/PhysRevD.85.094025
  [arXiv:1203.2654 [hep-ph]].
  %%CITATION = doi:10.1103/PhysRevD.85.094025;%%
  %437 citations counted in INSPIRE as of 06 Jul 2019
	
\bibitem{Pomarol:2013zra} 
  For example: A.~Pomarol and F.~Riva,
  %``Towards the Ultimate SM Fit to Close in on Higgs Physics,''
  JHEP {\bf 1401}, 151 (2014)
  doi:10.1007/JHEP01(2014)151
  [arXiv:1308.2803 [hep-ph]].
  %%CITATION = doi:10.1007/JHEP01(2014)151;%%
  %190 citations counted in INSPIRE as of 06 Jul 2019
	
	
%------------SMEFT and HEFT WW theorists-------------------------

\bibitem{Kilian:2014zja}
W.~Kilian, T.~Ohl, J.~Reuter, and M.~Sekulla,  
Phys. Rev. {\bf D91} (2015) 096007,
  [arXiv:1408.6207].

\bibitem{Brass:2018hfw}
S.~Brass, C.~Fleper, W.~Kilian, J.~Reuter, and M.~Sekulla, 
  Eur. Phys. J. {\bf C78} (2018), no.~11 931,
  [arXiv:1807.02512].

\bibitem{Gomez-Ambrosio:2018pnl}
R.~G\'omez-Ambrosio,  [arXiv:1809.04189].
	
	
	\bibitem{Espriu:2012ih}
D.~Espriu and B.~Yencho,  Phys. Rev. {\bf D87} (2013), no.~5 055017,
  [arXiv:1212.4158].

\bibitem{Espriu:2013fia}
D.~Espriu, F.~Mescia, and B.~Yencho,  Phys. Rev. {\bf D88} (2013) 055002,
  [arXiv:1307.2400].

\bibitem{Delgado:2013loa}
R.~L. Delgado, A.~Dobado, and F.~J. Llanes-Estrada,  
J. Phys. {\bf G41} (2014) 025002,
  [arXiv:1308.1629].

\bibitem{Delgado:2013hxa}
R.~L. Delgado, A.~Dobado, and F.~J. Llanes-Estrada,  
JHEP {\bf 02} (2014) 121,
  [arXiv:1311.5993].

\bibitem{Espriu:2014jya}
D.~Espriu and F.~Mescia,  Phys. Rev. {\bf D90} (2014), no.~1 015035,
  [ arXiv:1403.7386].

\bibitem{Delgado:2014jda}
R.~L. Delgado, A.~Dobado, M.~J. Herrero, and J.~J. Sanz-Cillero,
 JHEP {\bf 07} (2014) 149,
  [arXiv:1404.2866].

\bibitem{Delgado:2017cls}
R.~L. Delgado, A.~Dobado, D.~Espriu, C.~Garc{\'\i a-}Garcia, M.~J. Herrero,
  X.~Marcano, and J.~J. Sanz-Cillero,  
	JHEP {\bf 11}(2017) 098, 
	[arXiv:1707.04580].
	
	
%-----------WW in experiment-----------------------------------------

\bibitem{Aad:2014zda} 
  G.~Aad {\it et al.} [ATLAS Collaboration],
  %``Evidence for Electroweak Production of $W^{\pm}W^{\pm}jj$ in $pp$ Collisions at $\sqrt{s}=8$ TeV with the ATLAS Detector,''
  Phys.\ Rev.\ Lett.\  {\bf 113}, no. 14, 141803 (2014)
  doi:10.1103/PhysRevLett.113.141803
  [arXiv:1405.6241 [hep-ex]].
  %%CITATION = doi:10.1103/PhysRevLett.113.141803;%%
  %162 citations counted in INSPIRE as of 06 Jul 2019

\bibitem{Khachatryan:2014sta} 
  V.~Khachatryan {\it et al.} [CMS Collaboration],
  %``Study of vector boson scattering and search for new physics in events with two same-sign leptons and two jets,''
  Phys.\ Rev.\ Lett.\  {\bf 114}, no. 5, 051801 (2015)
  doi:10.1103/PhysRevLett.114.051801
  [arXiv:1410.6315 [hep-ex]].
  %%CITATION = doi:10.1103/PhysRevLett.114.051801;%%
  %111 citations counted in INSPIRE as of 06 Jul 2019

\bibitem{Sirunyan:2017ret} %55
  A.~M.~Sirunyan {\it et al.} [CMS Collaboration],
  %``Observation of electroweak production of same-sign W boson pairs in the two jet and two same-sign lepton final state in proton-proton collisions at $\sqrt{s} = $ 13 TeV,''
  Phys.\ Rev.\ Lett.\  {\bf 120}, no. 8, 081801 (2018)
  doi:10.1103/PhysRevLett.120.081801
  [arXiv:1709.05822 [hep-ex]].
  %%CITATION = doi:10.1103/PhysRevLett.120.081801;%%
  %49 citations counted in INSPIRE as of 06 Jul 2019

\bibitem{Aad:2016ett} 
  G.~Aad {\it et al.} [ATLAS Collaboration],
  %``Measurements of $W^\pm Z$ production cross sections in $pp$ collisions at $\sqrt{s} = 8$ TeV with the ATLAS detector and limits on anomalous gauge boson self-couplings,''
  Phys.\ Rev.\ D {\bf 93}, no. 9, 092004 (2016)
  doi:10.1103/PhysRevD.93.092004
  [arXiv:1603.02151 [hep-ex]].
  %%CITATION = doi:10.1103/PhysRevD.93.092004;%%
  %100 citations counted in INSPIRE as of 06 Jul 2019
	
\bibitem{Sirunyan:2017fvv} 
  A.~M.~Sirunyan {\it et al.} [CMS Collaboration],
  %``Measurement of vector boson scattering and constraints on anomalous quartic couplings from events with four leptons and two jets in proton-proton collisions at $\sqrt{s}=$ 13 TeV,''
  Phys.\ Lett.\ B {\bf 774}, 682 (2017)
  doi:10.1016/j.physletb.2017.10.020
  [arXiv:1708.02812 [hep-ex]].
  %%CITATION = doi:10.1016/j.physletb.2017.10.020;%%
  %24 citations counted in INSPIRE as of 06 Jul 2019
	


\bibitem{ATLAS-CONF-2018-030} 
  The ATLAS Collaboration, ATLAS-CONF-2018-030.
	%
\bibitem{Aad:2012twa} 
  G.~Aad {\it et al.} [ATLAS Collaboration],
  Eur.\ Phys.\ J.\ C {\bf 72}, 2173 (2012)
  doi:10.1140/epjc/s10052-012-2173-0
  [arXiv:1208.1390 [hep-ex]].
  %CITATION = doi:10.1140/epjc/s10052-012-2173-0;%%
  %135 citations counted in INSPIRE as of 06 Jul 2019		
	
\bibitem{Khachatryan:2016poo} 
  V.~Khachatryan {\it et al.} [CMS Collaboration],
  %``Measurement of the WZ production cross section in pp collisions at $\sqrt{s} = 7$ and 8 $\,\text{TeV}$ and search for anomalous triple gauge couplings at $\sqrt{s} = 8\,\text{TeV} $,''
  Eur.\ Phys.\ J.\ C {\bf 77}, no. 4, 236 (2017)
  doi:10.1140/epjc/s10052-017-4730-z
  [arXiv:1609.05721 [hep-ex]].
  %%CITATION = doi:10.1140/epjc/s10052-017-4730-z;%%
  %37 citations counted in INSPIRE as of 06 Jul 2019
	

\bibitem{Khachatryan:2016tgp} 
  V.~Khachatryan {\it et al.} [CMS Collaboration],
  %``Measurement of the WZ production cross section in pp collisions at $\sqrt(s) =$ 13 TeV,''
  Phys.\ Lett.\ B {\bf 766}, 268 (2017)
  doi:10.1016/j.physletb.2017.01.011
  [arXiv:1607.06943 [hep-ex]].
  %%CITATION = doi:10.1016/j.physletb.2017.01.011;%%
  %67 citations counted in INSPIRE as of 06 Jul 2019	
	
\bibitem{Green:2016trm} 
  D.~R.~Green, P.~Meade and M.~A.~Pleier,
  %``Multiboson interactions at the LHC,''
  Rev.\ Mod.\ Phys.\  {\bf 89}, no. 3, 035008 (2017)
  doi:10.1103/RevModPhys.89.035008
  [arXiv:1610.07572 [hep-ex]].
	
  %%CITATION = doi:10.1103/RevModPhys.89.035008;%%
  %22 citations counted in INSPIRE as of 08 Jul 2019
%\bibitem{Aaboud:2016yus} 
  %M.~Aaboud {\it et al.} [ATLAS Collaboration],
  %%``Measurement of the $W^{\pm}Z$ boson pair-production cross section in $pp$ collisions at $\sqrt{s}=13$ TeV with the ATLAS Detector,''
  %Phys.\ Lett.\ B {\bf 762}, 1 (2016)
  %doi:10.1016/j.physletb.2016.08.052
  %[arXiv:1606.04017 [hep-ex]].
  %%%CITATION = doi:10.1016/j.physletb.2016.08.052;%%	
	
	
%---------Chankowski---------------------------------------------------------------------------------------
\bibitem{chankowski} 
 P. Chankowski, lecture notes available at https://www.fuw.edu.pl/~chank/qftoei.html.

%--------CPT relations-----------------------------------------------------------------------------------

\bibitem{deRham:2017zjm} 
  C.~de Rham, S.~Melville, A.~J.~Tolley and S.~Y.~Zhou,
  %``UV complete me: Positivity Bounds for Particles with Spin,''
  JHEP {\bf 1803}, 011 (2018)
  doi:10.1007/JHEP03(2018)011
  [arXiv:1706.02712 [hep-th]]: the relations are discussed in the Appendix E.
  %%CITATION = doi:10.1007/JHEP03(2018)011;%%
  %25 citations counted in INSPIRE as of 07 Jul 2019
	
%-------Unitarity from gauge inv--------------------------------------------------------

\bibitem{Cornwall:1974km} 
  J.~M.~Cornwall, D.~N.~Levin and G.~Tiktopoulos,
  %``Derivation of Gauge Invariance from High-Energy Unitarity Bounds on the s Matrix,''
  Phys.\ Rev.\ D {\bf 10}, 1145 (1974)
  Erratum: [Phys.\ Rev.\ D {\bf 11}, 972 (1975)].
  doi:10.1103/PhysRevD.10.1145, 10.1103/PhysRevD.11.972
  %%CITATION = doi:10.1103/PhysRevD.10.1145, 10.1103/PhysRevD.11.972;%%
  %1198 citations counted in INSPIRE as of 07 Jul 2019
	
\bibitem{LlewellynSmith:1973yud} 
  C.~H.~Llewellyn Smith,
  %``High-Energy Behavior and Gauge Symmetry,''
  Phys.\ Lett.\  {\bf 46B}, 233 (1973).
  doi:10.1016/0370-2693(73)90692-8
  %%CITATION = doi:10.1016/0370-2693(73)90692-8;%%
  %467 citations counted in INSPIRE as of 07 Jul 2019

%--------Fermi-------------------------------------------------------------------------

\bibitem{Fermi:1933jpa} 
  E.~Fermi,
  %``Tentativo di una teoria dell'emissione dei raggi beta,''
  Ric.\ Sci.\  {\bf 4}, 491 (1933).
  %%CITATION = RISCA,4,491;%%
  %96 citations counted in INSPIRE as of 07 Jul 2019
%----------SMEFT bases--------------------------------------------------------------------

\bibitem{Falkowski:2015wza} 
  See the discussion in A.~Falkowski, B.~Fuks, K.~Mawatari, K.~Mimasu, F.~Riva and V.~Sanz,
  %``Rosetta: an operator basis translator for Standard Model effective field theory,''
  Eur.\ Phys.\ J.\ C {\bf 75}, no. 12, 583 (2015)
  doi:10.1140/epjc/s10052-015-3806-x
  [arXiv:1508.05895 [hep-ph]].
  %%CITATION = doi:10.1140/epjc/s10052-015-3806-x;%%
  %66 citations counted in INSPIRE as of 07 Jul 2019


%-----------chiral pert theory-----------------------------------------------------------------------

\bibitem{Pich:1998xt} %chiral
  A.~Pich,
  %``Effective field theory: Course,''
  hep-ph/9806303.
  %%CITATION = HEP-PH/9806303;%%
  %268 citations counted in INSPIRE as of 07 Jul 2019


\bibitem{Pich:1995bw} %chiral
  A.~Pich,
  %``Chiral perturbation theory,''
  Rept.\ Prog.\ Phys.\  {\bf 58}, 563 (1995)
  doi:10.1088/0034-4885/58/6/001
  [hep-ph/9502366].
  %%CITATION = doi:10.1088/0034-4885/58/6/001;%%
  %454 citations counted in INSPIRE as of 07 Jul 2019

\bibitem{Pich:2018ltt} 
  A.~Pich,
  %``Effective Field Theory with Nambu-Goldstone Modes,''
  arXiv:1804.05664 [hep-ph].
  %%CITATION = ARXIV:1804.05664;%%
  %12 citations counted in INSPIRE as of 07 Jul 2019

\bibitem{Scherer:2005ri} 
  S.~Scherer and M.~R.~Schindler,
  %``A Chiral perturbation theory primer,''
  hep-ph/0505265.
  %%CITATION = HEP-PH/0505265;%%
  %56 citations counted in INSPIRE as of 07 Jul 2019

%------field redefinitions-----------------------------------------------------------------------
\bibitem{Criado:2018sdb} 
  J.~C.~Criado and M.~PErez-Victoria,
  %``Field redefinitions in effective theories at higher orders,''
  JHEP {\bf 1903}, 038 (2019)
  doi:10.1007/JHEP03(2019)038
  [arXiv:1811.09413 [hep-ph]].
  %%CITATION = doi:10.1007/JHEP03(2019)038;%%
  %5 citations counted in INSPIRE as of 07 Jul 2019

%------CCWZ construction----------------------------------------------------------------------------

\bibitem{Coleman:1969sm} 
  S.~R.~Coleman, J.~Wess and B.~Zumino,
  %``Structure of phenomenological Lagrangians. 1.,''
  Phys.\ Rev.\  {\bf 177}, 2239 (1969).
  doi:10.1103/PhysRev.177.2239
  %%CITATION = doi:10.1103/PhysRev.177.2239;%%
  %1930 citations counted in INSPIRE as of 07 Jul 2019

\bibitem{Callan:1969sn} 
  C.~G.~Callan, Jr., S.~R.~Coleman, J.~Wess and B.~Zumino,
  %``Structure of phenomenological Lagrangians. 2.,''
  Phys.\ Rev.\  {\bf 177}, 2247 (1969).
  doi:10.1103/PhysRev.177.2247
  %%CITATION = doi:10.1103/PhysRev.177.2247;%%
  %1703 citations counted in INSPIRE as of 07 Jul 2019

%---------weinbergcountig---------------------------------------------------------------------

\bibitem{Weinberg:1978kz} 
  S.~Weinberg,
  %``Phenomenological Lagrangians,''
  Physica A {\bf 96}, no. 1-2, 327 (1979).
  doi:10.1016/0378-4371(79)90223-1
  %%CITATION = doi:10.1016/0378-4371(79)90223-1;%%
  %3244 citations counted in INSPIRE as of 07 Jul 2019

%------Xpt Op^4 Op^6-------------------------------------------------------------------------------

\bibitem{Gasser:1983yg} 
  J.~Gasser and H.~Leutwyler,
  %``Chiral Perturbation Theory to One Loop,''
  Annals Phys.\  {\bf 158}, 142 (1984).
  doi:10.1016/0003-4916(84)90242-2
  %%CITATION = doi:10.1016/0003-4916(84)90242-2;%%
  %4031 citations counted in INSPIRE as of 07 Jul 2019

\bibitem{Gasser:1984gg} 
  J.~Gasser and H.~Leutwyler,
  %``Chiral Perturbation Theory: Expansions in the Mass of the Strange Quark,''
  Nucl.\ Phys.\ B {\bf 250}, 465 (1985).
  doi:10.1016/0550-3213(85)90492-4
  %%CITATION = doi:10.1016/0550-3213(85)90492-4;%%
  %3761 citations counted in INSPIRE as of 07 Jul 2019

\bibitem{Bijnens:1999sh} 
  J.~Bijnens, G.~Colangelo and G.~Ecker,
  %``The Mesonic chiral Lagrangian of order p**6,''
  JHEP {\bf 9902}, 020 (1999)
  doi:10.1088/1126-6708/1999/02/020
  [hep-ph/9902437].
  %%CITATION = doi:10.1088/1126-6708/1999/02/020;%%
  %343 citations counted in INSPIRE as of 07 Jul 2019

\bibitem{Haefeli:2007ty} 
  C.~Haefeli, M.~A.~Ivanov, M.~Schmid and G.~Ecker,
  %``On the mesonic Lagrangian of order p**6 in chiral SU(2),''
  arXiv:0705.0576 [hep-ph].
  %%CITATION = ARXIV:0705.0576;%%
  %23 citations counted in INSPIRE as of 07 Jul 2019

\bibitem{Bijnens:2001bb}  
  J.~Bijnens, L.~Girlanda and P.~Talavera,
  %``The Anomalous chiral Lagrangian of order p**6,''
  Eur.\ Phys.\ J.\ C {\bf 23}, 539 (2002)
  doi:10.1007/s100520100887
  [hep-ph/0110400].
  %%CITATION = doi:10.1007/s100520100887;%%
  %125 citations counted in INSPIRE as of 07 Jul 2019

\bibitem{Ebertshauser:2001nj} 
  T.~Ebertshauser, H.~W.~Fearing and S.~Scherer,
  %``The Anomalous chiral perturbation theory meson Lagrangian to order p**6 revisited,''
  Phys.\ Rev.\ D {\bf 65}, 054033 (2002)
  doi:10.1103/PhysRevD.65.054033
  [hep-ph/0110261].
  %%CITATION = doi:10.1103/PhysRevD.65.054033;%%
  %71 citations counted in INSPIRE as of 07 Jul 2019

%---------NDA ------------------------------------------------------------------------------

\bibitem{Gavela:2016bzc} 
  B.~M.~Gavela, E.~E.~Jenkins, A.~V.~Manohar and L.~Merlo,
  %``Analysis of General Power Counting Rules in Effective Field Theory,''
  Eur.\ Phys.\ J.\ C {\bf 76}, no. 9, 485 (2016)
  doi:10.1140/epjc/s10052-016-4332-1
  [arXiv:1601.07551 [hep-ph]].
  %%CITATION = doi:10.1140/epjc/s10052-016-4332-1;%%
  %57 citations counted in INSPIRE as of 07 Jul 2019

%-------Custodial----------------------------------------------------------------------------

\bibitem{Appelquist:1980vg} 
  T.~Appelquist and C.~W.~Bernard,
  %``Strongly Interacting Higgs Bosons,''
  Phys.\ Rev.\ D {\bf 22}, 200 (1980).
  doi:10.1103/PhysRevD.22.200
  %%CITATION = doi:10.1103/PhysRevD.22.200;%%
  %674 citations counted in INSPIRE as of 07 Jul 2019
	
%---------EWXL-------------------------------------------------------------------------------

\bibitem{Dobado:1989ax}
A.~Dobado and M.~J. Herrero,  Phys. Lett. {\bf B228}
  (1989) 495--502.

\bibitem{Dobado:1989ue}
A.~Dobado and M.~J. Herrero,  Phys. Lett. {\bf B233} (1989) 505--511.

\bibitem{Dobado:1989gr}
A.~Dobado, M.~J. Herrero, and T.~N. Truong,  Phys. Lett. {\bf B235} (1990) 129.

\bibitem{Dobado:1990jy}
A.~Dobado, M.~J. Herrero, and J.~Terron,  
Z. Phys. {\bf C50} (1991) 205--220.

\bibitem{Dobado:1995qy}
A.~Dobado, M.~J. Herrero, J.~R. Pelaez, E.~Ruiz~Morales, and M.~T. Urdiales,  Phys. Lett. {\bf B352} (1995) 400--410,
  [hep-ph/9502309].

\bibitem{Dobado:1999xb}
A.~Dobado, M.~J. Herrero, J.~R. Pelaez, and E.~Ruiz~Morales,  Phys. Rev. {\bf D62} (2000) 055011,
  [hep-ph/9912224].

\bibitem{Alboteanu:2008my}
A.~Alboteanu, W.~Kilian, and J.~Reuter,  JHEP {\bf 11} (2008) 010,
  [arXiv:0806.4145].

%--------h/v reabsorbed in h wave funcion----------------------------------------------------------

\bibitem{Giudice:2007fh} 
  G.~F.~Giudice, C.~Grojean, A.~Pomarol and R.~Rattazzi,
  %``The Strongly-Interacting Light Higgs,''
  JHEP {\bf 0706}, 045 (2007)
  doi:10.1088/1126-6708/2007/06/045
  [hep-ph/0703164].
  %%CITATION = doi:10.1088/1126-6708/2007/06/045;%%
  %820 citations counted in INSPIRE as of 07 Jul 2019
%-------Technicolor --------------------------------------------------------------------
\bibitem{Susskind:1978ms} 
  L.~Susskind,
  %``Dynamics of Spontaneous Symmetry Breaking in the Weinberg-Salam Theory,''
  Phys.\ Rev.\ D {\bf 20}, 2619 (1979).
  doi:10.1103/PhysRevD.20.2619
  %%CITATION = doi:10.1103/PhysRevD.20.2619;%%
  %2642 citations counted in INSPIRE as of 07 Jul 2019

\bibitem{Dimopoulos:1979es} 
  S.~Dimopoulos and L.~Susskind,
  %``Mass Without Scalars,''
  Nucl.\ Phys.\ B {\bf 155}, 237 (1979).
  doi:10.1016/0550-3213(79)90364-X
  %%CITATION = doi:10.1016/0550-3213(79)90364-X;%%
  %1237 citations counted in INSPIRE as of 07 Jul 2019

\bibitem{Dimopoulos:1981xc} 
  S.~Dimopoulos and J.~Preskill,
  %``Massless Composites With Massive Constituents,''
  Nucl.\ Phys.\ B {\bf 199}, 206 (1982).
  doi:10.1016/0550-3213(82)90345-5
  %%CITATION = doi:10.1016/0550-3213(82)90345-5;%%
  %152 citations counted in INSPIRE as of 07 Jul 2019

%------CH---------------------------------------------------------------------------------

\bibitem{Contino:2010rs} 
  R.~Contino,
  %``The Higgs as a Composite Nambu-Goldstone Boson,''
  doi:10.1142/9789814327183\_0005
  arXiv:1005.4269 [hep-ph].
  %%CITATION = doi:10.1142/9789814327183_0005;%%
  %327 citations counted in INSPIRE as of 07 Jul 2019
	
\bibitem{Panico:2015jxa} 
  G.~Panico and A.~Wulzer,
  %``The Composite Nambu-Goldstone Higgs,''
  Lect.\ Notes Phys.\  {\bf 913}, pp.1 (2016)
  doi:10.1007/978-3-319-22617-0
  [arXiv:1506.01961 [hep-ph]].
  %%CITATION = doi:10.1007/978-3-319-22617-0;%%
  %266 citations counted in INSPIRE as of 07 Jul 2019


%--------Sigma decomp---------------------------------------------------------------------------
\bibitem{Alonso:2014wta} 
  R.~Alonso, I.~Brivio, B.~Gavela, L.~Merlo and S.~Rigolin,
  %``Sigma Decomposition,''
  JHEP {\bf 1412}, 034 (2014)
  doi:10.1007/JHEP12(2014)034
  [arXiv:1409.1589 [hep-ph]].
  %%CITATION = doi:10.1007/JHEP12(2014)034;%%
  %47 citations counted in INSPIRE as of 07 Jul 2019
	
%-------------testing non linearity --------------------------------
	
\bibitem{Brivio:2015kia}
I.~Brivio, M.~B. Gavela, L.~Merlo, K.~Mimasu, J.~M. No, R.~del Rey, and
  V.~Sanz, {\it {Non-Linear Higgs Portal to Dark Matter}},  JHEP {\bf 04}
  (2016) 141, [\href{http://xxx.lanl.gov/abs/1511.01099}{{\tt
  arXiv:1511.01099}}].
%
\bibitem{Brivio:2017ije}
I.~Brivio, M.~B. Gavela, L.~Merlo, K.~Mimasu, J.~M. No, R.~del Rey, and
  V.~Sanz,  Eur.
  Phys. J. {\bf C77} (2017), no.~8 572,
  [arXiv:1701.05379].	
	
%--------few per cent--------------------------------------------------------------------------

\bibitem{Butter:2016cvz} 
  A.~Butter, O.~J.~P.~Eboli, J.~Gonzalez-Fraile, M.~C.~Gonzalez-Garcia, T.~Plehn and M.~Rauch,
  %``The Gauge-Higgs Legacy of the LHC Run I,''
  JHEP {\bf 1607}, 152 (2016)
  doi:10.1007/JHEP07(2016)152
  [arXiv:1604.03105 [hep-ph]].
  %%CITATION = doi:10.1007/JHEP07(2016)152;%%
  %96 citations counted in INSPIRE as of 07 Jul 2019
	
%-------limits on dim-6---------------------------------------------------------------------
\bibitem{Falkowski:2016cxu} 
  A.~Falkowski, M.~Gonzalez-Alonso, A.~Greljo, D.~Marzocca and M.~Son,
  %``Anomalous Triple Gauge Couplings in the Effective Field Theory Approach at the LHC,''
  JHEP {\bf 1702}, 115 (2017)
  doi:10.1007/JHEP02(2017)115
  [arXiv:1609.06312 [hep-ph]].
  %%CITATION = doi:10.1007/JHEP02(2017)115;%%
  %64 citations counted in INSPIRE as of 07 Jul 2019
	
%-----positivity---------------------------------------------------------------------------

\bibitem{Adams:2006sv} 
  A.~Adams, N.~Arkani-Hamed, S.~Dubovsky, A.~Nicolis and R.~Rattazzi,
  %``Causality, analyticity and an IR obstruction to UV completion,''
  JHEP {\bf 0610}, 014 (2006)
  doi:10.1088/1126-6708/2006/10/014
  [hep-th/0602178].
  %%CITATION = doi:10.1088/1126-6708/2006/10/014;%%
  %516 citations counted in INSPIRE as of 07 Jul 2019

\bibitem{deRham:2017avq} 
  C.~de Rham, S.~Melville, A.~J.~Tolley and S.~Y.~Zhou,
  %``Positivity bounds for scalar field theories,''
  Phys.\ Rev.\ D {\bf 96}, no. 8, 081702 (2017)
  doi:10.1103/PhysRevD.96.081702
  [arXiv:1702.06134 [hep-th]].
  %%CITATION = doi:10.1103/PhysRevD.96.081702;%%
  %23 citations counted in INSPIRE as of 07 Jul 2019

%\bibitem{deRham:2017zjm} 
  %C.~de Rham, S.~Melville, A.~J.~Tolley and S.~Y.~Zhou,
  %%``UV complete me: Positivity Bounds for Particles with Spin,''
  %JHEP {\bf 1803}, 011 (2018)
  %doi:10.1007/JHEP03(2018)011
  %[arXiv:1706.02712 [hep-th]].
  %%%CITATION = doi:10.1007/JHEP03(2018)011;%%
  %%25 citations counted in INSPIRE as of 07 Jul 2019

\bibitem{Zhang:2018shp} 
  C.~Zhang and S.~Y.~Zhou,
  %``Positivity bounds on vector boson scattering at the LHC,''
  arXiv:1808.00010 [hep-ph].
  %%CITATION = ARXIV:1808.00010;%%
  %4 citations counted in INSPIRE as of 07 Jul 2019

%-------snow mass list of ops-----------------------------------------------------------------------

\bibitem{Degrande:2013rea} 
  C.~Degrande {\it et al.},
  %``Monte Carlo tools for studies of non-standard electroweak gauge boson interactions in multi-boson processes: A Snowmass White Paper,''
  arXiv:1309.7890 [hep-ph].
  %%CITATION = ARXIV:1309.7890;%%
  %36 citations counted in INSPIRE as of 07 Jul 2019

%--------vbfnlo------------------------------------------------------------------------------

\bibitem{Arnold:2008rz} 
  K.~Arnold {\it et al.},
  %``VBFNLO: A Parton level Monte Carlo for processes with electroweak bosons,''
  Comput.\ Phys.\ Commun.\  {\bf 180}, 1661 (2009)
  doi:10.1016/j.cpc.2009.03.006
  [arXiv:0811.4559 [hep-ph]];
  %%CITATION = doi:10.1016/j.cpc.2009.03.006;%%
  %462 citations counted in INSPIRE as of 07 Jul 2019
  J.~Baglio {\it et al.},
  %``VBFNLO: A Parton Level Monte Carlo for Processes with Electroweak Bosons -- Manual for Version 2.7.0,''
  arXiv:1107.4038 [hep-ph];
  %%CITATION = ARXIV:1107.4038;%%
  %180 citations counted in INSPIRE as of 07 Jul 2019
  J.~Baglio {\it et al.},
  %``Release Note - VBFNLO 2.7.0,''
  arXiv:1404.3940 [hep-ph].
  %%CITATION = ARXIV:1404.3940;%%
  %98 citations counted in INSPIRE as of 07 Jul 2019	

%-------Feynrules-------------------------------------------------------------------------------------

\bibitem{Christensen:2008py} 
  N.~D.~Christensen and C.~Duhr,
  %``FeynRules - Feynman rules made easy,''
  Comput.\ Phys.\ Commun.\  {\bf 180}, 1614 (2009)
  doi:10.1016/j.cpc.2009.02.018
  [arXiv:0806.4194 [hep-ph]];
  %%CITATION = doi:10.1016/j.cpc.2009.02.018;%%
  %697 citations counted in INSPIRE as of 07 Jul 2019
  A.~Alloul, N.~D.~Christensen, C.~Degrande, C.~Duhr and B.~Fuks,
  %``FeynRules  2.0 - A complete toolbox for tree-level phenomenology,''
  Comput.\ Phys.\ Commun.\  {\bf 185}, 2250 (2014)
  doi:10.1016/j.cpc.2014.04.012
  [arXiv:1310.1921 [hep-ph]].
  %%CITATION = doi:10.1016/j.cpc.2014.04.012;%%
  %1154 citations counted in INSPIRE as of 07 Jul 2019

%--------feyncalc-----------------------------------------------------------------------------

\bibitem{Mertig:1990an} 
  R.~Mertig, M.~Bohm and A.~Denner,
  %``FEYN CALC: Computer algebraic calculation of Feynman amplitudes,''
  Comput.\ Phys.\ Commun.\  {\bf 64}, 345 (1991).
  doi:10.1016/0010-4655(91)90130-D;
  %%CITATION = doi:10.1016/0010-4655(91)90130-D;%%
  %784 citations counted in INSPIRE as of 07 Jul 2019
  V.~Shtabovenko, R.~Mertig and F.~Orellana,
  %``New Developments in FeynCalc 9.0,''
  Comput.\ Phys.\ Commun.\  {\bf 207}, 432 (2016)
  doi:10.1016/j.cpc.2016.06.008
  [arXiv:1601.01167 [hep-ph]].
  %%CITATION = doi:10.1016/j.cpc.2016.06.008;%%
  %219 citations counted in INSPIRE as of 07 Jul 2019


%-------romao-----------------------------------------------------------------------------------

\bibitem{Romao:2016ien} 
  J.~C.~Romao,
  %``The need for the Higgs boson in the Standard Model,''
  arXiv:1603.04251 [hep-ph].
  %%CITATION = ARXIV:1603.04251;%%
%------non-interference -------------------------------------------------------------------------------

\bibitem{Azatov:2016sqh} 
  A.~Azatov, R.~Contino, C.~S.~Machado and F.~Riva,
  %``Helicity selection rules and noninterference for BSM amplitudes,''
  Phys.\ Rev.\ D {\bf 95}, no. 6, 065014 (2017)
  doi:10.1103/PhysRevD.95.065014
  [arXiv:1607.05236 [hep-ph]].
  %%CITATION = doi:10.1103/PhysRevD.95.065014;%%
  %51 citations counted in INSPIRE as of 07 Jul 2019

%-------madgraph----------------------------------------------------------

\bibitem{Alwall:2014hca} 
  J.~Alwall {\it et al.},
  %``The automated computation of tree-level and next-to-leading order differential cross sections, and their matching to parton shower simulations,''
  JHEP {\bf 1407}, 079 (2014)
  doi:10.1007/JHEP07(2014)079
  [arXiv:1405.0301 [hep-ph]].
  %%CITATION = doi:10.1007/JHEP07(2014)079;%%
  %3713 citations counted in INSPIRE as of 07 Jul 2019

%------ufo----------------------------------------------------------------

\bibitem{Degrande:2011ua} 
  C.~Degrande, C.~Duhr, B.~Fuks, D.~Grellscheid, O.~Mattelaer and T.~Reiter,
  %``UFO - The Universal FeynRules Output,''
  Comput.\ Phys.\ Commun.\  {\bf 183}, 1201 (2012)
  doi:10.1016/j.cpc.2012.01.022
  [arXiv:1108.2040 [hep-ph]].
  %%CITATION = doi:10.1016/j.cpc.2012.01.022;%%
  %644 citations counted in INSPIRE as of 07 Jul 2019
	
%---------pythia--------------------------------------

\bibitem{Sjostrand:2006za} 
  T.~Sjostrand, S.~Mrenna and P.~Z.~Skands,
  %``PYTHIA 6.4 Physics and Manual,''
  JHEP {\bf 0605}, 026 (2006)
  doi:10.1088/1126-6708/2006/05/026
  [hep-ph/0603175];
  %%CITATION = doi:10.1088/1126-6708/2006/05/026;%%
  %10206 citations counted in INSPIRE as of 07 Jul 2019
  T.~Sjöstrand {\it et al.},
  %``An Introduction to PYTHIA 8.2,''
  Comput.\ Phys.\ Commun.\  {\bf 191}, 159 (2015)
  doi:10.1016/j.cpc.2015.01.024
  [arXiv:1410.3012 [hep-ph]].
  %%CITATION = doi:10.1016/j.cpc.2015.01.024;%%
  %1675 citations counted in INSPIRE as of 07 Jul 2019
	
%-------Rpt---------------------------------------------------

\bibitem{Doroba:2012pd} 
  K.~Doroba, J.~Kalinowski, J.~Kuczmarski, S.~Pokorski, J.~Rosiek, M.~Szleper and S.~Tkaczyk,
  %``The $W_L W_L$ Scattering at the LHC: Improving the Selection Criteria,''
  Phys.\ Rev.\ D {\bf 86}, 036011 (2012)
  doi:10.1103/PhysRevD.86.036011
  [arXiv:1201.2768 [hep-ph]].
  %%CITATION = doi:10.1103/PhysRevD.86.036011;%%
  %29 citations counted in INSPIRE as of 07 Jul 2019
	
%------Mo1-------------------------------------------------------	
\bibitem{bib:M01}
  S. Todt, Ph.D. Thesis CERN-THESIS-2015-018, TU Dresden, 2015.

%----- lower limits more optimistic--------------------------------------
\bibitem{Degrande:2013yda} 
  C.~Degrande {\it et al.},
  %``Studies of Vector Boson Scattering And Triboson Production with DELPHES Parametrized Fast Simulation for Snowmass 2013,''
  arXiv:1309.7452 [physics.comp-ph].
  %%CITATION = ARXIV:1309.7452;%%
  %20 citations counted in INSPIRE as of 07 Jul 2019

%---------------madanalysis 5----------------------------

\bibitem{Conte:2012fm} 
  E.~Conte, B.~Fuks and G.~Serret,
  %``MadAnalysis 5, A User-Friendly Framework for Collider Phenomenology,''
  Comput.\ Phys.\ Commun.\  {\bf 184}, 222 (2013)
  doi:10.1016/j.cpc.2012.09.009
  [arXiv:1206.1599 [hep-ph]];
  %%CITATION = doi:10.1016/j.cpc.2012.09.009;%%
  %283 citations counted in INSPIRE as of 07 Jul 2019
  E.~Conte, B.~Dumont, B.~Fuks and C.~Wymant,
  %``Designing and recasting LHC analyses with MadAnalysis 5,''
  Eur.\ Phys.\ J.\ C {\bf 74}, no. 10, 3103 (2014)
  doi:10.1140/epjc/s10052-014-3103-0
  [arXiv:1405.3982 [hep-ph]];
  %%CITATION = doi:10.1140/epjc/s10052-014-3103-0;%%
  %123 citations counted in INSPIRE as of 07 Jul 2019
	B.~Dumont {\it et al.},
  %``Toward a public analysis database for LHC new physics searches using MADANALYSIS 5,''
  Eur.\ Phys.\ J.\ C {\bf 75}, no. 2, 56 (2015)
  doi:10.1140/epjc/s10052-014-3242-3
  [arXiv:1407.3278 [hep-ph]].
  %%CITATION = doi:10.1140/epjc/s10052-014-3242-3;%%
  %87 citations counted in INSPIRE as of 07 Jul 2019

%--------fastjet---------------------------------------------------

\bibitem{Cacciari:2011ma} 
  M.~Cacciari, G.~P.~Salam and G.~Soyez,
  %``FastJet User Manual,''
  Eur.\ Phys.\ J.\ C {\bf 72}, 1896 (2012)
  doi:10.1140/epjc/s10052-012-1896-2
  [arXiv:1111.6097 [hep-ph]].
  %%CITATION = doi:10.1140/epjc/s10052-012-1896-2;%%
  %2902 citations counted in INSPIRE as of 07 Jul 2019
	
%---- SO(5)/SO(4) CH mathing-------------------------------------	
	
\bibitem{Hierro:2015nna}
I.~M. Hierro, L.~Merlo, and S.~Rigolin,  JHEP {\bf 04} (2016) 016,
  [arXiv:1510.07899].
	

%--------------sigma model 	------------------------------------

\bibitem{Feruglio:2016zvt}
F.~Feruglio, B.~Gavela, K.~Kanshin, P.~A.~N. Machado, S.~Rigolin, and S.~Saa,  JHEP {\bf
  06} (2016) 038, [arXiv:1603.05668].
	
%-------traditional relation f<Lamda < 4pif -----------------------


\bibitem{Kaplan:1983fs}
D.~B. Kaplan and H.~Georgi,  Phys. Lett. {\bf B136} (1984) 183--186.
%--------w polarization----------------------------------------------

\bibitem{Ballestrero:2017bxn} 
  A.~Ballestrero, E.~Maina and G.~Pelliccioli,
  %``$W$ boson polarization in vector boson scattering at the LHC,''
  JHEP {\bf 1803}, 170 (2018)
  doi:10.1007/JHEP03(2018)170
  [arXiv:1710.09339 [hep-ph]].
  %%CITATION = doi:10.1007/JHEP03(2018)170;%%
  %9 citations counted in INSPIRE as of 07 Jul 2019
	
%------positivity bounds regions--------------------------------------------------------------

\bibitem{Bi:2019phv} 
  Q.~Bi, C.~Zhang and S.~Y.~Zhou,
  %``Positivity constraints on aQGC: carving out the physical parameter space,''
  JHEP {\bf 1906}, 137 (2019)
  doi:10.1007/JHEP06(2019)137
  [arXiv:1902.08977 [hep-ph]].
  %%CITATION = doi:10.1007/JHEP06(2019)137;%%
	
%%--------dim-6 in zz scattering-------------------------------------------------------------------------
%
%\bibitem{Gomez-Ambrosio:2018pnl} 
  %R.~Gomez-Ambrosio,
  %%``Studies of Dimension-Six EFT effects in Vector Boson Scattering,''
  %Eur.\ Phys.\ J.\ C {\bf 79}, no. 5, 389 (2019)
  %doi:10.1140/epjc/s10052-019-6893-2
  %[arXiv:1809.04189 [hep-ph]].
  %%%CITATION = doi:10.1140/epjc/s10052-019-6893-2;%%
  %%6 citations counted in INSPIRE as of 07 Jul 2019
	
%--------------------------------------------------------------------------------------------------

%%\bibitem{Sirunyan:2017fvv} 
  %%The ATLAS Collaboration, ATLAS-CONF-2018-030	 
  %%%``Measurement of vector boson scattering and constraints on anomalous quartic couplings from events with four leptons and two jets in proton-proton collisions at $\sqrt{s}=$ 13 TeV,''
  %%Phys.\ Lett.\ B {\bf 774}, 682 (2017)
  %%doi:10.1016/j.physletb.2017.10.020
  %%[arXiv:1708.02812 [hep-ex]].
  %%%%CITATION = doi:10.1016/j.physletb.2017.10.020;%%
  %%%24 citations counted in INSPIRE as of 06 Jul 2019
%%	
%%
%%
	%%
%%
%
	%%	%
  %%%57 citations counted in INSPIRE as of 06 Jul 2019
%%
%%
			%%
%%
				%%
%%
					%%
			%%
					%%
	%%
%%\bibitem{Aaboud:2016ffv}
%%{\bf ATLAS} Collaboration, M.~Aaboud {\em et.~al.}, {\it {Measurement of
  %%$W^{\pm}W^{\pm}$ vector-boson scattering and limits on anomalous quartic
  %%gauge couplings with the ATLAS detector}},  Phys. Rev. {\bf D96} (2017),
  %%no.~1 012007, [\href{http://xxx.lanl.gov/abs/1611.02428}{{\tt
  %%arXiv:1611.02428}}].
%
%
%%------------------------------------------------------------
%
%
%
%%\bibitem{Aad:2016ett}
%%{\bf ATLAS} Collaboration, G.~Aad {\em et.~al.}, {\it {Measurements of $W^\pm
  %%Z$ production cross sections in $pp$ collisions at $\sqrt{s} = 8$ TeV with
  %%the ATLAS detector and limits on anomalous gauge boson self-couplings}},
  %%Phys. Rev. {\bf D93} (2016), no.~9 092004,
  %%[\href{http://xxx.lanl.gov/abs/1603.02151}{{\tt arXiv:1603.02151}}].
%
%%\bibitem{Sirunyan:2017fvv}
%%{\bf CMS} Collaboration, A.~M. Sirunyan {\em et.~al.}, {\it {Measurement of
  %%vector boson scattering and constraints on anomalous quartic couplings from
  %%events with four leptons and two jets in proton-proton collisions at
  %%$\sqrt{s}=$ 13 TeV}},  Phys. Lett. {\bf B774} (2017) 682--705,
  %%[\href{http://xxx.lanl.gov/abs/1708.02812}{{\tt arXiv:1708.02812}}].
%
%%--------------------------------------------------------------------
%
%
%
%
%\bibitem{Aad:2012tfa}
%{\bf ATLAS} Collaboration, G.~Aad {\em et.~al.}, {\it {Observation of a New
  %Particle in the Search for the Standard Model Higgs Boson with the Atlas
  %Detector at the Lhc}},  Phys. Lett. {\bf B716} (2012) 1--29,
  %[\href{http://xxx.lanl.gov/abs/1207.7214}{{\tt arXiv:1207.7214}}].
%
%\bibitem{Chatrchyan:2012xdj}
%{\bf CMS} Collaboration, S.~Chatrchyan {\em et.~al.}, {\it {Observation of a
  %New Boson at a Mass of 125 GeV with the Cms Experiment at the Lhc}},  Phys.
  %Lett. {\bf B716} (2012) 30--61,
  %[\href{http://xxx.lanl.gov/abs/1207.7235}{{\tt arXiv:1207.7235}}].
%
%
%
%
%
%
%%
%\bibitem{Kaplan:1983fs}
%D.~B. Kaplan and H.~Georgi, {\it {$SU(2)$ $\times$ U(1) Breaking by Vacuum
  %Misalignment}},  Phys. Lett. {\bf B136} (1984) 183--186.
%
%\bibitem{Kaplan:1983sm}
%D.~B. Kaplan, H.~Georgi, and S.~Dimopoulos, {\it {Composite Higgs Scalars}},
  %Phys. Lett. {\bf B136} (1984) 187--190.
%
%\bibitem{Banks:1984gj}
%T.~Banks, {\it {Constraints on $SU(2)$ $\times$ U(1) Breaking by Vacuum
  %Misalignment}},  Nucl. Phys. {\bf B243} (1984) 125--130.
%
%\bibitem{Agashe:2004rs}
%K.~Agashe, R.~Contino, and A.~Pomarol, {\it {The Minimal Composite Higgs
  %Model}},  Nucl. Phys. {\bf B719} (2005) 165--187,
  %[\href{http://xxx.lanl.gov/abs/hep-ph/0412089}{{\tt hep-ph/0412089}}].
%
%\bibitem{Gripaios:2009pe}
%B.~Gripaios, A.~Pomarol, F.~Riva, and J.~Serra, {\it {Beyond the Minimal
  %Composite Higgs Model}},  JHEP {\bf 04} (2009) 070,
  %[\href{http://xxx.lanl.gov/abs/0902.1483}{{\tt arXiv:0902.1483}}].
%
%%\bibitem{Alonso:2014wta}
%%R.~Alonso, I.~Brivio, B.~Gavela, L.~Merlo, and S.~Rigolin, {\it {Sigma
  %%Decomposition}},  JHEP {\bf 12} (2014) 034,
  %%[\href{http://xxx.lanl.gov/abs/1409.1589}{{\tt arXiv:1409.1589}}].
%%
%\bibitem{Hierro:2015nna}
%I.~M. Hierro, L.~Merlo, and S.~Rigolin, {\it {Sigma Decomposition: the Cp-Odd
  %Lagrangian}},  JHEP {\bf 04} (2016) 016,
  %[\href{http://xxx.lanl.gov/abs/1510.07899}{{\tt arXiv:1510.07899}}].
%%
%\bibitem{Feruglio:2016zvt}
%F.~Feruglio, B.~Gavela, K.~Kanshin, P.~A.~N. Machado, S.~Rigolin, and S.~Saa,
  %{\it {The Minimal Linear Sigma Model for the Goldstone Higgs}},  JHEP {\bf
  %06} (2016) 038, [\href{http://xxx.lanl.gov/abs/1603.05668}{{\tt
  %arXiv:1603.05668}}].
%
%\bibitem{Gavela:2016vte}
%M.~B. Gavela, K.~Kanshin, P.~A.~N. Machado, and S.~Saa, {\it {The
  %linear-non-linear frontier for the Goldstone Higgs}},  Eur. Phys. J. {\bf
  %C76} (2016), no.~12 690, [\href{http://xxx.lanl.gov/abs/1610.08083}{{\tt
  %arXiv:1610.08083}}].
%
%\bibitem{Merlo:2017sun}
%L.~Merlo, F.~Pobbe, and S.~Rigolin, {\it {The Minimal Axion Minimal Linear
  %$\sigma$ Model}},  Eur. Phys. J. {\bf C78} (2018), no.~5 415,
  %[\href{http://xxx.lanl.gov/abs/1710.10500}{{\tt arXiv:1710.10500}}].
%
%\bibitem{Alonso-Gonzalez:2018vpc}
%J.~Alonso-González, L.~Merlo, F.~Pobbe, S.~Rigolin, and O.~Sumensari, {\it
  %{Testable Axion-Like Particles In The Minimal Linear $\sigma$ Model}},
  %\href{http://xxx.lanl.gov/abs/1807.08643}{{\tt arXiv:1807.08643}}.
%
%\bibitem{Halyo:1991pc}
%E.~Halyo, {\it {Technidilaton Or Higgs?}},  Mod. Phys. Lett. {\bf A8} (1993)
  %275--284.
%
%\bibitem{Goldberger:2008zz}
%W.~D. Goldberger, B.~Grinstein, and W.~Skiba, {\it {Distinguishing the Higgs
  %Boson from the Dilaton at the Large Hadron Collider}},  Phys. Rev. Lett. {\bf
  %100} (2008) 111802, [\href{http://xxx.lanl.gov/abs/0708.1463}{{\tt
  %arXiv:0708.1463}}].
%
%\bibitem{Hernandez-Leon:2017kea}
%P.~Hernandez-Leon and L.~Merlo, {\it {Distinguishing a Higgs-Like Dilaton
  %Scenario with a Complete Bosonic Effective Field Theory Basis}},  Phys. Rev.
  %{\bf D96} (2017), no.~7 075008,
  %[\href{http://xxx.lanl.gov/abs/1703.02064}{{\tt arXiv:1703.02064}}].
%
%%\bibitem{Appelquist:1980vg}
%%T.~Appelquist and C.~W. Bernard, {\it {Strongly Interacting Higgs Bosons}},
  %%Phys. Rev. {\bf D22} (1980) 200.
%
%\bibitem{Longhitano:1980iz}
%A.~C. Longhitano, {\it {Heavy Higgs Bosons in the Weinberg-Salam Model}},
  %Phys. Rev. {\bf D22} (1980) 1166.
%
%\bibitem{Longhitano:1980tm}
%A.~C. Longhitano, {\it {Low-Energy Impact of a Heavy Higgs Boson Sector}},
  %Nucl. Phys. {\bf B188} (1981) 118--154.
%
%%\bibitem{Gavela:2016bzc}
%%B.~M. Gavela, E.~E. Jenkins, A.~V. Manohar, and L.~Merlo, {\it {Analysis of
  %%General Power Counting Rules in Effective Field Theory}},  Eur. Phys. J. {\bf
  %%C76} (2016), no.~9 485, [\href{http://xxx.lanl.gov/abs/1601.07551}{{\tt
  %%arXiv:1601.07551}}].
%
%\bibitem{Brivio:2015kia}
%I.~Brivio, M.~B. Gavela, L.~Merlo, K.~Mimasu, J.~M. No, R.~del Rey, and
  %V.~Sanz, {\it {Non-Linear Higgs Portal to Dark Matter}},  JHEP {\bf 04}
  %(2016) 141, [\href{http://xxx.lanl.gov/abs/1511.01099}{{\tt
  %arXiv:1511.01099}}].
%%
%\bibitem{Brivio:2017ije}
%I.~Brivio, M.~B. Gavela, L.~Merlo, K.~Mimasu, J.~M. No, R.~del Rey, and
  %V.~Sanz, {\it {Alps Effective Field Theory and Collider Signatures}},  Eur.
  %Phys. J. {\bf C77} (2017), no.~8 572,
  %[\href{http://xxx.lanl.gov/abs/1701.05379}{{\tt arXiv:1701.05379}}].
%%XXXX
%
%
%%\bibitem{Aad:2014zda}
%%{\bf ATLAS} Collaboration, G.~Aad {\em et.~al.}, {\it {Evidence for Electroweak
  %%Production of $W^{\pm}W^{\pm}jj$ in $pp$ Collisions at $\sqrt{s}=8$ TeV with
  %%the ATLAS Detector}},  Phys. Rev. Lett. {\bf 113} (2014), no.~14 141803,
  %%[\href{http://xxx.lanl.gov/abs/1405.6241}{{\tt arXiv:1405.6241}}].
%
%%\bibitem{CMS:2014uib}
%%{\bf CMS} Collaboration, C.~Collaboration, {\it {Vector Boson Scattering in a
  %%Final State with Two Jets and Two Same-Sign Leptons}}, .
%%
%%%\bibitem{Khachatryan:2014sta}
%%%{\bf CMS} Collaboration, V.~Khachatryan {\em et.~al.}, {\it {Study of Vector
  %%%Boson Scattering and Search for New Physics in Events with Two Same-Sign
  %%%Leptons and Two Jets}},  Phys. Rev. Lett. {\bf 114} (2015), no.~5 051801,
  %%%[\href{http://xxx.lanl.gov/abs/1410.6315}{{\tt arXiv:1410.6315}}].
%%
%%\bibitem{Aaboud:2016ffv}
%%{\bf ATLAS} Collaboration, M.~Aaboud {\em et.~al.}, {\it {Measurement of
  %%$W^{\pm}W^{\pm}$ vector-boson scattering and limits on anomalous quartic
  %%gauge couplings with the ATLAS detector}},  Phys. Rev. {\bf D96} (2017),
  %%no.~1 012007, [\href{http://xxx.lanl.gov/abs/1611.02428}{{\tt
  %%arXiv:1611.02428}}].
%%
%%\bibitem{Aad:2016ett}
%%{\bf ATLAS} Collaboration, G.~Aad {\em et.~al.}, {\it {Measurements of $W^\pm
  %%Z$ production cross sections in $pp$ collisions at $\sqrt{s} = 8$ TeV with
  %%the ATLAS detector and limits on anomalous gauge boson self-couplings}},
  %%Phys. Rev. {\bf D93} (2016), no.~9 092004,
  %%[\href{http://xxx.lanl.gov/abs/1603.02151}{{\tt arXiv:1603.02151}}].
%%
%%\bibitem{Sirunyan:2017fvv}
%%{\bf CMS} Collaboration, A.~M. Sirunyan {\em et.~al.}, {\it {Measurement of
  %%vector boson scattering and constraints on anomalous quartic couplings from
  %%events with four leptons and two jets in proton-proton collisions at
  %%$\sqrt{s}=$ 13 TeV}},  Phys. Lett. {\bf B774} (2017) 682--705,
  %%[\href{http://xxx.lanl.gov/abs/1708.02812}{{\tt arXiv:1708.02812}}].
%
%%\bibitem{Sirunyan:2017ret}
%%{\bf CMS} Collaboration, A.~M. Sirunyan {\em et.~al.}, {\it {Observation of
  %%electroweak production of same-sign W boson pairs in the two jet and two
  %%same-sign lepton final state in proton-proton collisions at $\sqrt{s} = $ 13
  %%TeV}},  Phys. Rev. Lett. {\bf 120} (2018), no.~8 081801,
  %%[\href{http://xxx.lanl.gov/abs/1709.05822}{{\tt arXiv:1709.05822}}].
%
%\bibitem{Kilian:2014zja}
%W.~Kilian, T.~Ohl, J.~Reuter, and M.~Sekulla, {\it {High-Energy Vector Boson
  %Scattering After the Higgs Discovery}},  Phys. Rev. {\bf D91} (2015) 096007,
  %[\href{http://xxx.lanl.gov/abs/1408.6207}{{\tt arXiv:1408.6207}}].
%
%\bibitem{Kalinowski:2018oxd}
%J.~Kalinowski, P.~Kozów, S.~Pokorski, J.~Rosiek, M.~Szleper, and S.~Tkaczyk,
  %{\it {Same-sign WW scattering at the LHC: can we discover BSM effects before
  %discovering new states?}},  Eur. Phys. J. {\bf C78} (2018), no.~5 403,
  %[\href{http://xxx.lanl.gov/abs/1802.02366}{{\tt arXiv:1802.02366}}].
%
%\bibitem{Brass:2018hfw}
%S.~Brass, C.~Fleper, W.~Kilian, J.~Reuter, and M.~Sekulla, {\it {Transversal
  %Modes and Higgs Bosons in Electroweak Vector-Boson Scattering at the LHC}},
  %Eur. Phys. J. {\bf C78} (2018), no.~11 931,
  %[\href{http://xxx.lanl.gov/abs/1807.02512}{{\tt arXiv:1807.02512}}].
%
%%\bibitem{Gomez-Ambrosio:2018pnl}
%%R.~G\'omez-Ambrosio, {\it {Studies of Dimension-Six Eft Effects in Vector Boson
  %%Scattering}},  \href{http://xxx.lanl.gov/abs/1809.04189}{{\tt
  %%arXiv:1809.04189}}.
%
%\bibitem{Espriu:2012ih}
%D.~Espriu and B.~Yencho, {\it {Longitudinal WW scattering in light of the
  %“Higgs boson” discovery}},  Phys. Rev. {\bf D87} (2013), no.~5 055017,
  %[\href{http://xxx.lanl.gov/abs/1212.4158}{{\tt arXiv:1212.4158}}].
%
%\bibitem{Espriu:2013fia}
%D.~Espriu, F.~Mescia, and B.~Yencho, {\it {Radiative Corrections to Wl Wl
  %Scattering in Composite Higgs Models}},  Phys. Rev. {\bf D88} (2013) 055002,
  %[\href{http://xxx.lanl.gov/abs/1307.2400}{{\tt arXiv:1307.2400}}].
%
%\bibitem{Delgado:2013loa}
%R.~L. Delgado, A.~Dobado, and F.~J. Llanes-Estrada, {\it {Light ‘Higgs’,
  %yet strong interactions}},  J. Phys. {\bf G41} (2014) 025002,
  %[\href{http://xxx.lanl.gov/abs/1308.1629}{{\tt arXiv:1308.1629}}].
%
%\bibitem{Delgado:2013hxa}
%R.~L. Delgado, A.~Dobado, and F.~J. Llanes-Estrada, {\it {One-loop $W_LW_L$ and
  %$Z_LZ_L$ scattering from the electroweak Chiral Lagrangian with a light
  %Higgs-like scalar}},  JHEP {\bf 02} (2014) 121,
  %[\href{http://xxx.lanl.gov/abs/1311.5993}{{\tt arXiv:1311.5993}}].
%
%\bibitem{Espriu:2014jya}
%D.~Espriu and F.~Mescia, {\it {Unitarity and Causality Constraints in Composite
  %Higgs Models}},  Phys. Rev. {\bf D90} (2014), no.~1 015035,
  %[\href{http://xxx.lanl.gov/abs/1403.7386}{{\tt arXiv:1403.7386}}].
%
%\bibitem{Delgado:2014jda}
%R.~L. Delgado, A.~Dobado, M.~J. Herrero, and J.~J. Sanz-Cillero, {\it {One-loop
  %$\gamma\gamma \to$ W$_{L}^{+}$ W$_{L}^{-}$ and $\gamma\gamma \to$ Z$_{L}$
  %Z$_{L}$ from the Electroweak Chiral Lagrangian with a light Higgs-like
  %scalar}},  JHEP {\bf 07} (2014) 149,
  %[\href{http://xxx.lanl.gov/abs/1404.2866}{{\tt arXiv:1404.2866}}].
%
%\bibitem{Delgado:2017cls}
%R.~L. Delgado, A.~Dobado, D.~Espriu, C.~Garc{\'\i a-}Garcia, M.~J. Herrero,
  %X.~Marcano, and J.~J. Sanz-Cillero, {\it {Production of Vector Resonances at
  %the Lhc via Wz-Scattering: a Unitarized ECHL Analysis}},  JHEP {\bf 11}
  %(2017) 098, [\href{http://xxx.lanl.gov/abs/1707.04580}{{\tt
  %arXiv:1707.04580}}].
%
%\bibitem{Ballestrero:2009vw}
%A.~Ballestrero, G.~Bevilacqua, D.~Buarque~Franzosi, and E.~Maina, {\it {How
  %Well Can the Lhc Distinguish Between the Sm Light Higgs Scenario, a Composite
  %Higgs and the Higgsless Case Using Vv Scattering Channels?}},  JHEP {\bf 11}
  %(2009) 126, [\href{http://xxx.lanl.gov/abs/0909.3838}{{\tt
  %arXiv:0909.3838}}].
%
%\bibitem{BuarqueFranzosi:2017prc}
%D.~Buarque~Franzosi and P.~Ferrarese, {\it {Implications of Vector Boson
  %Scattering Unitarity in Composite Higgs Models}},  Phys. Rev. {\bf D96}
  %(2017), no.~5 055037, [\href{http://xxx.lanl.gov/abs/1705.02787}{{\tt
  %arXiv:1705.02787}}].
%
%%\bibitem{Panico:2015jxa}
%%G.~Panico and A.~Wulzer, {\it {The Composite Nambu-Goldstone Higgs}},  Lect.
  %%Notes Phys. {\bf 913} (2016) pp.1--316,
  %%[\href{http://xxx.lanl.gov/abs/1506.01961}{{\tt arXiv:1506.01961}}].
%
%\bibitem{Manohar:1983md}
%A.~Manohar and H.~Georgi, {\it {Chiral Quarks and the Nonrelativistic Quark
  %Model}},  Nucl. Phys. {\bf B234} (1984) 189--212.
%
%%\bibitem{Alwall:2014hca}
%%J.~Alwall, R.~Frederix, S.~Frixione, V.~Hirschi, F.~Maltoni, O.~Mattelaer,
  %%H.~S. Shao, T.~Stelzer, P.~Torrielli, and M.~Zaro, {\it {The Automated
  %%Computation of Tree-Level and Next-To-Leading Order Differential Cross
  %%Sections, and Their Matching to Parton Shower Simulations}},  JHEP {\bf 07}
  %%(2014) 079, [\href{http://xxx.lanl.gov/abs/1405.0301}{{\tt
  %%arXiv:1405.0301}}].
%%
%%\bibitem{Degrande:2011ua}
%%C.~Degrande, C.~Duhr, B.~Fuks, D.~Grellscheid, O.~Mattelaer, and T.~Reiter,
  %%{\it {Ufo - the Universal Feynrules Output}},  Comput. Phys. Commun. {\bf
  %%183} (2012) 1201--1214, [\href{http://xxx.lanl.gov/abs/1108.2040}{{\tt
  %%arXiv:1108.2040}}].
%
%\bibitem{Alloul:2013bka}
%A.~Alloul, N.~D. Christensen, C.~Degrande, C.~Duhr, and B.~Fuks, {\it
  %{FeynRules 2.0 - A complete toolbox for tree-level phenomenology}},  Comput.
  %Phys. Commun. {\bf 185} (2014) 2250--2300,
  %[\href{http://xxx.lanl.gov/abs/1310.1921}{{\tt arXiv:1310.1921}}].
  %
  %
%\bibitem{ww27paper}
%C.~Geetanjali, J.~Kalinowski, M.~Kaur, P.~Kozów, S.~Kaur, M.~Szleper, and
  %T.~Sławomir, {\it {In preparation}}.


\end{thebibliography}
%%%%%%%%%%%%%%%%%%%%%%%%%%%%

\end{document}